\documentclass[fleqn,twoside]{elsart}
\usepackage[letterpaper]{geometry}

\usepackage{amsmath}
\usepackage{amssymb}
\usepackage{graphicx}
\usepackage{epsfig,psfrag}
\usepackage{bbm} 

\newcommand{\ket}[1]{\left|#1\right\rangle}
\newcommand{\bra}[1]{\left\langle#1\right|}
\newcommand{\erw}[1]{\left\langle#1\right\rangle}
\def\Xint#1{\mathchoice
   {\XXint\displaystyle\textstyle{#1}}%
   {\XXint\textstyle\scriptstyle{#1}}%
   {\XXint\scriptstyle\scriptscriptstyle{#1}}%
   {\XXint\scriptscriptstyle\scriptscriptstyle{#1}}%
   \!\int}
\def\XXint#1#2#3{{\setbox0=\hbox{$#1{#2#3}{\int}$}
    \vcenter{\hbox{$#2#3$}}\kern-.5\wd0}}

\def\dashint{\Xint-}
\newcommand{\omegav}{\omega_{\sf v}} 
\newcommand{\sigmav}{\mbox{\boldmath$\sigma$}}
\newcommand{\omegavec}{\mbox{\boldmath$\omega$}}
\newcommand{\varepsilonv}{\mbox{\boldmath$\varepsilon$}}
\newcommand{\Omegav}{\mbox{\boldmath$\Omega$}}
\newcommand{\Dv}{\mbox{\boldmath$D$}}
\newcommand{\dv}{\mbox{\boldmath$d$}}
\newcommand{\calEv}{\mbox{\boldmath${\mathcal E}$}}

\numberwithin{equation}{section}   

\begin{document}

\begin{frontmatter}
\title{Coherent and Collective Quantum Optical Effects in Mesoscopic Systems}

 \author{Tobias Brandes}
 \ead{tobias.brandes@manchester.ac.uk}
 \address{School of Physics and Astronomy, The University of Manchester \\ PO Box 88, Manchester M60 1QD, UK}

\begin{abstract}
A review of coherent and collective quantum optical effects like superradiance and coherent population trapping in mesoscopic systems is presented.  Various new physical realizations of these phenomena are discussed, with a focus on their role for electronic transport and quantum dissipation in coupled nano-scale systems like quantum dots. A number of theoretical tools such as Master equations, polaron transformations, correlation functions, or level statistics are used to describe recent work on dissipative charge qubits (double quantum dots), the Dicke effect, phonon cavities, single  oscillators, dark states and adiabatic control in quantum transport, and large spin-boson models. The review attempts to establish connections between concepts from Mesoscopics (quantum transport, coherent scattering, quantum chaos), Quantum Optics (such as superradiance, dark states, boson cavities), and (in its last part) Quantum Information Theory.
\end{abstract}


\end{frontmatter}

\tableofcontents

\section{\bf Introduction}\label{section_introduction}

There is a growing interest in the transfer of concepts and methods between Quantum Optics and Condensed-Matter Physics. For example, well-known methods from Laser Physics like the control of quantum coherent superpositions or strong coupling of atoms to cavity photons have started to become feasible in artificial condensed-matter structures.  On the other hand, condensed matter concepts are used, e.g., in order to realize quantum phase transitions with  atoms in tunable optical lattices.  
The main direction of this Review is the one from Quantum Optics towards Condensed-Matter Physics, and to be more specific, towards mesoscopic systems such as artificial atoms (quantum dots). The primary subject therefore  are concepts, models, and methods which are originally mostly known in a quantum optical context, and the overall aim is to show how these appear and can be understood and implemented in Mesoscopics. Typical examples are the roles that (collective) spontaneous emission, coherent coupling to single boson modes, quantum cavities, dark resonances,  adiabatic steering etc. play for, e.g.,  electronic transport in low-dimensional systems such as (superconducting or semiconducting) charge qubits.

As is the case for Quantum Optics, quantum coherence is a very important (but not the only) ingredient of  physical phenomena in mesoscopic systems. Beside coherence, collective effects due to  interactions of electrons among themselves or with other degrees of freedom (such as phonons or photons) give rise to a plethora of intriguing many-body phenomena. At the same time, collective effects are also well-known in Quantum Optics. The laser is a good example for the realization of the paradigm of stimulated emission in a system with a large number of atoms, interacting through a radiation field. Another paradigm is spontaneous emission. As one of the most basic concepts of  quantum physics, it can be traced back to such early works as that of Albert Einstein in 1917. The corresponding realization of  spontaneous emission in a many-atom system (which will play a key role in this Review) is {\em superradiance}: this is the collective spontaneous emission of an initially excited ensemble of $N$ two-level systems interacting with a common photon field. As a function of time, this emission has the form of a very sudden peak on a short time scale $\sim 1/N$, with an abnormally large emission rate maximum $\sim N^2$. This effect was first proposed by Dicke in 1954, but it took nearly 20 years for the first experiments to confirm it in an optically pumped hydrogen fluoride gas. 

Outside Quantum Optics, Dicke superradiance has been known to appear in {\em condensed matter systems} for quite a while, with excitons and  electron-hole plasmas in semiconductors being the primary examples. In spite of the intriguing complexities involved, it is semiconductor quantum optics where physicists have probably been most successful so far in providing the condensed matter counterparts of genuine quantum optical effects. This indeed has led to a number of beautiful experiments such as the observation of Dicke superradiance from radiatively coupled exciton quantum wells.

On the other hand and quite surprisingly, the Dicke effect has been `re-discovered' relatively recently in the {\em electronic transport properties}  of a mesoscopic system in a theoretical work by Shabazyan and Raikh in 1994 on the tunneling of electrons through two coupled impurities. This  has been followed by a number of (still mostly theoretical) activities, where this effect is discussed in a new context and for physical systems that are completely different from their original counter-parts in Optics. For some of these (like quantum dots), the  analogies with the original optical systems seem to be fairly obvious at first sight, but in fact the mesoscopic `setup' (coupling to electron reservoirs, non-equilibrium etc.) brings in important new aspects and raises new questions. 

The purpose of the present Report is to give an overview over  quantum optical concepts and models (such as Dicke superradiance, adiabatic steering, single boson cavities) in Mesoscopics, with the main focus on their role for coherence and correlations in electronic scattering, in mesoscopic transport, quantum dissipation, and in such `genuine mesoscopic' fields as level statistics and quantum chaos. Most of the material covered here is theoretical, but there is an increasingly strong background of key experiments, only some of which are described here. The current rapid experimental and theoretical progress is also strongly driven by the desire to implement concepts from quantum information theory into real physical systems. It can therefore be expected that this field will still grow very much in the near future, and a Review, even if it is only on some special aspects of that field, might be helpful to those working or planning to work in this area.

A good deal of the theoretical models to be discussed here is motivated by experiments in mesoscopic systems, in particular on electronic transport in coupled, artificial  two-level systems such as semiconductor double quantum dots, or superconducting Cooper-pair boxes. Two examples in the semiconductor case are the control of spontaneous phonon emission, and single-qubit rotations. For the sake of definiteness, double quantum dots will be the primary example for two-level systems throughout many parts of this Review, but the reader should keep in mind that many of the  theoretical models can be translated (sometimes easily, sometimes probably not so easily) into other physical realizations.

Section \ref{section_transport} is devoted to electronic transport through double quantum dots and starts  with a short survey of experiments before moving on to a detailed theory part on models and methods, with more recent results on electron shot noise and time-dependent effects.
This is followed by a review of Dicke superradiance in section \ref{section_SR}, with applications such as entanglement in quantum dot arrays, and a section on dissipation effects in generic large-spin models that are of relevance to a large range of physical systems. 
Section \ref{section_spectral} starts with a brief analysis of the Dicke spectral line-shape effect and its mathematical structure, which  turns out to be very fruitful for understanding its wider implications for correlation functions and scattering matrices. This is  discussed in detail for the original Shabazyan-Raikh  and related models for tunneling and impurity scattering and concluded by a discussion of the effect in the ac-magneto-conductivity of  quantum wires. 

Section \ref{section_cavity} presents  electron transport through phonon cavities, and 
section \ref{section_oscillator} introduces single-mode quantum oscillator models, such as the Rabi-Hamiltonian, in the context of electronic transport. These models have started to play a great role in the description of mechanical and vibrational degrees of freedom in combination with transport in nanostructures, a topic that forms part of what can already safely been called a new area of  Mesoscopic Physics, i.e., nano-electromechanical systems.

Section \ref{section_dark} is devoted to the Dark Resonance effect and its spin-offs such as adiabatic transfer and rotations of quantum states. Dark resonances occur as quantum coherent `trapped' superpositions in three (or more) state systems that are driven by (at least) {\em two} time-dependent, monochromatic fields. Again, there are numerous applications of this effect in Laser Spectroscopy and Quantum Optics, ranging from laser cooling, population transfer up to  loss-free pulse propagation. In mesoscopic condensed-matter systems, experiments and theoretical schemes related to this effect have just started to appear which is why an introduction into this area should be quite useful.

Finally, section \ref{section_Dicke_Chaos} covers the Dicke superradiance model in its purest and, perhaps, most interesting one-boson mode version. It provides a discussion of an instability of the model, the precursors of which are related to a cross-over in its level statistics and its quantum-chaotic behavior. Exact solutions of this model have recently enlarged the class of systems for which entanglement close to a quantum phase transition can be discussed rigorously, which are briefly reviewed and compared with entanglement in the Dicke model.



\section{\bf Electronic Transport and Spontaneous Emission in Artificial Atoms (Two-Level Systems)} \label{section_transport}

Electronic transport is one of the most versatile and sensitive tools to explore the intriguing quantum properties of solid-state based systems. The quantum Hall effect \cite{KDP80}, with its fundamental  conductance unit $e^2/h$, gave a striking proof that `dirty' condensed matter systems indeed reveal beautiful `elementary' physics, and in fact was one of the first highlights of the new physics that by now  has established itself as the arena of mesoscopic phenomena. In fact,  electronic transport in the quantum regime can be considered as one of the central subjects of modern  Solid State Physics \cite{Kramer91,Kramer95,Kramer96,Dittrich,Ferry,Datta,Imry,Bastard,FukuyamaAndo}. Phase coherence of quantum states leads (or at least contributes) to effects such as, e.g., localization \cite{Kramer85,KM94} of electron wave functions, the quantization of the Hall resistance in two-dimensional electron gases \cite{Dittrich,Chakraborty,Janssen}, the famous conductance steps of quasi one-dimensional quantum wires or quantum point contacts \cite{Lan70,Bue86,Weeetal88,Whaetal88}, or Aharonov-Bohm like interference oscillations of the conductance of metallic rings or cylinders \cite{SS81}.

The technological and experimental advance has opened the test-ground for  a number of fundamental physical concepts related to the motion of electrons in lower dimensions. This has to be combined with a rising interest to observe, control and eventually utilize the two key  principles underlying our understanding of modern quantum devices: quantum superposition and quantum entanglement.

\subsection{Physical Systems and Experiments}

The most basic systems where quantum mechanical principles can be tested in electronic transport are two-level systems. These can naturally be described by a pseudo spin $1/2$ (single qubit) that refers either to the real electron spin or another degree of freedom that is described by a two-dimensional Hilbert space. The most successful experimental realizations so far have probably been {\em superconducting systems} based on either the charge or flux degree of freedom (the Review Article by Mahklin, Sch\"on and Shnirman \cite{MSS01} provides a good introduction). In 1999, the experiments by Nakamura, Pashkin and Tsai \cite{NPT99} in superconducting Cooper-pair boxes demonstrated controlled quantum mechanical oscillations for the first time in a condensed matter-based two-level system, with  more refined experiments following soon thereafter. These activities determine  a field which is still very much growing (and, needless to say, therefore cannot be treated in this Review in full detail). On of these examples at the time of writing this Review are the experiments by the Yale group on the coherent coupling of cavity photons to a Cooper-pair box, cf. section \ref{cavity_experiment}.

Furthermore, at least since the proposal by Loss and DiVincenzo in 1998 \cite{LD98}, there is a strong activity (still mostly theoretically) to test the huge potential of the {\em electron spin} for solid-state realizations of qubits and arrays of qubits. Fujisawa, Tokura and Hirayama \cite{FTH01} measured the spin-relaxation time in a single semiconductor quantum dot in the Coulomb blockade regime, where using a voltage pulse of fixed duration, the first excited and the ground state could be moved into and out of  a transport window between left and right chemical potential of the electron reservoirs. The resulting transient current revealed spin-flip relaxation times longer than a few $\mu$s for excited states whose spin differed from that of the ground state, whereas without spin-flip the relaxation times were much shorter (3 ns).

Charge relaxation due to spontaneous phonon emission in quantum dots is therefore in general much faster than spin-relaxation. In electron transport, spontaneous emission effects were first observed most prominently in experiments with semiconductor double quantum dots. These are discussed in some detail below, as the remainder of this section mainly deals with spontaneous emission effects in transport through two-level systems. The operation of a single charge-based qubit as realized in semiconductor double quantum dots was successfully demonstrated  by Hayashi and co-workers in 2003, an experiment which is discussed in section \ref{section_hayashi}.

\subsubsection{Quantum Dots}

Quantum dots are semiconductor structures containing a small number of electrons ($1 \sim 1000$) within a  region of space with typical sizes in the sub-micrometer range \cite{Ash96,DG98,Kouetal97,KM98,Steetal97}. Many properties of such systems can be investigated by transport, e.g. current-voltage measurements, if the dots are fabricated between contacts acting as source and drain for electrons which can enter or leave the dot. In contrast to real atoms, quantum dots are {\em open} systems with respect to the number of electrons $N$ which can easily be tuned with  external parameters such as gate voltages or magnetic fields. For example, by changing the size and the shape of the dot with external gate voltages, one can realize dots as artificial atoms, with the possibility to `scan through the periodic table' by adding one electron after the other within one and the same system. In fact, quantum effects such as discrete energy levels (atomic shell structure) and quantum chaos (as in nuclei) are observable in a controlled manner in quantum dots \cite{KM98}. Moreover, the experiments can be conducted in a regime which usually is not accessible to experiments with real atoms. For example, a singlet-triplet transition should occur in real helium atoms for magnetic fields such large as of the order of $10^5$T, as the they occur only in the vicinity of white dwarfs and pulsars \cite{BSD99}. In artificial atoms, which have a much larger size than real atoms, much smaller magnetic fields are sufficient to observe such effects \cite{MHW91,PG91}. 

Transport experiments  are very sensitive to energy scales down to a few micro electron volts. Traditionally, there are three effects which dominate transport through quantum dots: the tunnel effect, which is a quantum mechanical phenomenon where electrons can penetrate an electrostatic potential-barrier, the  charging effect which is due to the discreteness of the electron charge and known as Coulomb blockade effect, and size quantization due to the smallness of the dots, leading to discrete energies. Out of these three, the Coulomb blockade effect with its charging energy $U=e^2/2C$ for one additional electron is the most important and in fact sufficient to explain the simplest cases in the earlier experiments on quantum dots in terms of simple charging diagrams. There, the only `quantum' feature of quantum dots stems from the discreteness of the electron charge $e$, with the smallness of the dots providing the correspondingly small capacitances $C$ (and therefore sizable charging energies $U$), and the tunnel effect merely providing the contact between the dot and the outside world (i.e. the contact leads). On the theoretical side, this corresponds to a description of sequential tunneling in terms of simple rate equations, which was called `orthodox theory' for single electron charging effect in general, a good standard reference for which is provided by the volume on `Single  Charge Tunneling' edited by Grabert and Devoret \cite{Grabert}. 

As could be expected, a major thrust in quantum dot physics (starting in the 1990s) has been to go beyond this simple picture and to take a closer look at the above-mentioned effects. As for charge interaction and quantum size effects, this lead to detailed investigations of the internal structure of dots, with electron-electron correlations and spin effects playing a major role.  As for the tunnel effect, one can broadly speak of two main streams where either the `external' coupling of electrons between the dot and the reservoirs, or the coupling of  dots to other dots (coupled-dot systems) or to other external degrees of freedom (photons, phonons) is dealt with on a more serious level. The former case  with co-tunneling and the Kondo effect as the main key-words is intrinsically `solid state' physics, whereas the latter (in particular when it comes to two or more level systems interacting with bosons) has a number of analogies with Quantum Optics and is the main subject of this Review. One should bear in mind, however, that the distinction into two streams is a drastic simplification of what in reality is a very complex field of current research activities. 

Recent review articles on quantum dots are the ones by Reimann and Manninen \cite{RM02} on the electronic  structure of quantum dots, and the overview article on electronic structure and transport properties of quantum dots by  Tews \cite{Tews04}.

\begin{figure}[t]
\begin{center}
\includegraphics[width=0.4\textwidth]{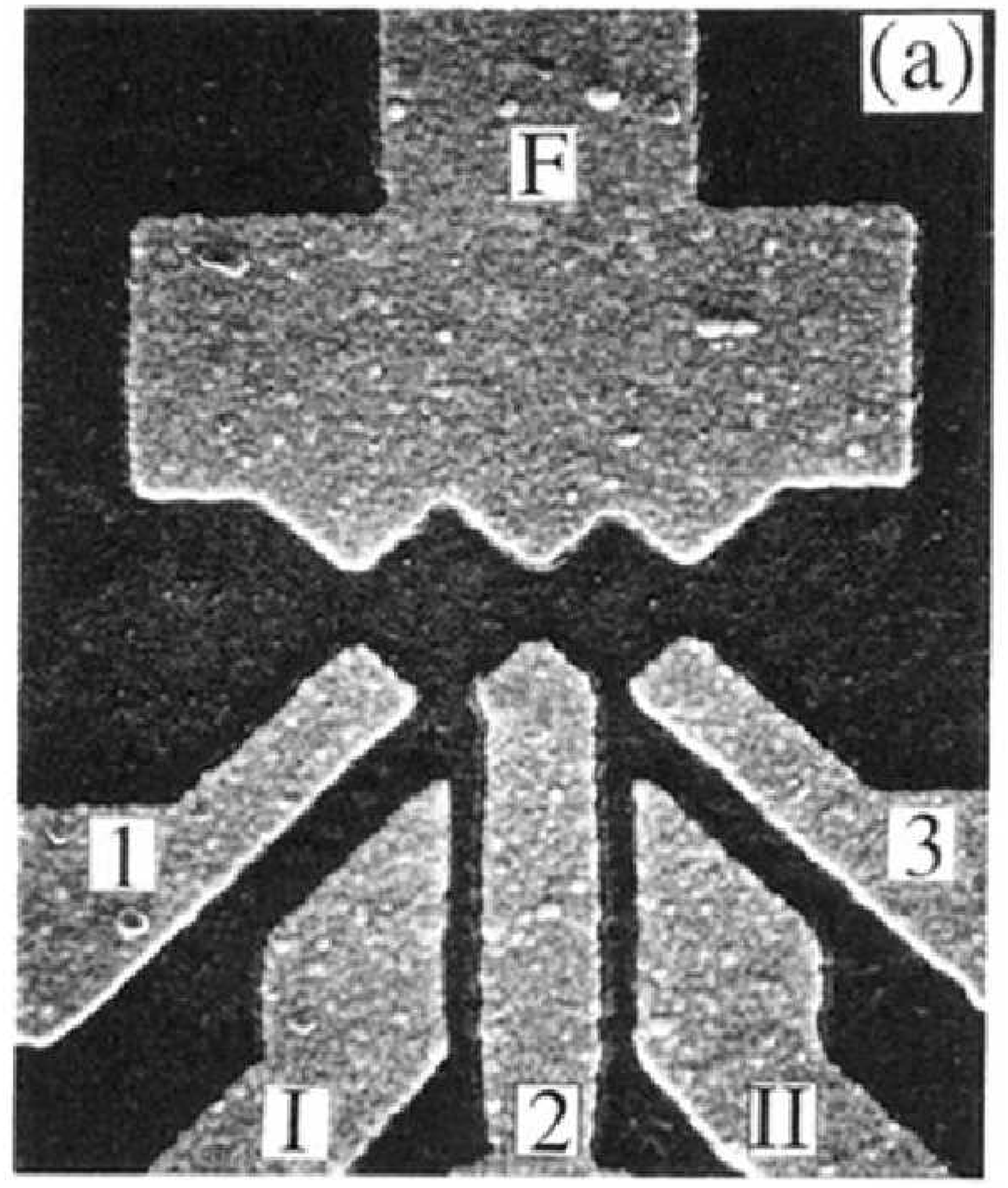}
\includegraphics[width=0.4\textwidth]{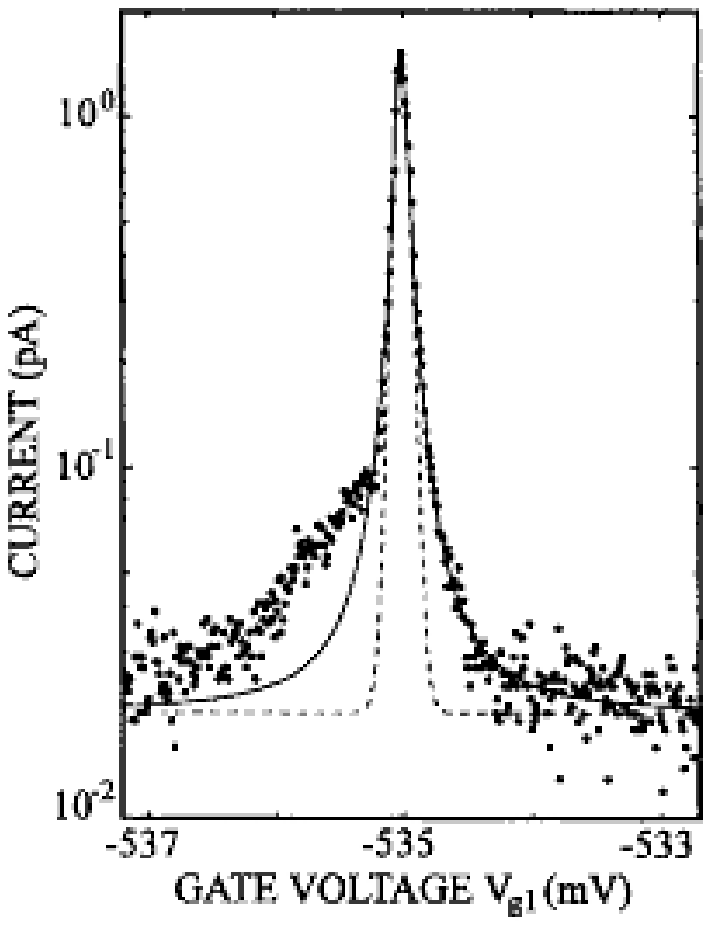}
\caption[]{{\bf Left:} Double quantum dot used in  the first experiment by van der Vaart and coworkers \cite{Vaartetal95} on resonant tunneling. Dimensions are 320 $\times$ 320  nm$^2$ (left dot) and 280 $\times$ 280  nm$^2$ (right dot). {\bf Right:} Resonant tunnel current through the double quantum dot \cite{Vaartetal95} (dots) as a function of inter-dot bias $\varepsilon$ at
source-drain voltage 400$\mu$V. Lorentzian fit (line) and fit $\sim \cosh^2 (2\varepsilon/k_BT )$ with $T=35$mK (dashed). From \cite{Vaartetal95}.
\label{vanderVaart_PRL1995_fig1.eps}}
\end{center}
\end{figure}

\subsubsection{Double Quantum Dots} \label{section_emission_experiments}
Coupling of two quantum dots leads to double quantum dots which in analogy with atomic and molecular physics sometimes are called `artificial molecules', although this terminology can be somewhat misleading: in the strong Coulomb blockade limit, double quantum dots are better described as two-level systems with controllable level-spacing and one additional transport electron, which rather suggests the analogy with a simple model for an {\em atom}, in particular if it comes to interaction with external fields such as photons or phonons. This view appears to be rather natural from a Quantum Optics point of view, too (cf. the classic book `Optical Resonance and Two-Level Atoms' by Allen and Eberly \cite{Allen}), and it furthermore fits with the terminology of quantum information technology, with the charge double dot (as in the experiment by Hayashi and co-workers) being the elementary one-qubit, cf. section \ref{section_hayashi}.

On the other hand, the distinction between the two regimes of  ionic-like bonding (weak tunneling between the two dots) and covalent bonding (strong tunneling) is often used in the literature; this also reflects the choice between two different starting points in the theoretical description, i.e.,  the basis of localized states  and the basis of delocalized (bonding and antibonding) states in the theory of the two-level system, as is discussed in section \ref{section_transport}.

Several groups have performed transport experiments with double quantum dots, with lateral structures offering experimental advantages over vertical dots with respect to their tunability of parameters. 
A recent overview  of the Delft and NTT experiments is given by van der Wiel, De Franceschi, Elzerman, Kouwenhoven, Fujisawa and Tarucha \cite{vdWetal03}, who review the stability diagram, linear and non-linear transport, resonant tunneling, and the influence of magnetic fields and microwave radiation on transport in lateral double quantum dots. 

As for the earlier double quantum dot experiments, van der Vaart and co-workers \cite{Vaartetal95}
investigated resonant tunneling in 1995 and found an asymmetry in the resonant line-shape that already hinted at physics beyond the simple elastic tunneling model, cf. Fig. (\ref{vanderVaart_PRL1995_fig1.eps}). Subsequently, Waugh and co-workers measured the tunnel-coupling induced splitting of the conductance peaks for double and triple quantum dots \cite{Wauetal96}. The Stuttgart group with Blick and co-workers explored the charging diagram for single-electron tunneling through a double quantum dot \cite{Blietal96}. Blick {\em et al.} later verified the coherent tunnel coupling  \cite{Blietal98b}, and  Rabi-oscillations (with millimeter continuous wave radiation \cite{Blietal98a}) in double dots.

\subsubsection{Resonant Tunneling and Phonon Emission in Double Quantum Dots}

\begin{figure}[t]
\includegraphics[width=0.5\textwidth]{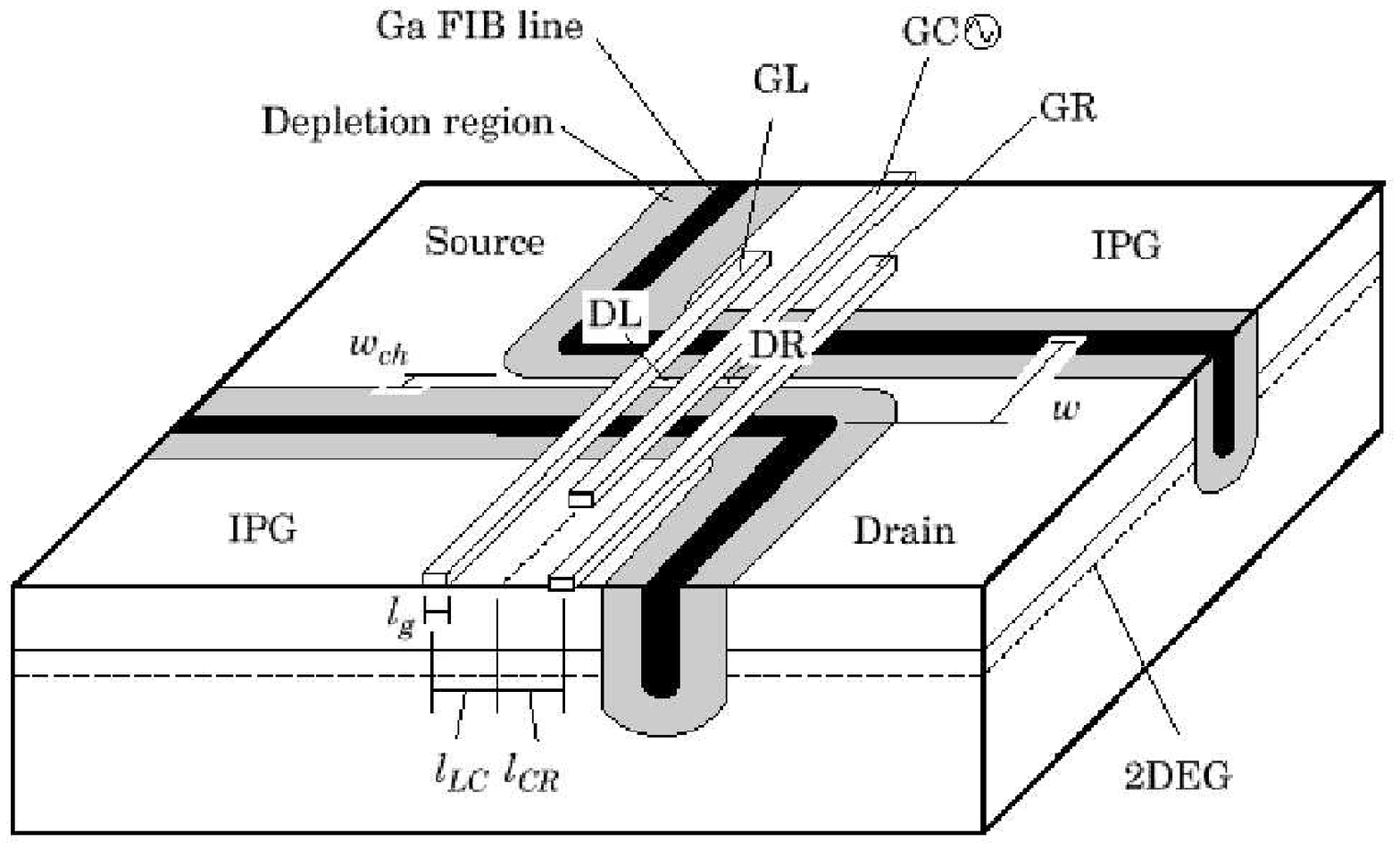}
\includegraphics[width=0.5\textwidth]{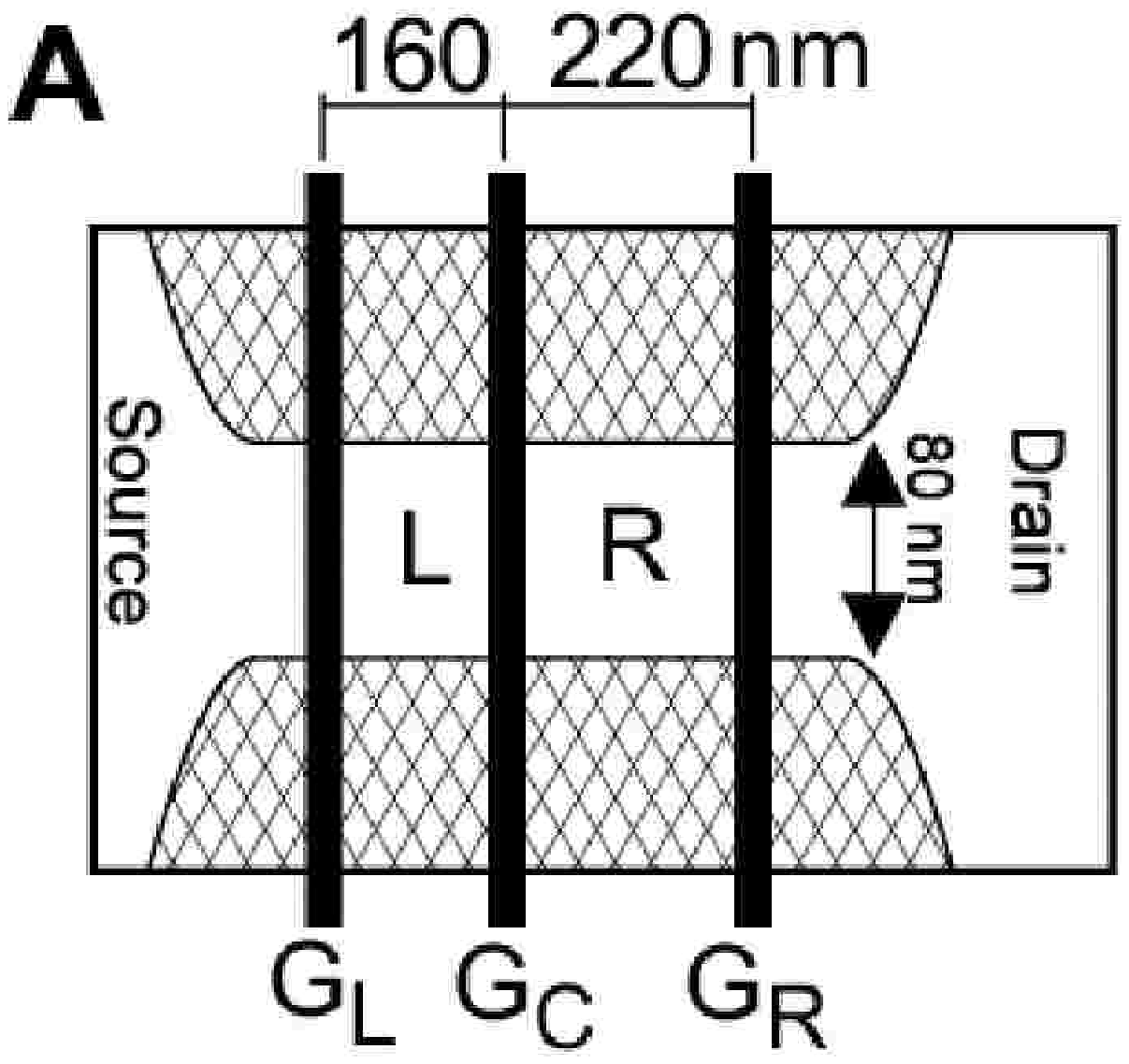}
\caption[]{\label{dotsample.eps}{\bf Left:} Schematic diagram of a `double gate single electron transistor'
by Fujisawa and Tarucha \cite{FT97}. The 2DEG is located 100 nm below the surface 
of an GaAs/AlGaAs modulation-doped heterostructure with mobility
$8\cdot 10^5$ cm$^2$ (Vs)$^{-1}$ and carrier concentration $3\cdot 10^{11}$ cm$^{-2}$ 
at 1.6 K in the dark and ungated. Ga focused ion beam implanted in-plane gates and Schottky
gates define the dot system. A double dot is formed by applying negative gate voltages to the
gates GL, GC, and GR. The structure can also be used for single-dot experiments, where
negative voltages are applied to GL and GC only. From \cite{FT97}.
{\bf Right:} Double quantum dots as used in the experiment by Fujisawa
and co-workers \cite{Fujetal98} (top view). Transport of electrons is through the narrow
channel that connects source and drain. The gates themselves have widths of 40 nm.
The two quantum dots contain approximately 15 (Left, L) and 25 (Right, R) electrons.
The charging energies are 4 meV (L) and 1 meV (R), the energy spacing for single
particle states in both dots is approximately 0.5 meV (L) and 0.25 meV (R). From \cite{Fujetal98}.}
\end{figure} 

Fujisawa and co-workers \cite{Fujetal98} performed a series of experiments on spontaneous emission of phonons in a lateral double quantum dot (similar experiments were performed with vertically coupled dots \cite{Taretal99}). Their device  was realized in a  GaAs/AlGaAs semiconductor heterostructure within the  two-dimensional electron gas \cite{FT97}.
Focused ion-beams were used to form in-plane gates which defined a narrow channel of tunable width. The channel itself was  connected to source and drain electron reservoirs and on top of it, three Schottky gates defined tunable tunnel barriers for electrons moving through the channel. The application of negative voltages to the left, central, and right Schottky gate defined two quantum dots (left $L$ and right $R$) which were coupled to each other, to the source, and to the drain.  The tunneling of electrons through the structure was sufficiently large in order to detect an electron current yet small enough to provide a well-defined number of electrons ($\sim 15$ and $\sim 25$) on the left and the right dot, respectively. The Coulomb charging energy ($\sim 4$ meV and $\sim 1$ meV) for placing an additional electron onto the dots was the largest energy scale, see Fig.(\ref{dotsample.eps}).

\begin{figure}[t]
\includegraphics[width=0.5\textwidth]{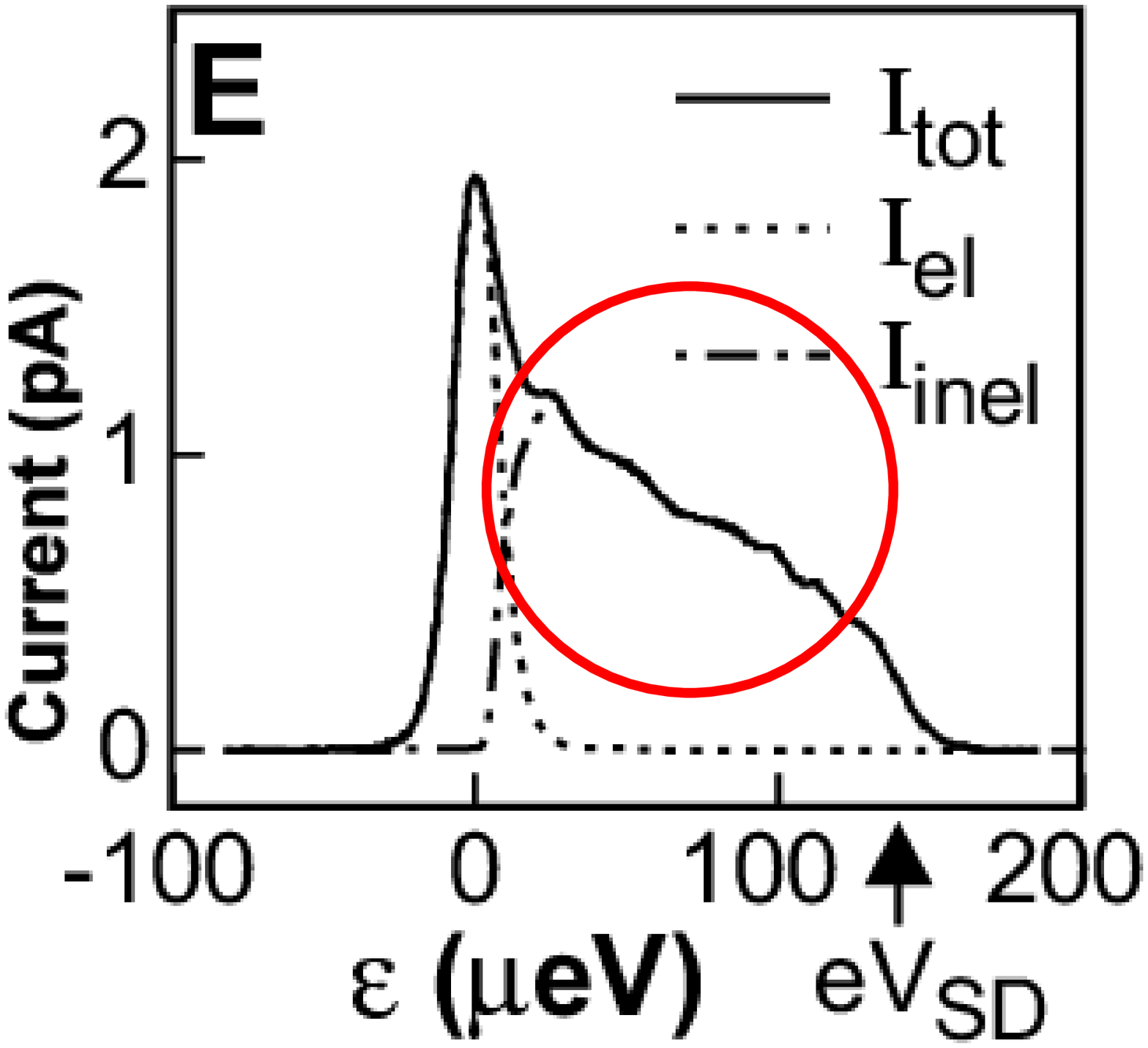}
\includegraphics[width=0.5\textwidth]{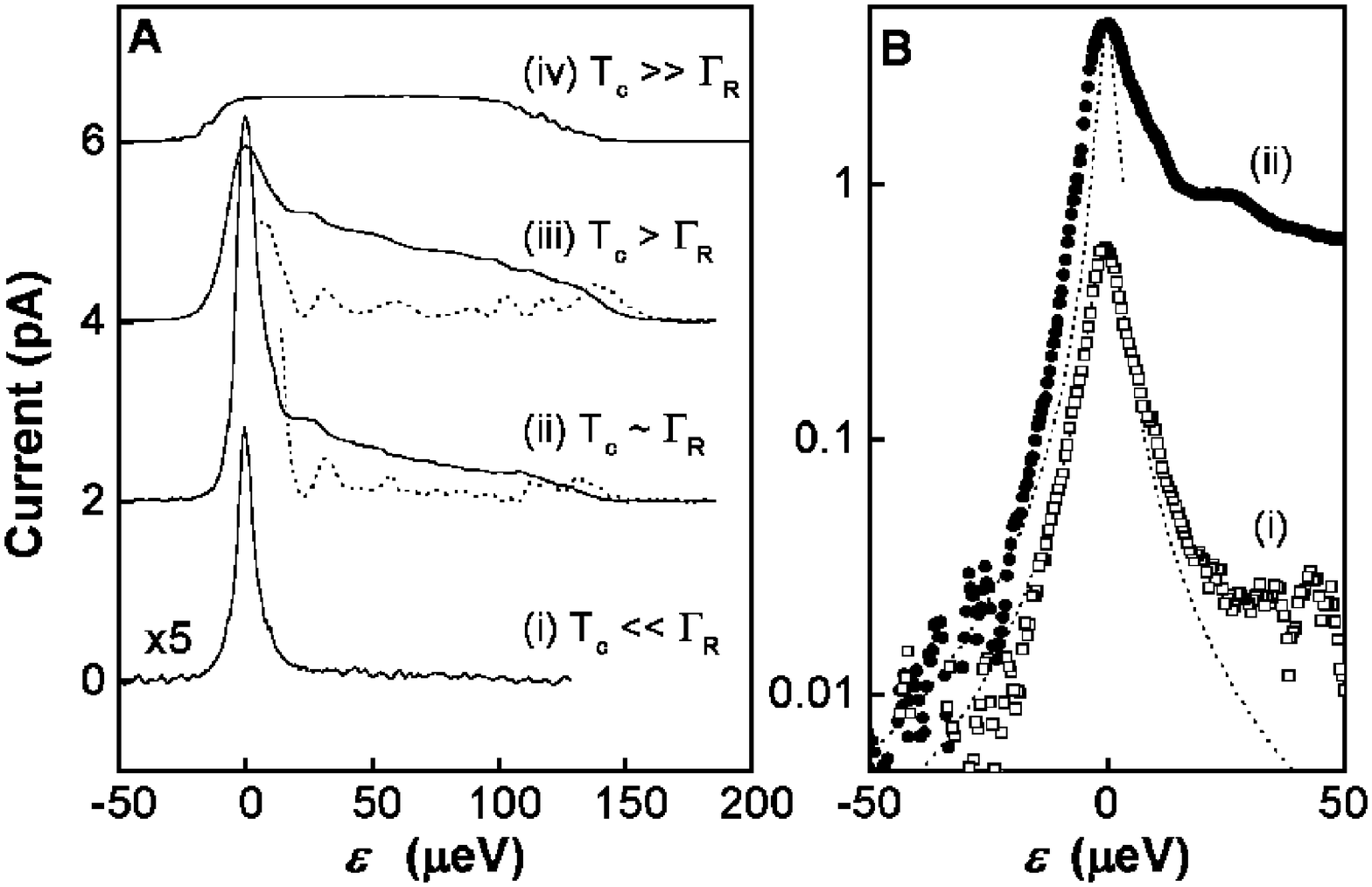}
\caption[]{\label{expcurrent1}{\bf Left:}  Current at temperature $T=23$ mK as a function of the energy
difference $\varepsilon$ in the experiment by Fujisawa
{\em et al.} \cite{Fujetal98}. The total measured current is decomposed into an elastic and 
an inelastic component\index{inelastic current}.
If the difference $\varepsilon$  between left and right dot energies
$E_L$ and $E_R$ is larger than the source-drain-voltage, tunneling is no longer
possible and the current drops to zero. The red circle marks the region of spontaneous 
emission, characterized by the large `shoulder' for $\varepsilon>0$ with an oscillation-like
structure on top of it.
{\bf Right:} Current at $T=23$ mK as a function of the energy
difference $\varepsilon$.
The curves in {\bf A} have an offset and are for different values of the coupling $T_c$ between the
dots and the rate $\Gamma_R$ for tunneling out into the drain region. The dotted curves are 
the negative derivatives of the currents with respect to energy $\varepsilon$ to enhance the
structure on the emission side of the current. {\bf B} shows curves (i) and (ii) from A in a 
double--logarithmic plot, where the dashed lines are Lorentzian fits. From \cite{Fujetal98}.
}
\end{figure}

By simultaneously tuning the gate voltages of the left and the right gate while keeping the central gate voltage constant, the double dot could switch between the three states $|0\rangle =|N_L,N_R\rangle$ (`empty state'), and $|L\rangle=|N_L+1,N_R \rangle$ and $|R\rangle =|N_L,N_R+1\rangle$ with only {\em one additional electron} either in the left or in the right dot (see the following section, where the model is explained in detail). The  experimental sophistication relied on being able  to maintain the state of the system within the Hilbert-space spanned by these states, and to vary the energy difference $\varepsilon=\varepsilon_L-\varepsilon_R$ of the dots without changing the other parameters such as the barrier transmission. The measured average spacing between single-particle states ($\sim 0.5$ and $\sim0.25$ meV) was a large energy scale compared to the scale on which $\varepsilon$ was varied. The largest value of  $\varepsilon$ was determined by the source-drain voltage which is around $0.14$ meV. The main outcomes of this experiment were the following: 
at low temperatures down to  $23$ mK, the stationary tunnel current $I$ as a function of $\varepsilon$ showed a resonant peak at $\varepsilon=0$ with a broad shoulder for $\varepsilon>0$ with oscillations in $\varepsilon$ on a scale of $\approx 20-30\mu$eV, see Fig.(\ref{expcurrent1}). As mentioned above, a similar asymmetry  had in fact  already been  observed in the first measurement of resonant tunneling through double quantum dots in 1995 by van der Vaart and co-workers \cite{Vaartetal95}, cf. Fig. (\ref{vanderVaart_PRL1995_fig1.eps}).

For larger temperatures $T$, the current measured by Fujisawa {\em et al.} increased stronger on the absorption side $\varepsilon<0$ than on the emission side. The data for larger $T$ could be reconstructed from the $23$ mK data by multiplication with the Einstein-Bose distribution $n(T)$ and $1+n(T)$ for emission and absorption, respectively. Furthermore, the functional form of the energy dependence of the current on the emission side was between $1/\varepsilon$ and $1/\varepsilon^2$. For larger distance between the left and right barrier ($600$ nm on a surface gate sample instead of $380$ nm for a focused ion beam sample), the period of the oscillations on the emission side appeared to become shorter, see Fig.(\ref{expcurrent1}).

From these experimental findings, Fujisawa {\em et al.} concluded that the {\em coupling to a bosonic environment} was of key importance in their experiment. To identify the microscopic mechanism of the spontaneous emission, they placed the double dot in different electromagnetic environments in order to test if a coupling to {\em photons} was responsible for these effects. Typical wavelengths in the  regime of relevant energies $\varepsilon$ are in the cm range for both photons and 2DEG plasmons. Placing the sample in microwave cavities of different sizes showed no effect on the spontaneous emission spectrum. Neither was there an effect by measuring different types of devices with different dimensions, which should change the coupling to plasmon. Instead, it was the coupling to {\em acoustic phonons} (optical phonons have too large energies in order to be relevant) which turned out to be the microscopic mechanism responsible for the emission spectrum. In fact, phonon energies in the relevant $\varepsilon$ regime correspond to wavelengths that roughly fit with the typical dimensions (a few 100 nm) of the double dot device used in the experiments.

\subsection{Transport Theory for  Dissipative Two-Level Systems} 

In the following, the dissipative double quantum dot as a model which is key to some of the following sections is introduced. It describes electron transport through two-level systems (coupled quantum dots) in the presence of a dissipative environment (phonons or other bosonic excitations).

\begin{figure}[t]
\includegraphics[width=0.5\textwidth]{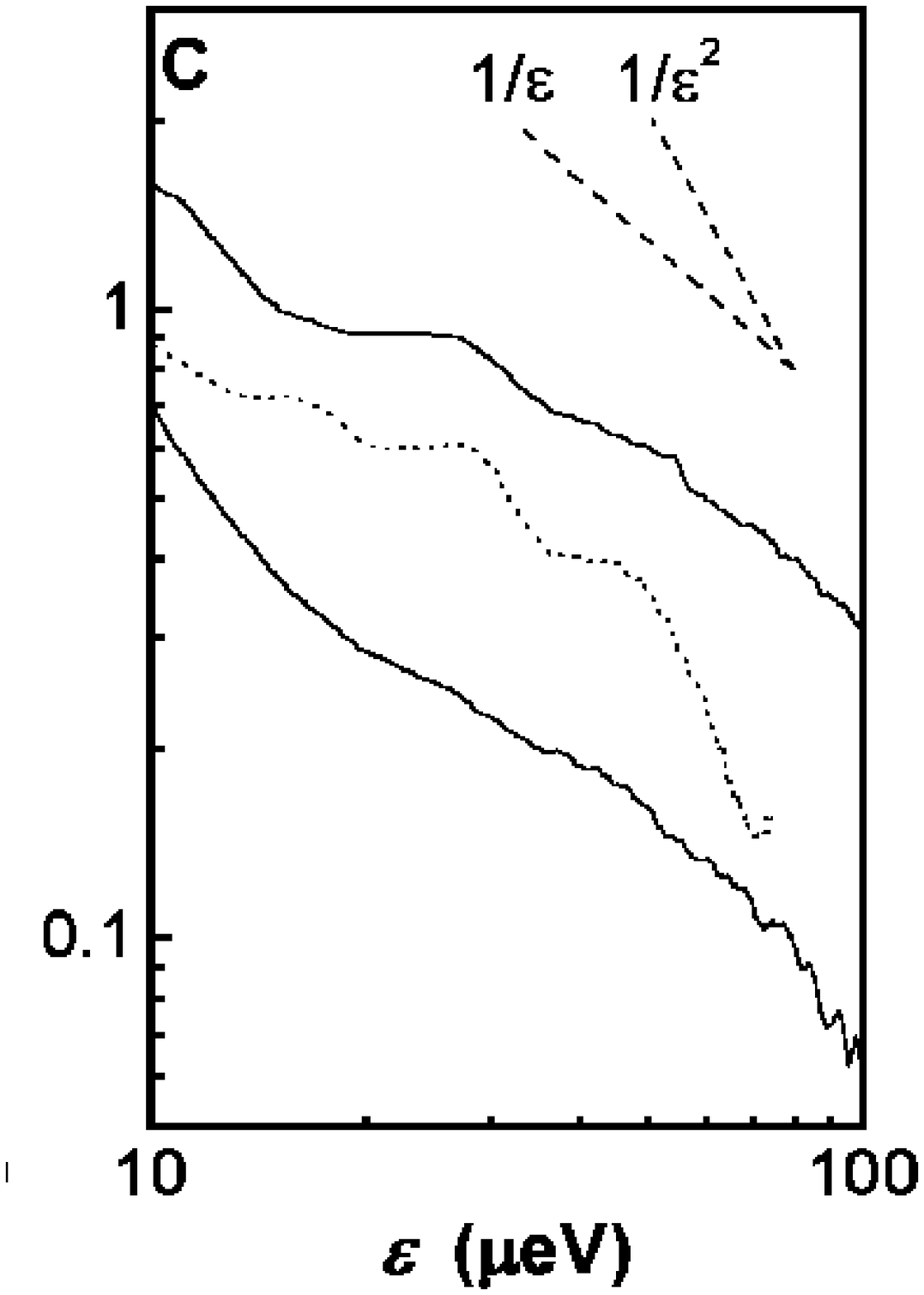}
\includegraphics[width=0.5\textwidth]{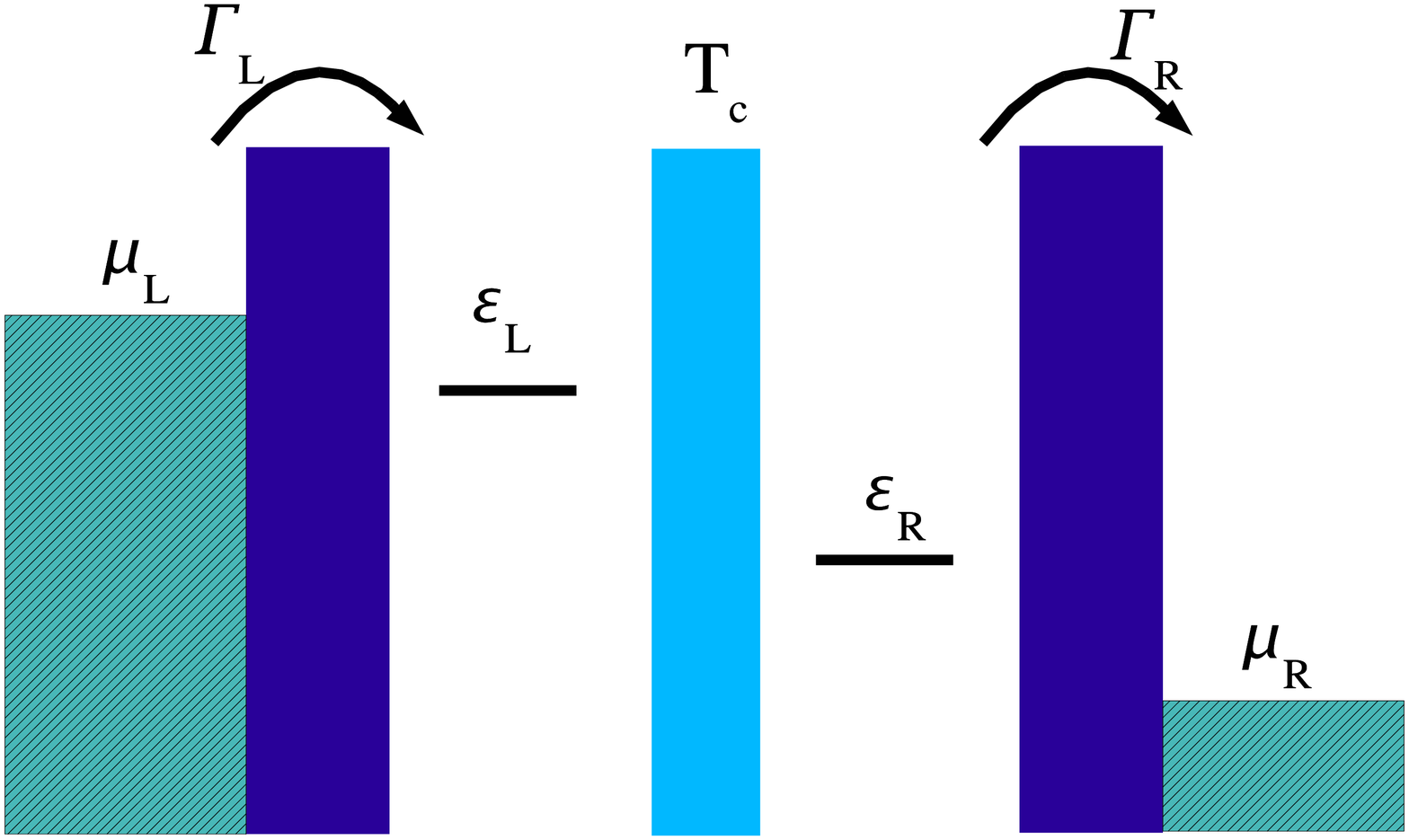}
\caption[]{\label{expcurrent3}{\bf Left:} Current on the emission side
$\varepsilon>0$ in the experiment by Fujisawa {\em et al.} \cite{Fujetal98}. The solid lines
correspond to data for different values of $T_c$. The dotted line represents data from a {\em surface gate sample} where the distance between left and right barriers is larger (600 nm). {\bf Right:} Double dot model consisting of left and right dot, coupled by a tunnel matrix element $T_c$. Left and right electron reservoirs with chemical potential $\mu_L$ ($\mu_R$) act as source and drain for electrons tunneling from left to right at rates $\Gamma_L$ and $\Gamma_R$. The energies $\varepsilon_L$ and $\varepsilon_R$  have to be understood as chemical potentials for the addition of one additional electron to the left and the right dots, respectively. The system is in the strong Coulomb blockade regime with only one additional electron allowed to enter the double dot. Phonons couple to the electronic density in both dots.}
\end{figure}

\subsubsection{Double Dot Model}\label{section_ddotmodel}
The possibly simplest model defines a double quantum dot as a composite system of two individual dots which for the sake of definiteness are called left and right dot (L and R) here and in the following,  and which are connected through a static tunnel barrier. The effective `qubit' Hilbert space $H^{(2)}\equiv {\rm  span}(|L\rangle,|R\rangle)$ of this system is assumed to be spanned by two many-body states $|L \rangle=|N_L+1,N_R \rangle$ and $|R \rangle=|N_L,N_R+1 \rangle$ with energies $\varepsilon_L$ and $\varepsilon_R$, corresponding to the lowest energy states for one additional electron in the left and the right dot. In contrast, the `empty' ground state $|0\rangle =|N_L,N_R\rangle$ has one electron less and $N_L$ electrons in the left and $N_R$ electrons in the right dot. Although this state plays a role in transport through the dot as discussed below, there are no superpositions between $|0\rangle$ and the states in $H^{(2)}$ (charge superselection rule). The left-right degree of freedom  in $H^{(2)}$ defines a `pseudospin' $1/2$ \cite{BK99} as described by Pauli matrices $\hat{\sigma}_z =  \hat{n}_L-\hat{n}_R$ and $\hat{\sigma}_x= \hat{p}+\hat{p}^{\dagger}$, which together with operators involving the empty state $|0\rangle$ form a closed operator algebra,
\begin{eqnarray}\label{operators}
\hat{n}_i&\equiv &|i\rangle \langle i|,
\quad  \hat{p}\equiv |L\rangle \langle R|,\quad 
 \hat{s}_i\equiv|0\rangle \langle i|, \quad
i=L,R.
\end{eqnarray}
Inter-dot tunneling between L and R is described by a single, real parameter which by convention is denoted as $T_c$ here and in the following.\footnote{This choice might be confusing to physicists working in  superconductivity, but has been used in much of the literature on double quantum dots which is why it is used here, too.}
The Hamiltonian of the double dot then reads
\begin{equation}\label{hdot}
{\mathcal H}_{\rm dot}\equiv \varepsilon_L\hat{n}_L+\varepsilon_R\hat{n}_R+T_c(\hat{p}+\hat{p}^{\dagger}),
\end{equation}
the eigenstates of which are readily obtained by diagonalizing the two-by two matrix 
\begin{eqnarray}
  {\mathcal H}_{\rm TLS}\equiv  \frac{\varepsilon}{2} \hat{\sigma}_z + T_c \hat{\sigma}_x=
\left(\begin{matrix}
 \frac{\varepsilon}{2} & T_c \\ T_c & - \frac{\varepsilon}{2}
\end{matrix}\right),
\quad \varepsilon\equiv \varepsilon_L - \varepsilon_R,
\end{eqnarray}
where here and in the following 
the trivial constant term $ \frac{1}{2}(\varepsilon_L + \varepsilon_R)$ is omitted. The  eigenstates  $|\pm\rangle$ and eigenvalues $\varepsilon_{\pm}$ of ${\mathcal H}_{\rm TLS}$ are 
\begin{eqnarray}
  \label{eq:twobytworesult}
|\pm \rangle &=& \frac{1}{N_{\pm}}\left[
\pm 2T_c |L\rangle + (\Delta \mp \varepsilon) |R\rangle \right],\quad 
N_{\pm}\equiv \sqrt{4T_c^2+(\Delta \mp \varepsilon)^2}\\
\varepsilon_{\pm}&=&\pm \frac{1}{2}\Delta,\quad \Delta\equiv\sqrt{\varepsilon^2+4T_c^2},
\end{eqnarray}
corresponding to hybridized wave functions, i.e. bonding and anti-bonding superpositions of the two, originally localized states $|L\rangle$ and $|R\rangle$. The corresponding eigenvalues $\varepsilon_{\pm}=\pm \frac{1}{2}\Delta$ of the double dot represent two energy surfaces over the $T_c$-$\varepsilon$ plane, with an avoided level crossing of splitting $\Delta$. For $\varepsilon=0$, one has $|\pm \rangle=(1/\sqrt{2})(\pm {\rm sign}(T_c)|L\rangle + |R\rangle )$ such that for the choice $T_c<0$  the  ground state $|-\rangle = (1/\sqrt{2}) (|L\rangle + |R\rangle )$ with energy $\varepsilon_-=-\frac{1}{2}\Delta$ is the {\em symmetric} superposition of $|L\rangle$ and $|R\rangle$. 

Electron transport through the double dot is introduced by connecting the left (right) dot to an electron reservoir in thermal equilibrium at chemical 
potential $\mu_L$ ($\mu_R$) with positive source-drain voltage $V_{SD}\equiv \mu_L-\mu_R$, inducing
tunneling of electrons from the left to the right.
One assumes that the ground state energies $\varepsilon_L$ of $|L\rangle$ and
$\varepsilon_R$ of $|R\rangle$ are in the window between source and drain energy, i.e.
$\mu_L>\varepsilon_L,\varepsilon_R>\mu_R$. 
Transport involves the state $|0\rangle$  and superpositions
within the two-dimensional Hilbert space $H^{(2)}\equiv {\rm  span}(|L\rangle,|R\rangle)$.
This restriction is physically justified under the following conditions: 
first, the source-drain voltage $V_{SD}$ has to be much smaller than the Coulomb charging energy $U_c$ to 
charge the double dot with more than one additional electron. Second,
many-body excited states outside $H^{(2)}$ can be neglected. 

The coupling to the electron
reservoirs ${\mathcal H}_{\rm res}$
is described by the usual tunnel Hamiltonian ${\mathcal H}_V$,
\begin{eqnarray}\label{Hresdef}
{\mathcal H}_{\rm res}=\sum_{k_i,i=L/R}\epsilon_{k_i}
c_{k_i}^{\dagger}c_{k_i},\quad
{\mathcal H}_V=\sum_{k_i{}}(V_k^i c_{k_i}^{\dagger}\hat{s}_i+H.c.),  \quad
\hat{s}_i\equiv|0 \rangle \langle i|,\quad i=L,R, 
\end{eqnarray}
where the $V_k^i$ couple to a continuum of channels $k$ in reservoir $i$. 
We note that the splitting of the whole electron system into reservoir and dot regions bears
some fundamental problems that are inherent in all descriptions that use the tunnel Hamiltonian
formalism 
\cite{Pra64,FraPhD,Fraetal99}.

Including the `empty' state $|0 \rangle=|N_L,N_R \rangle$, the completeness relation of the `open' double 
dot is now $\hat{1}=\hat{n}_0+\hat{n}_L+\hat{n}_R$.
In the above description,  spin polarization of the electrons has been assumed
so that only charge but no spin degrees of freedom are accounted for.
In the original `charge qubit' experiment \cite{Fujetal98}, a magnetic field between 1.6 and 2.4 T was
applied perpendicular to the dots in order to maximize the single-particle spacing and to 
spin polarize the electrons. The combination of both (real) spin and pseudo-spin degrees of freedom
was discussed recently by Borda, Zar\'{a}nd, Hofstetter, Halperin, and von Delft \cite{Boretal03} in the context of a $SU(4)$ Fermi liquid state and the Kondo effect in double quantum dots. 

Linear  coupling between the double dot and  bosonic modes (photons, phonons) is described by a Hamiltonian
\begin{eqnarray}\label{Hdp}
  {\mathcal H}_{\rm dp}=\sum_{\bf Q} \left(\alpha^L_{\bf Q} \hat{n}_L + \alpha^R_{\bf Q} \hat{n}_R 
+ \gamma_{\bf Q} \hat{p} + \gamma^*_{-{\bf Q}} \hat{p}^{\dagger}\right)\left(a_{-{\bf Q}} +a_{{\bf Q}}^{\dagger}\right),
\end{eqnarray}
where the coupling matrix elements $\alpha^L_{\bf Q}$, $\alpha^R_{\bf Q}$, and $\gamma_{\bf Q}$ 
and the frequency dispersions $\omega_{\bf Q}$ of the free boson Hamiltonian
\begin{eqnarray}
{\mathcal H}_{\rm B} =\sum_{\bf Q} \omega_{\bf Q} a_{\bf Q}^{\dagger}a_{\bf Q}  
\end{eqnarray}
have to be calculated from microscopic theories, cf. the following sections. The total Hamiltonian becomes
\begin{eqnarray}\label{Htotal}
  {\mathcal H} = {\mathcal H}_{\rm dot} + {\mathcal H}_{\rm dp}
+ {\mathcal H}_V
+ {\mathcal H}_{\rm B} + {\mathcal H}_{\rm res}
\end{eqnarray}
and generalizes the usual spin-boson model Hamiltonian ${\mathcal H}_{\rm SB}$ \cite{Legetal87,Weiss}
\begin{equation}\label{modelhamiltonian}
{\mathcal H}_{SB}= \Big[\frac{\varepsilon}{2}
+\sum_{\bf Q} \frac{g_{\bf Q}}{2} \left(a_{-\bf Q} + a^{\dagger}_{\bf
Q}\right)\Big]
\hat{\sigma}_z
+ T_c
\hat{\sigma}_x
+{\mathcal H}_B,\quad g_{\bf Q}\equiv \alpha^L_{\bf Q}-\alpha^R_{\bf Q}
\end{equation}
due to the additional coupling to the electron reservoirs (terms ${\mathcal H}_T + {\mathcal H}_{\rm res}$) and
the additional terms $\gamma_{\bf Q}$ in ${\mathcal H}_{\rm dp}$ which are 
off-diagonal in the localized basis $\{|L\rangle,|R\rangle\}$.
The usual spin-boson model ${\mathcal H}_{SB}$
corresponds to setting the off-diagonal-terms in \ Eq.~(\ref{Hdp}){} to zero, $\gamma_{\bf Q} =0$, whence
\begin{eqnarray}\label{newHdp}
{\mathcal H}_{\rm dp}=\sum_{\bf Q} \sum_{i=L,R}\alpha^i_{\bf Q} \hat{n}_i
\left(a_{-{\bf Q}} +a_{{\bf Q}}^{\dagger}\right),
\end{eqnarray}
which is used as electron-boson coupling Hamiltonian in the following. As the `dipole terms' $\gamma_{\bf Q}$ are proportional to the overlap of the wave functions  between the left and the right dot which itself determines the value of $T_c$, neglecting the $\gamma_{\bf Q}$ terms is argued to be justified for weak tunnel coupling $T_c$ \cite{GM88,Legetal87,BV03}. On the other hand, for larger $T_c$ these terms become more important, cf. section \ref{section_Keil_Schoeller}.

\subsubsection{Master Equation}
The easiest way to describe electron transport through quantum dots is to use rate equations
with tunnel rates calculated from the Hamiltonian \ Eq.~(\ref{Htotal}).
These equations have to be extended in order to account for coherences between the 
dots, i.e. the off-diagonal operators $\hat{p}$ and $\hat{p}^{\dagger}$ in \ Eq.~(\ref{hdot}).
This is similar to Quantum Optics where the optical Bloch equations for a two-level system \cite{Allen}
generalize the `diagonal' equations for the occupancies (Einstein equations). 
Gurvitz and Prager \cite{GP96,Gur98}, and Stoof and Nazarov \cite{SN96} have derived these equations for
double quantum dots in the limit of infinite source-drain voltage ($\mu_L\to \infty$, $\mu_R\to -\infty$), 
and for tunnel rates 
\begin{eqnarray}\label{tunnel_rates_def}
\Gamma_i\equiv 2\pi\sum_{k_i}|V_k^i|^2\delta(\varepsilon-\varepsilon_{k_i}),
\quad i=L/R,
\end{eqnarray}
assumed to be independent of energy, where the Born-Markov approximation
with respect to the electron reservoir coupling becomes exact. 
This limit, which is adopted throughout this Review,
is particularly useful for the discussion of coherent effects {\em within} the double dot 
system, as the role of the leads basically is to supply and carry away electrons, whereas
Kondo-type correlations between electrons in the leads and in the dots are completely suppressed.

Due to the coupling to bosons (the term ${\mathcal H}_{\rm dp}$ in  \ Eq.~(\ref{Htotal})), an exact 
calculation of the reduced density operator $\hat{\rho}(t)$ of the dot
is usually not possible, but one can invoke various approximation schemes,
the most common of which are perturbation theory in the inter-dot coupling 
$T_c$ (unitary polaron transformation),
and perturbation theory in the electron-boson coupling.

\subsubsection{Method 1: Polaron Transformation}\label{section_polaron}
The polaron transformation is a well-known method to solve problems 
where bosonic degrees of freedom couple to
a single localized state \cite{Mahan,GS88,WJW88,Jauho}. One defines a unitary transformation 
for all operators $\hat{O}$ in the Hamiltonian Eq.~(\ref{Htotal}),
\begin{eqnarray}
  \bar{O} \equiv e^S\hat{O}e^{-S},\quad S\equiv \sum_{i=L,R}\sum_{\bf Q}\hat{n}_i\left(
\frac{\alpha^i_{\bf Q}}{\omega_{\bf Q}}a^{\dagger}_{\bf Q} -
 \frac{\alpha^i_{-\bf Q}}{\omega_{\bf Q}}a_{\bf Q} \right),
\end{eqnarray}
which removes the electron-boson term \ Eq.~(\ref{newHdp}) and leads to the transformed
total Hamiltonian $\overline{\mathcal H}$,
\begin{eqnarray}\label{hamiltoniantrans}
  \overline{\mathcal H}&=&{\mathcal H}_0+ \overline{{\mathcal H}}_T
+\overline{{\mathcal H}}_V ,\quad
  {\mathcal H}_0\equiv \overline{\varepsilon_L}\hat{n}_L+\overline{\varepsilon_R}\hat{n}_R
+{\mathcal H}_{B}+{\mathcal H}_{\rm res}\nonumber\\
  \overline{{\mathcal H}}_T&\equiv&T_c(\hat{p}X+\hat{p}^{\dagger}X^{\dagger}),\quad
\overline{\varepsilon_i}\equiv \varepsilon_i-
\sum_{\bf{Q}}\frac{|\alpha^i_{\bf{Q}}|^2}{\omega_{\bf{Q}}}.
\end{eqnarray}
The energy difference $\varepsilon\equiv \overline{\varepsilon_L}-
\overline{\varepsilon_R}$ (using the same symbol for notational simplicity) is now renormalized
with the dot energies $\varepsilon_L$ and $\varepsilon_R$ renormalized 
to smaller values.
More important, however,  is the appearance of the factors $X$ and $X^{\dagger}$ in the
inter-dot coupling Hamiltonian $\overline{{\mathcal H}}_T$,
\begin{eqnarray}\label{Xdef}
  X \equiv \prod_{\bf{Q}}D_{\bf{Q}}\left(\frac{\alpha^L_{\bf Q}-\alpha^R_{\bf Q}}{\omega_{\bf Q}}\right),\quad
D_{\bf{Q}}(z) \equiv \exp \left(za^{\dagger}_{\bf{Q}}-z^*a_{\bf{Q}}^{\phantom{\dagger}} \right),
\end{eqnarray}
where $D_{\bf{Q}}(z)$ is the unitary displacement operator of a boson mode ${\bf Q}$.
The operation of $D_{\bf{Q}}(z)$ on the
vacuum of a boson field mode with creation operator $a_{\bf{Q}}^{\dagger}$ and ground state $|0\rangle_{\bf{Q}}$ 
creates a {\em coherent state} 
$|z\rangle_{\bf{Q}}=D_{\bf{Q}}(z)|0\rangle_{\bf{Q}}$ of that mode \cite{Walls}.

The Master equation can now be derived in the polaron-transformed frame, resulting into 
an explicit set of equations for the double dot expectation values,
\begin{eqnarray}\label{eom3new}
\frac{\partial}{\partial t}\langle n_L\rangle_t&=&-iT_c\left\{
\langle p \rangle_{t}-\langle p^{\dagger}\rangle_{t} \right\} 
+\Gamma_L\left[1-\langle n_L\rangle_{t} - \langle n_R\rangle_{t}\right]
\label{nLeq}\\
\frac{\partial}{\partial t}\langle
n_R\rangle_t&=&\phantom{-}iT_c \left\{ \langle
p\rangle_{t}-\langle p^{\dagger}\rangle_{t} \right\} 
-{\Gamma}_R\langle n_R\rangle_{t} 
\label{nReq}\\ 
\langle p\rangle_t&=& -\int_0^tdt'   e^{i\varepsilon(t-t')}
\left[ \left( \frac{\Gamma_R}{2}\langle {p}\rangle_{t'}
+ iT_c  \langle n_L\rangle_{t'} \right) C(t-t')- iT_c \langle
n_R\rangle_{t'} C^*(t-t') \right] \label{peqn}\\
\langle p^{\dagger}\rangle_t&=& -\int_0^tdt'   e^{-i\varepsilon(t-t')}
\left[ \left( \frac{\Gamma_R}{2} \langle {p}^{\dagger}\rangle_{t'}
- iT_c  \langle n_L\rangle_{t'} \right) C^*(t-t')+ iT_c \langle
n_R\rangle_{t'} C(t-t') \right], \label{pdaggereqn}
\end{eqnarray}
where the central quantity containing the coupling to the bosons is the 
equilibrium correlation function of the $X$ operators, \ Eq.~(\ref{Xdef}), for a 
boson density matrix $\rho_B$ in thermal equilibrium at inverse temperature
$\beta$,
\begin{eqnarray}\label{C_equi}
  C(t-t')&\equiv& {\rm Tr} \left( \rho_B X_t^{\phantom{\dagger}}X_{t'}^{\dagger} \right),\quad
\rho_B = \frac{e^{-\beta {\mathcal H}_B}}{{\rm Tr} e^{-\beta {\mathcal H}_B}}.
\end{eqnarray}
The function $C(t)$ can be evaluated explicitly and is expressed in terms
of the {\em boson spectral density} $J(\omega)$, 
\begin{eqnarray}\label{Ct}
   C(t)&\equiv&e^{-Q(t)},\quad Q(t)\equiv\int_0^{\infty}d\omega
\frac{J(\omega)}{\omega^2} \left[ \left(1- \cos \omega t\right)
\coth \left(\frac{\beta \omega}{2}\right) + i \sin \omega t
\right]\\
J(\omega)&\equiv& \sum_{\bf Q} \left| \alpha^L_{\bf Q}-\alpha^R_{\bf Q} \right|^2
\delta(\omega-\omega_{\bf Q}).\label{Jomega}
\end{eqnarray}


Details of the derivation of  Eq.~(\ref{eom3new})- \ Eq.~(\ref{pdaggereqn}) are given in Appendix \ref{appendix_pol}.
Several approximations have been used: first, the initial thermal density matrix $\bar{\chi}(0)$ of the total system at time $t=0$ in the polaron-transformed frame factorizes to lowest (zeroth) order in both $T_c$ and $V_k^i$ according to 
\begin{eqnarray}\label{factorization}
\bar{\chi}(0)\equiv \frac{\overline{e^{-\beta \mathcal{H}}}}{Z} \approx  \frac{e^{-\beta \mathcal{H}_0}}{Z_0}=
R_0 \otimes \rho_B \otimes {\rho}_{\rm dot},
\end{eqnarray}
where $R_0$ is the equilibrium density matrix of the electron reservoirs. 
Furthermore, for all times $t>0$ a decoupling approximation
\begin{eqnarray}\label{decoupling_app}
  \tilde{\chi}(t)\approx R_0 \otimes \rho_B \otimes \tilde{\rho}_{\rm dot}(t)
\end{eqnarray}
is used. The back-action on both electron reservoirs and the boson bath (which are assumed to stay in thermal equilibirum) is therefore neglected throughout. One then can factorize terms like $\langle
\hat{n}_LX_t^{\phantom{\dagger}}X_{t'}^{\dagger}\rangle_{t'}\approx \langle \hat{n}_L \rangle_{t'} 
 \langle X_t^{\phantom{\dagger}}X_{t'}^{\dagger}\rangle_B$ in the equation of the 
coherences $\langle \hat{p}\rangle_t$; these equations, however, are then no longer 
exact. In the original spin-boson problem ($\Gamma_{L/R}=0$), this amounts to second order perturbation 
theory in the inter-dot coupling $T_c$ \cite{Legetal87}, which is known to be equivalent to the 
so-called  non-interacting-blib-approximation (NIBA) \cite{Legetal87,Weiss}
of the dissipative spin-boson problem, whereas here the factorization also involves
the additional term $\frac{\Gamma_R}{2}\langle {p}\rangle_{t'}$ which
describes the broadening of the coherence $\langle \hat{p}\rangle_{t}$ due to 
electrons tunneling into the right reservoir. 

Finally, two additional terms 
in \ Eq.~(\ref{peqn}){} and (\ref{pdaggereqn}){}
describing the decay of an initial polarization of the system have been neglected. These terms in fact can be calculated exactly but they vanish in the stationary limit for long times $t\to \infty$.

\subsubsection{Method 2: Perturbation Theory in ${\mathcal H}_{dp}$}\label{section_perturbation}
An alternative way is a perturbation theory not in the inter-dot coupling $T_c$, but in the 
coupling ${\mathcal H}_{dp}$ to the boson system. Assuming  the boson system to be described
by a thermal equilibrium, standard second order perturbation theory and the Born-Markov 
approximation yield
\begin{eqnarray}\label{pequation}
  \frac{d}{dt}\langle p\rangle_t&=& \left(i\varepsilon-\frac{\Gamma_R}{2}-\gamma_p\right)\langle p\rangle_t
+ i T_c \left[\langle n_R\rangle_{t} -\langle n_L\rangle_{t}\right]
+ \gamma_+ \langle n_L\rangle_{t} - \gamma_- \langle n_R\rangle_{t},
\end{eqnarray}
with the correspondingly complex conjugated  equation for $\langle p^{\dagger}\rangle_t$, and the
equations for $\langle n_{L/R}\rangle_t$  identical to \ Eqs.~(\ref{nReq}), (\ref{nLeq}).
The rates $\gamma_p$ and $\gamma_{\pm}$ are defined as
\begin{eqnarray}\label{gammarates}
\gamma_p   &\equiv& \frac{1}{\Delta^2} \int_0^{\infty}\!\!dt \; (\varepsilon^2 
+ 4T_c^2 \cos \Delta t) \, \mbox{Re}\{K(t)\} \\
\gamma_+ &\equiv& \frac{T_c}{\Delta^2} \int_0^{\infty}\!\!dt \;
(\varepsilon \,(1\!-\!\cos \Delta t ) - i \Delta \sin \Delta t)\; K(t)\\
\gamma_- &\equiv& \frac{T_c}{\Delta^2} \int_0^{\infty}\!\!dt \;
(\varepsilon \,(1\!-\!\cos\Delta t) - i \Delta \sin\Delta t)\; K^*(t),
\end{eqnarray}
and the bosonic system enters solely via the correlation function
\begin{equation}\label{Kdefine}
K(t) = \int_0^{\infty}d\omega  J(\omega)
     [n_B(\omega) e^{i\omega t} + (1+n_B(\omega))  e^{-i \omega t}],
\end{equation}
where $n_B(\omega)=[e^{\beta \omega}-1]^{-1}$ is the Bose distribution 
at temperature $1/\beta$.
The explicit evaluation of Eq.~(\ref{gammarates})-(\ref{Kdefine}) leads to inelastic rates
\begin{eqnarray}\label{gammapdef}
\gamma_p\equiv 2\pi \frac{T_c^2}{\Delta^2} J(\Delta) \coth
\left(\beta \Delta /2\right),\quad 
\gamma_{\pm}\equiv-\frac{\varepsilon T_c}{\Delta^2} \frac{\pi}{2}
J(\Delta)  \coth \left(\beta \Delta /2\right)\mp
 \frac{T_c}{\Delta} \frac{\pi}{2}J(\Delta),
\end{eqnarray}
which completely determine dephasing and relaxation in the system. Some care has to be taken when evaluating
the rates,\ Eq.~(\ref{gammarates}), with the parametrized form
$ J(\omega) = 2\alpha \omega_{\rm ph}^{1-s} \omega^s e^{-\omega/\omega_c}$ for the boson spectral density in \ Eq.~(\ref{Kdefine}), cf. 
\ Eq.~(\ref{Jomegageneric}) in section \ref{section_spectraldensity}. In this case, it turns out that the Born-Markov approximation is in fact only meaningful and defined for $s\ge 1$. For $s<1$, this perturbation theory breaks down. In addition, the rates \ Eq.~(\ref{gammapdef}) acquire an additional term linear in the temperature $k_BT=1/\beta$ in the Ohmic case $s=1$, for which the rates explicitly read \cite{BV02}
\begin{eqnarray}
\label{raten-ohmsch}
\gamma_p &=& \frac{2\alpha \pi}{\Delta^2} \left( \frac{\varepsilon^2}{\beta}
 + 2 T_c^2 \Delta e^{-\Delta/\omega_c} \coth\left(\frac{\beta \Delta}{2}\right) \right) \\
\mbox{Re}\{\gamma_{\pm}\} &=& 2\alpha \frac{\pi T_c}{\Delta^2} 
\left( \frac{\varepsilon}{\beta} - \frac{\varepsilon}{2} \,\Delta\,
e^{-\Delta/\omega_c} \coth \left(\frac{\beta \Delta}{2}\right) \right.
\mp\left.\frac{\Delta^2}{2}e^{-\Delta/\omega_c}\right)\\
\mbox{Im}\{\gamma_{+}+\gamma_{-}\} &=& 4\alpha T_c\int_0^{\infty}d\omega\frac{\omega e^{-\omega/\omega_c}}
{\omega^2-\Delta^2}\left(1+\frac{2}{e^{\beta\omega}-1}\right)\\
\mbox{Im}\{\gamma_{+}-\gamma_{-}\} &=& -4\alpha\frac{\varepsilon T_c\omega_c}{\Delta^2}
\left[1-
\int_0^{\infty}\frac{d\omega}{\omega_c}\frac{\omega^2 e^{-\omega/\omega_c}}
{\omega^2-\Delta^2}\right].
\end{eqnarray}
The last two integrals can be evaluated approximately \cite{Gradstein} for small $\Delta/\omega_c$.
One finds that up to order $\Delta/\omega_c$,
\begin{eqnarray}
\mbox{Im}\{\gamma_{+}-\gamma_{-}\} &=& O\left(\frac{\Delta}{\omega_c}\right)\\
\mbox{Im}\{\gamma_{+}\} = \mbox{Im}\{\gamma_{-}\}
&=& 2\alpha T_c\left[ \ln\left(\frac{\beta\Delta}{2\pi}\right)
-\mbox{\rm Re}\Psi\left(\frac{i\beta\Delta}{2\pi}\right)\right.
\left.
-C-\ln\left(\frac{\Delta}{\omega_c}\right)\right]+O\left(\frac{\Delta}{\omega_c}\right).
\end{eqnarray}
Here,  $C=0.577216$ is the Euler number and $\Psi(x)$ is the logarithmic derivative of the Gamma function.
For the latter, one can use
\cite{Abramowitz} $\mbox{\rm Re}\Psi(iy)=\mbox{\rm Re}\Psi(1+iy)$ and
the expansions
\begin{eqnarray}
  \mbox{\rm Re}\Psi(iy)=\left\{
    \begin{array}{cc}
\ln y +\frac{1}{12y^2}+\frac{1}{120y^4}+\frac{1}{252y^6}+...,&y \to \infty\\
-C+y^2\sum_{n=1}^{\infty}n^{-1}\left(n^2+y^2\right)^{-1}, &|y|<\infty
    \end{array}\right.
\end{eqnarray}
The combination of the first (large arguments $y$) and the second expansion (small arguments $y$)
is  useful in numerical calculations.

\subsubsection{Matrix Formulation}
It is convenient to introduce the vectors ${\bf A}\equiv
(\hat{n}_L,\hat{n}_R,\hat{p},\hat{p}^{\dagger})$, ${\bf \Gamma}=\Gamma_L{\bf e}_1$ 
(${\bf e}_1,...,{\bf e}_4$ are unit vectors)
and a time-dependent matrix memory kernel ${M}$ in order to 
formally write the equations of motion (EOM) for the dot
as \cite{AB04}
\begin{equation}\label{expectation}
  \langle {\bf A}(t)\rangle   \!=\! \langle {\bf A}(0)\rangle
 + \int_{0}^{t}dt' \left\{{M}(t,t')\langle {\bf A}(t')\rangle  + {\bf \Gamma} \right\},
\end{equation}
where $\langle...\rangle\equiv {\rm Tr_{dot}} ...\hat{\rho}(t)$ and $\hat{\rho}(t)$ is the reduced density operator of the double dot.
This formulation is a particularly useful starting point for, e.g., the calculation of shot noise or
out-of-equilibrium situations like driven
double dots, where the bias $\varepsilon$ or the tunnel coupling $T_c$ are a function of time $t$
and consequently, the memory kernel $M$ is no longer time-translation invariant \cite{BAP04}, cf. sections 
\ref{section_noise} and \ref{section_AC}.

For constant $\varepsilon$ and $T_c$, Eq.~(\ref{expectation}) is easily 
solved by introducing the Laplace transformation $\hat{f}(z)=\int_{0}^{\infty}dt e^{-zt}f(t)$.
In $z$-space, one has
$\langle\hat{\bf A}(z)\rangle= [z-z\hat{M}(z)]^{-1}(\langle{\bf
A}(0)\rangle+{\bf \Gamma}/z)$ which serves as a starting point for the
analysis of stationary ($1/z$ coefficient in Laurent series for $z\rightarrow
0$) and non-stationary quantities. The memory kernel has a
block structure
\begin{equation}\label{block}
  z\hat{M}(z)=\left[
  \begin{matrix}
    -\hat{G} & \hat{T}\\ \hat{D}_{z}& \hat{\Sigma}_{z}
  \end{matrix}\right], \quad
 \hat{G}\equiv
\left(
  \begin{matrix}
\Gamma_L & \Gamma_L\\0&\Gamma_R
\end{matrix}\right),
\end{equation}
where $\hat{T} \equiv -iT_c(1-\sigma_x)$. 
The blocks $\hat{D}_{z}$ and $\hat{\Sigma}_{z}$
are determined by the equation of motion for the coherences 
$\langle\hat{p}\rangle=\langle\hat{p}^{\dagger }\rangle^*$ 
and contain the complete information on inelastic relaxation and dephasing of the system.

For weak boson coupling, the above  perturbation theory (PER, Method 2) 
in the correct basis of the hybridized states of the double dot
yields
\begin{equation}\label{PERb}
  \hat{D}^{\rm PER}=\hat{T}_c+
\left(
 \begin{matrix}
\gamma_+ &  - \gamma_-\\ \gamma_+ &  -\gamma_-
\end{matrix}\right),\quad
\hat{\Sigma}^{\rm PER}=
\left(
 \begin{matrix}
i\varepsilon-\gamma_p-\frac{\Gamma_R}{2}&0\\0& -i\varepsilon-\gamma_p-\frac{\Gamma_R}{2}
\end{matrix}\right).
\end{equation}
On the other hand,  for strong electron-boson coupling, the unitary transformation method (strong boson
coupling, POL, Method 1) with its integral equations \ Eq.~(\ref{peqn}), (\ref{pdaggereqn}),
yields  matrices in $z$-space of the form 
\begin{equation}\label{POLb}
   \hat{D}_{z}^{\rm POL}=iT_c
\left(\begin{matrix}
-1 & \frac{\hat{C}^*_{-\varepsilon}(z)}{\hat{C}_\varepsilon(z)}\\
1 &  -\frac{\hat{C}_{-\varepsilon}(z)}{\hat{C}^*_\varepsilon(z)}
\end{matrix}\right),\quad
\hat{\Sigma}_{z}^{\rm POL}=
\left(
 \begin{matrix}
z-1/C_\varepsilon(z)-\Gamma_R/2 &0\\0&
z-1/C_\varepsilon^{*}(z)-\Gamma_R/2
\end{matrix}\right),
\end{equation}
where
\begin{eqnarray}\label{C_eps_def}
  C_\varepsilon(z)&\equiv&\int_{0}^{\infty}dte^{-zt}e^{i\varepsilon t} C(t),\quad
C^{*}_\varepsilon(z)\equiv\int_{0}^{\infty}dte^{-zt}e^{-i\varepsilon t} C^{*}(t).
\end{eqnarray}
In contrast to the PER solution, where
$M(\tau)=M=z\hat{M}(z)$ is time-independent,
$M^{\rm POL}(\tau)$ is time-dependent and
$z\hat{M}(z)$ depends on $z$ in the POL approach.

\subsubsection{Stationary Current}
In the Master equation approach, 
the expectation value of the electron current through the double dot is obtained 
in a fairly straightforward  manner. One has to consider the average charge 
flowing through one of the three intersections, i.e.,  left lead/left dot, left dot/right dot, 
and right dot/right lead. This gives rise to the three corresponding electron currents
$I_L(t)$, $I_R(t)$, and the inter-dot current $I_{LR}(t)$. 
From the equations of motion, \ Eq.~(\ref{eom3new}), one recognizes that 
the temporal change of the occupancies $\langle n_{L/R}\rangle_t$ is due
to the sum of an `inter-dot' current $\propto T_c$ and a `lead-tunneling' part.
Specifically, the current from left to right through the left (right) tunnel barrier is \cite{BAP04}
\begin{eqnarray}\label{ILandR}
  I_L(t) &=& -e \Gamma_L\langle n_0\rangle_t = 
-e\Gamma_L\left[1-\langle n_L\rangle_{t} - \langle n_R\rangle_{t}\right],\quad
I_R(t) = -e \Gamma_R \langle n_R\rangle_{t},
\end{eqnarray}
and the inter-dot current is
\begin{eqnarray}\label{ILR}
  I_{LR}(t) &=& -ieT_c\left\{
\langle p \rangle_{t}-\langle p^{\dagger}\rangle_{t} \right\}
= -e \frac{\partial}{\partial t}\langle n_R\rangle_t +I_R(t)
= e \frac{\partial}{\partial t}\langle n_L\rangle_t  +I_L(t).
\end{eqnarray}
In the stationary case for times $t\to \infty$,
all the three currents are the same, $I_{LR}=I_R=I_L\equiv I$ and can be readily obtained from
the $1/z$ coefficient in the Laurent expansion of the Laplace transform $\hat{n}_R(z)$ of $\langle \hat{n}_R \rangle_t$ around $z=0$,
\begin{eqnarray}\label{currentstat}
  \langle \hat{I}\rangle_{t\to \infty} &=& 
  -e \lim_{z\to 0}\frac{\Gamma_R\Gamma_Lg_+(z)}
{\left[z+\Gamma_R+g_-(z)\right](z+\Gamma_L)+(z+\Gamma_R+\Gamma_L)g_+(z)}\\
 g_{+[-]}(z)&=&\pm iT_c({\bf e}_1-{\bf e}_2)
 \left[z-\hat{\Sigma}_z\right]^{-1}\hat{D}_z {\bf e}_{1[2]}.\label{gpmdefinition}
\end{eqnarray}
The explicit evaluation of the two-by-two blocks $\hat{D}_z$ and $\hat{\Sigma}_z$, 
cf. Eq. (\ref{block},\ref{PERb},\ref{POLb}), leads to 
\begin{eqnarray}\label{gpm1}
g^{\rm PER}_{\pm}(z)&\equiv&
  2T_c\frac{T_c(\gamma_p+ \Gamma_R/2+z)-\varepsilon\gamma_{\pm}}{(\gamma_p+ \Gamma_R/2+z)^2+\varepsilon
  ^2}\\
g^{\rm POL}_{+[-]}(z) &\equiv& T_c^2 \left[
\frac{C^{[*]}_{[-]\varepsilon}(z)}{1+\frac{\Gamma_R}{2}C_\varepsilon(z)}+
(C \leftrightarrow C^*)
\right].
\end{eqnarray}
In the expression for the current, \ Eq.~(\ref{currentstat}), 
the two `propagators' $g_\pm$ are summed up to infinite order in the inter-dot coupling $T_c$. 
For vanishing boson coupling, one has $g_+=g_-$, and the stationary current reduces to the 
Stoof-Nazarov expression \cite{SN96},
\begin{eqnarray}\label{current_SN96}
  \langle \hat{I}\rangle^{\rm SN}_{t\to \infty}=-e \frac{T_c^2\Gamma_R}
{\Gamma_R^2/4+\varepsilon^2+T_c^2(2+\Gamma_R/\Gamma_L)}.
\end{eqnarray}
The general expression  for the stationary current through double dots in POL \cite{BK99} reads
\footnote{This is the correct expression consistent with the definition \ Eq.~(\ref{tunnel_rates_def}){}
for the tunnel rates $\Gamma_{R/L}$, whereas the original version \cite{BK99} contained additional factors of
$2$ in $\Gamma_{R/L}$.}
\begin{eqnarray}\label{IPOL_final}
\langle I \rangle_{t\to
  \infty}&=&e T_c^2\frac{2\mbox{\rm Re } (C_{\varepsilon})+\Gamma_R|C_{\varepsilon}|^2}
  {|1+\Gamma_RC_{\varepsilon}/2|^2+2T_c^2B_{\varepsilon}}\\
B_{\varepsilon}&\equiv&
\mbox{\rm Re }\left\{(1+\Gamma_RC_{\varepsilon}/2)\left[
 \frac{C_{-\varepsilon}}{\Gamma_R}+\frac{C_{\varepsilon}^*}{\Gamma_L}\left(1+\frac{\Gamma_L}{\Gamma_R}\right)\right]\right\},
\quad C_\varepsilon \equiv \lim_{\delta\to 0}C(z=\varepsilon+i\delta).
\end{eqnarray}

\subsubsection{Boson Spectral Density $J(\omega)$}\label{section_spectraldensity}
The boson spectral density $J(\omega)=\sum_{\bf Q} \left| \alpha^L_{\bf Q}-\alpha^R_{\bf Q} \right|^2
\delta(\omega-\omega_{\bf Q})$, \ Eq.~(\ref{Jomega}){}, is the key quantity
entering into the theoretical description of dissipation within the framework of the 
spin-boson model, \ Eq.~(\ref{modelhamiltonian}). $J(\omega)$ determines the
inelastic rates $\gamma_p$ and $\gamma_{\pm}$, \ Eq.~(\ref{gammarates}) in the PER approach, and
the boson correlation function $C(t)$ via \ Eq.~(\ref{Ct}) in the POL approach. 

Models for $J(\omega)$ can be broadly divided into (A) phenomenological parametrizations, and (B)
microscopic models for specific forms of the electron-boson interaction (e.g., coupling to bulk phonons
or surface acoustic piezo-electric waves). 

\noindent (A) `Spin-Boson model parametrization' \cite{Weiss} in the exponentially damped power-law form
\begin{eqnarray}\label{Jomegageneric}
  J(\omega) = 2\alpha \omega_{\rm ph}^{1-s} \omega^s e^{-\omega/\omega_c},
\end{eqnarray}
where $0\le s\le 1$ corresponds to the sub-Ohmic, $s=1$ to the Ohmic, and $s>1$ to the super-Ohmic
case. The parameter $\omega_c$ is a high-frequency cut-off, and $\omega_{\rm ph}$ is a reference frequency
introduced in order to make the coupling parameter $\alpha$ dimensionless. The advantage of the generic
form \ Eq.~(\ref{Jomegageneric}){} is the vast amount of results in the quantum dissipation literature 
referring to it. Furthermore, this parametrization allows for an exact analytical expression of the boson correlation function $C(t)=\exp[-Q(t)]$, Eq.~(\ref{Ct}), for arbitrary temperatures $T=1/\beta$. Weiss \cite{Weiss} gives the explicit form of $Q(z)$ for complex times $z$, 
\begin{eqnarray}\label{Qz}
  Q(z) &=& 2\alpha \Gamma(s-1) \left(\frac{\omega_c}{\omega_{\rm ph}} \right)^{s-1} \Big\{
\left(1-(1+i\omega_c z)^{1-s}\right) + 2 (\beta\omega_c)^{1-s} \zeta\left(s-1,1+\frac{1}{\beta\omega_c}\right)\nonumber\\
&-& (\beta\omega_c)^{1-s} \left[ \zeta\left(s-1,1+\frac{1}{\beta\omega_c}+i\frac{z}{\beta}\right)
 + \zeta\left(s-1,1+\frac{1}{\beta\omega_c}-i\frac{z}{\beta}\right)\right] \Big\},
\end{eqnarray}
where $\zeta(z,q)$ is Riemann's generalized Zeta-function and $\Gamma(z)$ Euler's Gamma-function.

\noindent (B) Microscopic models naturally are more restricted towards specific situations but can yield
interesting insights into the dissipation mechanisms in the respective systems. 
%
Coupling of bulk acoustic phonons to the electron charge 
density in double quantum dots was assumed in \cite{BK99}, with the matrix elements
$\alpha_{\bf Q}^i=\lambda_{\bf Q}\int d^3 {\bf x} e^{i{\bf Q x}} \rho_i({\bf x})$ expressed in terms
of the local electron densities $\rho_i({\bf x}), i=L,R$ in the left and right dot. Assuming the 
electron density in both (isolated) dots described by the same
profile $\rho_{\rm dot}({\bf x})$ around the dot centers ${\bf x}_i$, one finds that the two coupling constants just differ by a phase factor,
\begin{eqnarray}\label{phasefactor}
  \alpha_{\bf Q}^{R}= \alpha_{\bf Q}^{L}e^{i{\bf Q d}},\quad {\bf{d}}={\bf{x}}_R-{\bf{x}}_L.
\end{eqnarray}
With ${\bf Q}=({\bf q}, q_z)$ and the vector ${\bf d}$  in the $x$-$y$ plane of lateral dots, one has
\begin{eqnarray}
  J(\omega) = \sum_{\bf{Q}} |\lambda_{\bf{Q}}|^2 
\hat{\rho}_{\rm dot}({\bf q}, q_z)|1-e^{i{\bf qd}}|^2
\delta(\omega-\omega_{\bf{Q}}).
\end{eqnarray}
The interference term $ |1-e^{i{\bf qd}}|^2$ is due to the lateral `double-slit' structure of the double
dot geometry interacting with three-dimensional acoustic waves; whether or not this interference
is washed out in $J(\omega)$ depends on the electron density profile and the 
details of the electron-phonon interaction. Analytical limits for $J(\omega)$ can be obtained
in the limit of infinitely sharp density profiles, i.e. $ \hat{\rho}_{\rm dot}({\bf q}, q_z)=1$: using
matrix elements
for piezoelectric and deformation potential phonons, one obtains \cite{BraHabil}
\begin{eqnarray}\label{Jmicro}
J_{\rm piezo}(\omega) 
&=& 2\alpha_{\rm piezo} \omega f\left(\frac{d \omega}{c}\right),\quad 
\quad J_{\rm def}(\omega) = \frac{2\alpha_{\rm def}}{\omega_{\rm ph}^2}\omega^3
f\left(\frac{d \omega}{c}\right),\quad 
f(x) \equiv \left(1-\frac{\sin x }{x} \right)
\\
2\alpha_{\rm piezo}&\equiv& \frac{P}{2\pi^2\hbar c^3 \rho_M},\quad 
\frac{2\alpha_{\rm def}}{\omega_{\rm ph}^2} = \frac{1}{\pi^2c^3}\frac{\Xi^2}{2\rho_Mc^2\hbar}.
\end{eqnarray}
For the piezoelectric interaction, the contributions from longitudinal and transversal phonons 
with dispersion $\omega_Q\equiv c|{\bf{Q}}|$ and  speed of sound $c=c_l,c_t$, respectively, were added here.  
Bruus, Flensberg and Smith 
\cite{BFS93} used a simplified angular average  $P=(eh_{14})^2({12}/{35}+{c_l 16}/{c_t 35} )$ in quantum wires
with the piezoelectric coupling denoted as $eh_{14}$. 
Furthermore, $\rho_M$ denotes the mass density of the crystal with volume $V$,
and $\Xi$ is the deformation potential. The contribution from bulk deformation potential phonons 
turns out to be small as compared with piezoelectric phonons where $2\alpha_{\rm piezo}\approx 0.05$.

Further microscopic models of the electron-phonon interaction in double-well potentials were done by Fedichkin and Fedorov \cite{fedichkin:032311} in their calculation of error rates in charge qubits. Furthermore, in a series of papers \cite{golovach:016601,khaetskii:195329,khaetskii:186802,khaetskii:125316,khaetskii:12639} Khaetskii and co-workers performed microscopic calculations for {\em spin} relaxation in quantum dots due to the interaction with phonons.

The forms \ Eq.~(\ref{Jmicro}){} for $J(\omega)$ represent examples of 
{\em structured} bosonic baths, where at least one additional energy scale
(in this case $\hbar c/d$, where $d$ is the distance between two dots and $c$ the speed of sound) enters
and leads to deviations from the exponentially damped power-law form  \ Eq.~(\ref{Jomegageneric}){}.
Note that the microscopic forms \ Eq.~(\ref{Jmicro}){} eventually also have a cut-off $\omega_c$ 
due to the finite extension of the electron density in the dots. In \cite{BV03} it was argued that the assumption of sharply localized positions between which the additional electron tunnels should be  justified by the strong intra-dot electron-electron repulsion. 
For $c/d \ll \omega \ll \omega_c$, the generic power-laws \ Eq.~(\ref{Jomegageneric}){}
match the piezo-electric case with $s=1$ (Ohmic) and the deformation potential case with $s=3$ (super-Ohmic). 
In the low-frequency limit, however, due to $f(x) = (1/6)x^2 + O(x^4)$ these exponents change
to $s=3$ and $s=5$, respectively.

A further phenomenological example for a boson spectral density for a
structured environment is the  Breit-Wigner form for a damped oscillator mode $\Omega$,
\begin{eqnarray}
  J(\omega) = \alpha\omega \frac{\Omega^4}{(\omega^2-\Omega^2)^2+4\omega^2\Gamma^2},
\end{eqnarray}
which was discussed recently by Thorwart, Paladino and Grifoni \cite{TPG04}, and
by Wilhelm, Kleff, and  von Delft \cite{WKD04}, who gave a comparison of 
the perturbative (Bloch-Redfield) and polaron (NIBA) method for the spin-boson model.

\subsubsection{$P(E)$-theory}\label{section_PEtheory}
The stationary current through double dots in POL, \ Eq.~(\ref{IPOL_final}){},
can be expanded to lowest order in the tunnel coupling $T_c$ and the rate $\Gamma_R$,
\begin{eqnarray}\label{IPE}
  \langle {I}\rangle_{t\to \infty}&\approx& 2 \pi e T_c^2 P(\varepsilon),\quad
 P(\varepsilon) \equiv \frac{1}{2\pi}\int_{-\infty}^{\infty}dt e^{i\varepsilon t}C(t).
\end{eqnarray}
The real quantity $P(\varepsilon)=\frac{1}{\pi}\mbox{\rm Re }C_{\varepsilon}$ is the 
probability density for inelastic tunneling from the left dot to the right dot 
with energy transfer $\varepsilon$ and plays the central role in the so-called $P(E)$-theory of
single electron tunneling in the presence of an electromagnetic environment 
\cite{Devetal90,Giretal90,IN91}. 

The function $P(\varepsilon)$ is normalized and obeys the detailed balance symmetry, 
\begin{eqnarray}\label{detailed_balance}
P(-\varepsilon)=\exp(-\varepsilon/k_BT)P(\varepsilon),   
\end{eqnarray}
but has to be derived for any specific realization of the dissipative environment. 
In the case of no phonon coupling, one has only elastic transitions and $P(\varepsilon) = \delta(\varepsilon)$. 
At zero temperature ($\beta\to\infty$), a simple perturbative expression for $P(\varepsilon)$ for arbitrary $J(\omega)$ can be found
by expanding $C(t)$, \ Eq.~(\ref{Ct}){}, to second order in the boson coupling,
$C(t) =1 -\int_0^{\infty}\frac{J(\omega)}{\omega^2}\left[1-e^{-i\omega t}\right]+ O(J^2)$ whence
$P(\varepsilon>0)=J(\varepsilon)/\varepsilon^2$. The resulting expression for the
inelastic current,
\begin{eqnarray}\label{Iinelastic}
 I_{\rm in}(\varepsilon) = -e 2 \pi  T_c^2 J(\varepsilon)/\varepsilon^2,  
\end{eqnarray}
is valid at $\varepsilon \gg \Gamma_R$ and is consistent with
an earlier result by Glazman and Matveev for inelastic tunneling through amorphous thin films
via pairs of impurities \cite{GM88}.

Aguado and Kouwenhoven \cite{AK00} have suggested to use tunable double quantum dots 
as {\em detectors of quantum noise} via  \ Eq.~(\ref{IPE}), where
the function $P(\varepsilon)$ in principle can be directly inferred from measurement of the 
current $I$ as a function of $\varepsilon\equiv\varepsilon_L=\varepsilon_R$. Deblock and coworkers \cite{Debetal03} have used very similar ideas to analyze their experiments on frequency dependent noise in a superconducting Josephson junction and a Cooper pair box, cf. section \ref{section_Deblock}.

Again, since
off-diagonal couplings $\gamma_{\bf Q}$ in the model, 
\ Eq.~(\ref{newHdp}){}, have not been taken into account, the information gained
on the environment by this method might not be complete. On the other hand, 
the $P(\varepsilon)$/spin-boson description takes into account arbitrary bosonic coupling strengths.
Furthermore, the underlying correlation function $C(t)$ can describe both equilibrium
and non-equilibrium situations. An example of the latter discussed in
\cite{AK00} is (shot) noise, i.e. fluctuations in the tunnel current 
through a quantum point contact that is capacitively coupled to a double quantum dot. 

For Ohmic dissipation $s=1$, at zero temperature absorption of energy from the environment is not possible and $P(\varepsilon)$ reads
\begin{eqnarray}\label{PET0}
P(\varepsilon) = \frac{\varepsilon^{2\alpha-1}}{\omega_c^{2\alpha}  \Gamma(2\alpha)} e^{-\varepsilon/\omega_c} \theta(\varepsilon),
\end{eqnarray}  
which is a Gamma distribution with parameter $g=2\alpha$. Another analytical solution for $P(\varepsilon)$ at finite temperatures is obtained at $\alpha=1/2$ \cite{VB04}, where the residue theorem yields
\begin{eqnarray}
P_{\alpha=\frac{1}{2}}(\varepsilon>0) &=& \frac{e^{-\varepsilon/\omega_c}}
{\omega_c \; \Gamma(1\!+\!1/\beta \omega_c)^2} \sum_{n=0}^{\infty} \, \frac{(-1)^n}{n!} \; 
\Gamma\Big(n+1+ \frac{2}{\beta\omega_c}\Big)\;
e^{-n \beta \varepsilon}, 
\end{eqnarray}
which at low temperatures, $k_BT\!=\!1/\beta\!\ll\!\omega_c$ can be approximated by a geometric series,
\begin{equation}\label{PEapprox12}
P_{\alpha=\frac{1}{2}}(\varepsilon) \approx \frac{e^{-\varepsilon/\omega_c}}
     {\omega_c \; \Gamma(1\!+\!1/\beta \omega_c)^2 \, (1+e^{-\beta \varepsilon})},
\end{equation}
with $P_{\alpha=\frac{1}{2}}(\varepsilon)$ following from \ Eq.~(\ref{detailed_balance}), and 
\begin{equation}
\label{eq_fourier-zero}
P_{\alpha=\frac{1}{2}}(\varepsilon=0) =  
\frac{\Gamma(1+2/\beta \omega_c)}
{2\omega_c  4^{1/\beta \omega_c}  \Gamma(1+1/\beta \omega_c)^2}.
\end{equation}

\subsubsection{Boson Shake-Up and Relation to X-Ray Singularity Problem}

Bascones, Herrero, Guinea and Sch\"on \cite{Basetal00} pointed out that electron tunneling through dots leads to excitations of electron-hole pairs in the adjacent electron reservoirs. These bosonic excitations 
possess an Ohmic spectral function $J(\omega)$ and for small $\omega$ therefore give the same  exponent $s=1$ as the piezoelectric spectral function, \ Eq.~(\ref{Jmicro}). Note, however, that this is only true  for the bulk case where the structure function $f\equiv 1$.


The appearance of a power-law singularity  in the inelastic tunneling probability $P(\epsilon)$, Eq. (\ref{PET0}), is well-known from the so-called X-ray singularity  problem. The latter belongs, together with the Kondo effect and the non Fermi-liquid effects in one-dimensional interacting electron systems (Tomonaga-Luttinger liquid) \cite{Mahan,Tom50,Voi95,Hal81a,Weh219}, to a class of problems in theoretical Solid State Physics that are essentially non-perturbative \cite{Doniach}. That is, simple perturbation theory in interaction parameters leads to logarithmic singularities which transform into power laws for Greens functions or other correlation functions after higher order re-summations, renormalization group methods, or approximation by exactly  solvable models.

X-ray transitions in metals are due to excitations of electrons from the metal ion core levels (e.g., the p-shells of sodium, magnesium, potassium) to the conduction band (absorption of photons), or the corresponding emission process with a transition of an electron from the conduction band to an empty ion core level, i.e. a recombination with an {\em core hole}. Energy conservation in a simple one-electron picture requires that for absorption there is a {\em threshold} energy (edge) $\hbar \omega_T=E_F+|E_c|$ for such processes, where $E_F$ is the Fermi energy and 
$E_c$ the core level energy, counted from the conduction band edge.
Following Mahan \cite{Mahan}, the core hole \index{core hole} interacts with the conduction band electron gas, which is described in an effective Wannier exciton picture by a Hamiltonian \cite{Mahan}
\begin{eqnarray}
  \label{eq:xrayhameffect}
  H = E_cd^{\dagger}d+\sum_{{\bf k}\sigma}\varepsilon_{\bf k}c^{\dagger}_{{\bf k}\sigma}c_{{\bf k}\sigma}
+\frac{1}{L^d}\sum_{{\bf kk'}\sigma}V_{\bf kk'}c^{\dagger}_{{\bf k}\sigma}c_{{\bf k'}\sigma}d^{\dagger}d.
\end{eqnarray}
Here, $d^{\dagger}$ denotes the creation operator of the core hole and $c_{\bf k}^{\dagger}$ the creation operator of a conduction band electron with Bloch wave vector ${\bf k}$ and spin $\sigma$, leading to an {\em algebraic singularity} in the core hole spectral function
\begin{eqnarray}
  \label{eq:xcorespec}
  A_h(\omega) &=& 2\Re e \int_0^{\infty}dt e^{i\omega t}\langle d(t)d^{\dagger}\rangle=
              \theta(\Omega)\frac{2\pi}{\Gamma(g)}\frac{e^{-\Omega}}
{\Omega^{1-g}},\quad \Omega = (\omega-\bar{\omega}_T)/\xi_0,
\end{eqnarray}
where  $\bar{\omega}_T$ is the (renormalized) photo-emission threshold energy,
and $\xi_0$ is a cutoff of the order of the Fermi energy. Here,
the dimensionless parameter $g$ for a three dimensional situation 
and for an interaction potential with $V_{\bf kk'}=V({\bf k-k'})$
is defined as \cite{Mahan}
\begin{eqnarray}
  \label{eq:gcoredef}
  g &=&\frac{m^2}{2\pi^2}\sum_{{\bf |q|}<2k_F}\frac{V(q)^2}{q},
\end{eqnarray}
where $m$ is the conduction band electron mass. The core hole spectral function $A_h(\omega)$ is thus strongly modified by the interaction with the electron gas: the sharp delta peak for the case of no interactions
becomes a power-law curve. The corresponding absorption step is obtained by integration of $A_h(\omega)$ \cite{Mahan}, it vanishes for non-zero $g$ when  approaching from above $\Omega\to 0^+$. This vanishing  of the absorption is called {\em orthogonality catastrophe}: the matrix elements for X-ray induced transitions in metals must depend on the overlap of two wave functions, i.e. the $N$-particle wave functions $|i\rangle$ and $|f\rangle$ before and after the appearance of the core hole, respectively. Here, $N$ is the number of electrons in the conduction band. A partial wave scattering analysis then shows that $|f\rangle$  (in the simplest case of s-wave scattering) can be considered as a Slater determinant composed of spherical waves $\propto \sin (kr+\delta)/kr$. The overlap of the two $N$-particle wave functions turns out to be  $  \langle f | i\rangle = N^{-\frac{1}{2}\alpha},\quad \alpha \equiv 2\frac{\delta^2}{\pi^2}.$ For large $N$, this overlap becomes very small though still finite for macroscopic numbers like $N\approx 10^{23}$ and $\alpha \approx 0.1$ \cite{Mahan}. The `catastrophe' of this effect consists in the fact that although all overlaps of initial and final {\em single particle} scattering waves are finite, the resulting {\em many-body} wave function overlap becomes arbitrarily small for large $N$. 
The fully dynamical theory takes into account the dynamical process of the excitations in the Fermion system that are induced by the sudden appearance of the core hole after absorption of an X-ray photon. In fact, these excitations are particle-hole pairs in the conduction band which can be regarded as bosons. For a spherically symmetric case, the X-ray problem can be solved exactly by a mapping to the Tomonaga model of interacting bosons in one dimension \cite{Mahan,SS69}. 

The analogy of inelastic tunneling through double quantum dots can be made by considering an additional electron initially in the left state $|L\rangle$ of the isolated dot. The operator $p^{\dagger} =|R\rangle\langle L|$ acts as a creation operator for an electron in the right dot or, alternatively and as there is only one additional electron in the double dot, $p^{\dagger}$ can be regarded as a creation operator for a {\em hole} in the left dot. The retarded hole Greens function 
\begin{equation}
  \label{eq:coregreen}
  G_p(t)=-i\theta(t)\langle p(t)p^{\dagger}\rangle=-i\theta(t)\langle pp^{\dagger}\rangle e^{i\varepsilon t}\langle X_tX^{\dagger}\rangle_0=-i\theta(t) e^{i\varepsilon t}C(t),
\end{equation}
is calculated in absence of tunneling, with the electron in the left dot at time $t\ge 0$ having excited its phonon cloud that already time-evolves according to the correlation function $C(t)$ for the phase factors $X$ stemming from the polaron transformation. The correlations in time can be translated into a frequency spectrum via the hole spectral function\cite{Mahan},
\begin{eqnarray}
  \label{eq:holespec}
  A_p(\omega) &=& -2\Im m G_p(\omega) = 2i \Im m \int_0^{\infty}dt e^{i\omega t}
e^{i\varepsilon t} C(t)=
2\pi P(\varepsilon+\omega),
\end{eqnarray}
using the detailed balance relation $C(t)= C^*(-t)$ and the definition of the inelastic tunneling probability, Eq. (\ref{IPE}). Comparison of  Eq.(\ref{PET0}), Eq.(\ref{eq:holespec}), and Eq.(\ref{eq:xcorespec}) shows that the spectral functions have identical form if one identifies the cut-offs $\xi_0=\omega_c$ and $\bar{\omega}_T$ with $-\varepsilon$, the only difference being the definition of the dimensionless coupling constant $g$. 

As pointed out by Mahan \cite{Mahan}, the power law behavior of Eq.(\ref{eq:holespec}) and 
Eq.(\ref{PET0}) is due to the logarithmic singular behavior of the function $Q(t)$  in $C(t) = \exp (-Q(t))$, Eq.(\ref{Ct}), which in turn results from an {\em infrared divergence} of the coupling function $J(\omega)/\omega^2$ for small $\omega$. This infrared divergence physically correspond to the generation of an infinite number of electron-hole pair excitations in the metal electron gas by the interaction with the core hole in the X-ray problem. In semiconductors, the (bulk) piezoelectric phonon coupling leads to the same kind of infrared divergence.

Again following Mahan \cite{Mahan4.3}, an alternative physical picture for the inelastic tunneling is obtained by considering the tunneling process from the point of view of the phonon and not from the electron (hole) system \cite{dualityusefulness}: a sudden tunnel event in which an electron tunnels from the left to the right dot appears as an additional energy term for the phonons, 
\begin{eqnarray}
  \label{eq:hamiltonsuddenly}
 \delta H_{\alpha\beta}&=&\sum_{\bf{Q}}\left(
  \alpha_{\bf{Q}}^L-\alpha_{\bf{Q}}^R\right)
  \left(a_{-\bf{Q}}+a^{\dagger}_{\bf{Q}}\right)
\end{eqnarray}
which is exactly the difference of the 
coupling energy before and after the tunnel event. This additional potential is linear in the phonon displacements $ a_{-\bf{Q}}+a^{\dagger}_{\bf{Q}}$ and `shakes up' the phonon systems in form of a dynamical displacement as expressed by the temporal correlation function $C(t) = \langle X_tX^{\dagger}\rangle$ of the unitary 
displacement operators, 
\begin{eqnarray}\label{transformedagain}
X&=&\prod_{\bf{Q}}D_{\bf{Q}}\left(\frac{\alpha_{\bf{Q}}-\beta_{\bf{Q}}}{\omega_{\bf{Q}}}\right),\quad
D_{\bf{Q}}(z)\equiv e^{za^{\dagger}_{\bf{Q}}-z^*a_{\bf{Q}}^{\phantom{\dagger}}}.
\end{eqnarray}

\subsubsection{Interference Oscillations in Current}\label{section_currentosc}

\begin{figure}[t]
\includegraphics[width=0.5\textwidth]{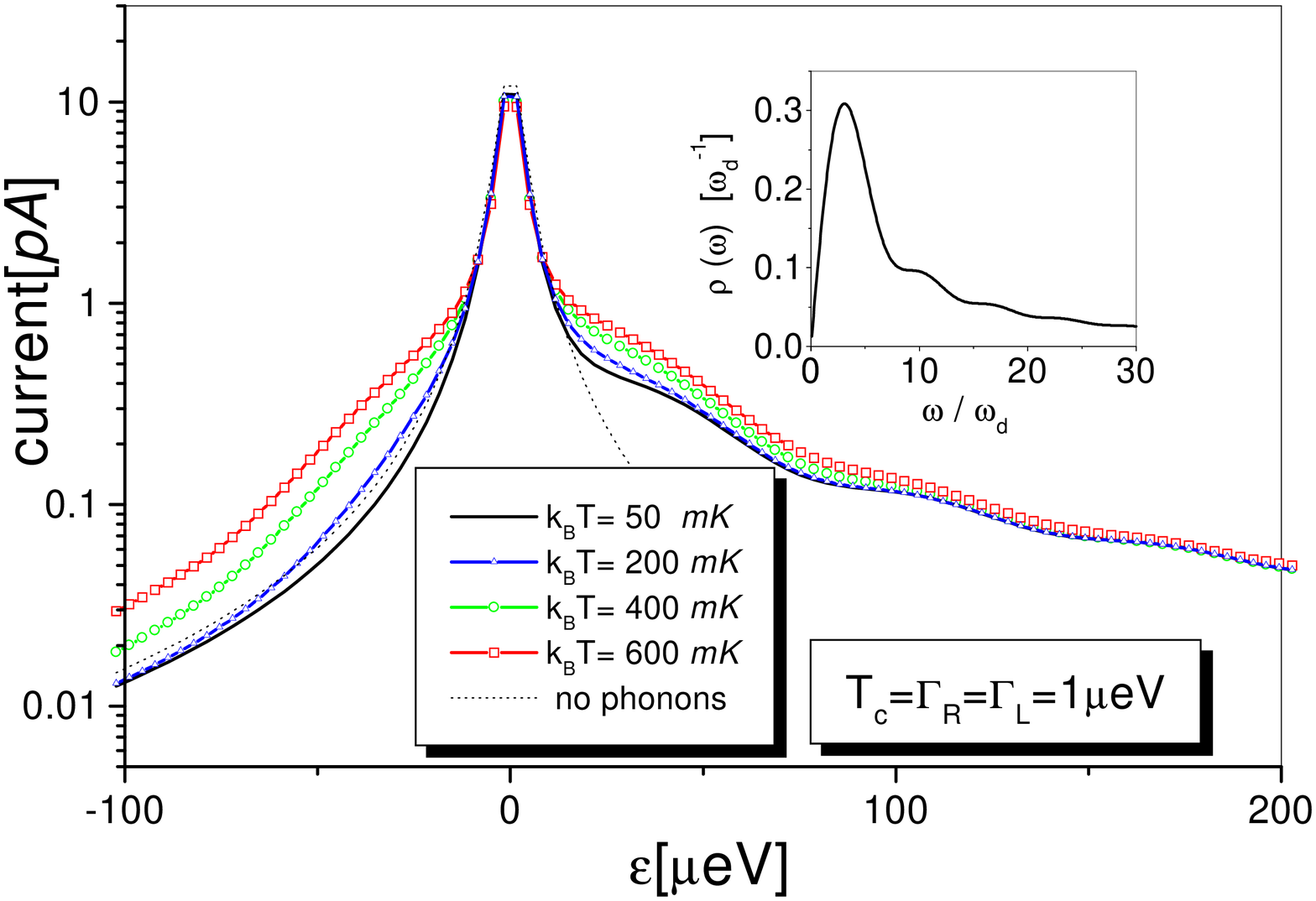}
\includegraphics[width=0.5\textwidth]{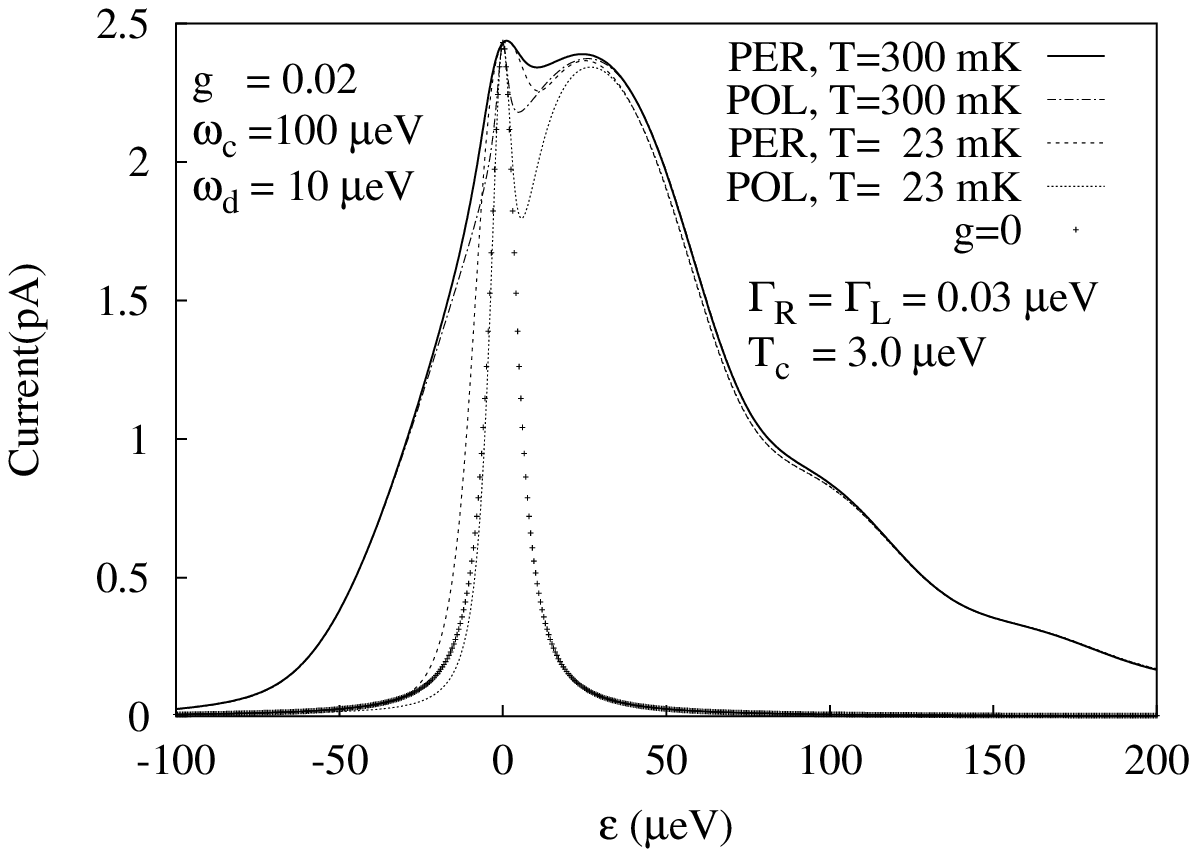}
\caption[]{\label{current_stat}Stationary tunnel current through double quantum dot, Eq.(\ref{currentstat}), as a function of the energy difference $\varepsilon$ between left and right dot ground state energies. {\bf Left:}  POL approach, \ Eq.~(\ref{IPOL_final}), dimensionless electron-phonon coupling parameter $\alpha=2g=0.025$.
Inset: $\rho(\omega)\equiv f(x)/x, x=\omega/\omega_d$ with $\omega_d\equiv c/d$, Eq.(\ref{Jmicro}). From \cite{BK99}. {\bf Right:} Comparison between POL and PER approach, from \cite{BV03}.}
\end{figure} 

The function $f(x)$ in the spectral density $J(\omega)$, \ Eq.~(\ref{Jmicro}), describes the interference oscillations in the electron-phonon matrix elements, cf. \ Eq.~(\ref{phasefactor}){}. These were   directly compared \cite{BK99} via \ Eq.~(\ref{Iinelastic}) with the oscillations in the current profile on the emission side at low temperature in the experiment by Fujisawa and co-workers \cite{Fujetal98}. Using parameters $d=200\cdot10^{-9}m$ and $c=5000 m/s$, the energy scale  $\hbar\omega_d\equiv \hbar c/d=
16.5\mu eV$ is in fact the scale on which the oscillations in \cite{Fujetal98} were observed. The corresponding stationary current was obtained from \ Eq.~(\ref{IPOL_final}){} by numerical evaluation of $C_\varepsilon$, \ Eq.~(\ref{C_eps_def}){}, with $C(t)$ split into  a zero-temperature and a finite temperature contribution (Appendix \ref{appendix_a}), cf. Fig. (\ref{current_stat}). At low temperatures, the broad oscillatory shoulder on the emission side $\varepsilon>0$  reflects the structure of the real part of $C_{\varepsilon}$. At higher temperatures, on the absorption side the current increases to larger values faster than on the emission side where the oscillations start to be smeared out. For $\varepsilon<0$ and larger temperature, a new  shoulder-like structure appears on the absorption side, a feature  similar to the one observed in the experiment \cite{Fujetal98}. The theoretical result \cite{BK99} for the inelastic current was based on the simple assumption of bulk  piezo-acoustic phonons and  was still at least a factor two smaller than the experimental one.  This might indicate that other phonon modes (such as surface acoustic phonons), or in fact higher order tunneling processes to and from the leads (co-tunneling) are important.

Another interesting observation was the scaling of the current a function of the ratio between temperature and energy $k_BT/|\varepsilon|$, re-confirming the equilibrium Bose-Einstein distribution for the phonon system. In analogy to the Einstein relations for emission and absorption, one defines the spontaneous emission rate $A(\varepsilon>0)\equiv[I(\varepsilon>0,T_0)-I_{el}(\varepsilon>0)]/e$, where $I_{el}(\varepsilon)$ is the elastic part of the current, i.e. the current for vanishing electron-phonon coupling $\alpha=0$. One introduces  similar definitions for the relative emission  $N$ and absorption $N^+$,
\begin{eqnarray}\label{abem}
N(\varepsilon>0,T)&\equiv&\left[I(\varepsilon,T)-I_{el}(\varepsilon)\right]/A(\varepsilon),\quad
N^+(\varepsilon<0,T)\equiv \left[I(\varepsilon,T)-I(\varepsilon,T_0)\right]/A(|\varepsilon|),
\end{eqnarray}
where  $T_0$ is the reference temperature. The numerical data for the stationary current  scaled well  to the Bose distribution function $n(x)=1/(e^x-1)$, i.e. $N(\varepsilon,T)=n(|\varepsilon|/k_BT)$ for absorption $\varepsilon<0$ and to $N^+(\varepsilon,T)=1+n(\varepsilon/k_BT)$ for emission $\varepsilon>0$
over an energy window $220 \mu eV>|\varepsilon|>20 \mu eV$ with a choice of $T_0 = 10$ mK. As in the experiment \cite{Fujetal98}, the analysis in terms of Einstein coefficients worked remarkably well \cite{BraHabil}.

A comparison between the perturbative (PER) and polaron transformation (POL) result for the stationary current,  was performed in \cite{BV03}. In both approaches, the currents \ Eq.~(\ref{currentstat}) are infinite sums of contributions from the two expressions $g_{\pm}(z)$, \ Eq.~(\ref{gpm1}), which were explicitly calculated. As PER works in the correct eigenstate base of the hybridized system (level splitting $\Delta$), whereas the energy scale $\varepsilon$ in POL is that of the two isolated dots (tunnel coupling $T_c=0$), one faces 
the general dilemma of two-level-boson Hamiltonians:  one either is in the correct base of the hybridized two-level system and perturbative in the boson coupling $\alpha$ (PER), or one starts from the `shifted oscillator' polaron picture that becomes correct only for $T_c=0$ (POL). The polaron (NIBA) approach does not coincide with standard damping theory \cite{SW90} because it does not incorporate the  square-root hybridization form of $\Delta= \sqrt{\varepsilon^2+4T_c^2}$ which is non-perturbative in $T_c$. However, it was argued in \cite{BV03} that for large $|\varepsilon| \gg T_c$, $\Delta \to |\varepsilon|$ whence POL and PER should coincide again and the polaron approach to work well even down to very low temperatures and small coupling constants $\alpha$. Fortunately, in the spontaneous emission regime of large positive $\varepsilon$ the agreement turned out to be very good indeed, cf. Fig. (\ref{current_stat}).

\subsubsection{Other Transport Theories for Coupled Quantum Dots, Co-Tunneling and Kondo Regime}

The amount of theoretical literature on transport through coupled quantum dots is huge and would  provide material for a detailed Review Article of its own, this being  yet another indication of the great interest researchers have taken in this topic. In the following, we therefore give only a relatively compact overview over parts of this field, which is still very much growing.

Inelastic tunneling through coupled impurities in disordered conductors was treated by Glazman and Matveev \cite{GM88} in a seminal work in 1988, which closely followed after the work of Glazman and Shekhter \cite{GS88} on resonant tunneling through an impurity level with arbitrary strong electron-phonon (polaron) coupling.  Raikh and Asenov \cite{RA92} later combined Hubbard and Coulomb correlations in their treatment of the Coulomb blockade for transport through coupled impurity levels and found step-like structures in the current voltage characteristics. 
References to earlier combined treatments of both the Coulomb blockade and the coherent coupling between coupled  dots can be found in the 1994 paper by Klimeck, Chen, and Datta \cite{KCD94}, who presented  a calculation of the linear conductance. 
Their prediction for a splitting of the conductance peaks both due to Coulomb interactions and the tunnel coupling was confirmed by exact digitalizations by Chen and coworkers \cite{Cheetal94}, and  by Niu, Liu, and Lin in a calculation with non-equilibrium Green's functions \cite{NLL95}, a technique also used by Zang, Birman, Bishop and Wang \cite{ZBBW97} in their theory of non-equilibrium transport and population inversion in double dots.
 In 1996, Pals and MacKinnon \cite{PM96} also used Green's functions and calculated the current through coherently coupled two-dot systems, and  Matveev, Glazman, and Baranger \cite{MGB96} gave a theory of the Coulomb blockade oscillations in double quantum dots.

The first systematic descriptions of transport through double quantum dots in terms of Master equations were given by Nazarov in 1993 \cite{Naz93}, and by Gurvitz and Prager \cite{GP96} and by Stoof and Nazarov \cite{SN96} in 1996, the latter including a time-dependent, driving  microwave field, cf. section \ref{section_AC}. These were later generalized to multiple-dot systems by Gurvitz \cite{Gur98} and by Wegewijs and Nazarov \cite{WN99}. Furthermore, Sun and Milburn \cite{SM99} applied the open system approach of Quantum Optics \cite{Carmichael} to current noise in resonant tunneling junctions and double dots \cite{SM00}, and 
Aono and Kawamura \cite{AK97} studied the stationary current and time-dependent current relaxation in double-dot systems, using Keldysh Green's functions.

Transport beyond the Master equation approach leads to co-tunneling (coherent transfer of two electrons) and  Kondo-physics, which again even only for double quantum dots has become such a large field that it cannot be reviewed here in detail at all. Pohjola, K\"onig, Salomaa, Schmid, Schoeller, and Sch\"on mapped a double dot  onto a single dot model with two levels and predicted a triple-peak structure in  the Kondo-regime of non-linear transport \cite{Pohetal97}, using a real-time renormalization group technique (see below and \cite{Pohetal99}), whereas Ivanov \cite{Iva97a} studied the Kondo effect in double quantum dots with the equation of motion method. Furthermore, Stafford, Kotlyar and Das Sarma \cite{SKS98}   calculated co-tunneling corrections to the persistent current through double dots embedded into an Aharonov-Bohm ring in an extension of the Hubbard model used earlier by Kotlyar and Das Sarma \cite{KDS97}.

The slave-boson mean field approximation was used by Georges and Meir \cite{GM99} and by Aono and Eto \cite{AE01} for the conductance,  and  for the nonlinear transport through  double quantum dots in the Kondo regime
by Aguado and Langreth \cite{AL00} and later by Orella, Lara, and Anda \cite{OLA02} who discussed nonlinear bistablity behavior. 
Motivated by experimental results by Jeong, Chang and Melloch \cite{JCM01}, Sun and Guo \cite{SG02} used a model with Coulomb interaction between the two dots and found a splitting of the  Kondo peaks in the conductance. 

\begin{figure}[t]
\begin{center}
\parbox{0.6\textwidth}{\includegraphics[width=0.5\textwidth]{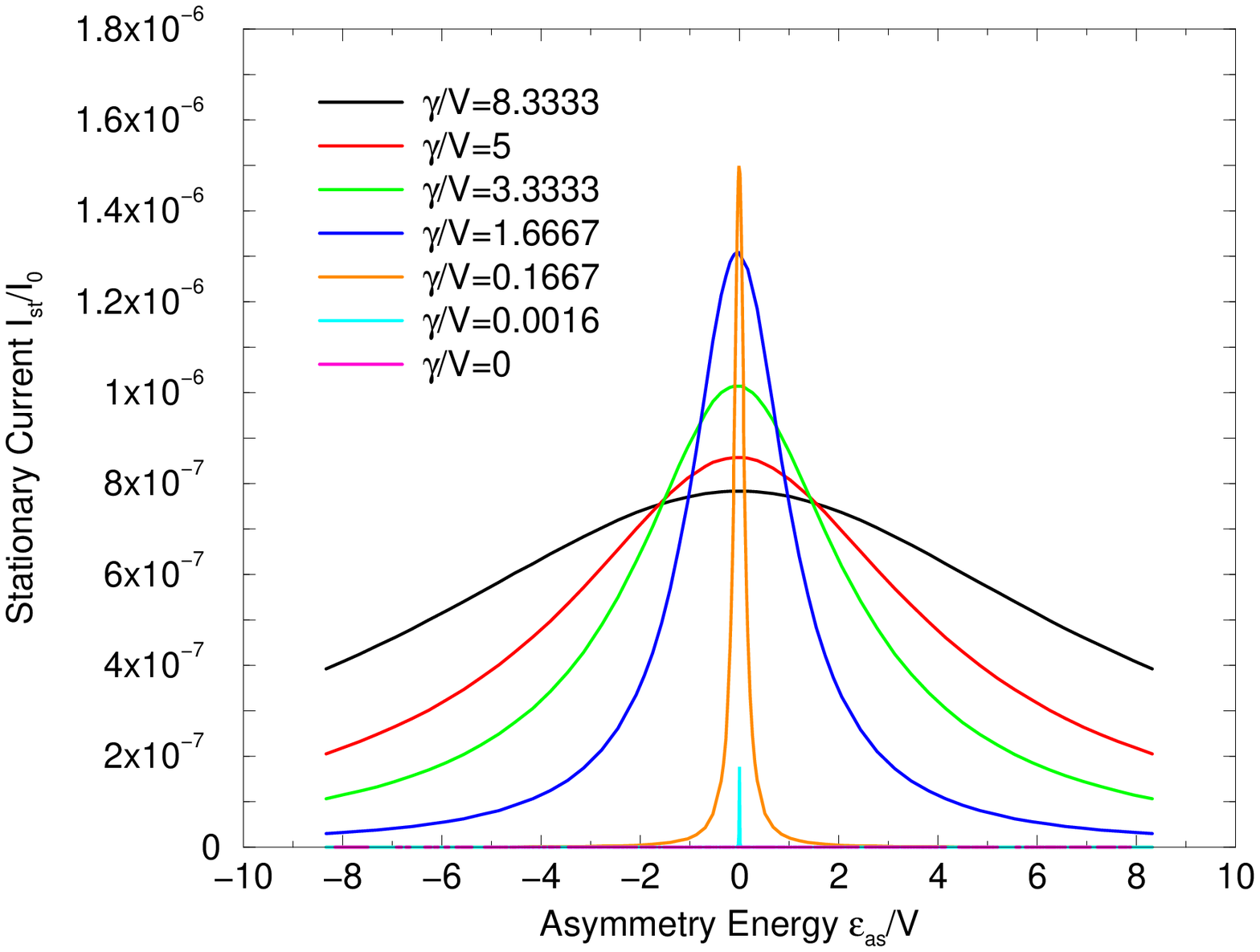}}
\parbox{0.2\textwidth}{ \begin{center}
\includegraphics[width=0.2\textwidth]{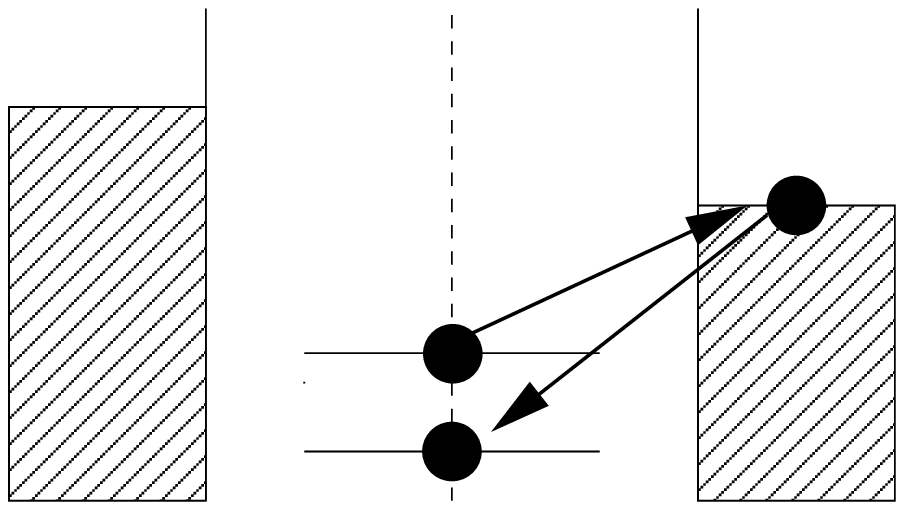}
\includegraphics[width=0.2\textwidth]{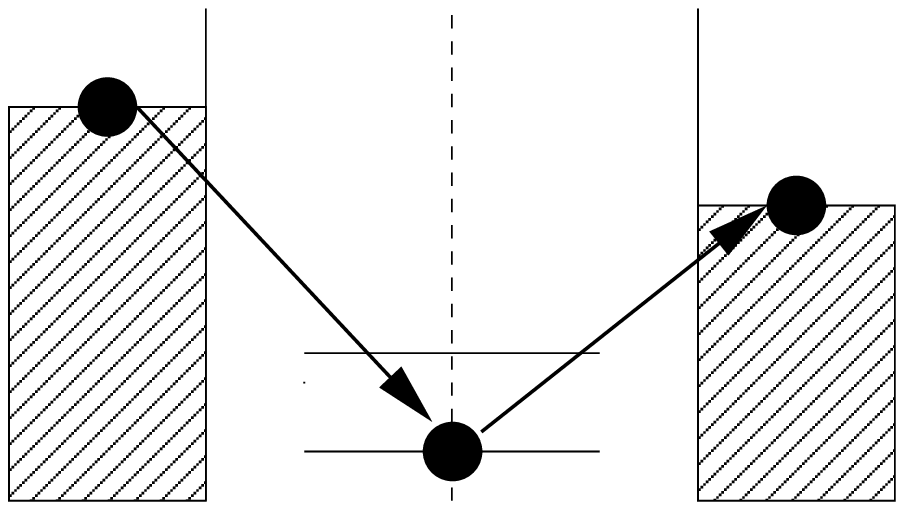}
\includegraphics[width=0.2\textwidth]{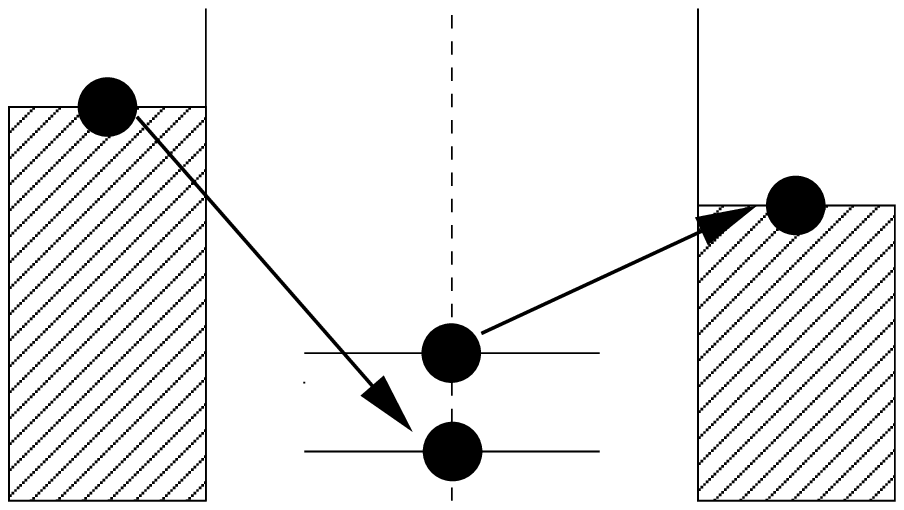}
\end{center}
}
\end{center}
\caption[]{\label{Hartmann_Wilhelm_PRB03_Fig4.eps}{\bf Left:} 
Stationary current in units $I_0=e\Gamma$ ($\Gamma=\Gamma_R=\Gamma_L=1$~GHz)  as a function of the inter-dot-bias $\varepsilon/2$ (in units of the finite bias voltage $V\equiv \mu_L-\mu_R=5.17 \mu$V) through double quantum dots in the co-tunneling regime after Hartmann and Wilhelm \cite{HW03}. Temperature $k_B T =140 $ mK; $\gamma$ denotes the tunnel coupling $T_c$ here. {\bf Right:} Co-tunneling processes contributing to the stationary current at finite bias voltage $V$. From \cite{HW03}.}
\end{figure}

Hartmann and Wilhelm \cite{HW03} calculated the co-tunneling contribution for transport at finite bias voltage $V$ in double quantum dots, starting from the basis of hybridized states, Eq. (\ref{eq:twobytworesult}), and performing a Schrieffer-Wolff transformation that took into account indirect transitions between final and initial dot states including one intermediate state, cf. Fig. (\ref{Hartmann_Wilhelm_PRB03_Fig4.eps}).  They then used the  transformed Hamiltonian in order to obtain the stationary current by means of the usual Bloch-Redfield (Master equation) method, by which they identified three transport regimes: no transport for the tunnel coupling $T_c<\varepsilon/2$, transport through both hybridized states for $\varepsilon<2T_c<\sqrt{V^2-\varepsilon^2}$, and transport through one of the hybridized states for $2T_c>\sqrt{V^2-\varepsilon^2}$, cf. Fig. (\ref{Hartmann_Wilhelm_PRB03_Fig4.eps}). Within the same formalism, they also analyzed dephasing and relaxation of charge, with the double dot regarded as a spin-boson problem with two  distinct baths (the electronic reservoirs) \cite{HW02}.

On the experimental side, co-tunneling and the Kondo regime in parallel transport through double quantum dots were studied by Holleitner and co-workers recently \cite{Holetal02,Holetal04}, whereas Rokhinson {\em et al.} \cite{Roketal03} used a Si double dot structure to analyze the effect of co-tunneling in the Coulomb blockade oscillation peaks of the conductance. 

Another interesting transport regime occurs in Aharonov-Bohm geometries, where electrons move through two {\em parallel}  quantum dots which are, for example, situated on the two arms of a mesoscopic ring `interferometer'. Marquardt and Bruder \cite{MB03} used $P(E)$ theory (section \ref{section_PEtheory}), in order to describe dephasing in such `which-path' interferometers, also cf. their paper and the Review by Hackenbroich \cite{Hac01} for further references.

\subsubsection{Real-Time Renormalization-Group (RTRG) Method} \label{section_Keil_Schoeller}

Keil and Schoeller \cite{KS02} calculated the stationary current through double quantum dots by using an
alternative  method that went beyond  perturbation theory (PER) and avoided the restrictions of the polaron
transformation method (POL). Their method allowed one to treat all three electron-phonon 
coupling parameters ($\alpha^L_{\bf Q}$, $\alpha^R_{\bf Q}$, and $\gamma_{\bf Q}$ in \ Eq.~(\ref{Hdp}))
on equal footing, and thereby to go beyond the spin-boson model which has  $\gamma_{\bf Q}=0$, \ Eq.~(\ref{newHdp}).
Furthermore, they avoided the somewhat unrealistic assumption of infinite bias voltage 
in the Gurvitz Master equation approach and kept $V=\mu_L-\mu_R$ at finite values. They also explicitly took into account a finite width $\sigma$ of the electron densities
in the left and the right dots,
\begin{eqnarray}\label{densityKS}
  \rho_{L/R}({\bf x}) = \left(\frac{1}{\pi \sigma^2}\right)^{3/2}e^{-({\bf x} - {\bf x}_{L/R})^2/\sigma^2},
\end{eqnarray}
which (as mentioned above) leads to a natural high-energy cutoff $D\equiv\hbar c/\sigma$, where $c$ is the speed of sound. 

The starting point of the RTRG method \cite{Schoeller99,KS02} was a set of two formally exact equations
for the time-dependent current $\langle \hat{I} \rangle(t)$ and the
reduced density matrix $\hat{\rho}(t)$ of the dot,
\begin{eqnarray}
  \langle \hat{I} \rangle(t)&=& 
\mbox{\rm Tr}_{\rm dot}
\left[\int_{0}^{t}dt' \Sigma_I(t-t')\hat{\rho}(t')\right],\quad
\frac{d}{dt}\hat{\rho}(t)+i\hat{L}_0\hat{\rho}(t)= \int_{0}^{t}dt' \Sigma(t-t')\hat{\rho}(t'),
\end{eqnarray}
with the operator for the current density between the left lead and the left dot,
\begin{eqnarray}
  \hat{I}\equiv ie\sum_{k_L} \left( (V_k^L)^*|0\rangle\langle L| c^{\dagger}_{k_L} -
 V_k^L|L\rangle\langle 0| c_{k_L} \right),
\end{eqnarray}
the free-time evolution Liouville super-operator $\hat{L}_0 \cdot \equiv [{\mathcal H}_0,\cdot]$ for
an effective dot Hamiltonian ${\mathcal H}_0$, and the two self-energy
operators $\Sigma_I$ and $\Sigma$ which described the coupling to the electron leads and the phonon bath. 
Here, ${\mathcal H}_0$ differs from the dot Hamiltonian ${\mathcal H}_{\rm dot}$, \ Eq.~(\ref{hdot}),
by a renormalized tunnel coupling $T_c^{\rm eff}=T_c - \alpha\omega_de^{-D/2\omega_d}\arctan D/2\omega_d$
and a renormalization $\propto \alpha$
of the energies of the states $|0\rangle$, $|L\rangle$, and $|R\rangle$, where again
$\alpha$ is the dimensionless electron-phonon coupling, and $\omega_d=c/d$ with $d$ the distance between 
the two dots.

Keil and Schoeller then generated renormalization group (RG) equations in the time-domain
by introducing a short-time cut-off $t_c$. By integrating out short time-scales, they derived a 
coupled set of differential equations for the Laplace transforms $\Sigma(z)$, $\Sigma_I(z)$, 
$\hat{L}_0$, and additional vertex operators which were defined in the diagrammatic expansion for the 
time-evolution of the total density matrix in the interaction picture. The RG scheme was perturbative
as it neglected multiple vertex operators, which however was justified for small coupling parameters $\alpha$. 

\begin{figure}[t]
\begin{center}
\includegraphics[width=0.4\textwidth]{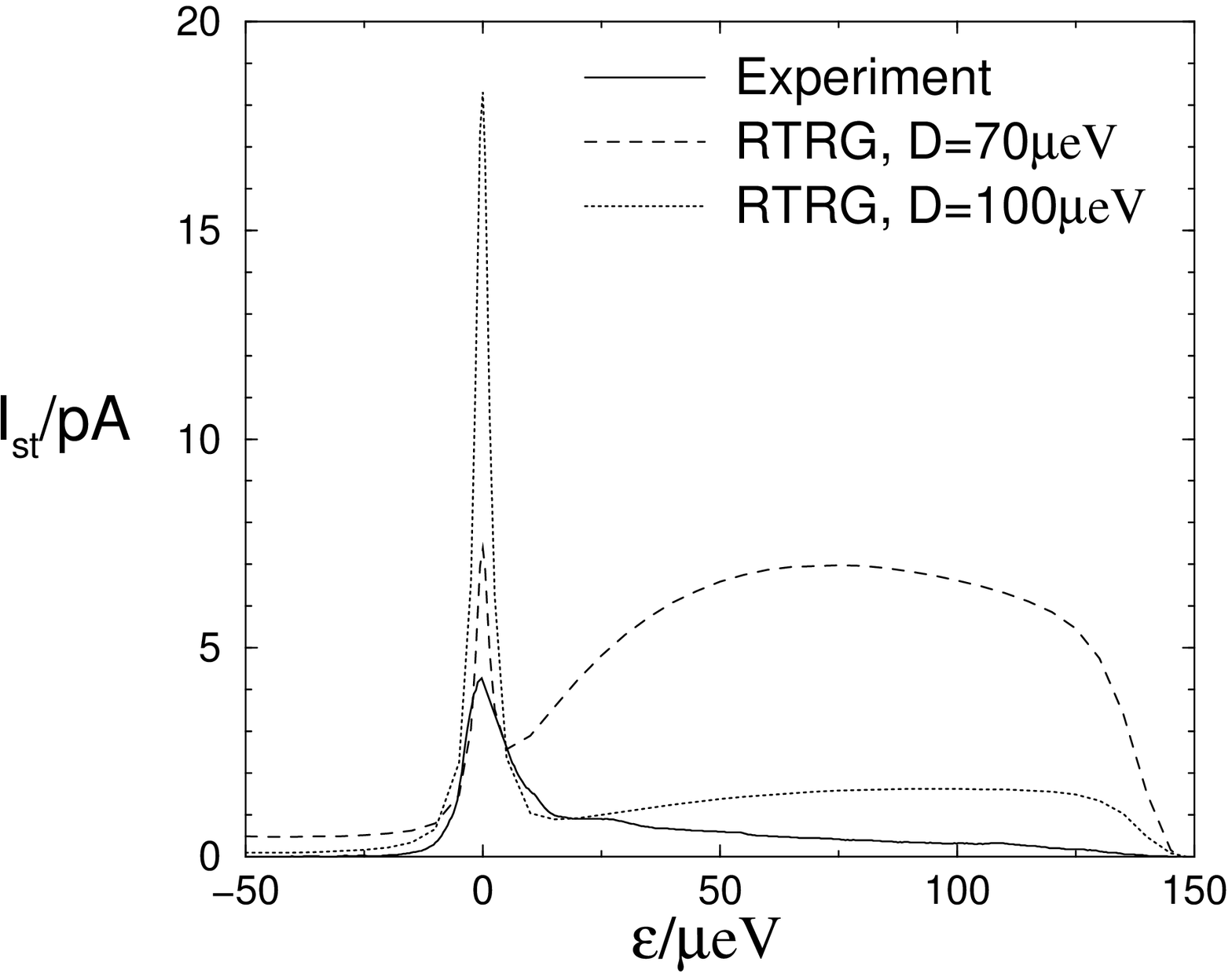}
\includegraphics[width=0.47\textwidth]{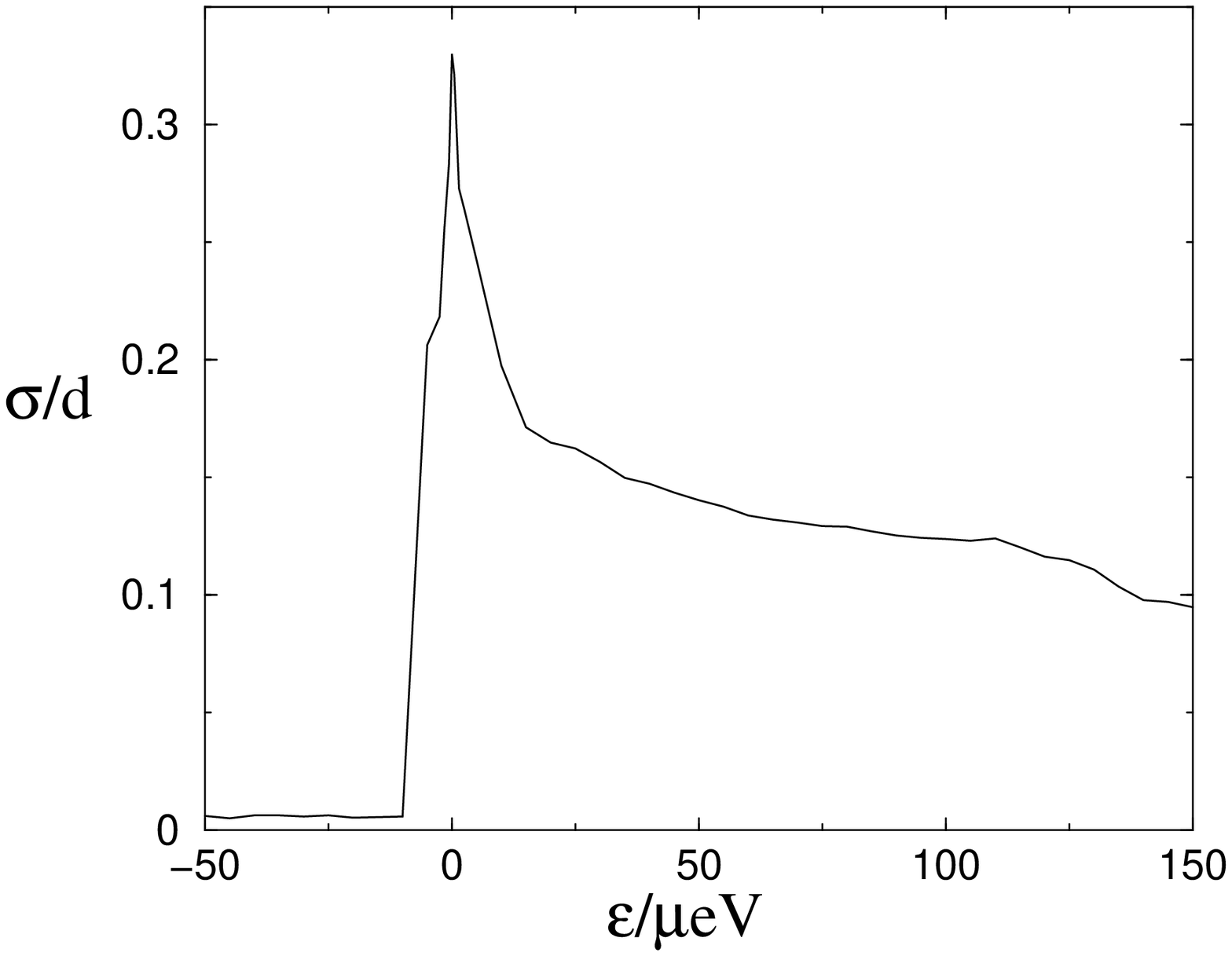}
\end{center}
\caption[]{\label{KS.eps}Left: 
stationary tunnel current through double quantum dots in comparison between
experiment \cite{Fujetal98} and RTRG method \cite{KS02}
with $T_c=0.375\mu$eV, $\Gamma_L=\Gamma_R=3.5\mu$eV, and cut-offs $D=70(100)\mu$eV, $D_{L}=D_R=1$meV. 
Right: fit of  $\varepsilon$-dependent electron density width $\sigma$ ($d$: dot distance).
From \cite{KS02}.}
\end{figure} 

A comparison between experimental data \cite{Fujetal98} and the RTRG calculations 
for the stationary current is shown in Fig. \ref{KS.eps} (left). With smaller cut-off $D$ (i.e. larger 
extension $\sigma$ of the electronic densities in the dots, \ Eq.~(\ref{densityKS}){}), the 
off-diagonal electron-phonon interaction (matrix elements $\gamma_{\bf Q}$) becomes more important 
and the inelastic current is increased. Keil and Schoeller explained the deviations 
from the experimental results by introducing an $\varepsilon$-dependence of $\sigma$, using
$\sigma$ as a fit-parameter for all $\varepsilon$  (Fig. \ref{KS.eps},
right) in order to match the experimental results. Larger energy separations  $\varepsilon>0$ then
imply   electron densities with sharper peaks.

\subsection{Shot-Noise and  Dissipation in the Open Spin-Boson Model}\label{section_noise}

Shot noise (quantum noise) of electrons has been recognized as a powerful tool in the analysis of electronic transport in mesoscopic systems for quite some time (cf. the chapter on noise in Imry's book \cite{Imry} on Mesoscopic Physics). Noise and fluctuations are also key theoretical concepts in Quantum Optics. Noise in mesoscopic conductors has been recently reviewed  by Blanter and B\"uttiker \cite{BB00}, and recent developments are presented in a volume on `Quantum Noise in Mesoscopic Physics' \cite{Naz03}. 

A large theoretical activity on the {\em detection of entanglement} in electron noise, or more generally in the full counting statistics of electrons, has revealed the usefulness of quantum noise for the purpose of quantum information processing in solids. For example, Burkard, Loss and Sukhorukov \cite{BLS00} theoretically demonstrated the possibility to detect entanglement in the bunching of spin singlets  and anti-bunching of spin triplets in an electron current passing a beam splitter. {\em Creation of entanglement} in solids has become a further and  widespread area of (so far) still mostly theoretical activities, ranging from the Loss-DiVincenzo proposal for spin-based qubits \cite{LD98}, superconducting systems \cite{MSS01}, semiconductor spintronics \cite{ALS02} up to entanglement of electron-hole pairs \cite{Benetal03}.

The spontaneous emission and, more generally, quantum dissipation effects discussed in the previous section for stationary transport have of course also a large impact on quantum noise.  Shimizu and Ueda \cite{SU92} investigated how dephasing and dissipation modifies quantum noise in mesoscopic conductors and found a suppression of noise by dissipative energy relaxation  processes. These authors furthermore investigated 
the effect of a bosonic bath on noise in a mesoscopic scatterer \cite{US93}.
Choi, Plastina, and Fazio \cite{CPF03} showed how to extract quantum coherence and the dephasing time $T_2$ from the frequency-dependent noise spectrum in a Cooper pair box. Elattari and Gurvitz \cite{EG02} calculated shot noise in coherent double quantum dots transport, and Mozyrsky and coworkers \cite{Mozetal02} derived an expression for the frequency-dependent noise spectrum  in a two-level quantum dot.

\subsubsection{Current and Charge Noise in Two-Level Systems}

Current noise is defined  by the power spectral density, a quantity sensitive to correlations between carriers,  
\begin{eqnarray}\label{SIdef}
\mathcal {S}_{I}(\omega)\equiv 2\int_{-\infty}^{\infty} d\tau
e^{i\omega\tau} \mathcal {S}_{I}(\tau)=\int_{-\infty}^{\infty}
d\tau e^{i\omega\tau}\langle
\{\Delta\hat{I}(\tau),\Delta\hat{I}(0)\} \rangle,
\end{eqnarray}
where  $ \Delta\hat{I}(t)\equiv \hat{I}(t)-\langle
\hat{I}(t)\rangle$ for the current operator $\hat{I}$. 
The Fano factor
\begin{eqnarray}
   \gamma\equiv\frac{\mathcal{S}_I(0)}{2qI}
\end{eqnarray}
quantifies deviations from the Poissonian noise, $\mathcal{S}_I(0)=2qI$ of
uncorrelated carriers with charge $q$.

The noise spectrum $\mathcal {S}_{I}(\omega)$ for electron transport through dissipative two-level systems was calculated  by Aguado and Brandes in \cite{AB04}, with examples for concrete realizations such as charge qubits in a Cooper pair box \cite{MSS01,CGNS02,CPF03} or the double  quantum dot system from the previous section. The theoretical treatment is basically identical in both cases: for the Josephson Quasiparticle Cycle  of
the superconducting single electron transistor (SSET) with charging energy $E_C\gg E_J$ (the Josephson coupling), only two charge states, $|2\rangle$ (one excess Cooper pair 
in the SSET) and $|0\rangle$ (no extra Cooper pair), are allowed. The consecutive quasiparticle events then  couple $|2\rangle$ and $|0\rangle$ with another state $|1\rangle$ through 
the cycle $|2\rangle\rightarrow |1\rangle\rightarrow
|0\rangle\Leftrightarrow |2\rangle$. Tunneling between $L$ and $R$ in the double dot system is analogous to coherent tunneling of a Cooper pair through one of the junctions, and tunneling to and from the double dot 
is analogous to the two quasiparticle events through the probe junction in the SSET \cite{CPF03}.

In Quantum Optics, the quantum regression theorem \cite{Carmichael} is a convenient tool to calculate temporal correlation functions within the framework of the Master equation.  Tunneling of particles to and from the two-level system requires  to relate the
reduced dynamics of the qubit to particle reservoir
operators like the current operator. In \cite{AB04}, this lead to an expression for the noise spectrum in terms of two contributions: the internal charge noise as obtained from the quantum regression theorem, and the current fluctuations in the particle reservoirs which were calculated by introducing an additional counting variable  $n$ for the number of particles having tunneled through the system. In fact, $S_I(\omega)$ in \ Eq.~(\ref{SIdef}) had to be calculated from the autocorrelations of the {\it total}
current $I$, i.e. particle plus displacement current\cite{BB00} under the 
current conservation condition. Left and right currents contribute to the total current as 
$I=a I_L + b I_R$, where $a$ and $b$ are capacitance coefficient ($a+b=1$) of the junctions
(Ramo-Shockley theorem), leading to an expression of $\mathcal {S}_{I}(\omega)$ in terms of
the spectra of particle
currents and the charge noise spectrum $S_{Q}(\omega)$, 
\begin{equation}
\label{fullnoise}
 S_I(\omega)=a S_{I_L}(\omega) + b
 S_{I_R}(\omega)-a b\omega^2S_{Q}(\omega)
\end{equation}
with $S_{Q}(\omega)$  defined as
\begin{eqnarray}
\label{charge}
  S_Q(\omega) &\equiv& \lim_{t\to \infty}
\int_{-\infty}^{\infty} d\tau e^{i\omega\tau}\langle
\{\hat{Q}(t),
\hat{Q}(t+\tau)\}\rangle = 2{\mbox Re} \left\{ \hat{f}(z=i\omega) +
\hat{f}(z=-i\omega)\right\},
\end{eqnarray}
where $\hat{Q}=\hat{n}_L+\hat{n}_R$ and $\hat{f}(z)$ is the
Laplace transform of
\begin{eqnarray}
\label{fequation}
  f(\tau) &=& \sum_{i,j=L,R}\langle \hat{n}_i(t)\hat{n}_j(t+\tau)\rangle
=
({\bf e}_1+{\bf e}_2)[{\bf C}_L(\tau)+{\bf C}_R(\tau)]\\
{\bf C}_\alpha(\tau)&\equiv& \langle \hat{n}_{\alpha}(t) {\bf A}(t+\tau) \rangle.
\end{eqnarray}
The equations of motions of  the charge correlation functions ${\bf C}_\alpha(\tau)$ \cite{AB04a}
(quantum regression theorem \cite{Carmichael}), 
\begin{eqnarray}\label{Cequation}
{\bf C}_i(\tau) &=& {\bf C}_i(0)+ \int_{0}^{\tau} d\tau'
\left\{{M}(\tau-\tau'){\bf C}_i(\tau') +  \langle n_{i} (t)
\rangle{\bf \Gamma}\right\},\quad {\bf \Gamma}=\Gamma_L {\bf e}_1,
\end{eqnarray}
are solved in terms of  the resolvent $[z-z\hat{M}(z)]^{-1}$, cf. Eq.
(\ref{block}).

The qubit dynamics was related with reservoir operators by introducing
a counting variable $n$ (number of electrons that have
tunneled through the right barrier \cite{MSS01,EG02}) 
and  expectation values, 
$O^{(n)}\equiv\sum_{i=0,L,R}{\rm Tr_{bath}} \langle n,i |\hat{O}\rho(t)|n,i\rangle$
with  $\langle \hat{O} \rangle=\sum_n O^{(n)}$. This lead to a system of equations of motion,
\begin{eqnarray}\label{generalized}
  \dot{n}_0^{(n)}&=&-\Gamma_L {n}_0^{(n)} + \Gamma_R {n}_R^{(n-1)},\quad
  \dot{n}_{L/R}^{(n)}= \pm\Gamma_{L/R} {n}_0^{(n)} \pm iT_c \left( p^{(n)}-
[p^{(n)}]^{\dagger}\right)
\end{eqnarray}
and similar equations for $p^{(n)}$ and $[p^{(n)}]^{\dagger}$, which
together with $P_n(t)=n_0^{(n)}(t)+n_L^{(n)}(t)+n_R^{(n)}(t)$ gave the
total probability of finding $n$ electrons in the collector by time
$t$. In particular, $I_R(t)=e\sum_n n\dot{P}_n(t)$ such that 
$S_{I_R}$ could be calculated from the Mac-Donald formula \cite{Mac48},
\begin{eqnarray}
  S_{I_R}(\omega)&=&2\omega e^2\int_0^{\infty} dt \sin(\omega t) \frac{d}{dt}
\left[ \langle n^2(t) \rangle - (t\langle I\rangle)^2 \right]
= 2eI\left\{1+\Gamma_R\left[\hat{n}_R(-i\omega) + \hat{n}_R(i\omega)\right]\right\}\nonumber\\
z\hat{n}_R(z)&=&\Gamma_Lg_+(z)/\left\{
[z+\Gamma_R+g_-(z)](z+\Gamma_L)+(z+\Gamma_R+\Gamma_L)g_+(z)
\right\},
\end{eqnarray}
where the $g_{+[-]}(z)$ are defined in \ Eq.~(\ref{gpmdefinition}).
In the {\em zero frequency limit} $z\rightarrow 0$, the result
\begin{eqnarray}
\label{shot}
  S_I(0) = 2eI \left(1+2\Gamma_R \frac{d}{dz}\left[z
  \hat{n}_R(z)\right]_{z=0}\right).
\end{eqnarray}
indicated the possibility  to investigate the shot noise of open dissipative two-level systems
for {\it arbitrary environments}.  In \cite{AB04}, it was pointed out that \ Eq.~(\ref{shot})
can not be written in the Khlus-Lesovik form $S_I(0)= 2 e^2 \int \frac{dE}{2\pi} t(E) [1-t(E)]$ 
with an effective transmission coefficient $t(E)$ as is the case for 
non-interacting mesoscopic conductors, cf. also \cite{SU92}.

\begin{figure}[t]
\begin{center}
\includegraphics[width=0.45\textwidth]{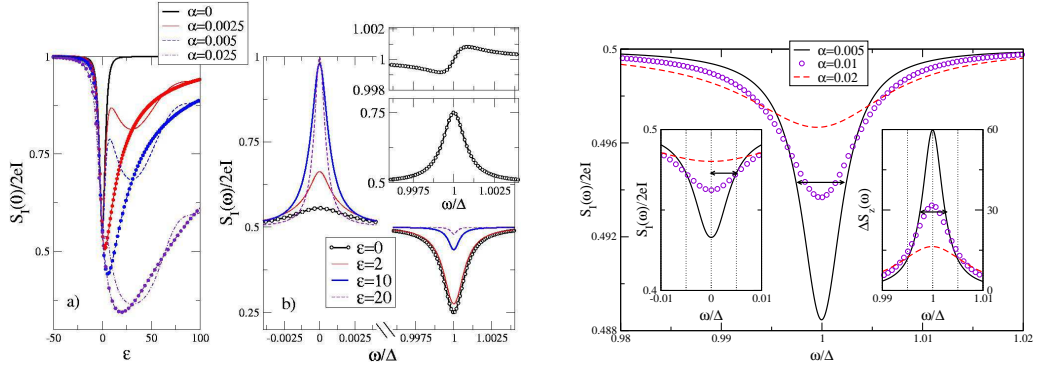}
\hspace{0.5cm}
\includegraphics[width=0.4\textwidth]{AB04_fig2.eps}
\end{center}
\caption[]{{\bf Left:}
a) Fano factor vs. $\varepsilon$ for different dissipative couplings $\alpha$ and
parameters $T_c=3$, $\Gamma=0.15$, $\omega_c=500$,
$\omega_d=10$, $k_BT=2$ (in $\mu eV$) corresponding to
typical experimental values~\cite{Fujetal98} in double quantum dots.
Lines: acoustic phonons, circles: generic Ohmic environment $\omega_d=0$.
b) Frequency dependent current noise ($\alpha=0$, $k_BT=0$, $\Gamma=0.01$).
Inset: (Top) Contribution to noise from particle currents $S_{I_R}(\omega)/2eI$. (Bottom) Charge noise contribution $\omega^2S_{Q}(\omega)/8eI$. $a=b=1/2$.
{\bf Right:} Effect of Ohmic dissipation on current noise near resonance
($\varepsilon=10$, $\Gamma=0.01$,
and  $\alpha=0.005, 0.01, 0.02$).
Inset: (Right)
pseudospin correlation function $S_z(\omega)$. Arrows indicate the calculated
relaxation rate $(\Gamma+\gamma_p)/\Delta\approx 0.005$ for $\alpha=0.005$.
(Left) 
low frequencies region near shot noise limit $\omega=0$. 
From \cite{AB04}.
\label{AB04.eps}}
\end{figure}

For $\alpha=0$, i.e. without coupling to the bosonic bath, \ Eq.~(\ref{gpm1}) yields
\begin{eqnarray}
g_{\pm}(z) = T_c^2 \frac{2z+\Gamma_R}{\left(z+\frac{\Gamma_R}{2}\right)^2+\varepsilon^2},  
\end{eqnarray}
which reproduces earlier results by  Elattari and Gurvitz \cite{EG02},
\begin{eqnarray}
\frac{d}{dz}\left[z \hat{n}_R(z)\right]_{z=0}=
-\frac{4T_c^2\Gamma_L}{\Gamma_R}
\frac{4\varepsilon^2(\Gamma_R-\Gamma_L)+3\Gamma_L\Gamma_R^2+\Gamma_R^3+8\Gamma_RT_c^2}
{\left[\Gamma_L\Gamma_R^2 + 4\Gamma_L\varepsilon^2
+4T_c^2(\Gamma_R+2\Gamma_L)\right]^2},
\end{eqnarray}
and similarly
one recovers the results for shot noise in the Cooper pair box obtained by Choi, Plastina and Fazio
\cite{CPF03}.
In particular, for $\alpha=0$ and
$\Gamma\equiv\Gamma_L=\Gamma_R$ ( left Fig. (\ref{AB04.eps}) a), solid line), the
smallest Fano factor has a minimum at $\varepsilon =0$ where quantum
coherence strongly suppresses noise with maximum suppression
($\gamma=1/5$) reached for $\Gamma =2\sqrt{2}T_c$. On the other hand, for large
$\varepsilon>0$ ($\varepsilon <0$) the charge becomes localized in
the right (left) level, and $S_I(0)$ is dominated by only one Poisson
process, the  noise of the right(left) barrier, and the Fano factor tends to unity,
$\gamma\rightarrow 1$.
For $\alpha\neq 0$, spontaneous
emission (for $\varepsilon >0$) reduces the noise well below the Poisson limit, with 
a maximal reduction  when the elastic and inelastic rates coincide, i.e.,
$\gamma_p=\Gamma_R$. 

On the other hand, for {\em finite  frequencies} $\omega$, $\gamma$ was found to have a peak around $\omega=0$ and a dip at the frequency
$\omega=\Delta$, where $\Delta=\sqrt{\varepsilon^2+4T_c^2}$ is the level splitting, which was shown to
directly reflect the resonance  of the subtracted charge noise $S_{Q}(\omega)$ around $\Delta$ 
(inset left Fig. (\ref{AB04.eps}) b), cf. Eq. (\ref{fullnoise}).
The dip in the high frequency noise at $\omega=\Delta$
is progressively destroyed (reduction of quantum
coherence) as $\varepsilon$ increases due to localization of the charge, or as the dissipation increases.

It was therefore argued \cite{AB04} that  $S_I(\omega)$ reveals the complete internal dissipative dynamics
of the two-level system, an argument that was supported by a calculation of the symmetrized pseudospin
correlation function 
\begin{eqnarray}
S_z(\omega)=1/2\int_{-\infty}^{\infty} d\omega e^{i\omega\tau}\langle\{\hat{\sigma}_z(\tau),\hat{\sigma}_z\}\rangle   
\end{eqnarray}
(right Fig. (\ref{AB04.eps}), right inset)
which is often used to investigate the dynamics of the spin-boson  problem \cite{Weiss} and which 
also shows the progressive damping of the coherent dynamics with increasing dissipation.
A further indication was the extraction of the {\em dephasing rate}
from the half-width of $S_I(\omega)$ around $\omega=\Delta$, and the {\em relaxation rate}
around $\omega=0$, indicated by the arrows and consistent with the relation between relaxation and dephasing time, $T_1 = \frac{1}{2}T_2$, in the underlying Markov approximation in the perturbative approach (PER). 

Results for the strong coupling (POL) regime were also discussed in \cite{AB04}, where
near $\omega=0$, POL and PER yielded nearly identical results for the noise $S_I(\omega)$
at very small $\alpha$, but with increasing $\alpha$ a 
cross-over to Poissonian noise near $\omega=0$ was found and interpreted as 
localized polaron formation. The delocalization-localization transition \cite{Legetal87,Weiss}
of the spin-boson model at $\alpha=1$  therefore also shows up in the shot noise near zero bias,
where the function $C_\varepsilon$ has a change in its analyticity.
A similar transition was found by Cedraschi and B\"uttiker in 
the suppression of the persistent current $I(|\varepsilon|)\propto
\mbox{\rm Im}C_{-|\varepsilon|}$ through
a strongly dissipative quantum ring containing  a quantum dot with bias $\varepsilon$
\cite{CB01}.

\subsubsection{Shot Noise Experiments}\label{section_Deblock}
Deblock, Onac, Gurevich, and Kouwenhoven \cite{Debetal03} measured the current noise spectrum in the frequencies range 6-90 GHz by using a superconductor-insulator-superconductor (SIS) tunnel junction, which converted a noise signal at frequency $\omega$ into a DC, photo-assisted quasi-particle tunnel current. They tested their on-chip noise analyzer for three situations: the first was a voltage ($V_J$) biased Josephson junction for $|eV_J|$ below twice the superconducting gap, leading to an AC current and (trivially) two delta-function noise peaks at frequencies $\omega=\pm 2eV_J/\hbar$. This measurement served to extract a trans-impedance $Z(\omega)$ of the system which was later used to analyze the data without additional fit parameters. The second case was a DC current ($I_J$) biased Josephson junction in the quasi-particle tunneling regime for  $|eV_J|$ above twice the superconducting gap . Using the same $Z(\omega)$,  good agreement with the experimental data was obtained with a {\em non-symmetrized} noise spectrum $S(\omega)=e I_J$, which is half the Poisson value,  $S(\omega)=2e I_J$.

Finally and most important, they used a Cooper pair box to confirm the peak in the spectral noise density as predicted by Choi, Plastina and Fazio \cite{CPF03}. The resonance of $S(\omega)$ appeared around the level splitting $\hbar\omega=\sqrt{4E_C(Q/e-1)^2+E_J^2}$, with $E_C$ the charging energy, $Q$ the charge in the box, and $E_J$ the Josephson coupling between the two states of $N$ and $N+1$ Cooper pairs in the box, thus again demonstrating the coherent quantum mechanical coupling between the two states.

\subsection{Time-Dependent Fields and Dissipation in Transport Through Coupled Quantum Dots}\label{section_AC}

\begin{figure}[t]
\includegraphics[width=0.45\textwidth]{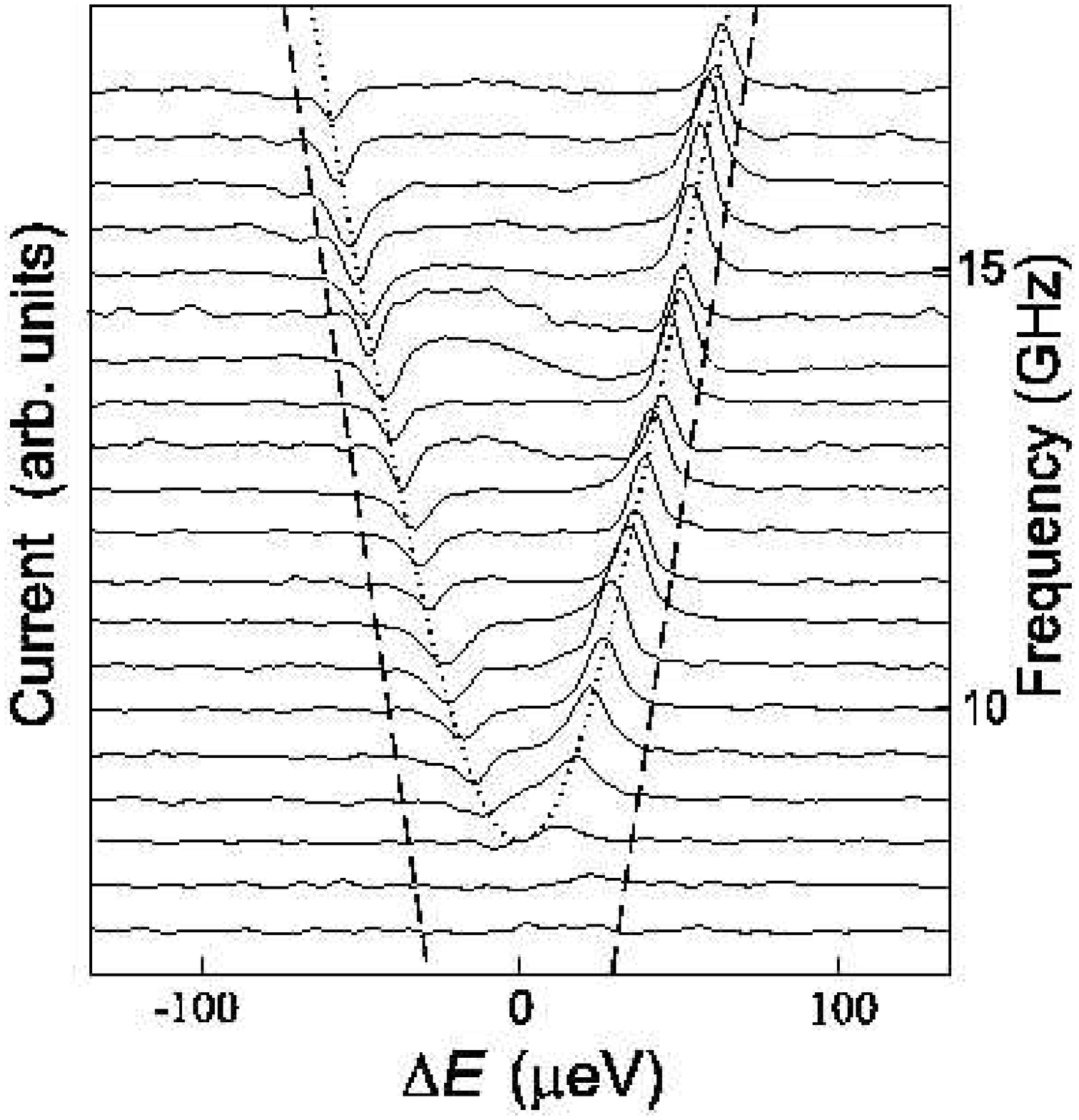}
\includegraphics[width=0.45\textwidth]{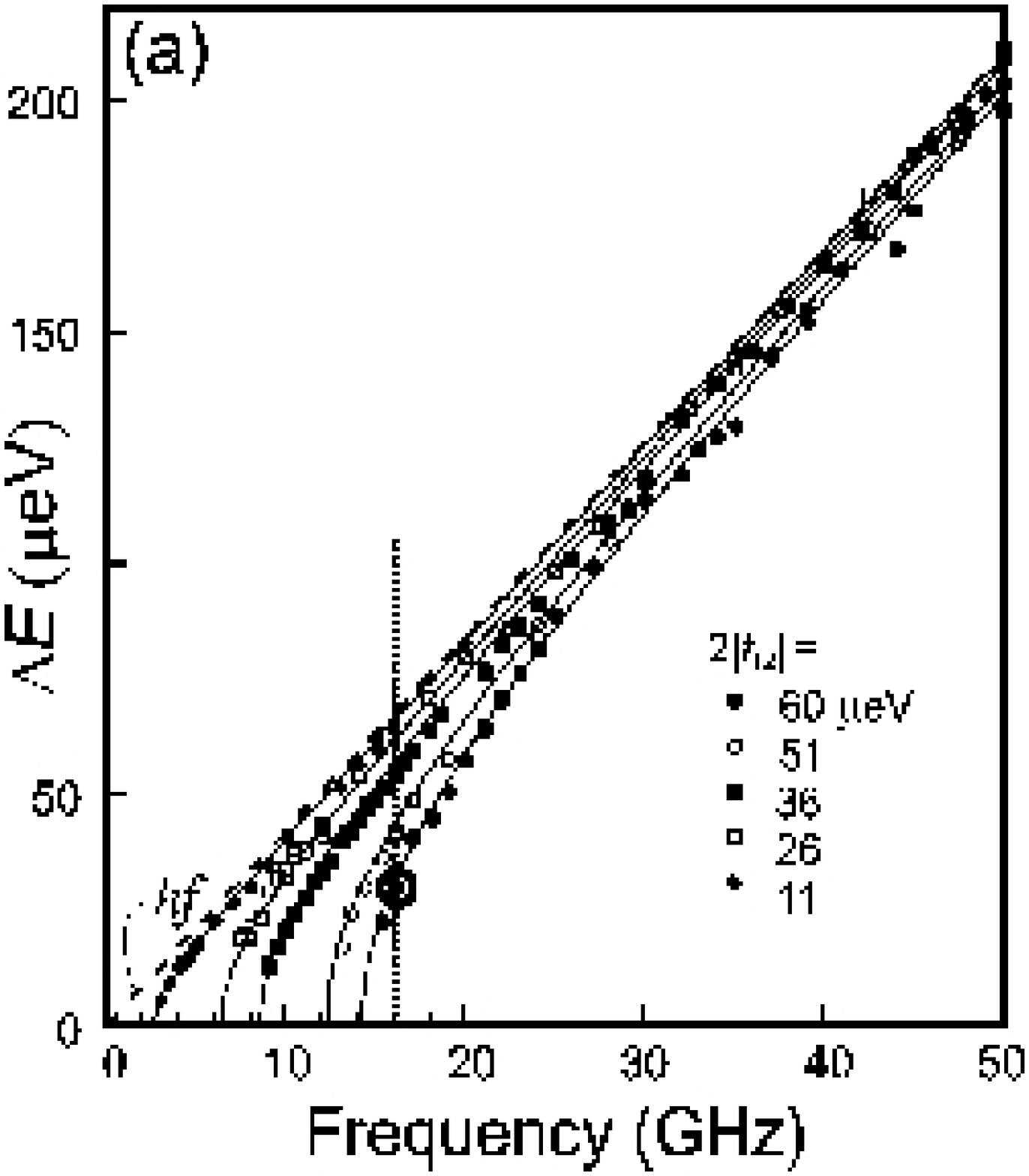}
\caption{\label{Oosterkamp3.eps}
{\bf Left:} Stationary current through a microwave irradiated double quantum dot (zero transport voltage)
for different microwave frequencies $f$ as a function of the 
energy difference $\varepsilon$ ($=\Delta E$ here) in the experiment by Oosterkamp and co-workers \cite{Oosetal98}. Positive or negative peaks occur whenever $hf$ matches the energy difference, $hf = \pm \Delta=\sqrt{\varepsilon^2+4T_c^2}$, between bonding and anti-bonding state in the double dot, cf. \ Eq.~(\ref{eq:twobytworesult}. From \cite{vdWetal03}).
{\bf Right:} The relation $\varepsilon (\equiv \Delta E) =\sqrt{(hf)^2-4T_c^2}$ is tested for various inter-dot coupling constants $T_c$ (denoted as $T$ in the picture). Inset shows the double dot sample. From \cite{Oosetal98}.}
\end{figure}

The interaction of two-level systems with light is one of the central paradigms of Quantum Optics; the study of transport  under irradiation with light therefore is a natural extension into the realm of quantum optical effects in mesoscopic transport through two-level systems as treated in this section. In the simplest of all cases, the light is not considered as a quantum object but as a simple monochromatic classical field with sinusoidal time-dependence, and one has to deal with time-dependent Hamiltonians. These systems are often called `ac-driven' and have received a lot of attention in the past. In the context of electronic transport and tunneling, this field has  recently been reviewed by Platero and Aguado \cite{PA04}. Furthermore, Grifoni and H\"anggi reviewed driven quantum tunneling in dissipative two-level and related systems \cite{GH98}.

An additional, time-dependent electric field in general is believed to give additional insight into the quantum dynamics of electrons, and in fact a large number of interesting phenomena like photo-sidebands, coherent suppression of tunneling, or zero-resistance states in the quantum Hall effect have been investigated. In this context, an essential point is the fact that the  field is not from the beginning treated as a perturbation (e.g., in linear response approximation), but is rather considered as inherent part of the system itself. By this, one has to deal with conditions of a {\em non-equilibrium system} under which the quantity of interest, e.g. a tunnel current or the screening of a static potential, has to be determined. 

For our purposes here, a simple distinction can be made between systems where the field is a simple, monochromatic  ac-field, or where it shows a more complicated time-dependence such as in the form of pulses with certain shapes. The latter case plays a mayor role in a variety of adiabatic phenomena such as charge pumping 
\cite{Grabert,Geretal90,Kouetal91,Potetal92,Swietal99,Bro98,PB01,CB02,MB01,RB01}, adiabatic control of state vectors \cite{Bonetal98,BRB01}, or operations relevant for quantum information processing in a condensed-matter
setting \cite{NPT99,MSS01,Ave98,Ave99,SLM01,EL01,TH02} and will be dealt with in section \ref{section_dark}. On the other hand,  a monochromatic time-variation is mostly discussed in the context of a high frequency regime and photo-excitations.

Theoretical approaches to ac-driven quantum dots comprise a large number of works that cannot be reviewed here, but cf. \cite{PA04}. Earlier works include, among others,  the papers by Bruder and Schoeller \cite{BS94}, Hettler and Schoeller \cite{HS95a}, Stafford and Wingreen \cite{SW96}, and Brune and coworkers \cite{BBS97}. The first systematic theory on transport through {\em double} quantum dots with ac-radiation in the strong Coulomb blockade regime was given by Stoof and Nazarov \cite{SN96}, which was later generalized to pumping of electrons and pulsed modulations by Hazelzet, Wegewijs, Stoof, and Nazarov \cite{HWSN01}. On the experimental side for double quantum dots, Oosterkamp and co-workers \cite{Oosetal98} used microwave excitations in order to probe the tunnel-coupling induced splitting into bonding and antibonding states, cf. Fig. (\ref{Oosterkamp3.eps}) and the Review by van der Wiel and coworkers \cite{vdWetal03}. Blick and co-workers \cite{Blietal98a} demonstrated Rabi-oscillations in double dots with ac-radiation, and later Holleitner and co-workers \cite{Holetal00} studied photo-assisted tunneling in double dots with an on-chip microwave source. 

Qin, Holleitner, Eberl and Blick \cite{QHEB01} furthermore reported the probing of bonding and anti-bonding states in double quantum dots with photon-assisted tunneling. They also found a remarkable combination of phonon and photon-assisted tunneling for microwave frequencies below 8 GHz. In their experiments, charging diagrams were measured as a function of the tunnel coupling $T_c$ and the inter-dot bias $\varepsilon$. At higher microwave frequencies $f=15$ and 20 GHz, the relation $\varepsilon=\sqrt{(hf)^2-4T_c^2}$ confirmed the coherent coupling of the two-level systems, cf. also  Fig. (\ref{Oosterkamp3.eps}). However, at $f=3$ GHz a modification of this square-root dependence was observed, and Qin and co-workers suggested a sequential process with photon absorption and a coherent coupling of the dot levels by a fundamental {\em phonon} frequency $f_{\rm ph}$, or alternatively a completely coherent coupling of phonon and photon, giving rise to `square-root laws' of the form \cite{QHEB01}
\begin{eqnarray}
  \varepsilon_{\rm seq} = \sqrt{(2hf_{\rm ph})^2 -4T_c^2} + 2hf,\quad 
  \varepsilon_{\rm coh} = \sqrt{(2hf_{\rm ph}+2hf)^2 -4T_c^2},
\end{eqnarray}
respectively, both of which were within the error bars of the experimental data. The phonon frequency $f_{\rm ph}\approx 10$ GHz was found to match well with the geometrical dimension of the double dot, which was argued to act like a phonon cavity for piezo-acoustic phonons, also cf. section \ref{section_cavity}. 

\subsubsection{Transport Model for Driven Two-Level System with Dissipation}
The combination of ac driving and transport in a {dissipative} two-level system (double quantum dot) was modeled by Brandes, Aguado, and Platero in \cite{BAP04} in a generalization of earlier work on closed, dissipative two-level systems with ac driving by Grifoni and H\"anggi \cite{GH98}, and coherently ac-driven double dots by Stoof and Nazarov \cite{SN96}. 
The model considered in \cite{BAP04}, 
\begin{eqnarray}\label{ddot_time}
  {\mathcal H}(t)= {\mathcal H}_{SB}(t)+{\mathcal H}_{res}+{\mathcal H}_{V}
\end{eqnarray}
was identical with the double-dot transport model in section \ref{section_ddotmodel}, but with a time-dependent
spin-boson part ${\mathcal H}_{SB}(t)$, cf. \ Eq.~(\ref{modelhamiltonian}), of which only the inter-dot bias $\varepsilon(t)$ was considered as time-varying according to 
\begin{eqnarray}
  \label{eq:epsdef1}
  \varepsilon(t)=\varepsilon + \Delta \sin (\Omega t),
\end{eqnarray}
where $\Omega$ is the angular frequency of an external electric field that leads to a symmetric modulation of the bias with amplitude $\Delta$. The question of whether or not the simple assumption \ Eq.~(\ref{eq:epsdef1}){} is sufficient in order to describe the effect of an ac-field is a non-trivial issue. Stoof and Nazarov \cite{SN96} argued that  Eq.~(\ref{eq:epsdef1}) describes the effect of a sinusoidal modulation of a gate voltage. Experimentally, however, the coupling to ac-fields is complicated, and in principle one has to expect additional modulations of other parameters such as the tunnel coupling $T_c$, or the tunnel rates $\Gamma_{L/R}$. From the quantum optical point of view, one would in fact  start from the bonding-antibonding basis $|\pm\rangle$, \ Eq.~(\ref{eq:twobytworesult}), and argue that the `light' induced transitions between them. A more precise model would involve detailed microscopic calculation of a) the electromagnetic field modes coupling into the system, e.g., its polarization, possible propagation effects etc., and b) the dipole (or higher if required) matrix elements for electron-photon coupling in the double-dot many-body system. Eventually, however, one would expect to recover models like \ Eq.~(\ref{ddot_time}), possibly with some modifications and microscopic expressions for the parameters.

In \cite{BAP04}, the evaluation of the photo-current through the dots under irradiation was performed within the Master equation approach in a generalization of the polaron transformation formalism developed in section \ref{section_polaron}. Using the boson correlation function, Eq.~(\ref{C_equi}), one then can define `polaron propagators' that incorporate the finite lifetime of the electron-boson quasi-particle due to tunneling out of the double dot at rate $\Gamma_R$, 
\begin{eqnarray}
  \label{eq:Ddef}
 \hat{D}_{\varepsilon}(z)&\equiv&\frac{\hat{C}_{\varepsilon}(z)}{1+\Gamma_R\hat{C}_{\varepsilon}(z)/2},\quad
 \hat{E}_{\varepsilon}(z)\equiv\frac{\hat{C}^*_{-\varepsilon}(z)}{1+\Gamma_R\hat{C}_{\varepsilon}(z)/2}\nonumber\\
 \hat{D}^*_{\varepsilon}(z)&\equiv&\frac{\hat{C}^*_{\varepsilon}(z)}{1+\Gamma_R\hat{C}^*_{\varepsilon}(z)/2},\quad
 \hat{E}^*_{\varepsilon}(z)\equiv\frac{\hat{C}_{-\varepsilon}(z)}{1+\Gamma_R\hat{C}^*_{\varepsilon}(z)/2},\nonumber\\
\end{eqnarray}
where $C_\varepsilon(z)$ etc. are defined in \ Eq.~(\ref{C_eps_def}) and the hat denotes the Laplace transformation. The propagators \ Eq.~(\ref{eq:Ddef}), transformed back into the time-domain, appear in closed equations for the occupancies $\langle n_{L/R} \rangle$,
\begin{eqnarray}
  \label{eq:eomLapl3}
  z\hat{n}_L(z)-\langle n_L \rangle_0 &=& - \int_{0}^{\infty}dt e^{-zt}
\left[ \langle n_L\rangle_{t} \hat{K}(z,t)-\langle n_R\rangle_{t} \hat{G}(z,t)\right]
+\Gamma_L\left[ \frac{1}{z} - \hat{n}_L(z)- \hat{n}_R(z) \right]
\nonumber\\
z\hat{n}_R(z)-\langle n_R \rangle_0 &=&  \int_{0}^{\infty}dt e^{-zt}
\left[ \langle n_L\rangle_{t} \hat{K}(z,t)-\langle n_R\rangle_{t} \hat{G}(z,t)\right]
-{\Gamma}_R \hat{n}_R(z)
\nonumber\\
\hat{K}(z,t)&\equiv&\int_{0}^{\infty}dt'e^{-zt'}
\left[T_c(t+t')T_c^*(t){D}_{\varepsilon}(t')+
T_c^*(t+t')T_c(t){D}^*_{\varepsilon}(t')\right]\nonumber\\
\hat{G}(z,t)&\equiv&\int_{0}^{\infty}dt'e^{-zt'}
\left[T_c(t+t')T_c^*(t){E}_{\varepsilon}(t')+
T_c^*(t+t')T_c(t){E}^*_{\varepsilon}(t')\right],
\end{eqnarray}
where the combination 
\begin{eqnarray}
    \label{eq:expdecomp1}
T_c(t+t')T_c^*(t') = T_c^2e^{i\int_t^{t+t'}ds {\Delta \sin(\Omega s)}}=
T_c^2\sum_{nn'}i^{n'-n}J_n\left(\frac{\Delta}{\Omega}\right)J_{n'}\left(\frac{\Delta}{\Omega}\right)
e^{-in\Omega t'} e^{-i(n-n')\Omega t}\nonumber
\end{eqnarray}
involves Bessel functions, as is typical for ac-tunneling problems. Decomposing into Fourier series, a closed set of equations for asymptotic, stationary quantities 
\begin{eqnarray}
\hat{n}_L^{asy}(z)=\sum_m \frac{\nu_m}{z+im\Omega},\quad
\hat{n}_R^{asy}(z)=\sum_m \frac{\mu_m}{z+im\Omega}
\end{eqnarray}
is then obtained in the form of an infinite system of linear equations for the Fourier coefficients $\nu_m$ and $\mu_m$,
\begin{eqnarray}
  \label{eq:fourier2}
 -iM\Omega
{\nu_M}&=&-\sum_{n}\left[\nu_nK_{M-n}(-iM\Omega)-\mu_n G_{M-n}(-iM\Omega)\right]
+\Gamma_L\left[\delta_{M,0}-\nu_M-\mu_M\right]\nonumber\\
\left[{\Gamma}_R -iM\Omega\right]{\mu_M}&=&\phantom{-}\sum_{n}\left[\nu_n K_{M-n}(-iM\Omega)-\mu_n G_{M-n}(-iM\Omega)\right],
\end{eqnarray}
which can be transformed into an infinite matrix equation that describes the contribution from all photo-side bands at frequencies $\pm n \Omega$  with coefficients given by the Fourier components of $\hat{K}(z,t)$ and
$\hat{G}(z,t)$,
\begin{eqnarray}\label{KGexplicit}
  {K}_m(-im'\Omega)&=& i^{-m}T_c^2\sum_n
\left[ J_n\left(\frac{\Delta}{\Omega}\right)J_{n-m}\left(\frac{\Delta}{\Omega}\right)
\hat{D}_{\varepsilon+(m'-n)\Omega}
+J_n\left(\frac{\Delta}{\Omega}\right)J_{n+m}\left(\frac{\Delta}{\Omega}\right)
\hat{D}^*_{\varepsilon-(m'+n)\Omega}\right]\nonumber\\
 {G}_m(-im'\Omega)&=& i^{-m}T_c^2\sum_n
\left[ J_n\left(\frac{\Delta}{\Omega}\right)J_{n-m}\left(\frac{\Delta}{\Omega}\right)
\hat{E}_{\varepsilon+(m'-n)\Omega}
+J_n\left(\frac{\Delta}{\Omega}\right)J_{n+m}\left(\frac{\Delta}{\Omega}\right)
\hat{E}^*_{\varepsilon-(m'+n)\Omega}\right].\nonumber\\
\end{eqnarray}
The closed expression for the stationary current $\bar{I}$ averaged over one period $\tau\equiv 2\pi/\Omega$,
\begin{eqnarray}\label{currentstationary}
\bar{I}=-e\Gamma_R
  \frac{K_0(0)-\sum_{n\ne 0}\left[
K_{-n}(0)/r_n+ G_{-n}(0)\right]\mu_n}
{\Gamma_R+K_0(0)/r_0+G_0(0)},
\end{eqnarray}
then is the starting point for a numerical and analytical analysis of ac-driven dissipative transport through two-level systems.

\subsubsection{Stationary Current}
In absence of the time-dependent (driving) part in $\varepsilon(t)$, i.e. for $\Delta=0$, one recovers the previous results for the stationary charge current, \ Eq.~(\ref{IPOL_final}). In the time-dependent case, the analysis is complicated by the fact that there are six energy scales, $T_c$, $\Omega$, $\varepsilon$, $\Gamma_{L/R}$, and 
$\omega_c$, the boson cut-off in the Ohmic $(s=1)$ boson spectral density $J(\omega)$, \ Eq.~(\ref{Jomegageneric}).

An expansion in lowest order of the tunnel coupling $T_c$ leads to the usual {\em Tien-Gordon result} \cite{TG63} for ac-driven tunneling, which in fact has often been used in the literature as the first starting point in the analysis of driven tunnel systems. This result is valid for 
\begin{eqnarray}
  T_c\sqrt{2+\frac{\Gamma_R}{\Gamma_L}}\ll \Omega,\Gamma_R,|\varepsilon+n\Omega|,\quad n=\pm 0,1,2,..,
\end{eqnarray}
a condition that one obtains when considering the expansion of the current in the un-driven case, \ Eq.~(\ref{current_SN96}) and which indicates that at the resonance points $\varepsilon=n\Omega$ such a  perturbation theory must break down, as is corroborated by numerical results. In this limit, one finds \cite{BAP04}
\begin{eqnarray}
  \label{eq:currentstatnew1}
  \overline{I}^{\rm TG}&\equiv&\sum_n J_n^2\left(\frac{{\Delta}}{\Omega}\right)
\left.\overline{I}_0\right|^{{\Delta} = 0}_{\varepsilon\to\varepsilon+n\Omega},
\end{eqnarray}
where the current in the driven system is expressed by a sum over  contributions of currents $\overline{I}_0$ in the un-driven case but evaluated at the side-band energies $\varepsilon+n\Omega$, weighted 
with squares of Bessel functions. 

A non-adiabatic limit is obtained for high frequencies,
\begin{eqnarray}
  \Omega\gg T_c,\varepsilon,\Gamma_R,\Gamma_L,
\end{eqnarray}
where Fourier components other than the central $n=0$ are neglected and
\begin{eqnarray}\label{currentfast}
 \bar{I} \approx \bar{I}^{\rm fast} \equiv \frac{-e\Gamma_R K_0(0) }{\Gamma_R+G_0(0)+K_0(0)\left[1+\Gamma_R/\Gamma_L\right]},
\end{eqnarray}
which in actual fact within lowest order of $T_c$ coincides with the Tien-Gordon expression, Eq.~(\ref{eq:currentstatnew1}). \ Eq.~(\ref{currentfast}) corresponds to a geometric series-like summation of an infinite number of terms $\propto T_c^2$ which is due to the integral equation structure of the underlying Master equation.

In order to systematically go beyond the Tien-Gordon approximation, Eq.~(\ref{eq:currentstatnew1}),  one has to
perform an expansion of the current in powers of $T_c^2$, which is cumbersome when done analytically but can be easily achieved by {\em truncating} the infinite matrix equation, Eq.~(\ref{eq:fourier2}), and solving it numerically. A third, analytical approximation discussed in \cite{BAP04} is based on results by Barata and Wreszinski \cite{BarataW00,Frasca} on higher order corrections to dynamical localization in a {\em closed} and  coherent driven two-level system. In that case, a third order correction of the tunnel coupling $T_c$ appears, 
\begin{eqnarray}\label{thirdorder}
  \delta T_c^{(3)}\equiv -\frac{2T_c^3}{\Omega^2}
 \sum_{n_1,n_2 \in Z} 
\frac{J_{2n_1+1}\left(\frac{\Delta}{\Omega}\right)J_{2n_2+1}\left(\frac{\Delta}{\Omega}\right)
J_{-2(n_1+n_2+1)}\left(\frac{\Delta}{\Omega}\right)}
{(2n_1+1)(2n_2+1)},
\end{eqnarray}
which can be used in order to define  a renormalized function $K_0^{(3)}(0)$,
\begin{eqnarray}
K_0^{(3)}(0)\equiv
\sum_n \left[ T_c J_n\left(\frac{{\Delta}}{\Omega}\right)  + \delta T_c^{(3)}\right] ^2
2 \mbox{\rm Re} D_{\varepsilon+n\Omega},
\end{eqnarray}
and $G_0^{(3)}(0)$ correspondingly that give rise to an expression for the tunnel current with the coupling between the dots  renormalized,
\begin{eqnarray}\label{currentthird}
  \bar{I}^{(3)} \equiv \frac{-e\Gamma_R K_0^{(3)}(0) }{\Gamma_R+G_0^{(3)}(0)+K_0^{(3)}(0)\left[1+\Gamma_R/\Gamma_L\right]}.
\end{eqnarray}

\begin{figure}[t]
  \includegraphics[width=0.45\columnwidth]{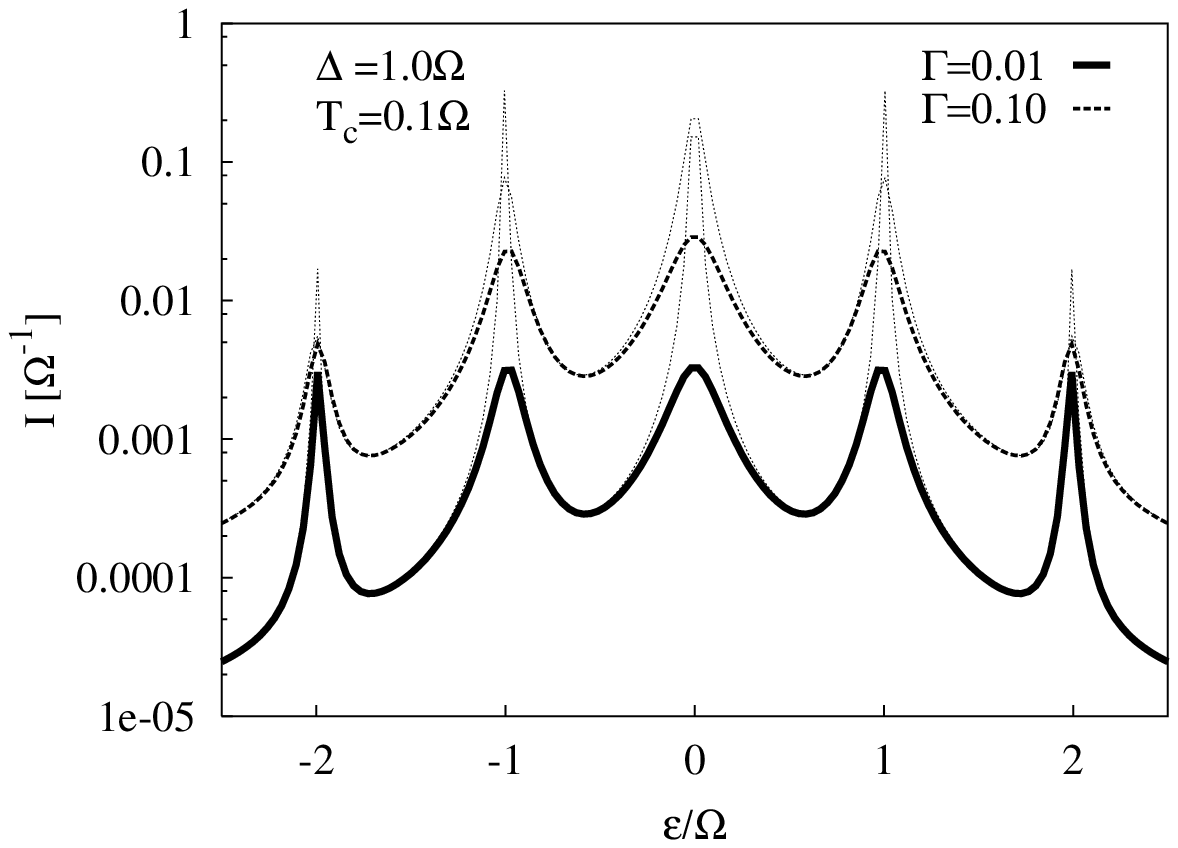}
  \includegraphics[width=0.45\columnwidth]{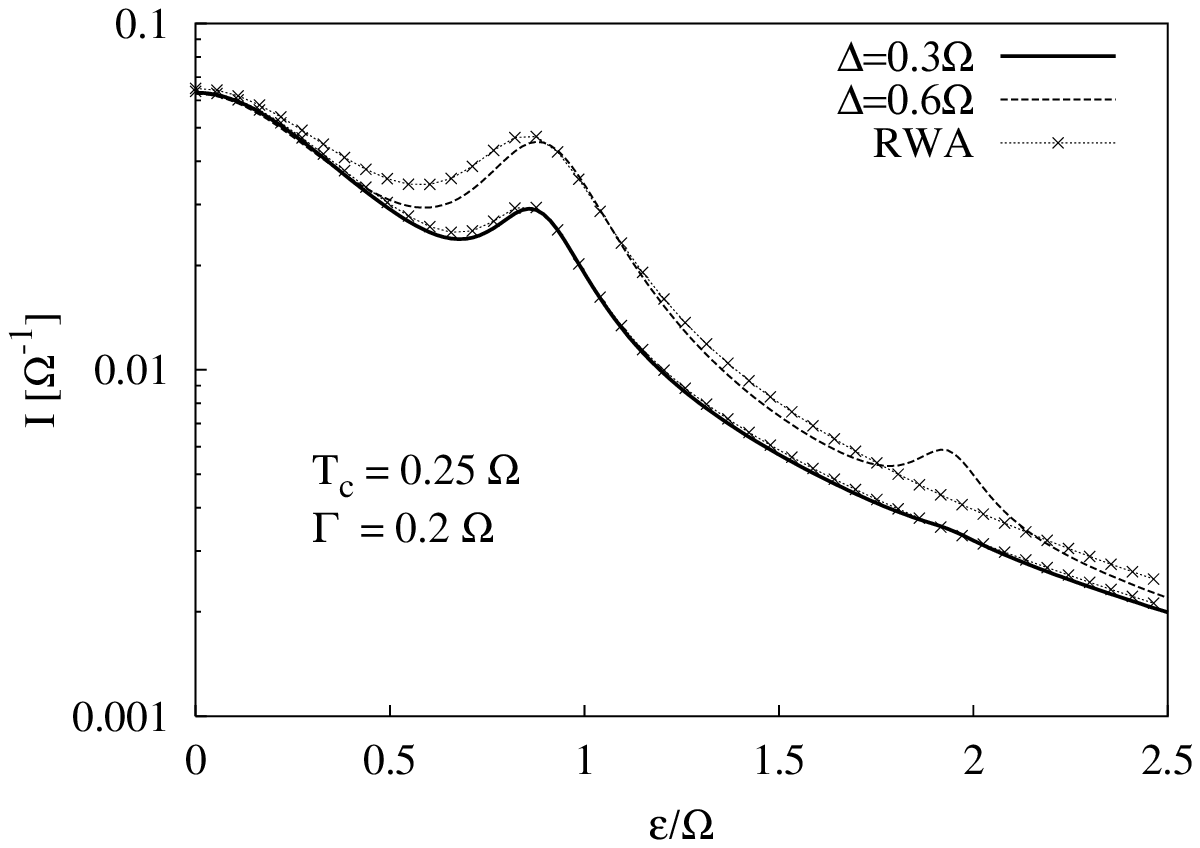}
\caption[]{\label{exact_tiengordon.eps}{\bf Left:}
Average current through double dot in Coulomb blockade regime with bias 
$\varepsilon+\Delta \sin \Omega t$. Coupling to left and right leads $\Gamma_L=\Gamma_R=\Gamma$.
Dotted lines indicate Tien-Gordon result, Eq.(\ref{eq:currentstatnew1}).
{\bf Right:} Comparison between RWA and exact result for first current side-peak.
From \cite{BAP04}.
}
\end{figure}

Fig. (\ref{exact_tiengordon.eps}), left,  shows results for the exact average stationary current and the Tien-Gordon expression, Eq.(\ref{eq:currentstatnew1}), in the coherent case $\alpha=0$ (no dissipation). Symmetric photo-side peaks appear at $\pm n \hbar \varepsilon$. The Tien-Gordon approximation overestimates the current close to these resonances, where terms of higher order in $T_c$ become important due to the non-linearity (in $T_c$) of the exact bonding and antibonding energies $\pm \sqrt{\varepsilon^2+4T_c^2}$ of the isolated two-level system.

Fig.(\ref{exact_tiengordon.eps}), right, compares the exact result with a Rotating Wave Approximation (RWA) for the first side-peak as obtained by Stoof and Nazarov \cite{SN96}, where in an interaction picture the fast-rotating terms with angular frequency $\pm \Omega$ are transformed away, and terms with higher rotation frequencies (such as $\pm 2\Omega$) are neglected. For smaller driving amplitude $\Delta$, the agreement is very good but becomes worse with increasing $\Delta$ when the position of the side-peak resonance point (which is independent of $\Delta$ in the RWA) starts to shift towards slightly larger values of the bias $\varepsilon$. For strong electric fields,  the RWA in fact is known to break down \cite{SanB04}. For example, the first corrections to the RWA in isolated two-level systems lead to the well-known Bloch-Siegert shift \cite{Allen} of the central resonance towards larger energies, which is consistent with the exact result in Fig. (\ref{exact_tiengordon.eps}).

\begin{figure}[t]
\begin{center}
  \includegraphics[width=0.475\columnwidth]{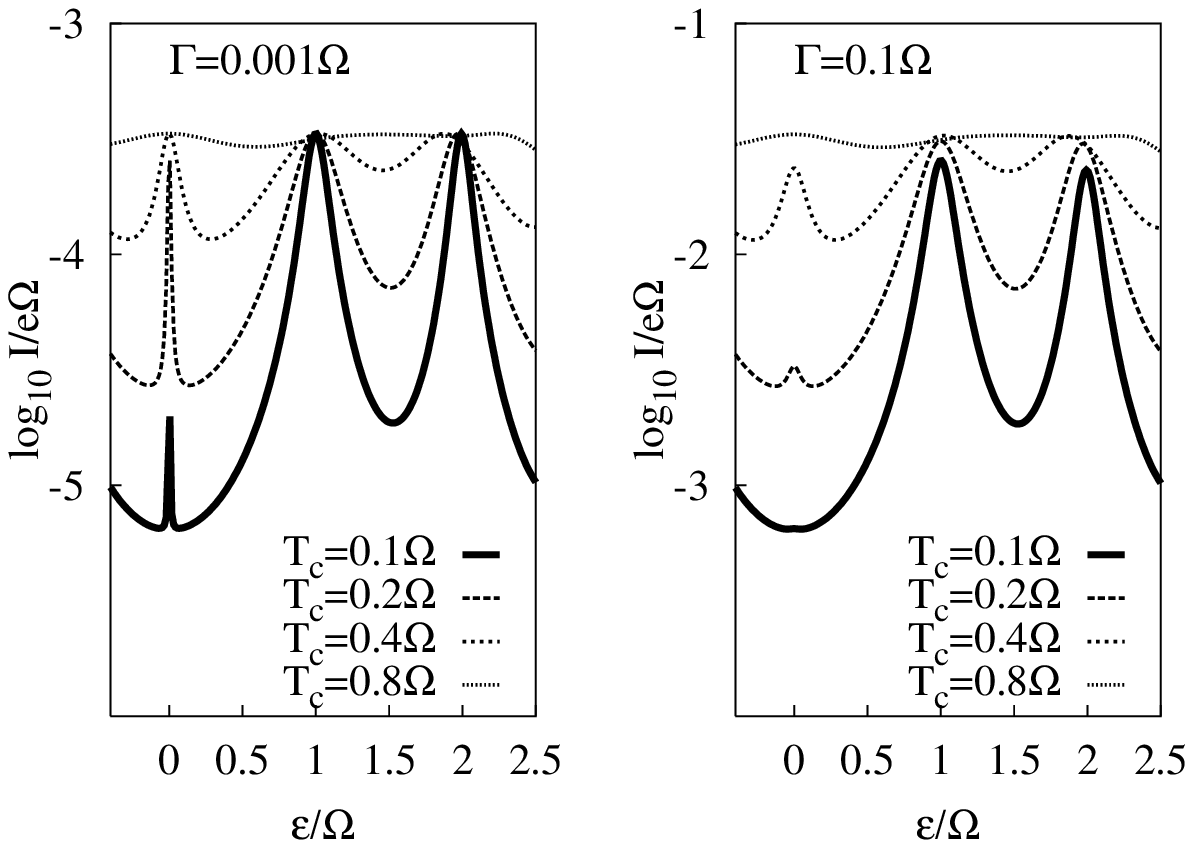}
  \includegraphics[width=0.475\columnwidth]{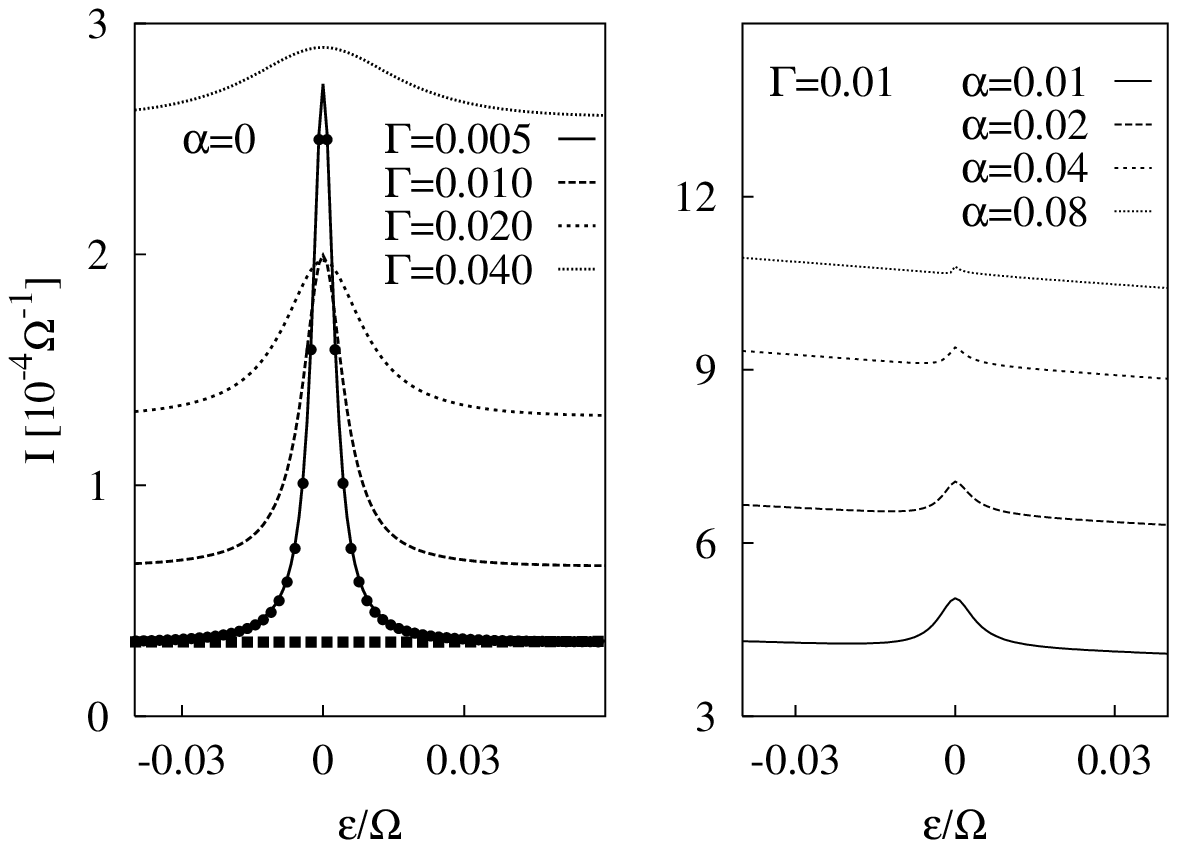}

a) \hfill b) \hfill c) \hfill d) \hfill
\end{center}
\caption[]{\label{dlregime.eps}a),b)
Average current for AC driving amplitude $\Delta=z_0 \Omega$ ($z_0$ first zero of Bessel function $J_0$) and various tunnel couplings $T_c$. Coupling to left and right leads $\Gamma_L=\Gamma_R=\Gamma$. c), d)  Central peak of average current through AC driven double quantum dot. Parameters $T_c=0.1$, $\Delta=z_0 \Omega$ (all rates in units of $\Omega$); c) coherent case $\alpha=0$ for different tunnel rates $\Gamma=\Gamma_L=\Gamma_R$, dots indicate third order results Eq.~(\ref{currentthird}), squares indicate the Tien-Gordon result Eq.~(\ref{eq:currentstatnew1}) for the case $\Gamma=0.005$, d) disappearance of central peak with increasing dissipation $\alpha$. From \cite{BAP04}.}
\end{figure}

Fig. (\ref{dlregime.eps}), a), b),  show  the average current for $\alpha=0$  in the dynamical localization  regime defined by $\Delta=z_0 \Omega$, where $z_0=2.4048...$ is the first zero of the Bessel function$J_0$. For this specific value of the ac-driving $\Delta$, to lowest order in $T_c$ the average current is strongly suppressed for $|\varepsilon|\lesssim \Omega$ as compared with the un-driven case $\Delta=0$. For small $T_c$, this suppression is  well-described by the Tien-Gordon expression (not shown ): since at $\Delta=z_0 \Omega$, the $n=0$ term in the sum Eq.~(\ref{eq:currentstatnew1}) is absent, the current is dominated by the shifted (un-driven) current contributions at bias $\varepsilon+n\Omega$ with $|n|\ge 1$, which however are very small due to the resonance shape of the un-driven current. 

Dynamical localization  (also called  Coherent Destruction of Tunneling) occurs in quantum system driven by a periodic electric field of a certain amplitude \cite{PA04,Groetal91}. An analysis in terms of Floquet states and energies shows that when two quasi-energies approach degeneracy, the time-scale for tunneling between the states diverges. For an isolated two-level system,   a monochromatic, sinusoidal field ${\varepsilon}(t)=\varepsilon + \Delta \sin(\Omega t)$, Eq.(\ref{eq:epsdef1}), leads to an effective renormalization of the coupling $T_c$ of the two levels, 
\begin{equation}
T_c \rightarrow T_{c,{\rm eff}} \equiv T_c J_0\left(\frac{\Delta}{\hbar\Omega}\right).
\label{CDT}
\end{equation}
which shows that at the first zero of the Bessel function $J_0$ the {effective tunnel splitting vanishes}, leading to a complete localization of the particle in the initial state.

The coherent suppression of the current is however {\em lifted} again very close to $\varepsilon=0$, where a small and sharp peak appears that becomes broader with increasing tunnel coupling $T_c$, but with its height being suppressed for increasing reservoir coupling $\Gamma$, cf. Fig. (\ref{dlregime.eps}) b). Fig. (\ref{dlregime.eps}) shows details for the central current peak around $\varepsilon=0$ at dynamical localization for coherent ($\alpha=0$, c) and incoherent ($\alpha>0$, d) tunneling. Again the Tien-Gordon description breaks down close to $\varepsilon=0$ where higher order terms in $T_c$ become important, in particular for $\varepsilon=0$ where the only relevant energy scale of the isolated two-level systems is $T_c$ itself. In contrast, the {\em third order approximation} Eq.~(\ref{currentthird}) reproduces very well the additional peak at $\varepsilon=0$, which indicates the importance of higher order terms in that regime. At $\varepsilon=0$, the charge between the two dots is strongly de-localized in the un-driven case, and this tunneling-induced quantum coherence persists into the strongly driven regime where its signature is a lifting of the dynamical localization close to $\varepsilon=0$. 

\begin{figure}[t]
\includegraphics[width=0.5\columnwidth]{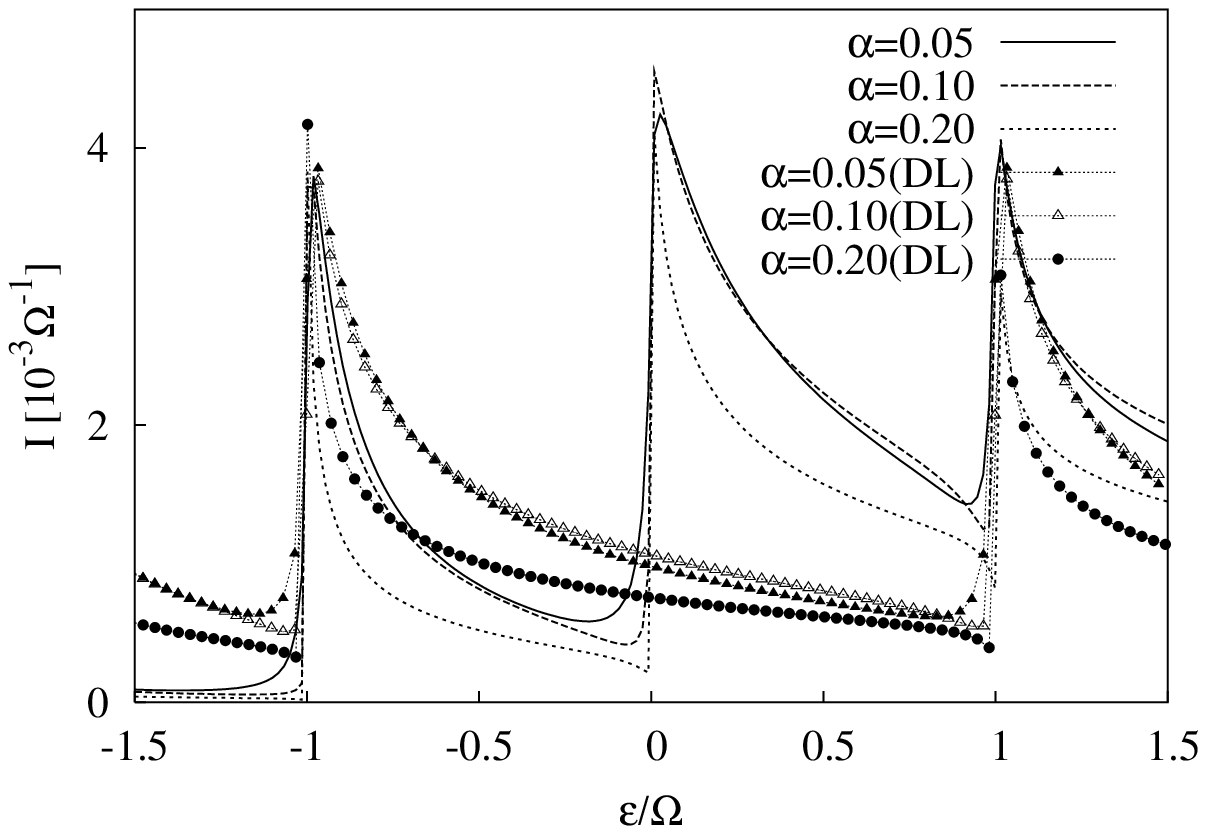}
\includegraphics[width=0.5\columnwidth]{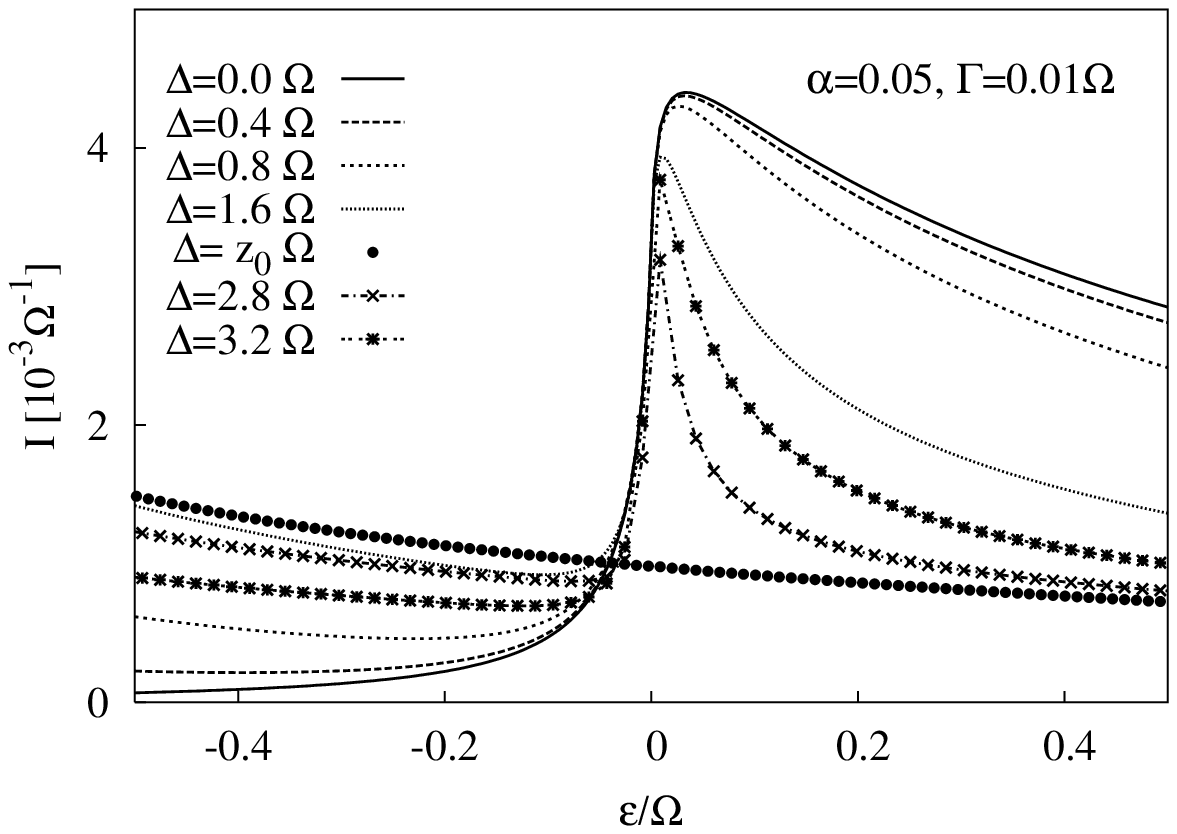}
\caption[]{\label{dissipation.eps}{\bf Left:} Average current through double dot in Coulomb blockade regime with bias $\varepsilon+\Delta \sin \Omega t$ for various Ohmic dissipation strengths $\alpha$ at zero temperature. Driving amplitude $\Delta=\Omega$ for lines without symbols, $\Delta=z_0 \Omega$ ($z_0$ first zero of Bessel function $J_0$) for lines with symbols. Tunnel coupling between dots $T_c=0.1\Omega$, bath cutoff $\omega_c=500 \Omega$, and lead tunnel rates $\Gamma_L=\Gamma_R=0.01\Omega$. {\bf Right:} Average current through driven double dot for various ac driving amplitudes $\Delta$ and fixed dissipation $\alpha=0.05$, tunnel coupling $T_c=0.1 \Omega$. From \cite{BAP04}. }
\end{figure}

Fig. (\ref{dissipation.eps}) shows the stationary current as a function of bias $\varepsilon$ for various Ohmic dissipation strengths $\alpha$ at zero temperature and finite ac driving amplitudes $\Delta$. For $\Delta=\Omega$, apart from the central resonant tunneling peak, side-bands at $\varepsilon= n\Omega$ appear which reproduce the asymmetry of the central peak around $\varepsilon=0$, cf. Fig. \ Eq.~(\ref{current_stat}) in section \ref{section_currentosc}. For ac driving amplitude $\Delta=z_0\Omega$ ($z_0$  the first zero of the Bessel function $J_0$) the current suppression strongly depends on the static bias $\varepsilon$: suppression occurs for $\varepsilon>0$ and, in general, {\em larger} values of the current for $\varepsilon<0$ as compared to the case of smaller ac amplitudes $\Delta$. A small driving amplitude $\Delta\lesssim 0.2$ nearly does not change the current at all. However,  the originally strongly asymmetric current curve is flattened out when $\Delta$ is tuned to larger values up to the dynamical localization value $\Delta=z_0\Omega$, where the ac field nearly completely destroys the strong asymmetry between the spontaneous emission ($\varepsilon>0$) and the absorption side ($\varepsilon<0$) of the current. The central $n=0$ photo-band is completely suppressed and the dominant contribution to the current stems from the $n=\pm 1$ bands. For $\varepsilon<0$, the current for $\Omega>|\varepsilon|$ is then due to photo-excitation of the electron into the first upper photo-sidebands and subsequent spontaneous emission of bosons of energy $E_1\equiv \Omega-|\varepsilon|$ to the bath. On the other hand, for $\Omega>\varepsilon>0$, photon emission blocks the current because  at $T=0$ there is no absorption of bosons from the bath. The remaining  photon absorption channel then leads to boson emission at an energy $E_2\equiv\Omega+\varepsilon$. This energy is larger as compared to the case for $\varepsilon<0$, $E_2>E_1$, and therefore has a smaller probability $P(E)\propto E^{2\alpha-1}e^{-E/\omega_c}$, cf. Eq.(\ref{PET0}), leading to a smaller current.


\section{\bf Superradiance, Large Spin Models, and Transport}\label{section_SR}
Whereas the previous section dealt with individual two-level systems and their interactions with boson and electron reservoirs, this section introduces the concept of collective effects in spontaneous emission (superradiance) and  various realizations of superradiance in mesoscopic systems. 

\subsection{Dicke Superradiance}\label{section_SRintro}
Superradiance is the collective  decay of an initially excited system of $N$ atoms due to spontaneous emission of photons. For large $N$, the emission as a function of time is not of the usual exponential form, but has the form of a very sudden peak (`flash') that occurs on a very short time-scale $\propto 1/N$ and has a  maximum $ \propto N^2$. The phenomenon of superradiance was predicted by Dicke in a seminal paper in 1954 \cite{Dic54}, shortly after his 1953 paper on spectral line-narrowing \cite{Dic53}. Both effects are  related to each other and referred to in the literature as `Dicke-effect'. In this section, we will discuss the Dicke superradiance effect only, whereas the Dicke spectral line effect will be dealt with in section \ref{section_spectral}.

The first observation of superradiance  in an optically pumped hydrogen fluoride gas by Skribanovitz, Herman, MacGillivray, and Feld \cite{Skretal73} was the starting point of considerable activities (both experimental and theoretical) since the 1970s on this collective, quantum optical effect, a good account of which is given in the text-books by Benedict {\em et al.} \cite{Benedict}, by Andreev {\em et al.} \cite{Andreev}, and the Review Article by Gross and Haroche \cite{GH82}. The wealth of physical concepts related to superradiance may in part have contributed to the quite recent revival of the effect, in particular in Solid State Physics. For example, coherent effects in semiconductors optics \cite{Ros98,Koch} have become accessible experimentally by ultrafast spectroscopy. Most prominently in semiconductor optics, the superradiance effect has been found in radiatively coupled quantum-well excitons \cite{Stretal96,Haaetal98,Stretal98} recently.

Conceptually, superradiance is the generalization of spontaneous emission from a single  to a many-body system, similar to the way that lasing is the extension of the concept of stimulated emission to a large ensemble of atoms. It has to be emphasized though, that superradiance and lasing are two different concepts which should not be confused. 

The clearest outline of the central idea behind superradiance is perhaps given in the introduction of Dicke's original paper. Let us take  a similar route here and first consider an excited atom, as described in the form of  a two-level system (ground and excited state), which can  decay due to spontaneous emission of photons. The decay rate $\Gamma$  is determined by the interaction of the atom with the light and can be calculated from the corresponding transition matrix elements ${{g}}_{{\bf Q}}$. Considering now {\em two} (instead of one) atoms at positions ${\bf r}_1$ and ${\bf r}_2$, this interaction (in dipole approximation with dipole moments $\hat{{d}}_1$ and  $\hat{{d}}_2$ of the two atoms) is proportional to the sums of terms 
$\hat{{d}}_1 e^{i({\bf Qr}_1)}$ and $\hat{{d}}_2 e^{i({\bf Qr}_2)}$, which 
interfere and thereby lead to a splitting of the spontaneous decay into a fast, `superradiant', decay channel, and a slow, `subradiant' decay channel. This splitting is called `Dicke-effect'.

Using Pauli matrices for two-level atoms, the Hamiltonian for two atoms interacting with the electromagnetic field reads \cite{Agarwal}
\begin{eqnarray}
  \label{eq:H2atoms}
  H&=&H_0+H_{\rm eph}+H_{\rm ph},\quad 
H_0\equiv \frac{1}{2}\omega_0\left(\hat{\sigma}_{z,1}+\hat{\sigma}_{z,2}\right),\quad 
H_{\rm ph}\equiv \sum_{\mathbf Qs}\omega_Qa^+_{{\bf Q}s}a_{{\bf Q}s}\\
H_{\rm eph}&\equiv&\sum_{\mathbf Qs}g_{\mathbf Qs}\left(a_{-{\bf Q}s} + a_{{\bf Q}s}^+\right)
\left[e^{i{\bf Q r}_1}\hat{\sigma}_{x,1}+e^{i{\bf Q r}_2}\hat{\sigma}_{x,2}\right], 
\end{eqnarray}
where the dipole operators are $\hat{{\bf d}}_i$ $={\bf d}\hat{\sigma}_{x,i}$, and $\hat{\sigma}_{z,i}$ and 
$\hat{\sigma}_{x,i}$ are the Pauli matrices in the $2\times 2$ space of the upper/lower level
$|\uparrow\rangle_i$,$|\downarrow\rangle_i$ of atom $i$,
\begin{eqnarray}\label{Pauli}
\hat{\sigma}_{x,i}\equiv  \begin{pmatrix} 0 & 1 \\ 1 & 0 \end{pmatrix}_{(i)},\quad
\hat{\sigma}_{y,i}\equiv \left( \begin{matrix} 0 & -i \\ i & 0 \end{matrix}\right)_{(i)},\quad
\hat{\sigma}_{z,i}\equiv \left( \begin{matrix} 1 & 0 \\ 0 & -1 \end{matrix}\right)_{(i)}.
\end{eqnarray}
Here, $\omega_0$ is the transition frequency between the upper and lower level. Furthermore, $\omega_Q=c|{\bf Q}|$ with the speed of light $c$,  and $a^+_{{\bf Q}s}$ creates a photon with wave vector ${\bf Q}$ and polarization $s$, and $g_{\mathbf Qs}=\tilde{{\bf g}}_{\mathbf Qs}{\bf d}$ is the coupling matrix element with $\tilde{{\bf g}}_{{\bf Q}s}=-i\left({2\pi cQ}/{V}\right)^{1/2}{\varepsilonv}_{{\bf Q}s}$ and light polarization vector ${\varepsilonv}_{{\bf Q},s}$ in a volume $V\to \infty$. The form \ Eq.~(\ref{eq:H2atoms}) of the Hamiltonian induces a preferential basis in the Hilbert space ${\mathcal{H}}_2=C^2 \otimes C^2$ of the two two-level systems, i.e. pseudo-spin singlet and triplet states according to 
\begin{eqnarray}
  \label{eq:dickebasistwoatoms}
  |S_0\rangle &\equiv&\frac{1}{\sqrt{2}}\left( |\uparrow\downarrow\rangle
-|\downarrow\uparrow\rangle\right)\nonumber\\
  |T_1\rangle &\equiv&|\uparrow\uparrow\rangle,\quad
  |T_0\rangle \equiv\frac{1}{\sqrt{2}}\left( |\uparrow\downarrow\rangle
+|\downarrow\uparrow\rangle\right),\quad
 |T_{-1}\rangle \equiv |\downarrow\downarrow\rangle,
\end{eqnarray}
which are a special example of $j=N/2$ Dicke states  (angular momentum states $|jm\rangle$ with $j=1$). Using this basis, one can easily calculate the non-zero matrix elements of $H_{\rm eph}$. Simple perturbation theory  (Fermi's Golden Rule) then yields  two transition rates $\Gamma_{\pm}$  for spontaneous emission of photons into a photon vacuum,
\begin{eqnarray}
  \label{eq:gammaplusminus}
  \Gamma_{\pm}(Q)=2\pi\sum_{{\bf Q}s}\frac{|g_{Qs}|^2}{2}
\left|1\pm \exp{[i{\bf Q}({\bf r}_2-{\bf r}_1)]}\right|^2
\delta(\omega_0-\omega_Q),\quad Q=\omega_0/c,
\end{eqnarray}
with superradiant decay at rate $\Gamma_{+}(Q)$ and subradiant decay at rate $\Gamma_{-}(Q)$. This splitting into two decay channels is the precursor of the more general case of $N$ radiators (ions, atoms,...), where the phenomenon is known as {\em Dicke superradiance}. Already for $N=2$, one can easily understand  how the time-dependence of the {\em collective} decay of radiators differs from the decay of single, independent  radiators:
The time dependence of the occupation probabilities ($T_1(t)$, $T_0(t)$, $T_{-1}(t)$, and $S_0(t)$) of the for states, Eq.(\ref{eq:dickebasistwoatoms}), are governed by  a set of rate equations
\begin{eqnarray}
  \label{eq:decaytimedep}
  \dot{T}_1&=&-(\Gamma_-+\Gamma_+)T_1,\quad
  \dot{S}_0=\Gamma_-(T_1-S_0)\nonumber\\
  \dot{T}_0&=&\Gamma_+(T_1-T_0),\quad
  \dot{T}_{-1}=\Gamma_-S_0+\Gamma_+T_0.
\end{eqnarray}
For simplicity, let us consider the case where the subradiant channel is completely suppressed, $\Gamma_-=0$ and $\Gamma_+=2\Gamma$, where $\Gamma$ is the emission rate of one individual radiator. This situation corresponds to the so-called {\em small--sample limit} $|{\bf Q}({\bf r}_2-{\bf r}_1)|\ll 1$ where the wave length of the emitted light is much larger than the distance between the two radiators. The rate equations can be easily solved \cite{GH82a},
\begin{eqnarray}
  \label{eq:solvetworad}
  T_1(t)&=&e^{-\Gamma_+t},\quad
  T_0(t)=\Gamma_+te^{-\Gamma_+t},\quad
  T_{-1}(t)=1-e^{-\Gamma_+t}(1+\Gamma_+t),
\end{eqnarray}
where initial conditions $T_1(0)=1$, $T_0(0)=S_0(0)=T_{-1}(0)=0$ were assumed. The total {\em coherent emission rate} $I_2(t)$ at time $t$ is the sum of the emission rates from $T_1$ and $T_0$ (the lowest level $T_{-1}$ does not radiate),
\begin{equation}
  \label{eq:ratetworad}
  I_2(t)=E_0\Gamma_+e^{-\Gamma_+t}(1+\Gamma_+t),\quad \Gamma_+=2\Gamma,
\end{equation}
where $E_0$ is a constant with dimension energy.
This has to be compared with the {\em incoherent sum} $2I_1(t)$ of the 
emission rates $I_1(t)$ from two independent radiators, which would give
$2I_1(t)=2E_0\Gamma e^{-\Gamma t}$. Not only does the superradiant decay have a rate $ \Gamma_+$ twice as large as in the incoherent case; the overall time-dependence is changed due to the term linear in $t$. For $N>2$ this change is even more drastic and leads to the superradiance `flash', see below. Note that energy conservation requires the total emitted energies to be the same in both the coherent and the incoherent case, i.e.
$  \int_0^{\infty}dt I_2(t)=\int_0^{\infty}dt 2I_1(t)\quad =2E_0$.

\subsubsection{N Atoms and the Dicke Effect}\label{section_Natoms}
The generalization of the interaction Hamiltonian in dipole approximation, \ Eq.~(\ref{eq:H2atoms}), to $N$ atoms at positions ${\bf r}_i$  with dipole moments $\hat{{\bf d}}_i={\bf d}\hat{\sigma}_{x,i}$  is given by the $-{\bf d E}$ interaction,
\begin{equation}\label{Nion}
H_{\rm eph}=\sum_{{\mathbf Qs}}{{g}}_{{\bf Q}s}
\left(a_{-{\bf Q}s} + a_{{\bf Q}s}^+\right)
\sum_{i=1}^N
{\hat{\sigma}}_{x,i}\e^{i{\bf Qr}_i}.
\end{equation}
In order to describe the full interaction of bound charges with the electromagnetic field, one has to add a self-energy term $H_{\rm self}$ on top of \ Eq.~(\ref{Nion}), as was first shown by Power and Zienau  and discussed in detail in the book by Agarwal \cite{Agarwal}. Alternatively, one can describe  the interaction in the ${\bf p A}$ gauge but then has to include electrostatic dipole-dipole interaction terms \cite{GH82}. Depending on various conditions, these terms are important for a realistic description of $N$-atom systems. They lead, together with the $N$-particle Lamb shifts from  the real parts in the Master equation \cite{GH82} derived from \ Eq.~(\ref{Nion}), to a modification of the `pure' superradiance scenario as  first discussed by Dicke. The `pure' superradiance case, however, is conceptually most transparent and relevant for a generalizations of collective decay to other types of interactions (e.g. with phonons). 

Assume a situation in which {\em all} the phase factors $e^{i{\bf Qr}_i}$ in \ Eq.~(\ref{Nion}){} are identical (say, unity for simplicity), for example when the maximal distance between any two atoms is much less than a typical wave length. Then, the coupling $H_{\rm eph}$ to the photon field does no longer depend on the individual coordinates of the atoms but  only on the {\em collective} pseudo-spin coordinate. One has to add up $N$ pseudo-spins $1/2$ to a single, large pseudo-spin which is described by angular momentum operators
\begin{eqnarray}
  \label{eq:angular}
  J_\alpha &\equiv& \frac{1}{2}\sum_{i=1}^N {\hat{\sigma}}_{\alpha,i},\quad \alpha=x,y,z\\
J_{\pm} &\equiv&  J_x\pm i J_y,\quad [J_z,J_{\pm}]=\pm J_{\pm},\quad [J_+,J_-]=2J_z
\end{eqnarray}
with angular momentum $j$ eigenstates which in this context are called Dicke states $| jm;\lambda \rangle$ defined via
\begin{eqnarray}
\label{eigenstates}
J^{2}  | jm;\lambda \rangle&=&  j(j+1)   | jm;\lambda \rangle,\quad
J_z  | jm;\lambda \rangle=  m   | jm;\lambda \rangle,
\end{eqnarray}                 
where $J^2=J_x^2+J_y^2+J_z^2$ is the total angular momentum squared and $j$ is sometimes called cooperation number \cite{Dic54}. Here, $\lambda$ denotes  additional quantum numbers apart from $j$ and $m$ which are necessary to completely label {\em all} the states of the $2^N$ dimensional Hilbert space $  {\mathcal{H}}_N=(C^2)^{\otimes N}$. For $N$ identical two-level systems, the additional  quantum numbers are provided by the permutation group $P_N$, as was shown by Arecchi, Courtens, Gilmore, and Thomas \cite{Areetal72}. The decomposition of the total Hilbert space ${\mathcal{H}}_N$  into irreducible representations ${\mathcal D}^j$ for angular momentum $j$ with dimension $2j+1$, and permutations of dimension ${\rm dim} \Gamma^{\lambda_1,\lambda_2}= {\lambda_1 +\lambda_2 \choose \lambda_2} - {\lambda_1 +\lambda_2 \choose \lambda_2-1}$, is reflected in the dimension formula \cite{Areetal72}
\begin{eqnarray}
  \sum_{2j=\lambda_1-\lambda_2\ge 0} {\rm dim} {\mathcal D}^j{\rm dim} \Gamma^{\lambda_1,\lambda_2}=2^N.
\end{eqnarray}
More generally, a system of $N$ atom $n$-level systems can be effectively described group-theoretically by the standard Young tableaux which characterize the irreducible representations of the group $S_N \otimes SU(n)$, a short summary of which is given in the paper by Keitel, Scully and S\"ussmann on triggered superradiance \cite{KSS92}.

In discussing superradiance, one usually considers a subspace of constant $j$ which is also invariant under permutation operations, omitting the labels $\lambda$ and thereby dealing with a constant angular momentum Hilbert space \cite{Areetal72}. The total Hamiltonian  for $N$ atoms interacting with the photon field then simply reads
\begin{equation}
  \label{eq:Htot}
  H_{\rm Dicke}=\omega_0J_z+\hat{A}J_x+H_{\rm ph},
\end{equation}
where $\hat{A}$ is a photon operator and $H_{\rm ph}$ the photon field Hamiltonian. Radiative transitions are due to transitions between Dicke states with the selection rule $m\to m\pm 1$, leaving $j$ constant. Using
\begin{eqnarray}\label{cjm_def}
  J_{\pm} | jm\rangle =c_{jm}^\pm  | j m\pm 1\rangle,\quad c_{jm}^\pm \equiv
\sqrt{j(j+1)-m(m\pm 1)},
\end{eqnarray}
and considering the initial state $|jj\rangle$ with $j=N/2$, spontaneous emission leads to a decrease of the quantum number $m$ step by step, with the corresponding emission intensity $I_{jm}$ from a state $| jm\rangle$ given by 
\begin{equation}
\label{emission}
I_{jm}=\hbar\omega_0\Gamma (j+m) (j-m+1),
\end{equation}
where $\Gamma$ is the spontaneous emission rate of one {\em single} atom from its excited state. Although this simplified argument only uses transition rates between pure states, it grasps the essential physics:  in the course of the spontaneous decay starting from the initial state, the intensity $I_{jm}$ reaches a maximum proportional to $N^{2}$ at the {\em Dicke peak} $m=0$, which is abnormally large in comparison with the intensity $N\Gamma \hbar\omega_0$ of the radiation of $N$  independently decaying radiators.

The time dependence of the emission peak can be obtained from a simple quasi-classical argument that regards the quantum number $m$ as a time-dependent, classical quantity. The energy of  an ensemble of identical atoms is $H_{0}= \omega _{0}J_{z}$. Equating the average energy loss rate, $-\hbar\omega _{0}\dot{m}(t)$, with the radiated intensity, Eq.~(\ref{emission}), one obtains an equation of motion for $m(t)$, 
\begin{eqnarray}\label{superradianceclassic}
\label{M(t)}
-\frac{d}{dt}m(t)&=&\Gamma \left[ j+m(t)\right)\left(j-m(t)+1\right]
\end{eqnarray}
The solution of this equation gives the hyperbolic secant solution to the superradiance problem, that is a time-dependent intensity 
\begin{equation}
\label{hyperbolic}
I(t)=I_{0}\frac{N^{2}}{2\cosh^{2}\left(N\Gamma [t-t_{d}]/2\right)},
\end{equation}
where the delay time $t_{d}$ depends on the initial condition at time $t=0$. As was discussed by Gross and Haroche \cite{GH82}, the quasi-classical description of the decay process holds if the system is prepared initially in a state $| jm_{0}\rangle$ with a large number of photons already emitted. If one starts from the totally inverted state $| jj\rangle$, the initial time evolution is dominated by strong quantum fluctuations (the phases of the single atoms are completely uncorrelated) which are not described by \ Eq.~(\ref{hyperbolic}).

The Dicke peak occurs on a short time scale $\sim 1/N$ and consists of photons with different wave vectors {\bf Q}. The mean number $N_{\bf {Q}}(t)$ of photons as a function of time can be calculated exactly in the small-sample limit of superradiance where the phase factors $e^{i{\bf Qr}_i}$ in \ Eq.~(\ref{Nion}) are neglected. For example, for the case of $N=2$ atoms, one finds \cite{Agarwal}
\begin{equation}
  \label{eq:2ionphotons}
  N_{\bf {Q}}(t\to\infty)=\frac{2|g_{\bf Q}|^2\left[(\omega_0-|{\bf Q}|c)^2+40\Gamma^2\right]}
{\left[(\omega_0-|{\bf Q}|c)^2+16\Gamma^2\right]\left[(\omega_0-|{\bf Q}|c)^2+4\Gamma^2\right]},
\end{equation}
where $\Gamma$ is the decay rate of one single atom. In reality, the small sample limit is never reached exactly, and instead of the collective operators $J_z$ and $J_{\pm}$, one introduces ${\bf Q}$-dependent operators $J_{\pm}({\bf Q})\equiv\sum_{i=1}^N J_{\pm}\exp{i({\bf Qr}_i)}$. An initial excitation with radiation in the form of a plane wave with wave vector ${\bf Q}$ then leads to a collective state $\propto J_{+}({\bf Q})^{p}|j,-j\rangle$ for some $p$. Subsequent spontaneous emission of photons with wave vector ${\bf Q}$ then is collective, conserves $j$ and decreases $m\to m-1$, while emission of photons with wave vector ${\bf Q}'\ne {\bf Q}$ can change $j$.

As a transient process, superradiance only occurs if the observation time scale $t$ is shorter than a {\em dephasing time scale} $\tau_{\phi}$ of processes that destroy phase coherence, and longer than the time $\tau $ which photons need to escape from the optical active region where the effect occurs \cite{Andreev}, such that recombination processes are unimportant. It is clear from the discussion so far that Dicke superradiance is a dissipative process and generalizes the Wigner-Weisskopf theory of spontaneous emission of a single atom to the many atom case. In this approach, the photon system itself is in its vacuum state throughout the time evolution. Any photon once emitted escapes from the system and thereby leaves  no possibility of re-absorption of photons. 

In the Master equation description of superradiance as a dissipative process, the back-action of the pseudo spin 
onto the  boson bath is usually disregarded. In a somewhat complementary approach, the dynamics of the photon field is treated on equal footing with the spin dynamics, and one has to solve the coupled Maxwell-Bloch equations, i.e. the Heisenberg equations of motion in some decoupling approximation, of the total system (spin + photon field). This approach is more suitable for the description of propagation effects for, e.g., an initial light pulse that excites the system, or re-absorption effects.

The condition 
\begin{equation}
\label{condition}
\tau \ll t \ll \tau_{\phi}, \Gamma ^{-1},
\end{equation}
determines the superradiant regime, together with the last inequality which involves $\Gamma^{-1}$, the time scale for the decay of an {\em individual} atom. For times $t$ much larger than the dephasing time $\tau_{\phi}$, there is a transition to the regime of {\em amplified spontaneous emission} \cite{Benedict}. In fact, the restriction Eq.~(\ref{condition}) of the {\em time-scale} for the  superradiant process can be seen in analogy to the restriction $l\ll L \ll L_{\phi}$ defining the {\em length scale} $L$ of a {\em mesoscopic system} where physics occurs between a microscopic (e.g. atomic) length scale $l$ and a dephasing length $L_{\phi}$ \cite{Imry}. Within this analogy, a superradiant system  can be called `mesoscopic in time'. 

\subsubsection{Sub- and Superradiance for $N=2$ Trapped Ions}\label{section_DeVB96}
DeVoe and Brewer \cite{DeVB96} measured the spontaneous emission rate of two ions as a function of the ion-ion distance in a trap of planar geometry which was strong enough to bring the ions (Ba$_{138}^+$) to a distance $d$ of the order of 1$\mu$m of each other.  The idea of their experiment was to determine $\Gamma_{\pm}(Q)$, Eq. (\ref{eq:gammaplusminus}), and to compare it to the spontaneous emission rate $\Gamma_{0}(Q)$ of a {\em single} ion within the same setup. This was done in a transient technique by exciting the ion molecule by a short laser pulse and recording the subsequent signal, i.e. the time of arrival of spontaneously emitted photons.

\begin{figure}[t]
\includegraphics[width=0.3\textwidth]{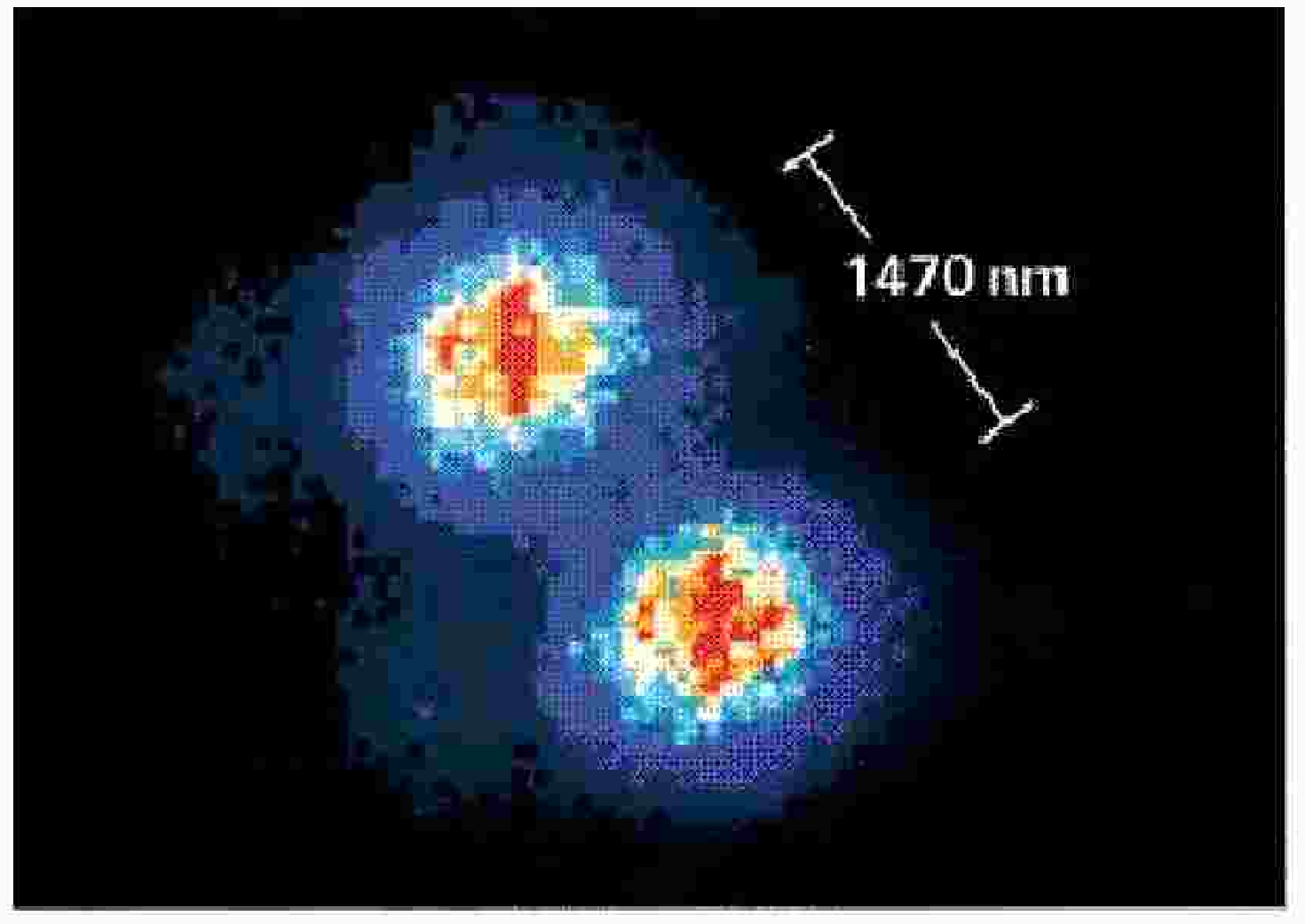}
\includegraphics[width=0.35\textwidth]{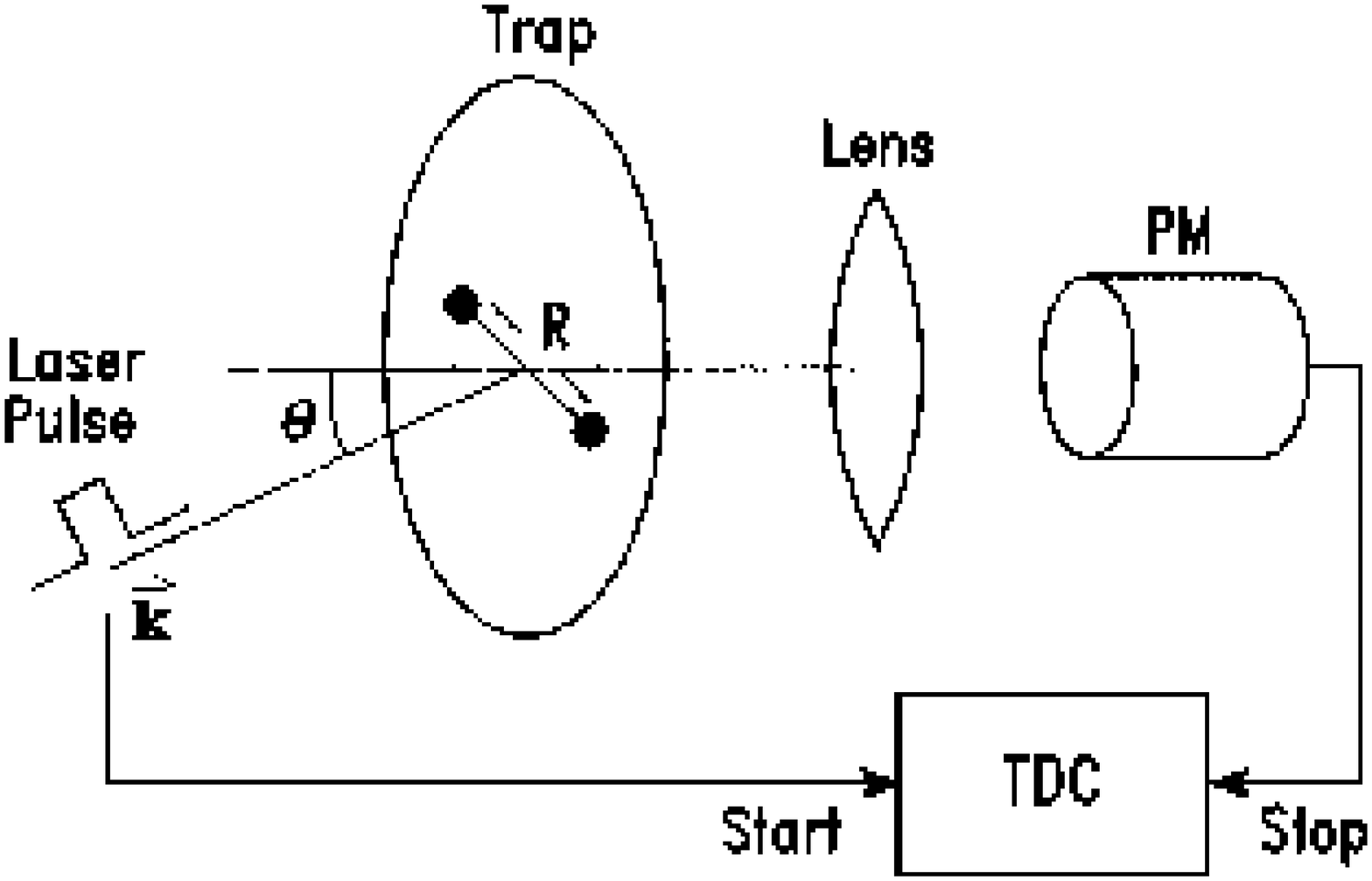}
\includegraphics[width=0.35\textwidth]{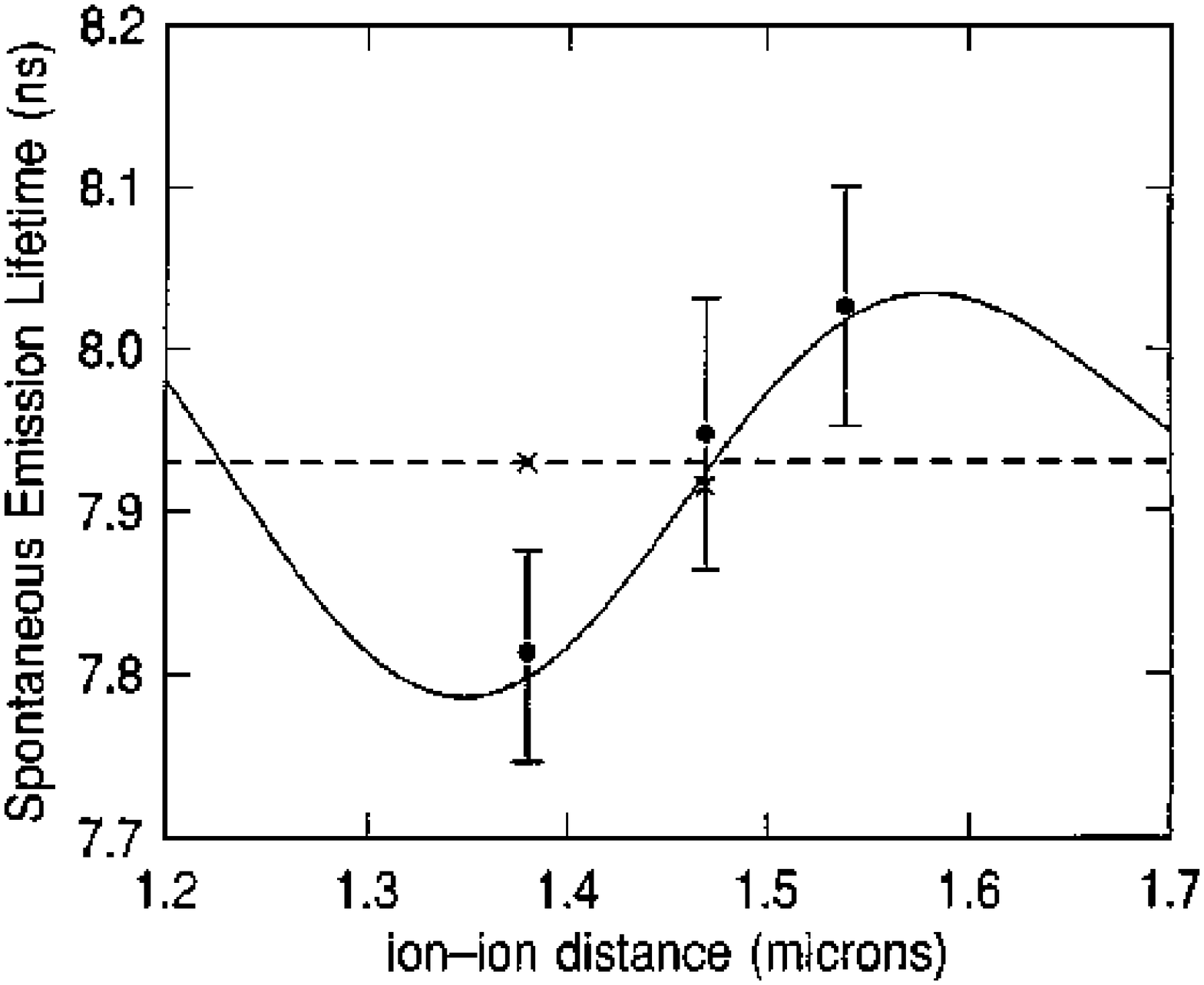}
\caption[]{\label{DeVB96fig3.eps}Double ion trap experiment by DeVoe and Brewer \cite{DeVB96}. The two-ion molecule is confined within a 80$\mu$m radius planar trap ({\bf left}) and excited with a laser pulse ({\bf center}). The time-to-digital converter (TDC) records the time of arrival of spontaneously emitted photons. {\bf Right:} Comparison of theory, Eq. (\ref{eq:subsuper}), and measured data for the identification of sub- and superradiance (Dicke effect). A laser beam excites the system at $t=0$; the start of the exciting pulse and the arrival of the spontaneous photons are recorded on a time to digital converter, which is fitted to an exponential decay. The dashed line indicates the life-time of a single ion in the same trap. Full circles with error bars are data for laser polarization perpendicular to the axis connecting the two ions, crosses are for parallel polarization. The points below the dashed line belong to the superradiant decay channel, whereas the points above the dashed line indicate belong to the subradiant channel. From \cite{DeVB96}.} 
\end{figure} 

It turned out that the best way to distinguish between the sub- and the superradiant decay channel was to choose the initial states of the system as the two states $S_0$ (singlet) and $T_0$ (triplet), which yield the subradiant and the superradiant emission rate, respectively. This was achieved by coherent excitation of the two-ion molecule, exciting dipole moments in the two ions with a phase difference of $0$ or $\pi$. Due to level degeneracy of the relevant $6^2P_{1/2}$ to $6^2S_{1/2}$ transition and due to loss of coherence because of  micro-motion Doppler shifts, the theoretical value for the factor $\alpha$ in the explicit form of the rates,
\begin{eqnarray}
  \label{eq:subsuper}
  \Gamma(Q)_{\pm}=\Gamma_0(Q)\left[1\pm\alpha\frac{\sin(Qd)}{(Qd)}\right],
\end{eqnarray}
 turned out to be
$\alpha = 0.33$. Diffraction limited images of the molecule, viewed through a window with a microscope,
provide the information on the distance between the ions \cite{DeVB96}. 

Measurements of the spontaneous rate $\Gamma$ at three different ion distances turned out
to be in good agreement with the (parameter free) theoretical prediction \cite{Benedict},
Eq.(\ref{eq:subsuper}). The data (statistical and systematic tests 
were performed) were averaged over a large number of runs.

\subsection{Dicke Effect in Quantum Dot Arrays}\label{section_Dicke_array}

The prominent role that phonon emission plays in transport through double quantum dots has been discussed in section \ref{section_transport}. {\em Collective} phonon emission effects and their impact on quantum transport were discussed for arrays of double quantum dots by Brandes, Inoue, and Shimizu \cite{BIS98}, and most explicitly for the case of $N=2$ double quantum dots interacting via a common phonon environment by Vorrath and Brandes \cite{VB03}. In the latter case, {\em charge} wave function entanglement occurs in a preferred formation  of either a  (pseudo) singlet or triplet configuration (depending on the internal level splittings of the dots and the coupling to electron reservoirs), which is  a realization of the Dicke effect in a stationary state of quantum transport.

\begin{figure}
\centering
\psfrag{Gl}{$\Gamma_L$}
\psfrag{Gr}{$\Gamma_R$}
\psfrag{Tc}{$T$}
\psfrag{e1}{$\varepsilon_L$}
\psfrag{e2}{$\varepsilon_R$}
\psfrag{Gl1}{$\Gamma_{L,1}$}
\psfrag{Gr1}{$\Gamma_{R,1}$}
\psfrag{Tc1}{$T_{1}$}
\psfrag{Gl2}{$\Gamma_{L,2}$}
\psfrag{Gr2}{$\Gamma_{R,2}$}
\psfrag{Tc2}{$T_{2}$}
\includegraphics[width=0.3\textwidth]{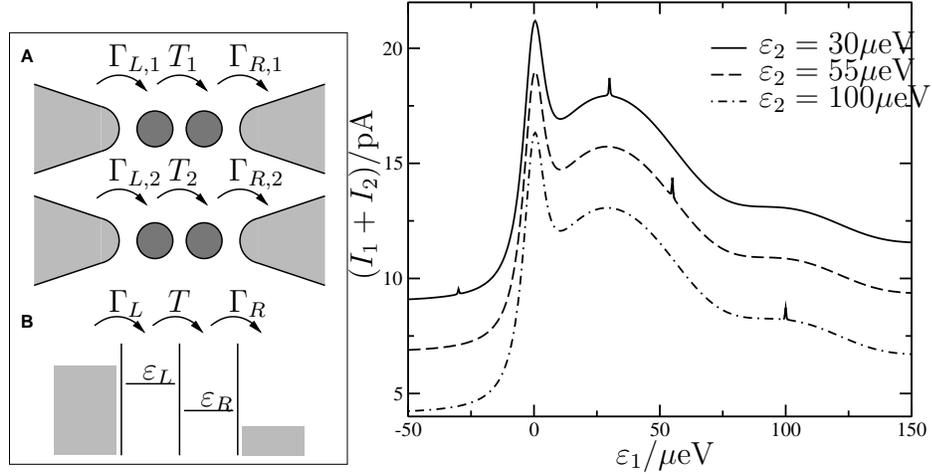}
\psfrag{I12}{\hspace*{-8mm}$(I_1+I_2)/$pA}
\psfrag{e1}{\hspace*{-5mm}$\varepsilon_1/\mu$eV}
\psfrag{eps30}{$\varepsilon_2=30\mu$eV}
\psfrag{eps55}{$\varepsilon_2=55\mu$eV}
\psfrag{eps100}{$\varepsilon_2=100\mu$eV}
\includegraphics[width=0.5\textwidth]{FI_sw.eps}
\caption{\label{VB_Fig1.eps} {\bf Left:} (A) $N=2$ charge qubit register with two double quantum dots coupled to independent electron leads. (B) Energy diagram of one individual double dot. {\bf Right:} Total current through two double quantum dots as a function of the bias $\varepsilon_1$,  parameters are $T_{c,1}\!=\!T_{c,2}\!=\!3\mu$eV, $\Gamma_{L,1}\!=\!\Gamma_{R,1}\!=\!  \Gamma_{L,2}\!=\!\Gamma_{R,2}\!=\!0.15\mu$eV, and for the spectral function $\alpha\!=\!0.005$, $T\!=\!23$mK, $\omega_d\!=\!10\mu$eV and $\omega_c\!=\!1$meV. From \cite{VB03}.}
\end{figure}

\subsubsection{Model and Master Equation for $N$  Double Quantum Dots}
The model \cite{VB03} describes  a `register' of $N$ double quantum dots, cf. Fig. \ref{VB_Fig1.eps}, coupled to independent left and right electron reservoirs as well as a common phonon reservoir, with the crucial assumption of the `small sample' limit, i.e., identical  electron-phonon matrix elements $\alpha_{\bf Q}^{L}$, $\alpha_{\bf Q}^{R}$  in the generalization of \ Eq.~(\ref{newHdp}){} to $N$ double quantum dots,
\begin{eqnarray}\label{newHdpN}
{\mathcal H}_{\rm dp}^N=\sum_{i=1}^N\sum_{\bf Q} 
\left(\alpha^L_{\bf Q} \hat{n}_{L,i} +\alpha^R_{\bf Q} \hat{n}_{R,i} \right)
\left(a_{-{\bf Q}} +a_{{\bf Q}}^{\dagger}\right),
\end{eqnarray}
where $\hat{n}_{L/R,i}$ refers to the number operator for the left/right level in the $i$th double dot.
Correspondingly, the other parts of the  Hamiltonians ${\mathcal H}_{\rm dot}$, \ Eq.~(\ref{hdot}){}, and ${\mathcal H}_{\rm res}$ and ${\mathcal H}_V$, \ Eq.~(\ref{Hresdef}){},  are generalized to their respective sums over the register index $i$. Ideally, a stacked layer of closely spaced double dots on top of each other would be a realization of the small sample limit. Phonon mediated collective effects between the members of the register should persist as long as a description in terms of a few  many-body states relevant for transport is possible. In \cite{VB03}, the Coulomb correlations between the double dots were neglected for simplicity.

The Master equation for the reduced density operator $\rho(t)$ of the register in Born-Markov approximation was derived in analogy to section \ref{section_transport}, 
\begin{equation}
\label{master_operator}
\begin{split}
&\dot{\rho} (t) 
  = i \; \sum_{i=1}^N \Big\{ \Big[\,\rho(t)\,, \varepsilon_{L,i} \, \hat{n}_{L,i} 
  + \varepsilon_{R,i} \, \hat{n}_{R,i} + T_{i} (\hat{p}_i + \hat{p}_i^{\dagger}) \Big] \\
&+ \frac{\Gamma_{L,i}}{2} \Big(
  2 \, \hat{s}_{L,i}^{\dagger}\,\rho(t)\,\hat{s}_{L,i}
  - \,\hat{s}_{L,i}\,\hat{s}_{L,i}^{\dagger}\,\rho(t)
  - \,\rho(t)\,\hat{s}_{L,i}\,\hat{s}_{L,i}^{\dagger} \Big) \\
&+ \frac{\Gamma_{R,i}}{2} \Big(
  2 \, \hat{s}_{R,i}\,\rho(t)\,\hat{s}_{R,i}^{\dagger}
  - \hat{s}_{R,i}^{\dagger}\,\hat{s}_{R,i}\,\rho(t)
  - \rho(t)\,\hat{s}_{R,i}^{\dagger}\,\hat{s}_{R,i} \Big) \Big\} \\
&-  \sum_{i,j} \, \Big\{ \Big[(\hat{n}_{L,i}\!-\!\hat{n}_{R,i}), \hat{A}_j\, \rho(t) \Big]
          \!-\! \Big[(\hat{n}_{L,i}\!-\!\hat{n}_{R,i}), \rho(t) \hat{A}_j^{\dagger} \Big] \Big\}\\
\hat{A}_j &\equiv  \frac{2 T_{j}}{\Delta_i^2} \,
      \Big( \,  2T_{j} \Gamma_{C,j} (\hat{n}_{L,j}\!-\!\hat{n}_{R,j}) 
    - \Gamma_{C,j}\, \varepsilon_j \, (\hat{p}_j + \hat{p}_j^{\dagger})
      + i\,  \Delta_j \Gamma_{S,j} (\hat{p}_j - \hat{p}_j^{\dagger})\, \Big),
\end{split}
\end{equation}
where $T_i$ is the tunnel coupling within double dot $i$, $\Delta_i=(\varepsilon_i^2+4 T_{i}^2)^{1/2}$, and
the inelastic rates are 
\begin{equation}
\begin{split}
\Gamma_{C,i} &\equiv 
    \frac{\pi}{2}\;J(\Delta_i)\,
      \coth \!\Big(\frac{\beta \Delta_i}{2}\Big), \quad
\Gamma_{S,i} &\equiv 
    -i \;\frac{\pi}{2}\;J(\Delta_i),
\end{split}
\end{equation}
with the spectral function $J(\omega)$ of the bosonic (phononic) environment, cf. \ Eq.~(\ref{Jomega}){}. The mixed terms $i\ne j$ in Eq.~(\ref{master_operator}) lead to collective effects (sub- and superradiance) in the stationary current through the system. The dimension of the density matrix scales as $9^N$ whence analytical solutions are cumbersome but were calculated for $N=2$ by Vorrath {\em et al.} \cite{VB03} for limiting cases. In general, for large $N$ even a numerical solution of  Master equations like \ Eq.~(\ref{master_operator}) becomes non-trivial. Special numerical techniques like Arnoldi iteration have been shown to be advantageous in this case \cite{FNJ04}. 

\subsubsection{Cross Coherence and Current Superradiance for $N=2$ Double Quantum Dots}
For $N=2$, the currents through double  dots $1$ and $2$ are expressed in terms of the matrix elements $\rho_{j\,i\,i'\,j'} = \;  {}_2 \!\bra{j} \otimes {}_1 \!\bra{i}\,\rho \,\ket{i'}_1 \otimes \ket{j'}_2$ ($ \quad i,j \in \{ L, R, 0\}$) of the density operator and read 
\begin{equation}
\label{current}
\begin{split}
I_1 &= - {2 eT_{1} } 
\mbox{Im} \big\{ \rho_{LRLL} + \rho_{RRLR} + \rho_{0RL0} \big\},\quad
I_2 &= - {2 eT_{2} } 
\mbox{Im} \big\{ \rho_{RLLL} + \rho_{RRRL} + \rho_{R00L} \big\},
\end{split}
\end{equation}
The numerical solution of Eq.~(\ref{master_operator}) yields the  stationary current as a function of the bias $\varepsilon_1$ in the first double dot while the bias $\varepsilon_2$ in the second is kept constant, as shown in Fig.~\ref{VB_Fig1.eps}. The overall shape of the current is very similar to the case of one individual double quantum dot (cf. section \ref{section_transport}), with its strong elastic peak around $\varepsilon_1=0$ and a broad inelastic shoulder for $\varepsilon_1>0$, but a new feature appears here in form of an additional peak at resonance $\varepsilon_1\!=\!\varepsilon_2$. This is due to the simultaneous coupling of both double dots to the same phonon environment, which induces an effective interaction between the two double quantum dots. The analysis of the effect starts from the observation that in spite of the large size of the  density matrix, this interaction is connected to six matrix elements (and their complex conjugates) only, i.e., $\rho_{RLLL}$, $\rho_{LRLL}$, $\rho_{RRLR}$, $\rho_{RRRL}$, and the two `cross coherence' matrix elements
\begin{equation}\label{crossdefinition}
\rho_{RLRL} = \langle p_1^{\dagger} p_2 \rangle,\quad 
\rho_{RRLL} = \langle p_1 p_2 \rangle.
\end{equation}
An approximate solution is then obtained by neglecting the cross interaction between the double dots in all but those six matrix elements, leading to an expression for the current {\em change} in double dot $1$, 
\begin{equation}
\label{Delta_current}
\Delta I_1 = \frac{2 e \,T_{1}\, \gamma_2}{\hbar \,\varepsilon_1}\,
 \Big( \textrm{Re}\langle p_1^{\dagger} p_2 \rangle 
  -\textrm{Re}\langle p_1 p_2 \rangle \Big),
\end{equation}
which is proportional to the (real parts of the) cross coherences. The increase of the current at $\varepsilon_1\!=\!\varepsilon_2$ is due to a corresponding peak of the $\langle p_1^{\dagger} p_2 \rangle$.

\begin{figure}
\centering
\psfrag{g}{$g$}
\psfrag{DI1}{\hspace*{-5mm}$\Delta I_1 /$pA}
\psfrag{T-S}{\hspace*{-5mm}$p_{T_0}-p_{S_0}$}
\psfrag{bilanz}{Rate equation}
\psfrag{num}{Master equation}
\includegraphics[width=0.4\textwidth]{F_bilanz_glg_sw.eps}
\psfrag{e1}{\hspace*{-4mm}$\varepsilon_1/\mu$eV}
\psfrag{I1}{\hspace*{-5mm}$I_1/$pA}
\psfrag{ee1}{\hspace*{-4mm}$\varepsilon_1/\mu$eV}
\psfrag{II1}{\hspace*{-10mm}
 $\langle \textrm{P}_{T_0} \rangle 
  - \langle \textrm{P}_{S_0}\rangle$}
\psfrag{Gr15}{$\Gamma_{R,2}= 0.15$}
\psfrag{Gr06}{$\Gamma_{R,2}= 0.06$}
\psfrag{Gr02}{$\Gamma_{R,2}= 0.02$}
\psfrag{Gr00}{$\Gamma_{R,2}= 0.00$}
\includegraphics[width=0.4\textwidth]{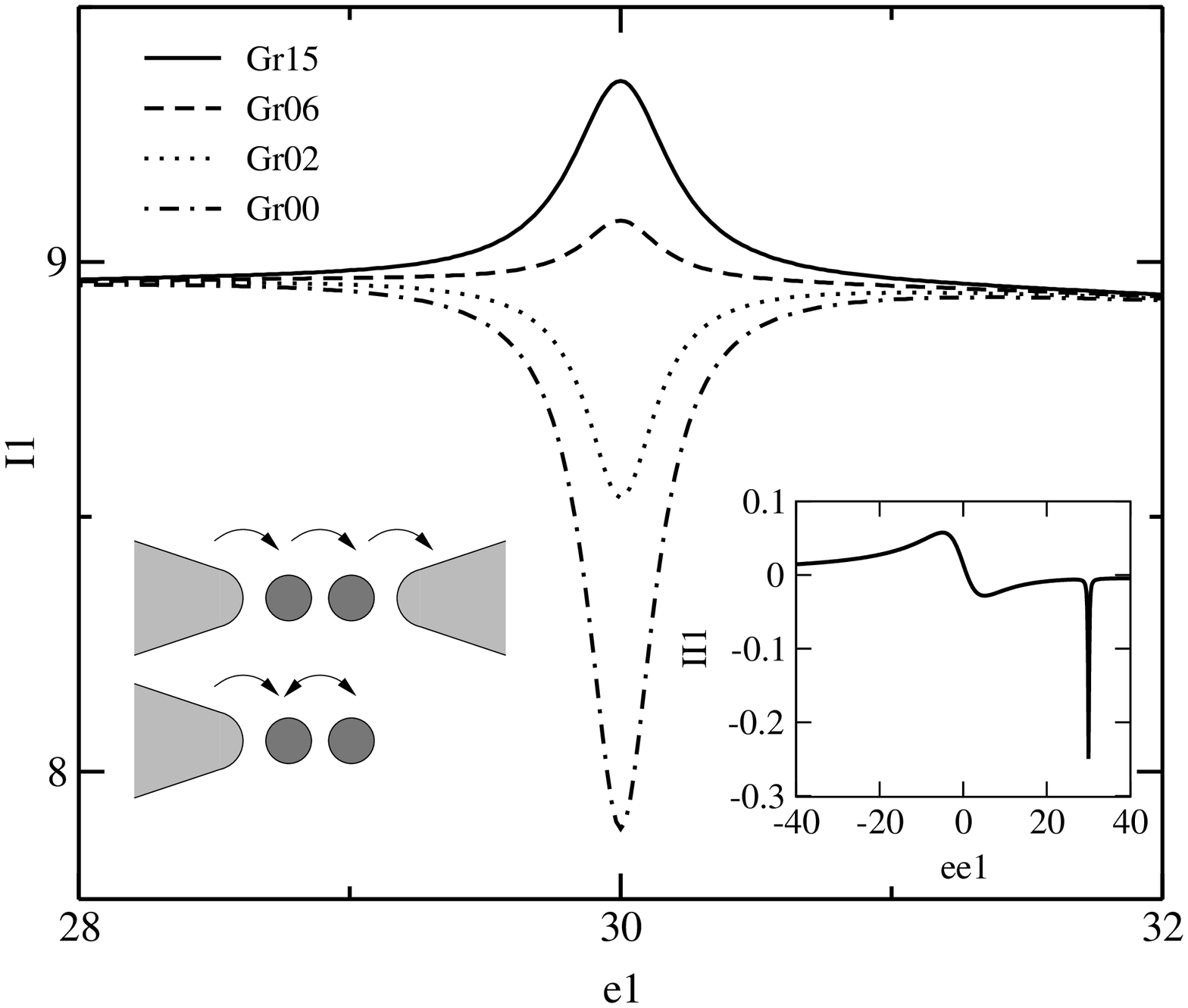}
\caption{\label{VB_Fig4.eps} {\bf Left:} Enhancement of the tunnel current $\Delta I_1$ at the resonance $\varepsilon_1\!=\!\varepsilon_2\!=\!30\mu$eV as a function of the dimensionless electron phonon coupling constant $g\equiv 2\alpha$. The additional current vanishes at $g\!\approx\!0.02$ when the tunnel rates to the double dot and between the dots become equal. The inset shows the difference in probabilities for triplet and singlet. {\bf Right:} Transition from an increased to a decreased
current through the first double quantum dot for different tunnel rates 
$\Gamma_{R,2}$ (in $\mu$eV) and  $\varepsilon_2\!=\!30\mu$eV.
The left inset shows schematically the set-up for $\Gamma_{R,2}\!=\!0$ and the
right inset gives the difference of triplet and singlet for the same  case. From \cite{VB03}.}
\end{figure}

A further analysis is possible by introducing pseudo singlet and triplet states, 
\begin{equation}
\label{basis_ts}
\begin{split}
\ket{T_+} &= \ket{L}_1 \ket{L}_2, \quad \ket{T_-} = \ket{R}_1  \ket{R}_2,\\
\ket{T_0} &= \frac{1}{\sqrt{2}} \,
      \Big(\ket{L}_1  \ket{R}_2 + \ket{R}_1 \ket{L}_2 \Big), \quad
\ket{S_0} &= \frac{1}{\sqrt{2}} \,
      \Big(\ket{L}_1  \ket{R}_2 - \ket{R}_1 \ket{L}_2 \Big).
\end{split}
\end{equation}
For $\varepsilon_1 \approx \varepsilon_2$, one has $\Delta I_1 \propto 2\textrm{Re} \langle p_1^{\dagger} p_2 \rangle$ $=\langle \textrm{P}_{T_0} \rangle - \langle \textrm{P}_{S_0}\rangle$, where $\textrm{P}$ is the projection operator on the triplet (singlet) state, and it follows that the current enhancement $\Delta I_1$ is due to an increased probability of finding the two electrons in a (pseudo) triplet rather than in a (pseudo) singlet state. This is in direct analogy to the $N=2$ Dicke effect for trapped ions as discussed above, with the difference that in the double dot system a third `empty' state $|0\rangle$ exists which allows current to flow through the system. One can use the singlet-triplet basis, \ Eq.~(\ref{basis_ts}){}, together with  five states $|00\rangle$, $|0L\rangle$,  $|L0\rangle$, $|0R\rangle$, and $|R0\rangle$ (indexes referring to the state of the first and the second double quantum dot) to derive nine coupled rate equations for the corresponding occupation probabilities \cite{VB03}. Assuming identical tunnel rates $\Gamma$ to all four leads, one obtains the inelastic current (for positive intra-dot bias $\varepsilon$) through the first double dot, as well as the triplet-single occupation difference,
\begin{equation} \label{Idouble1}
I_1     =  {e\Gamma} \frac{x (4x+1)}{9x^2+5x+1}, 
\quad p_{T_0} - p_{S_0} = - \frac{2x (x+2) (x-1)}{9x^3+23x^2+11x+2},\quad
\quad x=\nu / \Gamma,
\end{equation}
where $\nu\equiv 8\pi (T/\Delta)^2J(\Delta)$ is the inelastic decay rate within one double dot. This result is in excellent agreement with the  numerical solution of the full Master equation, cf. Fig.~(\ref{VB_Fig4.eps}). It explicitly demonstrates that  superradiance exists in arrays of artificial atoms, and can be probed as an enhanced current through the two double quantum dots at resonance $\varepsilon_1=\varepsilon_2$.

Tuning the individual tunnel rates can be used to generate current {\em sub}radiance, which occurs in a configuration where the two double quantum dots form a singlet state and electrons in the second double dot are prevented from tunneling into the right lead (inset of Fig.~\ref{VB_Fig4.eps}, right). The current peak $I_1$ then develops into a minimum as the tunneling rate $\Gamma_{R,2}$ is decreased to zero, which leads to an increased probability of the singlet state $\ket{S_0}$ and a {\em negative} cross coherence $\langle p_1^{\dagger} p_2 \rangle$ at resonance. 

A second configuration with fixed negative bias $\varepsilon_2\!<\!0$ in the second double dot can be used to generate a {\em current switch}. The resulting blockade of the second double dot can be lifted by the first one if the resonance condition $\varepsilon_1\!=\!-\varepsilon_2$ for the cross-coherence $\langle p_1 p_2 \rangle$ in Eq.~(\ref{Delta_current}) is fulfilled, when energy is transferred from the first to the second double dot, allowing electrons to tunnel from the left to the right in the second double dot.

\subsubsection{Oscillatory Superradiance for Large $N$}
Another extension of the Dicke model was discussed by Brandes, Inoue, and Shimizu in \cite{BIS98} for a superradiant `active region' of $N_i$ identical two-level systems  coupled to an `in' (left, $L$) and `out' (right, $R$) particle  reservoirs $H_{\rm res}$ by a tunnel Hamiltonian 
\begin{eqnarray}
  H_{T}=\sum_{ki}\left( t^L_k c^{\dagger}_{k,L}c_{i,\uparrow } + 
t^R_k c^{\dagger}_{k,R}c_{i,\downarrow } + H.c. \right),
\end{eqnarray}
where $c^{\dagger}_{k,\alpha }$ creates an electron in reservoir $\alpha$ and $c_{i,\sigma}^{\dagger}$ creates an electron in the upper/lower state $\sigma=(\uparrow, \downarrow) $ of the $i$-th two-level system. The real electron spin is assumed to play no role here and crucially, the tunnel matrix elements in $H_{T}$ are assumed to be $i$-independent.

In the extended Dicke model, the active region without electron reservoir coupling is assumed to be superradiant due to collective emission of bosons (photons, phonons). For a total number of $N\ll N_i$ electrons, this is the usual Dicke superradiance situation as described by a density operator in the basis of Dicke states $|JMN\rangle$, \ Eq.~(\ref{eigenstates}), where however additional quantum numbers other than $N$ (and represented by the index $\lambda$ in \ Eq.~(\ref{eigenstates})) are already neglected. Tunneling of electrons into and out of the active region  now provides a mechanism for pseudo-spin `pumping':  assuming large positive (negative) reservoir chemical potentials $\mu_L$ ($\mu_R$), electrons tunnel into the active region via upper levels $\sigma =\uparrow$, whereas they tunnel out of the active region from lower levels $\sigma =\downarrow$. 

Transitions between eigenstates of $H_{{\rm Dicke}}$, \ Eq.~(\ref{eq:Htot}), due to electron tunneling are described by rates  $\Gamma_{JMN\to J'M'N'}$, 
\begin{eqnarray}
\label{transitionrate}
\Gamma^{\alpha }_{JMN\to J'M'N'}&=&
 T^{\alpha }\left.
| \sum_{i\sigma }\langle J'M'N'| c^{\dagger}_{i\sigma }| JMN \rangle 
|^{2}\right.
\left.  f_{\alpha }(E_{J'M'N'}-E_{JMN})\right. \nonumber\\ 
&+& \left.T^{\alpha }
| \sum_{i\sigma }\langle J'M'N'| c^{\phantom{\dagger}}_{i\sigma }| JMN \rangle 
|^{2} \right.
\left.
\left[1-f_{\alpha }(E_{JMN}-E_{J'M'N'})\right]\right.\\
T^{\alpha }&\equiv&2\pi\sum_{k}t^{\alpha }_{k} (t^{\alpha }_{k})^{*}\delta (E-\varepsilon_{k}^{\alpha }), \quad \alpha={L/R},
\label{talpha}
\end{eqnarray}
where the  $\varepsilon_{k}^{\alpha }$ are single particle energies in reservoir $\alpha$ and the dependence of $T^{\alpha }$ on the energy difference $ \Delta E$ between final and initial state is neglected. This approach to tunneling of electrons through a region characterized by many-body states is in close analogy to (real) spin-dependent transport through quantum dots containing electrons interacting via Coulomb interaction as introduced by Weinmann and co-workers in their work on spin-blockade \cite{WHK95}. In the extended Dicke model here,  the role of the real electron spin is replaced by the (upper-lower level) pseudo spin, the total pseudo spin and its projection being denoted as  $J$ and $M$, respectively. 

Similar to spin-blockade related transport, the rates Eq.~(\ref{transitionrate}) are determined by the {\em Clebsch-Gordan coefficients} for adding or removing a single pseudo-up or down spin $j=1/2$ to the active region, 
\begin{eqnarray}
\label{Clebschapprox}
\left|\sum_{j\sigma }\langle J'M'N'| c^{\dagger}_{j\sigma }| JMN \rangle 
\right|^{2}&\approx& \delta_{N+1,N'} 
\left|\sum_{\sigma=\pm 1/2 }\langle J'M'| JM,j=\frac{1}{2} m=\sigma \rangle\right|^{2}\\
&=&
\frac{1}{2J+1}\left(
\delta _{J',J+1/2}\right.\left[
\delta _{M',M+1/2}(J+M+1)
+\delta _{M',M-1/2}(J-M+1)\right]\nonumber\\
&+& \delta _{J',J-1/2}\left[\delta _{M',M+1/2}(J-M)
+\left.\delta _{M',M-
1/2}(J+M)\right]\right).
\end{eqnarray}
where any further dependence on the specific form of the many-particle wave function in the active region is neglected and the proportionality factor is absorbed into the constant $T^{\alpha }$, Eq.~(\ref{talpha}).

\begin{figure}[t]
\begin{minipage}{\textwidth}
\parbox{0.4\textwidth}{\includegraphics[width=0.4\textwidth]{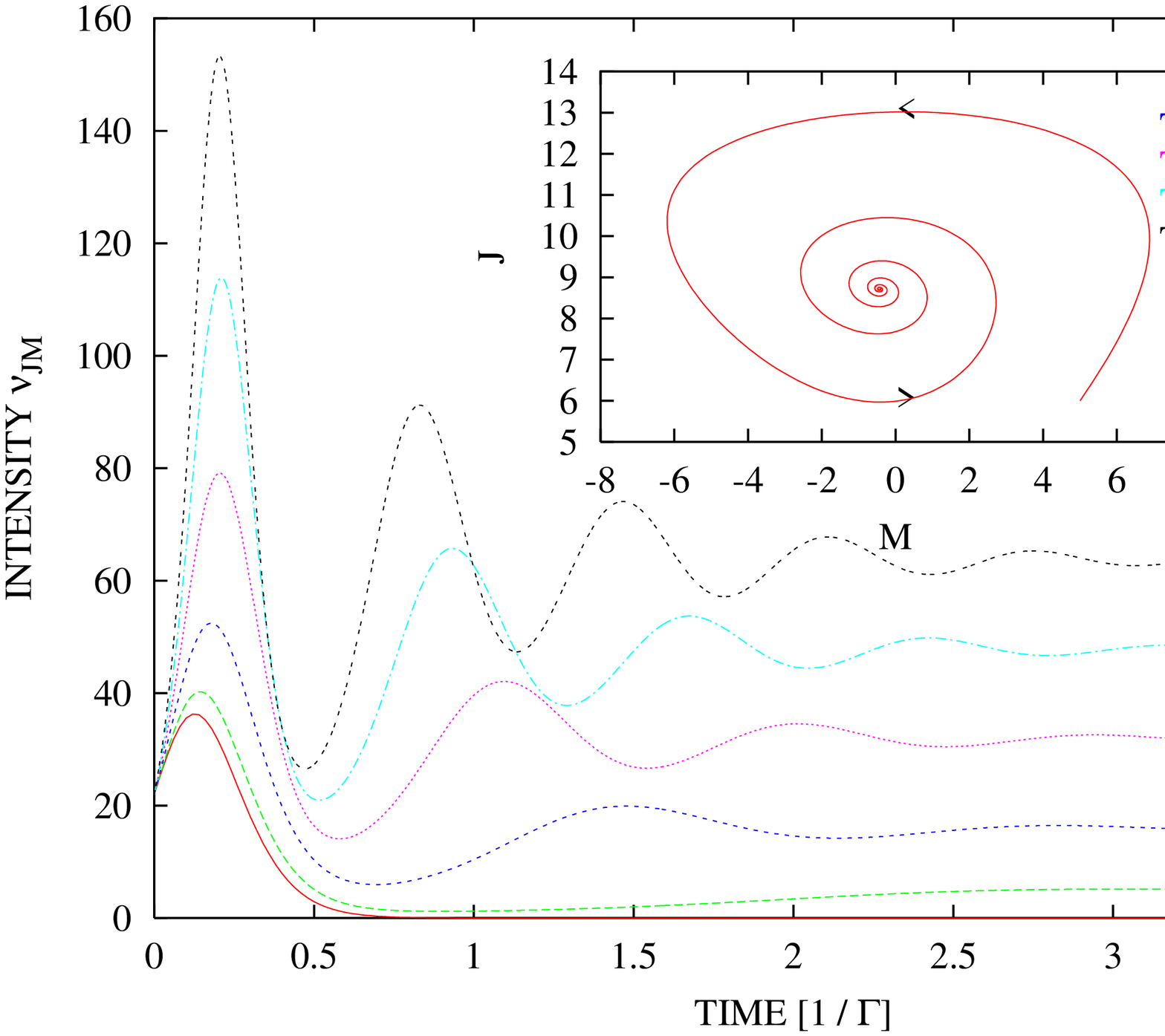}}
\hspace{0.2\textwidth}
\parbox{0.4\textwidth}{\includegraphics[width=0.4\textwidth]{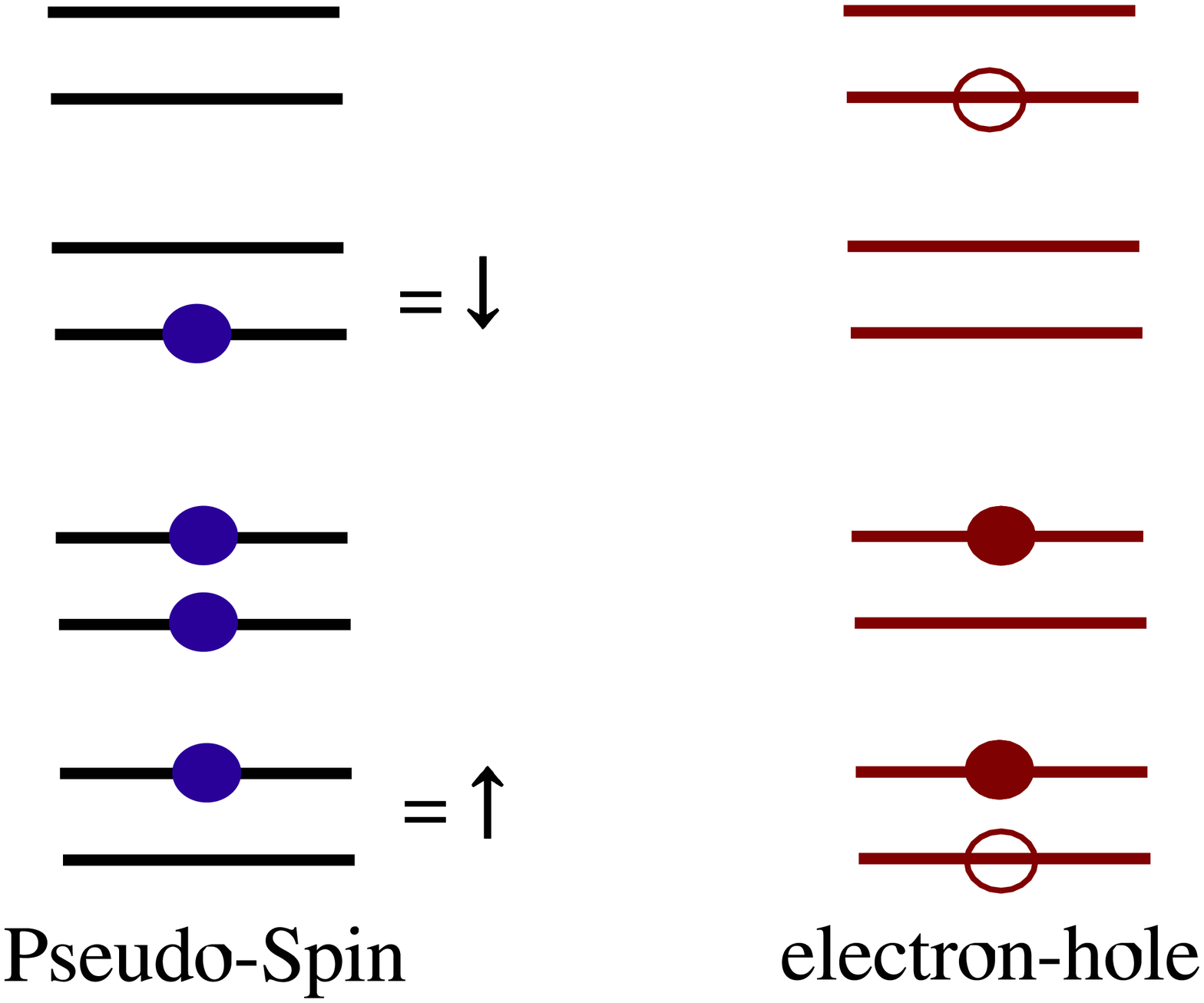}}
\end{minipage}
\caption[]{\label{BIS_Fig1.eps}{\bf Left:} Time evolution according to \ Eq.~(\ref{Dicke_rateeqn}) of the emission intensity $\nu_{JM}=I_{JM}/\Gamma \hbar\omega_{0}$ ($\Gamma$ spontaneous emission rate of single two-level system, $\omega_0$ level splitting) in a superradiant active region of two-level systems coupled to electron reservoirs with tunnel rates $T$. The Dicke peak is followed by oscillations, the angular frequency of which is approximately given by $\omega=\sqrt{2\Gamma T}$, \ Eq.~(\ref{SRoscanalytical}). 
Inset: $\langle J \rangle_{t}$ vs. $\langle M \rangle_{t}$ 
for $T=64$. {\bf Right:} Analogy between pseudo-spin and electron-hole model, cf. \ Eq.~(\ref{ehanalogy}){}. From \cite{BIS98}.}
\end{figure} 

Within these approximations, the dynamics of the active region is described in terms of rate equations for the probabilities $\rho (JMN)_{t}$ which are the diagonal elements of the reduced density operator at time $t$ in the basis of the Dicke states $|JM\rangle$ at a given number $N$ of electrons, 
\begin{eqnarray}\label{Dicke_rateeqn}
\frac{d}{dt}\rho (JMN)&=& -\frac{1}{\omega_0}\left[I_{JM}\rho (JMN)-I_{JM+1}\rho(JM+1N)\right]\nonumber\\
&+&\sum_{J'M'N'}\left[ \Gamma _{J'M'N'\to JMN}\rho (J'M'N')- \Gamma _{JMN\to J'M'N'}\rho (JMN)\right],
\end{eqnarray}
where $I_{JM}$ denotes the superradiant emission intensity, \ Eq.~(\ref{emission}). The rate equations \ Eq.~(\ref{Dicke_rateeqn}) can be either solved numerically or be used to derive an analytical solution in a quasi-classical approximation for large $J\gg 1$. In this limit, fluctuations of $M$ and $J$ are neglected and the probability distribution is entirely determine by the expectation values $J(t)$, $M(t)$ and $N(t)$ only \cite{GH82}, i.e $\rho (JMN)_{t}=\delta_{M,M(t)} \delta_{J,J(t)} \delta_{N,N(t)}$. Assuming identical tunnel matrix elements $T$ for in- and out-tunneling, one obtains a set of two differential equations  (the $N$ equation decouples),
\begin{eqnarray}\label{roughly}
\dot{M}(t)=-\Gamma\left[J(t)+M(t)\right]\left[J(t)-M(t)+1\right] + T,
\quad \dot{J}(t)=TM(t)/J(t)
\end{eqnarray}
which are governed by the two parameters $\Gamma $ and $T$, the emission rate and the tunnel rate, respectively. These equations have {\em damped oscillatory} solutions, cf. Fig. \ref{BIS_Fig1.eps}. In contrast to oscillatory superradiance in atomic systems \cite{Andreev}, the oscillations are not due to re-absorption of photons, but due to tunneling  of electrons into an active region characterized by a total pseudo spin $J$. Also in contrast to the original Dicke problem, $J$ is no longer conserved but develops a dynamics that is driven by the tunneling process which leads to a coupling of $J$ and $M$ as described by \ Eq.~(\ref{roughly}). The total number $N$ of electrons varies through single electron tunneling that changes the quantum numbers $J$ and $M$ and can lead to doubly occupied or empty single particle  levels. The change $\dot{J}$ of $J$ is proportional to $M$ itself, which follows from angular momentum addition rules (Clebsch-Gordan coefficients), whereas $M$ increases by electrons tunneling into the upper and out of the lower levels at the tunnel rate $T$. 

Instead of simple superradiant relaxation of the emission intensity (as described by \ Eq.~(\ref{hyperbolic})), the transient behavior is now determined by  a superradiant emission peak, followed by emission oscillations. In fact, after eliminating $J$ from Eq.~(\ref{roughly}) for large $J$, one obtains a single oscillator equation, 
\begin{eqnarray}\label{SRoscanalytical}
  \ddot{M}- 2\Gamma M\dot{M}+\omega ^{2}M=0,\quad \omega = \sqrt{2\Gamma T},
\end{eqnarray}
which  for $T>2\Gamma$ describes a harmonic oscillator with angular frequency $\omega$ and amplitude dependent damping. For smaller $T$, the oscillations are no longer visible and Eq.~(\ref{SRoscanalytical}) does no longer hold. For $T\to 0$, there is a smooth crossover to the usual Dicke peak, \ Eq.~(\ref{hyperbolic}),  with vanishing intensity at large times and without oscillations. 

In \cite{BIS98}, two physical systems were suggested for an experimental  realization of tunnel-driven, oscillatory superradiance. The first scenario described a  forward biased $pn$ junction in a system of electrons and holes in semiconductor quantum wells under strong perpendicular magnetic fields, the latter guaranteeing dispersion-less single electron levels $i$ with inter-band optical matrix elements diagonal in $i$. An initial optical or current excitation of the system was predicted \cite{BIS98} to lead to a superradiant peak of emitted light that would become strongly  enhanced if the  tunneling rate became higher. The correspondence with the pseudo-spin  model was established by mapping its four basic single particle states to the states of the electron-hole system;
\begin{eqnarray}\label{ehanalogy}
|{\rm empty}\rangle \to |0,h\rangle,\quad 
|\downarrow\rangle \to |0,0\rangle,
\quad |{\rm double}\rangle \to |e,0\rangle,\quad
|\uparrow\rangle \to |e,h\rangle,
\end{eqnarray}
i.e.,  the empty state becomes the state with and additional hole $|0,h\rangle$, the pseudo-spin down electron becomes the empty state $|0,0\rangle$, the doubly occupied state becomes the state with an additional electron $|e,0\rangle$, and the pseudo-spin up electron becomes the state $|e,h\rangle$ with an additional electron and hole which can radiatively decay. This realization, however, neglects Coulomb interactions that can lead  to strong correlations among electrons.  The kinetics of four-wave mixing for a two-dimensional magneto-plasma in strong magnetic fields  was calculated by Wu and Haug \cite{WH98} in a regime where  incoherent Coulomb scattering leads to  dephasing that increases with the  magnetic field. Still, the possibility  of collective quantum optical effects in the quantum Hall regime remains an open though not entirely new problem, since Landau-level lasing was suggested by Aoki already back in 1986 \cite{Aoki86}.

The second possible realization for superradiance with `electron pumping' was proposed \cite{BIS98} as an array of identical quantum dots with the ability to collectively radiate, with the dots having well-defined internal levels that allow transitions under emission of photons, cf. also the previous sections in \ref{section_Dicke_array}.

\subsection{Superradiance and Entanglement in other Quantum Dot Systems}

\subsubsection{Double Quantum Dot Excitons}\label{section_excitons}

\begin{figure}[t]
\includegraphics[width=0.5\textwidth]{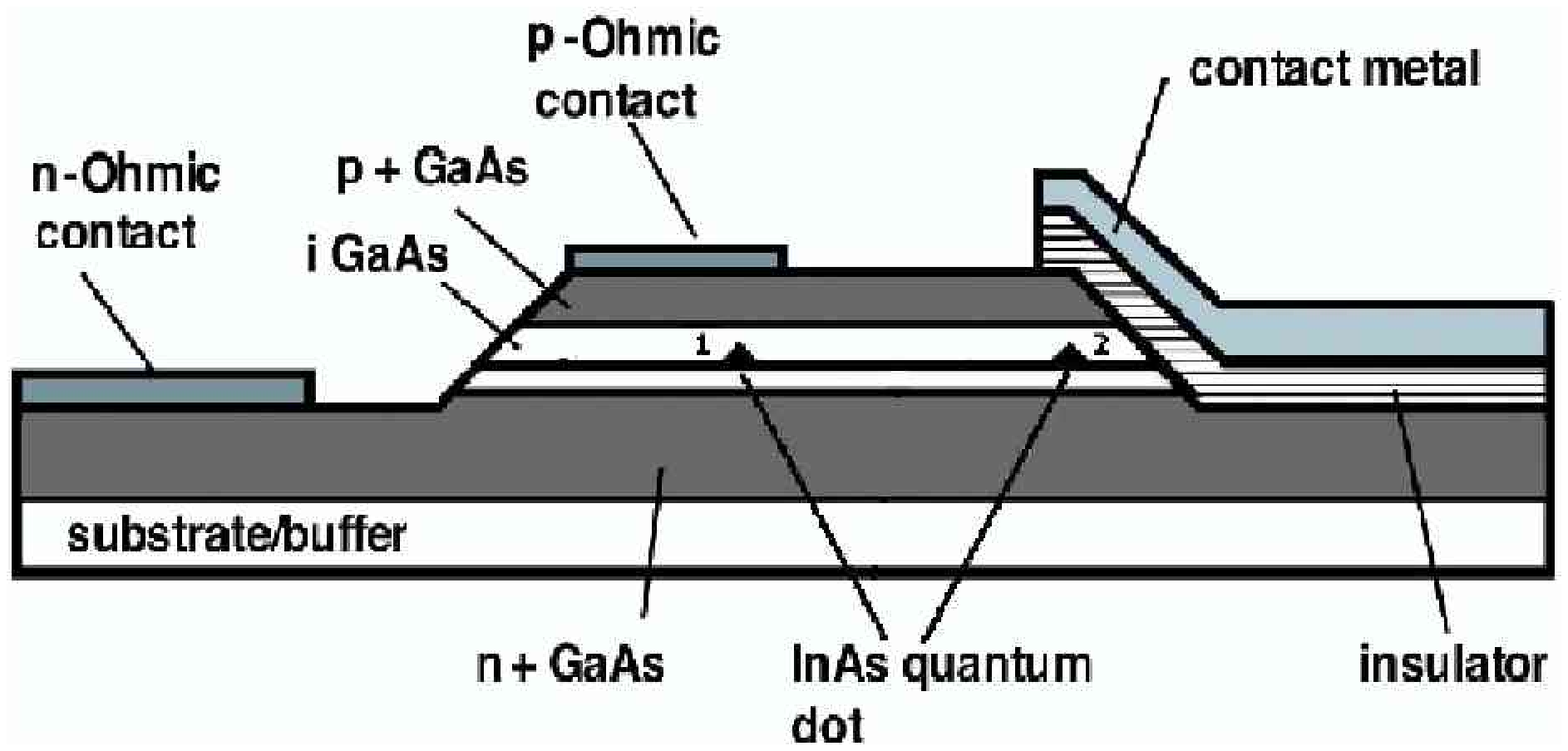}
\includegraphics[width=0.5\textwidth]{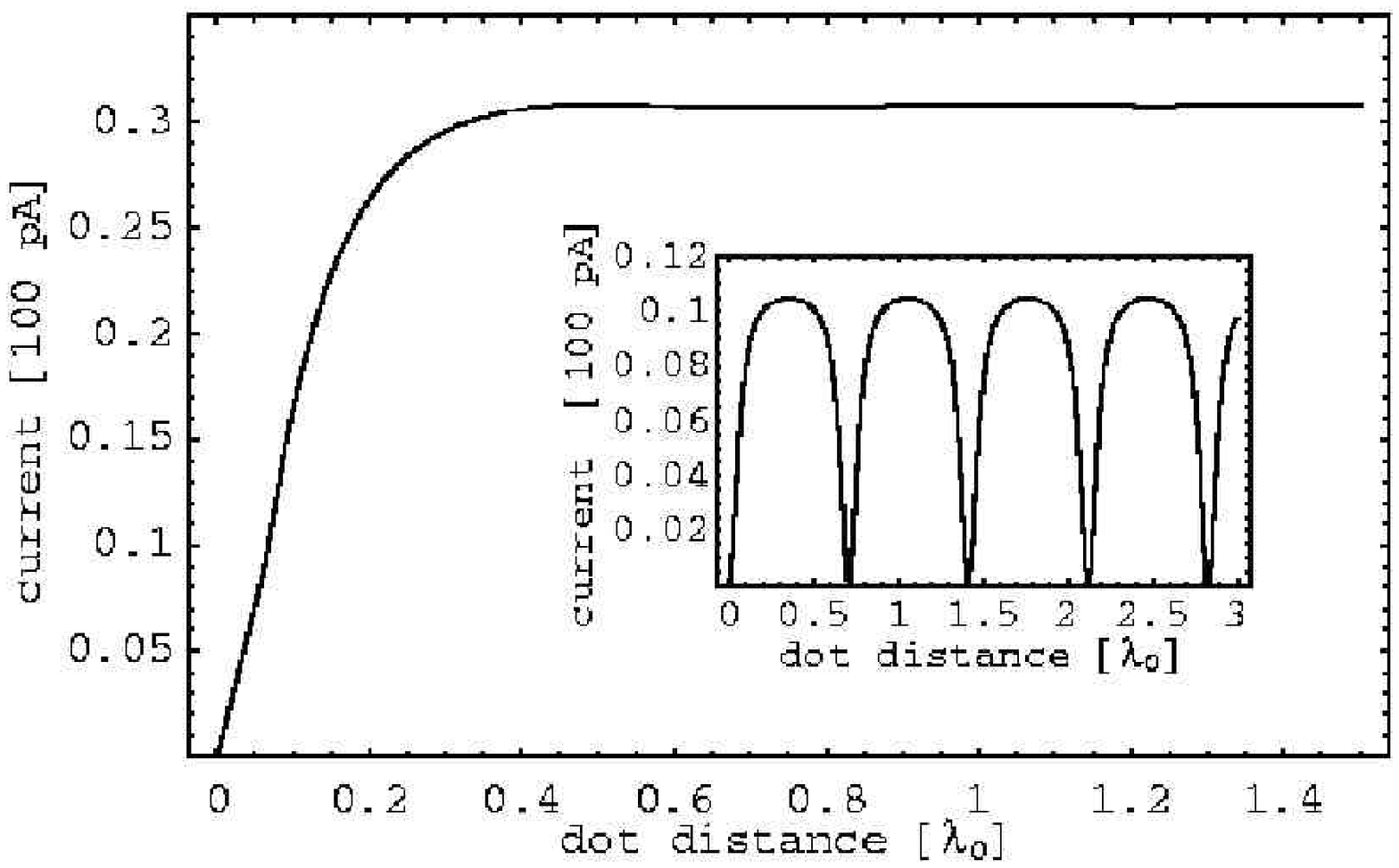}
\caption{\label{CCB03_fig1.eps}{\bf Left:} Double quantum dot exciton device structure suggested by Chen {\em et al.} \cite{CCB03}. {\bf Right:} Stationary current, \ Eq.~(\ref{current_chen}), and interference effect due to decay rates\ Eq.~(\ref{g2rates})  as a function of the dot distance (inset). 
From \cite{CCB03}.}
\end{figure}

Chen {\em et al.} further investigated the formal analogy, \ Eq.~(\ref{ehanalogy}),  between the pseudo-up/down spins and the electron-hole ($|eh\rangle$) and `empty' $|00\rangle$ states, and predicted superradiance in the electric current through excitonic double quantum dots \cite{CCB03}. Superradiant enhancement of excitonic decay in reduced dimensions is well-known, but {\em current} superradiance was  argued \cite{CCB03} to be an alternative tool for the detection of such collective effects. 
The main idea was to employ two spatially separated quantum dots ($1$ and $2$) which are radiatively coupled, cf. Fig. (\ref{CCB03_fig1.eps}), but with only dot $1$  being coupled to hole and electron reservoirs, which in fact is similar to the `Current Switch' configuration for the two double quantum dots considered above. Introducing the four states
\begin{eqnarray}
  \left| 0\right\rangle =\left|0,h;0,0\right\rangle, \left| U_{1}\right\rangle =\left|e,h;0,0\right\rangle , \left| U_{2}\right\rangle =\left|0,0;e,h\right\rangle, \left| D\right\rangle =\left|0,0;0,0\right\rangle ,
\end{eqnarray}
where $\left| 0,h;0,0\right\rangle $ denotes the state with one hole in dot 1, $\left| 0,0;0,0\right\rangle $ represents the ground state with no hole and electron in the quantum dots, and the exciton state $\left| e,h;0,0\right\rangle $ (in dot 1) can be converted to $\left|0,0;e,h\right\rangle $ (in dot 2) through  exciton-photon interactions. The latter was described by a Hamiltonian \cite{CBLC04}
\begin{eqnarray}
H_{I} &=&\sum_{\mathbf{k}}\frac{1}{\sqrt{2}}g\{D_{\mathbf{k}}b_{\mathbf{k}%
}[(1+e^{i\mathbf{k}\cdot \mathbf{r}})\left| S_{0}\right\rangle \left\langle
D\right| +(1-e^{i\mathbf{k}\cdot \mathbf{r}})\left| T_{0}\right\rangle \left\langle
D\right| ]+H.c.\},
\end{eqnarray}%
with super- and subradiant states as $\left| S_{0}\right\rangle =\frac{1}{%
\sqrt{2}}(\left| U_{1}\right\rangle -\left| U_{2}\right\rangle )$ and $%
\left| T_{0}\right\rangle =\frac{1}{\sqrt{2}}(\left| U_{1}\right\rangle
+\left| U_{2}\right\rangle )$, respectively. Furthermore, $b_{\mathbf{k}}$ is the photon operator, $gD_{\mathbf{k}}$ the coupling strength, $\mathbf{r}$ is the position vector between the two quantum dots, and  $g$ is a constant with the unit of the tunneling rate. Note that the dipole approximation was not used and the full $e^{i\mathbf{k}\cdot \mathbf{r}}$ terms kept in the Hamiltonian. The coupling of dot $1$ to the electron and hole reservoirs was described by the standard tunnel Hamiltonian 
\begin{equation}
H_{V}=\sum_{\mathbf{q}}(V_{\mathbf{q}}c_{\mathbf{q}}^{\dagger }\left|
0\right\rangle \left\langle U_{1}\right| +W_{\mathbf{q}}d_{\mathbf{q}%
}^{\dagger }\left| 0\right\rangle \left\langle D\right| +H.c.),
\end{equation}%
where $c_{\mathbf{q}}$ and $d_{\mathbf{q}}$ are the electron operators in the left and right reservoirs, respectively, giving rise to tunneling rates $\Gamma _{U}$ (electron reservoir) and $\Gamma _{D}$ (hole reservoir). The state $|e,0;0,0\rangle$ was argued to play no role for a dot configuration with thick tunnel barriers on the electron side. Equations of motion for the time-dependent occupation probabilities $n_{j}(t),$ $j=$ $0$, $D,$ $S_{0},T_{0}$ were then obtained in close analogy with Eq.~(\ref{eom3new}) and transformed into 
$z-$space,
\begin{eqnarray}
zn_{S_{0}/T_0}(z) &=&\mp ig\left[p_{S_{0},D}(z)-p_{D,S_{0}}(z)\right]+\Gamma _{U}\left[\frac{1}{z}%
-n_{S_{0}}(z)-n_{T_{0}}(z)-n_{D}(z)\right],  \\
zn_{D}(z) &=&-ig\left[p_{S_{0},D}(z)-p_{D,S_{0}}(z)+p_{T_{0},D}(z)-p_{D,T_{0}}(z)\right]-\frac{2\Gamma_{D}}{z}n_{D}(z)\\
p_{j_0,D}(z) &=&ig\gamma _{j}n_{j_{0}}(z)-\Gamma _{D}\gamma _{j}p_{j_{0},D}(z), \quad j = S,T,
\end{eqnarray}%
where $p_{S_{0},D}=p_{D,S_{0}}^{\ast }$ and $p_{T_{0},D}=p_{D,T_{0}}^{\ast } $ are off-diagonal matrix elements of the reduced density operator,  and a decoupling approximation similar to the one in  section \ref{section_polaron} was performed \cite{CBLC04a}. The  decay rates for superradiant and the subradiant channels,
\begin{equation}\label{g2rates}
g^{2}\gamma _{T/S}=\gamma _{0}\left(1\pm \frac{\sin (2\pi d/\lambda _{0})}{2\pi
d/\lambda _{0}}\right),
\end{equation}
depends on the ratio of inter-dot distance $d$ and the wave length $\lambda _{0}$ of the emitted light in an oscillatory form ($\gamma _{0}$ is the exciton decay rate in a single quantum dot). This is in close analogy to the ion trap experiment discussed in section \ref{section_DeVB96}, cf. \ Eq.~(\ref{eq:subsuper}). In the stationary limit, the current as defined by the temporal change of $\hat{n}_{D}(t)$ was obtained \cite{CCB03} as 
\begin{eqnarray}\label{current_chen}
  {\left\langle I\right\rangle }_{t\rightarrow \infty }=\frac{4g^{2}\gamma _{T}\gamma _{S}}{\gamma _{S}+\gamma _{T}[1+2\gamma
_{S}(g^{2}/\Gamma _{D}+g^{2}/\Gamma _{U}+\Gamma _{D})]},
\end{eqnarray}
which itself showed oscillations in $d/\lambda_0$ via \ Eq.~(\ref{g2rates}), cf. Fig. (\ref{CCB03_fig1.eps}), in close analogy to the two-ion case in section \ref{section_DeVB96}. The current is suppressed as the dot distance $d$ is much smaller than the wavelength $\lambda _{0}$. The emitted photon is reabsorbed immediately by the other dot and vice versa, with the current being blocked by this exchange process. The superradiant and the subradiant transport channels are in series in the limit where transport is determined by radiative decay, $g^2\gamma_{S/T}\ll \Gamma_{U/D}$, with $I\approx 4[1/g^{2}\gamma _{S}+1/g^{2}\gamma _{T}]^{-1}$.

\begin{figure}[t]
\includegraphics[width=0.5\textwidth]{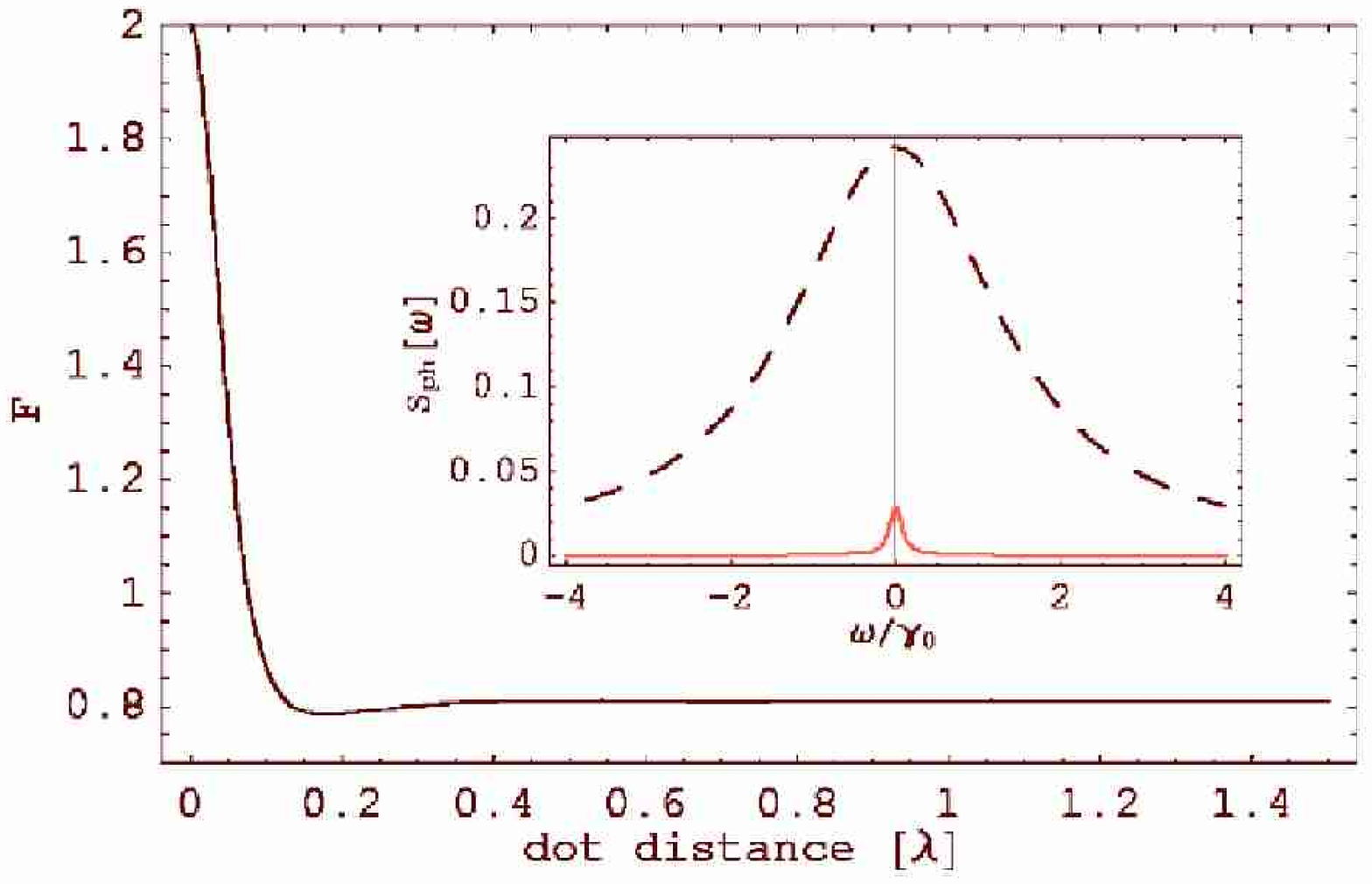}
\includegraphics[width=0.5\textwidth]{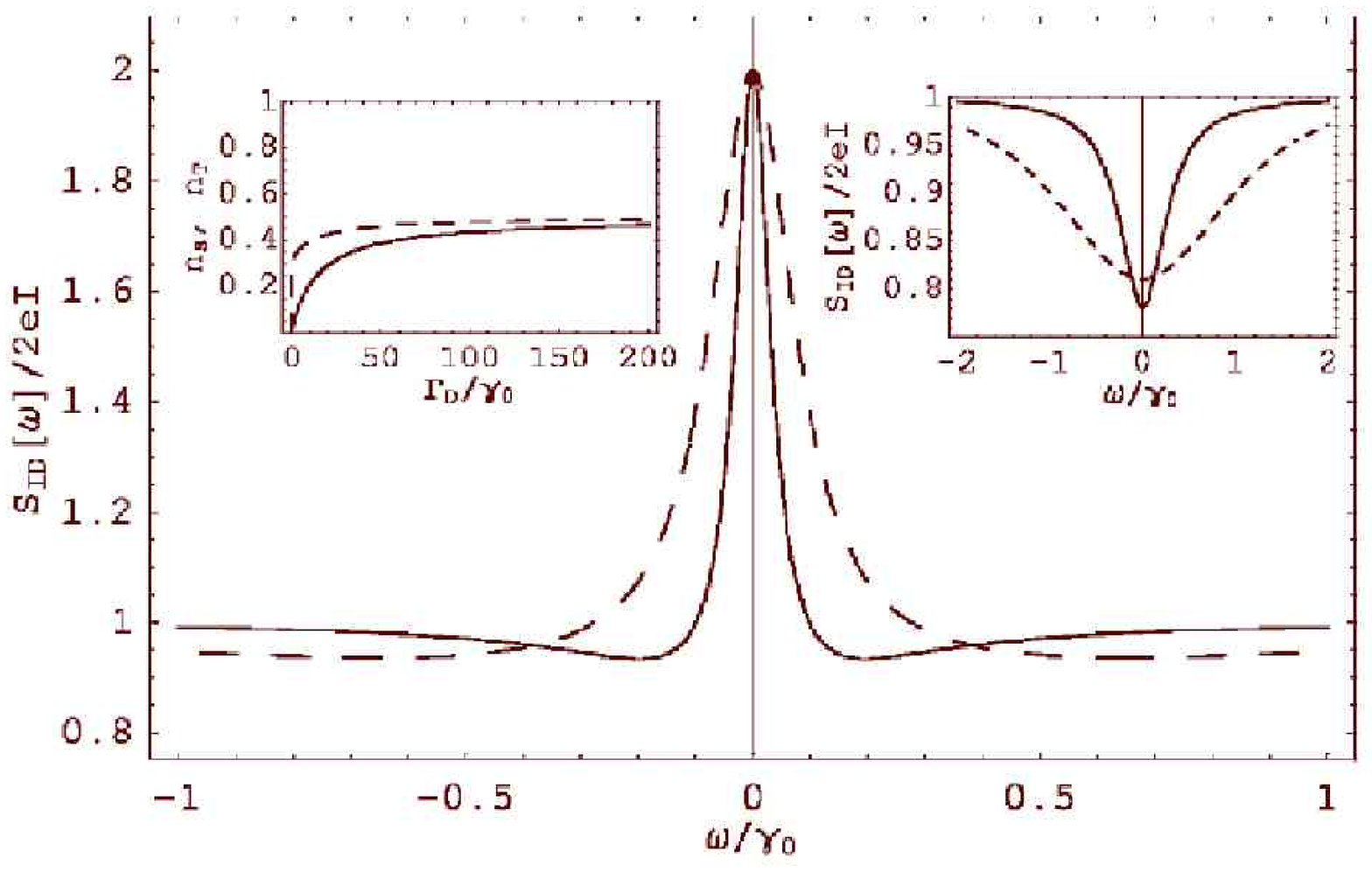}
\caption{\label{CBLC04_fig2.eps}{\bf Left:} Current noise Fano factor as a function of inter-dot distance. The vertical and horizontal units are $\frac{S_{I_{D}}(0)}{2eI}$ and $\protect\lambda $, respectively. Inset: Photon noise  $S_{ph}(\protect\omega )$ is equal to that in the  one-dot limit for $d\rightarrow \infty $ (dashed line), while it approaches \emph{zero noise} as $d=0.005\protect\lambda $ (red line). {\bf Right:} Effect of measurements on current noise \ $S_{ID}(\protect\omega )$
( ''maximum'' superradiance, $g^{2}\protect\gamma _{T}=2g^{2}\protect\gamma %
_{0}$ , $g^{2}\protect\gamma _{S}=0$). Solid and dashed lines correspond to $%
\Gamma _{D}=20$ $\protect\gamma _{0}$ and $\Gamma _{D}=$ $\protect\gamma %
_{0} $ , respectively. Right inset : the case of no superradiance. Left
inset : expectation value of the excited states $\left\langle
n_{S}\right\rangle $ and $\left\langle n_{T}\right\rangle $ as a function of 
$\Gamma _{D}$. From \cite{CBLC04}.}
\end{figure}

Chen {\em et al.} \cite{CCB03} suggested to include the double-dot system into a photon micro-cavity with strong electron-photon coupling. For a cavity of length $\lambda _{0}$, the three-dimensional version \ Eq.~(\ref{g2rates}){} for the two decay rates would then become
\begin{equation}
g^{2}\gamma _{cav,\pm }=\frac{\gamma _{0}}{\pi }\left| 1\pm e^{i2\pi d/(%
\sqrt{2}\lambda _{0})}\right| ^{2}.
\end{equation}%

A further property of the excitonic double dot system was  the fact that the interaction with the common photon field lead to emission-induced entanglement between the two dots. The maximum entangled state ($\left| S_{0}\right\rangle $) was reached as $d\ll\lambda _{0}$ which was checked by calculating the occupations $n_{S/T}$ in the stationary state. This entanglement was induced by the cooperative spontaneous decay which however can be controlled by, e.g.,  a voltage applied to a metallic gate that effectively tunes the band gap of dot 2 \cite{CCB03}.

Chen and co-workers also calculated  quantum noise in their electron-hole systems in close analogy to the formalism developed in section \ref{section_noise} for dissipative transport in double dots \cite{CBLC04}. The Fano factor $F$ in their model was enhanced by a factor of 2 for dot distances $d\ll \lambda$ (phonon wavelength) due to photon enhanced entanglement, cf. Fig. (\ref{CBLC04_fig2.eps}), with the  approximate expression 
\begin{equation}
F\equiv \frac{S_{ID}(0)}{2e\langle I\rangle }\approx 2-2g^{2}\gamma _{S}\left[\frac{1}{%
g^{2}\gamma _{T}}+3\left(\frac{1}{\Gamma _{D}}+\frac{1}{\Gamma _{U}}\right)+\frac{%
2\Gamma _{D}}{g^{2}}\right],
\end{equation}%
analogous to the result for current noise in the Cooper pair box by Choi {\em et al.} \cite{CPF03}.

In addition, one can compare  the current noise with the {\em photon} noise in the fluorescence spectrum, cf. inset of Fig. (\ref{CBLC04_fig2.eps}),  defined as 
\begin{eqnarray}
S_{ph}(\omega )&=&\frac{1}{\pi }\mbox{\rm Re}\int_{0}^{\infty }G^{(1)}[\tau
]e^{i\omega \tau }d\tau ,\\
G^{(1)}[\tau ] &\propto& \left| 1+e^{i2\pi d/\lambda }\right|
^{2}\left\langle p_{S_{0},D}(0)p_{D,S_{0}}(\tau )\right\rangle  
+\left| 1-e^{i2\pi d/\lambda }\right| ^{2}\left\langle
p_{T_{0},D}(0)p_{D,T_{0}}(\tau )\right\rangle 
\end{eqnarray}
and use the quantum regression theorem in order to calculate $G^{(1)}[\tau ]$. For small dot distances $d\ll \lambda$, the exciton does not decay and the  photon noise  approaches zero. 

A further interesting observation was the dependence of the current noise on the rate $\Gamma_D$ of hole tunneling, the effect of which can be thought of as a continuous measurement similar to the {\em quantum Zeno effect}.  Large $\Gamma_D$ turned out to narrow the noise spectrum $S_{ID}(\omega )$, cf. Fig. (\ref{CBLC04_fig2.eps}), right, and to localize the exciton in its excited state, with $\langle n_{S/T} \rangle$ approaching $\frac{1}{2}$.

\subsubsection{Nuclear Spins in Quantum Dots}

Eto, Ashiwa, and Murata \cite{EAM04} suggested a collective entanglement mechanism for nuclear spins by single electrons tunneling on and off  a quantum dot. In their model, they assumed a hyperfine interaction
\begin{eqnarray}\label{H_hf}
  H_{\rm hf} = 2 \sum_{k=1}^N \alpha_k {\bf S} \cdot {\bf I}_k
\end{eqnarray}
between $N$ nuclear spins ${\bf I}_k$ and the dot electron spin ${\bf S}$ with interaction constants $\alpha_k$ for nuclei at positions ${\bf r}_k$. They used a simple model for a double quantum dot with Zeeman-splitting $\delta E$ and on-site Coulomb repulsion $U$, cf. Fig. ({\ref{EAM04.eps}) left, where tunneling of an electron into the left dot leads to spin blockade when the additional spin is parallel to the one in the right dot. This spin blockade is lifted by a spin flip of the electron in the left dot under emission of a phonon, giving rise to a leakage electron current that can be measured. In lowest order perturbation theory in $H_{\rm hf}$ and the electron-phonon interaction $H_{\rm ep}$, the rate $\Gamma$ for an electron spin flip, e.g. from up to down, is a simple product of rates,
  \begin{eqnarray}\label{Gamma_hf}
    \Gamma= \gamma_{\rm ep} \gamma_{\rm hf},\quad 
\gamma_{\rm hf} = \left| \langle \Psi_{\rm i} \uparrow| H_{\rm hf} | \downarrow \Psi_{\rm f}\rangle \right|^2,
  \end{eqnarray}
where $\gamma_{\rm ep}$ is the rate for phonon emission (required for energy conservation), and the rate $\gamma_{\rm hf}$ depends on the state of the nuclear spins before/after the spin flip,  $\Psi_{\rm i/f}$. A simple approximation for the dynamics of the coupled electron-nuclei system is now found by following the temporal change of the nuclear state when  electron spins tunnel on and off: assuming identical $\alpha_k=\alpha$ for small dots, an initial state $\Psi^{(0)}=\sum_{jm;\lambda} c_{jm;\lambda}|jm;\lambda\rangle$ with random coefficients $c_{jm\lambda}$ in the basis of the collective states $|jm;\lambda\rangle$, cf. \ Eq.~(\ref{eigenstates}), is transformed into $\Psi^{(1)}=1/\sqrt{F^{(0)}}\sum_{jm;\lambda} c_{jm;\lambda}|jm\pm1;\lambda\rangle$, depending on whether the spin flip is up or down. The next electron spin transforms $\Psi^{(1)}$ into another collective spin state $\Psi^{(2)}$ of the nuclei, and so on, and recursive equations for the corresponding expansion coefficients $c_{jm;\lambda}^{(n)}$ and factors $F^{(n)}=\gamma_{\rm hf}^{(n)}/\alpha^2$ after $n$ flips are derived as 
  \begin{eqnarray}\label{cjm_recursion}
    c_{jm\mp 1;\lambda}^{(n)} &=& c^{\mp}_{jm} \sqrt{\frac{f^{(n-2)}}{f^{(n-2)}}} 
c_{jm;\lambda}^{(n-1)},\quad 
f^{(n)}\equiv\frac{1}{2^N}\sum_{jm}\frac{N!(2j+1) \left(c^{\mp}_{jm}\right)^2}{(N/2+j+1)!(N/2-j)!},
  \end{eqnarray}
where $c_{jm}^\pm \equiv \sqrt{j(j+1)-m(m\pm 1)}$, cf. \ Eq.~(\ref{cjm_def}), and $N$ is the total number of nuclei in the dot interacting with the electron spin. The total spin flip rate $\Gamma^{(n)}$, \ Eq.~(\ref{Gamma_hf}){},  after $n$ flips increases linearly with $n$, $\Gamma^{(n)}\approx n \Gamma^{(0)}$ for $1\ll n \ll N/2$, and saturates for $N/2\ll n$ as $\Gamma^{(n)}\approx (N/2) \Gamma^{(0)}$, a behavior which is reflected in the electronic current $I(t)$ through the dot, cf. Fig. ({\ref{EAM04.eps}) right.

\begin{figure}[t]
\begin{center}
\includegraphics[width=0.45\textwidth]{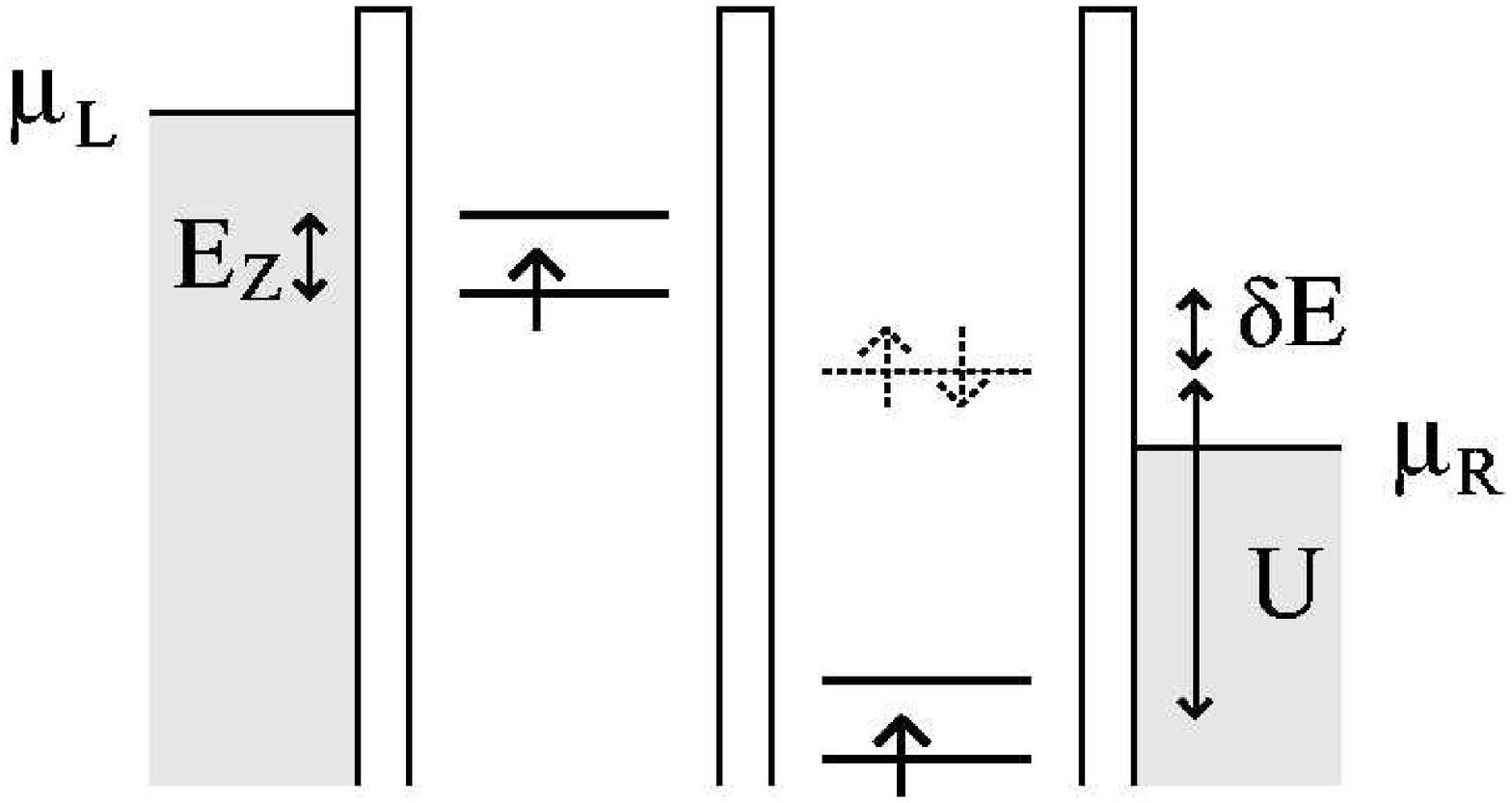} 
\includegraphics[width=0.45\textwidth]{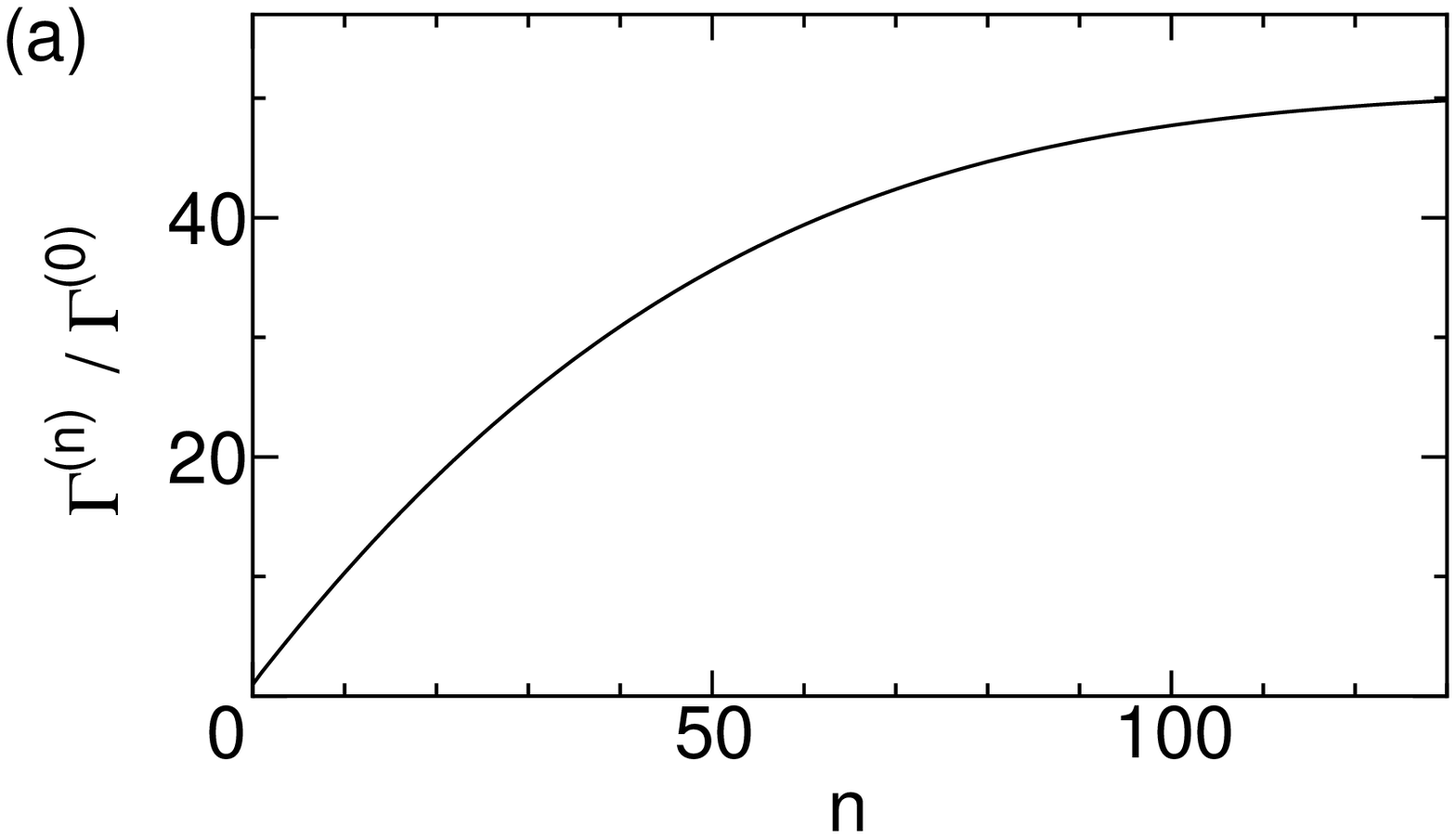} 
\end{center}
\caption{{\bf Left:} Spin blockade in double quantum dots in the  collective nuclear spin entanglement model by Eto, Ashiwa, and Murata \cite{EAM04}: the spin of an electron entering the left dot $L$ is parallel to that of the electron trapped in dot $R$. $E_Z$ is the Zeemann splitting and $U$ the on-site Coulomb interaction. The interaction via $H_{\rm hf}$, \ Eq.~(\ref{H_hf}), and  emission of phonons of energy  $\delta E$ leads to a leakage current through the dots. {\bf Right:} Spin-flip rate $\Gamma^{(n)}=\gamma_{\rm ep}\gamma_{\rm hf}^{(n)}$, $\gamma_{\rm hf}^{(n)}= \alpha^2 f^{(n)}/f^{(n-1)}$, Eq.~(\ref{cjm_recursion}){}, as a function of number $n$ of transported electrons accompanied by spin flip. From \cite{EAM04}.}
\label{EAM04.eps}
\end{figure}

\subsection{Large-Pseudo-Spin Models}


In the previous sections, the two-level system and its interaction with a dissipative environment played a prominent role. In fact, the famous spin-boson system \cite{Legetal87,Weiss} is one of the best studied models for quantum dissipation, and its importance has never been more obvious than in the light of experimental success in the generation of quantum superpositions and entanglement in noisy solid state environments, such as Cooper pair boxes \cite{NPT99} or semiconductor double  quantum dots \cite{Hayetal03}.
On the other hand, the cooperative phenomena (super and subradiance)  discussed above relied on  collective effects in combination with dissipation and therefore required a description in terms of spin-boson models for (pseudo) spin $j>\frac{1}{2}$.

\subsubsection{Collective Spins and  Dissipation}
There are several physical systems where dissipation of large spins plays a key role \cite{VB04}. Intrinsic spins greater than one half appear, e.g., in  the elements Gallium and Arsenic ({\em nuclear} spin of~$3/2$). Experiments by Kronm\"uller {\em et al.} \cite{Kroetal98,Kroetal99} and Smet {\em et al.} demonstrated the prominent role nuclear spins can have in the {\em quantum Hall effect}. 

Apel and Bychkov  \cite{AB99} discussed collective spin relaxation in quantum Hall systems due to spin-orbit interaction, $V_{SO}=-\frac{e\hbar}{2mc^2}{\bf S}\cdot {\bf E} \times {\bf p}$ for electrons with spin ${\bf S}=\frac{1}{2}{\sigmav}$ and momentum ${\bf p}$ in an electric field ${\bf E}$ due to piezo-electric lattice distortions. They used time-dependent perturbation theory in $V_{SO}$ for electrons in the lowest Landau level split by the Zeemann energy $\Delta$ with a free electronic Hamiltonian $H_0=-\Delta\sum_p(c^{\dagger}_{p\uparrow}c_{p\uparrow}-c^{\dagger}_{p\downarrow}c_{p\downarrow})$. Near filling factor $\nu=1$ and in Hartree-Fock approximation, they found the time-dependent spin relaxation of $\delta^z(t)=S^z(t=\infty)-S^z(t)$ given by 
\begin{eqnarray}\label{AB99_relax}
  \partial_t \delta^z(t) = -\frac{1}{\tau} \delta^z(t) [ \varepsilon + \delta^z(t)],\quad
\varepsilon\equiv \sqrt{4\bar{N}(1+\bar{N}) + (\nu-1)^2},
\end{eqnarray}
where the relaxation time $\tau$ depends on the dispersion of the collective spin-exciton modes, and $\bar{N}$ is the average phonon number. The form of the kinetic equation \ Eq.~(\ref{AB99_relax}) is identical to the one obtained from a simple non-interacting  model for (incoherent) excitonic relaxation. 

Nuclear spin relaxation was also studied by Apel and Bychkov \cite{ABy01} who generalized the Bloch equations  to higher spin. Furthermore, Maniv, Bychkov and Vagner \cite{MBV04} considered the hyperfine interaction, cf. \ Eq.~(\ref{H_hf}){}, and predicted a strong enhancement of the nuclear spin relaxation rate due to collective spin rotations of a single Skyrmion in a quantum Hall ferromagnet.

Another large-spin example is {\em molecular magnets} that contain a small number of metallic ions which couple magnetically, the most prominent examples, Mn${}_{12}$ and Fe${}_{8}$, being described by a spin of size~10~\cite{sessoli,wernsdorfer}. Chudnovsky and Garanin \cite{chudnovsky} considered a system of $N$ magnetic atoms (or molecules) with spin $S$,  in nearly degenerate situations where each magnetic atom is  described by an effective two-level system (pseudo-spin $\frac{1}{2}{\sigmav}_i$), giving rise to an effective Hamiltonian $H_{\rm eff}\equiv -\Delta J_x - W J_z$  with total pseudo-spin $J_\alpha= \frac{1}{2}\sum_{i=1}^N {{\sigma}}_{\alpha,i}$, $\alpha=x,y,z$, cf. Eq.~(\ref{eq:angular}). They considered the spin-photon \cite{chudnovsky} and spin-phonon \cite{CG04} interaction of the atoms in the small-sample limit of superradiance and described the dynamics for large total pseudo-spin $j\gg 1$  by the Landau-Lifshitz equation,
\begin{eqnarray}
  \dot{\bf n}= {\bf n} \times {\omegavec}_0 - \alpha {\bf n} \times ({\bf n} \times {\omegavec}_0),\quad
{\omegavec}_0 \equiv \Delta {\bf e}_x + W {\bf e}_z,\quad {\bf n}\equiv {\bf J}/j,
\end{eqnarray}
with the dimensionless damping coefficient $\alpha=j\Gamma_1/\sqrt{\Delta^2+W^2}$ proportional to $j=N/2$ indicating superradiance ($\Gamma_1$ is the relaxation rate for a single atom).

High-spin systems are also candidates for so-called `qu-dits' which are discussed in the context of quantum information processing and extend the standard qubit (spin $\frac{1}{2}$) to a higher-dimensional Hilbert space. Ensembles of two-state systems whose polarization is described by a large pseudo-spin can also be found in crystals and amorphous solids~\cite{wuerger}. Ahn and Mohanty have suggested collective effects of  two-level systems as a possible friction mechanism in micro-mechanical resonators~\cite{ahn}.

\subsubsection{Large-Spin-Boson Model: Weak Dissipation}
Vorrath \cite{VBK04} studied the \emph{large-spin-boson Hamiltonian} 
\begin{equation}
\label{eq_hamiltonian}
H = \varepsilon \, J_z + 2 T_c \, J_x 
    + J_z \sum_q \gamma_q \, (a_q^{\dagger}+a_{-q})
    + \sum_q \omega_q \, a_q^{\dagger} a_q,
\end{equation}
that generalizes the usual spin-boson Hamiltonian ${\mathcal H}_{SB}$, \ Eq.~(\ref{modelhamiltonian}), to arbitrary spin $j\ge \frac{1}{2}$. Various other generalization of the spin-boson Hamiltonian were already studied by other authors in dissipative tight-binding models for  multi-state systems \cite{Schmid83,GHM85,FZ85,Weiss}, or double-well potentials with additional, excited states \cite{TGH00,TGH01}.
In the form \ Eq.~(\ref{eq_hamiltonian}), the Hamiltonian $H$ at zero bias $\varepsilon=0$ is canonically equivalent to the Dicke model, 
\begin{equation}
\label{eq_dicke}
H_{\rm Dicke} = \omega_{\rm D} J_z + J_x \sum_q \gamma_q  
 (a_q^{\dagger}+a_{-q}) + \sum_q \omega_q \, a_q^{\dagger} a_q = e^{-i \pi/2  J_y} H  e^{i \pi/2  J_y}
\end{equation}
with the identification $\omega_{\rm D}=-2T_c$, also cf. Eq.~(\ref{eq:Htot}). 

\begin{figure}[t]
\centering
\psfrag{Jz}{\hspace*{-4mm} $\erw{J_z}$}
\psfrag{Jx}{\hspace*{-4mm} $\erw{J_x}$}
\psfrag{t}{\hspace*{-6mm} $t/ T_c^{-1}$}
\psfrag{jjz}{\footnotesize \hspace*{-2mm} $\erw{J_z}$}
\psfrag{jjx}{\footnotesize \hspace*{-2mm} $\erw{J_x}$}
\epsfig{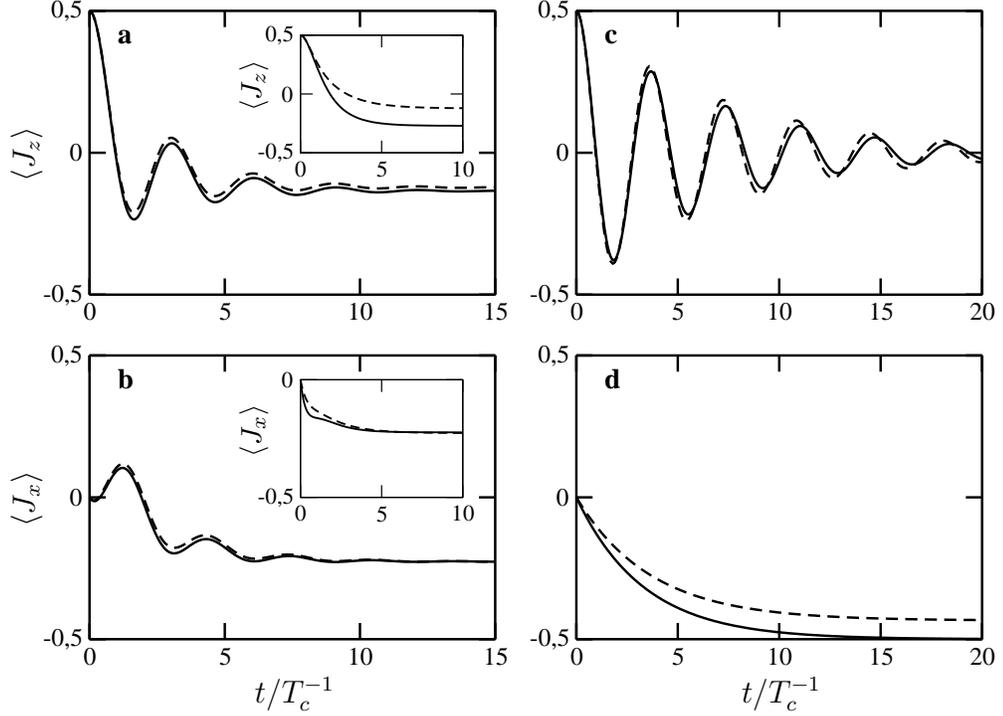}
\caption{\label{fig_spin} 
Dynamics of a spin 1/2 according to the Bloch equations (as derived from \ Eq.~(\ref{eq_master-final}),  
solid line) and the approximate solutions by Weiss (dashed line); a and b:
$\varepsilon=T_c$, $\alpha=0.05$ (inset: $\alpha=0.2$),
$\omega_c=50T_c$, and $k_BT=2T_c$ 
(dashed line: Eqs.~(21.132) and (21.134) of Ref.~\cite{Weiss}); 
c and d: $\varepsilon=0$, $\alpha=0.05$, $\omega_c=50T_c$, 
and $k_BT=0$ 
(dashed line: Eqs.~(21.172) and (21.173) of Ref.~\cite{Weiss}). From \cite{VBK04}.}
\end{figure}

Vorrath \cite{VBK04} derived the Master equation in second-order Born and Markov approximation for the reduced spin density matrix $\rho(t)$ corresponding to the Hamiltonian \ Eq.~(\ref{eq_hamiltonian}) as
\begin{equation}
\label{eq_master-final}
\begin{split}
\dot{\rho}(t) =  \; &i\, \big[ \rho(t) , \varepsilon J_z + 2 T_c\, J_x \big] 
- \frac{1}{\Delta^2} \, (\varepsilon^2 \, \Gamma + 4 T_c^2 \, \Gamma_c) \,
        \big[ J_z, J_z \, \rho(t)  \big] 
- \frac{2 T_c \, \varepsilon}{\Delta^2} \, (\Gamma - \Gamma_c) \,
        \big[ J_z, J_x \, \rho(t)  \big] \\
&+ \frac{2 T_c}{\Delta} \, \Gamma_s \,
        \big[ J_z, J_y \, \rho(t)  \big] 
+ \frac{1}{\Delta^2} \, (\varepsilon^2 \, \Gamma^* + 4 T_c^2 \, \Gamma_c^*) \,
        \big[ J_z,  \rho(t) \, J_z \big] \\
&+ \frac{2 T_c \, \varepsilon}{\Delta^2} \, (\Gamma^* - \Gamma_c^*) \,
        \big[ J_z,  \rho(t) \, J_x \big] 
- \frac{2 T_c}{\Delta} \, \Gamma_s^* \,
        \big[ J_z,  \rho(t) \, J_y \big],
\end{split}
\end{equation}
where an initial factorization condition was assumed and rates 
\begin{equation}
\label{eq_rates}
\begin{split}
\Gamma_c &= \frac{\pi}{2} \;J(\Delta) \,
             \coth\!\Big(\frac{\beta \Delta}{2}\Big)
 - \frac{i}{2} \; \dashint_0^{\infty} \! d\omega \; J(\omega) \;
 \Big( \frac{1}{\omega + \Delta} + \frac{1}{\omega - \Delta} \Big) ,\\
\Gamma_s &= \frac{1}{2} \; \dashint_0^{\infty} \! d\omega \; J(\omega) \,
 \coth\!\Big(\frac{\beta \omega}{2}\Big) \; 
 \Big( \frac{1}{\omega + \Delta} - \frac{1}{\omega - \Delta} \Big)
 - i \, \frac{\pi}{2} \;J(\Delta), \quad \Gamma\equiv \Gamma_c(\Delta\!\to\!0)
\end{split}
\end{equation}
with $\Delta=\sqrt{4T_c^2\!+\!\varepsilon^2}$ were defined.

For spin $j=\frac{1}{2}$ and a spectral density
\begin{eqnarray}
  J(\omega) = \sum_q|\gamma_q|^2\delta(\omega-\omega_q)=
2\alpha \omega_{\rm ph}^{1-s} \omega^s \exp(-\omega/\omega_c)
\end{eqnarray}
for weak Ohmic dissipation $s=1$ with $\alpha\ll 1$, cf. \ Eq.~(\ref{Jomegageneric}), Vorrath derived Bloch equations for the expectation values $\langle J_i\rangle$ as a function of time and compared the solutions with those obtained by Weiss within NIBA approximation for intermediate temperatures $k_BT=2T_c$, and with the solutions beyond NIBA \cite{Weiss} at zero temperature. The Born-Markov approximation turned out to correctly describe the spin-boson dynamics for weak dissipation at all temperatures, cf. Fig. ({\ref{fig_spin}), although systematic quantitative comparisons were not made. 

\begin{figure}[t]
\centering
\psfrag{t}{\hspace*{-4mm}$t/ T_c^{-1}$} 
\psfrag{g1}{\small $\alpha=0.0025$}
\psfrag{g2}{\small $\alpha=0.05$}
\psfrag{g3}{\small $\alpha=0.01$}
\psfrag{g4}{\small $\alpha=0.025$}
\epsfig{file=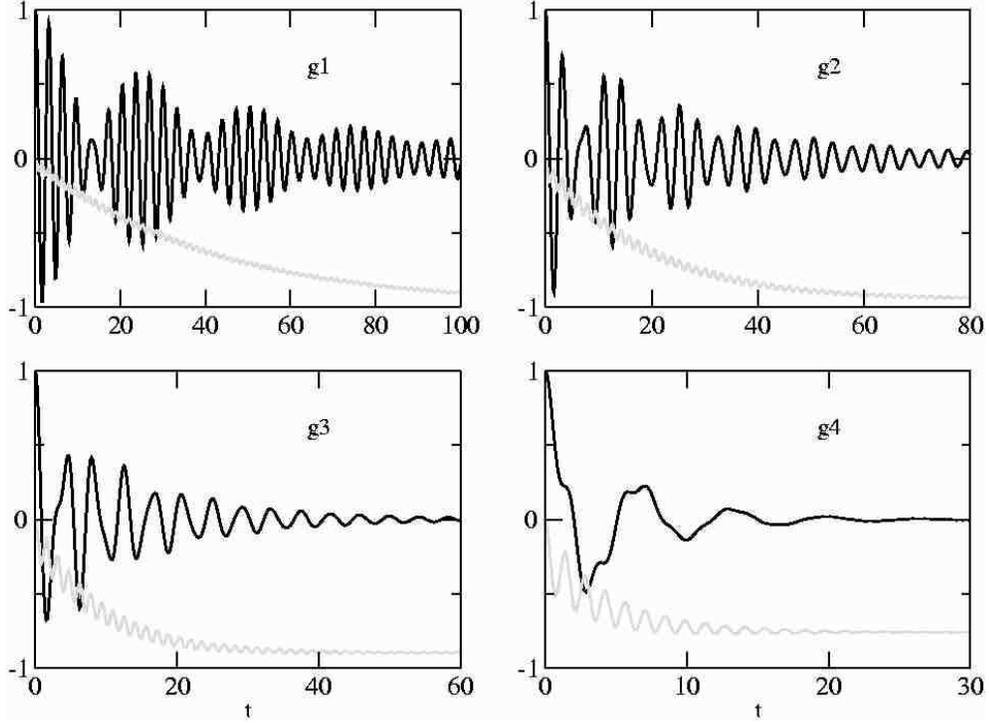,width=13cm}
\caption{ \label{fig_beats}
Time evolution of $J_z$ (black line) and $J_x$ (grey line) for different 
interaction strengths with the environment
($j=1$, $\varepsilon\!=\!0$, $\omega_c\!=\!50T_c$, and $k_BT\!=\!0$). From \cite{VBK04}.}
\end{figure}

For larger spin $j>1/2$ and $\varepsilon\ne 0$, the spin $z$-component $\langle J_z\rangle$ showed superradiant behavior in the form of increasingly faster, collective decay (as a function of time) with increasing spin $j$ \cite{VBK04}. Furthermore, a detailed analysis  for spin $j=1$ and bias $\varepsilon=0$ revealed {\em quantum beats} in the time-evolution of the spin, cf. Fig. ({\ref{fig_beats}), with similar beats occurring for higher spin. Non-resonant bosons lead to corrections of the energies for the spin eigenstates $\ket{+}$, $\ket{0}$, $\ket{-}$, which in second order perturbation theory are given by  
\begin{equation}
E_{\ket{\pm}}^{(2)} = - \frac{1}{2} \; \dashint_0^{\infty} \!d\omega \, 
J(\omega) \, \frac{1}{\omega\mp 2 T_c} \,, \qquad
E_{\ket{0}}^{(2)} = E_{\ket{+}}^{(2)} + E_{\ket{-}}^{(2)}, 
\end{equation}
thus leaving the eigenstates not equidistant any longer. The beat frequency 
\begin{equation}
\omega_b = \alpha \, \omega_c + \alpha \, T_c \Big[ \,
  e^{2T_c/\omega_c} \, {\rm Ei}\Big(\frac{-2T_c}{\omega_c}\Big) 
  -  e^{-2T_c/\omega_c} \, {\rm Ei}\Big(\frac{2T_c}{\omega_c}\Big) \, \Big],
\end{equation}
which for large cut-off $\omega_c\gg T_c$ is well approximated by $\omega_b=\alpha \omega_c$, was found to be in excellent agreement with numerical solutions of the Master equation.

\subsubsection{Large-Spin-Boson Model: Strong Dissipation}
For the large-spin-boson model in the regime of strong dissipation, Vorrath and Brandes \cite{VB04}  used perturbation theory for $H=H_0+V$, Eq.~(\ref{eq_hamiltonian}), with respect to the tunneling part $V=2 T_c J_x$. Using a polaron transformation, cf. section \ref{section_polaron}, one obtains
\begin{equation}
\label{eq_polaron-hamiltonian}
\bar{H}_0 =  e^{\sigma J_z} H_0 \, e^{-\sigma J_z} 
 = \varepsilon J_z - \kappa J_z^2 + \sum_q \omega_q \, a_q^{\dagger} a_q,\quad
\sigma \equiv \sum_q \, \frac{\gamma_q}{\omega_q}\,(a_q^{\dagger}-a_{-q}),
\end{equation}
with $\kappa = \sum_q \frac{|\gamma_q|^2}{\omega_q}=2 \alpha \omega_c$
for Ohmic dissipation. Spin and boson  subsystems become independent and can be treated separately. A new non-trivial term, $- \kappa J_z^2$, appears in the spin part of the transformed Hamiltonian~(\ref{eq_polaron-hamiltonian}). In the spin-boson model with spin $j=1/2$, this term is constant and has no physical consequences, whereas for larger spins $j>1/2$ it dominates the properties of the system.
The eigenenergies~$E_m$ of the spin subsystem directly follow from $\bar{H}_0$ as 
\begin{eqnarray}
 E_m = \varepsilon m- \kappa m^2, \quad -j \le m \le j, 
\end{eqnarray}
whereas in the transformed picture the tunnel term 
\begin{equation}
\label{eq_V_polaron}
\bar{V} = T_c \, (J_+ \, X + J_- \, X^{\dagger}), \quad 
X = e^{\sigma},
\end{equation}
now contains the unitary boson displacement (`shake-up') operators $X$, cf. section \ref{section_polaron}.  The Markov approximation is applied by assuming that the memory time of the environment corresponding to the width of the correlation  function $C(t)$ is the shortest time-scale in the problem, which however is {\em not} identical to the replacement of $C(t)$ by a Delta-function.

The Master equation for the spin density operator is calculated in the basis of the Dicke states $|jm\rangle$, 
\begin{equation}
\label{eq_master}
\begin{split}
\dot{\tilde{\rho}}&_{m,m}(t) = 2 \pi T_c^2 \Big[
 - {c_{jm}^{\,-}}^2 
   P \big(\varepsilon\!-\!\kappa(2m\!-\!1)\big) 
   \tilde{\rho}_{m,m}(t) 
-  {c_{jm}^{\,+}}^2 
   P \big(\!-\!\varepsilon\!+\!\kappa(2m\!+\!1)\big) 
   \tilde{\rho}_{m,m}(t) \\
&+  {c_{cm}^{\,-}}^2 
   P \big(\!-\!\varepsilon\!+\!\kappa(2m\!-\!1)\big) \;
   \tilde{\rho}_{m-1,m-1}(t) 
+  {c_{jm}^{\,+}}^2 
   P \big(\varepsilon\!-\!\kappa(2m\!+\!1)\big) 
   \tilde{\rho}_{m+1,m+1}(t) \Big],
\end{split}
\end{equation}
where in addition to the superradiant factors $c_{jm}^\pm$, \ Eq.~(\ref{cjm_def}),  the rate $P(E)$ for inelastic transitions due to boson emission or absorption from the dissipative environment appears, cf. \ Eq.~(\ref{IPE}) in section \ref{section_spectraldensity}. Here, $E$ is the energy difference $E=\varepsilon-\kappa(2m\pm 1)$ between the Dicke states $|jm\rangle$ and $|jm \pm 1\rangle$. The range of validity  of \ Eq.~(\ref{eq_master}){} is restricted to that of the NIBA (strong couplings, $\alpha\!\ge\!1$ at zero temperature and intermediate to strong couplings at finite temperatures).

For spin $j=1/2$, one recovers the usual results of the spin-one-half boson model \cite{Weiss}, with the Master equation predicting an exponential relaxation  of $J_z$ to the equilibrium value $\langle J_z \rangle_{\infty}=(P(-\varepsilon)\!-\!P(\varepsilon))/2(P(-\varepsilon)\!+\!P(\varepsilon))$ with relaxation rate 
$\gamma \!=\! 2 \pi T_c^2 (P(-\varepsilon)\!+\!P(\varepsilon))$: 1) for the zero temperature version of $P(\varepsilon)$, \ Eq.~(\ref{PET0}), the spin remains in its initial state at zero bias, $\varepsilon = 0$, which 
is the well-known localization phenomenon of the spin-boson model at $\alpha\!\ge\!1$~\cite{chakravarty,bray}. For a finite bias $\varepsilon\ne 0$, the system relaxes with rate
\begin{equation}
\gamma_{T=0} = \frac{2 \pi \, T_c^2 \, \varepsilon^{2\alpha-1}}
{\omega_c^{2\alpha} \, \Gamma(2\alpha)} \; e^{-\varepsilon/\omega_c},
\end{equation}
which agrees in leading order in $\varepsilon/\omega_c$ with the relaxation rate of the spin-boson model~\cite{Legetal87}. 2) at $\alpha \!=\! 1/2$, the spin-boson model has an exact solution and corresponds to the Toulouse limit of the anisotropic Kondo model. Using the analytic expression  \ Eq.~(\ref{PEapprox12}), at zero bias the relaxation rate is 
\begin{equation}
\gamma_{\varepsilon=0} = 4^{-1/\beta\omega_c} \;
\frac{2 \pi \, T_c^2 \, \Gamma(1\!+\!2/\beta\omega_c)}
{\omega_c \, |\Gamma(1\!+\!1/\beta\omega_c)|^2},
\end{equation}
which correctly converges to the zero temperature result of the spin-boson model, $\gamma \!=\! 2\pi T_c^2/\omega_c$, cf.~\cite{SWeiss90}.

\begin{figure}[t]
\includegraphics[width=0.45\textwidth]{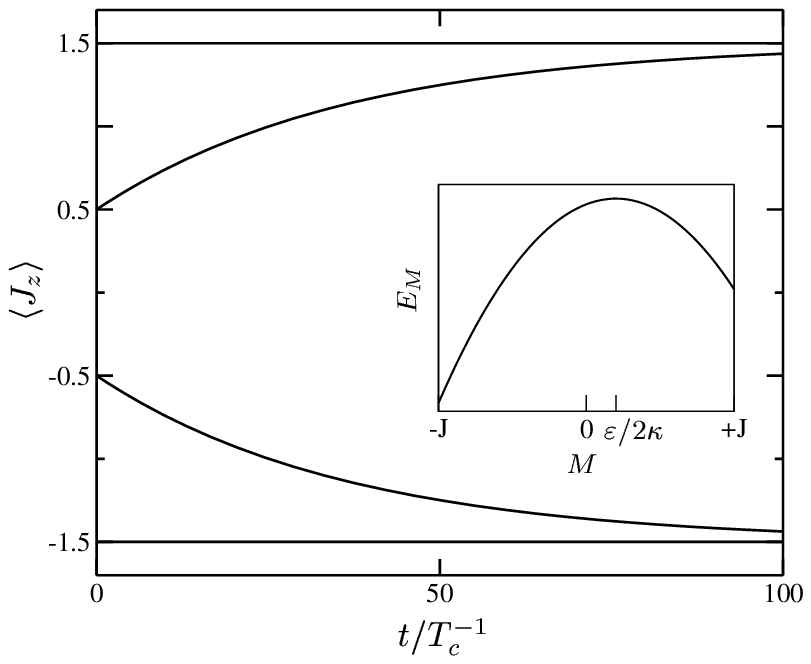}
\includegraphics[width=0.45\textwidth]{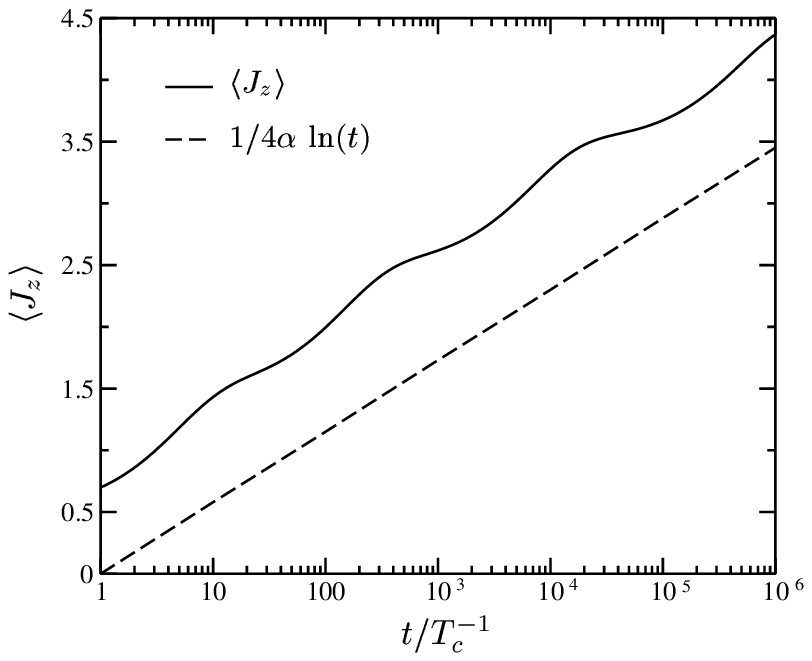}
\caption{ \label{VB04_Fig1.eps} {\bf Left:} Time-dependence of  $J_z$ for a spin $3/2$ with different initial values ($\varepsilon\!=\!0$, $\alpha\!=\!1$, $\omega_c\!=\!50T_c$, and $k_BT\!=\!0$). Inset: eigenenergies of the unperturbed system. {\bf Right:} Logarithmic plot of the relaxation of a spin of size $j\!=\!4.5$ with initial value $\langle J_z \rangle_0\!=\!1/2$ ($\varepsilon\!=\!0$, $\alpha\!=\!1$, $\omega_c\!=\!50T_c$, and $k_BT\!=\!0$)
and the approximation~(\ref{eq_log-relax}) with $C\!=\!0$. From \cite{VB04}.}
\end{figure}

While spins $j=1/2$ are localized for $\alpha\!\ge\!1$ and $\varepsilon=0$, spins $j>1/2$ relax towards one of the two  energy minima of $\bar{H}_0$, i.e. the polarized states $|j,\pm j\rangle$ on the inverted parabola, cf. Fig.~\ref{VB04_Fig1.eps}, depending on the initial spin value. 
For an initial value $|jm_0\rangle$ on the ascending branch, $m_0 > 0$, the Master equation describes transitions $m \to m+1$ at rate
\begin{equation}
\label{eq_rate}
\Gamma_{m \to m+1} =  2 \pi T_c^2 \; {c_{jm}^{\,+}}^2 \;   P \big(\!-\!\varepsilon\!+\!\kappa(2m\!+\!1)\big),
\end{equation}
with at zero temperature, \ Eq.~(\ref{PET0}), obeys $\Gamma_{m \to m+1}\ll \Gamma_{m-1 \to m}$ such that each transition happens much slower than the previous one. As a consequence, the transition $m-1 \to m$ is  completed before the next transition, $m \to m+1$, becomes effective. The spin therefore cascades down to its equilibrium value $+j$, with the time $t(m)$ needed to relax to a state $|jm \rangle$ approximately being independent of the initial state and only governed by the last transition. An estimate is obtained from Eq.~(\ref{eq_rate}),
\begin{equation}
t(m) \approx \frac{1}{2 \pi T_c^2  {c_{jm}^{-}}^2 
   P \big(-\varepsilon+\kappa(2m-1)\big)} .
\end{equation}
Furthermore, for Ohmic dissipation one can derive an approximation for 
\begin{equation}
\label{eq_log-relax}
\langle J_z \rangle \approx \frac{1}{4\alpha} \ln(t) + C,
\end{equation}
where all other parameters are absorbed in the constant~$C$. The logarithmic relaxation, cf. Fig.~\ref{VB04_Fig1.eps}, is due to the exponential cut-off in the boson spectral density $J(\omega)=2\alpha \omega e^{-\omega/\omega_c}$, cf. \ Eq.~(\ref{Jomegageneric}){}. Other forms of $J(\omega)$ will therefore lead to other time-dependences of large-spin relaxation in the strong coupling limit.

At finite temperatures, the spin can also absorb energy from the environment and transitions in both directions $m \leftrightarrow m + 1$ are possible.  Due to the detailed balance relation, \ Eq.~(\ref{detailed_balance}){}, the absorption rate is much smaller than the emission rate and does not deviate much from  the zero temperature `ultra-slow radiance' behavior \cite{VB04}.

\subsubsection{Collective Decoherence of Qubit Registers}

Reina, Quiroga and Johnson \cite{RQJ02} considered an exactly solvable spin-boson model where the coupling to a number of $L$  spin-$\frac{1}{2}$s (qubits) to a bath of boson modes $q$ is via the individual spin-$z$ components $\sigma_z^n$ only,
\begin{eqnarray}\label{H_Johnson}
  H=\sum_{n=1}^L \varepsilon_n\sigma_z^n + 
\sum_{q}\sum_{n=1}^L\sigma_z^n\left( g_q^n a^{\dagger}_q +  {g_q^n}^* a^{\phantom{\dagger}}_q\right)
+ \sum_q \omega_q a^{\dagger}_q a^{\phantom{\dagger}}_q.
\end{eqnarray}
The absence of  tunneling term with coupling to $\sigma_x^n$ makes this model somewhat unrealistic from the point of view of applications to real physical situations. Models like \ Eq.~(\ref{H_Johnson}) are sometimes called `pure dephasing models' in the literature; the fact that they can be solved exactly makes them attractive for, e.g. illustrating the temporal decay of off-diagonal elements of a density matrix due to dissipative environment coupling. 

The exact solution  is accomplished by  calculating the time evolution operator, which is given by $U_I(t)$ $=\hat{T}\exp[-i\int_0^t dt' H_I(t') ]$ in the interaction picture with respect to the free spin and boson part of Eq.~(\ref{H_Johnson}), where care has to be taken to properly take into account the time-ordering operation $\hat{T}$, cf. the discussion in \cite{RQJ02}. A case of particular interest occurs for `collective decoherence' where all the coupling constant $g_q^n\equiv g_q$ are identical and the matrix elements $\rho_{i_n,j_n}(t)$ of the reduced density operator  of the spin systems evolve as
\begin{eqnarray}
  \rho_{i_n,j_n}(t) &=& 
\exp \Big\{ i \Theta(t) \left[ \left( \sum_m i_m \right)^2 -  \left( \sum_m j_m \right)^2 \right]
- \Gamma(t) \left[ \sum_m (i_m-j_m)\right]^2 \Big\}
\rho_{i_n,j_n}(0) \label{rho_Johnson}\\
\Theta(t)&\equiv& \int_{0}^{\infty}d\omega \frac{J(\omega)}{\omega^2} \left[ \omega t - \sin \omega t \right],\quad \Gamma(t) \equiv  \int_{0}^{\infty}d\omega \frac{J(\omega)}{\omega^2} \left(1-\cos \omega t \right)
\cosh \left(\frac{\beta\omega}{2}\right).
\end{eqnarray}
The result \ Eq.~(\ref{rho_Johnson}) reflects the abelian nature of the dephasing: the time evolution of the density operator only consists in a multiplication of the initial density operator with an exponential factor $\exp \{...\}$ which, however, itself depends on the coordinates $i_n,j_n$. Note that the real `decay rate' $ \Gamma(t)$ is given by the real part of the function $Q(t)$ in Eq.~(\ref{Ct}), $\Gamma(t) = \mbox{\rm Re } Q(t)$, which therefore with the explicit expression\ Eq.~(\ref{Qz}) can be calculated analytically. 

Reina, Quiroga and Johnson \cite{RQJ02} discussed the case of $L=1$ and $L=2$ qubits for $s=1$ (Ohmic) and $s=3$ (super-Ohmic) dissipation in detail. An interesting result is the fact that, apart from the diagonal elements of the density operator \ Eq.~(\ref{rho_Johnson}), the matrix elements $\langle \uparrow\downarrow| \rho(t)| \downarrow \uparrow \rangle$ and $\langle \downarrow\uparrow| \rho(t)| \uparrow \downarrow \rangle$
do not decay at all. For identical $g_q^n$, the interaction part of the Hamiltonian \ Eq.~(\ref{H_Johnson}) in fact gives zero on the states $|\uparrow\downarrow\rangle$ and $|\downarrow\uparrow\rangle$. In contrast, the matrix elements $\langle \uparrow\uparrow| \rho(t)| \downarrow \downarrow \rangle$  (and  correspondingly $\langle \downarrow\downarrow| \rho(t)| \uparrow \uparrow \rangle$) decohere fast according to
\begin{eqnarray}
  \langle \uparrow\uparrow| \rho(t)| \downarrow \downarrow \rangle  = 
\exp\left[-4\Gamma(t)\right] \langle \uparrow\uparrow| \rho(0)| \downarrow \downarrow \rangle, 
\end{eqnarray}
which has been called `super-decoherence' in an earlier paper by Palma, Suominen and Ekert \cite{PSE96}.

\subsubsection{Superradiance in Arrays of Cooper Pair Boxes}
Rodrigues, Gy\"orffy, and Spiller \cite{RGS04} proposed a model for collective effects in the Cooper-pair tunnel current in an array of  Cooper-pair boxes coupled to a large, superconducting reservoir with BCS Hamiltonian $\mathcal{H}_{\rm BCS}$. They started from a total Hamiltonian
\begin{eqnarray}
  \mathcal{H}&=& \mathcal{H}_{\rm array} + \mathcal{H}_{\rm BCS} + \mathcal{H}_T\\
\mathcal{H}_{\rm array}&=& \sum_{i=1}^{l_B}E_i^{\rm ch}\sigma_i^z,\quad
\mathcal{H}_T=-\sum_{k,i}T_{k,i}(\sigma_i^+ c_{-k\downarrow}c_{k\uparrow} + H.c. ),
\end{eqnarray}
where in $\mathcal{H}_{\rm array}$ and the tunneling Hamiltonian $\mathcal{H}_T$  each of the $l_b$ Cooper-pair boxes was described as a two-level system with pseudo spin $\frac{1}{2}$. For $i$-independent charging energies $E_i^{\rm ch}$ and tunnel matrix elements $T_{k,i}$, the Cooper-pair array was described by a collective spin $S_b$ of size $l_b/2$. In addition, they wrote the BCS Hamiltonian using $k$-dependent Nambu spins $\sigma_k^+=c^{\dagger}_{k\uparrow}c^{\dagger}_{-k\downarrow}$ etc. and assumed a strong coupling limit in which the dispersion of the particle energies $\xi_k$ (counted from the chemical potential) and the pairing field $\Delta_k$ became negligible, giving rise to a spin representation $   \mathcal{H}_{\rm BCS} \approx 2\xi S_r^z - S_r^+ \Delta - S_r^- \Delta^*$ (the total spin $S_r=l_r/2\to \infty$ represented half the degeneracy in the reservoir),  which they checked to reproduce standard BCS results. 

Neglecting furthermore the $k$-dependence of the $T_{k,i}$ allowed them to work with an effective Hamiltonian with two large spins representing the box-array ($b$) and the reservoir ($r$),
\begin{eqnarray}
  \mathcal{H}_{\rm eff} = E^{\rm ch} S_b^z + 2 \xi S_r^z - S_r^+ \Delta - S_r^- \Delta^*
- T(S_b^+ S_r^- + S_r^+ S_b^-).
\end{eqnarray}
Using lowest order time-dependent perturbation theory in $\mathcal{H}_T$, they then calculated the expectation value of the tunnel current operator $\hat{I}\equiv T(S_b^+S_r^--S_b^-S_r^+)$ for various initial conditions (number and coherent states for array and reservoir). They found a current proportional to the {\em square} of the number $l_b$ of boxes in the array which demonstrated the Dicke superradiance effect in their system. In a further calculation, Rodrigues {\em et al.} also made explicit predictions for a collective,  time-dependent quantum revival effect in analogy with the quantum optical revivals in the Jaynes-Cummings model \cite{Walls}.


\section{\bf Dicke Effect and Spectral Line-shapes}\label{section_spectral}

\subsection{Introduction}

The original Dicke effect as predicted by Dicke in 1953 \cite{Dic53} is a phenomenon that occurs in the line shapes  of absorption spectra in a gas of atoms. Line shapes for the absorption of light with wave vector ${\bf k}$ are subject to Doppler broadening due to frequency shifts ${\bf kv}$, where ${\bf v}$ is the velocity of an individual atom. Dicke showed that velocity-changing collisions of the radiating atoms with the atoms of a (non-radiating) {buffer gas} can lead  to a substantial {\em narrowing } of the spectral line shape in the form of a very sharp peak on top of a broad line shape, centered around the transition frequency of the atom, cf. Fig. \ref{dickespec.eps}. The {\em incoherent} coupling of {two} independent relaxation channels (spontaneous emission and velocity-changing collisions) leads to a splitting into two combined decay channels for the whole system, cf. section \ref{section_collision}. This phenomenon is somewhat in analogy with the {\em coherent} coupling of two (real) energy levels (level repulsion), leading to the formation of a bonding and an anti-bonding state. The difference is that the Dicke effect is related to the splitting of decay rates (`imaginary energies'), and not real energies, into a large (fast, superradiant) and a small (slow, subradiant) decay rate. In fact, the splitting into two decay modes can be considered as a precursor of the  phenomenon of Dicke superradiance \cite{Dic54}, where a symmetric mode of $N$ radiators gives rise to an abnormally large decay on a time scale $1/N$.

From the theoretical point of view, spectral line-shapes are determined by poles of  correlation functions in the complex frequency plane. The poles are eigenvalues of a collision matrix which, for the simplest case of only two poles, belong to symmetric and anti-symmetric eigenmodes. As a function of an external parameter (e.g. the pressure of an atomic gas), these poles can move through the lower frequency half-plane, whereby the spectral line-shape becomes a superposition of a strongly broadened and a strongly sharpened peak. 

\begin{figure}[t]
\includegraphics[width=0.5\textwidth]{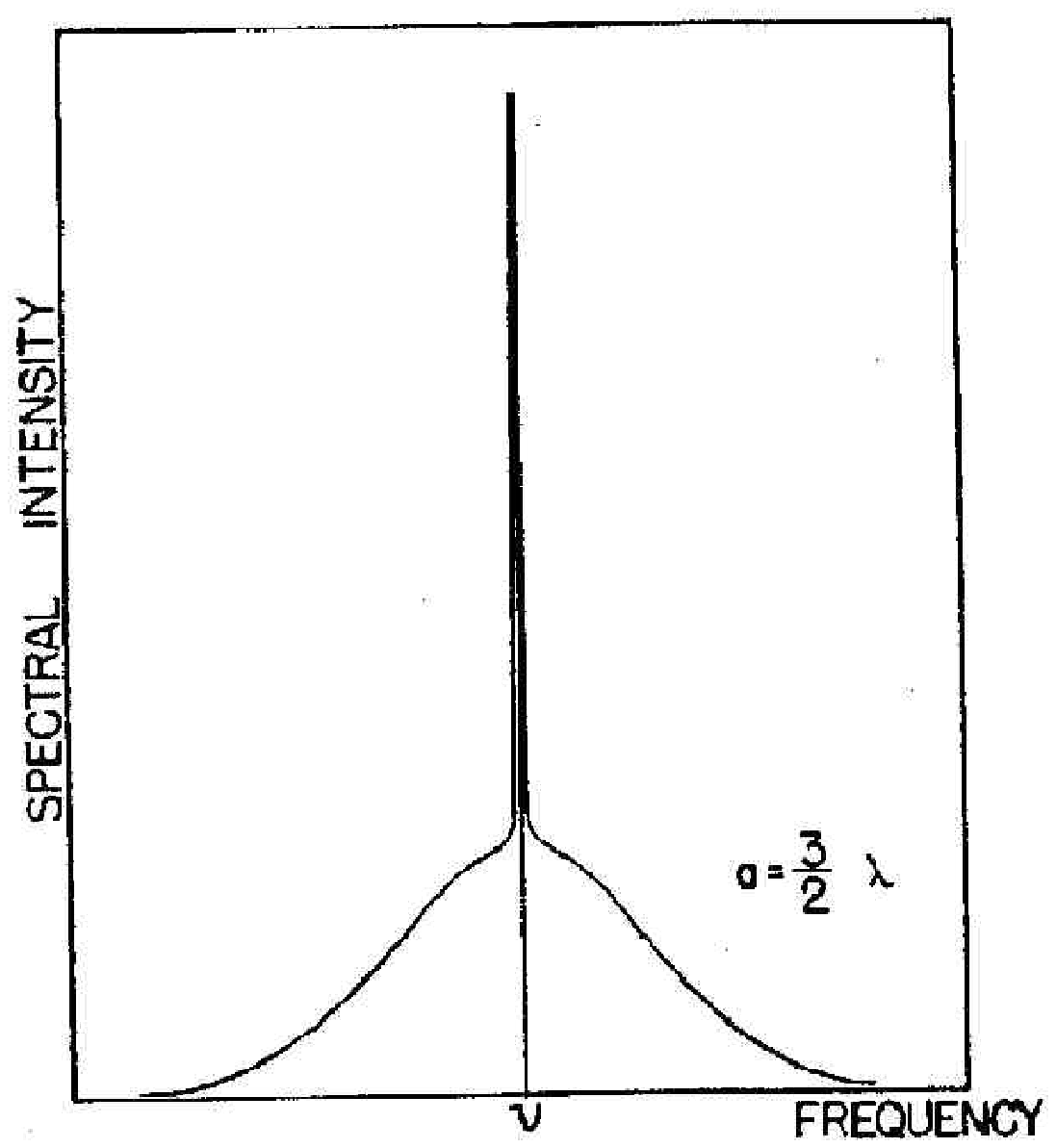}
\includegraphics[width=0.5\textwidth]{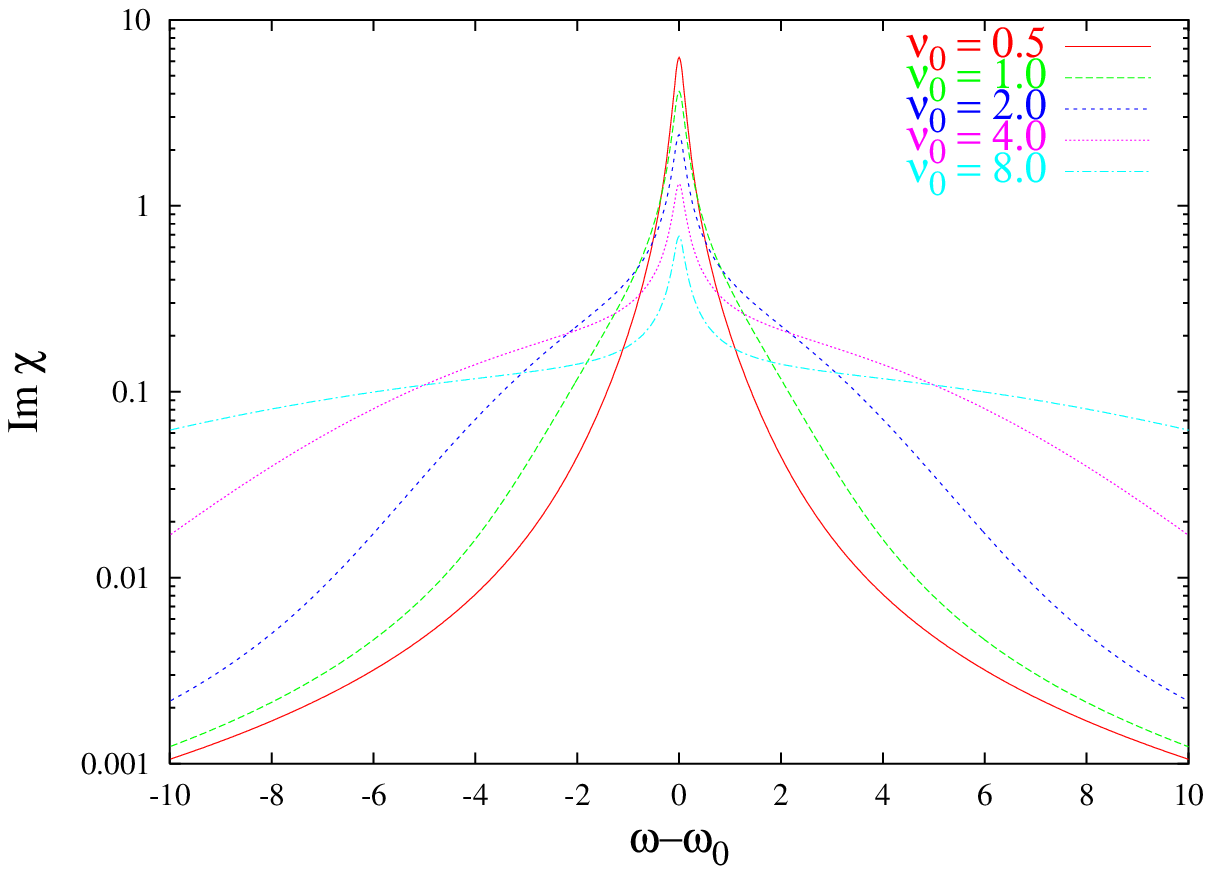}
\caption[]{\label{dickespec.eps}{\bf Left:} Line narrowing due to collisions of a Doppler--broadened spectral
line in the original 1953 Dicke paper \cite{Dic53}. The radiating gas is modeled
within a one-dimensional box of width $a$; $\lambda$ is the light wavelength. From \cite{Dic53}. {\bf Right:}
Imaginary part of the polarizability $\chi(k,\omega)$
in units of $d^2/\Gamma$ ($d$: dipole moment)
around $\omega=\omega_0$ for a one-dimensional model (see text). All frequencies are in units of the
collision rate $\Gamma$ of the radiating atoms with the atoms of the buffer gas.
The spontaneous emission rate $\gamma=0.1$; 
the atom mass $M$ and the light wave vector $k$ enter into the frequency
$\nu_0=kp_0/M$ which determines the width of the momentum distribution. For broad distributions (larger $\nu_0$), the sharp `Dicke-peak' appears on top of the Doppler-broadened line-shape.}
\end{figure} 

From an abstract point of view, the Dicke effect has its roots in the properties of eigenvalues and eigenvectors of matrices of a special form. Consider a $N\times N$ matrix 
\begin{eqnarray}\label{spectral_matrix}
  A=    \begin{pmatrix}
      1 & q & q & q & ...\\
      q & 1 & q & q & ...\\
      q & q & 1 & q & ...\\
      q & q & q & 1 & ...\\
      ...  
    \end{pmatrix},
\end{eqnarray}
representing a coupling among $N$ objects $i=1,...,N$ with identical real coupling strengths $A_{i\ne j}=q$ and unity `self-coupling'. The eigenspace of $A$ is spanned by a single, `superradiant' and symmetric eigenvector $(1,1,1,...,1)$ with eigenvalue $1+q(N-1)$, and $N-1$ degenerate, `subradiant'  eigenvectors, e.g.  $(1,-1,0,...,0)$, $(1,0,-1,...,0)$ etc. with eigenvalue $1-q$. This splitting into sub- and superradiant subspaces is a very generic feature due to the high symmetry of $A$, and has important consequences for physical systems where such symmetries play a role.

Although the Dicke spectral line effect has been known and experimentally verified for a long time in atomic systems \cite{Benedict,Andreev,Ber75}, only recently predictions were made for it to occur in transport and scattering properties of mesoscopic systems such as  for resonant electron tunneling  via two impurities \cite{SR94}, resonant scattering  in a strong magnetic field \cite{SU98}, or the emission from disordered mesoscopic systems \cite{SRV00}. This section represents an introduction to the effect by an explicit calculation of the collision-induced narrowing of the polarizability $\chi(\omega)$ of an atomic gas (as was considered in Dicke's 1953 paper), a short overview over the results from the seminal paper by Shabazyan and Raikh, and recent work related to the effect.

\subsection{Atomic Line Shapes and Collision Effects}\label{section_collision}
The Dicke effect (line narrowing due to collisions), its experimental consequences and the conditions under which it can been observed have been  reviewed   by Berman \cite{Ber75}. Here, we provide a short  review of the theoretical aspects of the original effect by using the Boltzmann equation for a gas of two-level atoms of mass $M$ as described by a one-particle density matrix, defined as a trace of the statistical operator $\rho$,
\begin{equation}
  \label{eq:oneparticledensity}
  \rho_{\sigma\sigma'}({\bf r}_1,{\bf r}_2;t)\equiv 
{\rm Tr} \left(\rho \Psi_{\sigma'}^+({\bf r}_2t)\Psi_{\sigma}({\bf r}_1t)\right),
\end{equation}
where the field operator $\Psi_{\sigma}^+({\bf r}_2)$ creates an atom at position ${\bf r}_2$ with the upper level ($\sigma = \uparrow$) or the lower level ($\sigma = \downarrow$) occupied. The `spin'-index $\sigma$ thus denotes the internal degree of freedom of the atom. 
An electric field ${\bf E}({\bf x},t)$ now gives rise to dipole transitions within an atom at position ${\bf x}$. If the corresponding matrix element is denoted as ${\bf d}$ (for simplicity we set ${\bf d}\equiv{\bf d}_{\uparrow\downarrow}\equiv {\bf d}_{\downarrow\uparrow}$), and the transition frequency is $\omega_0$, the  Hamiltonian
of the system in second quantization is 
\begin{eqnarray}
  \label{eq:gashamil}
  H&=&\sum_{\sigma=\pm}\int d^3{\bf x}  \Psi^+_{\sigma}({\bf x})
\left[ \sigma\frac{\omega_0}{2}-\frac{\Delta}{2M} \right]
\Psi_{\sigma}({\bf x}) \nonumber\\
&+&\int d^3{\bf x}({\bf dE}({\bf x},t))\left[ \Psi^+_{\uparrow}({\bf x})
 \Psi_{\downarrow}({\bf x}) +
\Psi^+_{\downarrow}({\bf x}) \Psi_{\uparrow}({\bf x})\right],
\end{eqnarray}
where $\Delta$ is the Laplacian and we have set $\hbar=1$. The quantum-mechanical distribution function
\begin{equation}
  \label{eq:wigner}
  f({\bf p}, {\bf r}, t) =\frac{1}{(2\pi)^3}\int d^3{\bf r}'
e^{-i{\bf pr'}}\rho({\bf r},{\bf r'};t)
\end{equation}
with ${\bf r} = ({\bf r}_1 + {\bf r}_2)/2$ and 
${\bf r}' = {\bf r}_1 - {\bf r}_2$ obeys an equation of motion as derived from the Heisenberg equations of the field operators $\Psi_{\sigma}({\bf x})$,  
\begin{eqnarray}
  \label{eq:eqnofmotion}
  \left(\frac{\partial}{\partial t}-i\omega_0+{\bf v_p \nabla_r}\right)f_{\downarrow\uparrow}
({\bf p}, {\bf r}, t) 
&=&i{\bf dE}({\bf r},t)\left[
f_{\uparrow\uparrow}({\bf p}, {\bf r}, t)- 
f_{\downarrow\downarrow}({\bf p}, {\bf r}, t)\right],\nonumber\\
  \left(\frac{\partial}{\partial t}+i\omega_0+{\bf v_p \nabla_r}\right)f_{\uparrow\downarrow}
({\bf p}, {\bf r}, t) 
&=&-i{\bf dE}({\bf r},t)\left[
f_{\uparrow\uparrow}({\bf p}, {\bf r}, t)- 
f_{\downarrow\downarrow}({\bf p}, {\bf r}, t)\right]
\end{eqnarray}
with ${\bf v_p}={\bf p}/M$
and corresponding equations for $f_{\uparrow\uparrow}$ and $f_{\downarrow\downarrow}$. The electric field ${\bf E}({\bf x},t)$ has been assumed to spatially vary on a length scale which is much larger than the de-Broglie wave length of the atoms; apart from this Eq.(\ref{eq:eqnofmotion}) is exact.

The Dicke effect has its origin in collisions of the atoms with a buffer gas. These collisions are assumed to change only the momentum ${\bf p}$ of the atoms and not their internal degree of freedom $\sigma$. Furthermore, the buffer gas is optically inactive; a situation that in a condensed matter setting would correspond to elastic scattering of electrons at impurities in electronic systems like metals or semiconductors. In the theoretical description of these scattering events, one introduces a {\em collision term} 
\begin{eqnarray}
  \label{eq:collterm}
 {\mathcal{L}}[f_{\sigma,\sigma'}]({\bf p}, {\bf r}, t)\equiv 
&-&\int d{\bf p}' W({\bf p},{\bf p}')
\left[f_{\sigma,\sigma'}({\bf p}, {\bf r}, t) - f_{\sigma,\sigma'}({\bf p}', {\bf r}, t)\right]
\end{eqnarray}
on the r.h.s. of the kinetic equation Eq.(\ref{eq:eqnofmotion}), where $W({\bf p},{\bf p}')$ is the probability for scattering from ${\bf p}$ to ${\bf p}'$, which can be calculated in second order perturbation theory (Fermi's Golden rule) from a scattering potential. Furthermore, the spontaneous decay due to spontaneous emission of light from the upper level of the atoms leads to a decay of the polarization at a rate $\gamma$. This dissipative process is introduced as an additional collision term for  $f_{{\uparrow\downarrow}}$ and $f_{{\downarrow\uparrow}}$
\begin{eqnarray}
{\mathcal L}'[f_{{\downarrow\uparrow}}]=-\gamma f_{{\downarrow\uparrow}},\quad
{\mathcal L}'[f_{{\uparrow\downarrow}}]=-\gamma f_{{\uparrow\downarrow}}.
\end{eqnarray}
The {\em polarization} of the atom gas 
\begin{equation}
  \label{eq:defpolarization}
  {\bf P}({\bf r},t)={\bf d} \int d{\bf p}\left[f_{\uparrow\downarrow}({\bf p}, {\bf r}, t)
+ f_{\downarrow\uparrow}({\bf p}, {\bf r}, t)\right]
\end{equation}
is obtained in {\em linear response} to the electric field, i.e., the occupation probabilities of the upper and lower level are assumed to be constant in time and space,
$f_{\uparrow\uparrow}({\bf p}, {\bf r}, t)- 
f_{\downarrow\downarrow}({\bf p}, {\bf r}, t)$ $=N({\bf p})$.
The resulting equation of motion for $f_{\uparrow\downarrow}$ then becomes
\begin{eqnarray}
  \label{eq:eqnofmotion2}
  \left(\frac{\partial}{\partial t}+i\omega_0+\gamma+{\bf v_p \nabla_r}\right))f_{\uparrow\downarrow}
({\bf p}, {\bf r}, t)
&=&-i{\bf dE}({\bf r},t)N({\bf p})+ {\mathcal{L}}[f_{\uparrow\downarrow}]({\bf p}, {\bf r}, t),
\end{eqnarray}
which is a linearized {\em Boltzmann equation} for the distribution
function $f_{\uparrow\downarrow}$. 
Dicke originally considered the scattering processes in a one-dimensional model: atoms bouncing back and forth within a one-dimensional container \cite{Dic53}, a situation that easily allows one to understand the line narrowing from Eq.(\ref{eq:eqnofmotion2}). Due to energy conservation, $W(p,p')\propto \delta(p^2-p^{'2})$, which one can write as 
\begin{equation}
  \label{eq:Wppdef}
  W(p,p')=\Gamma(p)[\delta(p-p')+\delta(p+p')],
\end{equation}
where $\Gamma(p)=\Gamma(-p)$ is a scattering rate with dimension $1/$time. In the collision integral, only the back-scattering term remains, i.e.,
\begin{eqnarray}
  \label{eq:collterm2}
  {\mathcal{L}}[f_{\sigma,\sigma'}]( p, r, t)&\equiv &
-\int d{p}' \Gamma(p)\delta(p+p')
\left[f_{\sigma,\sigma'}({p},{r},t) - f_{\sigma,\sigma'}({-p}, {r}, t)\right]\nonumber\\
&=&-\Gamma(p)\left[f_{\sigma,\sigma'}({p},{r},t) - f_{\sigma,\sigma'}({-p}, {r}, t)\right].
\end{eqnarray}
The solution of Eq.(\ref{eq:eqnofmotion2}) is easily obtained in Fourier-space where
$\partial_t\to-i\omega$ and $\partial_r\to ik$;
\begin{eqnarray}
  \label{eq:twolevelfourier}
\left(-i\omega+i\omega_0+\gamma+\Gamma(p)+iv_pk\right)f_{\uparrow\downarrow}(p,k,\omega)\nonumber\\
&-&\Gamma(p)f_{\uparrow\downarrow}(-p,k,\omega)=-i{dE}({q},\omega)N({p}).
\end{eqnarray}
This can be solved by writing a second equation for $f_{\uparrow\downarrow}(-p,k,\omega)$
by simply changing $p\to -p$. The result is a two-by-two system of equations
for $f_{\uparrow\downarrow}(p)$ and $f_{\uparrow\downarrow}(-p)$ (omitting all other variables for the moment),
\begin{eqnarray}
  \label{eq:twolevelfourier2}
&  &
\left(
  \begin{array}[c]{cc}
-i\Omega_p+\Gamma(p)    & -\Gamma(p)\\
-\Gamma(p) & -i\Omega_{-p}+\Gamma(p)
  \end{array}
\right)
\left(
  \begin{array}[c]{c}
f_{\uparrow\downarrow}(p)\\
f_{\uparrow\downarrow}(-p)
\end{array}
\right)
=
\left(
  \begin{array}[c]{c}
g(p)\\
g(-p)
\end{array}
\right),\nonumber\\
& &
\end{eqnarray}
with the abbreviations
$g(p)\equiv -i{dE}({k},\omega)N({p})$ and $\Omega_p\equiv \omega-\omega_0-v_pk+i\gamma$. 
Note that the velocity $v_p$ is an odd function of $p$, $v_p\equiv p/M=-v_{-p}$.
Inverting the two-by-two matrix yields 
\begin{eqnarray}
  \label{eq:fpkwresult}
  f_{\uparrow\downarrow}(p,k,\omega) &=&i{dE}({k},\omega)N({p})
\frac{-i(\omega-\omega_0+v_pk+i\gamma)+2\Gamma(p)}
{[\omega-\omega_+(p,k)][\omega-\omega_-(p,k)]},
\end{eqnarray}
where the two poles $\omega_{\pm}(p,k)$ in the denominator of Eq.(\ref{eq:fpkwresult}) are given by
\begin{eqnarray}
\label{zerofpkresult}
\omega_{\pm}(p,k)\equiv \omega_0-i\gamma-i\left(\Gamma(p)\pm\sqrt{\Gamma(p)^2-v_p^2k^2}\right),
\end{eqnarray}
and the result for $f_{\downarrow\uparrow}(p,k,\omega) $ is obtained from Eq.(\ref{eq:fpkwresult}) by changing $\omega_0\to -\omega_0$ and $N(p)\to -N(p)$.  
Using these results, one can now express a linear relation between the Fourier
transform of the  polarization, Eq.(\ref{eq:defpolarization}), and the electric
field $E(k,\omega)$,
\begin{eqnarray}
  \label{eq:linearpol}
  P(k,\omega)&=&\chi(k,\omega)E(k,\omega)\nonumber\\
\chi(k,\omega) &=& d^2\int dp N(p) \frac
{\omega-\omega_0+v_pk+i\gamma+2i\Gamma(p)}
{[\omega-\omega_+(p,k)][\omega-\omega_-(p,k)]}-(\omega_0\to -\omega_0).
\end{eqnarray}
The spectral line shape is determined by the {\em polarizability} $\chi(k,\omega)$, the form of which in turn depends on the position of the poles $\omega_{\pm}(p,k)$ in the complex $\omega$-plane. It is useful to consider  two limiting cases:
\begin{enumerate} 
\item the collision-less limit $\Gamma^2(p)\ll v_p^2k^2$, cf. Fig.(\ref{zero.eps}) : in this case,
$\omega_{\pm}(p,k) \approx \omega_0\pm v_pk -i\gamma$. The line-width is determined
by the broadening through spontaneous emission $\gamma$ and is shifted from the
central position $\omega_0$ by the {\em Doppler-shifts} $\pm v_pk$. Note that
the final result for the polarizability still involves an integration over the
distribution function $N(p)$ and therefore depends on the occupations
of the upper and lower levels. This leads to the final {\em Doppler broadening }
due to the Doppler-shifts $\pm v_pk$.
\item the {\em Dicke-limit} $\Gamma^2(p)\gg v_p^2k^2$, cf. Fig.(\ref{zero.eps}), is a more interesting case, where in the square-root in the two poles the Doppler-broadening can be neglected and 
\begin{eqnarray}
  \label{eq:twopolesdicke}
  \omega_+&=&\omega_0-i\gamma-2i\Gamma(p),\quad
  \omega_-=\omega_0-i\gamma.
\end{eqnarray}
The first pole $\omega_+$ corresponds to a broad resonance of width $\gamma+2\Gamma(p)$, the second pole $\omega_-$ corresponds to a resonance whose width is solely determined by the `natural' line-width $\gamma$, i.e. a resonance which is no longer Doppler-broadened.
\end{enumerate} 
The splitting into two qualitatively different decay channels is the key feature of the Dicke effect. We have already encountered it in the emission of light from a two-ion system, section \ref{section_DeVB96}, where the spontaneous decay split into one fast (superradiant) and one slow (subradiant) channel.
\begin{figure}[t]
\begin{center}
\includegraphics[width=0.6\textwidth]{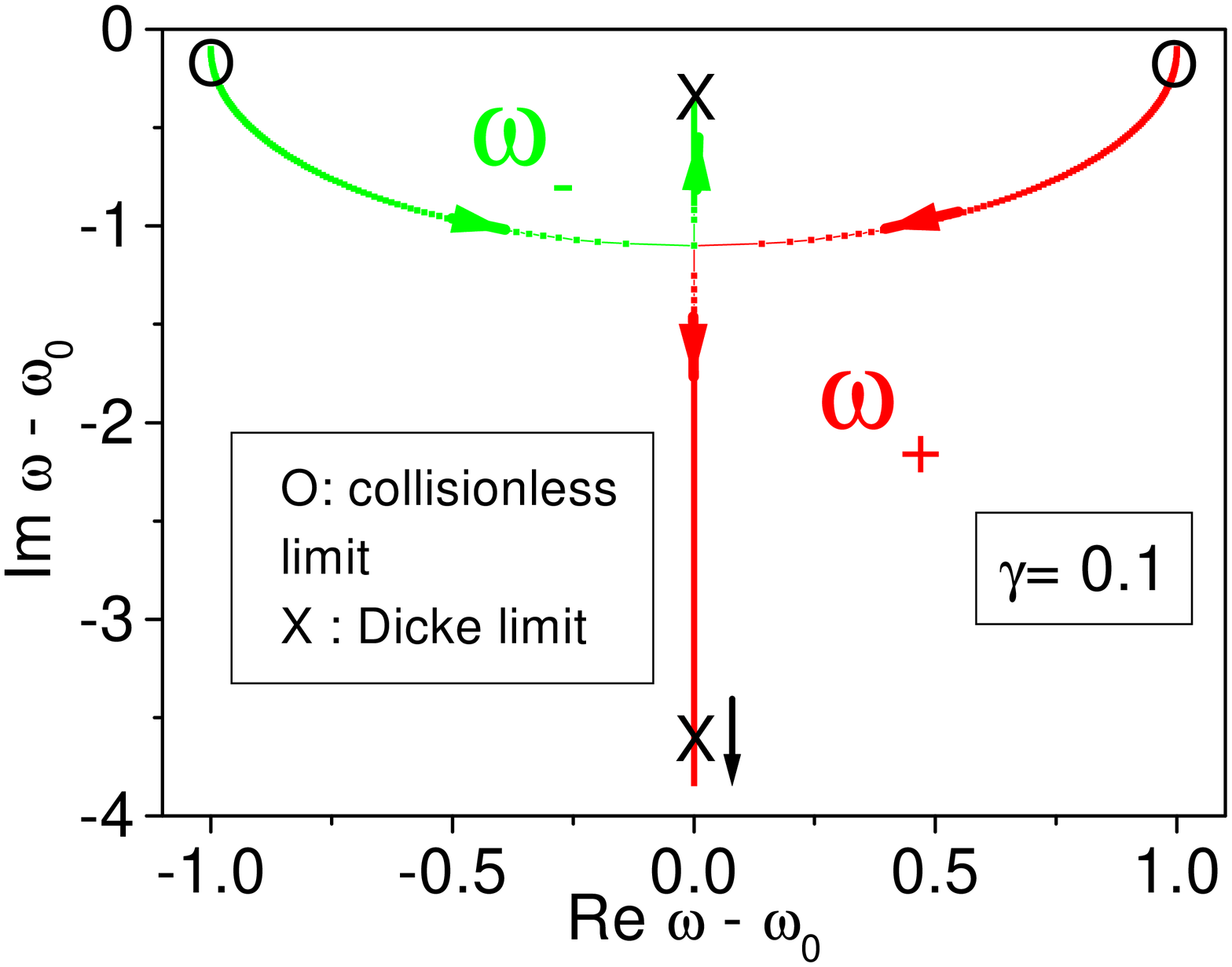}
\end{center}
\caption[]{\label{zero.eps}Zeros $\omega_{\pm}-\omega_0$ according to
Eq.(\ref{zerofpkresult}) appearing in the
distribution function Eq.(\ref{eq:fpkwresult}) and the polarizability
Eq.(\ref{eq:linearpol}). The real and imaginary part of the frequencies
are in units of the Doppler shift $v_pk$ which is fixed here. The
two curves are plots parametric in the elastic collision rate $\Gamma(p)$; the arrows
indicate the direction of increasing $\Gamma(p)$. For $\Gamma(p)\gg|v_p k|$, both
curves approach the {\em Dicke limit } Eq.(\ref{eq:twopolesdicke}), where
the imaginary part of $\omega_--\omega_0$ becomes the negative of $\gamma$, and
the imaginary part of $\omega_+-\omega_0$ flows to minus infinity.}
\end{figure} 
In fact, in the Dicke-limit the polarizability is given by a sum of 
the two resonances $\omega_{\pm}$: from Eq.(\ref{eq:linearpol}), one obtains
\begin{eqnarray}
  \label{twopolesexpand}
\chi(k,\omega) &=& d^2\int dp N(p) \frac
{\omega-\omega_0+v_pk+i\gamma+2i\Gamma(p)}
{\omega_+-\omega_-}\times\nonumber\\
&\times&\left[\frac{1}{\omega-\omega_+}-\frac{1}{\omega-\omega_-}\right]
-(\omega_0\to -\omega_0)
\end{eqnarray}
In the Dicke limit, this becomes
\begin{eqnarray}
\chi(k,\omega)&\approx& 
 d^2\int dp  \frac
{N(p)}
{-2i\Gamma(p)}\left[\frac{v_pk}{\omega-\omega_+}-\frac{2i\Gamma(p)}{\omega-\omega_-}\right]
-(\omega_0\to -\omega_0).
\end{eqnarray}
The two resonances thus correspond to an {\em anti-symmetric term} $v_pk/(\omega-\omega_+)$ 
(odd function of $p$) and a {\em symmetric term } $2i\Gamma(p)/(\omega-\omega_-)$ 
(even function of $p$).
Note that the anti-symmetric term gives no contribution to $\chi(k,\omega)$ for even
distribution $N(p)=N(-p)$. Still, the appearance of a definite type
of symmetry together with each type of resonance is typical for the Dicke effect and has 
its origin in the coupling of the two components $f(p)$ and $f(-p)$ in the matrix equation
Eq.(\ref{eq:twolevelfourier2}). The latter can be re-written 
(again considering only the component $f_{\uparrow\downarrow}$),
\begin{eqnarray}
  \label{eq:suggestive}
  (A-\lambda 1)\left(
  \begin{array}[c]{c}
f_{\uparrow\downarrow}(p)\\
f_{\uparrow\downarrow}(-p)
\end{array}
\right)
=
-ig(p)
\left(
  \begin{array}[c]{c}
1\\
1
\end{array}
\right),\quad A\equiv
\left(\begin{array}[c]{cc}
v_pk-i\Gamma(p)    & i\Gamma(p)\\
i\Gamma(p) & -v_pk-i\Gamma(p)
  \end{array}\right)
\end{eqnarray}
where  $N(p)=N(-p)$ and $\lambda\equiv \omega-\omega_0+i\gamma$. In the limit $\Gamma(p)\gg |v_pk|$, the matrix $A$ has the 
eigenvectors $(1,1)$ and $(1,-1)$ with eigenvalues $0$ and $-2i\Gamma(p)$, respectively.
For the symmetric eigenvector $(1,1)$ , the effect of the collisions is therefore annihilated to zero, and this eigenvector solves  Eq.(\ref{eq:suggestive}) with 
$ -\lambda f_{\uparrow\downarrow}(p)=-ig(p)$,
meaning 
\begin{eqnarray}
  \label{eq:solvesugg1}
  f_{\uparrow\downarrow}(p)=\frac{dE(k,\omega)N(p)}
{\omega-\omega_0+i\gamma}.
\end{eqnarray}
This agrees with the previous result Eq.(\ref{eq:fpkwresult}) in the 
Dicke limit $\Gamma(p)\gg |v_pk|$: the collision broadening 
has disappeared and the line is determined by the remaining natural
line width $\gamma$. 
%
It is instructive to discuss a quantitative numerical example, using a Gaussian distribution function $N(p)=f_{\uparrow\uparrow}-f_{\downarrow\downarrow}= -(2\pi p_0)^{-1/2}e^{-{p^2}/{2p_0^2}}$.
The imaginary part $\chi_1^{''}(k,\omega)$ of the first term in the polarizability Eq.(\ref{eq:linearpol}),
\begin{equation}
  \chi_1^{''}(k,\omega)\equiv d^2\Im {\rm m}\int dp N(p)\frac 
{\omega-\omega_0+v_pk+i\gamma+2i\Gamma(p)}
{[\omega-\omega_+(p,k)][\omega-\omega_-(p,k)]},
\end{equation}
corresponds to the resonance around $\omega\approx \omega_0$. 
The result (which requires one numerical integration) for  constant $\Gamma(p)=\Gamma$ is shown in Fig.(\ref{dickespec.eps} right) for different widths $\nu_0\equiv p_0k/M$ of the distribution $N(p)$. For a sharp momentum distribution (small $\nu_0$), the line-width is determined by the spontaneous emission rate $\gamma$ and there is basically no Doppler-broadening (Dicke-limit). In the opposite case of a broad momentum distribution, the form of the line  is determined by a sharp peak of width $\sim\gamma$ on top of a broad curve of width $\sim v_0$, which reflects the appearance of the {\em two poles} $\omega_+$ and $\omega_-$ in $\chi(k,\omega)$. 

\subsection{Spectral Function for Tunneling Via Two Impurity Levels}
The appearance of the Dicke effect in a spectral function for electronic states was  first found by Shahbazyan and Raikh in their paper from 1994 \cite{SR94}. They considered two-channel resonant tunneling of electrons through two impurities (localized states) coupled to electron reservoirs, cf. Fig.(\ref{SR94fig12.eps}).  If Coulomb interactions among the electrons are neglected, the {\em conductance} of the whole system can then be expressed by its scattering properties \cite{Lan70,Bue86,Jauho12}. 

We follow the discussion of Shahbazyan and Ulloa who later generalized this problem to the case of scattering properties in a strong magnetic field \cite{SU98}.
The starting point for the analysis of the conductance is the spectral function of the system, which can be expressed by the imaginary part of the retarded Green's function \cite{Mahan,Doniach}. For the case of two energy levels $\varepsilon_1$ and $\varepsilon_2$ that are assumed to belong to two spatially separated localized impurity states, the spectral function is defined by a two-by-two matrix in the Hilbert space of the two localized states,
\begin{eqnarray}\label{ShabS}
S(\omega)=-\frac{1}{\pi}\Im {\rm m}\frac{1}{2}{\rm Tr}\frac{1}{\omega -\hat{\varepsilon}+i\hat{W}}.
\end{eqnarray}
Scattering between the localized states $i\to |{\bf k}\rangle \to j$ is possible via virtual transitions to extended states (plane waves $|{\bf k}\rangle$) of the electron reservoir. Then, $\hat{\varepsilon}$ is diagonal in the $\varepsilon_i$, and $\hat{W}$ is a self-energy operator that describes the possibility of transitions between localized levels $i$ and $j$ via extended states with wave vector ${\bf k}$. In second order perturbation theory, the self-energy operator $\hat{W}$ is given by
\begin{equation}
W_{ij}=\pi \sum_{\bf k}t_{i{\bf k}}t_{{\bf k}j}\delta(\omega-E_{\bf k}),
\end{equation} 
where $\hbar=1$ and the dependence on $\omega$ of $\hat{W}$ is no longer indicated. The quantities $t_{i{\bf k}}$ are overlaps between the localized states $i$ and the plane waves $|{\bf k}\rangle$, their dependence on the impurity position ${\bf r}_i$ is given by the phase factor from the plane wave at the position of the impurity, i.e. $t_{i{\bf k}}\propto \exp(i{\bf kr}_i)$. 
\begin{figure}[tb]
\includegraphics[width=0.5\textwidth]{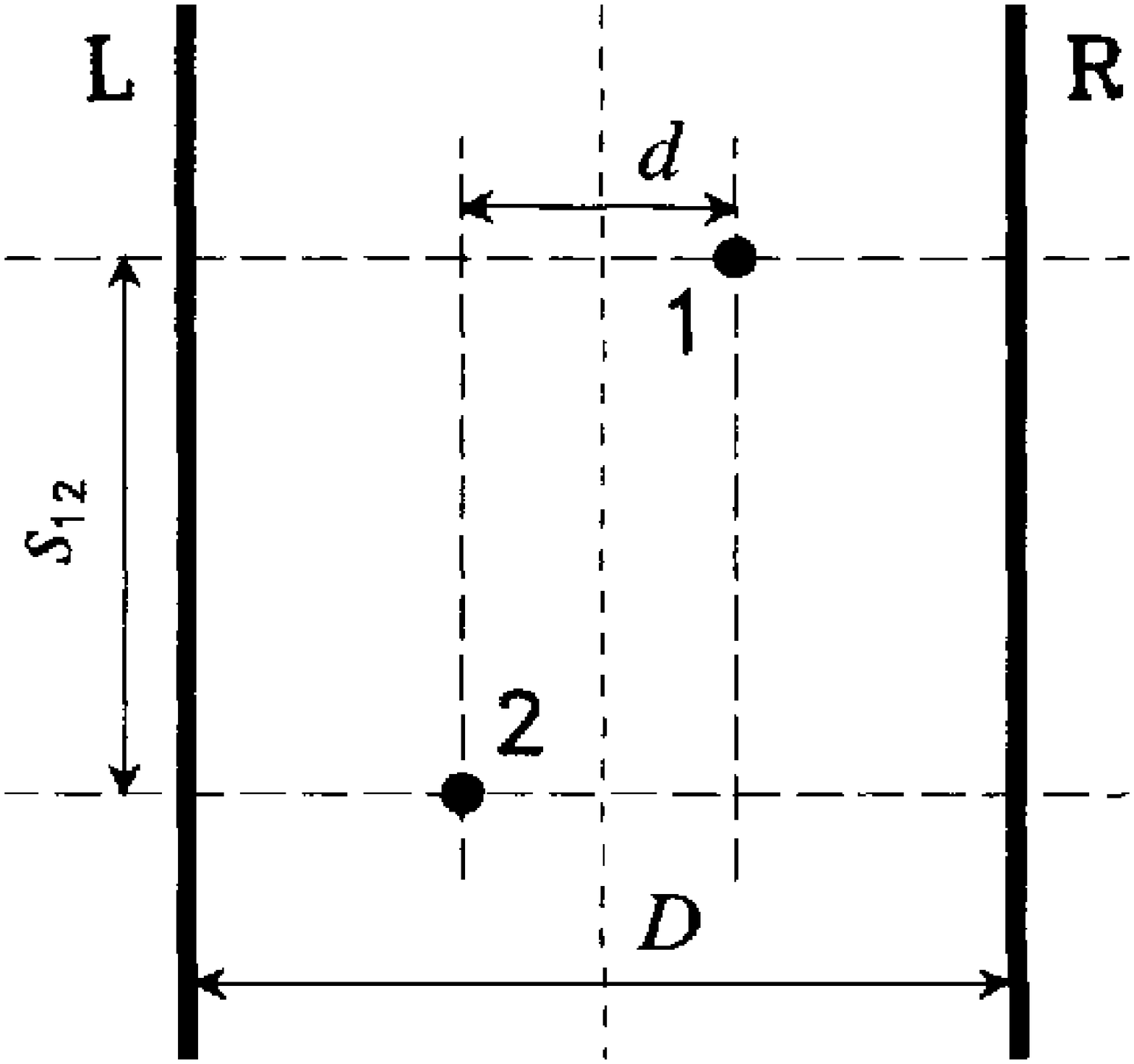}
\includegraphics[width=0.5\textwidth]{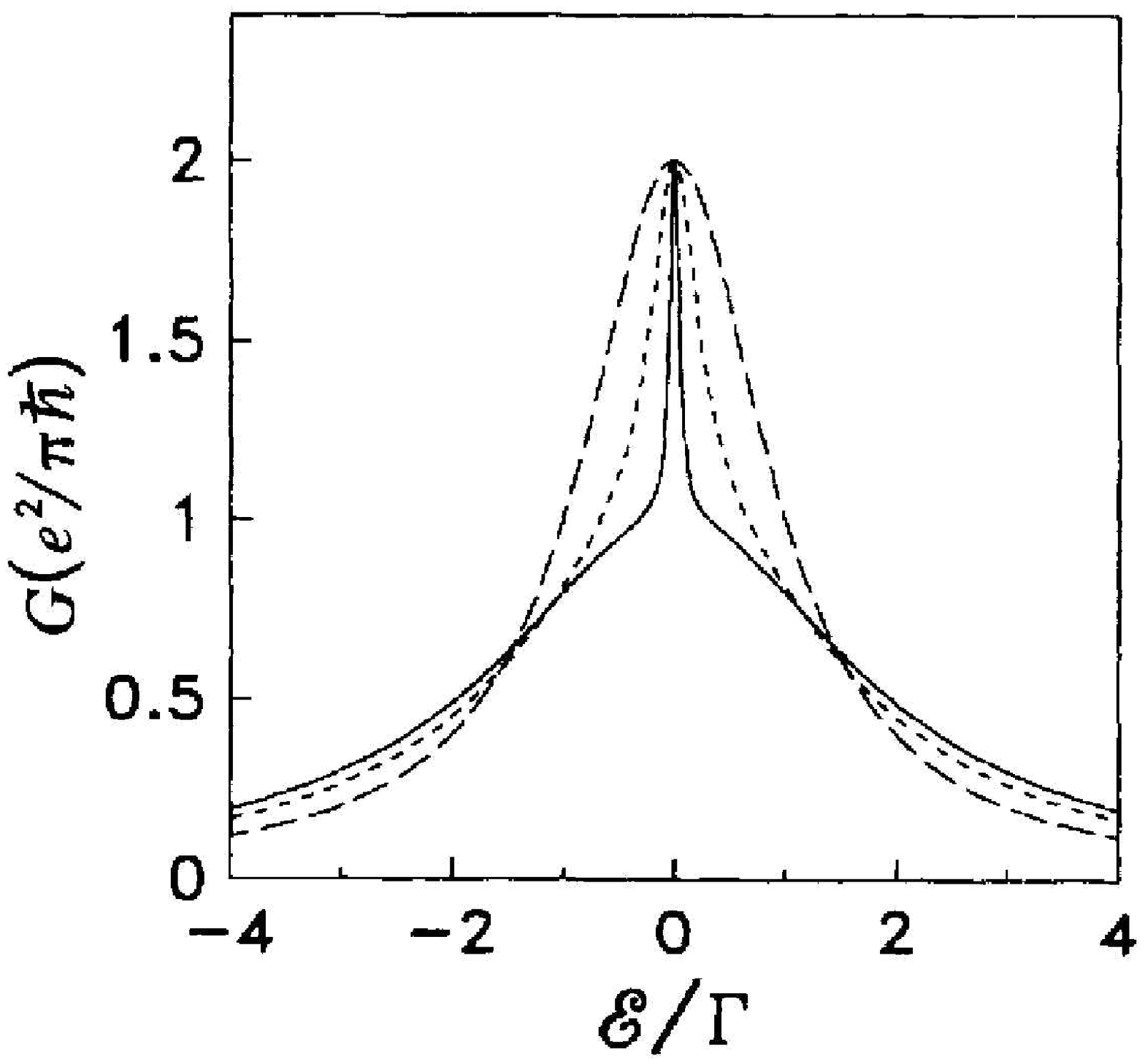}
\caption[]{\label{SR94fig12.eps}Resonant tunneling through two impurity levels, from
Shahbazyan and Raikh \cite{SR94}. {\bf Left:} Tunnel junction with two resonant impurities
1 and 2 in a distance $d$ in horizontal and distance $s_{12}$ in vertical direction. {\bf Right:}
linear conductance for identical impurity levels $E$ as a function of ${ \mathcal{E}}=E_F-E$, where
$E_F$ is the Fermi energy of the tunneling electron. The characteristic  shape of 
the spectral function Eq.(\ref{ShabS1}), as known from the Dicke effect,
appears here in the conductance with increasing parameter $q=0$, $q=0.75$, $q=0.95$, cf. 
Eq.(\ref{shabJ0}). $\Gamma$ is the
tunneling rate through the left and the right barrier. From \cite{SR94}.}
\end{figure} 
Note that this spatial dependence of the matrix element $t_{i{\bf k}}$ is similar to the relation $\alpha^L_{\bf Q}\propto \exp(i{\bf Qr}_L)$, $\alpha_{\bf Q}^R\propto \exp(i{\bf Qr}_R)$ that lead to the interference in the matrix elements for electron-phonon coupling in double quantum dots as discussed in section \ref{section_transport}, cf. Eq. (\ref{phasefactor}). 
The non-diagonal elements $W_{12}$ can be shown to be oscillating functions of the impurity distance $r_{ij}$,
\begin{equation}
\label{shabJ0}
W_{12}=q\sqrt{W_1W_2},\quad q = J_0(r_{12}k_F),
\end{equation}
where $k_F$ is the Fermi wave vector and $J_0$ the Bessel function that results from an angular integral in the plane of the two impurities.  If the diagonal elements $W_{11}$ and $W_{22}$ and both energies are assumed to be identical, $\varepsilon_1=\varepsilon_2=\varepsilon$ and $W_{11}=W_{22}=W$, one has
\begin{eqnarray}\label{ShabS1}
\hat{W}&=&W\left(
  \begin{array}[c]{cc}
1 & q\\
q & 1 
  \end{array}
          \right),\quad 
S(\omega) =\frac{1}{2\pi}\left[ \frac{W_-}{(\omega-\varepsilon)^2+W_-^2}
+\frac{W_+}{(\omega-\varepsilon)^2+W_+^2} \right],
\end{eqnarray}
with $W_{\pm}=(1\pm q)W$.
This spectral function consists of a superposition of two Lorentzians (one narrow line with width $W_-$, corresponding to a subradiant channel, and one broad line with width $W_+$, corresponding to a superradiant channel) and therefore represents another example of the Dicke spectral line effect discussed in the previous section. Furthermore, this splitting is analogous to the spitting of a radiating decay channel of two coupled radiators as discussed in section \ref{section_SRintro}. If the parameter $q$ is small, $q\ll 1$, one has $W_+\approx W_-\approx W$ and the spectral function is a simple Lorentzian if width $W$. The crossover to the Dicke regime with the splitting into a sharp and a broad part of $S(\omega)$ is thus governed by $q=J_0(r_{12}k_F)$ and therefore by the ratio of the distance of the impurities to the Fermi wavelength of the electron. This again shows that the effect is due to interference. 
The two localized impurity states are coupled by the continuum of plane waves. As for their scattering properties, they have to be considered as a {single quantum mechanical entity}, as long as their distance is of the same order or smaller than the wavelength of the scattering electrons. In this case, the (linear) conductance $G(E_F)$ for resonant tunneling shows the typical feature of the Dicke effect as a function of the energy  $E_F$ of a tunneling electron: as $G(E_F)$ is determined by the spectral function $S(\omega)$ \cite{Jauho12}, the Dicke peak becomes directly visible in the  conductance, cf. Fig. (\ref{SR94fig12.eps}). If the energies $\varepsilon_1$ and $\varepsilon_2$ of the two impurity levels differ from each other, the resonant peak even shows a more complex behavior; as a function of the parameter $q$ there is a crossover to a sharp transmission minimum \cite{SR94}.

Several authors have built upon the 1994 paper by Shahbazyan and Raikh  and found features in electronic transport which in one way or the other are related to the Dicke effect. Shahbazyan and Ulloa \cite{SU98} studied the Dicke effect for resonant scattering in a strong magnetic field, using an exact solution for the density density of states in the lowest Landau level as calculated from a zero-dimensional field-theory \cite{SU97}. Furthermore, Kubala and K\"onig studied a generalization  including an Aharonov-Bohm flux $\varphi$ in a ring-geometry connected to  left and right leads with resonant scattering through single  electronic levels $\varepsilon_1$ and $\varepsilon_2$ of two embedded quantum dots \cite{KK02,KK03}. They calculated  the transmission $T(\omega)$ through the ring within the usual Meir-Wingreen formalism \cite{MW92},
\begin{eqnarray}
  T(\omega)={\rm Tr} \left[{\bf{G}}^a(\omega) {\bf \Gamma}^R {\bf G}^r(\omega) {\bf \Gamma}^L\right],
\quad {\bf{G}}^r(\omega) \equiv 
\begin{pmatrix}
 \omega-\varepsilon_1+i\frac{\Gamma}{2} & i\frac{\Gamma}{2} \cos \frac{\varphi}{2} \\
  i\frac{\Gamma}{2}  \cos \frac{\varphi}{2}       & \omega-\varepsilon_2+i\frac{\Gamma}{2}  
\end{pmatrix}^{-1} 
\end{eqnarray}
with $2\times 2$ matrices  ${\bf{\Gamma}}^{L/R}$ for the coupling to the leads and retarded and advanced Green's functions, ${\bf{G}}^{r/a}(\omega)$.
The imaginary part of ${\bf{G}}^r(\omega)^{-1}$  contains the sum of the tunnel rates, $\Gamma=\Gamma_L+\Gamma_R$ and  again has `Dicke' form , cf. \ Eq.~(\ref{spectral_matrix}), leading to  a sharp suppression of transport around $\varepsilon_1=\varepsilon_2=0$.

\subsection{Cooperative Light Emission From Disordered Conductors}
A large part of mesoscopic physics deals with universal properties of disordered, coherent electronic systems. The related topic of `random lasing' has  attracted a lot of attention recently; some shorter Review Articles by Hackenbroich and Haake, and by Apalkov, Raikh and Shapiro can be found in \cite{Weh283}. 

Shabazyan, Raikh, and Vardeny \cite{SRV00} studied a related problem, i.e.,  {\em spontaneous} cooperative emission from a {disordered} mesoscopic  system, motivated by experimental evidence  for collective excitonic light emission from strongly disordered polymers. 
They used a completely classical description of superradiance as a collective phenomenon, which was in the spirit of a generalization of the classical description of spontaneous emission from a single, classical radiator. The microscopic, quantum mechanical description of superradiance in fact is two-fold and can be performed following  two alternative schemes \cite{GH82}: in the {\em Schr\"odinger picture}, a Master equation for the reduced density operator of the electronic system is derived. The degrees of freedom of the electromagnetic field are regarded as dissipative bath leading to spontaneous emission; they are integrated out whence the coupling to the electromagnetic field basically enters as one single parameter (the decay rate of a single radiator). On the other hand, in the  {\em Heisenberg picture}, the equations of motion for the field operators of the polarization, occupation numbers, and the polarization are derived, and  the electromagnetic field is dealt with on a classical level by using  Maxwell's equations. This second approach is in particular useful in order to study the classical aspects of superradiance, and furthermore additional aspects like propagation effects for the electromagnetic field etc. Both alternatives are valid (though entirely different) routes towards cooperative emission (superradiance), cf. the discussion in the Review article by Gross and Haroche \cite{GH82}.

The starting point in the work by Shabazyan, Raikh, and Vardeny was a system of $N\gg 1$ classical harmonic oscillators of charge $e$, mass $m$, dipole orientation $\textbf{n}_i$ at random positions $\textbf{r}_i$ and with random frequencies $\omega_i$, interacting via their common radiation field ${\bf E} (\textbf{r},t)$. The equations of motions for the oscillator displacements,
\begin{eqnarray}\label{SRV1}
  \ddot{u}_i(t) + \omega_i^2 u_i(t) = \frac{e}{m}\textbf{n}_i {\bf E} (\textbf{r}_i,t),
\end{eqnarray}
are closed by using the wave equation for the electric field,
\begin{eqnarray}\label{SRV2}
  \Delta \textbf{E}(\textbf{r},t) -\frac{1}{c^2}\ddot{\textbf{E}}(\textbf{r},t) = \frac{4\pi}{c^2}\dot{\textbf{J}}(\textbf{r},t),\quad
\textbf{J}(\textbf{r},t)\equiv e \sum_i \textbf{n}_i \dot{u}_i(t) \delta(\textbf{r} - \textbf{r}_i),
\end{eqnarray}
where the source term (the macroscopic polarization)  is again determined by the oscillator displacements. 
As mentioned above, the combination of the two sets of equations, \ Eq.~(\ref{SRV1}) and (\ref{SRV2}), for light-matter interaction constitutes the `classical' approach to superradiance and is complementary to the completely quantum mechanical approach based on collective spontaneous emission as presented in the introduction to section \ref{section_SRintro}. 

After Laplace transforming and expansion into eigenmodes of the field, one arrives at a simple set of linear equations for the (re-scaled) oscillator displacements $v_i(i\omega)$,
\begin{eqnarray}\label{SRVeigen1}
  (\omega_i-\omega)v_i+\frac{1}{\tau}\sum_j (\beta_{ij} + i \alpha _{ij}) v_j = -\frac{i}{2}e^{-i\phi_i},
\end{eqnarray}
where $\alpha_{ij}$ and $\beta_{ij}$ are the imaginary and the real part of the effective interaction matrix elements between the radiators as mediated by the electric field, $\tau$ is the radiative life time of an individual oscillator, and $\phi_i$ the initial oscillator phases. 

\begin{figure}[t]
\includegraphics[width=0.5\textwidth]{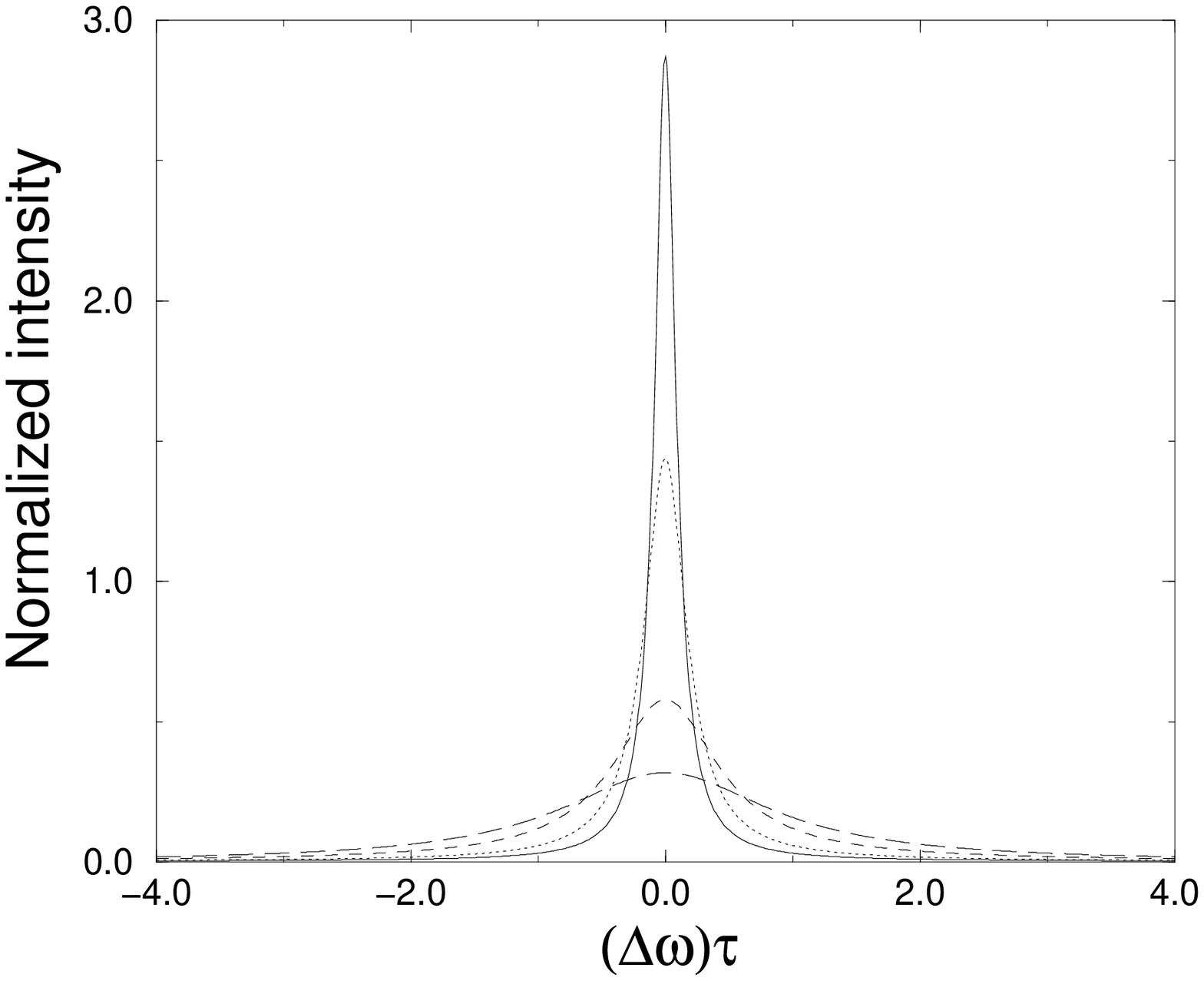}
\includegraphics[width=0.4\textwidth]{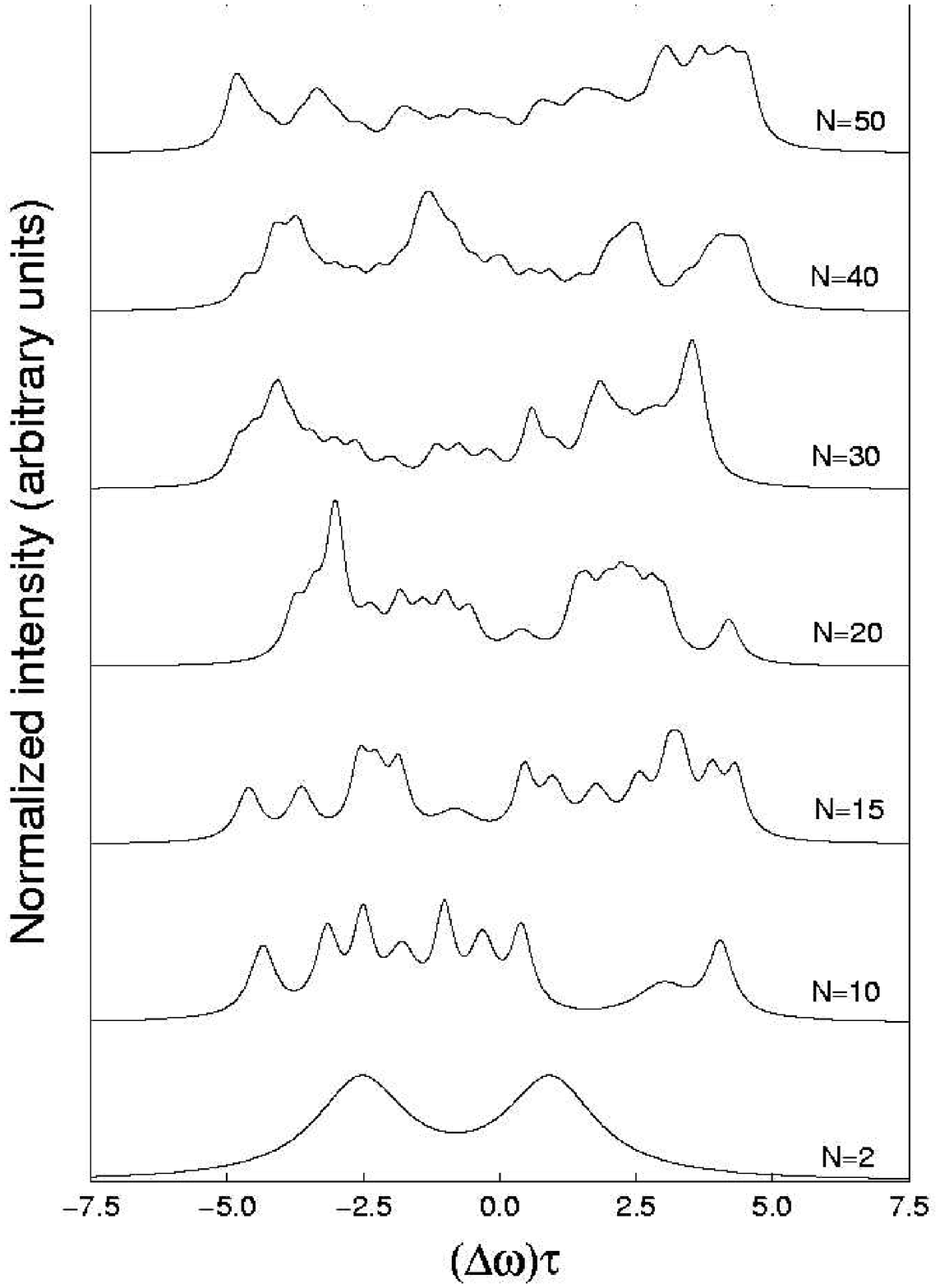}
\caption[]{\label{SRV.eps} Mesoscopic superradiance from disordered systems after Shahbazyan, Raikh, and Vardeny \cite{SRV00}. {\bf Left:} Spectral intensity of $N$ identical oscillators, Eq.\ (\ref{subsuperspectral}),  vs. 
$\Delta\omega=\omega-\omega_0$ for $N=10$, $\alpha=0$ (long-dashed line), $\alpha=0.5$ (dashed line), $\alpha=0.8$ (dotted line), and $\alpha=0.9$ (solid line). {\bf Right:} Spectral intensity for several sets of random oscillator frequencies $\omega_i$ with $\Omega\tau=5.0$ and $\alpha=0.8$. From \cite{SRV00}.}
\end{figure} 

The limit of pure Dicke superradiance then follows from neglecting the dephasing terms $\beta_{ij}$ (which are due to effective dipole-dipole interactions) and by setting 
\begin{eqnarray}
  \alpha_{ij}=\alpha \textbf{n}_i\textbf{n}_j,\quad \alpha_{ii}=1,
\end{eqnarray}
with $(1-\alpha)\sim L^2/\lambda_0^2\ll 1$, where $L$ is the system size and $\lambda_0$ the wavelength corresponding to the central oscillator frequency $\omega_0$. For identical oscillators $\omega_i=\omega_0$ and identical dipole moment orientations $\textbf{n}_i$, the eigenvalue problem \ Eq.~(\ref{SRVeigen1}) is identical to the one for the matrix $A$ discussed in the introduction, cf. \ Eq.~(\ref{spectral_matrix}), and one obtains an emission spectrum
\begin{eqnarray}\label{subsuperspectral}
  I(\omega) \propto \left[ \frac{(N-1)(1-\alpha)/\tau}{(\omega_0-\omega)^2+(1-\alpha)^2/\tau^2}
+ \frac{(1-\alpha+\alpha N)/\tau}{(\omega_0-\omega)^2+(1-\alpha+\alpha N)^2/\tau^2}\right],
\end{eqnarray}
which is a superposition of a wide Lorentzian (corresponding to the single superradiant mode) and a narrow Lorentzian (corresponding to $N-1$ subradiant modes), cf. Fig.(\ref{SRV.eps}).

Disorder in the orientations $\textbf{n}_i$ alone was shown to have no qualitative effect on the emission spectrum. For frequencies $\omega_i$ randomly distributed in an interval $(\omega_0-\Omega, \omega_0+\Omega)$, however, striking {\em mesoscopic} features appear in $I(\omega)$. Instead of the naively expected smearing of the sharp (subradiant) Dicke peak \ Eq.~(\ref{subsuperspectral}), the coupling of the oscillators leads to a multitude of sharp peaks in $I(\omega)$, cf. Fig. \ref{SRV.eps}. Shabazyan, Raikh, and Vardeny showed that the splitting into a single superradiant and $N-1$ subradiant modes persists even in the disordered case, as long  as the {\em mean frequency spacing} $\Omega/N$ is much smaller than the inverse life time (individual oscillator line-width) $\tau^{-1}$. The precise form of $I(\omega)$ depends on the specific (random) choice of the  $\omega_i$ and is therefore a `fingerprint' of the frequency distribution. On the other hand, a detailed analysis showed that some universal features of $I(\omega)$ depend on $\tau$, $\Omega$, $L$, and $N$ only. In particular, in an intermediate regime $\Omega\tau \lesssim N \lesssim \Omega\tau (1-\alpha)^{-1/2}$, the width of the many peaks was shown to decrease with increasing $N$, with the system behaving as a `point sample', whereas for larger $N$ a cross-over occurs into a regime with peaks becoming broader with increasing $N$.

\subsection{ac-Drude Conductivity of Quantum Wires} 
The Dicke spectral line effect also appears in the ac conductivity $\sigma(\omega)$ of quantum wires in a magnetic field, which is yet another example of electronic transport in a mesoscopic system. Quantum wires are electronic systems where the motion of electrons
is confined in two perpendicular direction of space and free in the  third \cite{Dittrich,Ferry,Datta,Imry,Jauho,Koch}. In the presence of impurity scattering and when only the two lowest subbands of the wire are occupied, the absorptive part of $\sigma(\omega)$ shows (as a function of $\omega$) the Dicke effect in analogy to the spectral line narrowing discussed above. The parameter that drives the effect is the magnetic field $B$. Impurity back-scattering becomes more and more suppressed with increasing $B$, which leads to a crossover in $\sigma(\omega)$ from a broad Lorentzian to a very sharp and high peak on top of a broad Lorentzian. This is due to inter-subband scattering, by which the transport rates for the two subbands become coupled and split into one fast and one slow mode, corresponding to the superradiant and the subradiant channel in the 
superradiance problem.

A model for this effect takes a quantum wire in $x$-direction within a quantum well in the $x$-$y$-plane under a magnetic field in $z$ direction, cf. Fig.(\ref{wiretransport.eps}) left, with the wire defined by a harmonic confinement potential of frequency $\omega_0$. The single electron eigenstates $|nk\rangle$ with eigenenergies $\varepsilon_{nk}$ of the clean system (no impurities, Landau gauge) have two quantum numbers $n$ (Landau band) and $k$ (momentum in direction of the wire) \cite{Janssen}. In the Drude model for the conductivity $\sigma(\omega)$ of the wire, quantum interference effects and localization of electrons  are disregarded, and the electronic transport is determined by the average electron scattering rate at the impurities \cite{Jaeckle,Jauho1,MLB92,Bra96a}. The {\em memory function formalism} by G\"otze and W\"olffle \cite{GW72,Brenig} is an alternative to a calculation of  $\sigma(\omega)$ via the Boltzmann equation, as has been done by, e.g.,  Bruus, Flensberg and Smith 
\cite{BFS93} or Akera and Ando \cite{AA90} for $\omega=0$.
\begin{figure}[t]
\includegraphics[width=0.5\textwidth]{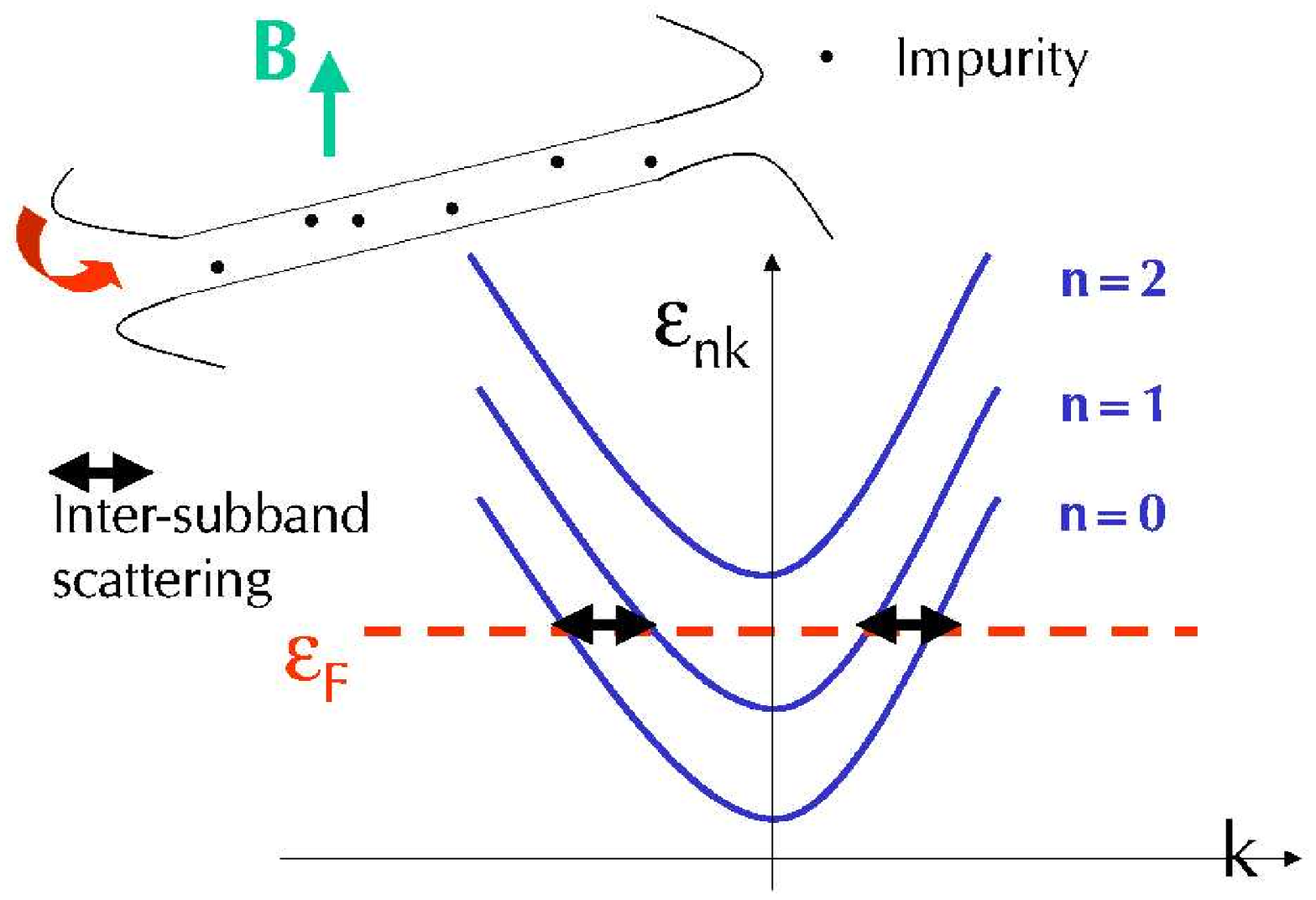}
\includegraphics[width=0.5\textwidth]{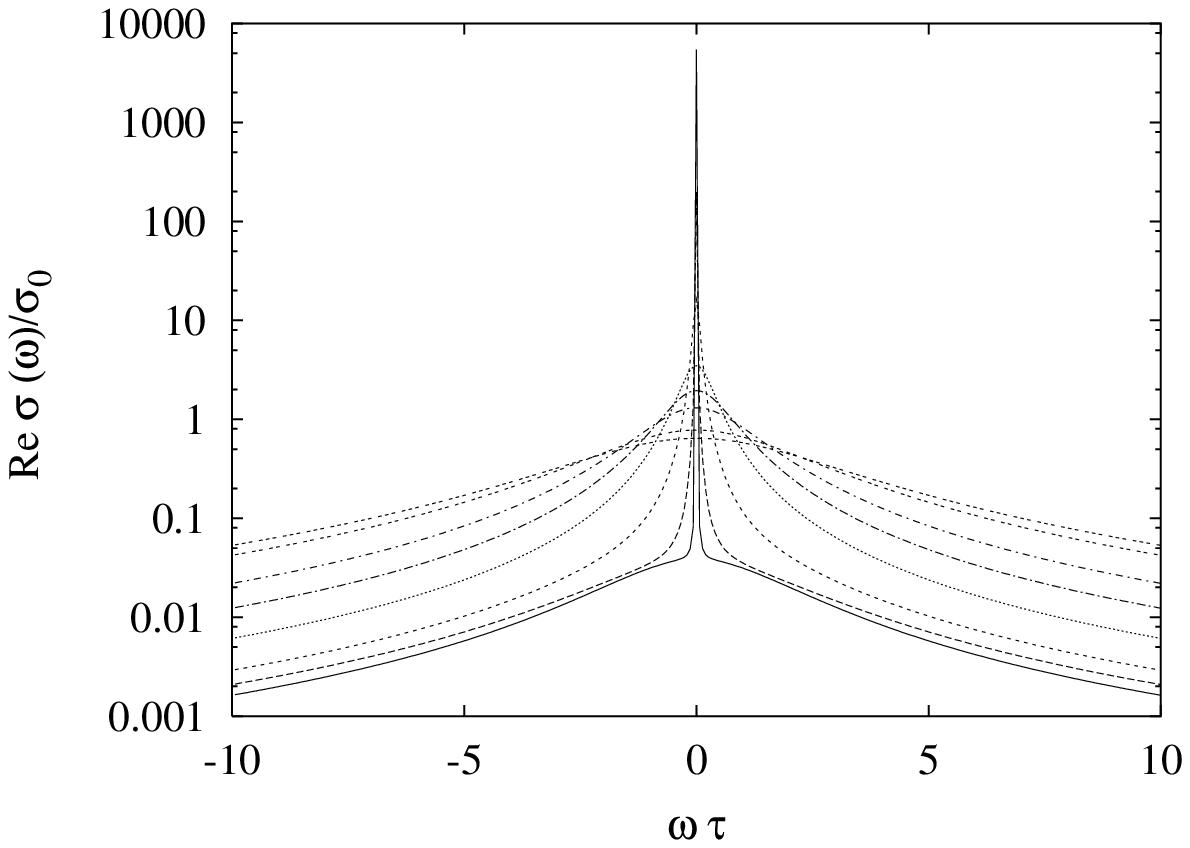}
\caption[]{\label{wiretransport.eps}{\bf Left:} Subband dispersion $\varepsilon_{nk}$ of a wire
(upper left) in a magnetic field $B$. {\bf Right:} Real part of the frequency-dependent Drude 
conductivity of a two--channel quantum wire in a magnetic field, Eq. (\ref{eq:condlongeq}),
in units of $\sigma_0:=e^2sv_{F_0}\tau/\pi$ (s=1 for spin--polarized electrons).
Different curves are for $\omega_c/\omega_0$ = $0,0.4,0.8,1.2,1.6,2.0,2.4,2.8$,
where $\omega_0$ is the frequency of the harmonic confinement potential, and
$\omega_c=eB/m$ the cyclotron frequency for magnetic field $B$.}
\end{figure} 
One of the  advantages of the memory function formalism is that non-trivial interaction effects can in principle be incorporated  into the formalism by  interaction dependent correlation functions. This in particular is useful to combine exact results, e.g. for  correlation functions of interacting one-dimensional systems, with a perturbative description of impurity scattering \cite{AR82,Sol79}. Such effects, however, are neglected in the calculation below, and the final result can be shown to coincide with the one obtained from the Boltzmann equation in the limit of zero temperature and small frequencies $\omega$.

The Hamiltonian of the wire is given by 
\begin{equation}
\label{hamiltonian_drude}
H=\sum_{nk}\varepsilon _{nk}c_{nk}^{+}c_{nk}+\frac{1}{L_s}\sum_{nmkq}V_{nm}(q)c^{+}_{nk}c_{mk+q},
\end{equation}
where $L_{s}$ is the length of the wire, $c_{nk}$ the electron creation operator for band $n$, and $V_{nm}(q)$ the matrix element for impurity scattering with momentum transfer $q$ from a state with quantum number $nk$ to a state $mk+q$. To simplify the notation, the spin index $\sigma $ in the operators $c_{nk\sigma }^{(+)}$ has not been written out explicitly. The scattering potential is assumed to be spin-independent and summation over the spin is included in all $k,k'$--sums.
The linear  response of an electronic system to a monochromatic  electric field ${\bf E}({\bf x})\cos (\omega t)$ in general is governed by a non-local conductivity tensor $\sigma({\bf x},{\bf x'},\omega)$. Many electronic transport properties of quantum wires (many-subband quasi one-dimensional systems) have to be discussed in terms of the {\em conductance} $\Gamma$ (the inverse resistance) \cite{Lan70,Bue86,BraPhD,FD93,BPT93,BWK93} rather than the conductivity, although the former is related to the latter in special cases \cite{FL81,SS88,MK89,KM89,VMK89}. 
The conductance is regarded as the proper transport property to explain, e.g., phenomena like step-like features in the electronic transport properties, i.e. a quantization of $\Gamma$ in multiples of $2e^2/h$ \cite{Weh219}. This and other phenomena like localization due to disorder \cite{KM94} in general exist due to phase coherence \cite{SAI90,CS86,ZB92}. In presence of phase breaking processes, a crossover to a regime that can be described by a Drude-like theory is expected even for one-dimensional systems when their length $L_s$ becomes larger than the distance $L_{\phi}$ over which phase coherence is maintained. In this case, the conductivity $\sigma(\omega)$ becomes a meaningful quantity. Furthermore, the conductivity as physical quantity in quantum wires is also used to describe {\em deviations} from ideal, unperturbed situations, e.g. deviations from conductance plateaus due to scattering processes where a low order (sometimes renormalized) perturbation theory \cite{KB96a} is possible. 

The homogeneous conductivity as a function of complex frequency $z$ is expressed in terms of the current-current correlation function \cite{GW72,Brenig},
\begin{equation}
\label{sigma}
\sigma (z)=-i\frac{e^{2}}{z}\left(\chi(z)-\frac{n_{e}}{m^{*}}\right),
\end{equation}
where
\begin{equation}
\label{zubarev}
-\chi(z)=
\langle\langle \hat{j};\hat{j} \rangle\rangle_{z}\equiv -iL_s\int_{0}^{\infty} dt e^{izt}\langle 
[\hat{j}(t),\hat{j}(0)]\rangle_{0}
\end{equation}
is the (Zubarev) correlation function of the $q=0$ component of the mass current density operator $\hat{j}=\hat{j}(q=0)$. Furthermore, $n_{e}$ is the electron density, $-e<0$ the electron charge and $m^{*}$ its conduction band mass. The multichannel wire  is described as a set of quasi one-dimensional subbands (channels) $n=1,...,N_{c}$ of dispersion $\varepsilon _{nk}$ and corresponding electron velocities $v_{nk}=\partial\varepsilon _{nk} / 
\partial k$ (we set $\hbar =1$). The current in the total system is the sum of the currents of all channels, 
\begin{equation}
\label{currentoperator}
\hat{j}=\frac{1}{L_{s}}\sum_{n,k}v_{nk}c_{nk}^{+}c_{nk}\equiv \sum _n j_n,
\end{equation}
which allows one to write the conductivity as
\begin{equation}
\label{sigmageneral}
\sigma (z)=-i\frac{e^{2}}{z}\left(\sum_{n,m}\chi_{nm}(z)-
\frac{n_{e}}{m^{*}}\right), \quad 
\chi_{nm}(z)\equiv -\langle\langle j_{n},j_{m}\rangle\rangle_{z}
\end{equation}
in terms of a matrix of current-current correlation functions. The total number of electrons $N_{e}$ is given by $N_{e}=\sum_{n,|k|<k_{n}}$. Here, the Fermi momentum $k_{n}$ in subband $n$ is related to the Fermi energy $\varepsilon_F$ as $\varepsilon_{nk}=\varepsilon _{F}, k=k_n$, which in turn is determined by the total number of electrons  via $N_{e}=\sum_{n,|k|<k_{n}}$ and the magnetic field dependent band structure $\varepsilon_{nk}$.
One has 
\begin{equation}
\label{chi0}
\frac{n_{e}}{m^{*}}\equiv \sum_{nm}\chi^{0}_{nm}, \quad \chi_{nm}^{0}\equiv 
\delta _{nm}\frac{s}{\pi}v_{n}, \quad v_{n}=v_{nk=k_{n}}=k_{n}/m^{*},
\quad\mbox{\rm{$s$ spin degeneracy}},
\end{equation}
where $v_{n}$ is the Fermi velocity in subband $n$ and the sum in Eq. (\ref{chi0}) runs over all occupied subbands. In Appendix (\ref{Appmemory}), a multichannel version of the memory function formalism \cite{GW72} is used to find the expression for the frequency dependent conductivity $\sigma(\omega)$ at zero temperature $T=0$ and small excitations $\hbar \omega$ around the Fermi surface, i.e. frequencies $\omega\ll |\varepsilon_F-\varepsilon_{n=0,k=0}|/\hbar$. In the following, excitations much smaller than the inter-subband distance $\hbar \omega_B = \hbar \sqrt{\omega_0^2+\omega_c^2}$ are assumed, where $\omega_c = |e|B/m^*c$ is the cyclotron frequency for magnetic field $B$. An estimate for the relevant frequency range for $B=0$ is $\hbar \omega_0=1$ meV, i.e. $\omega_0=1500$ GHz, and frequencies $\omega$ from $0-100$ GHz $\ll \omega_0$  are in the microwave spectroscopy regime.

The general expression for $\sigma(\omega)$ is given in Appendix (\ref{Appmemory}), Eqs. (\ref{Lnmarray}), (\ref{eq:condl}), together with (\ref{chi0}). In the case where the two lowest subbands $n=0$ and $n=1$ are occupied, the expression for the conductivity is
\begin{eqnarray}
  \label{eq:2channelcond}
  \sigma(z)=ie^2\frac{ z\frac{s}{\pi}(v_0+v_1)+i\left[
\frac{v_0}{v_1}L_{11}+\frac{v_1}{v_0}L_{00}-2L_{01}\right]} 
{\left(z+i\frac{\pi L_{00}}{sv_0} \right) \left(z+i\frac{\pi L_{11}}{sv_1} \right)
+\frac{\pi^2 L_{01}^2}{s^2v_0v_1}}
\end{eqnarray}
with 
\begin{eqnarray}
  \label{eq:Liidef}
  L_{00}&=&\frac{s}{\pi}L_s\frac{v_0}{v_1}\left( |V_{01}(k_0-k_1)|^2+|V_{01}(k_0+k_1)|^2+
\frac{2s}{\pi}V_{00}(2k_0)^2\right)\nonumber\\
  L_{11}&=&\frac{s}{\pi}L_s\frac{v_1}{v_0}\left( |V_{01}(k_0-k_1)|^2+|V_{01}(k_0+k_1)|^2+
\frac{2s}{\pi}V_{11}(2k_1)^2\right)\nonumber\\
  L_{01}&=&\frac{s}{\pi}L_s\left( |V_{01}(k_0+k_1)|^2-|V_{01}(k_0-k_1)|^2\right).
\end{eqnarray}
Here, $s=2$ if the electrons are taken as spin degenerate, and $s=1$ if the electrons are
assumed to be spin-polarized.
%
%
In lowest order perturbation theory (Born approximation) in the scattering off random impurities, it is sufficient to know the impurity averaged square of the matrix element 
\begin{eqnarray}
  \label{eq:imaveragematrix}
  \overline{|V_{nn'}(k-k')|^2}=n_i^{2D}\sum_{{\bf q}}|u({\bf q})|^2
|\langle nk|e^{-i{\bf qx}}|n'k'\rangle |^2,
\end{eqnarray}
that enters into the expressions $L_{ij}$ in Eq.(\ref{eq:Liidef}). Here, $u({\bf q})$ is the two-dimensional Fourier transform of the static potential of a single impurity potential $u(x,y)$. All impurities are assumed to be identical scatterers and distributed with a concentration $n_i^{2D}$ per area $L^2$. Finite quantum well thickness corrections (form factors) are neglected here for simplicity. The averaged matrix elements are calculated in Appendix (\ref{Appmatrix}) for Delta-scatterers, where the Fourier component $|u({\bf q})|^2\equiv V_0^2$ is a constant. The dependence on the magnetic field is only through the ratio $\beta\equiv (\omega_c/\omega_0)^2$. 
We express the scattering matrix elements by the scattering rate $\tau^{-1}$ without
magnetic field,
\begin{eqnarray}
  \label{eq:taubzerodefhere}
  \tau^{-1}&\equiv  n_i^{2D}V_0^2m^*/\sqrt{4\pi}\hbar^3,
\end{eqnarray}
where in comparison with \cite{BFS93},  $\tau^{-1}$ is defined with an additional factor of $1/\sqrt{4\pi}$ for convenience.

The frequency dependence of the real part of the conductivity, Eq.(\ref{eq:condlongeq}), is shown in Fig. (\ref{wiretransport.eps}) for the Fermi energy fixed between the bands $n=1$ and $n=2$, i.e. $\varepsilon_F=2\hbar\omega_B$. The real part $\Re {\rm e} \sigma(\omega)$ has a Lorentzian shape for small magnetic fields. For increasing magnetic field, i.e. larger $\omega_c/\omega_0$, this shape develops into a very sharp Lorentzian on top
of a broad Lorentzian, indicating that one of the two poles $z_{\pm}$ in $\sigma(z)$ approaches zero which again is the {\em Dicke effect} as discussed above. Here, in the Dicke limit the subradiant pole is zero and has no small finite imaginary part, since scattering processes other than impurity 
scattering is not included, which is  in contrast to Eq.(\ref{eq:twopolesdicke}), where spontaneous emission at a rate $\gamma$ lead to a finite imaginary part $-i\gamma$ in both zeros.

\begin{figure}[t]
\begin{center}
\includegraphics[width=0.25\textwidth]{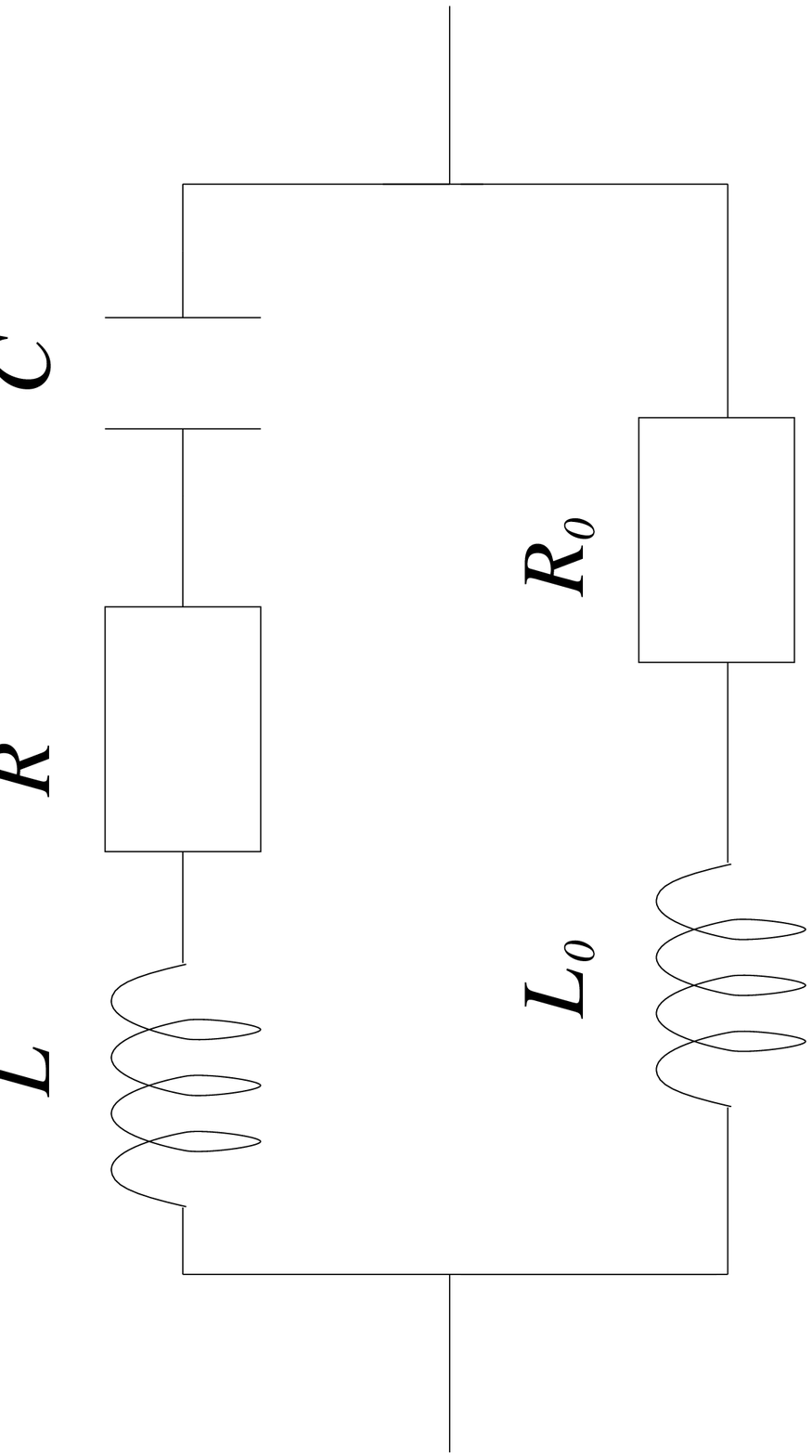}
\includegraphics[width=0.65\textwidth]{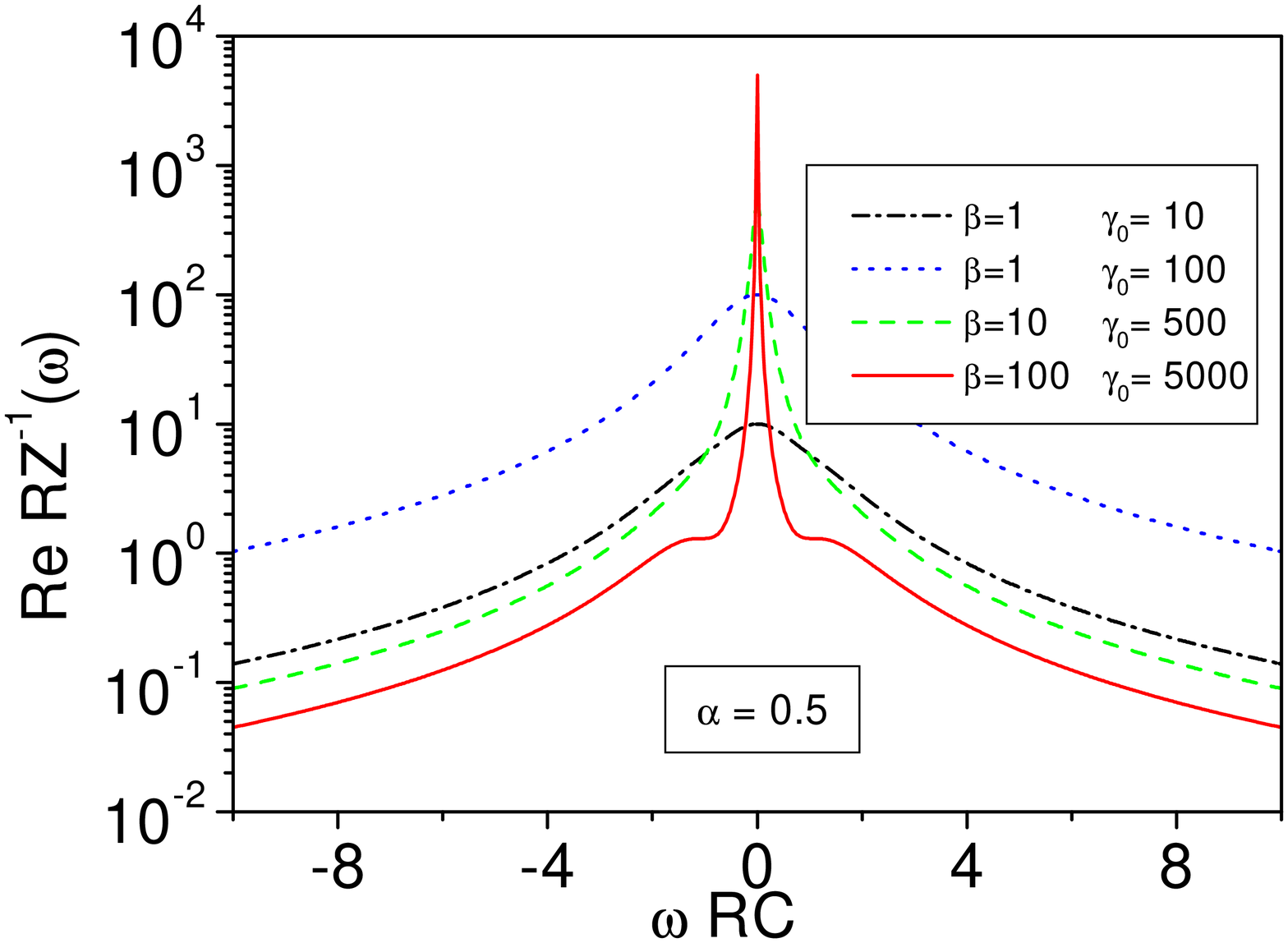}
\end{center}
\caption[]{\label{circuit.eps}
{\bf Left:} Classical circuit to simulate the Dicke effect in a two-subband quantum wire. {\bf Right:} Real part of the inverse impedance $Z^{-1}(\omega)$ (conductance in units of $R$) for a classical circuit, Fig.(\ref{circuit.eps}), that simulates the Dicke effect in a two-subband quantum wire, Fig.(\ref{wiretransport.eps}).}
\end{figure} 

The two poles  \index{poles in complex plane}of $\sigma(z)$ determine the width of $\Re {\rm e}\sigma(\omega)$. For large magnetic fields $B$,  one can neglect the terms which are not due to intersubband {\em forward} scattering  in $\tilde{L}_{00}$, $\tilde{L}_{11}$, $\tilde{L}_{01}$ , and
\begin{eqnarray}
  \label{eq:l00approx}
  \tilde{L}_{01}&\approx& \frac{-s}{L_s\pi}|V_{01}(k_0-k_1)|^2,\quad
  \tilde{L}_{00}= \frac{v_0}{v_1}\tilde{L}_{01},\quad
  \tilde{L}_{11}= \frac{v_1}{v_0}\tilde{L}_{01}.
\end{eqnarray}
The quadratic equation that determines the poles of $\sigma(z)$ then has the solutions
\begin{eqnarray}
  \label{eq:poleslimit}
  z_{-}=0,\quad z_+=\frac{-i|V_{01}(k_0-k_1)|^2}{L_s}\left(\frac{1}{v_0} 
+\frac{1}{v_1} \right).
\end{eqnarray}
In this limit one of the poles becomes zero, corresponding to the very sharp peak in $\Re {\rm e}\sigma(\omega)$. 

This analysis demonstrates that the coupling of the two subbands by the intersubband impurity scattering is essential for the appearance of the Dicke effect in this example of electronic transport. Furthermore, for large magnetic fields, backscattering with momentum transfer $2k_0$, $2k_1$, and $k_0+k_1$ from one side to the other side of the wire becomes largely suppressed due to the exponential dependence of the matrix elements on the square of the momentum transfer, cf. Eq.(\ref{eq:sqrt3etc}), (\ref{eq:Vfinalarray}). With increasing magnetic fields, such scattering processes become much weaker than intersubband forward scattering, i.e. scattering between the bands $n=0$ and $n=1$. This absence of backward scattering, of course, leads to a larger DC conductivity. 

In the Dicke-limit Eq.(\ref{eq:l00approx}), simple algebraic manipulations lead to an expression for $\sigma(z)$ with the Fermi velocities $v_0$ and $v_1$ in subband $n=0$ and $n=1$,
\begin{eqnarray}
  \label{eq:sigmasimple}
  \sigma(z)&\approx&ie^2\frac{s}{\pi}\left( \frac{v_+}{z-z_+} +\frac{v_-}{z-z_-}\right)\nonumber\\
v_+&\equiv &(v_0-v_1)\frac{v_0/v_1-1}{v_0/v_1+1},\quad v_-\equiv \frac{4v_0v_1}{v_0+v_1}.
\end{eqnarray}
The conductivity then becomes a sum of two contributions from the `superradiant'mode corresponding to $z_+$ and the `subradiant' mode corresponding to $z_-$. Note that these modes are superpositions of contributions  from both subbands $n=0$ and $n=1$. 

Another observation is the fact that it is possible to simulate the behavior of $\Re {\rm e} \sigma(\omega)$ as a function of $\omega$ by a {\em classical electrical circuit} composed of two impedances in parallel: this circuit consists of one huge inductance $L_0$ which is in series with a small resistance $R_0$, the whole being in parallel with a small inductance $L$, a large resistance $R$, and a capacitance $C$ in series. Such classical circuits were in fact used in the past to simulate the ac transport properties of  more complicated systems such as mesoscopic tunnel barriers \cite{BWK93,CSK98}. The complex impedance
\begin{eqnarray}
  \label{eq:impedance}
  Z^{-1}(\omega)=\frac{i\omega C}{1+i\omega R C-\omega^2 L C}+
\frac{1}{R_0 + i\omega L_0}
\end{eqnarray}
contains the time scale $RC$ and the three parameters 
\begin{eqnarray}
  \label{eq:impedance1}
\alpha\equiv L/R^2C,\quad \beta \equiv  L_0/RR_0C,\quad \gamma_0\equiv R/R_0,
\end{eqnarray}
by which a  fit that qualitatively compares well with $\Re {\rm e} \sigma(\omega)$ can be achieved. Note that the case $\beta/\alpha\equiv L_0R_0/LR\gg 1$ together with $\gamma_0\equiv R/R_0\gg 1$ sets very drastic conditions for the possible ratios $L_0/L$ and $R_0/R$, if one tried to simulate $\Re {\rm e}\sigma(\omega)$ by a classical circuit in real experiments. 

Checking the range of frequencies where the effect could be
observed experimentally, one recognizes from Figs. (\ref{wiretransport.eps}) that $\omega$ has to be varied such that $0.1 \lesssim \omega \tau \lesssim 5$ in order to scan the characteristic shape of the Dicke peak. Impurity scattering times for AlGaAs/GaAs heterostructures are between $3.8\cdot 10^{-12}$ s and $3.8 \cdot 10^{-10}$ s for mobilities between $10^5-10^7$ cm$^2/$Vs, cf. \cite{Ferry}. A scattering time of $10^{-11}$ s requires frequencies of $\omega\approx$100 GHz for $\omega\tau \approx 1$, which is consistent with the requirement of $\omega$ being much smaller than the effective confinement frequency ($\omega_0=1500$ Ghz for $\hbar\omega_0=1$ meV). An experimental check of the Dicke effect in quantum wires under magnetic fields would therefore require microwave absorption experiments, i.e. determination of $\Re {\rm e}\sigma(\omega)$ in relatively long wires. The above calculation applies for the case where the two lowest subbands are occupied. Temperatures $T$ should be much lower than the subband-distance energy $\hbar \omega_B$, because thermal excitation of carriers would smear the effect. For $\hbar \omega_B$ of the order of a few meV, $T$ should be of the order of a few Kelvin or less. The Dicke peak appears for magnetic fields such that $\omega_c/\omega_0$ becomes of the order and larger than unity. For convenience, we note that the cyclotron energy in GaAs is $\hbar\omega_c$[meV] = $1.728 B$[T].


\section{\bf Phonon Cavities and Single Electron Tunneling} \label{section_cavity}
Optics deals with light, acoustics deals with sound. Optics has an
underlying microscopic theory that is linear both in its classical 
(electrodynamics) and quantum version (quantum electrodynamics), whereas in acoustics the 
linearity is an approximation: sound is based on matter-matter interaction which is non-linear.

Propagation of waves in media can be controlled by boundary conditions, material
properties and geometry. On the optics side, photonic crystals  or photon cavities are 
examples where the solutions to Maxwell's equations are `designed' 
in order to achieve a specific purpose (refractive properties, 
confinement of single photons etc). In a similar way, vibrational properties of matter can be
controlled (sound insulation being an example for classical sound-waves). The theoretical
framework is elasticity theory, the simplest model being
an isotropic material with a displacement field $\textbf{u}(\textbf{r})$ obeying a
`generalized wave equation' 
\begin{eqnarray}\label{waveequation}
  \frac{\partial^2}{\partial t^2} \textbf{u}(\textbf{r},t)=
c_t^2{\bf \nabla}^2 \textbf{u}(\textbf{r},t) + (c_l^2-c_t^2) {\bf \nabla}
\left({\bf \nabla} \cdot \textbf{u}(\textbf{r},t)\right),
\end{eqnarray}
with the transversal ($c_t$)and longitudinal ($c_l$)  sound velocities entering as parameters.

Given the importance of electron-phonon interactions as a dissipation mechanism in
single electron tunneling, it is natural to ask how to control these interactions. In quantum optics,
the controlled enhancement or reduction of spontaneous emissions of {\em photons} from atoms
defines the primary goal of cavity quantum electrodynamics (cavity QED). As for phonons, one obvious
approach towards control of phonon-induced dephasing therefore is to build  the electronic system
into a {\em phonon cavity}.  

Regarding the possible combinations of phonon cavities and electronic transport, one can broadly distinguish between two classes of cavities: 1. `natural' phonon cavities, where the modification of 
the vibrational properties of the system  comes `for free'
and goes hand in hand with the modification (as compared to the bulk) of the electronic properties. 
Recent examples are carbon nanotubes \cite{DE00}, or individual molecules \cite{Paretal00,JGA00,Reietal02,BS01,GK02};
2. `artificial' phonon cavities, which are the subject of the rest of this section and where the electronic system (2DEG, single or double quantum dot) is embedded into a  nanostructure whose  phononic properties are modified by additional fabrication steps such as under-etching and material removal \cite{Blietal00}.

\subsection{Lamb-Wave Cavity Model}\label{section_thinplate}

\begin{figure}[t]
\includegraphics[width=0.5\textwidth]{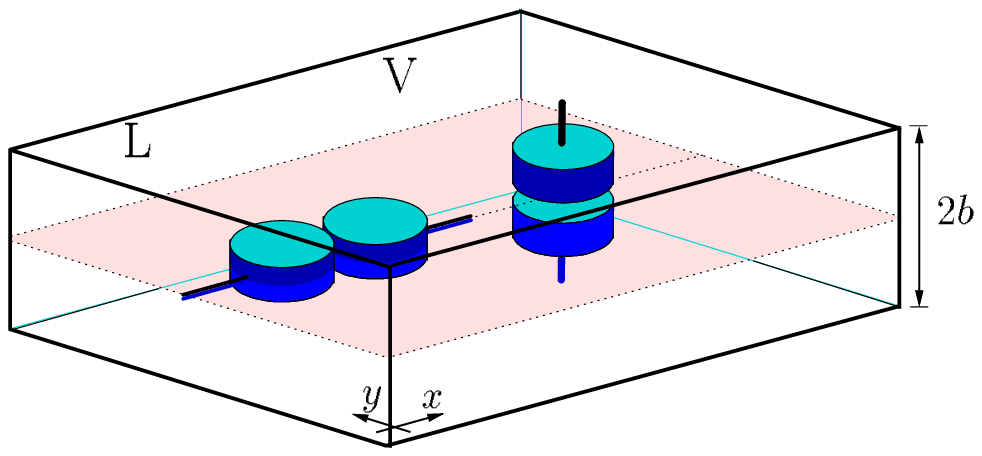}
\includegraphics[width=0.5\textwidth]{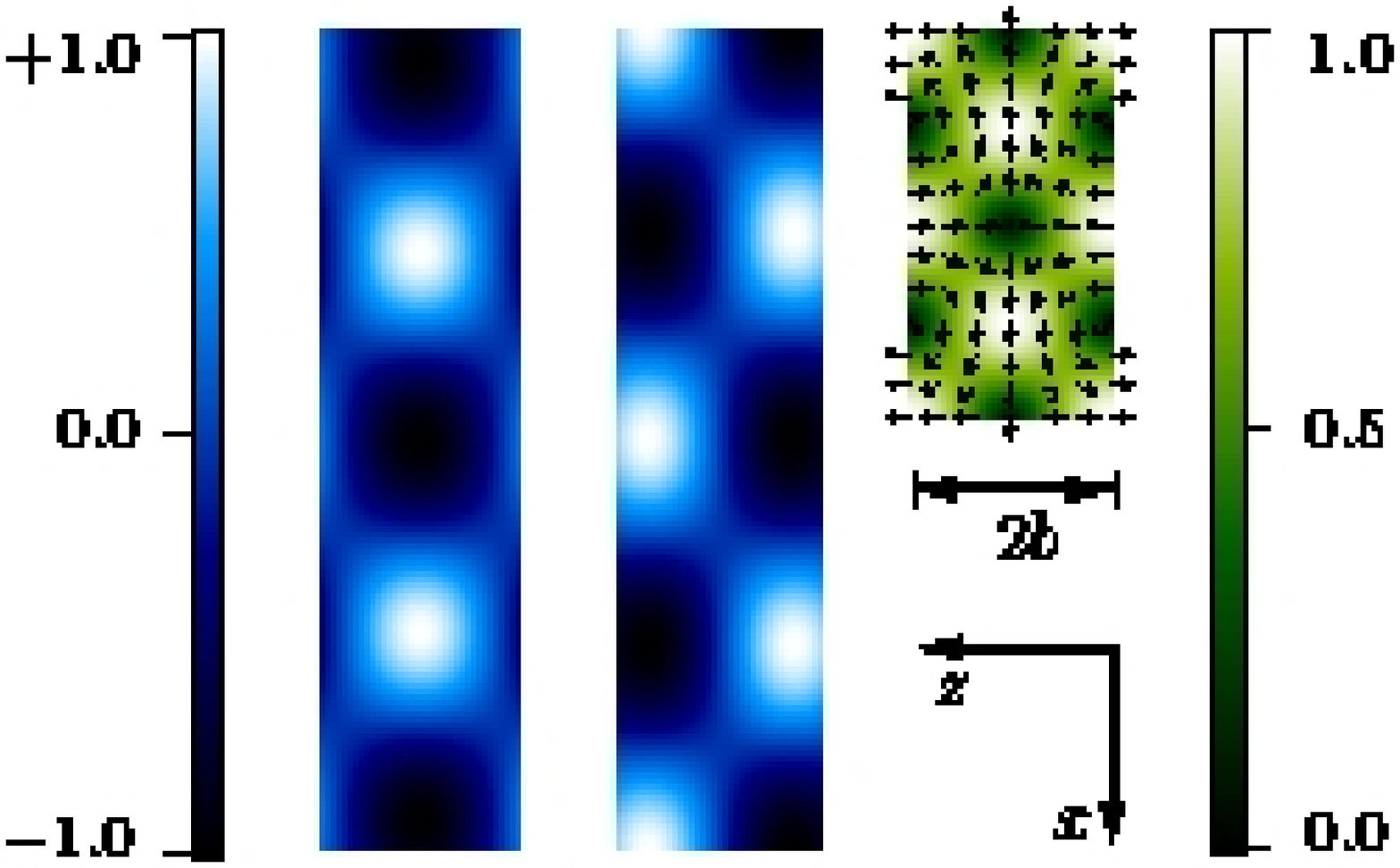}
\caption[]{\label{DBK02_fig1.eps}{\bf Left:}
  Scheme for {\em lateral} (L) and {\em vertical} (V) configurations of a
  double quantum dot in a phonon nano-cavity. {\bf Right:} Deformation potential
  induced by dilatational (left) and flexural modes (center) at
  $q_{\|}b=\pi/2$ ($n=2$ subbands), and displacement field (right)
  ${\bf u}(x,z)$ of $n=0$ dilatational mode at $\Delta=\hbar \omega_0$.
  Greyscale: moduli of deformation potentials (left) and displacement
  fields (right) (arb.  units). From \cite{DBK02}.
}
\end{figure}  

The simplest phonon cavity model is a homogeneous, two-dimensional thin plate (slab) of thickness $2b$. Debald and co-workers \cite{DBK02} used this model and calculated the transport current through a double quantum dot in various configurations, cf. Fig.(\ref{DBK02_fig1.eps}) left, where it turned out that phonon cavity effects strongly determined the electronic properties of the dots. 

The phonons were described by a displacement field $\textbf{u}(\bf{r})$, cf. \ Eq.~(\ref{waveequation}), which was determined by the vibrational modes of the slab \cite{Landau7}. These modes (Lamb waves) were classified according to the symmetry of their displacement fields with respect to the slab's mid-plane. Dilatational modes yield a symmetric elongation and compression, whereas flexural modes
yield an anti-symmetric field and a periodic bending of the slab, cf. Fig. (\ref{DBK02_fig1.eps}), right.
The third mode family consists of vertically polarized shear waves but turned out to be less
important because these waves do not coupled to charges via the deformation potential (see below).

\subsubsection{Phonon Confinement and Nano-mechanical `Fingerprints'}
Debald {\em et al.} \cite{DBK02} showed that the confinement due to the finite plate thickness leads to phonon quantization into subbands. The corresponding phonon dispersion relation was determined from the  Rayleigh-Lamb equations, 
\begin{eqnarray} \label{rl1}
\frac{\tan q_{t,n} b}{\tan q_{l,n} b} &=& -\left[ \frac{4 q_{\|}^2
    q_{l,n} q_{t,n}}{(q_{\|}^2  - q_{t,n}^2)^2} \right]^{\pm 1},\quad
 \omega_{n,q_{\|}}^2 = c_l^2 (q_{\|}^2 + q_{l,n}^2) =
c_t^2 (q_{\|}^2  + q_{t,n}^2), 
\end{eqnarray}
where the exponents $\pm 1$ correspond to dilatational and flexural modes, respectively. For each in-plane component $\bf{q}_{\|}$ of the wave vector one obtains infinitely many subbands (label $n$) which correspond to a discrete set of transversal wave vectors in the direction of the confinement. The two sound velocities $c_l$ and $c_t$ in the elastic medium are associated with longitudinal and transversal wave propagation and give rise to two sets of transversal wave vectors, $q_{l,n}$ and $q_{t,n}$.

\begin{figure}[t]
\begin{center}
\includegraphics[width=0.7\textwidth]{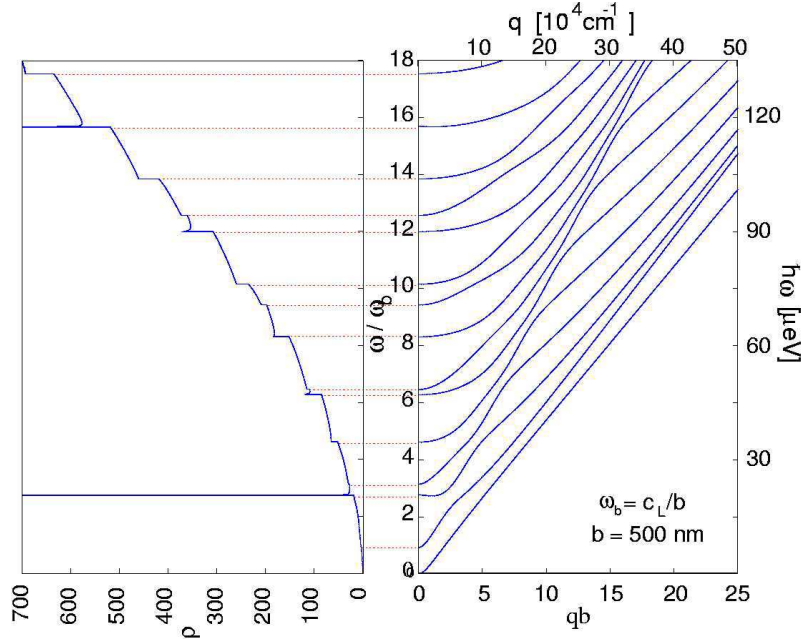}
\end{center}
\caption{Density of states ({\bf left}) and dispersion relation ({\bf right}) of
typical cavity phonons. The characteristic energy is given by $\hbar \omega_b =
\hbar c_l / b$ with the longitudinal speed of sound $c_l$ and the cavity width
$2b$. For a GaAs planar cavity of width $2b=1\,\mu$m, one has 
$\hbar \omega_b=7.5\,\mu$eV. The minimum in the dispersion of the third subband
leads to a van Hove singularity in the phonon DOS at $\hbar \omegav \approx
2.5 \, \hbar \omega_b$. From \cite{VDKB02}.} 
\label{VDKB02_fig3.eps}
\end{figure}

Examples of a cavity phonon dispersion relation and the corresponding phononic density of 
states $\rho(\omega)$ are shown in Fig. ({\ref{VDKB02_fig3.eps}) for flexural modes. 
As a particularly striking feature,  {\em phononic van Hove singularities} appear at
angular frequencies that correspond to 
a minimum in the dispersion relation $\omega_{n,q_{\|}}$ for finite ${q}_{\|}$. 
These zero phonon group velocities (with preceding {\em negative} phonon group velocities
for smaller $q$ in the corresponding subband) are due to the complicated
non-linear structure of the Rayleigh-Lamb equations for the planar cavity.
They occur in an irregular sequence 
that can be considered as a characteristic `fingerprint' of the
mechanically confined  nanostructure.

\subsubsection{Inelastic Scattering Rates in Double Quantum Dots}
Mechanical confinement effects in nano-structures modify the electron-phonon interaction in phonon cavities. Since double quantum dots are sensitive detectors of quantum noise ( \cite{AK00}, cf. section \ref{section_spectraldensity}), the boson spectral density $J(\omega)$,  and  via $I_{\rm in}= - e 2\pi T_c^2 J(\varepsilon)/\varepsilon^2$ (cf. \ Eq.~(\ref{Iinelastic})) the inelastic current through a double dot `detector', is strongly modified due to
phonon confinement.

In analogy to the bulk phonon case, one can define bosonic  spectral density for a confined slab geometry in the vertical and lateral  configurations,
\begin{eqnarray}\label{Jslab}
J_{\rm vertical}(\omega) &=&
  \sum_{\textbf{q}_{\|},n}\!
{|\lambda_{\rm dp}^{\rm flex}(\textbf{q}_{\|}, n)|^2}
4 \sin^2 \left( \frac{q_{l,n}d}{2}\right) 
\delta(\omega - \omega_{n,q_{\|}}),\\
J_{\rm lateral}(\omega) &=&
  \sum_{\textbf{q}_{\|},n}\!
{|\lambda_{\rm dp}^{\rm dil}(\textbf{q}_{\|}, n)|^2}
\left|e^{i\textbf{q}_{\|} \textbf{d}}-1 \right|^2  \cos^2 \left( \frac{q_{l,n}d}{2}\right) 
\delta(\omega -\omega_{n,q_{\|}}),
\end{eqnarray}
where again the vector ${\textbf d}$ connects the two dots, and the electron density is assumed to be sharply peaked near the dot centers which are located symmetrically within the slab. Here, the matrix elements for the deformation potential (DP) interaction are given by
\begin{equation}
\label{lambdadp}
\lambda_{\rm dp}^{\rm dil/flex}(q_{\|},n) = B_n^{\rm dp}(q_{\|}) ( q_{t,n}^2 -
q_{\|}^2)(q_{l,n}^2 + q_{\|}^2) {\rm tsc }{q_{t,n} b},\quad B_n^{\rm dp} \equiv F_n(\hbar\Xi^2/2\rho_M\,
\omega_{n,q_{\|}}A)^{1/2},
\end{equation}
where tsc ${x} = \sin x$ or $\cos x$ for
dilatational and flexural modes, respectively, $\Xi$ is the deformation potential, $\rho_M$ the mass density, $A$ the area of the slab, and $F_n$ the normalization constant for the $n^{\rm th}$ eigenmode \cite{DBK02}. Similar expressions can be derived for the piezo-electric potential \cite{DBK02}. Three observations can be made with respect to the properties of the spectral densities, \ Eq.~(\ref{Jslab}):

1. In the vertical geometry only flexural phonons, and in the lateral geometry only dilatational phonons couple to the dots via the deformation potential. This is a consequence of the symmetry of the modes and the corresponding electron-phonon interaction vertices. For the piezo-electric interaction, this symmetry is actually reversed, where vertical (lateral) dots only couple to dilatational (flexural) phonons \cite{DBK02}.

2. The deformation potential \ Eq.~(\ref{lambdadp}) {\em vanishes} for $\textbf{q}_{\|}=q_{t,n}$. For this value of $\textbf{q}_{\|}$, the divergence of the displacement field $\textbf{u}({\bf r})$ is zero, cf. Fig. (\ref{DBK02_fig1.eps}). From the Rayleigh-Lamb equations, \ Eq.~(\ref{rl1}), one obtains the corresponding smallest energy in, e.g., the lateral configuration (dilatational phonons) as
\begin{eqnarray}\label{vanishenergy}
  \hbar \omega_0 = \frac{\pi}{\sqrt{2}}\frac{\hbar c_t}{b}.
\end{eqnarray}

3. The quantization into {\em subbands} and the {\em van-Hove singularities} in the bare phonon density of states $\rho(\omega) = \sum_{\textbf{q}_{\|},n}\delta(\omega - \omega_{n,q_{\|}})$, have to appear in the 
spectral densities \ Eq.~(\ref{Jslab}) as well.

\begin{figure}[t] 
  \begin{center}
\includegraphics[width=0.7\textwidth]{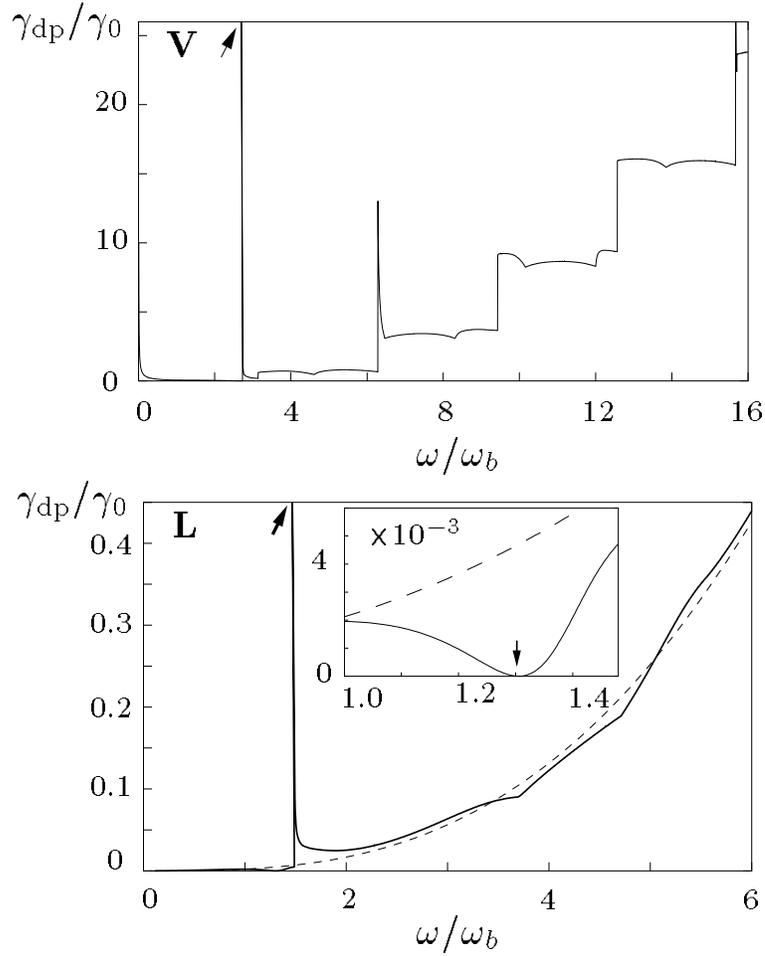}
\caption{Inelastic phonon emission rate $\gamma_{\rm dp}(\omega)$ of 
  vertical (V) and lateral (L) double dots in a phonon cavity of width
  $2b$ due to deformation potential.  Phonon-subband quantization
  effects appear on an energy scale $\hbar \omega_b = \hbar c_l/b$
  with the longitudinal speed of sound $c_l$; $\gamma_0$ nominal
  scattering rate (see text). Coupling to {\em flexural} (top) and
  {\em dilatational} modes (bottom, dashed: bulk rate).  Inset:
  Suppression of $\gamma_{\rm dp}(\omega)$ from slab phonons at
  $\omega = \omega_0$ (arrow). From \cite{DBK02}.}
\label{DBK02_fig3.eps}
\end{center}
\end{figure}

All these features are confirmed by numerical calculations. In Fig. (\ref{DBK02_fig3.eps}), the inelastic electron-phonon scattering rates
\begin{eqnarray}\label{gammadp}
  \gamma_{\rm dp}(\omega) \equiv 2\pi T_c^2 \frac{J(\omega)}{\omega^2}.
\end{eqnarray}
for the deformation potential coupling in the vertical (V) and lateral (L) configuration are shown in units of a nominal scattering rate  
$\gamma_0\equiv T_c^2 \Xi^2/\hbar\rho c_l^4 b$ for $b=5d$. The van-Hove singularities appear up 
as singularities in the inelastic rate in both cases. The phonon-subband quantization appears
as a staircase for the flexural modes (V), and as cusps for the dilatational modes (L). In the latter
case, the overall form of the curve is (apart from the singularity) quite close to 
the bulk scattering rate.  The most striking feature there, however, is the
suppression of the inelastic rate for small $\omega$ and  its {\em complete   vanishing} at
the energy $\hbar\omega_0$, \ Eq.~(\ref{vanishenergy}). Near $\omega_0$, the remaining contribution of
the $n=0$-subband mode is drastically suppressed as compared with bulk phonons.

\subsubsection{Suppression of Dephasing}
In \cite{DBK02}, it was argued that  the properties 1.-3. discussed above 
are generic features due to the slab geometry. In particular, a similar  vanishing of the inelastic rate occurs for  piezo-electric (PZ) coupling to phonons, where the angular dependence is reversed as compared to the deformation potential case. As a result, one can `switch off' the coupling to dilatational phonons either for PZ scattering in the vertical configuration, or for DP scattering in the lateral configuration at a certain energy $\hbar \omega_0$. The electron-phonon scattering is then mediated by the remaining, other interaction mechanism that couples the electrons to the flexural modes. Since the ratio $\gamma_{\rm pz}/\gamma_{\rm dp} \propto b^2$, for very thin plates (small $b$) the DP interaction dominates and the proper choice to `switch off' the scattering would be the lateral configuration, with a small contribution remaining if the material is piezo-electric, and vice versa. 

If the level-splitting $\Delta$ of a dissipative two-level system was tuned to a dissipation-free point, $\Delta=\hbar\omega_0$, this would in fact constitute a `dissipation-free manifold' for one-qubit rotations, for example in the parameter space $(\varepsilon,T_c)$ of two hybridized states with $\Delta= \sqrt{\varepsilon^2 +4T_c^2}=\hbar\omega_0$, cf. \ Eq.~(\ref{eq:twobytworesult}) and the discussion in section \ref{section_adiabaticrotation}.

The Golden-Rule type calculation of the inelastic rates, \ Eq.~(\ref{gammadp}), however, neglects 
4-th and higher order terms in the coupling constant (virtual processes) that can lead to a small
but finite phonon-induced dephasing rate even at $\Delta=\hbar\omega_0$, not to speak
of other dephasing mechanisms such as spontaneous
emission of photons (although negligible with respect to the phonon
contribution in second order \cite{FujetalTaretal}), or 
plasmons and electron-hole pair excitations in nearby  leads.

Rather than the suppression, the {\em enhancement} (van-Hove singularities) of the electron-phonon coupling in nano-cavities actually seems to be relevant to experiments with quantum dots in phonon-cavities as discussed in  section \ref{section_blick}.

\subsection{Surface Acoustic Wave Cavity Model}

Vorrath and co-workers \cite{VDKB02} discussed another phonon cavity model based on ideas by Kouwenhoven and van der Wiel \cite{KvdW99}, who suggested to place a double quantum dot between two arms of a surface acoustic wave (SAW) inter-digitated transducer.

The model is defined for a surface of a semiconductor heterostructure with an infinite lattice of  metallic stripes (spacing $l_0$, infinite length), with two coupled quantum dots located at a distance $z_0$ beneath the surface at the interface of the heterostructure,  cf. Fig. (\ref{VDKB02_fig1.eps}), left. Surface acoustic waves propagate along the surface of a medium while their typical penetration depth into the medium is of the order of one wavelength. In piezo-electric materials like GaAs, their displacement field  generates an electric potential that dominates the interaction with electrons. As the piezo-electric potential of the wave has to meet the electric boundary conditions at the interface between the medium and the air, the electron-phonon interaction strongly depends on the electric properties of the surface. The  (connected) metallic stripes give rise to an additional boundary condition for the potential
\begin{equation}
\label{bcondition}
\varphi(x\!=\!n l_0, y) = \mbox{const.}, \qquad n\in \mathbbm{Z},
\end{equation}
where the width of the stripes is neglected. 

Surface waves propagate as plane waves with wave vector $\mathbf{q}=(q_x,q_y)$ along the surface.
Considering only standing waves in $x$-direction and traveling waves in $y$-direction, 
the displacement field is given by \begin{equation}
\label{displacement}
\mathbf{w_q}(\mathbf{r},t) = C e^{i(q_y y -\omega t)} \left(
\begin{array}{c}
  a(q,z) \cos(\alpha) \cos(q_x x) \\
  i a(q,z) \sin(\alpha) \sin(q_x x) \\
  -b(q,z) \sin(q_x x)
\end{array} \right),
\end{equation}
where the functions $a(q,z)$ and $b(q,z)$ describe the decay of the SAW amplitude with depth $z$ of the medium and $\alpha$ is the angle between the $x$-axis and the wave vector $\mathbf{q}$. The corresponding piezo-electric potential is 
\begin{equation}
\label{piezo}
\varphi_{\bf q}(\mathbf{r},t) = - C \frac{e_{14}}{\varepsilon_0 \varepsilon} \;
    \big(\!\cos^2(\alpha) - \sin^2(\alpha)\big)\, 
    f(qz) \sin(q_x x) \, e^{i(q_y y -\omega t)},
\end{equation}
with $e_{14}$ the piezo-electric stress constant, $\varepsilon_0$ the dielectric constant, and $\varepsilon$ the relative permittivity of the medium. The function $f(qz)$ describes the decay in $z$-direction  and follows from the boundary condition for the electric field on the surface. 
Assuming a non-conducting surface together with the boundary condition Eq. (\ref{bcondition}), one obtains the restriction 
\begin{equation}
\label{qx}
q_x = m \, \frac{\pi}{l_0}, \qquad m \in \mathbbm{N}.
\end{equation}
Vorrath {\em et al.} derived the corresponding electron-phonon interaction potential as
\begin{eqnarray}\label{H_SAW}
  V_{\rm ep}({\bf r}) = \sum_{\mathbf{q}} \left[-e\varphi_{\bf q}(\mathbf{r},t=0)\right]
   \Big( b_{q_x, q_y}+ b_{q_x,-q_y}^{\dagger} \Big),
\end{eqnarray}
where $b_{q_x, q_y}$ is the phonon annihilation operator for the mode $(q_x, q_y)$, $-e$ the electron charge, and $\varphi_{\bf q}$ the piezo-electric potential, \ Eq.~(\ref{piezo}), where the normalization constant $C$ in (\ref{displacement}) is defined  as $C = \frac{1}{L} \sqrt{\frac{\hbar}{\rho \lambda v}}$ ($\rho$ is the density of the medium, $\lambda$ a material parameter and $v$ the velocity of the SAW) and does not depend on the wave vector $\mathbf{q}$ but on the quantization area $L^2$.

\begin{figure}[t]
\centering
\psfrag{0}{$0$}
\psfrag{L}{$l_0$}
\psfrag{x}{$x$}
\psfrag{y}{$y$}
\psfrag{[010]}{\hspace*{-16mm}[010]}
\psfrag{[100]}{\hspace*{-16mm}[100]}
\psfrag{x}{\hspace*{-3mm}$\omega / \omega_v$}
\psfrag{rho}{\hspace*{-6mm}$J(\omega)N\omega_v/\omega^2$}
\includegraphics[width=0.4\textwidth]{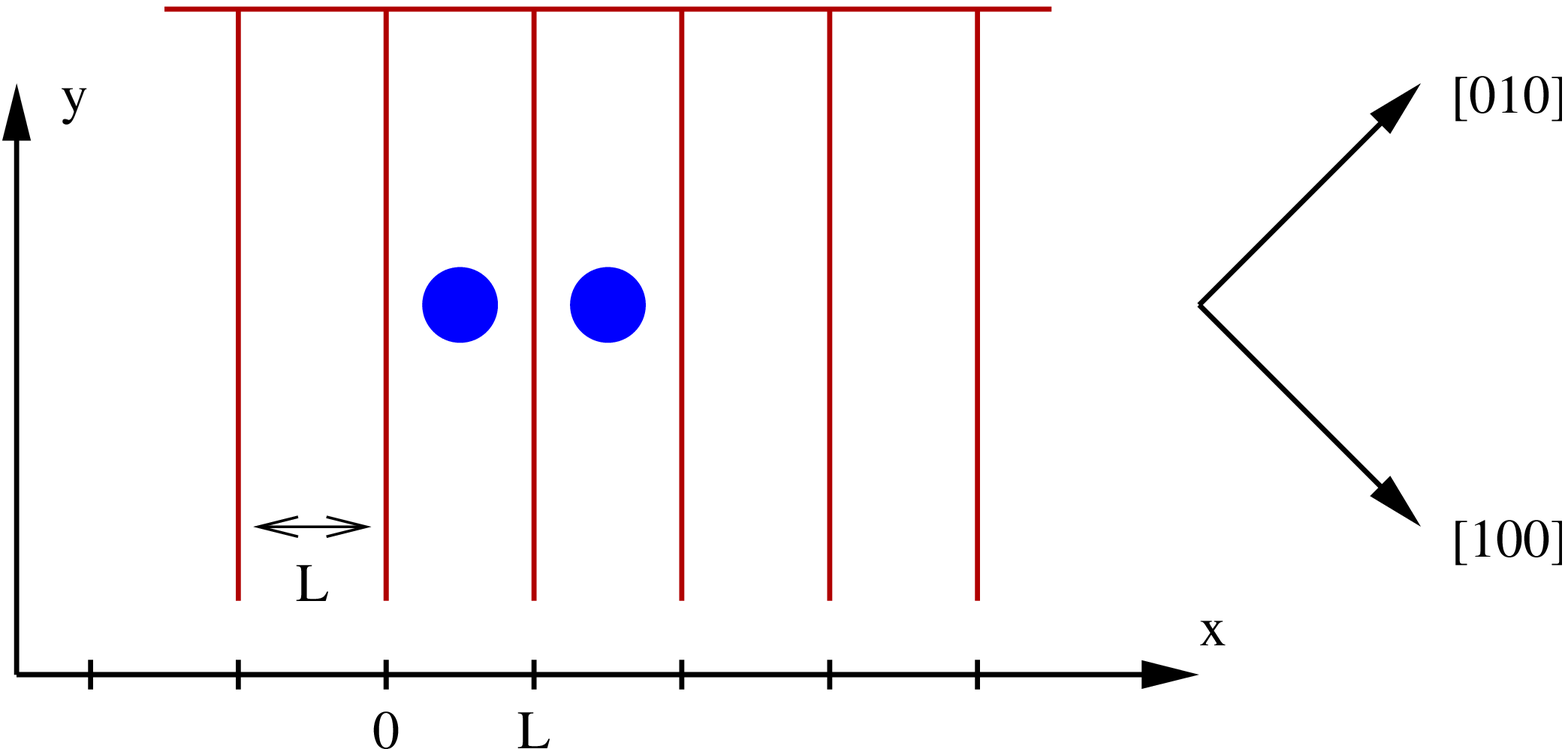}
\includegraphics[width=0.4\textwidth]{VDKB02_fig2.eps}
\caption{{\bf Left:} Surface Acoustic Wave Cavity Model (top view) with  quantum dots beneath the surface between metal stripes. Crystal axes include an angle of 45 degrees with the stripes. {\bf Right:} 
Spectral phonon density  for  finite system-length~$L=Nl_0$, \ Eq.~(\ref{J_SAW}). The quantum dots are $z_0=100$ nm beneath the surface and their distance is equal to the spacing of the metallic stripes $l_0=250$ nm. Material parameters are taken for GaAs. The frequency $\omega_v$ corresponds to an energy of 22 $\mu$eV. From \cite{VDKB02}.}
\label{VDKB02_fig1.eps}
\end{figure}

The boson spectral density $J(\omega)$
corresponding to interaction with SAW modes in \ Eq.~(\ref{H_SAW}) was calculated as \cite{VDKB02}
\begin{eqnarray}\label{J_SAW}
  J_{\rm SAW}(\omega) = \frac{1}{N \omega_v}
 \frac{4}{\pi^2 \hbar  \rho \lambda v^3}
\left( \frac{e e_{14}}{\varepsilon_0 \varepsilon} \right)^2
{\omega_v}^2 f^2(\omega z_0/v) 
\sum_{m=1,3,...}^{m<\omega/\omega_v}
\Big[ 2 m^2 \left( \frac{\omega_v}{\omega} \right)^2 -1\Big]^2,
\end{eqnarray}
where a (finite) system-length $L$ was given in units of $N$  spacings, $L=Nl_0$, and the typical frequency scale $\omega_v\equiv \pi v/l_0 $  was introduced. 
Fig. (\ref{VDKB02_fig1.eps}) right shows numerical examples of $J_{\rm SAW}(\omega)$, from which the corresponding inelastic current through the double dots again is given by $I_{\rm in}(\varepsilon) = -e 2 \pi  T^2_c J(\varepsilon)/\varepsilon^2$,  \ Eq.~(\ref{Iinelastic}), in lowest order of the tunnel coupling $T_c$.
For energies smaller than $\hbar \omega_v$ the lowest standing wave mode can not be excited and consequently the inelastic current exhibits a gap in that energy region. The excitation of higher modes manifests itself in steps in the inelastic current at $\omega/\omega_v\!=\!1,3,5,\ldots$ . Furthermore, at $\omega/\omega_v\!=\!\sqrt{2}$ the spectral density  vanishes, because the SAW would be emitted along the crystal axes without any piezo-electric interaction in that direction.

The scaling of $J_{\rm SAW}(\omega)$ with the inverse of the system length $L$ is due to the fact that the energy $\hbar \omega_q$ of one phonon is distributed over the whole sample. By increasing $L$, the amplitude of the displacement, the piezo-electric potential, and therewith the interaction strength is decreased and finally vanishes. For traveling waves, this effect is canceled by an increasing number of modes within each interval of energy. In the cavity model however,  the standing wave modes are independent of the system-size and therefore the boson spectral density $J_{\rm SAW}(\omega)$ is completely suppressed by the metallic stripes in the limit $L\to \infty$.

\subsection{Experiments on Electron Tunneling in Suspended Nano-Structures}\label{section_blick}

Weig and co-workers \cite{Hoeetal04} performed transport experiments with single quantum dots embedded into a phonon cavity that was produced as a freestanding, 130 nm thin  GaAs/AlGaAs membrane. The technique of embedding and controlling a two-dimensional electron gas into a suspended semiconductor structure was pioneered by Blick and co-workers \cite{Blietal00}.

\begin{figure}[t]
\centering
\includegraphics[width=0.4\textwidth]{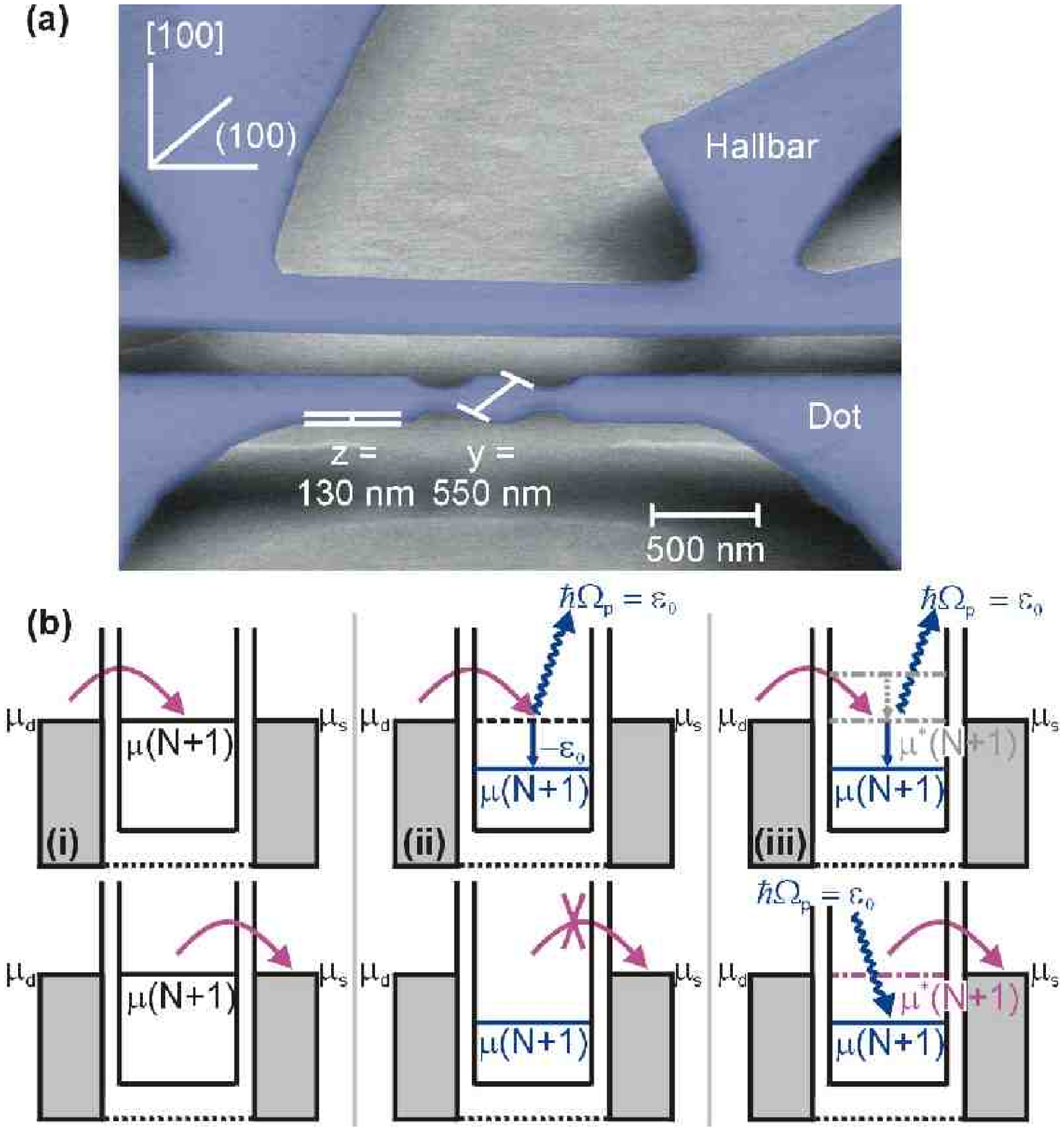}
\includegraphics[width=0.5\textwidth]{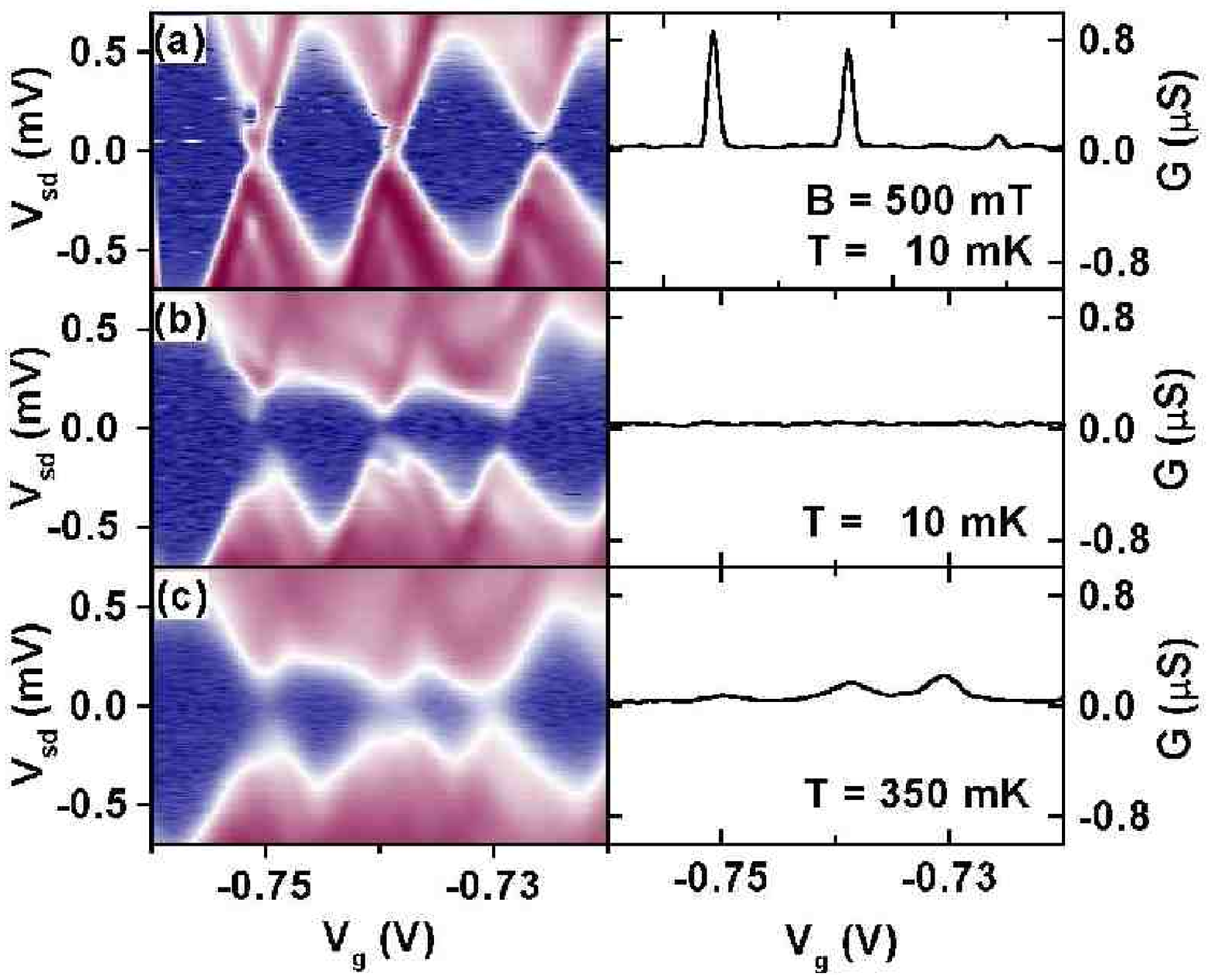}
\caption{{\bf Left:} 
(a) Suspended quantum dot cavity and Hall-bar formed in a $130$\,nm thin GaAs/AlGaAs membrane.
(b) Level diagrams for single electron tunneling: (i) In the orthodox model electrons sequentially
tunnel through the dot, if the  
chemical potential $\mu(N+1)$ is aligned between the reservoirs. (ii)
Tunneling into the phonon  
cavity  results in  the excitation of a cavity phonon with energy
$\hbar \Omega_{\rm ph}$, leading to a level 
mismatch $\epsilon_0$ and thus to `phonon blockade'. (iii) Single
electron tunneling
is re-established by a higher lying electronic state $\mu^*(N+1)$ 
which re-absorbs the phonon.
{\bf Right:}Transport spectrum of suspended single quantum dot 
and zero bias conductance: (a) Single electron
resonances at electron temperature $100$\,mK and a 
perpendicular magnetic field of $500$\,mT. 
(b) At zero magnetic field conductance is suppressed for bias
voltages below $100\,\mu$V due to phonon excitation.
(c) The conductance pattern at $350$\,mK shows that phonon
blockade starts to be lifted because of thermal broadening of the
Fermi function supplying empty states in the reservoirs. From \cite{Hoeetal04}.} \label{Hohetal_fig1}
\end{figure}

The phonon cavity, Fig. (\ref{Hohetal_fig1}) left, was produced by completely removing the layer beneath the membrane, and the quantum dot was formed by two constrictions on the membrane. A negative gate voltage $V_g$ applied to the nearby in-plane gate electrode created tunnel barriers for the dot and controlled the dot electrochemical potential $\mu(N+1)$. Standard Coulomb diamond diagrams \cite{Grabert,Tews04} as a function of $V_g$ and $V_{sd}$, the source-drain voltage, were used to analyse the linear and non-linear transport through the dots, cf. Fig. (\ref{Hohetal_fig1}) right. At a finite perpendicular magnetic field $B=500$ mT and an electron temperature $T_e=100$ mK, conventional Coulomb blockade (CB) was observed in the form of CB oscillation peaks as a function of $V_g$ in the conductance $G$, and an electron number of $N\approx 1400$ was deduced. 

A novel feature was found for zero magnetic field in the form of a complete suppression of the linear conductance over several CB oscillation peaks, and the opening of an energy  gap $\varepsilon_0$ between the CB diamonds, Fig. (\ref{Hohetal_fig1}), right b), which resulted into a blockade of transport that could only be overcome by either increasing $V_{sd}$ or the temperature $T$, Fig. (\ref{Hohetal_fig1}), right c).

\begin{figure}[t]
\centering
\includegraphics[width=0.45\textwidth]{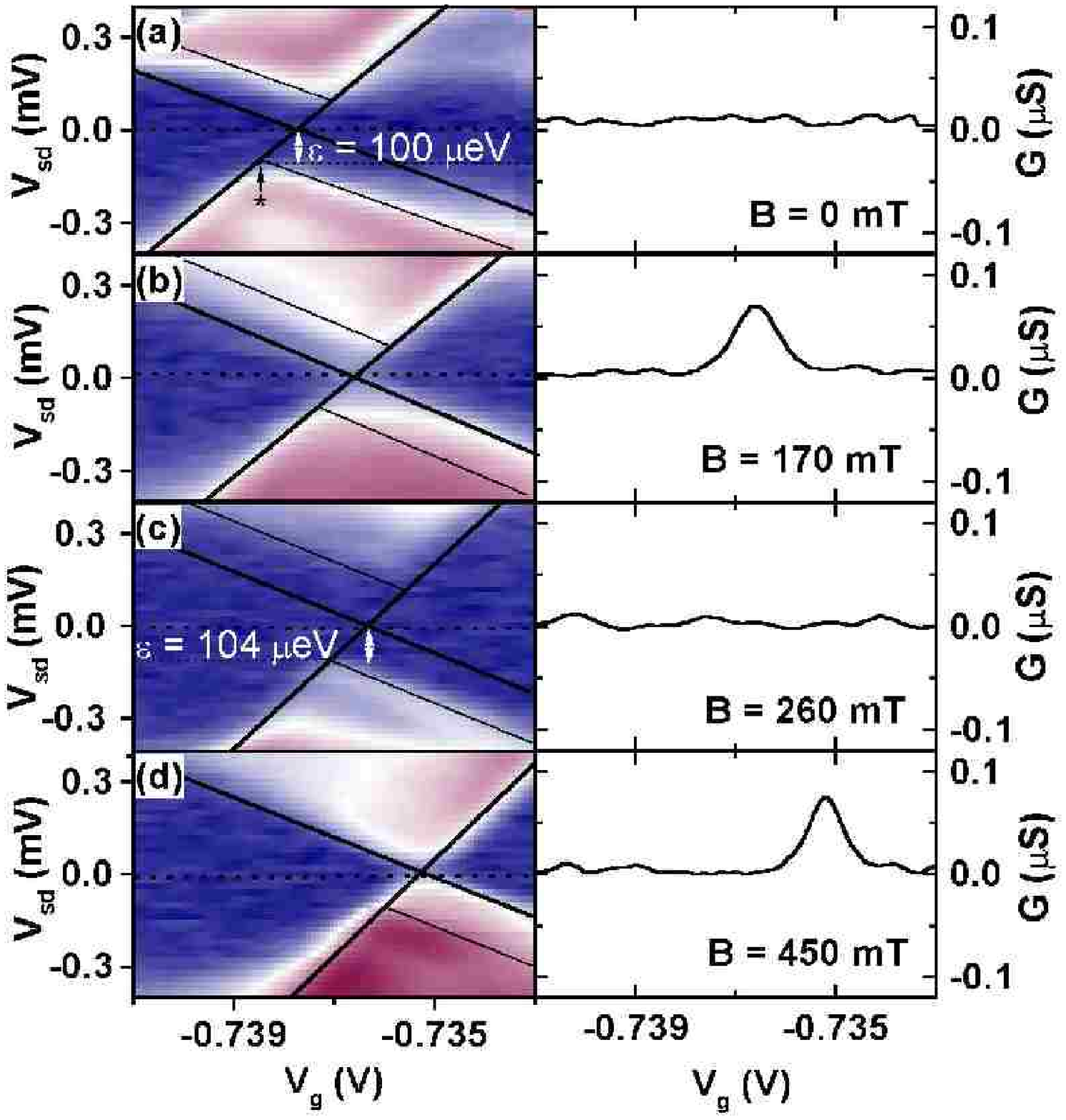}
\includegraphics[width=0.45\textwidth]{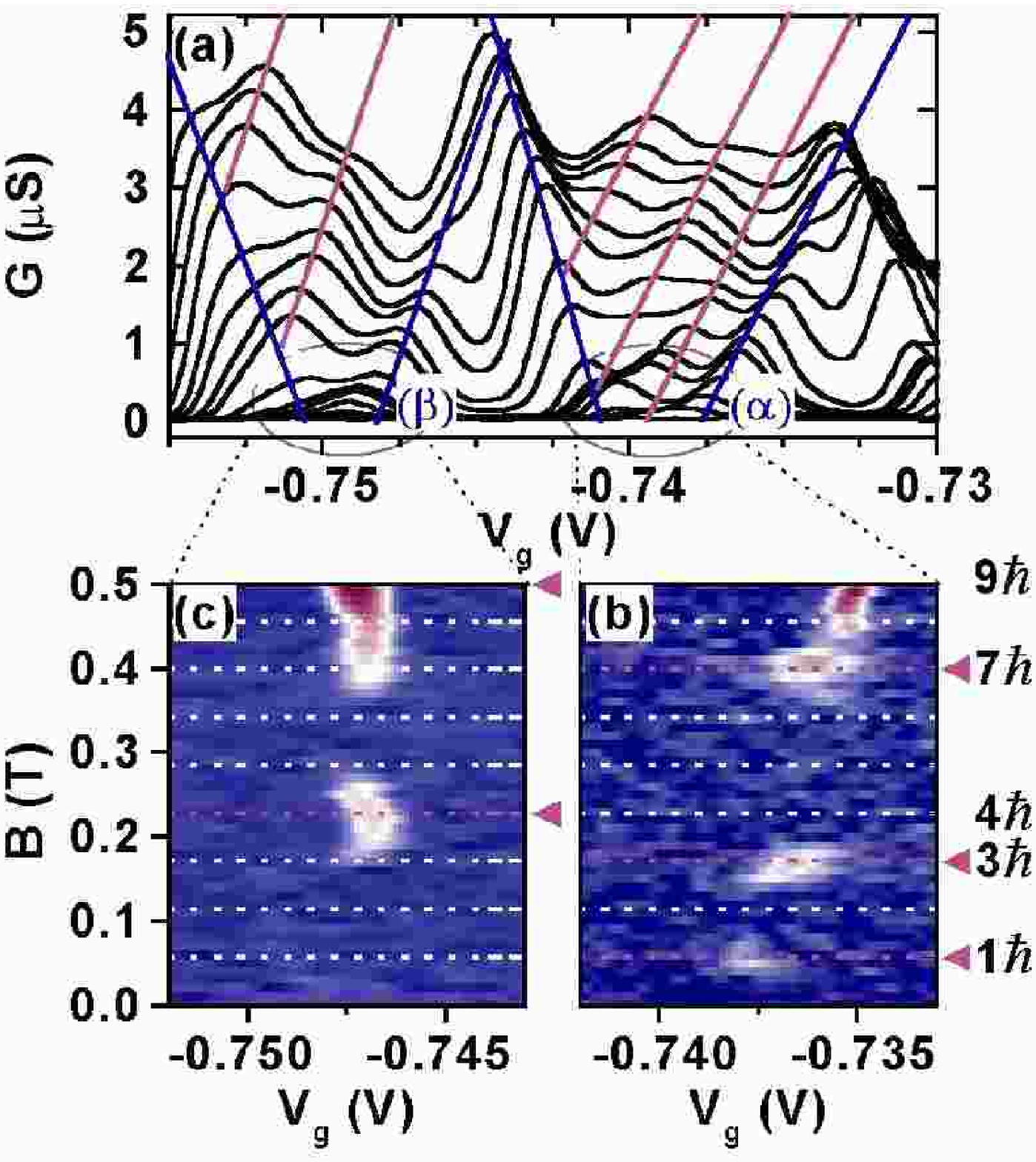}
\caption{{\bf Left:}
Transport spectrum for (a) $B = 0$\,mT, (b) $170$\,mT, (c) $260$\,mT,
and (d) $450$\,mT. 
The line plots
give the zero bias trace. At certain magnetic fields (b,d)
excited quantum dot states with higher magnetic momentum are brought
into resonance with the cavity phonon re-enabling single electron
tunneling. Otherwise (a,c) transport is suppressed due to phonon
blockade with an excitation barrier of around $100\,\mu$eV.
{\bf Right:} (a) Line plot of conductance resonances $\alpha$ and $\beta$ at $B=0$ and 
different source-drain bias voltages between $0\,\mu$V and
$-800\,\mu$V. Blue lines follow the ground states, while red lines
mark excited states. 
b) Zero bias conductance for
resonance $\alpha$ plotted against gate voltage $V_{\rm g}$ and
magnetic field $B$. 
Finite conductance
appears for $57$\,mT, $170$\,mT, and $400$\,mT. (c) Similar plot for
resonance $\beta$ (blue:
$0.02\,\mu$S, red: $2\,\mu$S): Non-zero conductance is found for
$230$\,mT and $510$\,mT. From \cite{Hoeetal04}.
}
\label{Hohetal_fig3.eps}
\end{figure}

A simple model \cite{Hoeetal04} was developed along the lines of single electron transport in molecular transistors, where similar energy gaps in transport through $C_{60}$ molecules were observed by Park and co-workers \cite{Paretal00}.  Fig. (\ref{Hohetal_fig1}) left b) compares the situation of conventional Coulomb blockade (i) with Coulomb blockade in a suspended phonon cavity (ii), where electron tunneling excites a localized cavity phonon with  energy $\hbar \Omega_{\rm ph}$ that goes along with a drop of the chemical potential $\mu(N+1)$ of the dot, leading to a blockade (`phonon blockade') of single electron tunneling. The energy gap $\varepsilon_0=100\mu$eV  was found to compare well with the phonon energy $\hbar \Omega_{\rm ph}$ in the thin plate model of section \ref{section_thinplate} corresponding to the lowest van-Hove singularity, where electron-phonon coupling is expected to be strongly enhanced. In an analogy to the M\"ossbauer effect for $\gamma$-radiation emitting nuclei in solids, the `recoil' of the tunneling electron is taken up by the crystal as a whole, if the dot is produced in a usual, non-freestanding matrix, with the resulting transport being elastic (case i).  On the other hand, a freestanding phonon cavity picks up the recoil energy of the tunneling electron , with the electron relaxing to a new ground state trapped below the chemical potentials of the leads (case ii). To re-establish single electron tunneling (case iii), the cavity phonon has to be re-absorbed such that (similar to Rabi oscillations in a two-level system) the electron can tunnel out again via a higher lying electronic state with chemical potential $\mu^*(N+1)$.

Weig and co-workers presented additional data on the vanishing and re-appearance of the `phonon blockade' effect at different magnetic fields, cf. Fig.(\ref{Hohetal_fig3.eps}), left, by tuning excited states with angular momentum $l\hbar$, $l=1,2,...$ in resonance and thereby lifting the `phonon blockade', cf.  (b) and (d). Furthermore, Fig.(\ref{Hohetal_fig3.eps}) right,  shows conductance traces for various bias voltages and zero magnetic field $B=0$, (a), and the linear conductance near two CB peaks at finite $B$, (b) and (c). The energies of the excited states in (a) matched the number of discrete magnetic fields in (b) and (c) which was a further indication for the lifting of the `phonon blockade' by excited states.


\section{\bf Single Oscillators in Quantum Transport}\label{section_oscillator}

Single bosonic modes play  a key role in the modeling for the interaction of matter with photons or phonons in confined geometries. A prime example with respect to matter-light interaction is cavity quantum electrodynamics where the coupling between atoms and photons is used in order to, e.g., transfer quantum coherence from light to matter (control of tunneling by electromagnetic fields \cite{NS96}) and vice versa \cite{UWK96,SU97,SU99}. The discussions in section \ref{section_transport} and \ref{section_cavity} have made it clear that {\em phonons} interacting with  electrons in confined geometries can give rise to what might be called `semiconductor phonon cavity QED' in analogy with semiconductor cavity  quantum electrodynamics \cite{Yamamoto}. Furthermore, the newly emerging field of nano-mechanics shows that vibrational properties of mesoscopic systems give rise to new and surprising electronic transport phenomena such as `shuttling' in movable nano-structures  \cite{Goretal98,WZ99,AWZB01,AM02}, cf. section \ref{section_shuttle}. More or less closely related topics are single-phonon physics, the quantization of the thermal conductance, displacement detection, and macroscopic superposition and tunneling of mechanical states, which are topics covered in a recent Review article by Blencowe \cite{Blen04} on quantum electromechanical systems.

The following two  subsections present  models in which one of the fundamental models in Quantum Optics is adapted to electronic transport. It is probably fair to say that the Rabi Hamiltonian \cite{Rabi37},
\begin{eqnarray}\label{H_Rabi}
  {\mathcal H}_{\rm Rabi} = \frac{\varepsilon}{2}\sigma_z + g \sigma_x (a^{\dagger} + a) + \Omega a^{\dagger}a,
\end{eqnarray}
is the simplest and at the same time the best studied model  for the interaction of matter with light \cite{Allen}, where `matter' is represented by the most elementary quantum object, i.e., a (Pseudo) spin $\frac{1}{2}$.  Section \ref{section_delta} deals with a single boson model in one of the `classic' areas of mesoscopic physics, i.e., the transmission coefficient for the motion of (quasi) one-dimensional, non-interacting electrons in a scattering potential. Section \ref{section_boson} then presents the opposite extreme of electron transport in the strong Coulomb blockade regime, where the limit of one single boson mode in the open spin-boson model, similar to the double quantum dots from section \ref{section_transport}, is discussed. Section \ref{section_shuttle} gives a very brief introduction into non-linear boson coupling and electron shuttling, and section \ref{cavity_experiment} shortly discussed recent experimental results on a realization of \ Eq.~(\ref{H_Rabi}) with Cooper pair boxes.

\subsection{Transmission Coefficient of a Dynamical Impurity, Fano Resonances}\label{section_delta}
An exactly solvable mesoscopic scattering model for the transmission of electrons through a barrier in presence of coupling to a boson (photon or phonon) mode was discussed by Brandes and Robinson in \cite{BR02}. The model describes a {\em single} electron of mass $m$ in one dimension that interacts with a delta-barrier, the coupling strength of which is itself is a dynamical quantity,
\begin{eqnarray}\label{Hamiltonian_delta}
  H=\frac{p^2}{2m}+\delta(x)\left\{ g_0+g_1[a^{\dagger}+a]\right\}+\Omega a^{\dagger}a.
\end{eqnarray}
Here, $a^{\dagger}$ creates a boson of frequency $\Omega$ 
and $g_1[a^{\dagger}+a]$ is a dynamical contribution added to the static coupling constant $g_0$. The constant zero point energy is omitted since it merely  shifts the energy scale by $\Omega/2$. 
The lattice version of this model was  originally introduced by Gelfand, Schmitt-Rink and Levi \cite{GSL89} in 1989 in the study of tunneling in presence of Einstein phonons of frequency $\Omega$. Their model had the form of a one-dimensional tight binding Hamiltonian, 
\begin{eqnarray}
  H_{ij}^{\rm GSR}=-t_{ij} + \delta_{ij}\left[V_{0i}+V_1\delta_{i0}(a+a^{\dagger})\right],
\end{eqnarray}
and they used a continued-fraction expansion which lead to  singularities (cusps and infinite slopes) in the transmission coefficient as a function of energy, similar to the results in the continuous model discussed below. Lopez-Castillo, Tannous, and Jay-Gerin \cite{LTJ90} compared these results shortly afterwards with those from a corresponding time-dependent classical Hamiltonian,
\begin{eqnarray}
  H_{ij}^{\rm LTJ}=-t_{ij} + \delta_{ij}\left[V_{0i}+V_1\delta_{i0}\sin \Omega t \right],
\end{eqnarray}
and found very similar features. The time-dependent, classical  version of the continuous model Hamiltonian, \ Eq.~(\ref{Hamiltonian_delta}), reads 
\begin{eqnarray}\label{Hamiltonianc}
  H_{\rm cl}(t)=\frac{p^2}{2m}+\delta(x)\left\{ g_0+2g_1\cos(\Omega t)
\right\}.
\end{eqnarray}
and is obtained as the interaction picture Hamiltonian of Eq.(\ref{Hamiltonian_delta}) 
with respect to $H_B=\Omega a^{\dagger}a$,
after replacing the boson operators by $a^{\dagger}=a=1$. 

In its  time-dependent version, Eq.(\ref{Hamiltonianc}) was used as a model for scattering in quasi-one-dimensional quantum wires by Bagwell \cite{Bag90}, who found Fano-type resonances in the transmission coefficient as a function of the energy of an incident electron. It soon turned out that the scattering properties of this Hamiltonian are  quite intriguing as they very much depend on the relative sign and strength of the two coupling parameters $g_0$ and $g_1$. Bagwell and Lake \cite{BL92} furthermore studied the interplay between evanescent modes and quasi-bound states in quasi one-dimensional scattering. Very recently, Martinez and Reichl \cite{MR01}, and Kim, Park, Sim and Schomerus investigated the behavior of the transmission amplitude of a one-dimensional  time-dependent impurity potential in the complex energy plane \cite{KPSS03}.

One should mention that in contrast to the classical, time-dependent  Eq.(\ref{Hamiltonianc}), one immediate pitfall of the quantum model \ Eq.~(\ref{Hamiltonian_delta}) is the fact that its many-electron, second quantized counterpart is non-trivial: even without electron-electron interactions, the coupling of the Fermi sea to a common boson mode induces effective interactions among the electrons, and one has to deal with a non-trivial correlation problem.

\subsubsection{Transmission Coefficient}
In the comparison between the  peculiarities of the quantum version Eq. (\ref{Hamiltonian_delta}) with those of the classical model $H_{\rm cl}(t)$, \ Eq.~(\ref{Hamiltonianc}), it turns out that beside transmission zeroes, there are points of perfect transparency in the Fano resonance that only appear in the `quantum' model $H$ but not in $H_{\rm cl}$. In order to calculate the transmission coefficient, the total wave function $|\Psi\rangle$ of the coupled electron-boson system is expanded in the oscillator basis $\{ |n\rangle \}$ as
\begin{eqnarray}\label{totalwave}
  \langle x|\Psi\rangle=\sum_{n=0}^{\infty}\psi_n(x) |n\rangle,
\end{eqnarray}
with wave function coefficients $\psi_n(x)$ depending on the position $x$ of the electron. One solves the stationary Schr\"odinger equation at total energy $E>0$, implying a scattering condition for the electron part of the wave function in demanding that there is no electron incident from the right. For $x\ne 0$, the $\psi_n(x)$ are superpositions of plane waves if $E$ is above the threshold for the $n$-th boson energy, 
\begin{eqnarray}\label{waveprop}
 \psi_n(x<0)&=&a_ne^{ik_nx}+b_ne^{-ik_nx},\quad 
 \psi_n(x>0)=t_ne^{ik_nx},\quad k_n\equiv \sqrt{E-n\Omega},\quad
E>n\Omega,
\end{eqnarray}
whereas normalizable evanescent modes occur if $E$ is below the threshold,
\begin{eqnarray}\label{waveevan}
 \psi_n(x<0)&=&b_ne^{\kappa_nx},\quad
 \psi_n(x>0)=t_ne^{-\kappa_nx},\quad \kappa_n\equiv \sqrt{n\Omega-E},\quad
E<n\Omega,
\end{eqnarray}
where one sets $\hbar = 2m =1$. Imposing the condition that the boson is in its ground state for an electron incoming from the left, $  a_n=\delta_{n,0}$, and setting the corresponding amplitude $A=A_0$ to unity, one obtains
$a_n+b_n=t_n$ for all $n$ from the continuity of $\psi_n(x)$ at $x=0$, whereas the jump in derivative of $\psi_n(x)$ across the delta barrier leads to a recursion relation for the transmission amplitudes $t_n$,
\begin{eqnarray}\label{recursion}
  g_1\sqrt{n}t_{n-1}+(g_0 -2i\gamma_n)t_n + g_1 \sqrt{n+1} t_{n+1}&=&-2i\gamma_n \delta_{n,0},
\end{eqnarray}
where the $\gamma_n$ are real (imaginary) above (below) the boson energy $n\Omega$,
\begin{eqnarray}
\gamma_n&=& k_n \theta(E-n\Omega)+i\kappa_n\theta(n\Omega-E).  
\end{eqnarray}
The total transmission coefficient $T(E)$ is then obtained from the sum over all {\em propagating} modes,
\begin{eqnarray}\label{transmission}
  T(E) = \sum_{n=0}^{[E/\Omega]}\frac{k_n(E)}{k_0(E)}|t_n(E)|^2,
\end{eqnarray}
where the sum runs up to the largest $n$ such that $k_n$ remains real. Although Eq.(\ref{transmission}) is a finite sum, its evaluation requires the solution of the {\em infinite} recursion relation Eq.(\ref{recursion}) due to the fact  that the propagating modes are coupled to all evanescent modes. The transmission amplitudes can be determined from the linear matrix equation 
\begin{eqnarray}\label{matrix}
M{\bf t} &=& {\bf a},\quad {\bf t}=(t_0,t_1,t_2,...),\quad
{\bf a}= (-2i\gamma_0,0,0,...)\nonumber\\
M&=&\begin{array}{r}
    \left(
    \begin{array}{ccccc}
        g_0-2i\gamma_0 & \sqrt{1}g_1 & 0 &    \\
        \sqrt{1}g_1 & g_0-2i\gamma_1 & \sqrt{2}g_1 & 0   \\
        0 & \sqrt{2}g_1 & g_0-2i\gamma_2 & \ddots   \\
          & 0  & \ddots & \ddots 
    \end{array}\right).
\end{array}
\end{eqnarray}
Numerically, this is easily solved by truncation of the matrix $M$.  Alternatively, one can solve Eq.(\ref{matrix}) recursively which actually is numerically more efficient. In particular, the result for the zero-channel transmission amplitude $t_0(E)$ can be written in a very intuitive form:  defining the `Greens function' $G_0(E)$ by
\begin{eqnarray}\label{G0def}
  G_0(E) \equiv [-2i\gamma_0(E) +g_0]^{-1},
\end{eqnarray}
one writes $t_0(E)$ with the help of a recursively defined 
`self energy' $\Sigma^{(N)}(E)$,
\begin{eqnarray}\label{selfenergy}
  t_0(E) &=& \frac{-2i\gamma_0(E)}{G^{-1}_0(E)-\Sigma^{(1)}(E)},
\quad  \Sigma^{(N)}(E) = \frac{Ng_1^2}{G^{-1}_0(E-N\Omega)- \Sigma^{(N+1)}(E)},
\end{eqnarray}
and by using $\gamma_n(E)=\gamma_0(E-n\Omega)$ the self energy $\Sigma^{(1)}(E)$ can be represented as a continued fraction 
\begin{eqnarray}\label{sigmacontinued}
  \Sigma^{(1)}(E) &=&
\frac{g_1^2}{\displaystyle G^{-1}_0(E-\Omega) - \frac{2g_1^2}{\displaystyle G^{-1}_0(E-2\Omega) -
\frac{3g_1^2}{\displaystyle G^{-1}_0(E-3\Omega) - \frac{4g_1^2}{\displaystyle \ddots}}}}\,,
\end{eqnarray}
which also demonstrates that $t_0(E)$ depends on $g_1$ only through the square $g_1^2$. 

Truncating the matrix $M$ to a $N\times N$ matrix corresponds to the approximation that sets $\Sigma^{(N)}(E)\equiv 0$ and recursively solves Eq. (\ref{selfenergy}) for $\Sigma^{(N-1)}(E)$ down to $\Sigma^{(1)}(E)$. In the simplest approximation, truncating at $N=2$ one obtains
\begin{eqnarray}\label{t0approx}
  t_{0,{N=2}}(E) &=& \frac{-2i\gamma_0(E)}{G^{-1}_0(E)-\Sigma^{(1)}_{N=2}(E)}
=\frac{-2i\gamma_0(E)}{-2i\gamma_0(E)+g_0 -\frac{\displaystyle g_1^2}{\displaystyle -2i\gamma_1(E)+g_0}},
\end{eqnarray}
from which an interesting observation can be made with respect to the stability of the recursion for large coupling constants $g_1$: the truncation at $N+1$ is only consistent if the truncated self energy $\Sigma^{(N)}(E)$ is a small correction to the inverse `free propagator', $Ng_1^2/|G^{-1}_0(E-N\Omega)|<  |G^{-1}_0(E-(N-1)\Omega)|$,
which by use of Eq. (\ref{G0def}) at large $N$ implies $Ng_1^2<4N\Omega$ or $g_1<2\sqrt{\Omega}$. In \cite{BR02}, it was argued that the tridiagonal form of the matrix, Eq. (\ref{matrix}),  implies that the recursion method is perturbative in the coupling $g_1$, and it was conjectured that for $g_1$ above the critical value, the perturbation based  on the oscillator basis $\{ |n\rangle \}$  should break down, similar to other numerical  approaches that start from a weak coupling regime in single boson Hamiltonians, such as the standard Rabi Hamiltonian \cite{rabi1}, \ Eq.~(\ref{H_Rabi}){}.

\subsubsection{Comparison to the Classical Case}
\begin{figure}[t]
\includegraphics[width=0.5\textwidth]{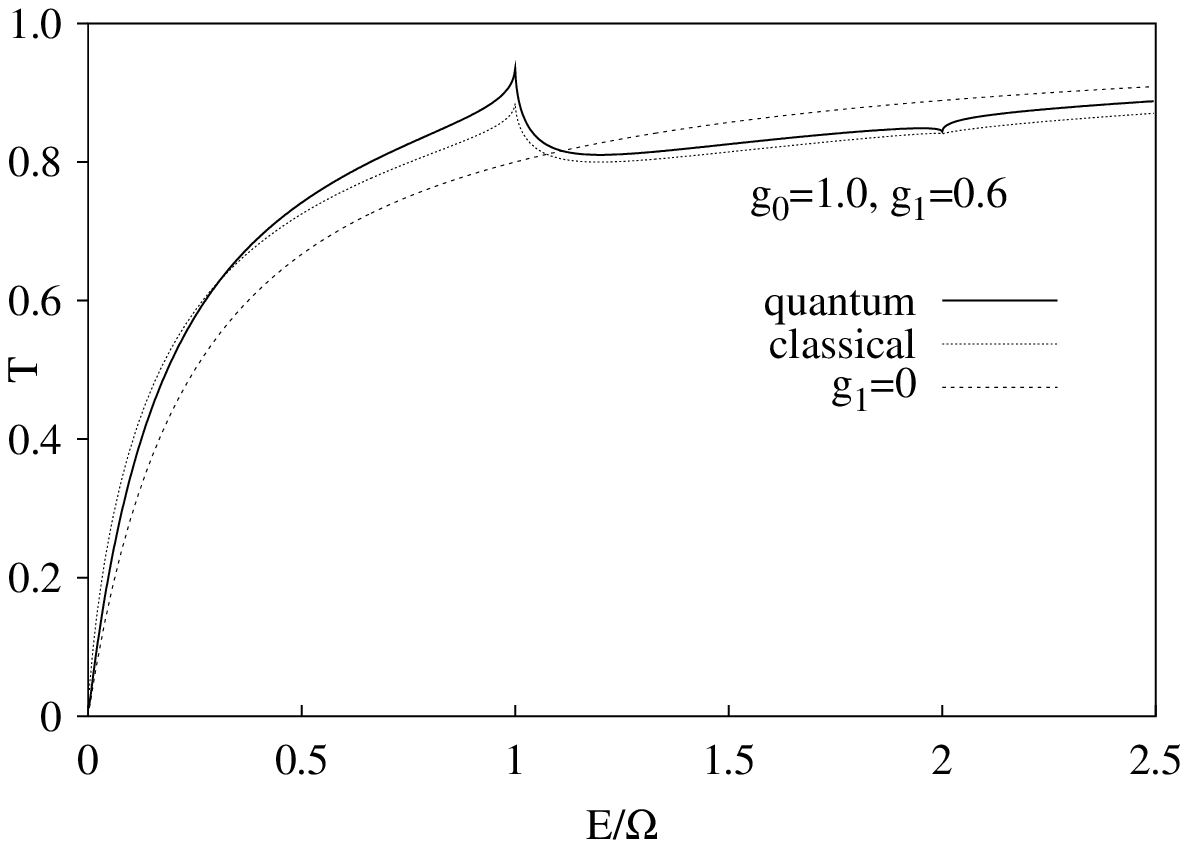}
\includegraphics[width=0.5\textwidth]{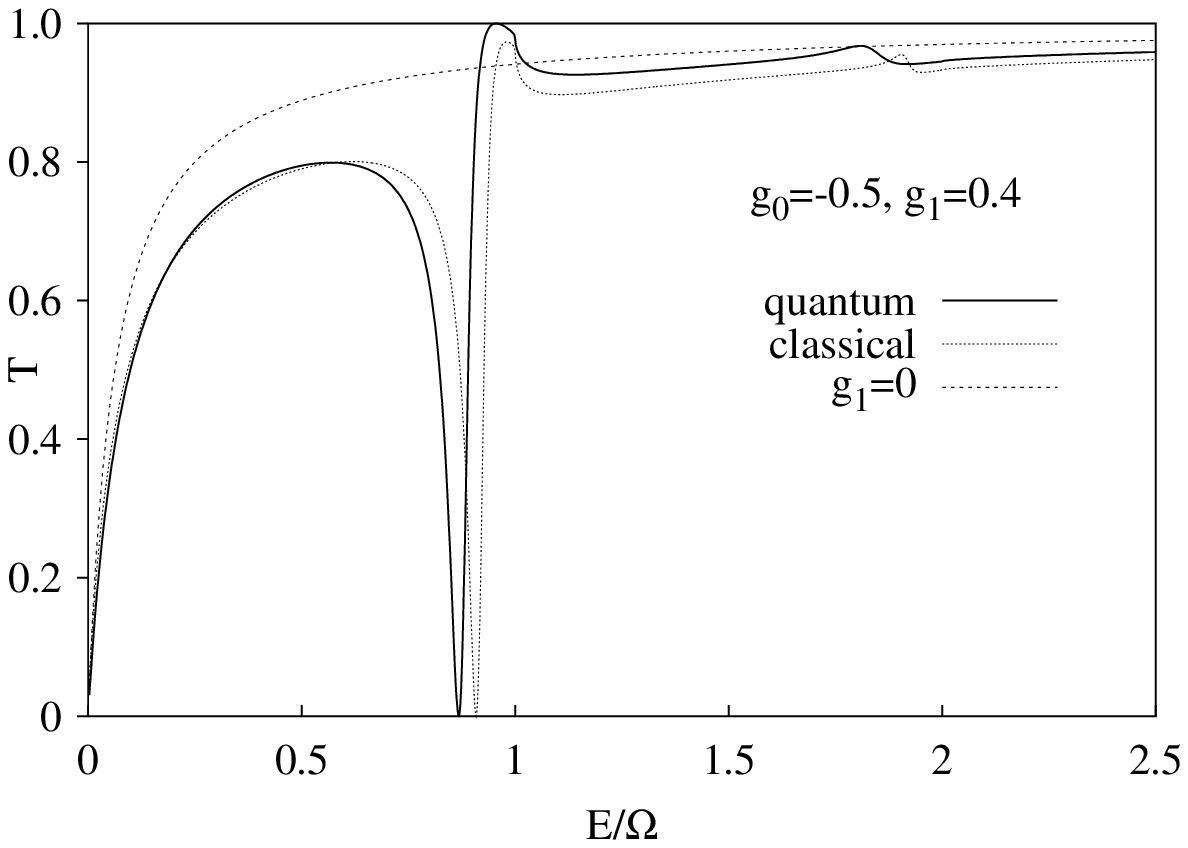}
\caption[]{\label{delta1.eps}Transmission coefficient through a dynamical 
one-dimensional delta barrier
with repulsive ($g_0>0$, {\bf left}) and attractive ($g_0<0$, {\bf right}) static part, cf. Eq. (\ref{Hamiltonian_delta}) and
(\ref{Hamiltonianc}). 
$E$ is the energy of the incident particle. From \cite{BR02}.}
\end{figure} 

A recursion relation corresponding to Eq. (\ref{recursion}) for the classical time-dependent
Hamiltonian, Eq. (\ref{Hamiltonianc}), was derived (and discussed) by Bagwell and Lake \cite{BL92} as
\begin{eqnarray}
  g_1 t_{n-1}+(g_0 -2i\gamma_n)t_n + g_1 t_{n+1}&=&-2i\gamma_n \delta_{n,0},\quad n=0,\pm 1,\pm 2,...,
\end{eqnarray}
where $t_n$ is the coefficient of the time-dependent electron wave function in photon side-band $n$, and $n$ runs through positive {\em and negative} integers $n$. In further contrast to the recursion relation Eq. (\ref{recursion}), the factors  $\sqrt{n}$ and $\sqrt{n+1}$ multiplying the coupling constant $g_1$ do not appear in the classical case. This latter fact is an important difference to the quantum case where these terms lead to the factors $N$ that multiply $g_1^2$ in the self energies $\Sigma^{(N)}(E)$, Eq. (\ref{selfenergy}), and  eventually to the breakdown of the perturbative approach for large $g_1$ in the quantum case. 
 
A continued fraction representation of $t_0(E)$ for the classical case was derived  by Martinez and Reichl \cite{MR01}, and the corresponding matrix defining the transmission amplitudes ${\bf t}_{\rm cl}=(...,t_{-2},t_{-1},t_0,t_1,t_2,...)$ is the infinite tridiagonal matrix $M_{\rm cl}$ with $ g_0-i\gamma_{n}$
on the diagonal and $g_1$ on the lower and upper diagonals,
\begin{eqnarray}\label{matrixcl}
M_{\rm cl}&=&\begin{array}{r}
    \left(
    \begin{array}{ccccc}
\ddots  & \ddots & 0 & &\\
      \ddots &  g_0-2i\gamma_{-1} & g_1 & 0 &    \\
   0  &   g_1 & g_0-2i\gamma_0 & g_1 & 0   \\
     &   0 & g_1 & g_0-2i\gamma_1 & \ddots   \\
      &    & 0  & \ddots & \ddots 
    \end{array}\right).
\end{array}
\end{eqnarray}
Following \cite{BR02}, Fig. (\ref{delta1.eps}) presents a comparison between the transmission coefficient $T(E)$, Eq. (\ref{transmission}), for the quantum  and the classical barrier. In the repulsive case with $0<g_1<g_0$, the dynamical part of the barrier  is only  a weak perturbation to the unperturbed $(g_1=0)$ case. Additional structures (cusps) appear at the boson (photo side-band) energies $n\Omega$ although the overall $T(E)$-curve resembles the $(g_1=0)$ case.
\begin{figure}[t]
\centering
\includegraphics[width=0.45\textwidth]{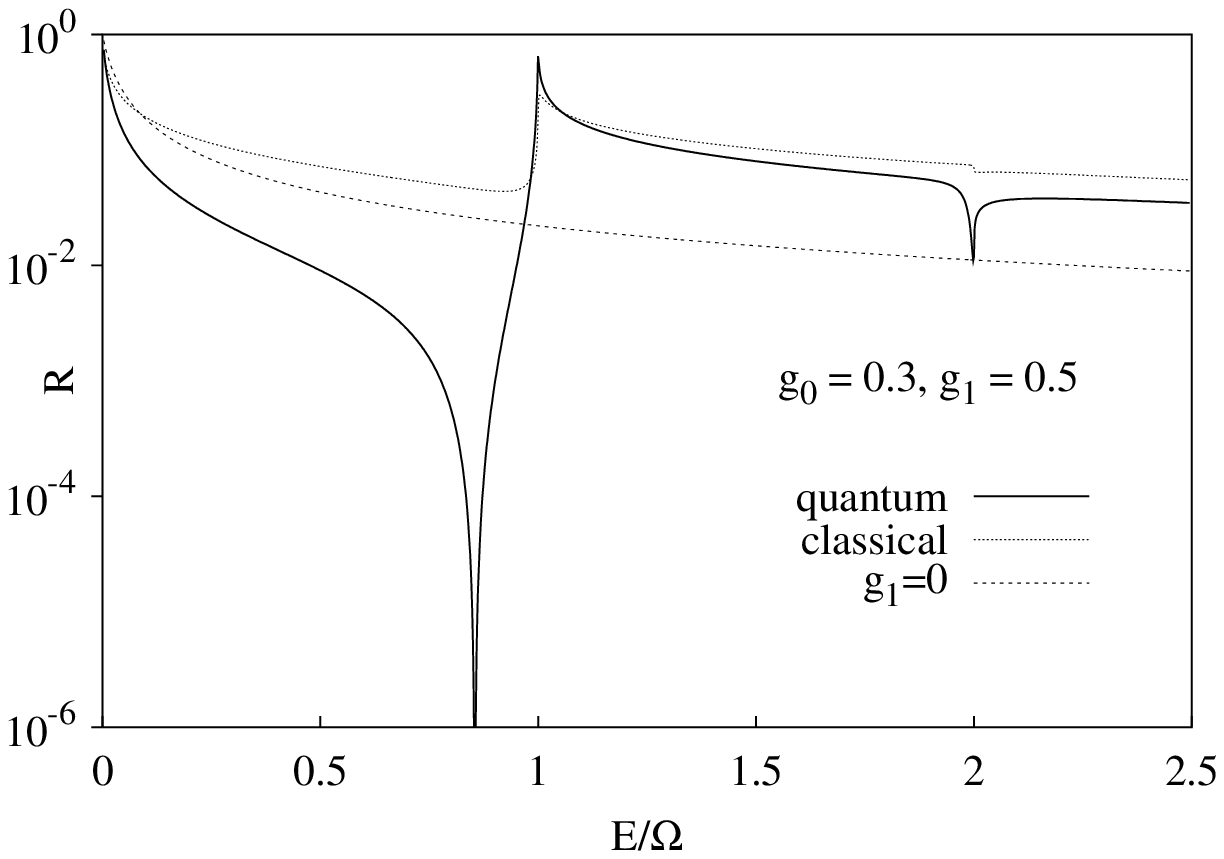}
\includegraphics[width=0.45\textwidth]{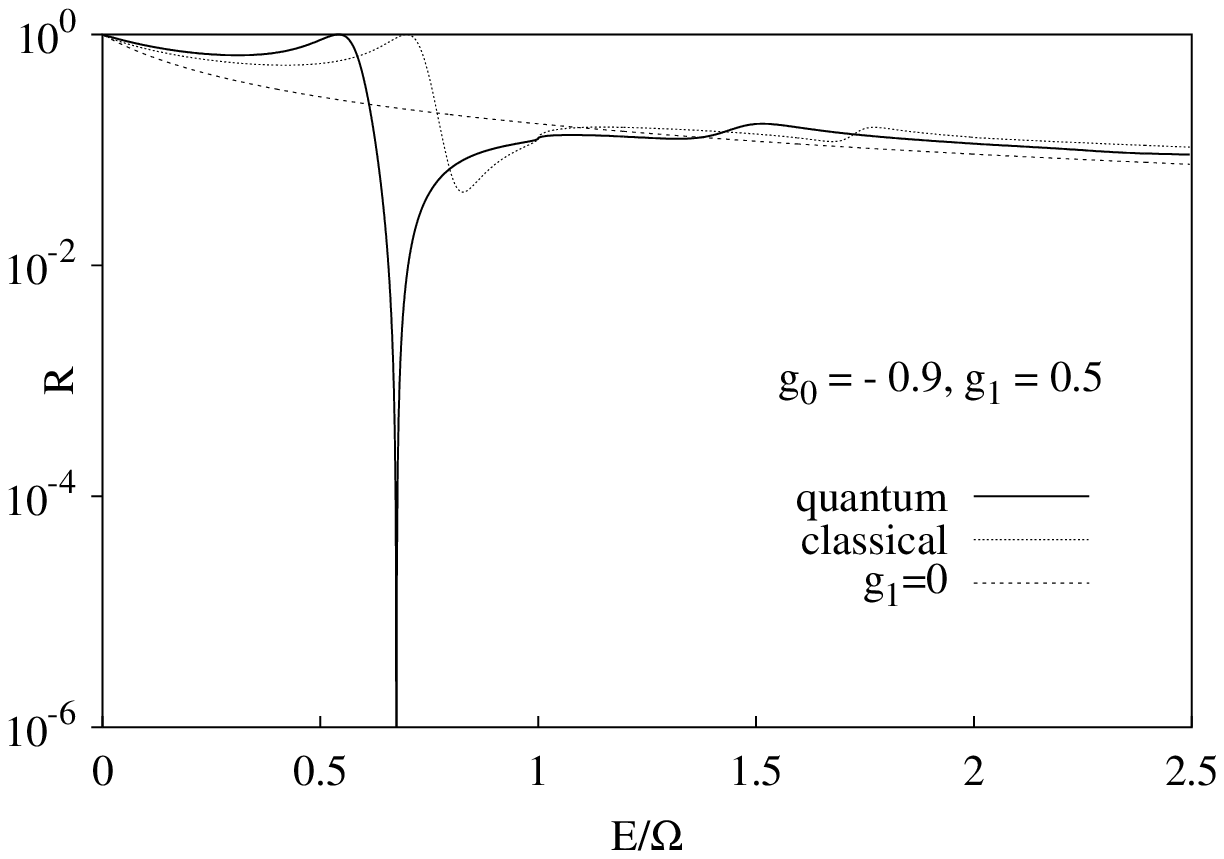}
\caption[]{\label{delta3.eps}Logarithmic plot of reflection coefficient $R\equiv 1-T$ for dynamical delta
barrier with static repulsive ($g_0>0$, {\bf left}) and attractive ($g_0<0$, {\bf right}) core.  From \cite{BR02}.}
\end{figure}

The more interesting case occurs for barriers with an attractive static part, $g_0<0$ (Fig. (\ref{delta1.eps}), right). A Fano type resonance appears below the first threshold $E=\Omega$ where the transmission coefficient has a zero in both the classical and the quantum case. In the classical case, this is a well-known phenomenon \cite{BL92}: the transmission zero  for weak coupling (small $g_1$) shows up when the Fano resonance condition
\begin{eqnarray}\label{Fano1}
 2\kappa_1(E)+g_0=0 
\end{eqnarray}
is fulfilled. There, the energy of the electron in the first side channel ($n=1$) coincides with the bound state of the attractive delta barrier potential, $E-\Omega = -g_0^2/4$. In the {\em quantum case}, the self energy in Eq.(\ref{selfenergy}) diverges at the zeros of $T(E)$, 
\begin{eqnarray}\label{Fano1a}
  [{\Sigma^{(1)}(E)}]^{-1}=0.
\end{eqnarray}
For $g_1\to 0$, ${\Sigma^{(1)}}(E)\to \Sigma^{(1)}_{N=2}(E)= g_1^2/(2\kappa_1(E)+g_0)$, cf. Eq.(\ref{t0approx}), and the two conditions Eq.(\ref{Fano1}) and Eq.(\ref{Fano1a}) coincide.

The most interesting feature in the scattering properties of the dynamical quantum barrier however is the appearance of an energy  close to the  first channel $(n=1)$ threshold where {\em perfect transmission} $T(E)=1$ occurs. This is clearly visible in the vanishing of the reflection coefficient, $R(E)\equiv 1-T(E)$, in the logarithmic plot Fig. (\ref{delta3.eps}). For a repulsive static part, $g_0=0.3$, this occurs at an energy below the energy where the reflection coefficient comes close to unity, and above that energy if the static part is attractive ($g_0=-0.9$). On the other hand, in the classical case the reflection coefficient never reaches zero in neither the repulsive nor the attractive case. This contrast  becomes even more obvious in the two-dimensional plot  where the zeros in $R$ correspond to `ridges' in the $g_0$-$E$ plane, cf. Fig. (\ref{delta4.eps}).

\begin{figure}[t]
\psfrag{E}{\hspace*{-8mm}$E/\Omega$}
\psfrag{g}{$g_0$}
\psfrag{20}{$20$}
\psfrag{1}{\hspace*{-1mm}$1$}
\psfrag{0.5}{$0.5$}
\psfrag{-1}{\hspace*{-3mm}$-1$}
\psfrag{0}{\hspace*{-1mm}$0$}
\includegraphics[width=0.95\textwidth]{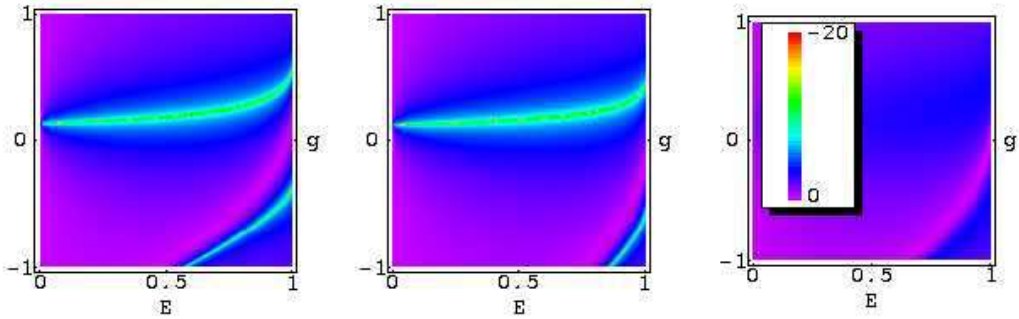}
\caption[]{\label{delta4.eps}Density plot of $\ln R$ (reflection coefficient) for the 
quantum delta barrier at $g_1=0.5$.  Exact solution from Eq. (\ref{selfenergy}) ({\bf left}),
from the $N=2$ truncation Eq. (\ref{t0approx}) ({\bf center}), and from the classical model Eq.(\ref{Hamiltonianc})
({\bf right}).
The light `ridges' correspond to curves of perfect transmission $T$, cf. Eqs. (\ref{perfecta}) and 
Eqs. (\ref{perfectb}). From \cite{BR02}.}
\end{figure} 
Perfect transparency ($R=1-T=0$) can be understood by considering the transmission amplitude $t_0(E)$ which determines the total transmission below the first side-band threshold. Recalling that $t_0(E)=-2ik_0/(-2ik_0+g_0-\Sigma^{(1)}(E))$, in the quantum case the transmission coefficient becomes unity when
\begin{eqnarray}\label{perfect}
  g_0-\Sigma^{(1)}(E)=0.
\end{eqnarray}
The exact continued fraction expression for the self energy, Eq.(\ref{sigmacontinued}), then implies that for $0<E<\Omega$, $\Sigma^{(1)}(E)$ is real because $G_0^{-1}(E-n\Omega)=2\sqrt{n\Omega-E}+g_0$ is real
for $n\ge 1$. The condition Eq.(\ref{perfect}) then means that the self energy  renormalizes the static part $g_0$ of the scattering potential to exactly zero.
  
This renormalization was analysed in \cite{BR02} for small $g_1$ with the perturbative expression corresponding to truncating the matrix $M$, Eq.(\ref{matrix}), to a two-by-two matrix. The perfect transparency condition Eq.(\ref{perfect}) then becomes 
\begin{eqnarray}
  \label{perfecta}
  g_0 - \frac{g_1^2}{2\kappa_1(E)+g_0}=0,\quad 0<E<\Omega,\quad (N=2 \mbox{ truncation.}),
\end{eqnarray}
which determines the position of the perfect transmission energy. The solution of the quadratic Eq.(\ref{perfecta}) defines two curves in the $E$-$g_0$-plane with perfect transmission for $0<E<\Omega$,
\begin{eqnarray}\label{perfectb}
  g_0=-\sqrt{\Omega-E}\pm \sqrt{\Omega-E+g_1^2},
\end{eqnarray}
which  can be clearly identified in the logarithmic density plots of the reflection coefficient $R=1-T$, cf. Fig. (\ref{delta4.eps}). The $N=2$ approximation to the transmission amplitude, Eq.(\ref{t0approx}), thus turns out to reproduce these features quite well even at moderate coupling constants $g_1$.

The above results are consistent with general properties of resonance line shapes in quasi-one-dimensional scattering as reviewed by N\"ockel and Stone \cite{NS94}. The boson mode in the Hamiltonian $H$, \ Eq.~(\ref{Hamiltonian_delta}), can be regarded as representing a simple harmonic oscillator confinement potential $V_{\rm osc}(y)$ in transversal direction $y$ of the quantum wire and thus giving rise to sub-band quantization of the transmission. The above truncation at $N=2$ corresponds to the two-channel approximation in the Feshbach approach \cite{NS94}. Furthermore, from this picture the difference between the transmissions in the  quantum and the classical (time-dependent) case, \ Eq.~(\ref{Hamiltonian_delta}){} and \ Eq.~(\ref{Hamiltonianc}), becomes clear: in the quantum case, one has inversion symmetry of the potentials $\delta(x)$ and $V_{\rm osc}(y)$ which was shown to imply that there are energies for which the transmission $T$ goes to zero {\em and} unity near the Fano resonance. In the classical case, this inversion symmetry is broken and the zero reflection point, $R=0$, $T=1$, is lost.

\subsection{Rabi Hamiltonian and Beyond: Transport Through Quantum Dots Coupled to Single Oscillator Modes}\label{section_boson}

One obtains a `transport version' of the Rabi Hamiltonian, \ Eq.~(\ref{H_Rabi}),  when one allows the particle (electron) number on the two-level atom  to fluctuate. This situation usually cannot be achieved in atomic physics unless one ionizes the atom. On the other hand, the restriction of fixed particle number  can easily be lifted, e.g.,  in the solid state by tunnel-coupling to particle reservoirs. The Cooper pair box or in fact the double-dot model (which formed a central part in section \ref{section_transport}) is therefore  a natural candidate for a `transport Rabi Hamiltonian'. Using the double-dot version, the Hamiltonian reads 
\begin{eqnarray}\label{H_Rabidot}
  {\mathcal H} &=& {\mathcal H}_{\rm dot} + {\mathcal H}_{\rm dp}
+ {\mathcal H}_V
+ {\mathcal H}_{\rm B} + {\mathcal H}_{\rm res},\quad 
{\mathcal H}_{\rm dot}=\varepsilon_L\hat{n}_L+\varepsilon_R\hat{n}_R+T_c(\hat{p}+\hat{p}^{\dagger})\nonumber\\
{\mathcal H}_{\rm dp}&=&\left(\alpha^L \hat{n}_L + \alpha^R \hat{n}_R 
+ \gamma \hat{p} + \gamma^*\hat{p}^{\dagger}\right)\left(a +a^{\dagger}\right),\quad {\mathcal H}_V=\sum_{k_i{},i=L/R}(V_k^i c_{k_i}^{\dagger}|0 \rangle \langle i|+H.c.)\nonumber\\
 {\mathcal H}_{\rm res}&=&\sum_{k_i,i=L/R}\epsilon_{k_i}
c_{k_i}^{\dagger}c_{k_i},\quad
 \quad {\mathcal H}_{\rm B} = \omega a^{\dagger} a,
\end{eqnarray}
which is the direct generalization of \ Eq.~(\ref{Htotal}) to a single boson mode $a^{\dagger}$ and was studied by Brandes and Lambert in \cite{BL03}. In contrast to the multi-mode boson version in section \ref{section_transport}, in the one-mode version \ Eq.~(\ref{H_Rabidot}) the boson degree of freedom  is not regarded as a dissipative bath, but treated on equal footing with the electronic degrees of freedom. 

The transport Master equation for the reduced density operator $\rho(t)$ of the systems (dot + boson) reads,
\begin{eqnarray}\label{master1}
\frac{d}{dt}\rho(t) &=& -i\left[{\mathcal H}_{\rm dot} + {\mathcal H}_{\rm dp}+ {\mathcal H}_{\rm B}
,\rho(t)\right]-\frac{\Gamma_L}{2}\left(s_L s_L^\dagger \rho(t) - 2s_L^\dagger
\rho(t)s_L + \rho(t)s_L s_L^\dagger\right)\nonumber\\
&-&\frac{\Gamma_R}{2}\left(s_R^\dagger s_R \rho(t) - 2s_R
\rho(t)s_R^\dagger + \rho(t)s_R^\dagger s_R\right)-
\frac{\gamma_b}{2}\left(2a \rho a^\dagger - a^\dagger a \rho - \rho
a^\dagger a\right),\nonumber
\end{eqnarray}
where $s_L = |0\rangle \langle L|$, $s_R = |0\rangle \langle R|$, and again only the three states `empty', `left', and `right' are involved in the description of the double-dot, where tunneling to the right and from the left electron reservoir in the infinite bias limit $\mu_L-\mu_R\to \infty$ occurs at rates $\Gamma_{L/R}$. 
 A further damping term of the  bosonic system at  rate $\gamma_b$ in \ Eq.~(\ref{master1}) describes photon or phonon cavity losses in Lindblad-form \cite{KMT97} and is crucial for the numerical stability in the stationary limit.

\begin{figure}[t]
\includegraphics[width=0.5\textwidth]{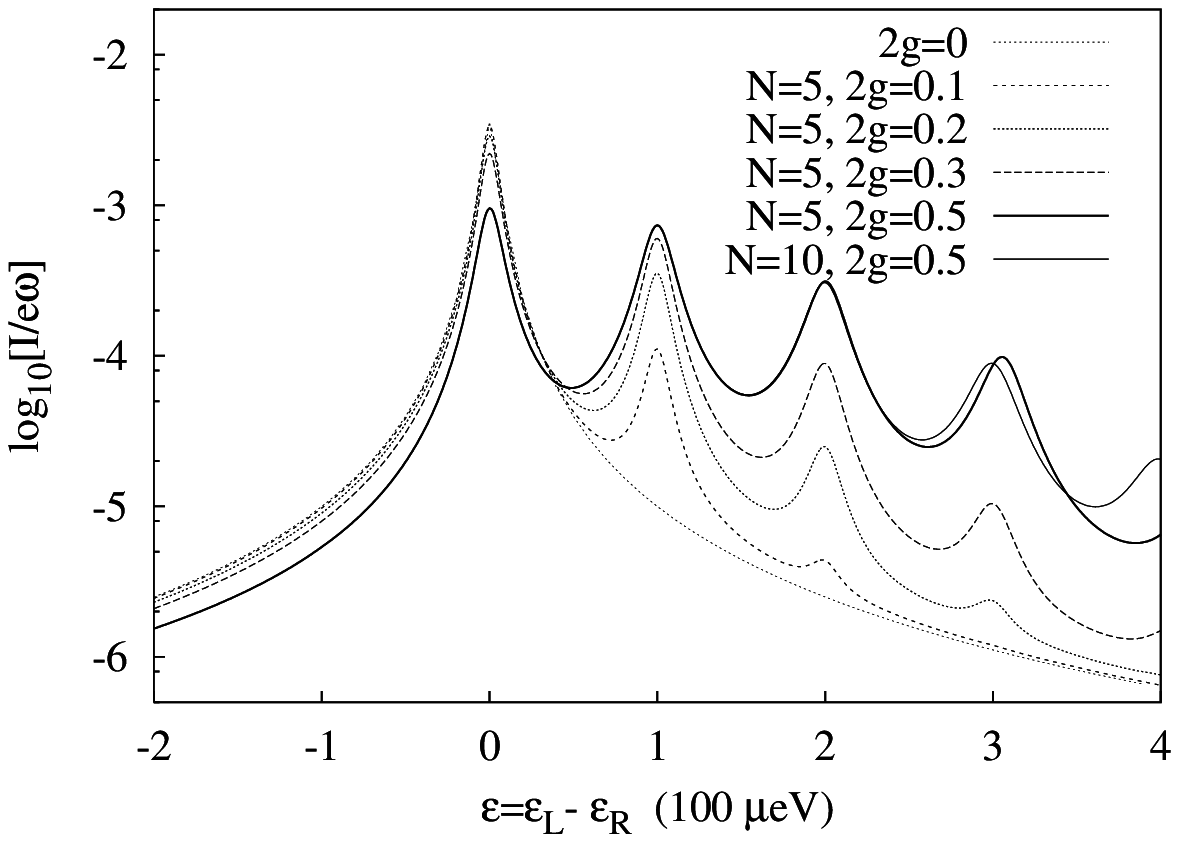}
\includegraphics[width=0.5\textwidth]{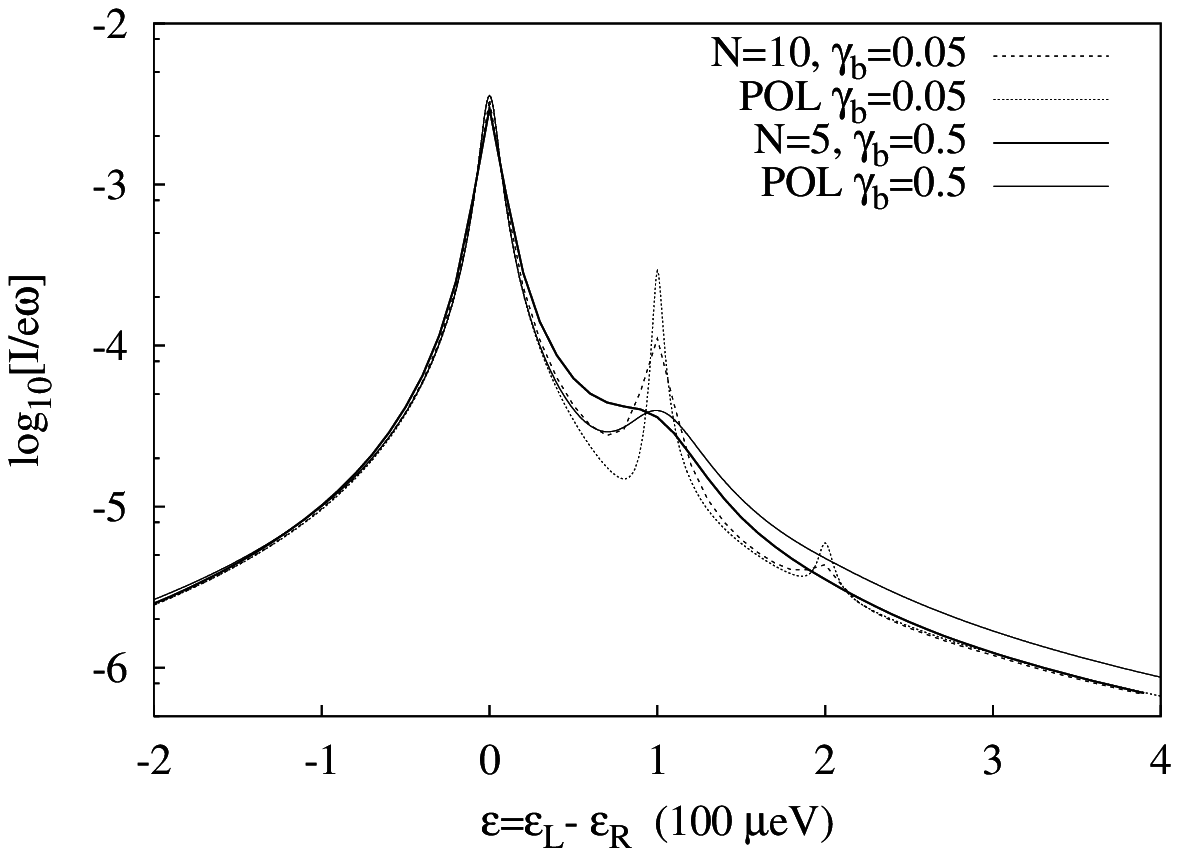}
\caption{{\bf Left:} Stationary current in `transport Rabi Hamiltonian' (double quantum dot coupled to  single boson mode) with $\Gamma_L=\Gamma_R=0.1$, $T_c=0.01$, $\gamma_b=0.05$ and boson coupling varying $2g$. $N$: number of boson states in truncated Hilbert space. {\bf Right:} Approximate (POL) and numerical results. From \cite{BL03}.}
\label{BL02_Fig1.eps}
\end{figure}

In order to numerically solve the system of linear equations resulting from taking matrix elements of Eq. (\ref{master1}) in the boson number state basis, the bosonic Hilbert space has to be truncated at a finite number $N$ of boson states which leaves the total number of equations at $5 N^2 + 10  N + 5$. The numerical solution becomes
a standard inversion of a fully occupied matrix and is easily achieved for $N$ up to 20 on a PC, whereas for larger $N$ more advanced methods like Arnoldi iteration in Krylov subspaces \cite{FNJ04} are more efficient.

\subsubsection{Stationary Current}

The stationary current, cf. \ Eq.~(\ref{ILandR}) and (\ref{ILR}), for various  boson couplings $\alpha^L=-\alpha^R=2g$, $\gamma=0$, is shown in Fig. \ref{BL02_Fig1.eps}, left. Resonances appear at multiples $\varepsilon\equiv \varepsilon_L-\varepsilon_R = n \omega>0$ similar to photo-assisted tunneling (cf. section \ref{section_AC}) , but in contrast to those only for positive $\varepsilon$ because on the absorption side of the profile ($\varepsilon<0$) the damped boson $(\gamma_b>0)$ relaxes  to its ground state. Analytical expression for the stationary current can be obtained when the polaron transformation method (POL) from section \ref{section_polaron} and the corresponding result for the current,\ Eq.~(\ref{IPOL_final}), is used together with an expression for the boson correlation function $C(t)$ in presence of damping. The latter can be calculated from a Master equation for a damped boson mode $\rho_{\rm B}(t)$,
\begin{eqnarray}
  \frac{d}{dt}\rho_{\rm B}(t) &=& -i[\omega a^{\dagger}a,\rho_{\rm B}] - \frac{\gamma_b}{2}\left( 2a\rho_{\rm B} a^{\dagger} -
a^{\dagger} a \rho_{\rm B} - \rho_{\rm B} a^{\dagger} a \right),
\end{eqnarray}
and leads to 
\begin{eqnarray}
  C(t) = \exp\left\{-|\xi|^2\left( 1- e^{-\left(\frac{\gamma_b}{2}+i\omega\right) t}\right)\right\},
\quad \xi=\frac{4g}{\omega}.
\end{eqnarray}
The analytical results compare quite well with the numerics for small coupling constants $g$ as shown in Fig. \ref{BL02_Fig1.eps}, right.

\subsubsection{Boson Distribution, Wigner Representation}\label{Wigner}
The stationary state of the boson mode is obtained from the total density matrix by tracing out the electronic degrees of freedom, $\rho_b \equiv \lim_{t\to \infty} {\rm Tr_{\rm dot}}\rho(t)$. With an electron current  flowing through the dot and interacting with the boson mode, this will not be a thermodynamic equilibrium state but a non-equilibrium state that is controllable by the system parameters, such as $\varepsilon$ and the tunnel rates $\Gamma_{L/R}$.  

For $\varepsilon\ll 0$, an approximate solution for $\rho_b$ is obtained by noting that the electron is predominantly localized in the left dot and one can approximate the operator $\sigma_z=| L\rangle \langle L| - |R\rangle \langle R|$ by its expectation value $\langle \sigma_z \rangle =1$ whence the boson system is effectively described by
\begin{eqnarray}
  H_{\rm eff} = 2g(a+a^{\dagger}) + \omega a^{\dagger}a,
\end{eqnarray}
a shifted harmonic oscillator with ground state  $|GS\rangle =|-2g/\omega\rangle$ ($|z\rangle$ denotes a coherent state, $a|z\rangle = z |z\rangle$), which can easily be seen by introducing new operators $b\equiv a+2g/\omega$ whence $H_{\rm eff} =  \omega b^{\dagger}b-4g^2/\omega$ 
and $b|GS\rangle = 0$. It follows that $ \rho_b  \approx |z\rangle \langle z|,\quad z=-2g/\omega$ and that the occupation probability $p_n\equiv (\rho_b)_{nn}$ is given by a Poisson distribution, $p_n =|\langle n|GS\rangle|^2 |z|^{2n}e^{-|z|^2}/n!$, which is well confirmed by numerical results \cite{BL03}.

In the general case of  arbitrary $\varepsilon$, one has to obtain $\rho_b$ numerically. A useful way to represent the boson state is the Wigner representation, which is a representation in position ($x$) and momentum ($p$) space of the harmonic oscillator, where 
\begin{eqnarray}\label{xpdef}
  x=\frac{(a+a^\dagger)}{\sqrt{2}},\quad p=\frac{i(-a +
a^\dagger)}{\sqrt{2}}
\end{eqnarray}
and the Wigner function is defined as \cite{CG69}
\begin{eqnarray}
  W(x,p)\equiv \frac{1}{\pi}{\rm Tr}\left(\rho_b D(2\alpha) U_0 \right),  \quad \alpha=\frac{x+ip}{\sqrt{2}},
\end{eqnarray}
where $D(\alpha)\equiv \exp[\alpha a^{\dagger} - \alpha^* a]$ is a unitary displacement operator and $U_0\equiv \exp[i\pi a^{\dagger} a]$ is the parity operator for the boson \cite{CV98}.
$W(x,p)$ is known to be a symmetric Gaussian for a pure coherent boson state, and a symmetric Gaussian multiplied with a polynomial for a  pure number state \cite{Walls}. Using the number state basis $\{|n\rangle \}$ and the matrix elements ($ m\ge n$),
\begin{eqnarray}\label{Laguerre}
\langle m | D(\alpha) | n\rangle &=&
\langle m| D^{\dagger}(\alpha)|n\rangle^*
=(-1)^{m-n} \langle m| D(\alpha)|n\rangle^* \\
&=& \sqrt{\frac{n!}{m!}}\alpha^{m-n}e^{-\frac{1}{2}|\alpha|^2}L_{n}^{m-n}\left(|\alpha|^2\right),
\end{eqnarray}
where $L_{n}^{m-n}$ is a Laguerre polynomial and $\alpha=({x+ip})/{\sqrt{2}}$, one obtains $W(x,p)$ directly from the matrix elements $\langle n|\rho_b| m\rangle$. 
\begin{figure}[t]
\begin{center}
\psfrag{eps}{$\varepsilon$}
\psfrag{=}{$=$}
\psfrag{0.00}{ $0.00$}
\psfrag{0.25}{ $0.25$}
\psfrag{0.50}{ $0.50$}
\psfrag{0.75}{ $0.75$}
\psfrag{1.00}{ $1.00$}
\psfrag{1.25}{ $1.25$}
\psfrag{1.50}{ $1.50$}
\psfrag{1.75}{ $1.75$}
\psfrag{2.00}{ $2.00$}
\psfrag{2.25}{ $2.25$}
\includegraphics[width=0.75\textwidth]{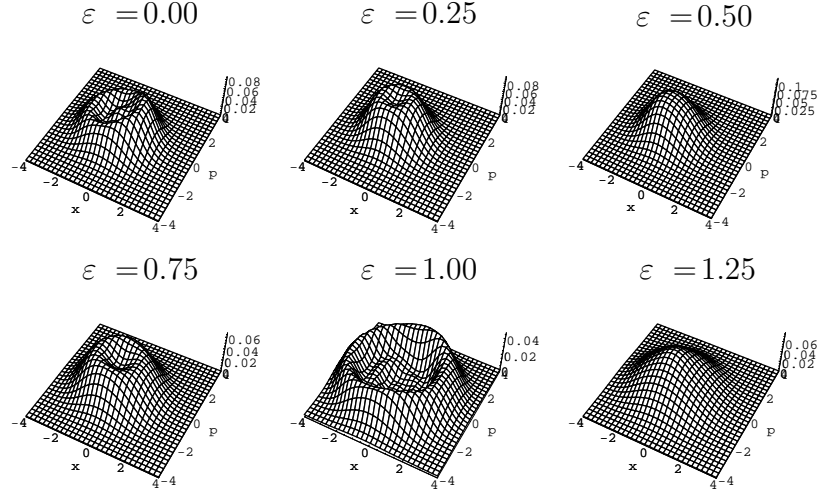}
\caption[]{\label{BL02_Fig6.eps}Wigner distribution functions for the single bosonic mode in the `transport Rabi model'. Parameters are $\Gamma_L=\Gamma_R=T_c=0.1$, $\gamma_b =0.005$, $g=0.2$, $N=20$. Stationary current resonances occur at $\varepsilon=0.0, 1.0, 2.0,...$. From \cite{BL03}.}
\end{center}
\end{figure}
As shown in Fig. (\ref{BL02_Fig6.eps}), $W(x,p)$ closely resembles a Gaussian between two resonance energies $\varepsilon=n\omega$, whereas close  to the resonance energies, the distribution spreads out in rings around the origin, which is  consistent with the increased Fock state occupation numbers. Additional calculations \cite{BL03} show that the  position and momentum variances also increase at these energies. The resonances at $\varepsilon=n\omega>0$ correspond to the emission of bosons by the electron as it tunnels through the dot.

\subsection{Non-linear Couplings, Nano-Electromechanics,  and Shuttle Effects}\label{section_shuttle}
Single oscillators play a central role in the  emerging field of nano-electromechanics, where vibrational (mechanical) and electronic degrees of freedom are strongly coupled to each other, leading to novel transport regimes. Single electron shuttling was introduced by Gorelik and co-workers \cite{Goretal98} as a mechanism to transfer charge  by a cyclic loading and unloading of a metallic grain oscillating between two electrodes. Weiss and Zwerger \cite{WZ99} used a Master equation  in order to combined the Coulomb blockade effect with shuttling, a method used later by Erbe, Weiss, Zwerger and Blick \cite{AWZB01} to compare with experimental data in a `quantum bell'. 

One important and novel ingredient in quantum shuttles is a non-linear dependence of the matrix element for electron tunneling on the oscillator coordinate $x$. Armour and MacKinnon \cite{AM02} introduced a {\em three-dot} model with a central dot oscillating between two other dots that are connected to external electron reservoirs. 
A generic Hamiltonian for a {\em one-dot} nano-mechanical single electron transistor, used by several groups, combines a single `resonator' (oscillator) mode with the resonant level model (no electron spin included), 
\begin{eqnarray}\label{H_shuttle}
   {\mathcal H} &=& {\mathcal H}_{\rm dot} + {\mathcal H}_{\rm osc}
+ {\mathcal H}_V
+ {\mathcal H}_{\rm B} + {\mathcal H}_{\rm res} \nonumber\\
{\mathcal H}_{\rm dot}&\equiv& (\varepsilon_0-eE x) c^{\dagger}c,\quad {\mathcal H}_{\rm osc}
=\frac{p^2}{2m}+\frac{m\omega^2x^2}{2},\quad 
{\mathcal H}_V=\sum_{k_i{},i=L/R}(V_{k_i}(x) c_{k_i}^{\dagger}c + H.c.),
\end{eqnarray}
where ${\mathcal H}_{\rm res}=\sum_{k_i,i=L/R}\epsilon_{k_i} c_{k_i}^{\dagger}c_{k_i}$ describes leads on the left and right side, ${\mathcal H}_{\rm B}$ is a dissipative bath coupled to the oscillator, $E$ is the inner electric field, and the $x$-dependence of the left and right tunnel matrix element $V_{k_i}(x)$ is assumed to be exponential, 
\begin{eqnarray}
  V_{k_L}(x)= V_{k_L} e^{-x/\lambda},\quad  V_{k_R}(x)= V_{k_R} e^{x/\lambda}.
\end{eqnarray}
The analysis of the Hamiltonian, Eq.~(\ref{H_shuttle}), is complicated by the fact that it contains a number of length  and energy scales: $\lambda$ is the electron tunneling length, $x_0\equiv\sqrt{\hbar/m\omega}$ is the amplitude of oscillator zero-point fluctuations, and $d\equiv eE/m\omega^2$ is the ponderomotive shift of the oscillator by the field $E$. Furthermore, $V_{\rm sd}\equiv \mu_L-\mu_R$ gives the source-drain bias between left and right reservoir, $\hbar \omega$ the oscillator energy, and $\Gamma_{L,R}$ are bare left and  right tunnel rates derived from ${\mathcal H}_V$. In addition, the bath $ {\mathcal H}_{\rm B}$ introduces a damping rate $\gamma$ and temperature $T$ (which in principle can differ from the temperature of the leads).

\begin{figure}[t]
\begin{center}
\includegraphics[width=0.7\textwidth]{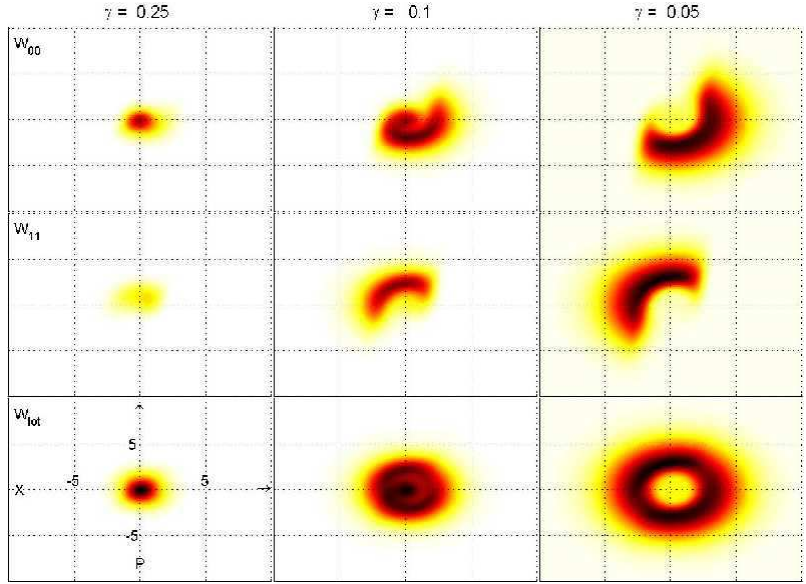}
\caption[]{\label{NDJ_Fig1.eps}Wigner distributions  in the phase-space analysis of the  `quantum shuttle' (single electron plus resonator) model \ Eq.~(\ref{H_shuttle}) by Novotn\'{y}  and co-workers \cite{NDJ04}, showing the transition from tunneling (strong damping $\gamma$) to shuttling (small $\gamma$). The latter regime is indicated by the half-moon shapes of the charge-resolved $W_{00}$ (upper row describing an empty level when the oscillator goes from right to left) and  $W_{11}$ (middle row, describing an electron shuttled from left to right), whereas $W_{\rm tot}=W_{00}+W_{11}$ (lower row) corresponds to the total oscillator state. From  \cite{NDJ04}.}
\end{center}
\end{figure} 

In transport regimes where single oscillator modes are of primary importance, methods from Quantum Optics like phase space representations and  Master equations are obviously relevant theoretical tools. 
Novotn\'{y}, Donarini, and Jauho \cite{NDJ04} used the numerical solution of a Master equation corresponding to \ Eq.~(\ref{H_shuttle}) in a truncated oscillator basis and in the limit $V_{sd}\to \infty$, including oscillator damping in Lindblad-form at rate $\gamma$. They used a Wigner function representation (cf. section \ref{Wigner}) for the discharged and charged oscillator states in order to clearly identify a tunneling-to-shuttling crossover that occurred when tuning from strong to weak damping $\gamma$ in a `quantum regime' defined by $\lambda\sim x_0$, cf. Fig. (\ref{NDJ_Fig1.eps}). This crossover was further analysed by a calculation of zero-frequency shot noise in a subsequent paper \cite{NDFJ04}. 

Fedorets, Gorelik, Shekter, and Jonson \cite{FGSJ04} studied the regime $\lambda \gg x_0$  in an analytical treatment of two coupled equations of motion for the Wigner functions $W_{\pm}(x,p)$ corresponding to the sum and difference of the `empty dot' and 'occupied dot' density matrix elements. Using polar coordinate in  phase space, $x = A\sin \varphi$ and $p=A \cos \varphi$, they found a stationary solution $W_{+}(A)$ for the oscillator state that reflected the instability towards shuttling when the dissipation was weak enough, which was consistent with the numerical results for $W_{\rm tot}$ in \cite{NDJ04}, cf. Fig. (\ref{NDJ_Fig1.eps}).  In their analysis, they furthermore distinguished between a classical regime for large fields, $E\gg E_q$, and a quantum regime for fields $E\ll E_q$ below a certain field $E_q$. 

Armour, Blencowe, and Zhang \cite{ABZ04} on the other hand used a  Master equation by essentially treating the bosonic mode as a classical harmonic oscillator in the regime of large source-drain voltage, $V_{\rm sd}\gg k_BT, \hbar \omega$. Their description involved Poisson-brackets similar to the `mixed quantum classical ensembles' used by Kantorovich \cite{Kan02}. They found an effective temperature and intrinsic damping caused by the action of the tunneling electrons on the resonator, similar to Mozyrsky and Martin \cite{MM03} who derived an effective friction coefficient by comparison with the Caldeira-Leggett model.

\subsection{Superconducting Cavity-QED Experiments}\label{cavity_experiment}
The Yale group successfully demonstrated the coherent coupling between single photons and a superconducting Cooper-pair box in experiments by Wallraff {\em et al.} \cite{Waletal04} and Schuster {\em et al.} \cite{Schuetal04}. They fabricated  the two-junction Cooper-pair box onto a silicon chip between the walls of a quasi-one-dimensional `on chip'  wave-guide resonator (transmission line cavity for the photons). Blais, Huang, Wallraff, Girvin, and Schoelkopf \cite{Blaetal04} described the Cooper-pair box in the two-level charge regime limit by the usual two-level Hamiltonian,
\begin{eqnarray}
  H_{\rm CPB} \equiv -\frac{\varepsilon}{2}\bar{\sigma}_z - \frac{E_J\cos (\pi \Phi/\Phi_0)}{2}\bar{\sigma}_x,\quad \varepsilon\equiv 4 E_C(1-2n_g),
\end{eqnarray}
with Pauli matrices in the basis of the island-eigenstates with $N$ and $N+1$ Cooper-pairs, $n_g$ the dimensionless, voltage-tunable  polarization charge, and $E_J\cos (\pi \Phi/\Phi_0)$ the flux $(\Phi)$-tunable Josephson energy. At the charge degeneracy point, $n_g=1/2$, they showed that their system (without dissipation) could be described by the Rabi Hamiltonian, \ Eq.~(\ref{H_Rabi}),  which they approximated by the Jaynes-Cummings model in the rotating wave approximation (RWA), 
\begin{eqnarray}\label{H_JC}
  H_{\rm JC} \equiv \hbar \omega_r\left(a^{\dagger}a+\frac{1}{2}\right)
+\frac{\hbar\Omega}{2}\sigma_z+\hbar g \left(a^{\dagger}\sigma_-+a\sigma_+\right),
\end{eqnarray}
where $\omega_r$ is the resonator angular frequency, $g$ the coupling constant that determines the Rabi frequency $\nu_{\rm Rabi}=g/\pi$, and $\Omega=E_J\cos (\pi \Phi/\Phi_0)$ the energy spitting in the basis of the eigenstates of $H_{\rm CPB}$. 

In the experiments \cite{Waletal04} at low temperatures $T<100$ mK with photon occupations $n\equiv \langle a^{\dagger}a\rangle<0.06$, the frequency-dependent transmission spectrum of a probe beam through the coupled resonator clearly showed two peaks split by the Rabi frequency $\nu_{\rm Rabi}\approx 11.6$ MHz, as expected from \ Eq.~(\ref{H_JC})  at resonance $\Delta\equiv \Omega-\omega_r=0$ with $\nu_r=\omega_r/2\pi = 6.04$ GHz. In this regime, $g\gg \kappa,\gamma$ was strong enough to treat the cavity and the qubit decay (at rates $\kappa$ and $\gamma$, respectively) perturbatively, and weak enough $\nu_{\rm Rabi}/\nu_r\ll 1$ to use the Jaynes-Cummings instead of the Rabi Hamiltonian. 

Another interesting case was tested for the `dispersive regime' of large detuning $\Delta$ with $g/\Delta\ll 1$, where the Hamiltonian \ Eq.~(\ref{H_JC}) to second order in $g$ can be  approximated through a unitary transformation $U=\exp\left[(g/\Delta)(a\sigma^+ - a^{\dagger} \sigma_-)\right]$ as
\begin{eqnarray}\label{H_U}
  H_U = \hbar \left[\omega_r  +\frac{g^2}{\Delta}\sigma_z\right]a^{\dagger}a
+\frac{\hbar}{2}\left[ \Omega + \frac{g^2}{\Delta}\right] \sigma_z.
\end{eqnarray}
The frequency shift $\pm g^2/\Delta$ could be identified in the phase shifts of a transmitted microwave at a fixed frequency by tuning the gate charge $n_g$ and the flux $\Phi$. In a second experiment \cite{Schuetal04}, Schuster {\em et al.} verified the ac-Stark shift (term $\pm n g^2/\Delta$ in \ Eq.~(\ref{H_U})) by measuring the qubit level separation as a function of the microwave power and thereby the photon number $n$.

The coupling of a single harmonic oscillator and a superconducting qubit was furthermore achieved by the Delft group in experiments by Chiorescu {\em et al.} \cite{Chietal04}, who integrated a flux qubit into a larger SQUID, the latter providing the oscillator mode. Their system had an interaction term $\lambda\sigma_z(a^{\dagger}+a)$, leading to Rabi-oscillations in the SQUID switching when microwave pulses were applied. Goorden, Thorwart and Grifoni gave an analysis of the related driven two-level system with coupling to a detector and a dissipative environment \cite{GTG04}.


\section{\bf Dark States and Adiabatic Control in Electron Transport}\label{section_dark}
One of the most remarkable features of quantum systems is the possibility to modify their physical properties by creating quantum superpositions (linear combinations of states). The simplest and most basic quantum system where that is possible is the two-level system,  which is central to so many areas of physics.  It plays a major role in the modeling of light-matter interactions, as is for example reviewed in the classical textbook by Allen and Eberly \cite{Allen} on `Optical Resonance and Two-Level Atoms'. Shortly after that book was published, the discovery of dark states in {\em three-level systems} sparked an enormous amount of activities, leading to the establishment of several new branches of Quantum Optics and Laser Spectroscopy, such as laser cooling and adiabatic population transfer.

This section reviews recent theoretical activities on dark resonance effects  and the associated adiabatic transfer schemes in the solid state. After a short introduction to coherent population trapping,  dark resonances and their use for control of electron transport are discussed, before we come back to two-level systems (qubits) in the context of  dissipative adiabatic transfer.

\subsection{Coherent Population Trapping (CPT)}
The first observation of dark states by Alzetta, Gozzini, Moi, and Orriols \cite{Alzetal76} in 1976 occurred in the form of a black line across the fluorescence path of a multi-mode dye laser beam through a sodium vapor cell. The three-level system there consisted of the two Zeeman-splitted $3^2S_{1/2}$ ground state hyper-fine levels, coupled by simultaneous application of two near-resonant monochromatic radiation fields to an excited $3^2P_{1/2}$ state.  A magnetic field $H$ with a spatial gradient then matched the ground-state level splitting $\delta\nu(H)$ with a frequency difference between  two laser modes at the position of the black line. The theoretical treatment in the same year by Arimondo and Orriols \cite{AO76}, and (independently) Whitley and Stroud \cite{WS76}, laid the foundations for explaining  the trapping of dissipative, driven three-level systems in a superposition of the two splitted ground-states which is  decoupled from the light and therefore `dark'. 

\subsubsection{Coherent Population Trapping Model}
Dark states, coherent population trapping, and related phenomena in Quantum Optics are reviewed by Arimondo 
in \cite{Arimondo96}. The basic physical effect is quite simple and can be explained in a model
of three states $|0\rangle$, $|1\rangle$,  $|2\rangle$ 
driven by two classical, monochromatic fields 
\begin{eqnarray}\label{fielddef}
{\bf E}_j(t)=\calEv_j\cos(\omega_j t+\varphi_j), \quad j=1,2,   
\end{eqnarray}
with angular frequencies $\omega_i$ and phases $\varphi_i$, cf.
Fig. (\ref{threelevel.eps}).
In the $\Lambda$-configuration, by convention $\varepsilon_1,\varepsilon_2 < \varepsilon_0$, although
other notations are used in the literature as well. 
The two fields couple to the two transitions $|0\rangle \leftrightarrow |1\rangle$ and
$|0\rangle \leftrightarrow |2\rangle$ and are {\em detuned} off the two excitation energies
by $\hbar\delta_1\equiv\varepsilon_0-\varepsilon_1-\hbar\omega_1$
and $\hbar\delta_2\equiv\varepsilon_0-\varepsilon_2-\hbar\omega_2$. The Hamiltonian in dipole approximation 
with coupling to dipole moment operators ${\dv}_i$ is 
\begin{eqnarray}\label{H3levelnonRWA}
H(t)&=&H_0-\sum_{j=0}^2{\dv}_j {\bf E}_j(t),\quad  
H_0\equiv\sum_{j=0}^2\varepsilon_j |j\rangle \langle j|,
\end{eqnarray}
which is often replaced by  adopting the  rotating wave approximation (RWA) 
by neglecting counter-rotating terms in Eq.~(\ref{H3levelnonRWA}),
\begin{eqnarray}\label{H3level}
H_{\rm RWA}(t)&=&H_0+H_I(t),\quad  
H_I(t)=-\sum_{j=1,2}\frac{\hbar\Omega_j}{2}e^{-i(\omega_j t+\varphi_j)}|0\rangle\langle j|
+ h.c.,
\end{eqnarray}
where the  {\em Rabi frequencies} 
\begin{eqnarray}
\Omega_j\equiv \frac{1}{\hbar}\langle 0| \calEv_j{\dv}_j |j\rangle,\quad  (j=1,2) 
\end{eqnarray}
define the coupling strength to the electric field.

\subsubsection{Dark States}
\begin{figure}[t]
\includegraphics[width=0.5\textwidth]{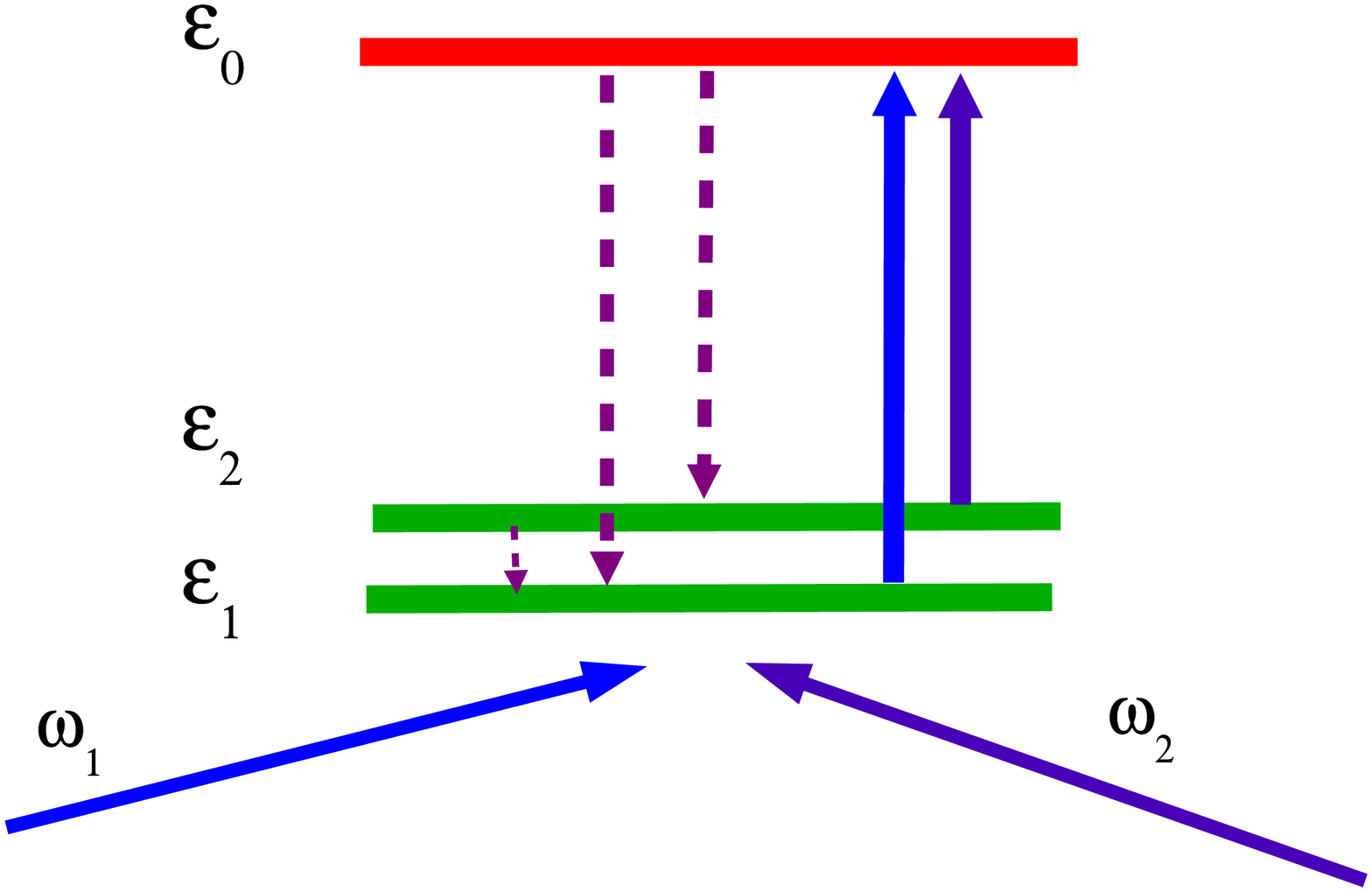}
\includegraphics[width=0.5\textwidth]{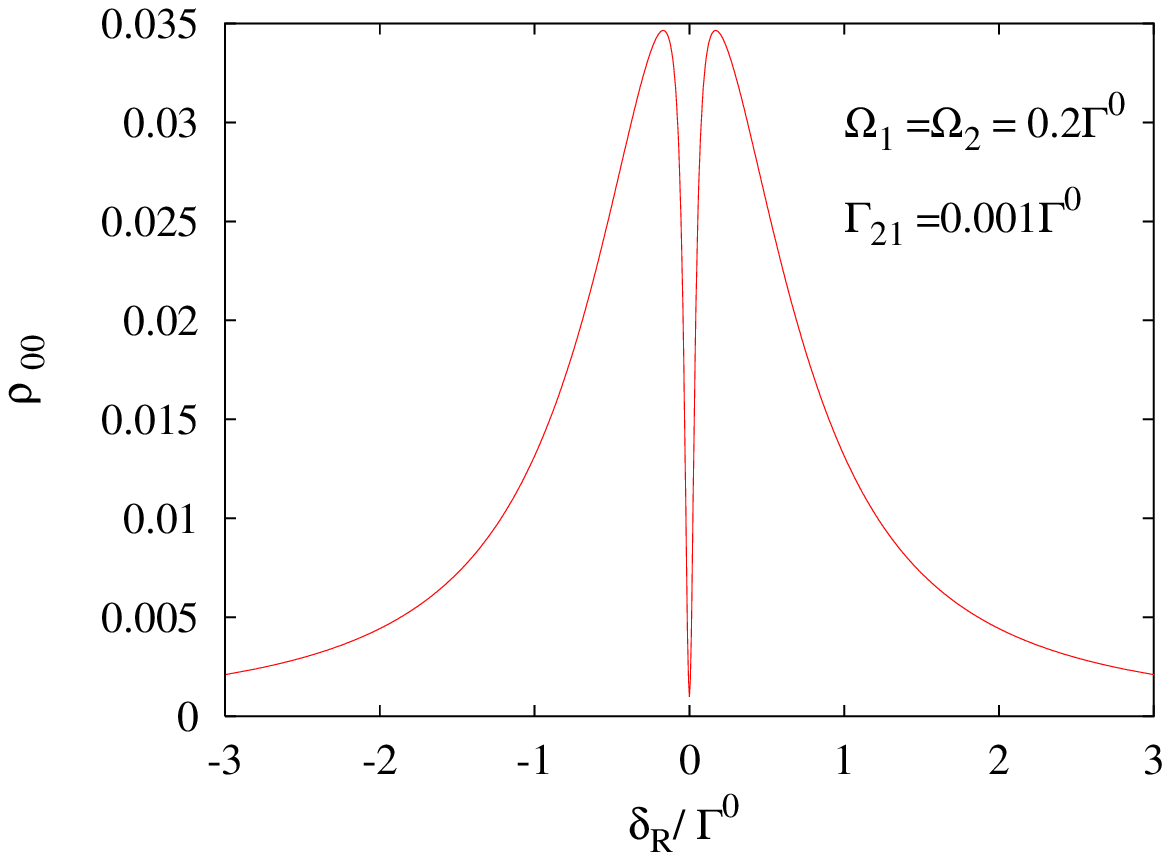}
\caption[]{\label{threelevel.eps} {\bf Left:} Three-level system under irradiation. Dashed lines indicate decay due to spontaneous emission of photons. {\bf Right:} Stationary occupation of the upper level $|0\rangle$. $\Omega_1$ and $\Omega_2$ denote the Rabi frequencies corresponding to both radiation fields, $\Gamma^0$ is the decay rate of the upper level, $\Gamma_{21}=2\gamma_p$ is the decay rate of level $|2\rangle$. From \cite{Bra00}.}
\end{figure} 

The dark state is obtained by 
a simple rotation of the basis triple $\{ |0\rangle,|1\rangle,|2\rangle \}$ 
within the ${\rm span}\{(|1\rangle,|2\rangle\}$-subspace into ($\hbar=1$)
\begin{eqnarray}
  |0\rangle &,&\quad |NC\rangle(t) \equiv \cos \theta |1\rangle - e^{i\phi(t)} 
\sin \theta |2\rangle,\quad
|C\rangle(t) \equiv \sin \theta |1\rangle + e^{i\phi(t)} \cos \theta |2\rangle,\\
\phi(t)&\equiv& (\omega_1-\omega_2)t+\varphi_2 -\varphi_1,\quad 
\cos \theta \equiv \Omega_2/\sqrt{\Omega_1^2+\Omega_2^2},\quad
\sin \theta \equiv \Omega_1/\sqrt{\Omega_1^2+\Omega_2^2}.
\end{eqnarray}
In the interaction picture with respect to $H_0$ (denoted by $\tilde{}$ as usual),
the time-dependence of $\tilde{|NC\rangle}(t)$ is governed by 
$\tilde{H}_I(t)=-\frac{1}{2}\sum_i\Omega_ie^{i(\delta_it-\varphi_i)}|0\rangle\langle i| + h.c.$, and a simple
calculation yields
\begin{eqnarray}
  \delta_R\equiv \delta_2-\delta_1= \varepsilon_1+\omega_1-\varepsilon_2-\omega_2=0
\rightarrow \tilde{H}_I(t)\tilde{|NC\rangle}(t)=0,
\end{eqnarray}
which means that at {\em Raman resonance} $\delta_R=0$ the {\em dark state} (non-coupling state) 
$|NC\rangle(t)$ completely `decouples from the light', i.e. once the system is in the (time-dependent) superposition  $|NC\rangle(t)$, it can no longer be excited into the state $|0\rangle$. 
In contrast, the 
coupled state $|C\rangle(t)$, which is orthogonal to $|NC\rangle(t)$,  can be excited and couples to the light. 
Note that for $\delta_R=0$, in the interaction picture  $\tilde{|NC\rangle}(t)=e^{-i\varepsilon_1t}
[cos \theta |1\rangle - e^{i(\varphi_2-\varphi_1)} \sin \theta |2\rangle]$
is in fact time-independent apart from the trivial phase factor $e^{-i\varepsilon_1t}$.

\subsubsection{Dissipation, RWA,  and Dark Resonances}
The presence of a dissipative environment requires a description of CPT in terms of a density operator rather than in terms of pure states. In fact, dissipation plays a central role in achieving a stable trapping of the three-level system into the dark state: spontaneous decay from the upper level ($|0\rangle$ in the $\Lambda$ configuration) into the two lower levels $|1\rangle$ and $|2\rangle$ leads to a re-shuffling of the level occupancies. The continuous pumping of electrons  into $|0\rangle$ is only from the coupled state $|C\rangle$, whereas spontaneous emission leads to transitions into  {\em both} the coupled state $|C\rangle$ and the dark state $|NC\rangle$. The combination of this one-sided pumping with the two-fold decay  drives the system into the dark state as a stationary `dead-end'. Dissipation in the form of spontaneous decay, in combination with the time-dependent pumping by the two external fields, is the driving force for CPT to evolve but also leads to decoherence of the $|NC\rangle$ superposition within the $|1\rangle$-$|2\rangle$ subspace. A quantitative analysis starts from the stationary solution of the Master equation for the matrix elements $\rho_{ij}(t)$ of the reduced density operator of the three-level system,
\begin{eqnarray}
    \dot{\rho}_{00} &=& -\Gamma^0\,\rho_{00}+i{\Omega_1}\cos(\omega_1t)\,\rho_{01}+i{\Omega_2}\cos(\omega_2t)\,\rho_{02}-i{\Omega_1}\cos(\omega_1t)\,\rho_{10}-i{\Omega_2}\cos(\omega_2t)\,\rho_{20}\\
    \dot{\rho}_{11} &=& \alpha_1{\Gamma^0}\,\rho_{00} + 2\gamma_p{\rho}_{22}
-i{\Omega_1}\cos(\omega_1t)\,\rho_{01}+i\Omega_1\cos(\omega_1t)\,\rho_{10} \\
    \dot{\rho}_{22} &=& \alpha_2{\Gamma^0}\,\rho_{00} - 2\gamma_p{\rho}_{22} -i{\Omega_2}\cos(\omega_2t)\,\rho_{02}+ i\Omega_2\cos(\omega_2t)\,\rho_{20} \\
    \dot{\rho}_{01} &=& -\left({\Gamma^0}/{2}+i\omega_{01}\right)\,\rho_{01}+ i\Omega_1\,\cos(\omega_1t)\,(\rho_{00}-\rho_{11})-i\Omega_2\cos(\omega_2t)\,\rho_{21} \\
   \dot{\rho}_{02} &=& -\left({\Gamma^0}/{2}+i\omega_{02}\right)\,\rho_{02}+i\Omega_2\,\cos(\omega_2t)\,(\rho_{00}-\rho_{22})-i\Omega_1\cos(\omega_1t)\,{\rho_{21}} \\
    \dot{\rho}_{21} &=& -(\gamma_p+i\omega_{21})\,\rho_{21}+i\Omega_1\,\cos(\omega_1t)\,\rho_{20}-i{\Omega_2}\,\cos(\omega_2t)\,\rho_{01},
\end{eqnarray}
with $\rho_{ij}^*=\rho_{ji}$ and $\hbar\omega_{ij}=E_i-E_j$.
The decay of the excited state $\Gamma^0$ into $|1\rangle$ ($|2\rangle$) occurs at rates $\alpha_1\Gamma^0$  ($\alpha_2\Gamma^0$, $\alpha_1+\alpha_2=1$), whereas $|2\rangle$ relaxes into $|1\rangle$ at twice the dephasing rate $\gamma{_p}$ within the Born-Markov approximation, cf. \ Eq.~(\ref{gammapdef}){}.

In the rotating wave approximation (RWA), the coupling to the external fields and the damping are assumed to be small and the system is close to resonance, 
\begin{equation}\label{rwa condition}
  \Omega_1,\,\Omega_2,\,\Gamma^0,\,\gamma_p,\,|\omega_1-\omega_{01}|,\,|\omega_2-\omega_{02}|\ll\omega_1\sim\omega_2.
\end{equation} 
Introducing the unitary operator $ U(t)\equiv {\rm diag}(1,e^{-i(\omega_1-\omega_2)t},e^{-i\omega_1 t})$
in the  eigenstate basis \{$|1\rangle$, $|2\rangle$, $|0\rangle$\},
one obtains transformed quantities in a rotated frame as 
$\rho_U\equiv U^{\dagger} \rho U$, $H_U(t)= -i\hbar U^+\frac{\partial U}{\partial t}+U^+H(t)U$, with
($E_1=0$ for convenience)
\begin{equation}
 {H}_U(t)=\hbar\left[
  \begin{array}{ccc}
    0 & 0 & \frac{\Omega_1}{2}(1+e^{-i2\omega_1 t}) \\
    0 & -\delta_1+\delta_2 &  \frac{\Omega_2}{2}(1+e^{-i2\omega_2 t})\\
    \frac{\Omega_1}{2}(1+e^{i2\omega_1 t}) & \frac{\Omega_2}{2}(1+e^{i2\omega_2 t}) & -\delta_1 
  \end{array}
\right]. \nonumber \label{the transformed hamiltonian}
\end{equation}
In the RWA, one neglects the oscillating part of $H_U(t)$  which is replaced by an effective {\em time-independent} Hamiltonian \cite{SB04}
\begin{equation}
 \tilde{H}_{\rm RWA}=\hbar\left[
  \begin{array}{ccc}
    0 & 0 & \frac{\Omega_1}{2} \\
    0 & -\delta_1+\delta_2 &  \frac{\Omega_2}{2}\\
    \frac{\Omega_1}{2} & \frac{\Omega_2}{2} & -\delta_1 
  \end{array}
\right] 
\end{equation}
that governs the equations of motion of the density operator in the RWA. Alternatively, one can start from the RWA Hamiltonian \ Eq.~(\ref{H3level}){} and derive the corresponding  equations of motions for the density matrix by directly transforming away the fast dependencies in the time evolution, see \cite{Arimondo96} and below. Corrections to coherent population trapping due to counter-rotating terms were investigated by Sanchez and Brandes \cite{SB04} in a systematic truncation scheme beyond the RWA.

The best insight into the phenomenon comes from plotting the stationary matrix elements of the density matrix as a function of the `Raman' detuning $\delta_R$, i.e. the difference of the relative detunings of the external light frequencies from the two transition frequencies, cf. Fig. (\ref{threelevel.eps}). This can be achieved, e.g., by fixing the second frequency $\omega_2$ exactly on resonance such that $\delta_2=0$ and varying $\omega_1=-\delta_R$. The population $\hat{\rho}_{00}$ of the upper level then shows a typical resonance shape, i.e. it increases coming from large $|\delta_R|$ towards the center $\delta_R=0$. Shortly before the resonance condition for the first light source, i.e. $\delta_R=0$, is reached, the population drastically decreases in the form of a very sharp anti-resonance, up to a vanishing  $\hat{\rho}_{00}$ for $\delta_R=0$.  For $\delta_R=0$, the population (i.e. all the electrons in the ensemble of three-level systems) is trapped in the dark superposition $|NC \rangle$ that cannot be brought back to the excited state $|0\rangle$. The dephasing rate $\gamma_p$ and the two Rabi frequencies determine the small half-width $\delta_{1/2}$ of the anti-resonance \cite{Arimondo96},
\begin{equation}
\delta_{1/2}\approx \gamma_p+\frac{|\Omega_R|^2}{2\Gamma^0},
\end{equation}
where $\Omega_R\equiv (\Omega_1^2+\Omega_2^2)^{1/2}$.

\subsection{Dark Resonance Current Switch}\label{sectionswitch}
\begin{figure}[t]
\begin{center}
\includegraphics[width=0.5\textwidth]{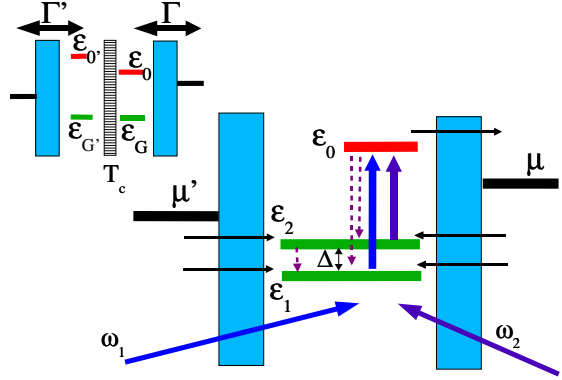}
\end{center}
\caption[]{\label{ddot_switch.eps} Level scheme for CPT in two coupled quantum dots in Coulomb blockade regime. Two tunnel coupled ground-states $|G\rangle$ and $|G'\rangle$ (small inset) form states $|1\rangle$ and $|2\rangle$ from which an electron is pumped to the excited state $|0\rangle$ by two light sources of frequency $\omega_1$ and $\omega_2$. Relaxation by acoustic phonon emission is indicated by dashed arrows.  From \cite{BR00}.}
\end{figure} 

A new transport mechanism based on the coherent population trapping effect in tunnel-coupled quantum dots was suggested by Brandes and Renzoni in \cite{BR00}. The original proposal  with two coupled quantum dots in the strong Coulomb blockade regime is actually very close to three-level systems in atoms, with the additional possibility to test the effect (and its modifications) in electronic transport, cf. Fig. \ref{ddot_switch.eps}. The dark state  appears in the form of a sharp anti-resonance in the {\em stationary current} through a double dot as a function of the Raman detuning $\delta_R$, i.e. the detuning difference of the two classical laser (or microwave) fields. The half-width of the anti-resonance can then be used to extract valuable information, such as the relaxation and dephasing times of tunnel coupled dot-ground state superpositions, from transport experiments.

\subsubsection{Model}
The model is defined by a double quantum dot in the strong Coulomb blockade regime with two tunnel-coupled ground states $|G\rangle$ and $|G'\rangle$ (see Fig. \ref{ddot_switch.eps}, inset) which hybridize via tunnel coupling $T_c$ into states $|1\rangle$ and $|2\rangle$ with energy difference $\Delta\equiv\varepsilon_2-\varepsilon_1=(\varepsilon^2+4T_c^2)^{1/2}$. The excited state $|0\rangle$ is assumed to have an unchanged number of electrons and is realized in the right dot.  The energy of the first excited level $|0'\rangle$ of the other (left) dot is assumed to be off resonance for transitions to and from the two ground states, and the hybridization of $|0'\rangle$ with $|0\rangle$ can be neglected for $|\varepsilon_{0'}-\varepsilon_0|\gg T_c$. The two ac-fields pump electrons into the excited level $|0\rangle$ such that transport becomes possible if both dots are connected to electron reservoirs. Again, the Coulomb charging energy $U$ is assumed to be  so large that states with two additional electrons can be neglected (typical values are  1meV $\lesssim U \lesssim$ 4meV in lateral double dots \cite{FujetalTaretal}).
Furthermore, the chemical potentials $\mu$ and $\mu'$ are tuned to values slightly above $\varepsilon_2$; this excludes the co-tunneling like re-entrant resonant tunneling process that can exist in three-level dots and has been discussed by Kuznetsov and co-workers \cite{Kuzetal96}.

In the dipole and rotating wave approximation, the coupling to the time-dependent fields is described by 
\begin{eqnarray}\label{VAL}
V_{\rm AL}(t) &=& -{\frac{\hbar}{2}}\left[
\Omega_P e^{-i\omega_Pt} |0\rangle\langle 1| +
\Omega_S e^{-i\omega_St} |0\rangle\langle 2|\right] + H.c.,
\end{eqnarray}
where for later convenience we have already introduced the pump(P)-Stoke(S) notation for the two (Rabi) frequencies $\omega_P$ and $\omega_S$ ($\Omega_P$ and $\Omega_S$).

The decay rate $\Gamma^0$  is primarily due to acoustic phonon coupling. The branching ratios $\alpha_1=1-\alpha_2=(\Delta+\varepsilon)^2/[(\Delta+\varepsilon)^2+4T_c^2]$ can be calculated using  the eigenstates \ Eq.~(\ref{eq:twobytworesult}).  Furthermore, $|2\rangle$ decays into $|1\rangle$ at the rate $2\gamma_p$, where $\gamma_p$ is the (dephasing) rate for the decay of the `coherence' (density matrix element $\rho_{12}$) within the Born-Markov approximation, cf. \ Eq.~(\ref{gammapdef}){}.

If the chemical potentials $\mu$ and $\mu'$ are as indicated in Fig.(\ref{ddot_switch.eps}), electron tunneling occurs by {\em in}-tunneling that changes $|E\rangle$ into $|G\rangle$ at a rate $\Gamma$, and $|E\rangle$ into $|G'\rangle$ at the rate $\Gamma'$, whereas {\em out}-tunneling from $|G\rangle$ and $|G'\rangle$  is Pauli blocked. The corresponding rates $\gamma_1$ and $\gamma_2$ for tunneling into the hybridized states $|1\rangle$ and $|2\rangle$ are $\gamma_{1,2}=[(\Delta \pm \varepsilon )^2 \Gamma +4T_c^2 \Gamma']/ [(\Delta \pm \varepsilon )^2 +4T_c^2 ]$. On the other hand, electrons can leave the dots only by tunneling {\em out of} the state $|0\rangle$ (but not in) at the rate $\Gamma$ into the right lead (negligible hybridization of $|0\rangle$ with $|0'\rangle$ was assumed). Setting $\Gamma=\Gamma'$ for simplicity in the following and denoting the `empty state' by $e$ (the symbol $0$ is used already for the excited three-level state), the resulting density-matrix equations then are given by \cite{BRB01}
\begin{eqnarray}
\dot{\rho}_{1,1} &=& \alpha_1\Gamma^0 {\rho}_{0,0}+\Gamma {\rho}_{e,e} 
+2\gamma_p \rho_{2,2}+ {\rm Im}[\Omega_P \tilde{\rho}_{1,0}] \label{3leveleom11}\\
\dot{\rho}_{2,2} &=& \alpha_2\Gamma^0 {\rho}_{0,0}+\Gamma {\rho}_{e,e}
 -2\gamma_p \rho_{2,2}+ {\rm Im}[\Omega_S\tilde{\rho}_{2,0}] \\
\dot{\rho}_{0,0} &=& -(\Gamma+\Gamma^0) \rho_{0,0} -
{\rm Im}[\Omega_P \tilde{\rho}_{1,0}]
- {\rm Im}[\Omega_S\tilde{\rho}_{2,0}] \\
\dot{\rho}_{e,e} &=& -2\Gamma\rho_{e,e} +\Gamma \rho_{0,0}\\
\dot{\tilde{\rho}}_{1,0} &=& -\left[
\frac{1}{2}\left( \Gamma+\Gamma^0\right) + i\delta_P\right]
\tilde{\rho}_{1,0}
+{i\over 2}\Omega_P (\rho_{0,0}-\rho_{1,1})
 -{i\over 2}\Omega_S \tilde{\rho}_{1,2}\\
\dot{\tilde{\rho}}_{2,0} &=& 
-\left[ \frac{1}{2} \left( \Gamma+\Gamma^0\right) +i
\delta_S\right] \tilde{\rho}_{2,0}
+{i\over 2}\Omega_S (\rho_{0,0}-\rho_{2,2})
-{i\over 2}\Omega_P \tilde{\rho}_{1,2}^{*}\\
\dot{\tilde{\rho}}_{1,2} &=& -\left( \gamma_p +i\delta_R\right)
\tilde{\rho}_{1,2}
+{i\over 2}\Omega_P \tilde{\rho}_{0,2}
-{i\over 2}\Omega_S \tilde{\rho}_{1,0}\label{3leveleom12},
\end{eqnarray}
where $\tilde{\rho}_{0j}=\tilde{\rho}_{j0}^*=\rho_{0j}e^{i\omega_j t}$ are the slowly-varying off-diagonal matrix elements of the reduced density operator of the double dot.

\subsubsection{Stationary Current}
\begin{figure}[t]
\includegraphics[width=0.45\textwidth]{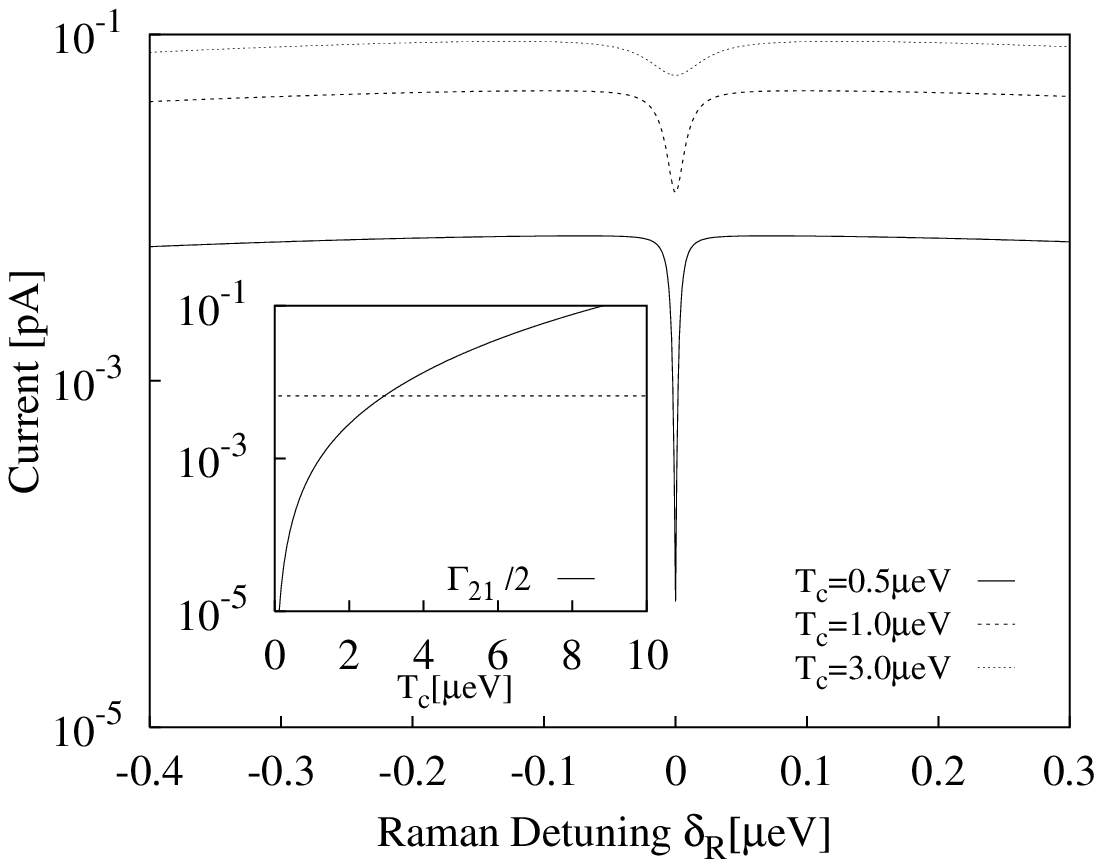}
\includegraphics[width=0.45\textwidth]{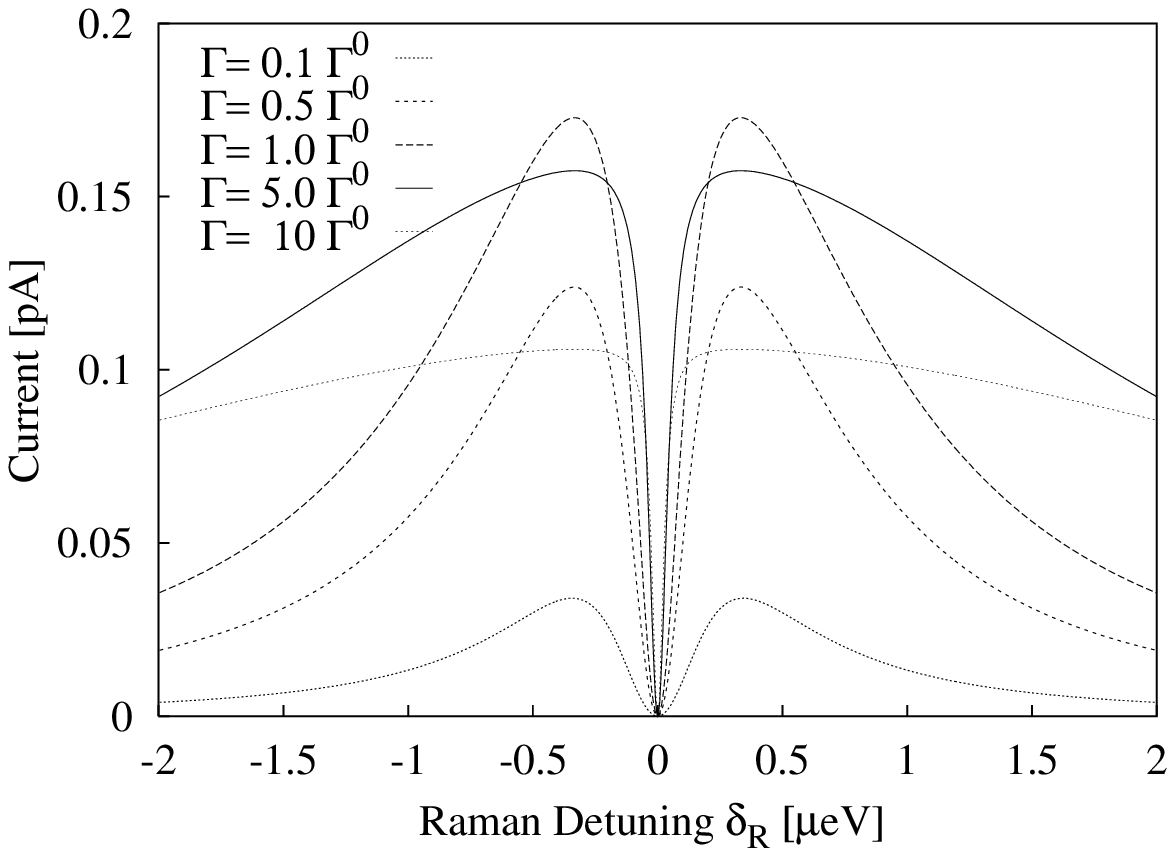}
\caption[]{\label{cu_CPT.eps} {\bf Left:} Tunnel current anti-resonance through double dot system  with ground-state energy difference $\varepsilon=10\mu$eV.  Rabi frequencies $\Omega_1=\Omega_2$, parameters are $\Omega_R=(\Omega_1^2+\Omega_2^2)^{1/2}=0.2\Gamma^0$, tunnel rates $\Gamma=\Gamma'=\Gamma^0=10^9$s$^{-1}$, where $\Gamma^0$ is the relaxation rate due to acoustic phonon emission from $|0\rangle$. Inset: Dephasing rate $\gamma_p=2\pi \frac{T_c^2}{\Delta^2} J_{\rm piezo}(\Delta)$ (in $\mu eV/\hbar$) with $\Delta=(\varepsilon^2+4T_c^2)^{1/2}$, $\hbar c/d=20 \mu$eV, and $\alpha_{\rm piezo}=0.025$, \ Eq.~(\ref{Jmicro}). Dashed line: crossover at $\gamma_p=|\Omega_R|^2/2[\Gamma^0+\Gamma]$, cf. Eq.(\ref{delta12}). {\bf Right:} Current for fixed $T_c=1\mu$eV, $\Omega_R=1.0\Gamma^0$,  and different tunnel rates $\Gamma=\Gamma'$. From \cite{BR00}.}
\end{figure} 

The  solution for the density operator is used to obtain the   electric current 
\begin{equation}\label{defcurrentCPT}
I(t)=-e\Gamma  [\rho_{0,0}(t)-\rho_{e,e}(t)],
\end{equation}
as the net flow of electrons with charge $-e<0$ through the dot. For the stationary case, Fig. (\ref{cu_CPT.eps}) shows  $I$ as a function of the Raman detuning $\delta_R$ for constant $\Omega_1=\Omega_2$ at zero temperature \cite{BR00}. Close to $\delta_R=0$, the overall Lorentzian profile shows the typical CPT  anti-resonance. The half-width $\delta_{1/2}$ of the current anti-resonance is given by ($\Omega_R\equiv(\Omega_1^2+\Omega_2^2)^{1/2} $)
\begin{equation}\label{delta12}
\delta_{1/2}\approx \gamma_p+\frac{|\Omega_R|^2}{2[\Gamma^0+\Gamma]},\quad \varepsilon=0,
\end{equation}
which shows that $\delta_{1/2}$ increases with the dephasing rate $\gamma_p$. In Fig. (\ref{cu_CPT.eps}), $\gamma_p=2\pi \frac{T_c^2}{\Delta^2} J_{\rm piezo}(\Delta)$ is completely due to spontaneous emission of piezo-electric phonons for zero temperature (Eq.~(\ref{gammapdef}) with $\beta\to\infty$ and \ Eq.~(\ref{Jmicro}) with $\alpha_{\rm piezo}=0.025$). For fixed coupling strength to the time-dependent fields and increasing tunnel coupling $T_c$, $\gamma_p$ increases whence the anti-resonance becomes broader and finally disappears for large $T_c$. The  vanishing of the anti-resonance sets in for $\gamma_p \gtrsim |\Omega_R|^2/2[\Gamma^0+\Gamma]$, cf. the inset of Fig. (\ref{cu_CPT.eps}). 
On the other hand, with {\em increasing} elastic tunneling $\Gamma$ out of the dot, the current increases until an overall maximal value is reached at $\Gamma\approx \Gamma^0$, cf. Fig. (\ref{cu_CPT.eps}) (right); $I(\delta_R)$ decreases again and becomes very broad if the elastic tunneling becomes much faster than the inelastic relaxation $\Gamma^0$, and with increasing $\Gamma$ the center anti-resonance then becomes sharper and sharper, its half-width $\delta_{1/2}$ approaching the limit $\gamma_p$, Eq. (\ref{delta12}).

The three-level dark resonance therefore acts as an `optical switch' based on an optical double-resonance. The dark state thus created is protected deeply below the Fermi sea of the contact reservoirs by the Pauli principle and the Coulomb blockade. In \cite{BR00} it was furthermore pointed out that the CPT transport mechanism  differs physically from other transport effects in AC-driven systems (e.g., coherent 
destruction of tunneling \cite{GrossmannWagner,GH98}, tunneling through photo-sidebands \cite{BruderSchoellerInarrea}, or coherent pumping of electrons \cite{SW96,WS99}) that depend on an additional time-dependent phase 
that electrons pick up while tunneling, with dissipation being a disturbance rather then necessary for those effects to occur. In contrast, the CPT effect in dots requires incoherent relaxation (phonon emission) in order to trap the system in the dark state.

\subsection{Adiabatic Transfer of Quantum States}

A remarkable feature of coherent population trapping is the possibility to control and indeed rotate the dark state $|NC\rangle$ into arbitrary superpositions within the qubit ${\rm span}(|1\rangle,|2\rangle)$. This can be achieved by slow, adiabatic variation of the two Rabi frequencies $\Omega_1$ and
$\Omega_2$, where the adiabatic theorem guarantees that a given state follows the slow variation of parameters in the Hamiltonian. 

\subsubsection{Adiabatic Transfer and STIRAP}
Bergmann and co-workers \cite{Beretal98} have developed this technique for three-level systems, where it is called `stimulated Raman adiabatic passage' (STIRAP) and has found widespread application in coherent control and the adiabatic transfer of populations (in the ensemble sense) from one state into another.

To be specific, consider the two Rabi frequencies in \ Eq.~(\ref{VAL}) as time-dependent parameters in pump (P) and Stoke (S) pulse form,
\begin{eqnarray}
\label{eq:pulse}
\Omega_P(t)&=&\Omega^0 \sin \theta e^{-(t-\tau)^2/T^2},\quad
\Omega_S(t)=\Omega^0 \left( e^{-t^2/T^2} + \cos \theta 
e^{-(t-\tau)^2/T^2}\right),
\end{eqnarray}
where $\tau$ and $T$ are the pulse delay and pulse duration, respectively, and the Gaussian form in Eq.~(\ref{eq:pulse}) has been chosen for convenience. With this choice, the Stokes  pulse $S$ (field ${\bf E}_j(t)$ in \ Eq.~(\ref{fielddef}) with time-dependent amplitude $|\calEv_2(t)| \propto \Omega_S(t) $) first couples $|2\rangle$ to $|0\rangle$, before a second pulse (the pump pulse $P$),
partially overlapping with $S$, couples $|1\rangle$ to $|0\rangle$ \cite{marte,vita,BRB01}. For large times $t\gtrsim T$, an initial state $|\Psi_{in}\rangle=|1\rangle$ is then adiabatically transformed into a superposition,
\begin{eqnarray}
  |\Psi_{in}\rangle=|1\rangle \to |\Psi_f\rangle=\cos\theta |1\rangle - \sin\theta |2\rangle.
\end{eqnarray}
The requirement for this to happen is that during the whole process the Raman resonance condition, $\delta_R\equiv\varepsilon_1+\omega_1-\varepsilon_2-\omega_2=0$, is preserved and therefore dark states are adiabatically transformed into dark states. The pulse sequence \ Eq.~(\ref{eq:pulse}) is called `counter-intuitive': transferring population out of $|1\rangle$ is achieved by first `pumping' the $|2\rangle-|0\rangle$ transition and not the 
$|1\rangle-|0\rangle$ transition.

\subsubsection{STIRAP and Transport in Double Quantum Dots}

\begin{figure}[t]
\begin{center}
\includegraphics[width=0.3\textwidth]{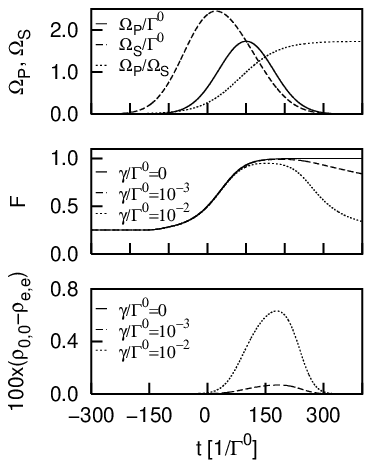}
\includegraphics[width=0.3\textwidth]{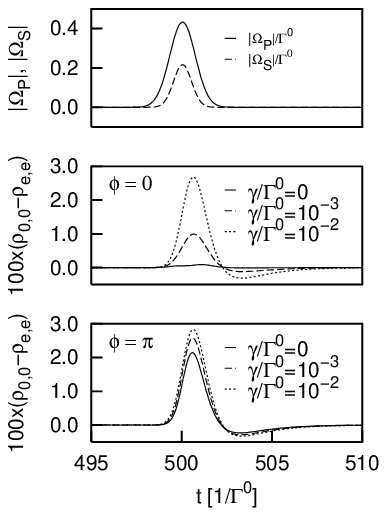}
\end{center}
\caption[]{\label{stirap_fig2.eps} {\bf Left:} Rabi frequencies, fidelity and electric current as a function of the
interaction time for STIRAP in double quantum dots. $\Omega^0 = 2\Gamma^o$, $\Gamma= \Gamma^0/3$, $\theta=\pi/3$, $\alpha_1=\alpha_2=1/2$, $T=\tau=100/\Gamma^0$.
 {\bf Right:} Second probe pulses Rabi frequencies, \ Eq.~(\ref{eq:probe}){}, and current pulse $I(t)$ in double pulse scheme. $\Omega_p = 0.5\Gamma^o$, $T_p=1/\Gamma^0$ and $\Delta t = 500/\Gamma^0$.  From \cite{BRB01}.}
\end{figure} 

A realization of STIRAP rotations with microwave pulses in double quantum dots was suggested by Brandes, Renzoni, and Blick in \cite{BRB01}, where in addition a scheme to determine the dephasing rate $\gamma_p$ from time-dependent transport measurements was developed. Two-source microwave techniques have in fact been used to experimentally investigate ground and excited states in single quantum dots already \cite{Qinetal01}.

The STIRAP-transport model adopted in \cite{BRB01} is an extension of the `current switch' model \cite{BR00}  of section \ref{sectionswitch} to time-dependent pulses, \ Eq.~(\ref{eq:pulse}). The time-dependent current $I(t)$, \ Eq.~(\ref{defcurrentCPT}), is calculated by numerical solution of \ Eq.~(\ref{3leveleom11}-\ref{3leveleom12}), which together with the preparation {\em fidelity} $F(t)\equiv \langle \Psi_f | \rho(t)  \Psi_f| \rangle$ of the final state $|\Psi_f\rangle=\cos\theta |1\rangle - \sin\theta |2\rangle$ and the pulse form \ Eq.~(\ref{eq:pulse}) is shown in Fig. (\ref{stirap_fig2.eps}), left,  for different values of $\gamma_p$. For $\gamma_p=0$, one obtains fidelity one because the STIRAP pulses prepare the double dot in the desired superposition $|\Psi_f\rangle$. The current through the dot is zero for $\gamma_p=0$, when the dark state is stable and no electrons can be excited to state $|0\rangle$. For finite $\gamma_p>0$, the dark state decoheres and leads to a finite current pulse $I(t)$ which increases with increasing $\gamma_p$. As this indicator current is very weak, a more sensitive detection scheme was suggested \cite{BRB01} in the form of a double-pulse sequence, where the two `preparation' pulses \ Eq.~(\ref{eq:pulse}) are applied  simultaneously at a second, later time $\Delta t>0$ 
\begin{eqnarray}
\label{eq:probe}
\Omega_P^{\rm probe}(t)&=&\Omega_p \sin \theta e^{-(t-\Delta t)^2/T_p^2},\quad \Omega_S^{\rm probe}(t)=\Omega_p \cos (\theta + \phi) e^{-(t-\Delta t)^2/T_p^2},
\end{eqnarray}
with the ratios of their amplitudes chosen to correspond to $|\psi_f\rangle$ ($\phi=0$) or to its orthogonal state ($\phi=\pi$). For $\gamma_p=0$ and $\phi=0$, nothing happens as the dot stays in the state $|\psi_f\rangle$ and the subsequent application of the probe pulses \ Eq.~(\ref{eq:probe}) does not produce any current through the dot. For $\phi=\pi$, however, the probe pulses are in anti-phase with the ground state superposition and a large current follows. For nonzero  $\gamma_p\ne 0$, the superposition decays into a mixture on a time scale $1/\gamma_p$, and the application of the probe pulses results in a current through the dot both for $\phi=0$ and $\phi=\pi$. The larger  $\gamma_p$, the less sensitive is the current to the relative phase $\phi$ of the probe pulses which gives rise to the definition of the {\em contrast} 
\begin{equation}
C=\frac{I_{max}(\phi=\pi)-I_{max}(\phi=0)}{I_{max}(\phi=\pi)+I_{max}(\phi=0)}
\end{equation}
as a measure to extract $\gamma_p$ from a transport experiment in coupled dots.

\subsubsection{Quantum Dot Excitons}
Hohenester and co-workers \cite{Hohetal00} proposed a STIRAP scheme in two coupled quantum dots with two hole states ($|L\rangle$ and $|R\rangle$) in the valence band and one electron state $|e\rangle$ in the conduction band of a $p-i$ semiconductor double-quantum well structure. In their scheme, an external gate voltage and the Coulomb blockade guaranteed population of the double dot with one additional hole only. An external electric field leads to localization of the hole wave functions in the left ($L$) or the right ($R$) dot, whereas the electron wave function spread across both dots due to the smaller electron mass. Coulomb interactions between electrons and holes were taken into account by exact diagonalization in the Fock-Darwin single particle basis \cite{Hohetal00,THM02}, from which one could clearly identify a three-level system with the two low-energy states $|L\rangle$ and $|R\rangle$ and the excited (correlated) charged-exciton state $|X^+\rangle$. The STIRAP process was then realized within the usual two-pulse (pump-Stoke) configuration, allowing adiabatic population transfer between $|L\rangle$ and $|R\rangle$. Troiani, Molinari and Hohenester subsequently extended this scheme by taking into account the spin-degree of freedom in order to realize an optical qubit gate \cite{TMH03}, cf. \ref{section_KR}.

\subsection{Higher-dimensional Hilbert Spaces, Geometrical Phase}

A generalization of adiabatic schemes based on STIRAP is obtained in Hilbert spaces of  dimension $d\ge 4$. Unanyan, Shore and Bergmann \cite{USB99} used STIRAP in four-level systems and established a relation to non-Abelian geometrical phase factors. Duan, Cirac, and Zoller \cite{DCZ01} showed how the universal set of one- and two-qubit quantum gates can be realized by adiabatic variation of three independent  Rabi frequencies. Faoro, Siewert, and Fazio \cite{FSF03} applied this scheme to a network of superconducting Josephson junctions, with three fluxes varied cyclically, and gave explicit expressions for non-Abelian holonomies. Somewhat closer to the original STIRAP scheme, Kis and Renzoni used four-level systems and a double-STIRAP process to directly construct the operator for adiabatic rotations around a given axis of arbitrary one-qubits. 

The underlying physics of these schemes is again quite simple and can best be formulated in a geometric fashion. One extends the  Hamiltonian in the dipole and rotating wave approximation, \ Eq.~(\ref{H3level}){}, from $d=3$ ($|0\rangle,|1\rangle,|2\rangle$) to $d=N+1>3$ levels ($|0\rangle,|1\rangle,...,|N\rangle$), with $N$ classical monochromatic ac-fields with  Rabi frequencies $\Omega_i$ and all frequencies $\omega_i=(E_0-E_i)$ on resonance. In the interaction picture, one then obtains  a {\em time-independent} (for constant $\Omega_i$) interaction Hamiltonian,
\begin{eqnarray}\label{Hgeo}
  H_I&=&-\frac{\hbar}{2}\left( |0\rangle \langle {\Omegav}| + |{\Omegav} \rangle \langle 0|\right),\quad
 |{\Omegav}\rangle \equiv \sum_{i=1}^N
\Omega_i^* |i\rangle,
\end{eqnarray}
where the vector ${\Omegav}\equiv (\Omega_1,...,\Omega_N)$ contains the (complex) Rabi frequencies. 
The interaction Hamiltonian, \ Eq.~(\ref{Hgeo}), immediately gives rise to an $N-1$ dimensional subspace of dark states $|{\Dv}\rangle$ with $H_I|{\Dv}\rangle=0$, defined by the  $N-1$ dimensional manifold of vectors ${\Dv}$ that are orthogonal to ${\Omegav}$ and therefore have $\langle {\Omegav}|{\Dv}\rangle=0$. In this language, one clearly understands that there are no dark states for $N=1$ (two-level system), one dark state for $N=2$ (three-level system), two linearly independent dark states in $N=3$ (four-level system) etc. 

\subsubsection{Double STIRAP and  SU(2) Qubit-Rotations}\label{section_KR}
The above-mentioned STIRAP  schemes now start from making ${\Omegav}$ time-dependent by allowing slow, adiabatic variations such that $|\dot{{\Omegav}}|/|{{\Omegav}}|\ll |{{\Omegav}}|$. In the first step of their four-level double-STIRAP scheme ($N=3$ in \ Eq.~(\ref{Hgeo}){}), Kis and Renzoni \cite{KR02} chose a parametrization ${\Omegav} \equiv(\Omega_P \cos \chi,\Omega_P e^{-i\eta} \sin \chi, \Omega_S)$ with fixed $\chi,\eta$ and (`counter-intuitive') Stoke ($\Omega_S$) and delayed pump ($\Omega_P$) pulse, cf.  \ Eq.~(\ref{eq:pulse}) for $\theta=\pi/2$. The dark subspace is spanned by the constant state $|NC_1\rangle\equiv-\sin \chi |1\rangle + e^{i\eta} \cos \chi |2\rangle$ and the orthogonal and slowly moving $|NC_2\rangle \propto \Omega_S |C_1\rangle - \Omega_P |3\rangle$, where the bright state $ |C_1\rangle = \cos \chi |1\rangle + e^{i\eta} \sin \chi |2\rangle$. An initial one-qubit state $|\Psi_{\rm in}\rangle\in {\rm span}(|1\rangle, |2\rangle)$ is now decomposed into its orthogonal components along $|NC_1\rangle$ and $|C_1\rangle$, with only the latter component slowly dragged along with $|NC_2\rangle$ (which for large times becomes $-|3\rangle$), 
\begin{eqnarray}\label{psiinter}
  |\Psi_{\rm in}\rangle \to |\Psi_{\rm inter}\rangle=\langle NC_1|\Psi_{\rm in}\rangle |NC_1\rangle
-\langle C_1|\Psi_{\rm in}\rangle |3\rangle.
\end{eqnarray}
The {\em second} STIRAP step now has ${\Omegav} \equiv(\Omega_S \cos \chi,\Omega_S e^{-i\eta} \sin \chi, \Omega_Pe^{-i\delta})$ and a dark subspace spanned by $|NC_1\rangle$ and the slowly moving $|NC'_2\rangle\propto 
\Omega_Pe^{-i\delta}|C_1\rangle -\Omega_S|3\rangle$. The intermediate state $|\Psi_{\rm inter}\rangle$, \ Eq.~(\ref{psiinter}), lies in the new dark subspace. With the  roles of $\Omega_S$ and $\Omega_P$ now exchanged, $|NC'_2\rangle$ moves from $-|3\rangle$ into $e^{-i\delta}|C_1\rangle$ whereas $|NC_1\rangle$ remains constant, and therefore
\begin{eqnarray}
  |\Psi_{\rm inter}\rangle \to \langle NC_1|\Psi_{\rm in}\rangle |NC_1\rangle
+ e^{-i\delta}\langle C_1|\Psi_{\rm in}\rangle|C_1\rangle=e^{-i\delta/2}\exp
\left(-i\frac{\delta}{2}\mathbf{n\hat{\sigma}}\right) |\Psi_{\rm in}\rangle,
\end{eqnarray}
which apart from the overall phase factor $e^{-i\delta/2}$ is 
an SU(2) rotation of the initial qubit $|\Psi_{\rm in}\rangle$ about the unit vector
$\mathbf{n}=(\sin 2\chi \cos \eta, \sin 2\chi \sin \eta, \cos 2\chi)$ through the angle $\delta$ 
($\mathbf{\hat{\sigma}}$ is the vector of the Pauli matrices), as can be
checked by direct calculation. Note that the STIRAP directions ${\Omegav}$ are chosen such that
the constant dark state has $\langle NC_1|\mathbf{\hat{\sigma}}|NC_1\rangle\propto \mathbf{n}$ and thereby
defines the rotation axis. The scheme is robust against fluctuations of the pulse shapes and areas $\Omega_P(t)$ and $\Omega_S(t)$.

Troiani, Molinari and Hohenester \cite{TMH03} showed how the Kis-Renzoni scheme can be utilized to achieve not only one-qubit rotations, but also conditional (two-qubit) gates in coupled quantum dots with both orbital and spin degree of freedom. Their scheme allows to rotate a spin qubit into an orbital qubit, and in addition to perform a controlled NOT by utilizing STIRAP and charge-charge interactions. 
Kis and Paspalakis \cite{KP04} suggested qubit rotations in three-level SQUIDs interacting with two non-adiabatic microwave pulses. An initial state is again split into its components along the $|NC\rangle,|C\rangle$ basis belonging to ${\Omegav}=(\Omega(t)\cos \varphi, \Omega(t)e^{i\eta} \sin \varphi)$, leading to a two-level system defined in the subspace orthogonal to the dark state $|NC\rangle$, where switching on and off of $\Omega(t)$ leads to the desired rotation in the form of a standard Rabi rotation. Paspalakis and co-workers furthermore suggested adiabatic passage to achieve entanglement between two three-level SQUIDs, an various other applications of adiabatic rotations in double dots and SQUIDS \cite{Paspalakis}. Along similar lines, Chen, Piermarocchi, Sham, Gammon, and Steel \cite{Chenetal04} devised an adiabatic qubit rotation for a single spin in a quantum dot.

Thanopulos, Kr\'{a}l, and Shapiro \cite{TKS04} used  a generalized double STIRAP for adiabatic population transfer between an initial and a final {\em wave packet} composed of $n$ nearly degenerate states $|k^{(')}\rangle$.  A Rabi frequency vector ${\Omegav}_0$ links the excited (`parking') state $|0\rangle$ with $n$ non-degenerate auxiliary states, which in turn are linked to the $|k\rangle$ states by $n$ linearly independent Rabi frequency vectors ${\Omegav}_k$ ($k=1,..,n$). A choice ${\Omegav}_0\propto \sum_{k=1}^n a_k {\Omegav}_k$ with appropriate $a_k$ now allows to utilize the single dark state of the system to rotate between the wave packets via $|0\rangle$.

\subsubsection{Non-Abelian Holonomies}
The Hamiltonian, \ Eq.~(\ref{Hgeo}), can alternatively be regarded as a part of some given Hamiltonian 
$H=\varepsilon_0 |0\rangle \langle 0| + H_I$ (in the Schr\"odinger picture), where all the energies of levels
$|1\rangle,...|N\rangle$ are the same (and set to zero for simplicity). This second, more general interpretation actually gives rise to many generalizations of adiabatic schemes beyond atomic physics. 

Since the subspace spanned by the $N-1$ dark states $|\Dv\rangle$ of $H$ is degenerate, cyclic adiabatic variation of ${\Omegav}$ gives rise to  Wilczek-Zee non-Abelian holonomies \cite{WZ84}, that generalize the Abelian Berry phase to a {\em matrix phase} that involves superpositions of the degenerate eigenstates. Within that subspace, the usual dynamical time-evolution of states is then replaced by `geometric evolutions' that can be used, e.g. for quantum computation ({\em holonomic quantum computation}).  

Duan, Cirac, and Zoller \cite{DCZ01} constructed the gate $U_1=e^{i\phi_1|2\rangle \langle 2|}$  with $\Omegav=(0,\Omega\sin \frac{\theta}{2}e^{i\varphi},-\Omega\cos \frac{\theta}{2})$ in the notation of Eq.~(\ref{Hgeo}), with a dark state $|\Dv\rangle = \cos \frac{\theta}{2}|2\rangle + \sin \frac{\theta}{2}e^{i\varphi} |3\rangle$ and $\theta, \varphi$ cyclically varied (starting and ending with $\theta=0$), giving the Berry phase $\phi_1=\oint\sin \theta d\theta d\varphi=\oint d\Omega$, which is the solid  angle swept by the vector into $(\theta,\varphi)$ direction. Similarly, they constructed the gate $U_2 = e^{i\phi_2\sigma_y}$, $\sigma_y \equiv i(|2\rangle \langle 1| -|1\rangle \langle 2|)$, using  $\Omegav=\Omega(\sin\theta \cos \varphi, \sin \theta \sin \varphi, \cos \theta)$ with the two degenerate dark states, 
\begin{eqnarray}
  |\Dv_1\rangle=\cos \theta(\cos \varphi |1\rangle + \sin \varphi |2\rangle) - \sin \theta |3\rangle, \quad |\Dv_2\rangle=\cos \varphi |2\rangle - \sin\varphi |1\rangle,
\end{eqnarray}
with again $\phi_2=\oint d\Omega$ the solid  angle swept by $(\theta,\varphi)$.

Following the non-Abelian holonomy schemes by Unanyan, Shore and Bergmann \cite{USB99}, and  Duan, Cirac, and Zoller \cite{DCZ01}, an application to {\em networks of superconducting Josephson junctions} with variable SQUID loop Josephson couplings $J_i(\bar{\Phi}_i), i=1,2,3$, was suggested by Faoro, Siewert, and Fazio \cite{FSF03}. These authors considered a Hamiltonian of the type Eq.~(\ref{Hgeo}),
\begin{eqnarray}
  H = \delta E_C |0\rangle +\frac{1}{2}\sum_{i=1}^3\left(J_i(\bar{\Phi}_i) |0\rangle \langle i| +H.c.\right),
\end{eqnarray}
where $\bar{\Phi}_i$ is the external magnetic flux through loop $i$ in units of the flux quantum $hc/2e$, and  $\delta E_C$ is the energy difference between the three degenerate charge states $|i\rangle$ (corresponding to one excess Cooper pair on island $i$) and the state $|0\rangle$ with  one excess Cooper pair on a forth superconducting island. They considered the dark subspace spanned by $|\Dv_1\rangle = -J_2|1\rangle + J_1|2\rangle$ and $|\Dv_2\rangle = -J_3(J_1^*|1\rangle + J_2^*|2\rangle)+(|J_1|^2+|J_2|^2)|3\rangle$, and unitary transformations $U_\gamma$ on a closed loop $\gamma$ in that subspace as 
\begin{eqnarray}
  U_\gamma = {\rm P} \exp \oint_\gamma \sum_{j}A_jd\bar{\Phi}_j,\quad 
(A_j)_{\alpha\beta}=\langle \Dv _\alpha |\frac{\partial}{\partial \bar{\Phi}_j}|\Dv _\beta \rangle,\quad
\alpha=1,2,
\end{eqnarray}
with the path ordering symbol ${\rm P}$. Choosing appropriate loops $\gamma$ in the parameter space of the three fluxes $\bar{\Phi}_j$ then yields transformations corresponding to charge pumping, one-qubit gates, or two-qubit gates by coupling two qubits via Josephson junctions.

\subsubsection{Quantum Adiabatic Pumping through Triple Dots}\label{section_triple_dot}
Renzoni and Brandes suggested {\em quantum adiabatic following} as a mechanism for {\em charge pumping} in strongly Coulomb-blocked systems \cite{RB01}, in a regime that is opposite to adiabatic quantum pumping in non- or weakly interacting systems. In the latter case, which by itself is a relatively new area of mesoscopic transport \cite{Bro98}, parametric change of the scattering matrix leads to adiabatic pumping of charges through mesoscopic scatterers which typically are in the metallic regime. Concepts from metallic systems, such as mesoscopic fluctuations \cite{ZSP99,aleiner},  symmetries \cite{aleiner}, or resonances \cite{WWG00}, are then generalized to the time-dependent case. Experiments in large quantum dots \cite{SMCG99} have demonstrated the feasibility of such `adiabatic quantum electron pumps'. 
\begin{figure}[t]
\begin{center}
\includegraphics[width=0.35\textwidth]{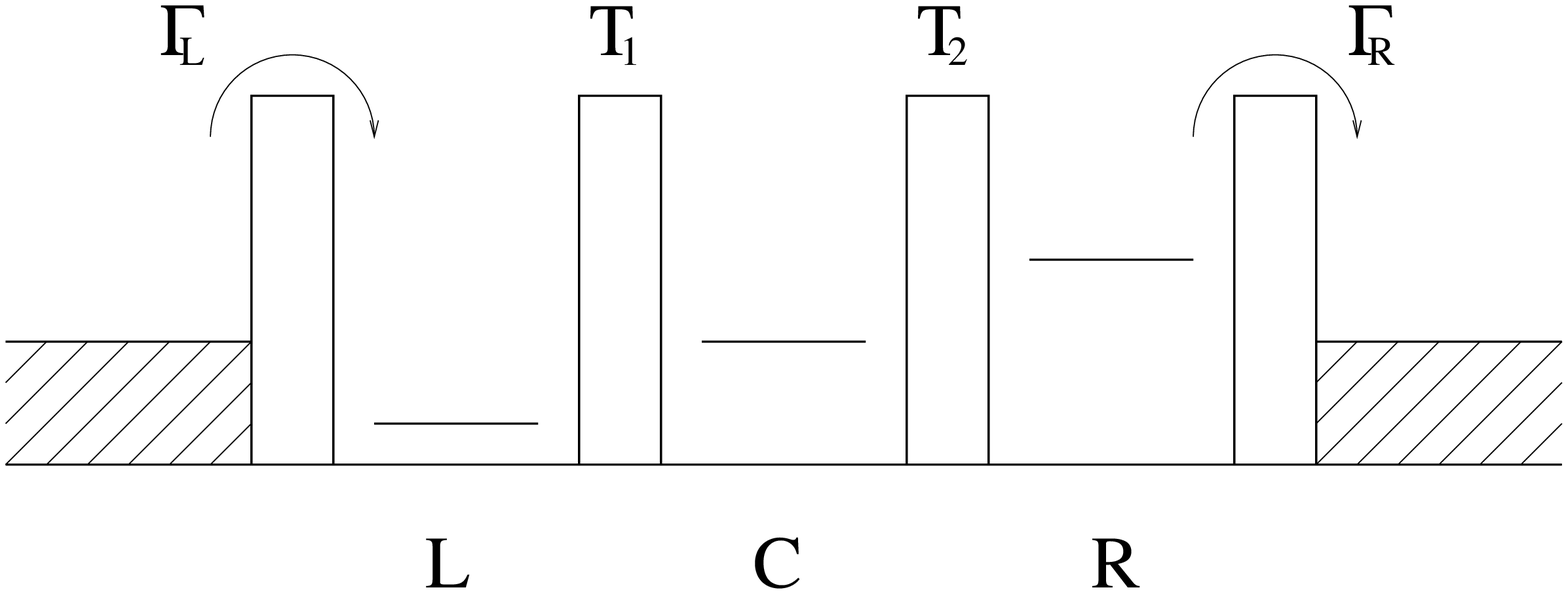}
\includegraphics[width=0.6\textwidth]{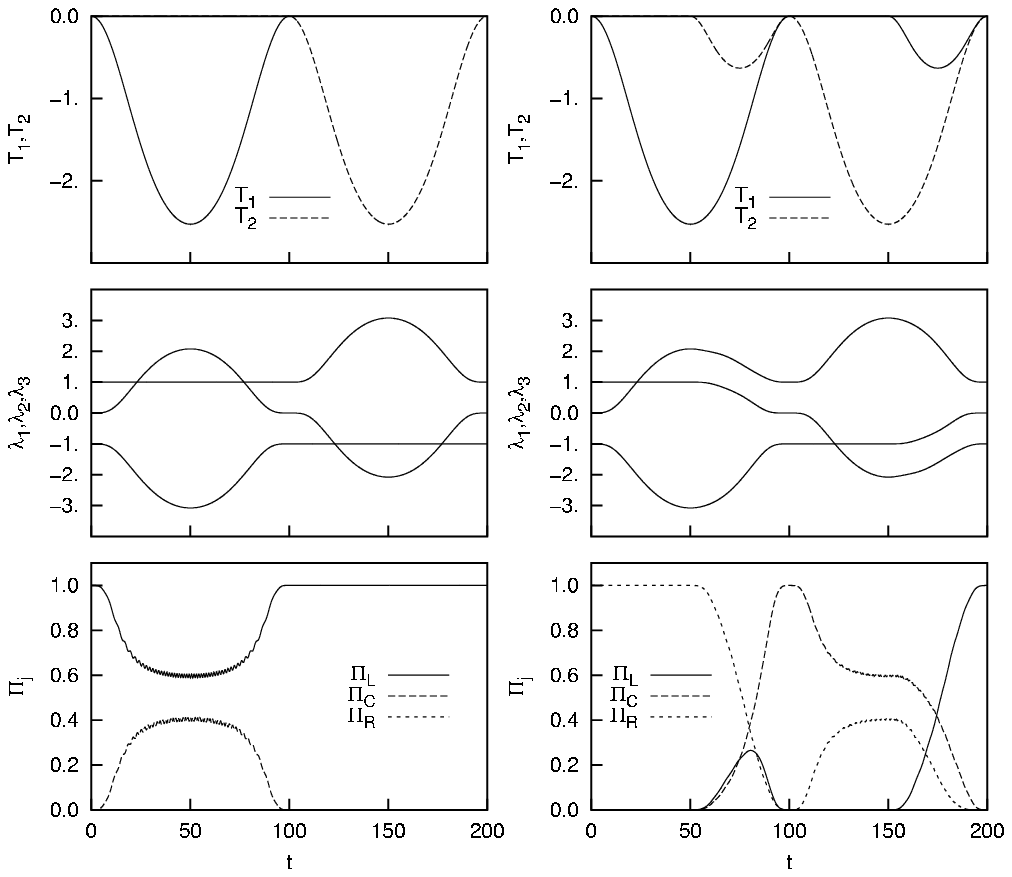}
\end{center}
\caption[]{\label{steering_fig1.eps} {\bf Left:} Triple dot for quantum adiabatic charge pumping with time-dependent tunnel couplings $T_1$, $T_2$.  {\bf Right:} Transfer of an electron from the right dot to the left dot. Tunnel-coupling pulse sequence (top) and corresponding time evolution of the adiabatic energy eigenvalues (center) and populations $\Pi_{\alpha}=c^{*}_{\alpha} c_{\alpha}$ ($\alpha=L,C,R$, bottom), as determined by numerically solving the Schr\"odinger equation, \ Eq.~(\ref{eq:steering}). In the left panel, only one of the tunnel barriers is open at a time. From \cite{RB01}.}
\end{figure} 
In contrast, the triple dot system considered in \cite{RB01} is closer to the original idea of adiabatic following in atomic physics, and extends the concept of (classical) adiabatic transfer in single electron devices \cite{pumpexperiment,Grabert} (such as single electron turnstiles) to  the (quantum) adiabatic control of the wave function itself. The Hamiltonian
\begin{eqnarray}\label{H_threedot}
H(t)&=&\sum_{\alpha=L,C,R}\left[\varepsilon_\alpha|\alpha\rangle\langle \alpha|\right]+
\hbar T_1 (t)\left[ |L\rangle\langle C| + |C\rangle\langle L| \right] +
\hbar T_2 (t) \left[ |C\rangle\langle R|+ |R\rangle\langle C| \right]~,
\label{ham}
\end{eqnarray}
describes the left, right, and central dot, cf. Fig. (\ref{steering_fig1.eps}), in a four dimensional Hilbert space with basis $\{ |0\rangle, |L\rangle,|C\rangle,|R\rangle \}$, where again $|0\rangle$ denotes  the `empty' state and the time-dependence of the (real) $T_i,i=1,2$ is slow.

For degenerate dot ground states $E_C=E_R=E_L=0$, $H(t)$ is of the form Eq.~(\ref{Hgeo}) with $N=2$, $|\Omegav\rangle =-2T_1(t) |L\rangle - 2T_2(t) |R\rangle$, and the central dot state $|C\rangle$ corresponding to $|0\rangle$ in Eq.~(\ref{Hgeo}). Adiabatic holonomies as the ones discussed above then correspond to rotations in  the one-dimensional subspace spanned by the single dark state of $H(t)$, 
\begin{eqnarray}
  |\Dv\rangle=\frac{1}{\sqrt{T_1^2+T_2^2}}(T_2|L\rangle - T_1|R\rangle),
\end{eqnarray}
which shows that by adiabatic variation of $T_1$ and $T_2$, the state of the triple dot can be rotated from, e.g., an electron in the right dot to an electron in the left dot, without intermittent occupation of the central dot at any time. 

For {\em non-zero} ground state energies  $E_R=-E_L>0$ (and $E_C=0$), $H(t)$ no longer has the form Eq.~(\ref{Hgeo}) due to the free part $H_0\equiv \sum_{\alpha=L,C,R}\left[\varepsilon_\alpha|\alpha\rangle\langle \alpha|\right]$ that gives rise to a dynamical phase (or alternatively, in the interaction picture with respect to $H_0$, multiplication of the $T_i$ by fast oscillating phase factors $e^{i(E_C-E_R)t}, e^{i(E_C-E_L)t}$). Still, adiabatic transfer with $H(t)$ is possible for parametric time-dependence of the $T_i(t)$.  The system state follows the adiabatic evolution of its eigenvalues,  Fig. (\ref{steering_fig1.eps}), right. An electron can be transferred adiabatically from the right to the left dot, using the double pulse sequence for $T_1$ and $T_2$ as shown in  Fig. (\ref{steering_fig1.eps}), right. A long $T_1$ pulse, which alone would produce a pair of level crossings, is followed by a shorter $T_2$ pulse which changes the second level crossing into an  {\em anti-crossing}, so that the electron is adiabatically transferred to the center dot. For the transfer from the center dot to the left one the role of $T_1$ and $T_2$ are exchanged which  results in the transfer of the additional electron to the left dot. In this picture, the above case $E_C=E_R=E_L=0$ with the three adiabatic eigenvalues $0$ and $\pm \sqrt{T_1^2+T_2^2}$ corresponds to following the adiabatic rotation of the dark state $|\Dv\rangle$ along the (constant) zero eigenvalue. This was used by Greentree, Cole, Hamilton, and Hollenberg \cite{GCHH04} in their triple dot system and their extension to multi-dot systems. 

The coherent time evolution of the isolated triple-dot is governed by the Schr\"odinger equation
\begin{eqnarray}\label{eq:steering}
|\psi (t)\rangle &=& c_L (t) \exp[-iE_Lt/\hbar] |L\rangle +
                 c_C (t) \exp[-iE_Ct/\hbar] |C\rangle +   c_R (t) \exp[-iE_Rt/\hbar] |R\rangle
\\
\dot{c}_L (t) &=& -i T_1(t) c_C(t) \exp[-i(E_C-E_L)t/\hbar]\nonumber\\
\dot{c}_C (t) &=& -i T_1(t) c_L(t) \exp[-i(E_L-E_C)t/\hbar] -i T_2(t)c_R(t) \exp[-i(E_R-E_C)t/\hbar]\nonumber\\
\dot{c}_R (t) &=& -i T_2(t) c_C(t) \exp[-i(E_C-E_R)t/\hbar]~\nonumber,
\end{eqnarray}
but the coupling to left and right leads can be easily incorporated within a Master equation description \cite{RB01}. Transport from, e.g., the left to the right lead is then followed by a charge leakage to the right lead at the tunnel rate $\Gamma_R$. At the same time, charge tunnels at rate $\Gamma_L$ from the left lead into the left dot whence there is a net charge transport through the triple dot which after the tunnel-couplings sequence (including a `leakage time' of the order of $1/{\rm min}(\Gamma_L,\Gamma_R)$) is returned to the initial state with (almost) the whole charge in the left dot. 

The whole adiabatic transfer scheme relies on the existence of pairs of level crossings and anti-crossings, with a level {\em crossing} corresponding to $T_1$ or $T_2$ becoming zero. If the tunnel rates are kept at non-zero values $T_i<0$ all the time, the previous degeneracies at the level crossings are lifted and the crossings become anti-crossings. In this case,  the transfer mechanism across these points is Landau-Zener tunneling, whereas outside the `nearly crossings' the dynamics remains adiabatic. In the very extreme case of arbitrarily {\em slow} tuning of (never vanishing) $T_i(t)$, the Landau-Zener tunneling becomes exponentially small and there is no transfer of charge at all any longer. 

Estimates for experimentally relevant parameters given in \cite{RB01} assume the ground state energy difference $\hbar \omega_0$ between two adjacent dots to fulfill $\hbar \omega_0 \ll U, \Delta$, where $U$ is the Coulomb charging energy and $\Delta$ the single particle level spacing within a single dot. For $\hbar\omega_0\sim 0.1$meV, one has operation frequencies as $\nu \sim 10^8$s$^{-1}$, with the temperature smearing of the Fermi distribution being negligible if $k_BT \ll \hbar \omega_0 \sim 1K$, thus defining an operation window $h \nu,k_BT \ll \hbar \omega_0 \ll U, \Delta$.

\subsection{Quantum Dissipation and Adiabatic Rotations}
Dissipation clearly has a strong impact on the dynamics of quantum systems. This is very obvious for the `usual' dynamical time-evolution, for example in the damping of quantum mechanical coherent oscillations of charge qubits as first observed by  Nakamura, Pashkin, and Tsai \cite{NPT99} in superconductors and by Hayashi and co-workers in semiconducting quantum dots (\cite{Hayetal03}, see below). Decoherence also occurs during adiabatic rotations and can theoretically be dealt with in the usual quantum dissipative framework, i.e., using Master equations, spin-boson models, path integrals etc. 

Loss and DiVincenzo \cite{LD98} introduced quantum gates based on the electron spin in quantum dots. Dephasing of spin degrees of freedom due to spin-orbit coupling or the coupling to nuclear spins is expected to be much weaker than dephasing of charge superpositions, but spin and charge can become coupled during switching operations  whereby charge dephasing also influences spin-based qubits. Adiabatic quantum computation with Cooper pairs, including adiabatic controlled-NOT gates, was proposed by Averin \cite{Ave98,Ave99}. For adiabatic one- or two-qubit operations, one has a close analogy with the dissipative Landau-Zener-problem (cf. the Review by Grifoni and H\"anggi \cite{GH98} for further references). Dephasing in geometrical quantum operations was discussed by  Nazir, Spiller, and  Munro \cite{NSM02}.

\subsubsection{Dissipative Adiabatic Rotations in Quantum Dots}\label{section_adiabaticrotation}
Brandes and Vorrath \cite{BV02} investigated the role of dissipation for  one- and two-qubit adiabatic state rotations in the double dot model from section \ref{section_transport}. Coherent adiabatic transfer without dissipation in the charge qubit span$(\{|L\rangle, |R\rangle \}$ is described by the time-dependent Hamiltonian
\begin{eqnarray}\label{H0define}
  H_0^{(1)}(t) &=& \frac{\varepsilon(t)}{2}\sigma_z + T_c(t) \sigma_x, \quad
\varepsilon(t)=\varepsilon_0+\varepsilon_1 \cos \Omega t,\quad 
T_c(t)=-T_c \exp [-(t - t_0)^2/\tau^2],
\end{eqnarray}
corresponding to a change of the bias $\varepsilon(t)$ with a simultaneous switching  of the tunnel coupling $T_c(t)$ between the dots. If the rotation is slow, $\Omega,\tau^{-1},t_0^{-1}\ll \Delta/\hbar$, an initial ground-state $|L\rangle$ of the system is rotated into the instantaneous superposition $|-\rangle$, Eq. (\ref{eq:twobytworesult}). The time $t$ is a parameter in approximate expectation values like 
\begin{eqnarray}
  \langle \sigma_z \rangle_{\rm ad} = -\varepsilon(t)/\Delta(t)
\end{eqnarray}
(this is Crisp's solution for the adiabatic following of an atom in a near resonance light pulse \cite{Crisp73,Allen}), where $\Delta(t)$ is the time-dependent
adiabatic level splitting between $|-\rangle$ and $|+\rangle$, Eq. (\ref{eq:twobytworesult}). The exact numerical solution for $\langle \sigma_z \rangle(t)$  exhibits the expected quantum mechanical oscillations with frequency $\Delta(t)/\hbar$ around the adiabatic value, which are strongest when the tunnel coupling is fully switched on, cf. Fig. (\ref{2surface}), right.

\begin{figure}[t]
\includegraphics[width=0.3\textwidth]{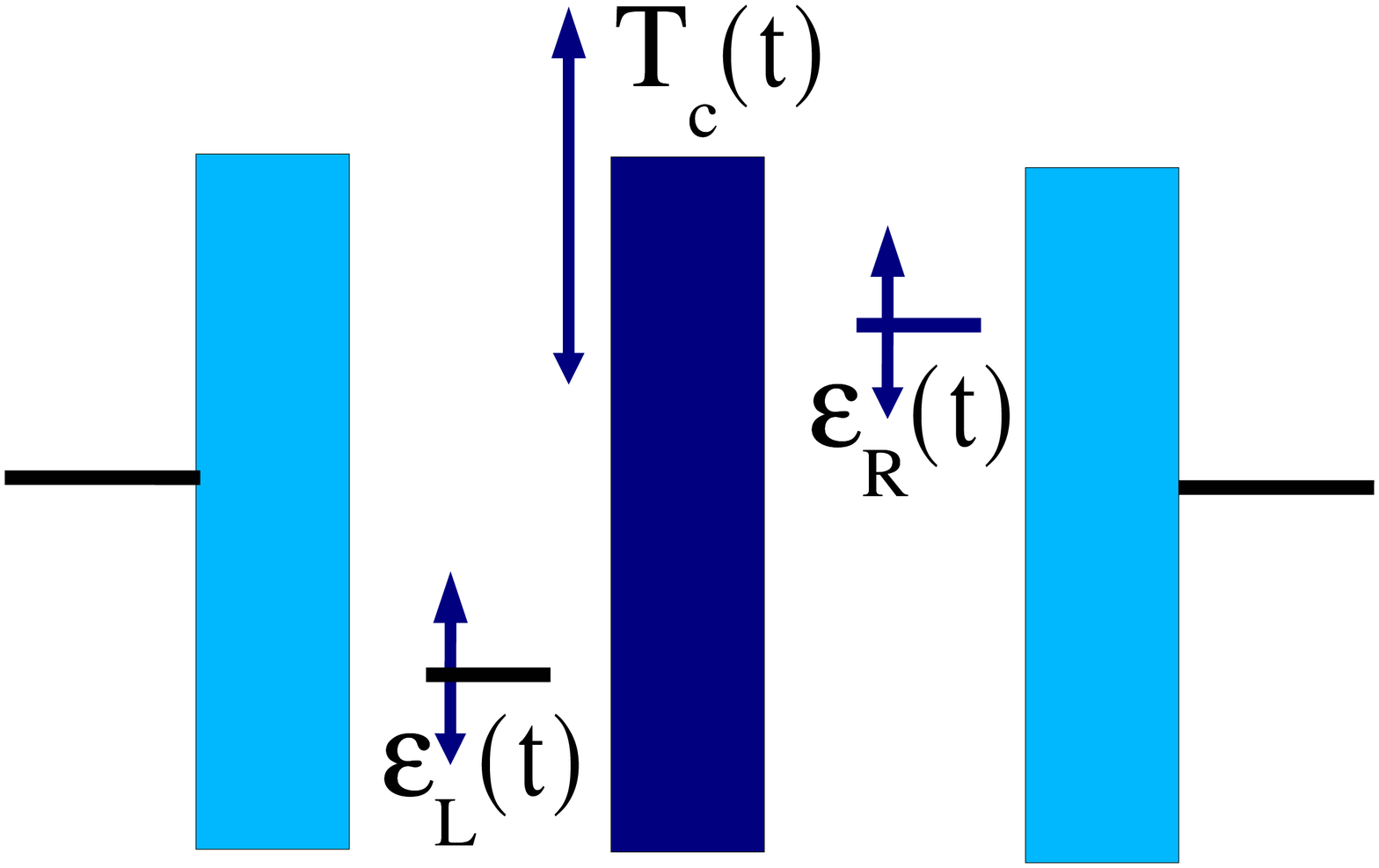}
\includegraphics[width=0.3\textwidth]{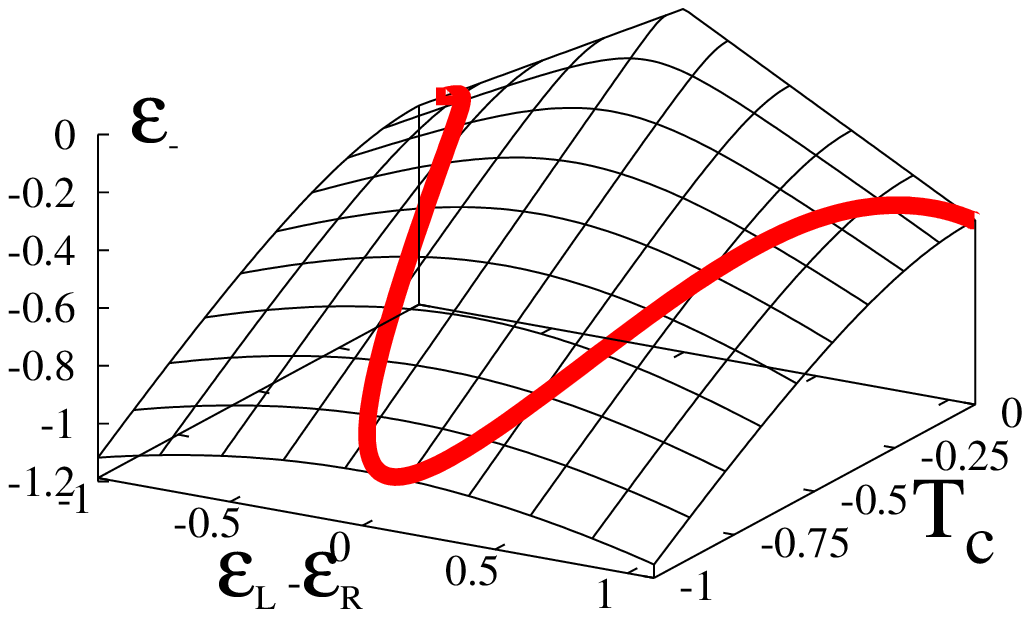}
\includegraphics[width=0.4\textwidth]{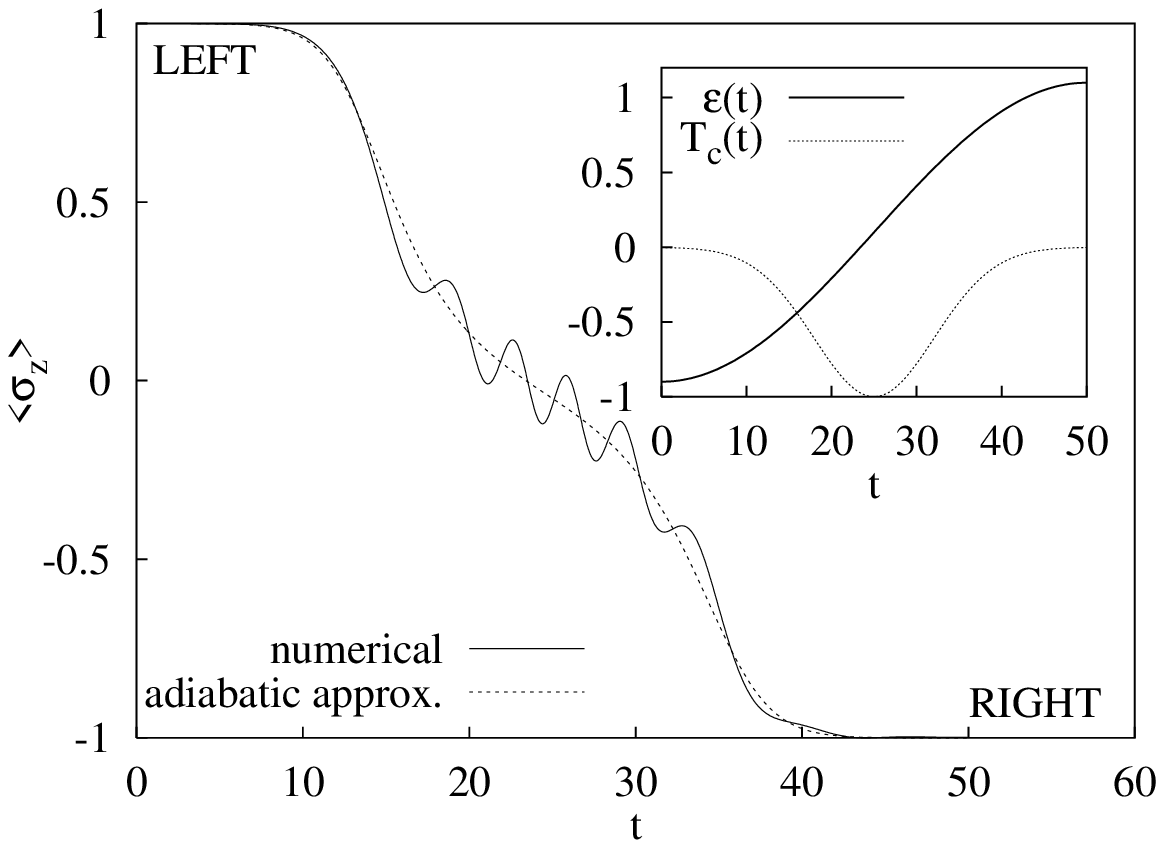}
\caption[]{\label{2surface}{\bf Left:} Double dot  with time-dependent energy level difference $\varepsilon(t)=\varepsilon_L(t)-\varepsilon_R(t)$ and tunnel matrix element $T_c(t)$, connected to electron reservoirs. {\bf Center:} Surface of the lower energy eigenvalue  $\varepsilon_-$ of the two-level Hamiltonian $H_0^{(1)}(t)$, Eq. (\ref{H0define}). To adiabatically transfer an electron from the left to the right dot, $\varepsilon$ and 
$T_c$ are varied as a function of time as in Eq. (\ref{H0define}), corresponding to the curve on the $\varepsilon_-$ surface. {\bf Right:} Inversion $\langle \sigma_z \rangle$ for zero dissipation in a two-level system, Eq. (\ref{H0define}),  with time-dependent tunnel matrix element $T_c(t)$ and energy splitting $\varepsilon(t)$, cf inset. Energies (times) are in units of  the amplitude $T_c$ ($\hbar/T_c$) in Eq. (\ref{H0define}); the other parameters are $t_0=25$, $\Omega= \pi/(2t_0)$, $\tau=10$, $\varepsilon_0=0.1$, $\varepsilon_1=-1$. Dotted line: adiabatic approximation $\langle \sigma_z \rangle_{\rm ad} = -\varepsilon(t)/\Delta(t)$ (see text). From \cite{BV02}}
\end{figure} 

In \cite{BV02}, the influence of dissipation on this adiabatic rotation was described using the spin-boson Hamiltonian, \ Eq.~(\ref{modelhamiltonian}). Generalizing the strong coupling (POL) approach, section \ref{section_polaron}, to time-dependent $\varepsilon$ and $T_c$, one obtains an integro-differential equation \cite{GH98}
\begin{eqnarray}\label{polaronsigmaz}
  \frac{\partial}{\partial t}\langle \sigma_z \rangle_t 
&=&-\int_0^t dt' \sum_{\pm}
 \left[1\pm\langle \sigma_z \rangle_{t'}\right]
(\pm )2T_c(t)T_c(t') {\rm Re } 
\left\{e^{\pm i\int_{t'}^t ds \varepsilon(s)} C(t-t')\right\},
\end{eqnarray}
which can be solved by standard numerical techniques and compared with the perturbative (PER) approach, section \ref{section_perturbation}. Due to Landau-Zener tunneling from the adiabatic ground state $|-\rangle$ to the excited state $|+\rangle$, there is always a finite albeit small probability $P_L$ for the electron to remain in the left dot even in absence of dissipation. Introducing the deviation $\delta\langle \sigma_z\rangle_t \equiv\langle \sigma_z\rangle_t +1$ from the ideal,  non-dissipative value $-1$ of $\langle \sigma_z \rangle$ after the rotation, one can discuss the trade-off between too fast (Landau-Zener transitions become stronger), and too slow swap operations where inelastic transitions to the excited level will have sufficient time to destroy the coherent transfer.

Due to  the time-dependence of $\varepsilon$ and $T_c$, one has to go beyond the simple Bloch equation description of decoherence \cite{SS01} by introducing  a unitary transformation of  the original Hamiltonian (\ref{modelhamiltonian}) into the $|\pm\rangle$ basis, again considering the time $t$ as a slowly varying parameter \cite{GH98,BV02}. The deviation $\delta \langle \sigma_z \rangle_t$ of the inversion due to the coupling to the bosons is then given by
\begin{eqnarray}\label{deltazanalytic}
     \delta \langle \sigma_z \rangle_t &=& 2\int_0^{\infty}d\omega J(\omega) \Big\{
n_B(\omega) f(\omega,t) + \left[1+n_B(\omega)\right] f(-\omega,t)\Big\} \nonumber\\
f(\omega,t) &\equiv&\left|\int_0^t dt' \frac{T_c(t')}{\Delta(t')}
e^{-i\int_{0}^{t'}ds \left[\Delta(s)-\omega\right]}\right|^2,
\end{eqnarray}
where $n_B(\omega)$ is the Bose equilibrium distribution at inverse temperature $\beta$ and $J(\omega)$ the boson spectral density, cf. section \ref{section_spectraldensity}. Analytical results are obtained for exact `Rabi rotations', 
\begin{eqnarray}\label{harmonicpulses}
  T_c(t) &=&-\frac{\Delta}{2}\sin \Omega t,\quad \varepsilon(t) = -\Delta \cos \Omega t,
\end{eqnarray}
for which one has an exact solution in absence of dissipation, determined by an ellipse in the $\varepsilon,T_c$ plane with {\em constant} excitation energy $\Delta=\sqrt{\varepsilon^2+4T_c(t)^2}$ to the excited state. For a pulse of length $t_f=\pi/\Omega$, one obtains 
$  f\left(\omega,t_f\right)  \to \frac{c}{\Omega}\delta(\Delta-\omega)$ with $c = \frac{\pi^3J_{3/2}(\pi)}{4\sqrt{2}}$ in the adiabatic limit $\Omega/\Delta\to 0$ and thereby
\begin{eqnarray}\label{sigmazanalytic}
  \delta \langle \sigma_z \rangle_{f}&\approx& 
1-\left[ \left(\frac{\Delta}{\omega_R}\right)^2 +
\left(\frac{\Omega}{\omega_R}\right)^2 \cos\left(\frac{\pi \omega_R}{\Omega}\right) \right]
+ 4.94 \frac{J(\Delta)/\Omega}{\exp({\beta\Delta})-1},\quad{\Omega}\ll{\Delta}, 
\end{eqnarray}
where $\omega_R\equiv\sqrt{\Omega^2+\Delta^2}$. The inversion change $\delta \langle \sigma_z \rangle_{f}$ as a function of the pulse frequency $\Omega$ is shown in Fig. (\ref{ddotzener_figure4.eps}). The $1/\Omega$ dependence  of the dissipative contribution  to $\delta \langle \sigma_z \rangle_{f}$ is clearly visible at small $\Omega$, indicating that for too long pulse duration the electron swap remains incomplete due to incoherent dissipation. On the other hand, if the pulse duration is too short (larger $\Omega$), the oscillatory
coherent contribution from $\langle \sigma_z \rangle_f^{\rm Rabi}$ dominates.

\begin{figure}[t]
\includegraphics[width=0.5\textwidth]{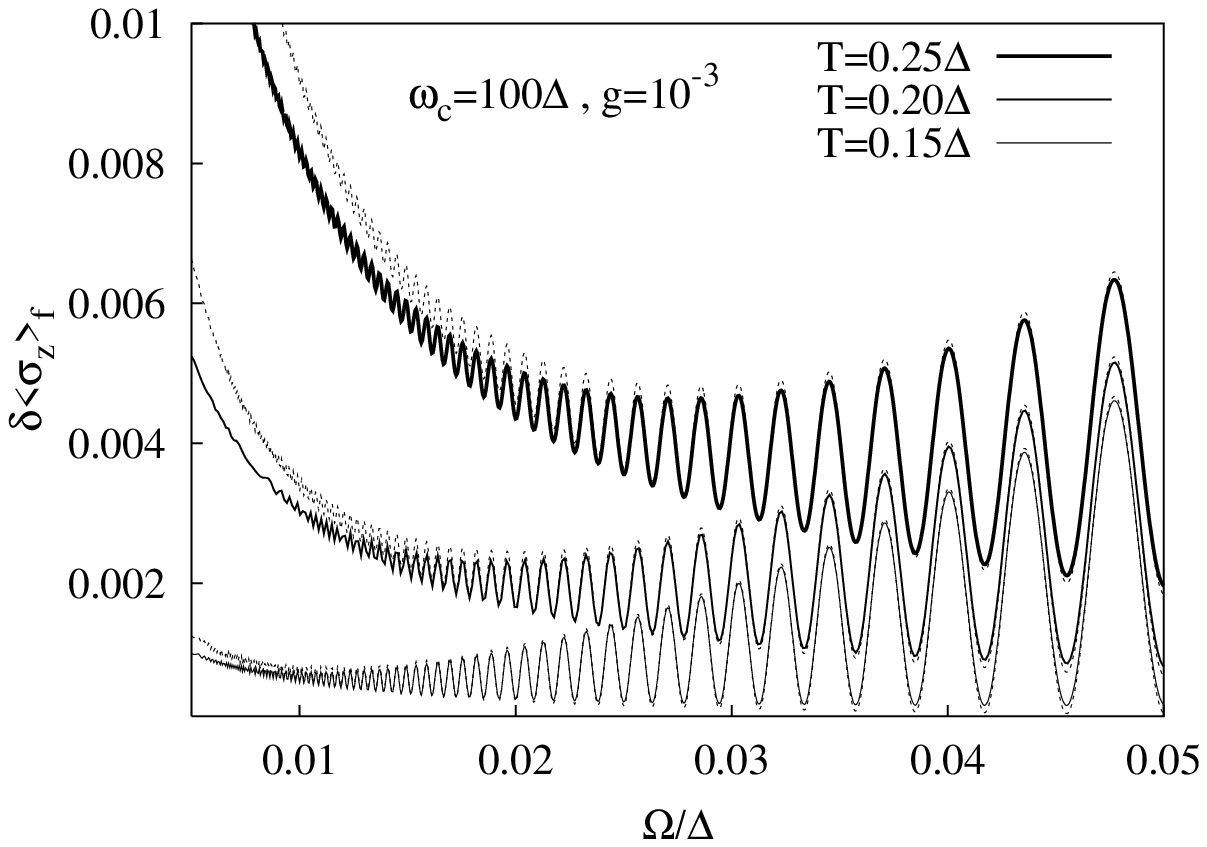}
\includegraphics[width=0.5\textwidth]{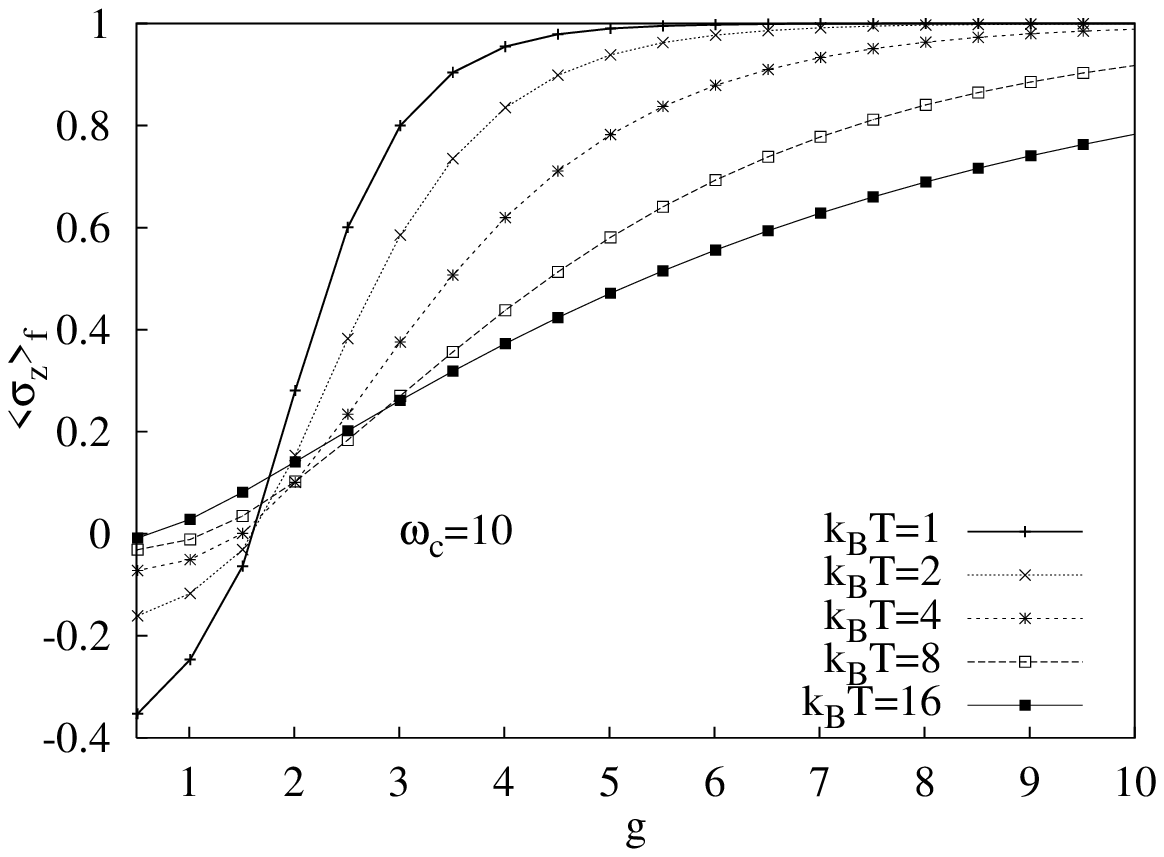}
\caption{\label{ddotzener_figure4.eps}{\bf Left:} 
Inversion change $\delta\langle \sigma_z\rangle_f$ after time $t_f=\pi/\Omega$ for 
sinusoidal pulses Eq.(\ref{harmonicpulses}) as 
obtained from the Master equation, cf. section \ref{section_perturbation}, with Ohmic dissipation,
\ Eq.~(\ref{Jomegageneric}) with $s=1$. 
Dotted curves correspond to the analytical prediction, Eq. (\ref{sigmazanalytic}). {\bf Right:} Inversion $\langle \sigma_z\rangle$ for strong electron-boson coupling (POL regime, Ohmic dissipation) 
after application of the pulse $T_c(t)$ and $\varepsilon(t)$, Eq. (\ref{H0define}), with crossover at $g\equiv2\alpha\approx 2$ where the temperature dependence changes. From \cite{BV02}.}
\end{figure}
The  Rabi rotation, Eq. (\ref{harmonicpulses}), keeps the energy difference to the excited state $|+\rangle$ constant throughout the adiabatic rotation. If therefore $\Delta$ is chosen to coincide with a zero of $J(\omega)$ (as occurs for phononic cavities, cf. section \ref{section_cavity}), the dissipative contribution to \ Eq.~(\ref{sigmazanalytic}){} vanishes and one obtains a {\em decoherence-free manifold} in the parameter space of the system. 

Another interesting case is the zero temperature limit of the weak coupling form \ Eq.~(\ref{deltazanalytic}), where only the  term with $f(-\omega,t_f)$ remains due to the small, but finite probability for spontaneous emission {\em during} Landau-Zener transitions from $|-\rangle$ to $|+\rangle$ \cite{GH98,BV02}. On the other hand, for strong electron-boson coupling there is a cross-over in the temperature dependence of the POL result, \ Eq.~(\ref{polaronsigmaz}){}, for $\langle \sigma_z \rangle_{f}$, cf. Fig. (\ref{ddotzener_figure4.eps}), right: for couplings $\alpha\lesssim 1$ (Ohmic dissipation $s=1$, \ Eq.~(\ref{Jomegageneric})), a temperature increase leads to an increase of $\langle \sigma_z \rangle_{f}$, which is as in
the weak coupling case. However, above $\alpha\gtrsim 1$, the temperature dependence changes in that
larger $k_BT$ lead to smaller values of  $\langle \sigma_z \rangle_{f}$ because the system tends to remain localized  in the left dot state $|L\rangle$ and no tunneling to the right state $|R\rangle$ occurs. In this regime, higher temperatures destroy the localization and lead to smaller $\langle \sigma_z \rangle_{f}$, which is consistent with the transition ($\alpha=1$) in the dissipative two-level dynamics \cite{Legetal87} for static bias and tunnel coupling.

\subsubsection{Adiabatic Quantum Pumping}\label{section_dotpump}
A combination of adiabatic rotations and electron transport in the above scheme was used \cite{BV02} to
extract the inversion $\delta\langle \sigma_z \rangle_f$, \ Eq.~(\ref{sigmazanalytic}), from the average
current $\langle I \rangle$ pumped through the system. The pumping cycle separates 
the quantum mechanical time evolution of the two-level system  from a merely `classical' decharging and charging process. An additional electron in the left dot and an adiabatic rotation of the parameters $(\varepsilon(t),T_c(t))$ is performed in the `Safe Haven' of the Coulomb- and the Pauli-blockade \cite{BR00} with the left and right 
energy levels of the two dots well below the chemical potentials $\mu=\mu_L=\mu_R$ of the leads. The cycle continues with closed tunnel barrier $T_c=0$ and increasing $\varepsilon_R(t)$ such that the two dots then are still in a superposition of the left and the right state. The subsequent lifting of the right level above the chemical potential of the right leads constitutes a measurement of that superposition (collapse of the wave-function): the electron is either in the right dot (with a high probability $1-\frac{1}{2}\delta\langle \sigma_z \rangle_f$) and tunnels out, or the electron is in the left dot (and nothing happens because the left level is still below $\mu$ and the system is Coulomb blocked). For tunnel rates $\Gamma_R,\Gamma_L\gg t_{\rm cycle}^{-1}$,  the precise value of  $\Gamma_R,\Gamma_L$ and the precise shape of the $\varepsilon(t)$-pulse for $t_f<t<t_{\rm cycle}$  has no effect on the total charge transferred within one cycle. With the probability to transfer one electron from the left to the right in one cycle being $1-\frac{1}{2}\delta\langle \sigma_z \rangle_f$, on the average an electron current 
\begin{eqnarray}\label{pumpcurrent}
  \langle I \rangle = -e \left[1-\frac{1}{2}\delta\langle \sigma_z \rangle_f\right]{t_{\rm cycle}}^{-1}
\end{eqnarray}
then flows from left to right. This scheme, with its pulse-like changes of the parameters $\varepsilon,T_c$ and 
the leads acting only as classical measurement devices, has great similarities with the scheme used in the
Nakamura {\em et al.} \cite{NPT99} interference experiment in a superconducting Cooper pair box, and the pumping sequence by Hayashi {\em et al.} \cite{Hayetal03} in their one-qubit interference experiment in double quantum dots, cf. the next section.

\subsubsection{Experiments in One-Qubit Double Quantum Dots}\label{section_hayashi}

The NTT group with Hayashi, Fujisawa, Cheong, Jeong, and Hirayama \cite{Hayetal03} successfully realized coherent time evolution of superposition states in a single charge qubit based on semiconductor double quantum dots. Similar to the experiment in superconducting charge qubits by Nakamura {\em et al.} \cite{NPT99} and to the pumping scheme of section \ref{section_dotpump}, they used a pulse technique to switch the source-drain voltage from large bias $V_{SD}$ (electrons can tunnel in) to zero bias (the double dot is isolated), cf. Fig. (\ref{Hayetal03_fig1.eps}) left. At the same time, the inter-dot bias $\varepsilon$ was also switched, giving rise to a time-dependent Hamiltonian
\begin{eqnarray}\label{H_Hayashi}
  H(t) = \frac{\varepsilon(t)}{2}\sigma_z + T_c \sigma_x
\end{eqnarray}
which described the isolated double quantum dot. Coulomb blockade prevented other electrons to enter the system in the isolated phase (d), Fig. (\ref{Hayetal03_fig1.eps}) left, with the coherent time-evolution of the system for $\varepsilon=0$ only disturbed by inelastic processes such as phonon coupling, or co-tunneling processes. Restoring a large bias $V_{SD}$ after the pulse time $t_p$, Fig. (\ref{Hayetal03_fig1.eps} left e), provided a strong measurement and, since repeated many times at frequency $f_{\rm rep}=100$ MHz, a read-out of the charge state of the system in the form of an electric current $I_p$.

\begin{figure}[t]
\begin{center}
\includegraphics[width=0.45\textwidth]{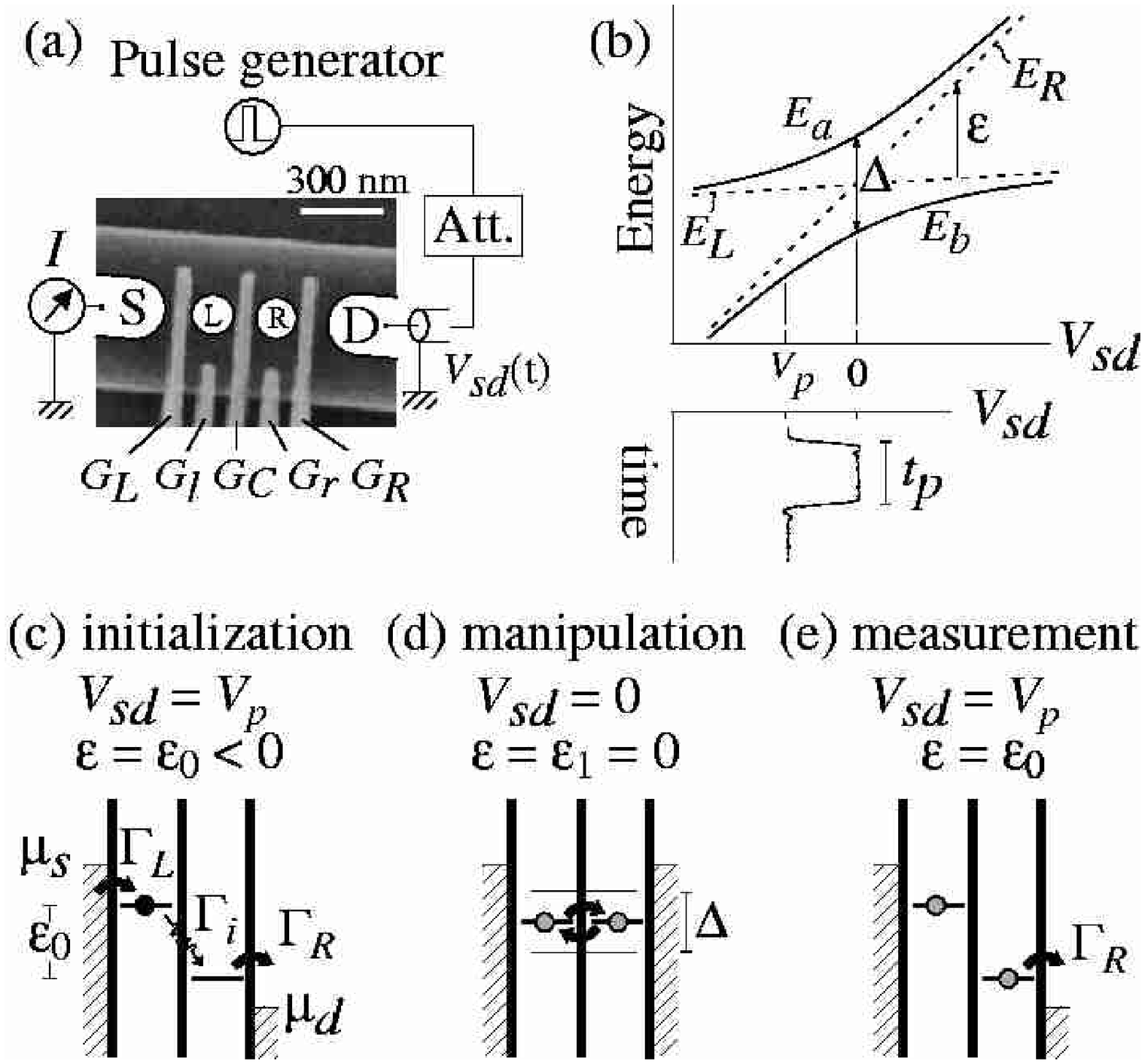}
\includegraphics[width=0.45\textwidth]{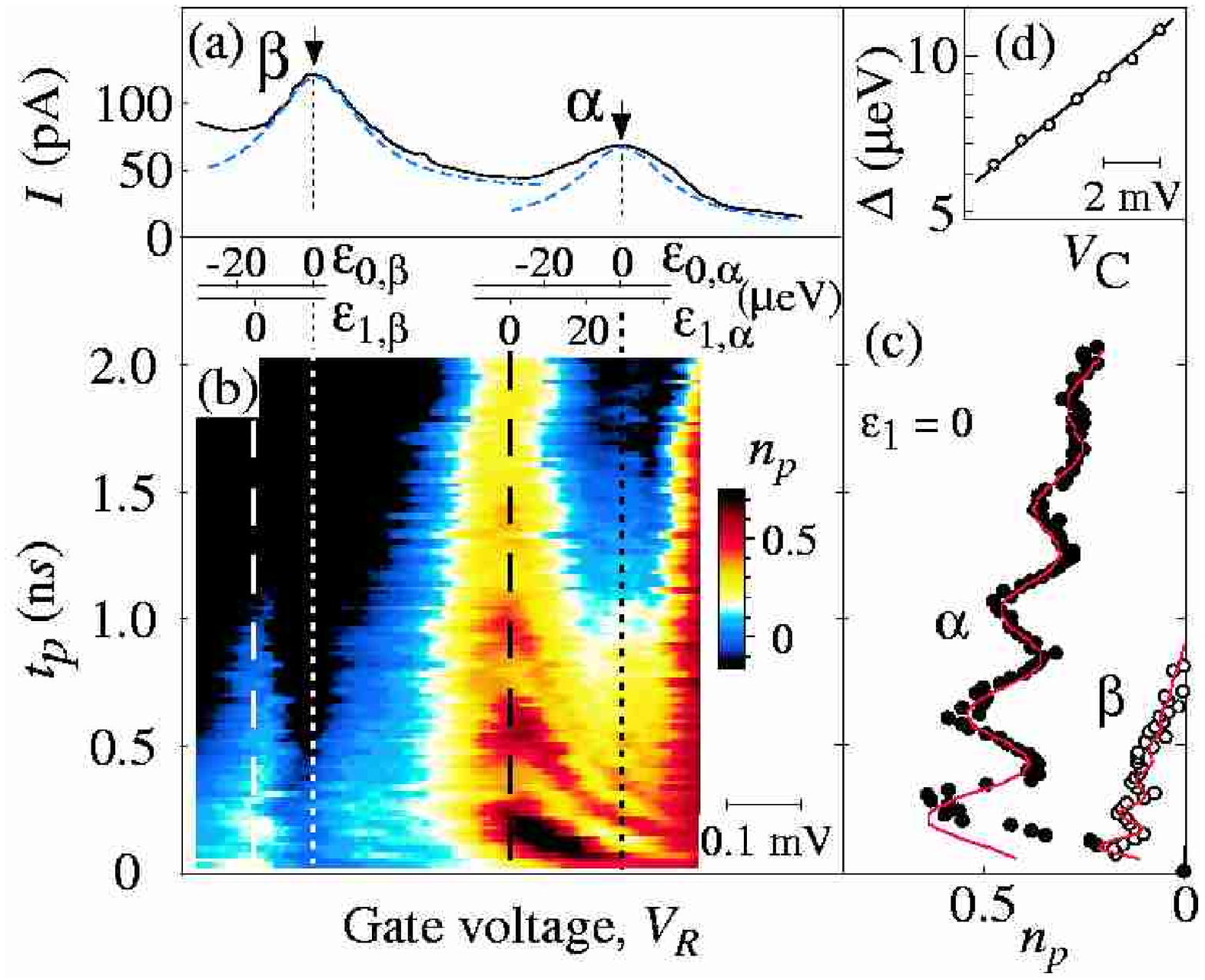}
\end{center}
\caption{\label{Hayetal03_fig1.eps}{\bf Left:} (a) Double dot in the experiment by Hayashi {\em et al.} \cite{Hayetal03} with tunable source-drain voltage $V_{SD}(t)$, energy splitting $\varepsilon(t)$, and tunnel coupling $T_c (=\Delta/2$ in \cite{Hayetal03}), giving rise to the time-dependent Hamiltonian, \ Eq.~(\ref{H_Hayashi}), and the sequence (c-e) with quantum mechanical oscillations between left (L) and right (R) dot, (d). {\bf Right:} Non-linear current profile as a function of $\varepsilon$ near the two resonance peaks, $\alpha$ and $\beta$. (b) Mean dot occupancy $n_p\equiv I_p/ef_{\rm rep}$ as a function of $V_R$ (inter-dot bias $\varepsilon$) and pulse duration $t_p$. (c) The main result: coherent oscillations in the two two-level systems, $\alpha$ and $\beta$. (d) Central gate voltage dependence of tunnel coupling $\Delta=2T_c$. From \cite{Hayetal03}.}
\end{figure}

The measurements were carried out at two resonant tunneling peaks $\alpha$ and $\beta$, cf. Fig. (\ref{Hayetal03_fig1.eps} right a), each corresponding to an effective two-level system as realized within the many-electron ($N_L\sim N_R \sim 25$)  double-dot (charging energy $E_c \sim 1.3$ meV) at electron temperatures $T_e \sim 100$ mK and a magnetic field 0.5 T. The  curves of $n_p\equiv I_p/ef_{\rm rep}$ as a function of pulse length $t_p$, Fig. (\ref{Hayetal03_fig1.eps} right c), were extracted from the $t_p$-$V_R$-diagram, Fig. (\ref{Hayetal03_fig1.eps} right b), where the gate voltage effectively tuned the bias $\varepsilon$ during the coherent time-evolution phase of the isolated double dot. From these, decoherence times $T_2$ where extracted using a fit $n_p(t_p)$ with an exponentially damped cosine function. Hayashi and co-workers then discussed three possible dephasing mechanism: first, background charge fluctuations and gate voltage noise was held responsible for random fluctuations of $\varepsilon$, leading to strong dephasing for large $\varepsilon$. Second, co-tunneling rates where found to be comparable to the fitted $T_2^{-1}$ rates for large tunnel couplings $\Gamma=\Gamma_R\sim\Gamma_L\sim 30\mu$eV, but to have a minor effect at $\Gamma\sim 13\mu$eV. Third,  dephasing rates $\gamma_p$ from electron-phonon coupling were found to play a major role for lattice temperatures above 100 mK, where the boson spectral density with Ohmic dissipation ($s=1$) and a coupling parameter $g=2\alpha=0.03$ was used to calculate $\gamma_p$ according to \ Eq.~(\ref{gammapdef}).

\subsubsection{Dissipative Quantum Oscillations in Double Dots}

\begin{figure}[t]
\includegraphics[width=0.7\columnwidth]{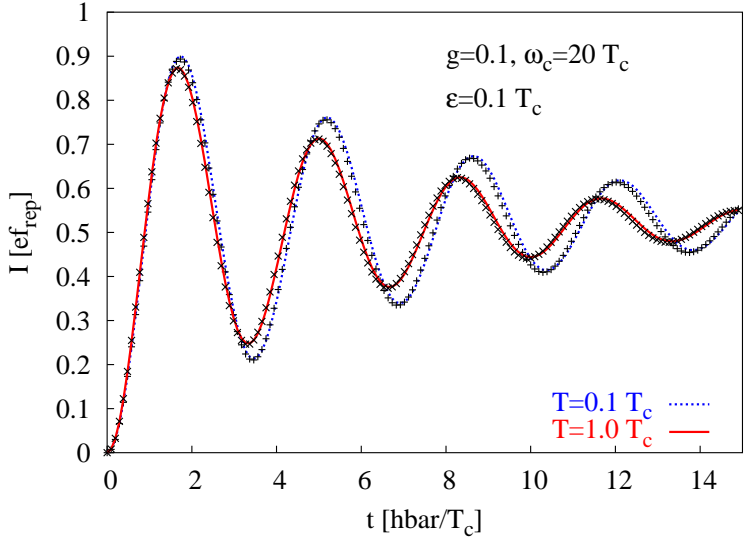}
\caption[]{\label{ntteps1}Calculated average current due to quantum oscillations of an electron in a double quantum dot after time $t$, starting in the left dot at $t=0$.  Ohmic dissipation at temperature $T$, coupling parameter $g\equiv2\alpha=0.1$, and cutoff frequency $\omega_c$. The curves represent the numerical solutions of the Bloch equations whereas the crosses correspond to the analytical
solutions Eq. (\ref{npanalytic}). 
}
\end{figure} 

The damped oscillations of $n_p(t_p)$, as observed  in the experiment by Hayashi {\em et al.}, also follow  from analytical calculations for the time evolution of the double dot system from the Master equation with weak dissipation, cf. section (\ref{section_perturbation}). For the isolated dot, one has to set $\Gamma_R=\Gamma_L=0$ and the initial condition $\langle n_L\rangle_0=1$, $\langle p\rangle_0=0$ in  Eq.~(\ref{nLeq}),  Eq.~(\ref{nReq}), Eq.~(\ref{pequation}), which by Laplace transformation yield 
\begin{eqnarray}\label{laplace_nR}
  \hat{n}_{R}(z)&=&\frac{(z+\gamma_p)
(2T_c^2-2T_c\Im\gamma_+)-2\varepsilon T_c\Re\gamma_+}{
z\left[z\left\{(z+\gamma_p)^2+\varepsilon^2\right\} 
-2\varepsilon T_c\Re(\gamma_++\gamma_-)
\right.+\left.(z+\gamma_p)\left\{4T_c^2 -2T_c \Im(\gamma_++\gamma_-)\right\}
\right]},
\end{eqnarray}
with the rates $\gamma_p$ and $\gamma_{\pm}$ defined in  Eq.~(\ref{gammapdef}). The zeroes of the denominator in \ Eq.~(\ref{laplace_nR}) to first order in the dimensionless coupling constant $\alpha$ are
\begin{eqnarray}
  z_0&=&0,\quad z_1=-\Gamma_p,\quad z_{\pm}= -\frac{\Gamma_p}{2}-\gamma_1
\pm i E\\
\Gamma_p&\equiv& 2\pi\frac{T_c^2}{\Delta^2}J(\Delta)\coth \frac{\beta\Delta}{2},\quad
\gamma_1\equiv\delta_{s,1}\frac{2\alpha\pi\varepsilon^2}{\beta\Delta^2},\quad
E\equiv \left[\Delta-\frac{T_c}{\Delta}\mbox{Im}(\gamma_++\gamma_-)\right]
\end{eqnarray}
with $\beta=1/(k_BT)$ and again $\Delta\equiv \sqrt{\varepsilon^2+4T_c^2}$ as the  level splitting of the double dot. Note that there is a temperature dependent renormalization (Lamb shift) of the level splitting from the term $-(T_c/\Delta)\mbox{Im}(\gamma_++\gamma_-)$ in the energy $E$ which determines the period of the oscillations. 
By simply Laplace back-transforming \ Eq.~(\ref{laplace_nR}), an explicit solution to lowest order in $\alpha$ is obtained for $\langle n_R\rangle_t \equiv n_p(t))$,
\begin{eqnarray}\label{npanalytic}
\langle n_R\rangle_t&\approx&
\frac{2T_c^2}{E^2}\left\{\kappa
+
\left[\kappa\frac{\gamma_1}{\Gamma_p}
-\frac{\varepsilon\mbox{Re}\gamma_+}{T_c\Gamma_p}
\right]\left(1-e^{-\Gamma_p t}\right)\right.\\
&-&\left.e^{-(\frac{\Gamma_p}{2}+\gamma_1)t}\left[
\left(\frac{\kappa\Gamma_p}{2E}-\frac{\varepsilon\mbox{Re}\gamma_+}{ET_c} \right) \sin Et 
+\kappa\cos Et \right]\right\},\quad
\kappa\equiv1-\frac{\mbox{Im}\gamma_+}{T_c}.
\end{eqnarray}
As shown in Fig. (\ref{ntteps1}), this  perfectly agrees with the numerical solution of the Master equation which should be called Bloch equation in this context as only two levels are involved, cf. Eq.~(\ref{nLeq}),  Eq.~(\ref{nReq}), Eq.~(\ref{pequation}). 

Non-Markovian corrections to this Born-Markov theory have been calculated recently by Loss and DiVincenzo \cite{LD03}.

\subsubsection{Charge Shelving and Adiabatic Fast Passage}
Greentree, Hamilton, and Green \cite{GHG04} suggested a pumping scheme with bias spectroscopy similar to the optical Autler-Townes experiment. They considered  a three-level Hamiltonian where the right state $|R\rangle$ of a double well $|L\rangle,|R\rangle$ qubit is coupled to an additional probe-state, $|p\rangle$, 
\begin{eqnarray}
  H(t) = \varepsilon_p(t) |p\rangle \langle p| - T_c (|L\rangle \langle R| + |L\rangle \langle R|) -  T_p(t) (|R\rangle \langle p| + |p\rangle \langle R|),
\end{eqnarray}
cf. Eq.~(\ref{H_threedot}) for the triple dot in section \ref{section_triple_dot}. Similar to the time-dependent variation of $\varepsilon(t)$ and $T_c(t)$ in the double dot system, \ Eq.~(\ref{H0define}){}, they demonstrated pumping in the form of {\em charge shelving} by linearly increasing $\varepsilon_p(t)$ and simultaneously switching the tunnel coupling $T_p(t)$ on and off at fixed $T_c$: an initially anti-symmetric eigenstate $\Psi(t=0)=(1/\sqrt{2})(|L\rangle-|R\rangle)$ of the $|L\rangle,|R\rangle$ qubit is driven into $|p\rangle$, the population of which can adiabatically approach unity on a short time scale of a few $T_c^{-1}$. On the other hand, for the initially symmetric state $\Psi(t=0)=(1/\sqrt{2})(|L\rangle+|R\rangle)$ with lower energy, the final population of $|p\rangle$ is very small. Therefore, the third state $|p\rangle$ provides a read-out for the qubit which is reversible in absence of dissipation such that an electron  can be pumped back into the anti-symmetric state and thereby reset the qubit.

\subsubsection{Spin Qubit Swaps}
The adiabatic swapping model in section \ref{section_adiabaticrotation} can also be applied to study decoherence due to charge dephasing in spin-based two-qubit systems, where spin and charge become coupled during switching operations \cite{LD98,BLD99}. An example is the Loss-DiVincenzo proposal for quantum operations with spin states of coupled single-electron quantum dots \cite{LD98}. Thorwart and H\"anggi \cite{TH02} discussed dissipation and decoherence in quantum XOR gates within a numerical scheme, predicting gate fidelities to be very sensitive to the dissipative bath coupling constant, but only weakly on temperature. Recently, Requist, Schliemann,  Abanov, and Loss calculated corrections to adiabaticity due to double occupancy errors of two quantum dot spin-qubits \cite{RSAL04}.

Schliemann, Loss, and MacDonald \cite{SLM01} suggested a swap operation  where two electrons with spin  are localized on two coupled quantum dots $A$ and $B$, giving rise to a basis of six states, with four basis vectors with the two electrons on different dots (spin singlet and triplets),
\begin{eqnarray}
|S_1\rangle &\equiv&2^{-1/2}(c_{A\uparrow}^{\dagger}c_{B\downarrow}^{\dagger}
-c_{A\downarrow}^{\dagger}c_{B\uparrow}^{\dagger})|0\rangle\nonumber\\
|T^{-1}\rangle&\equiv&c_{A\downarrow}^{\dagger}c_{B\downarrow}^{\dagger}|0\rangle,\quad
|T^{1}\rangle \equiv c_{A\uparrow}^{\dagger}c_{B\uparrow}^{\dagger}|0\rangle,\quad
|T^0\rangle \equiv 2^{-1/2}(c_{A\uparrow}^{\dagger}c_{B\downarrow}^{\dagger}
+c_{A\downarrow}^{\dagger}c_{B\uparrow}^{\dagger})|0\rangle, 
\end{eqnarray}
and two states with two electrons on dot $A$ (`left') or dot $B$ (`right'), 
\begin{eqnarray}\label{LRdefine}
  |L\rangle &\equiv& c^{\dagger}_{A\uparrow}c^{\dagger}_{A\downarrow}|0\rangle
= 2^{-1/2}\left[ |S_2\rangle + |S_3\rangle \right],\quad
  |R\rangle \equiv c^{\dagger}_{B\uparrow}c^{\dagger}_{B\downarrow}|0\rangle
= 2^{-1/2} \left[ |S_2\rangle - |S_3\rangle \right],
\end{eqnarray}
which are superpositions of two spin singlets $|S_{2,3}\rangle \equiv 2^{-1/2} (c_{A\uparrow}^{\dagger}c_{A\downarrow}^{\dagger} \pm c_{B\uparrow}^{\dagger}c_{B\downarrow}^{\dagger})|0\rangle $
that differ in their orbital wave function.

\begin{figure}[t]
\begin{center}
\includegraphics[width=0.7\textwidth]{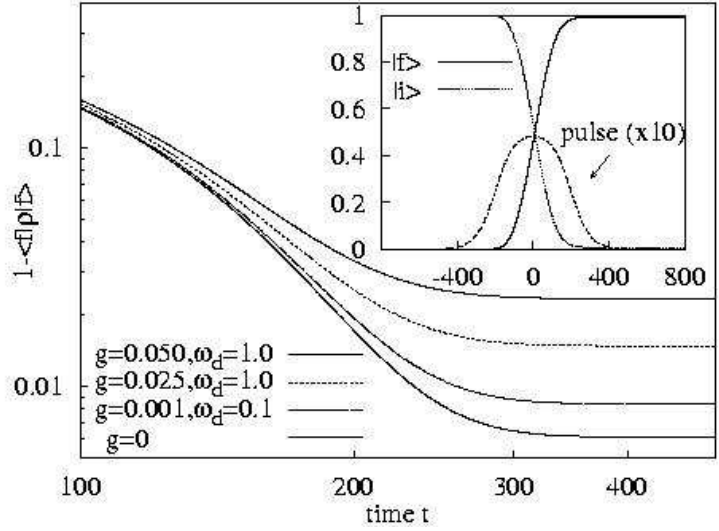}
\end{center}
\caption{\label{ddotzener_figure7.eps}$1-\langle f |\rho(t)| f \rangle$ (fidelity) for a two-electron double dot qubit swap from initial state $|i\rangle$ to final state $|f\rangle$, Eq.~(\ref{swapdefinition}), as a function of time (in units of $U_H/\hbar$, $U_H$: on-site Coulomb repulsion); electron-phonon spectral density \ Eq.~(\ref{Jmicro}) with coupling $g=2 \alpha_{\rm piezo}$ and $\omega_d=c/d$. Inset: $\langle i |\rho(t)| i \rangle$ and $\langle f |\rho(t)| f \rangle$ for $g=0$. From \cite{BV02}.}
\end{figure}
 During a swap operation from an initial state $|i\rangle$ to a final state $|f\rangle$,
\begin{eqnarray}\label{swapdefinition}
|i\rangle \equiv  \frac{1}{\sqrt{2}}\left[|T^0\rangle + |S_1\rangle\right]
\to |f\rangle \equiv \frac{1}{\sqrt{2}}\left[|T^0\rangle - |S_1\rangle\right],
\end{eqnarray}
(which can be achieved \cite{LD98,BLD99} by an adiabatically opening and then closing of 
the tunnel barrier between the two dots as a function of time), 
{\em charge} decoherence occurs for intermediate, doubly occupied states in  span$\{|L\rangle,|R\rangle\}$ (= span$\{|S_2\rangle,|S_3\rangle\}$) which leads out of the subspace span$\{|S_1\rangle,|T^0\rangle\}$. 
Piezo-electric phonons then couple to the electron charge and incoherently mix states in the singlet sector which leads to a loss of fidelity of the swap operation. This process can be described in a four-dimensional Hilbert space ${\mathcal H}^{(4)}$, spanned by the three singlets $|S_j\rangle$ and the triplet $|T^0\rangle\equiv |0\rangle$,
with a time-dependent Hamiltonian 
\begin{eqnarray}\label{H02define}
  H_0^{(2)}(t)&=&\sum_{j=0}^3\varepsilon_j |j\rangle \langle j|
+ T(t) \left[ |1\rangle \langle 2| +  |2\rangle \langle 1| \right],
\end{eqnarray}
where $\varepsilon_j$ denotes the energies of the spin singlet states, $\varepsilon_1=\varepsilon_0$, 
$\varepsilon_2=\varepsilon_0+U_H$, $\varepsilon_3=\varepsilon_0+U_H-2X$ with
the spin triplet energy $\varepsilon_0$, the on-site Coulomb repulsion $U_H>0$, 
the exchange term $X>0$, and the 
time-dependent tunnel coupling element between the dots $T(t)$. The total Hamiltonian in presence of bosons coupling to the charge degree of freedom, $H^{(2)}(t) = H_0^{(2)}(t)+ \frac{1}{2}\sigma_z \hat{A} + H_B$, then 
has exactly the same form as in the one-qubit case, but with the free Hamiltonian $H_0^{(1)}(t)$ replaced by $H_0^{(2)}(t)$, a new coupling constant $\bar{g}_{\bf Q}$ \cite{BV02},  
and $\sigma_z \equiv |L\rangle\langle L|-|R\rangle\langle R|$ now referring to the two-particle states Eq. (\ref{LRdefine}). With the restriction $|T(t)| \ll U_H, 2X$, inelastic transitions  are determined by the dynamics in the subspace spanned by the states $|2\rangle$ and $|3\rangle$ and admixtures from $|1\rangle $  through the hybridization between $|1\rangle$ and $|2\rangle$ can be neglected. Within the Born-Markov approximation, the adiabatic rates then depend on the  energy difference $\Delta=2X$ between $|2\rangle$ and $|3\rangle$ only. As $\Delta$  remains constant throughout the operation, this again means that charge dissipation to second order can be switched off in phonon cavities when $\Delta$ is tuned to a `gap' energy $\hbar \omega_0$, cf. section \ref{section_cavity}.

Results for the fidelity $\langle f|\rho(t)|f\rangle$ are shown in Fig. \ref{ddotzener_figure7.eps}, where a pulse \cite{SLM01}
\begin{eqnarray}
  T(t) = \frac{T_0}{1+\cosh(t/\tau)/\cosh(T/2\tau)}
\end{eqnarray}
with $T_0= 0.05$, $T=400$, 
$\tau = 50$ was chosen, together with $X=0.5$ and a temperature $1/\beta=0.1$ (in units of $U_H$).
Even in absence of dissipation, the non-adiabacity of the operation results in a finite value
of $1-\langle f|\rho(t)|f\rangle$ after the swap \cite{SLM01}. The electron-phonon interaction, modeled with a spectral density $J_{\rm piezo}(\omega)$ as in \ Eq.~(\ref{Jmicro}) with different coupling parameters $g\equiv 2\alpha_{\rm piezo}$,  acts when charge between the dots is moved during the opening of the tunneling barrier. Consequently, the two states $|2\rangle$ and $|3\rangle$ become mixed incoherently, leading to a finite, irreversible occupation probability of the energetic lower state $|3\rangle$ even after the pulse operation. Spontaneous emission of phonons occurring during the slow swap leads to a dephasing rate $\Gamma\approx \pi g X/\hbar $. In this case, even relatively small values of $g$ can lead to a considerable fidelity loss of the operation.


\section{\bf Large Spin-Boson (Single Mode Dicke) Models, Chaos, and Entanglement}\label{section_Dicke_Chaos}
Most of the material treated in this Review so far dealt with the appearance of  quantum optical effects in  electronic transport properties of mesoscopic systems. A central topic  was the interaction between matter and light, and more specifically the interaction between bosonic (phonons, photons) and fermionic degrees of freedom, where the latter sometimes corresponded to single electrons, or were represented by `pseudo-spins' such as in two-level systems  and charge qubits. Many of the theoretical models that were presented in the previous sections  investigated these interactions within a wider context (e.g., with coupling to other electron reservoirs in order to describe transport), which often  required additional approximations in order to make any progress, even in a completely numerical treatment.

Sometimes, a much `cleaner' theoretical set-up can be achieved by going back to some of the original quantum optical Hamiltonians, with the goal to look at them with a `mesoscopic eye'. This program has been followed by a (seemingly growing) number of theorists, probably  motivated by (at least partly) some of the following reasons: - the realization that quantum optical concepts are useful in other areas of physics as well, - the experimental success in Quantum Optics and related areas such as Bose-Einstein condensation, - the possibility to study `fundamental' problems (measurement process, entanglement, quantum chaos) in conceptually very simple systems. Mainly driven by this last motivation, the final section of this Review therefore presents an overview over newer results on one important  class of models  from Quantum Optics, the single-mode Dicke superradiance model (and some of its allies), and their relation to ideas from quantum information theory (entanglement), quantum chaos and Mesoscopics (level statistics, scaling), as well as the old question of the quantum-classical crossover. 

\subsection{Single-Mode Superradiance Model}
The single-mode Dicke model describes the interaction of $N$ two-level systems with a bosonic mode of angular frequency $\omega$,
\begin{eqnarray}\label{Dicke_first}
  H_{\rm Dicke} = \frac{\omega_0}{2} \sum_{i=1}^{N} \hat{\sigma}_{z,i} + \frac{\lambda}{\sqrt{N}}\sum_{i=1}^{N}  \hat{\sigma}_{x,i}
\left(a^{\dagger} + a\right)+\omega a^{\dagger}a,
\end{eqnarray}
where $\omega_0$ is the transition angular frequency between the upper and lower level, cf. \ Eq.~(\ref{Pauli}), and the factor $1/\sqrt{N}$ is due to the dipole matrix element containing a factor $1/\sqrt{V}$, where $V$ is the volume  of the boson cavity and one works at constant density $\rho=N/V$, absorbing the factor $\sqrt{\rho}$ into the coupling matrix element. Crucially, the coupling constant $\lambda$ to the bosonic mode does not depend on the atom index $i$.  The interaction term in the one-mode Hamiltonian \ Eq.~(\ref{Dicke_first})  in fact is a special case of the multi-mode interaction, \ Eq.~(\ref{Nion}) for one single mode $({\bf Q}s)$, where the dependence on the phase factors $e^{i{\bf Q r}_i}$ is neglected. One can then introduce collective atomic operators (angular momentum operators), 
\begin{eqnarray}
  \label{eq:angular1}
  J_\alpha &\equiv& \frac{1}{2}\sum_{i=1}^N {\hat{\sigma}}_{\alpha,i},\quad \alpha=x,y,z;\quad 
J_{\pm} \equiv  J_x\pm i J_y,\quad [J_z,J_{\pm}]=\pm J_{\pm},\quad [J_+,J_-]=2J_z,
\end{eqnarray}
cf. \ Eq.~(\ref{eq:angular}). 

In cavity quantum electrodynamics, this model  describes collective light-matter interactions in a photon cavity. On the transport side, possible candidates for experimental systems would be arrays of excitonic quantum dots (the case $N=2$ would correspond to the system treated in section \ref{section_excitons}), and electrons in several quantum dots interacting with single phonon modes. An example of the latter is the `phonon cavity quantum dynamics' of nano-electromechanical systems, cf. section \ref{section_cavity}.


\subsubsection{Hamiltonians}
For the rest of this section, we consider the $j=N/2$ subspace of the $2^N$ dimensional total atomic Hilbertspace ${\mathcal{H}}_N=(C^2)^{\otimes N}$, which is spanned by Dicke states
$| jm\rangle$ with maximum total angular momentum $j=N/2$, cf. the discussion in section \ref{section_Natoms}. In terms of the collective operators \ Eq.~(\ref{eq:angular1}), the single-mode Dicke Hamiltonian then reads 
\begin{eqnarray}\label{Dicke1}
  H_{\rm Dicke} = \omega_0 J_z + \frac{\lambda}{\sqrt{2j}}\left(a^{\dagger} + a\right) 
(J_+ + J_-)+\omega a^{\dagger}a, 
\end{eqnarray}
which is the generalization of the Rabi Hamiltonian
\begin{eqnarray}\label{Dicke_Rabi}
  H_{\rm Rabi} = \frac{\omega_0}{2}  \hat{\sigma}_z + {\lambda}\left(a^{\dagger} + a\right)  \hat{\sigma}_x
+\omega a^{\dagger}a
\end{eqnarray}
to $j=N/2>1/2$. In Quantum Optics, the Dicke Hamiltonian is often considered within the rotating wave approximation (RWA),
\begin{eqnarray}\label{Dicke_RWA}
   H_{\rm Dicke}^{\rm RWA} = \omega_0 J_z + \frac{\lambda}{\sqrt{2j}}\left(a^{\dagger}J_- + a J_+\right) 
+\omega a^{\dagger}a,
\end{eqnarray}
which in comparison with the full Hamiltonian $H_{\rm Dicke}$ does not contain the `counter-rotating' terms
$a^{\dagger} J_+$ and $a J_-$, and which is the generalization of the Jaynes-Cumming Hamiltonian
\begin{eqnarray}\label{Jaynes-Cummings}
  H_{\rm Jaynes-Cummings} = \frac{\omega_0}{2}  \hat{\sigma}_z + {\lambda}\left(a^{\dagger} \hat{\sigma}_- + a \hat{\sigma}_+\right) +\omega a^{\dagger}a
\end{eqnarray}
to $j=N/2>1/2$. The absence of counter-rotating terms makes the RWA-Hamiltonian integrable and therefore has dramatic consequences when it comes to the discussion of quantum chaos. The RWA-form $H_{\rm Dicke}^{\rm RWA}$ conserves the {\em excitation number operator} $\hat{N}_{\rm ex}$, whereas the full Dicke Hamiltonian only conserves the {\em parity operator} $\hat{\Pi}$, both of which are defined \cite{EB03} as 
\begin{eqnarray}\label{Pi_definition}
  \hat{\Pi}\equiv \exp\left[ i \pi \hat{N}_{\rm ex} \right],\quad
 \hat{N}_{\rm ex} \equiv a^{\dagger}a + J_z + j.
\end{eqnarray}
The meaning of these operators can be most easily understood in the analogy of the spin-boson Hamiltonian with a single particle on a  two-dimensional lattice, cf. Fig. \ref{Dicke_lattice.eps}, where  each point represents a basis vector $|n\rangle \otimes |jm\rangle$ with $|n\rangle$ representing the number states, $a^{\dagger}a |n\rangle = n |n\rangle$, and $|jm\rangle$ the Dicke states. The lattice is finite in `$m$' direction, but infinite in the `$n$' direction. For the full Dicke  Hamiltonian  $H_{\rm Dicke}$, the interaction conserves the parity $\hat{\Pi}$, and states with an even total excitation number $n + m +j$ interact only with other even states, whereas odd states interact only with odd states.  This has the effect of dividing the total lattice into a motion of the particle on one of the two inter-weaved sub-lattices, which corresponds to the two different parity sectors. On the other hand, the RWA version $H_{\rm Dicke}^{\rm RWA}$ induces an even more drastic  splitting of the total Hilbert space into an (infinite) number of finite-dimensional subspaces that are characterized by the excitation number $\hat{N}$. In the lattice picture, this corresponds to independent clusters joined in the direction $\searrow$, cf. Fig. \ref{Dicke_lattice.eps}.

\begin{figure}[t]
\includegraphics[width=0.5\textwidth]{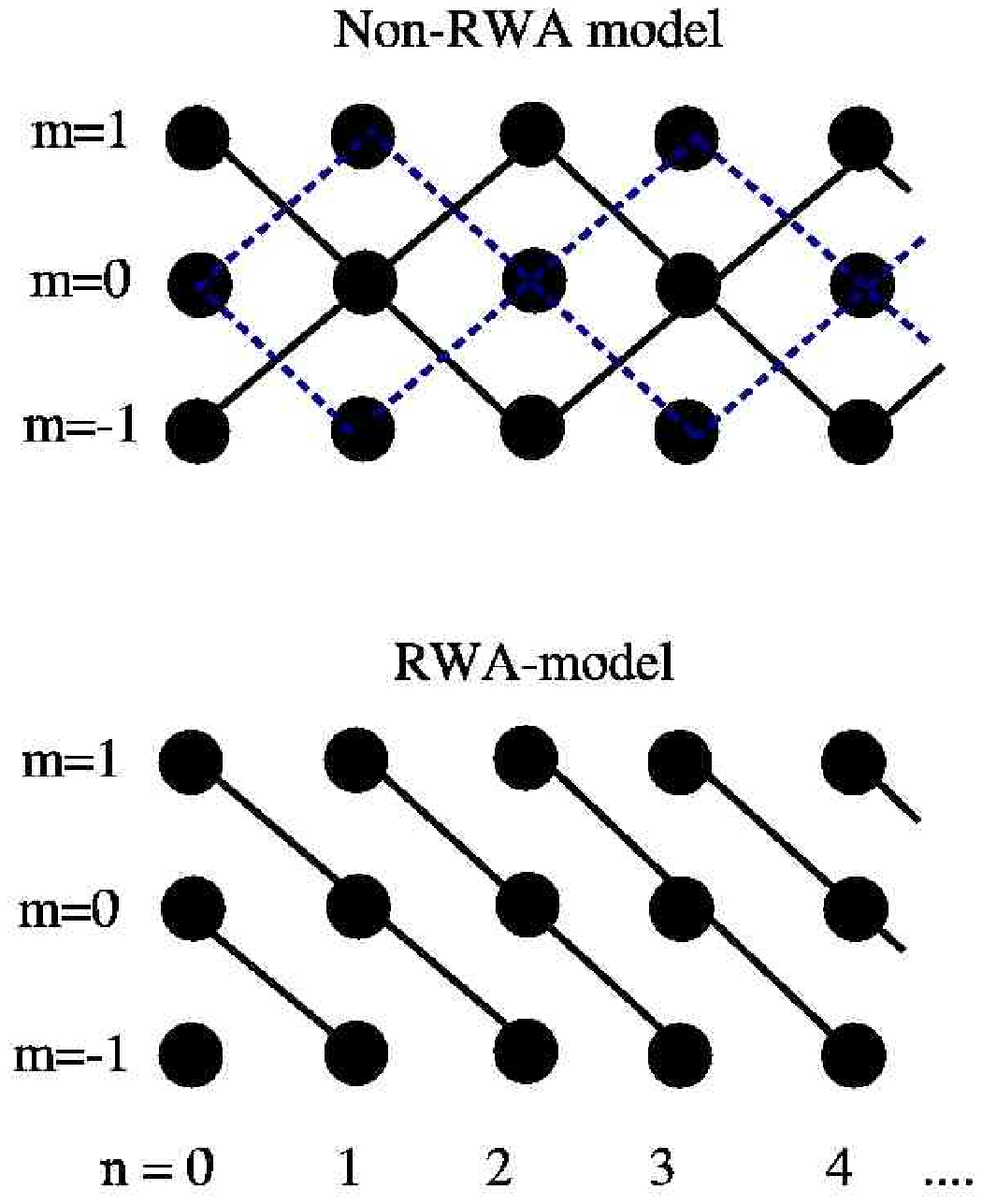}
\includegraphics[width=0.5\textwidth]{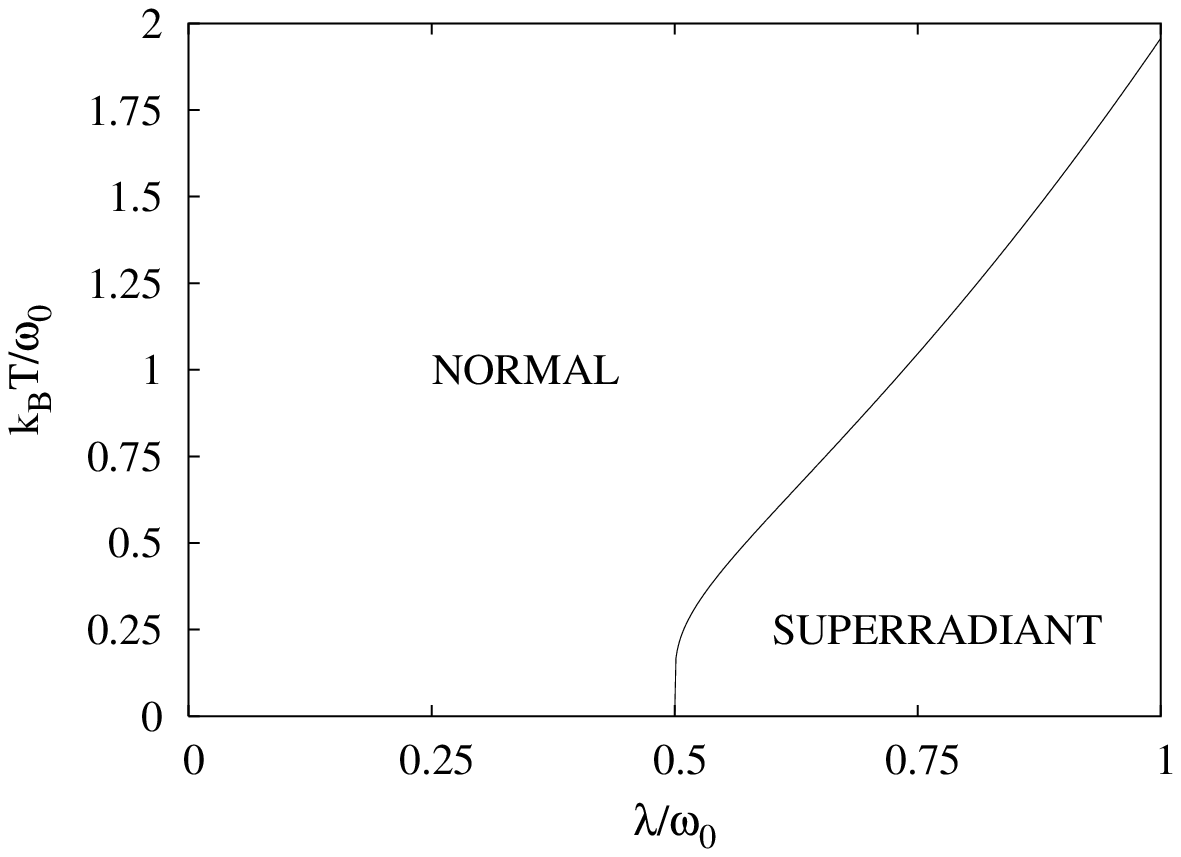}
\caption[]{\label{Dicke_lattice.eps}{\bf Left:} Lattice analogy for Dicke model in (non) rotating wave approximation, the case shown here is for $j=1$. {\bf Right:} Phase diagram for the Dicke Hamiltonian $H_{\rm Dicke}$, \ Eq.~(\ref{Dicke_first}), in the thermodynamic limit $j\to \infty$.}
\end{figure} 

\subsubsection{Phase Transition}
The phase transition for the  RWA Dicke model, \ Eq.~(\ref{Dicke_RWA}), was first   rigorously derived in 1973 by Hepp and Lieb \cite{HL73} who used spectral properties of  finite matrices derived from the model. At the same time, Wang and Hioe gave \cite{WH73}  a more transparent (though less rigorous) proof using bosonic coherent state. A simple generalisation for the non-RWA version, \ Eq.~(\ref{Dicke_first}), was soon given by Hepp and Lieb \cite{HL73a}, and by Carmichael, Gardiner, and Walls \cite{CGW73} who started from the canonical partition function $Z(N,T)\equiv {\rm Tr} \exp (-\beta H_{\rm Dicke})$, $\beta=1/k_BT$, and traced out the field as in \cite{WH73},
\begin{eqnarray}
  Z(N,T) &=& \int \frac{d^2\alpha}{\pi}e^{-\beta|\alpha|^2} \left[{\rm Tr} \exp \left\{-\beta 
\left( \frac{\omega_0}{2} \hat{\sigma}_z + \frac{\lambda}{\sqrt{N}} (\alpha+\alpha^*)  \hat{\sigma}_x\right) \right\}\right]^N\nonumber\\
&=&\int_{0}^{\infty}dr r \int_{0}^{2\pi} \frac{d \theta}{\pi} e^{-\beta r^2}
 \left[2 \cosh  \left\{\frac{\beta \omega_0}{2} \left(
1+ \frac{16\lambda^2r^2 \cos^2 \theta}{\omega_0^2 N} \right)^{1/2} \right\}\right]^N,
\end{eqnarray}
where the boson frequency $\omega$ has been set to unity. This integral is evaluated asymptotically using the method of steepest descents, from which the phase diagram in the thermodynamic limit $N\to \infty$ follows: for $\lambda < \sqrt{\omega \omega_0}/2$, the system is in the `normal' phase with a free energy $f(T)$ per particle given by
\begin{eqnarray}
  -f_{\rm n} (T) = \beta^{-1}\ln [2 \cosh (\frac{1}{2} \beta \omega_0)],
\end{eqnarray}
which is just the free energy of a non-interacting two-level system. For $\lambda > \sqrt{\omega \omega_0}/2$, however, there is a critical temperature $T_c$ given by $\omega \omega_0/4\lambda^2 = \tanh (\frac{1}{2}\omega_0 /k_BT_c)$ below which the system in a `superradiant' state with a free energy per particle given by
\begin{eqnarray}\label{f_SR}
  - f_{\rm SR}(T) &=&
\beta ^{-1} \ln \left[2 \cosh \left (4 \frac{\beta}{\omega} \lambda^2 x \right) \right] - 4 \frac{\lambda^2}{\omega} x^2 + \frac{\omega \omega_0^2}{16\lambda^2},
\quad 2x = \tanh (4\beta \lambda^2 x) > 0.
\end{eqnarray}
The self-consistent equation for $x$ in \ Eq.~(\ref{f_SR}) indicates that this phase transition is of mean-field type. At the phase boundary in the phase diagram, Fig. \ref{Dicke_lattice.eps}, the system changes discontinuously between  the normal phase, where the boson occupation per particle is zero, and the superradiant phase that has a macroscopic boson occupancy \cite{WH73},
\begin{eqnarray}
\lim_{N\to \infty} \frac{1}{N} \langle a^{\dagger} a \rangle _{\rm n} = 0,\quad
\lim_{N\to \infty}\frac{1}{N} \langle a^{\dagger} a \rangle _{\rm RS} = 4 \frac{\lambda^2}{\omega^2} x^2 -\frac{\omega_0^2}{16\lambda^2},
\end{eqnarray}
where again $x$ is given by the positive solution of $2x = \tanh (4\beta \lambda^2 x)$. At zero temperature $(\beta\to \infty)$, this solution is $x=\frac{1}{2}$, and one easily obtains quantities like the ground state energy
\begin{eqnarray}
  E^0_{\rm n} = -\omega_0/2,\quad  E^0_{\rm SR} = -\frac{\lambda^2}{\omega}-\frac{\omega \omega_0^2}{16\lambda^2}
\end{eqnarray}
and other quantities at $T=0$ from the finite-$T$ results in the thermodynamic limit $N\to \infty$. 

As mentioned above, the original derivations of the thermodynamic properties for the infinite-$N$ Dicke model were first made for the RWA model. Hebb and Lieb \cite{HL73a}, and Carmichael, Gardiner, and Walls \cite{CGW73} in fact showed that in the limit $N\to\infty$, the thermodynamic properties of the non-RWA model $H_{\rm Dicke}$ are obtained by simply using the expressions obtained from  the RWA model $H_{\rm Dicke}^{\rm RWA}$ and doubling the coupling constant, $\lambda\to 2\lambda$. Consequently, the phase transition in $H_{\rm Dicke}^{\rm RWA}$ occurs at a critical coupling $\lambda_c^{\rm RWA}=\sqrt{\omega\omega_0}$ that is twice as large as $\lambda_c=\sqrt{\omega\omega_0}/2$ in the non-RWA model $H_{\rm Dicke}$. A heuristic argument for the factor two is the doubling of interacting vertices in $H_{\rm Dicke}$ as compared to the RWA model. A more recent comparison between RWA and non-RWA, in particular with respect to the integrability of the Dicke model, is given in \cite{EB03a}.

\subsubsection{Effective Hamiltonians and Finite-$N$ Results}
Emary and Brandes \cite{EB03,EB03a} studied the one-mode Dicke model $H_{\rm Dicke}$, \ Eq.~(\ref{Dicke_first}),  at arbitrary $N=2j$ but at zero temperature $T=0$ with the aim to relate quantum chaotic behavior as obtained from the spectrum of  $H_{\rm Dicke}$ at finite $N$ to the transition for $N\to \infty$. In the terminology of statistical mechanics, the transition at $T=0$ is only driven by quantum (and not thermal) fluctuations and thus is a quantum phase transition, although one of a special kind: the absence of any intrinsic, physical length scale in the model makes it exactly solvable. The phase transition in fact can be related to an instability for $N\to \infty$ of the quadratic form describing the interaction of two bosonic modes, one of which represents the original photon mode $a^{\dagger}$ whereas the other represents the spin $j$. This is formalized by the Holstein-Primakoff representation of the angular momentum operators in terms of a single bosonic mode $b^{\dagger}$,
\begin{eqnarray}\label{Primakoff}
  J_+ = b^\dagger \sqrt{2j - b^\dagger b},\quad   J_- = \sqrt{2j - b^\dagger b}~ b,\quad 
  J_z = {b^\dagger b - j},
\end{eqnarray}
which are inserted into $H_{\rm Dicke}$ and then expanded for large $j$. Hillery and Mlodinow \cite{HM84} used this method in their analysis of the RWA form, $H_{\rm Dicke}^{\rm RWA}$ \ Eq.~(\ref{Dicke_RWA}){}, in the normal phase. For a general survey on boson realizations of Lie algebras, cf. the review by Klein and Marshalek \cite{KM91}. 

A very suitable method for the case of the Dicke Hamiltonian  is to introduce position and momentum operators for the two bosonic modes \cite{EB03a},
\begin{eqnarray}\label{xypp}
x = \frac{1}{\sqrt{2 \omega}}({a^\dagger + a}),\quad 
p_x =i\sqrt{\frac{\omega}{2}}({a^\dagger - a}), \quad 
y =\frac{1}{\sqrt{2 \omega_0}}({b^\dagger + b}), \quad 
p_y = i\sqrt{\frac{\omega_0}{2}}({b^\dagger - b}),
\end{eqnarray}
which turns out to be particularly useful when discussing properties of the ground state wave function, and which leads to a Hamiltonian describing the {\em normal phase for $N\to \infty$}, 
\begin{equation}\label{H_normal_Dicke}
H^{(1)} = \frac{1}{2}
\left\{ \omega^2 x^2 + p_x^2 + \omega_0^2 y^2 + p_y^2 
+ 4 \lambda \sqrt{\omega \omega_0}~ x y - \omega_0 - \omega\right\}
-j \omega_0.
\end{equation}
This is easily diagonalized and leads to 
\begin{eqnarray}
H^{(1)}  &=&  \varepsilon^{(1)}_- c_1^\dagger c_1 
+ \varepsilon^{(1)}_+  c^\dagger_2 c_2 
+\frac{1}{2} \left(\varepsilon^{(1)}_+ + \varepsilon^{(1)}_- -\omega - \omega_0\right) - j \omega_0\nonumber\\
\left[\varepsilon_{\pm}^{(1)}\right]^2 &=& \frac{1}{2}\left\{ \omega^2 + \omega_0^2 \pm \sqrt{({\omega_0^2 - \omega^2})^2 + 16 \lambda^2 \omega \omega_0}\right\},
\label{lcepspm}
\end{eqnarray}
with two excitation energies for the two new, collective bosonic modes $1$ and $2$. The excitation energy $\varepsilon_-$ is real only for $\lambda \le \lambda_c \equiv \sqrt{\omega \omega_0}/2$ which indicates the transition: $H^{(1)}$ remains valid in the normal phase but becomes invalid in the superradiant phase. 

The ground-state wave function of $H^{(1)}$, \ Eq.~(\ref{H_normal_Dicke}), is a simple product of two harmonic oscillator wave functions which in the $x$-$y$ representation reads
\begin{eqnarray}\label{Psi_normal}
  \Psi^{(1)}_G({x,y}) &=& 
  G_-( x\cos \gamma - y\sin \gamma)
  G_+(x\sin \gamma + y\cos \gamma)\nonumber\\
\gamma&=& \frac{1}{2}\arctan \frac{4 \lambda \sqrt{\omega \omega_0}}{\omega_0^2 - \omega^2},\quad
G_{\pm}(q) \equiv (\varepsilon^{(1)}_{\pm}/\pi)^{1/4} e^{-\varepsilon_{\pm}^{(1)}q^2/2}.
\end{eqnarray}
Close below the critical point $\lambda_c$, the excitation energy $\varepsilon_-=\varepsilon_-^{(1)}$ 
vanishes as 
\begin{eqnarray}\label{varepsilon_scaling}
  \varepsilon_- \propto |\lambda-\lambda_c|^{z \nu},
\end{eqnarray}
with the dynamical exponent $z=2$ and the `localization length' exponent $\nu =1/4$ describing the divergence of the characteristic length $\xi \equiv {\varepsilon_-^{-1/z}}\propto |\lambda-\lambda_c|^{-\nu}$ in the oscillator wave function $G_-$, and the same exponents when approaching from above $\lambda_c$. At the critical point of the coupling constant $\lambda=\lambda_c$, $\xi$ becomes infinite and the Gaussian wave function, \ Eq.~(\ref{Psi_normal}), is infinitely stretched along the $x=-y$ line in the $x$-$y$ plane. This is consistent with the results for the ground state wave function as obtained from a numerical diagonalization for the finite $j=N/2$ Dicke Hamiltonian $H_{\rm Dicke}$, \ Eq.~(\ref{Dicke_first}), as shown in the $x$-$y$ representation in Fig. \ref{EB03a_Fig12.eps} for $j=5$: the wave function starts as a single lobe centered at the origin for low coupling.  As the coupling increases, the two modes start mixing, leading to a stretching of the  single-peaked wave function, which then {\em splits} into two lobes as the coupling is increased above approximately $\lambda_c$.   The two lobes move away from each other in their respective quadrants of the $x$-$y$ plane when further increasing $\lambda$ above $\lambda_c$. 

\begin{figure}[t]
  \centerline{
    \includegraphics[clip=true,width=0.35\textwidth]{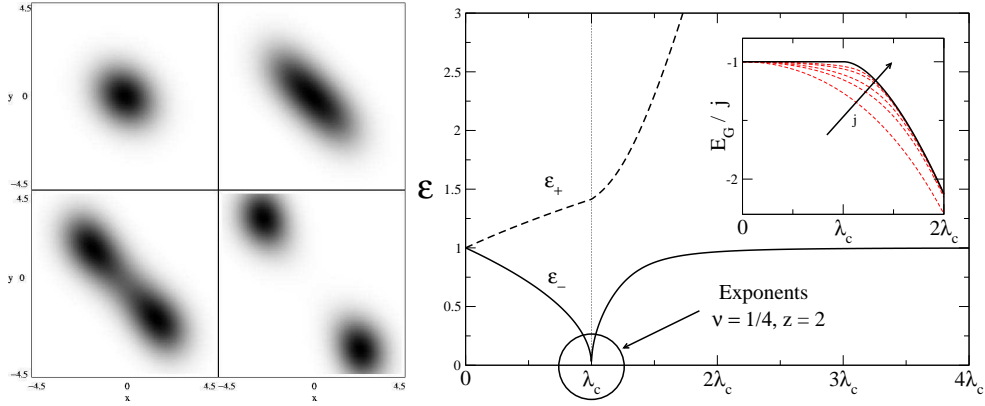}
    \includegraphics[clip=true,width=0.5\textwidth]{EB03_Fig1.eps}}
  \caption{
  \label{EB03a_Fig12.eps}{\bf Left:}
    Modulus of the ground-state wave function 
    $\psi({x,y})$ 
    of the Dicke Hamiltonian in the abstract $x$-$y$ representation  
    for finite $j=5$, at couplings of 
    $\lambda / \lambda_c$ = 0.2, 0.5, 0.6, 0.7.  Black corresponds 
    to $\mathrm{Max}|\psi|$ and white corresponds to zero.  The 
    Hamiltonian is resonant $\omega =\omega_0=1$; $\lambda_c=0.5$. From \cite{EB03a}.
{\bf Right:} Excitation energies $\varepsilon_\pm$ in the thermodynamic limit.  Inset:
scaled ground-state energy, $E_G /j$, in the thermodynamic limit
(solid line) and at various finite values of $j=1/2,1,3/2,3,5$ 
(dashed lines).  From \cite{EB03}. }
\end{figure}

For large but finite $j$, the ground-state with $\lambda>\lambda_c$ is a coherent superposition of two wave function lobes that are macroscopically separated in the $x$-$y$ plane. For $j\to \infty$, i.e. in the thermodynamic limit, the macroscopic separation becomes so large that this Schr\"odinger cat is `split into two halves'. It was shown in \cite{EB03a} that this superradiant regime is described by {\em two} equivalent effective Hamiltonians $H^{(2)}$, each describing the low-energy excitations in the frame of reference of one of the lobes. For any finite $j$, the ground-state obeys the parity symmetry $\hat{\Pi}$, \ Eq.~(\ref{Pi_definition}){}, meaning that the wave function is always invariant under a rotation of $\pi$ in the $x$-$y$ plane. For $j\to \infty$, the ground-state is two-fold degenerate, the system choses to sit in one of the lobes that is `super-selected' whereby the parity symmetry of $H_{\rm Dicke}$ is spontaneously broken. Recently, Frasca \cite{Fra04} discussed the Schr\"odinger cat and the $N\to \infty$ limit of the Dicke model in the context of decoherence.

\begin{figure}[t]
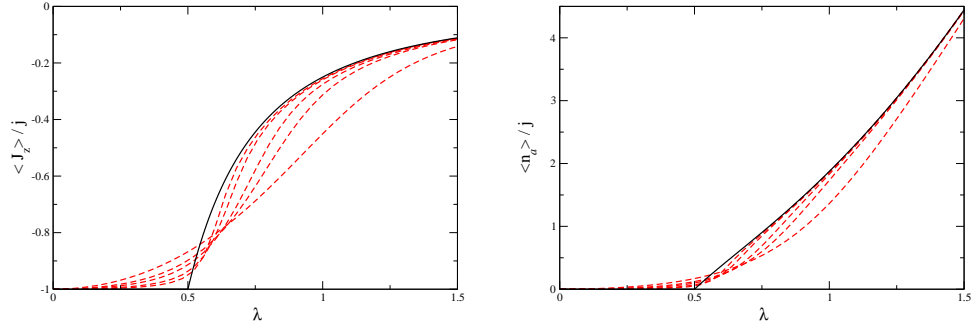

  \centerline{
    \includegraphics[clip=true,width=0.4\textwidth]{EB03a_Fig4a.eps}
     ~~~
    \includegraphics[clip=true,width=0.4\textwidth]{EB03a_Fig4b.eps}}
  \caption{\label{EB03a_Fig4.eps}
    Scaled atomic inversion ({\bf left}) and mean photon number ({\bf right})
    of the Dicke Hamiltonian as a 
    function of coupling $\lambda$.
    Solid lines denote results in the thermodynamic limit, whereas
    dashed lines correspond to the results for 
    various finite values of $j=\frac{1}{2},1,\frac{3}{2},3,5$.
    The Hamiltonian is resonant: $\omega = \omega_0 = 1$, 
    $\lambda_c=0.5$. From \cite{EB03a}.
   }
\end{figure}

The effective Hamiltonians $H^{(2)}$ for the superradiant phase are obtained by using the Holstein-Primakoff transformation, \ Eq.~(\ref{Primakoff}){}, and a canonical transformation that displaces the two bosonic modes, thereby taking into account the macroscopic displacement and occupation of the field ($a^{\dagger}$) and the field ($b^{\dagger}$) mode,
\begin{eqnarray}\label{abshift}
a^\dagger \rightarrow c^\dagger \pm  \sqrt{\alpha},\quad b^\dagger \rightarrow d^\dagger \mp \sqrt{\beta}.
\end{eqnarray}
Here, the upper and lower sign refer to the two equivalent Hamiltonians that describe the system for $j\to \infty$, with $\alpha$ and $\beta$ to be determined by expanding the  canonically transformed Dicke Hamiltonian for large $j$, retaining only up to quadratic terms in the new bosonic operators $c^{(\dagger)}$ and $d^{(\dagger)}$. Elimination of {\em linear} terms in these operators then leads to two equations \cite{EB03a},
\begin{eqnarray}\label{abeqs}
  2\lambda \sqrt{\frac{\beta (2j-\beta)}{2j}}-\omega\sqrt{\alpha}=0,\quad
\left[\frac{4\lambda^2}{\omega j }(j-\beta) - \omega_0\right]\sqrt{\beta}=0,
\end{eqnarray}
with trivial solutions $\alpha=\beta=0$ that recover the normal phase (the Hamiltonian $H^{(1)}$), and non-trivial solutions determining the superradiant $H^{(2)}$, which after some further transformations is brought into diagonal form,
\begin{eqnarray}
H^{(2)} &=& \varepsilon^{(2)}_- e_{1}^\dagger e_{1} 
+ \varepsilon^{(2)}_+  e^\dagger_{2} e_{2}
-j\left\{
    \frac{2 \lambda^2}{\omega}+ \frac{\omega_0^2 \omega}{8 \lambda^2}  
  \right\}+\frac{1}{2} 
\left({
  \varepsilon^{(2)}_+ + \varepsilon^{(2)}_- 
  -\frac{\omega_0}{2 \mu}({1+\mu}) - \omega 
  -\frac{2\lambda^2}{\omega}({1-\mu})
 }\right)\nonumber\\
2\left[{\varepsilon_\pm^{(2)}}\right]^2 &=&
  \frac{\omega_0^2}{\mu^2} +  \omega^2
  \pm \sqrt{\left[ \frac{\omega_0^2}{\mu^2} - \omega^2 \right]^2 
  + 4\omega^2 \omega_0^2},\quad \mu \equiv \frac{\omega \omega_0}{4 \lambda^2} 
             = \frac{\lambda_c^2}{\lambda^2} .
\end{eqnarray}
The values of $\alpha$ and $\beta$ as determined from \ Eq.~(\ref{abeqs}){} are the same for both signs in \ Eq.~(\ref{abshift}) and related to the atomic inversion and the mean photon number 
\begin{eqnarray}\label{alphabetarelated}
\langle{J_z}\rangle/j = \beta/j-1,\quad  \langle{a^\dagger a}\rangle/j = \alpha/j,
\end{eqnarray}
where the brackets refer to ground state expectation values. One thereby obtains two exactly equivalent Hamiltonians $H^{(2)}$, which are valid for $\lambda\ge \lambda_c$ such that  the excitation energy $\varepsilon_-^{(2)}$ remains real.

As the Hamiltonians $H^{(1)}$ and $H^{(2)}$ are in diagonal form, they present the exact analytical solution for the Dicke one-mode model at arbitrary coupling strength in the limit $j\to \infty$ and allow one to derive exact results for the spectrum, expectation values, wave function properties, entanglement etc. that can be compared to their respective finite-$j$ counter-parts as obtained from numerical diagonalizations. Examples of such a comparison are shown for the ground state energy $E_G$ and the excitation energies $\varepsilon_{\pm}$,  Fig. (\ref{EB03a_Fig12.eps}), and for  the atomic inversion $\langle J_z \rangle$ and the photon number $\langle a^{\dagger} a \rangle$ in Fig. (\ref{EB03a_Fig4.eps}).

\subsubsection{Level Statistics}  

The nearest-neighbor level spacing distribution $P(S)$ for level spacings $S_n=E_{n+1}-E_n$ at finite $j$ was obtained in \cite{EB03a} by direct numerical diagonalization of  $H_{\rm Dicke}$, \ Eq.~(\ref{Dicke_first}). Signatures of the $T=0$ normal-superradiant phase transition for $j\to \infty$ can be related to a cross-over  from the Poissonian distribution, $P_{\rm P}(S)=\exp(-S)$ at $\lambda<\lambda_c$, to the Wigner-Dyson distribution, $P_{\rm W}(S) = \pi S/2 \exp(-\pi S^2/4)$, in the finite-$j$ level statistics, cf. Fig (\ref{EB03a_Fig9.eps}).  At low $j\le 3$, however, it should be noted that the $P(S)$ do not correspond to any of the universal random matrix theory ensembles but are rather non-generic distributions, an example being the `picket-fence' character of $P(S)$ for the Rabi Hamiltonian $j=1/2$, \ Eq.~(\ref{Dicke_Rabi}). The cross-over as a function of $\lambda$ becomes sharper for larger $j$, and one might regard any deviations from a sharp transition as `finite-size' effects, i.e., deviations from the $j\to \infty$ limit. This interpretation, however, is somewhat misleading because in this limit the system, although going through a phase transition at $\lambda=\lambda_c$, remains integrable. 

\begin{figure}[t]
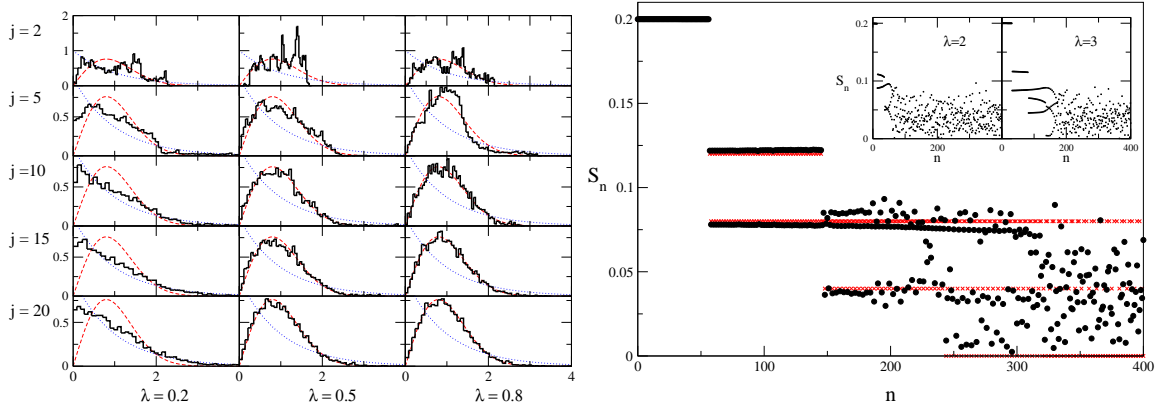

  \centerline {\includegraphics[clip=true,width=0.5\columnwidth]{EB03a_Fig9.eps}
  \includegraphics[clip=true,width=0.5\columnwidth]{EB03a_Fig11.eps} }
  \caption{\label{EB03a_Fig9.eps}
    {\bf Left:} Nearest-neighbor distributions  $P(S)$ for the 
    Dicke Hamiltonian, for different couplings $\lambda$ and pseudo-spin $j$, and comparison with 
    the universal Poissonian (dots) to Wigner (dashes) 
    distributions. {\bf Right:}   Nearest-neighbor spacing 
    $S_n=E_{n+1}-E_n$ vs. eigenvalue 
    number $n$ plot for $j=5$ with $\lambda =4$.
    Horizontal crosses: results for the integrable 
    $\lambda \rightarrow \infty$ Hamiltonian.  Inset: $j=5$ results
    with $\lambda =2$ and $\lambda=3$. Results shown are for
    $\omega = \omega_0 = 1$, $\lambda_c=0.5$. From \cite{EB03}.}
\end{figure}

The cross-over in the level statistics of the Dicke model is also consistent with the bifurcation of the ground wave function into a macroscopic superposition, cf. Fig. (\ref{EB03a_Fig9.eps}), left. This can be regarded as a transition from a localized, quasi-integrable regime for $\lambda<\Lambda_c$ (corresponding to  Poissonian level statistics), to a delocalized, chaotic regime for $\lambda>\Lambda_c$ (corresponding to Wigner-Dyson statistics). 

Another peculiarity of the spectrum is the close co-existence of very regular and very irregular parts at fixed, finite $j$ and $\lambda$ as a function of the level index  $n$, cf. Fig. (\ref{EB03a_Fig9.eps}), right. The regular part of the nearest-neighbor spacings $S_n$ can be compared with the integrable {\em strong coupling limit} $\lambda\to \infty$ of the model $H_{\rm Dicke}$, in which the term $\omega_0 J_z$ becomes a negligible perturbation and the system corresponds to a shifted harmonic oscillator. For $\lambda \gg \lambda_c$, the spectrum becomes very regular and close to the exact $\lambda\to \infty$ limit at low energies, whereas outside this region the spectrum is very irregular and described by the Wigner-Dyson distribution.

\begin{figure}[t]
  \begin{center}
   \includegraphics[width=0.8\textwidth]{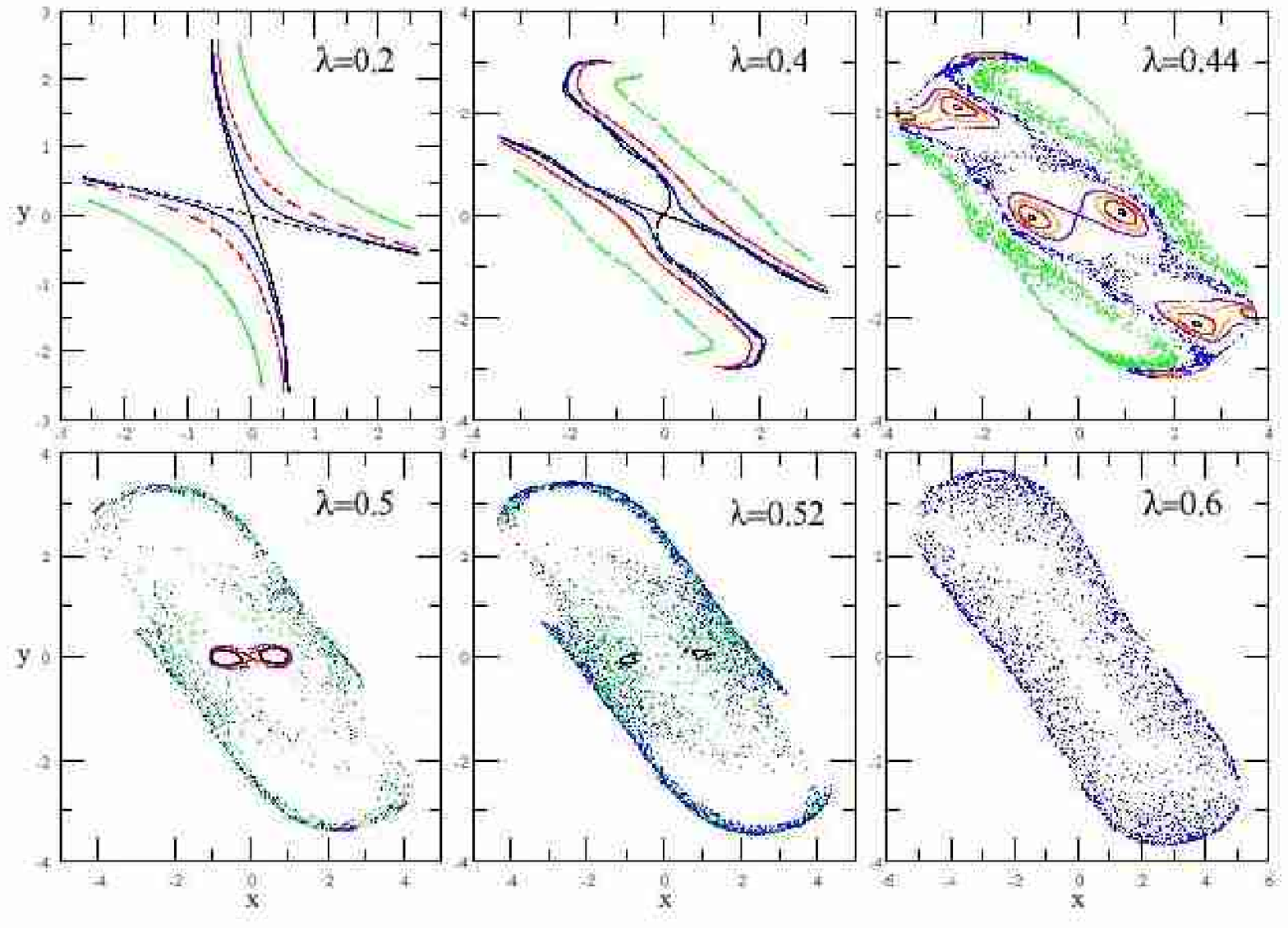} 
   \includegraphics[width=0.7\textwidth]{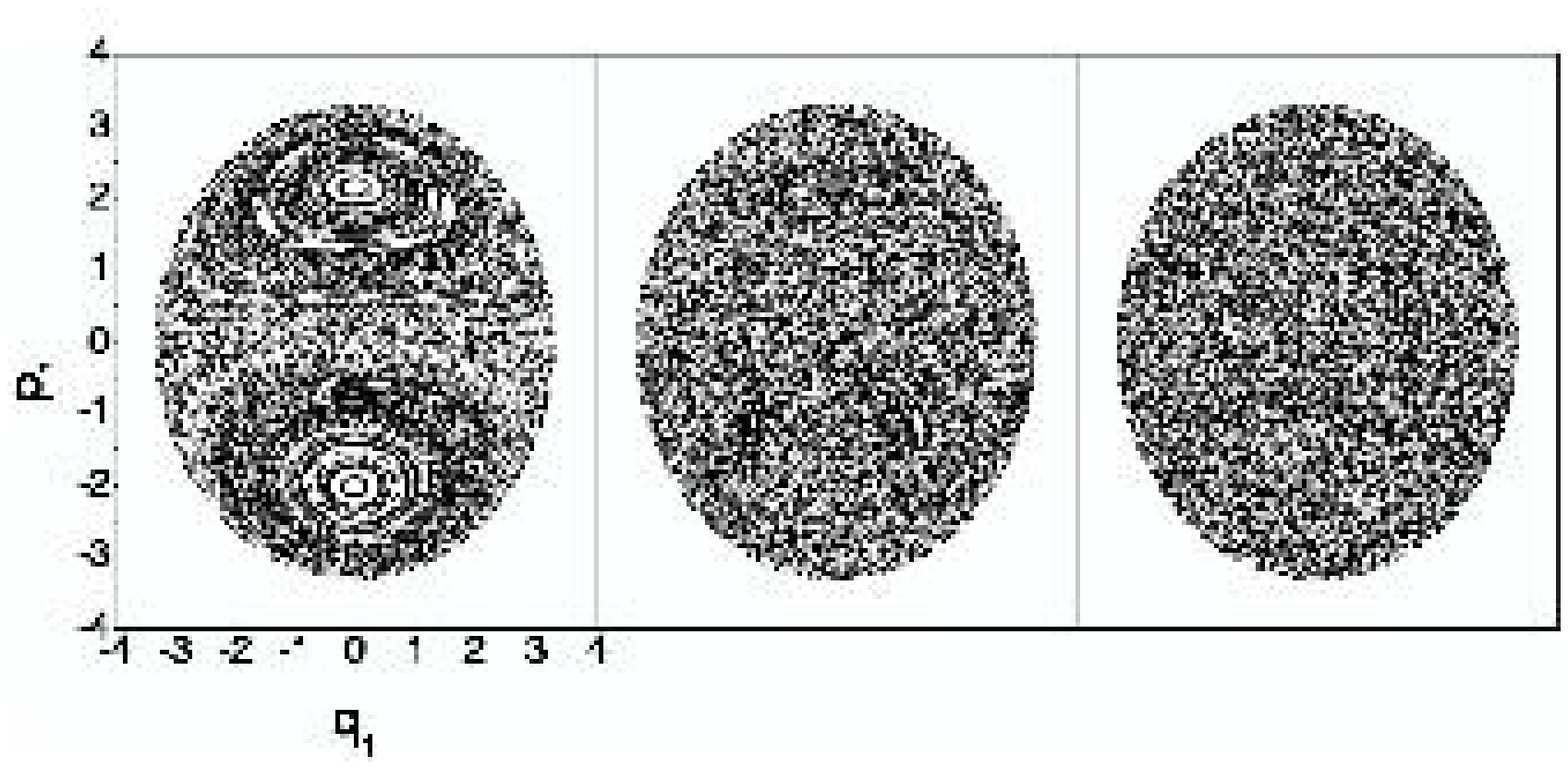}
\end{center}
  \caption{\label{EB03a_Fig14.eps} {\bf Top:} 
    Poincar\'{e} sections for the classical Dicke Model for a sequence 
    of increasing couplings, with $j=5$ and $E=-3$. From \cite{EB03a}. {\bf Bottom:} Poincar\'{e} sections as calculated by Hou and Hu \cite{HH04}. From \cite{HH04}.}
\end{figure}

\subsubsection{Semi-Classical Model and Chaos}
Emary and Brandes \cite{EB03a} derived a classical Hamiltonian from the Dicke model $H_{\rm Dicke}$ in bosonic form, using the Holstein-Primakoff transformation,  Eq.~(\ref{Primakoff}), and a subsequent replacement of position and momentum operators $x$,$y$,$p_x$,$p_y$, \ Eq.~(\ref{xypp}), by classical variables. The resulting classical model,
\begin{eqnarray}\label{Dicke_classical}
  H_{\rm Dicke}^{\rm cl} &=& \frac{1}{2}\left(p_x^2 + p_y^2 \right) + U(x,y,p_y),\nonumber\\
U(x,y,p_y)&\equiv& \frac{1}{2}\left(\omega^2 x^2 + \omega_0^2 y^2 -\omega-\omega_0\right) -j\omega_0 
+ 2 \lambda \sqrt{\omega\omega_0}xy \sqrt{1-\frac{\omega_0^2y^2+p_y^2-\omega_0}{4j\omega_0}},
\end{eqnarray}
described the motion of a single particle in a two-dimensional, {\em momentum-dependent} potential $U(x,y,p_y)$. In the limit $j\to \infty$, the square-root non-linearity in \ Eq.~(\ref{Dicke_classical}) vanishes, and by diagonalization one finds the same symmetry-breaking phase transition as for the quantum model $H_{\rm Dicke}$. For finite $j$, a stability analysis of Hamilton's equations from $H_{\rm Dicke}^{\rm cl}$ yields a fixed point $x=y=p_x=p_y=0$ in phase space that is stable in the `localized regime' $\lambda<\lambda_c/\sqrt{1+/4j}$, where again $\lambda_c=\sqrt{\omega\omega_0}/2$ is the critical coupling found in the quantum model. Two other fixed points with $p_x=p_y=0$ exist in the `delocalized regime' in the $x$-$y$ plane at points $(\pm x_0,\mp y_0)$ which are stable for $\lambda>\lambda_c/\sqrt{1+/4j}$ and correspond to the two lobes of the Schr\"odinger cat ground state superposition in the superradiant regime of the quantum model.

Poincar\'{e} sections for $H_{\rm Dicke}^{\rm cl}$ with $p_x=0$ and $p_y>0$ fixed by the total energy $E$ are shown in Fig.(\ref{EB03a_Fig14.eps}), left. At low $\lambda$, the Poincar\'e sections consist of a series of regular, periodic orbits. Approaching the critical coupling, the character of the periodic orbits changes and a number of chaotic trajectories emerges.  Increasing the coupling further results in the break up of the remaining periodic orbits and the whole phase space becomes chaotic. This transition to chaos in the classical system mirrors very closely the transition in the quantum system. Hou and Hu \cite{HH04} recently confirmed these findings in a calculation of Poincar\'{e} sections through the $y$-$p_y$ plane for $H_{\rm Dicke}^{\rm cl}$ with $x=0$ fixed, cf. Fig.(\ref{EB03a_Fig14.eps}).

\subsection{Phase Transitions in Generalized Dicke Models}
A generalized form of the one-mode Dicke model was considered in a work \cite{EB04} that shed further light on the instability of large-spin boson Hamiltonians in the thermodynamic limit $j\to \infty$. The generic model 
\begin{eqnarray}\label{Hamiltonian_generic}
  H = \omega a^{\dagger} a + \left ({\bf \Omega } + a^{\dagger} {\bf \Lambda }
+ a {\bf \Lambda }^{\dagger} \right) {\bf J},
\end{eqnarray}
describes the simplest coupling between the Heisenberg-Weyl (harmonic oscillator) algebra ($1,a,a^{\dagger}$) and the  algebra of the angular momentum (spin $j$) operators  $J_x=\frac{1}{2}(J_++J_-)$, $J_y=\frac{1}{2}(J_+-J_-)$, $J_z$, with $ {\bf J}=(J_x,J_y,J_z)$ and the three-dimensional coupling constant vectors ${\bf \Omega }$ and 
${\bf \Lambda }$, the latter being in general complex. This generic form contains a number of special, well-known cases, cf. Table (\ref{special_generic}).

\begin{table}[b]
 \caption{\label{special_generic}Special cases of the generic spin-boson Hamiltonian, \ Eq.~(\ref{Hamiltonian_generic}){}. The three-dimensional unit vectors are ${\bf e}_\alpha$, $\alpha=x,y,z$. }
   \begin{tabular}{cll}
   \hline \hline
   Model & ${\bf \Omega}$      & ${\bf\Lambda } $ \\
   \hline
   Rabi ($j=1/2$), Dicke ($j\ge 1/2$)         & $\omega_0 {\bf e}_z$ & $\lambda {\bf e}_x$ \\
  Jaynes-Cummings ($j=1/2$), RWA-Dicke ($j\ge 1/2$)  & $\omega_0 {\bf e}_z$ & $\lambda ({\bf e}_x-i {\bf e}_y )$ \\
  abelian (`one mode dephasing') & $\omega_0 {\bf e}_\alpha$ &  $\lambda {\bf e}_\alpha$, $\alpha=x,y,z$ \\
 one mode (large) spin-boson & $\varepsilon {\bf e}_z + 2 T_c {\bf e}_x$ & $ \lambda {\bf e}_z$\\ 
\hline \hline
  \end{tabular}
\end{table}

\begin{figure}[t]
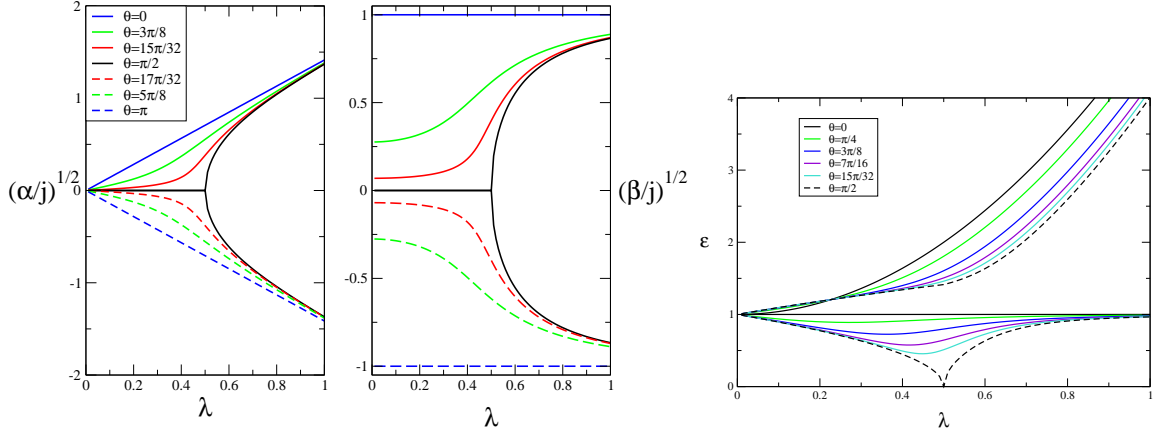

  \centerline{
    \includegraphics[clip=true,width=0.6\columnwidth]{EB04_Fig1.eps}
    \includegraphics[clip=true,width=0.4\columnwidth]{EB04_Fig3.eps}
  }
  \caption{{\bf Left:} The two displacement parameters $\sqrt{\alpha}$ 
           and $\sqrt{\beta}$ 
           as a function
           of the coupling $\lambda$ for various different angles $\theta$ in 
           the class of Hamiltonians $H_\theta$, \ Eq.~(\ref{oham}){}.  {\bf Right:} 
Excitation energies of $H_\theta$. From \cite{EB04}.
           \label{EB04_Fig1.eps}
          }
\end{figure}

The class of models discussed in \cite{EB04} was for real coupling vectors $  {\bf \Lambda }={\bf \Lambda }^{\dagger}$ and ${\bf \Omega}$, simplifying the most general case \ Eq.~(\ref{Hamiltonian_generic}){} which in general has three real, linearly independent three-dimensional coupling constant vectors. In the $x$-$z$ plane, the generalized one-mode Dicke models then are defined as
\begin{eqnarray}
  H_{\theta} = \omega a^\dagger a + \Omega ({ J_x \cos \theta + J_z \sin \theta})
  + \frac{2 \lambda}{\sqrt{2j}}\left({a^\dagger + a}\right) J_x,
 \label{oham}
\end{eqnarray}
which for fixed frequencies $\omega,\Omega$, and coupling constant $\lambda$ are characterized by the angle  $\theta$ that can be restricted to $0\le \theta \le \pi$. Again employing the Holstein-Primakoff representation of the angular momentum operators, \ Eq.~(\ref{Primakoff}){}, shifting the oscillator modes $a \rightarrow a \pm \sqrt{\alpha}$ and $b \rightarrow b \mp \sqrt{\beta}$ (cf. \ Eq.~(\ref{abshift}){} with  $\alpha$ and $\beta$ of $O(j)$), and proceeding to the thermodynamic limit $j\to \infty$ yields an effective Hamiltonian with terms up to quadratic order in the bosonic operators. As before in the treatment of  $H_{\rm Dicke}$,  the linear terms  can be eliminated, which yields an equation for $\beta$,
\begin{eqnarray}
  \frac{4 \lambda^2}{\omega} \frac{j-\beta}{j}\sqrt{\beta}
  -\Omega\sin\theta \sqrt{\beta}
  +\Omega\cos\theta \frac{j-\beta}{\sqrt{2j -\beta}}
  =0,
  \label{beta_solution}
\end{eqnarray}
and $\sqrt{\alpha} = ({2 \lambda}/{\omega})[{(2j-\beta)\beta}/{2j}]^{1/2}$, leading to a Hamiltonian that can be diagonalized after a Bogoliubov transformation of the bosonic operators.

\begin{figure}[t]
  \centerline{
    \includegraphics[clip=true,width=0.7\columnwidth]{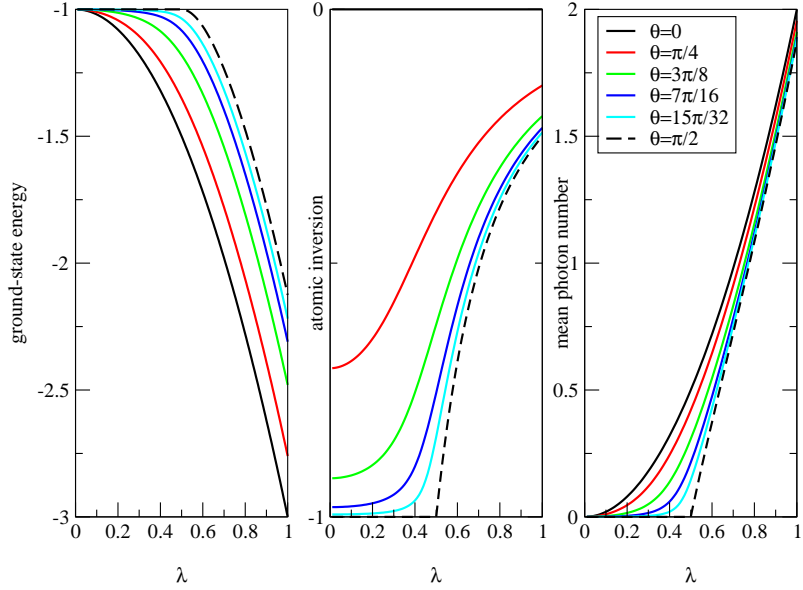}
  }
  \caption{{\bf From left:} Ground-state energy, atomic inversion and
           mean photon number of the ground state
           as a function 
           of the coupling $\lambda$ for various different angles $\theta$.
           The Hamiltonian is  on scaled resonance,
           $\omega = \Omega=1$, $\lambda_c=0.5$. From \cite{EB04}.
           \label{EB04_Fig5.eps}
          }
\end{figure}

Again, the parameters $\alpha$ and $\beta$ are related to the atomic inversion and mean field occupation, \ Eq.~(\ref{alphabetarelated}), although not all solutions of \ Eq.~(\ref{beta_solution}){} are physically valid \cite{EB04}. In particular, it turns out that a co-existence of {\em two} physically valid solutions, corresponding to both the upper and the lower sign in the shifted bosonic operators, \ Eq.~(\ref{abshift}){}, exist only for the special case $\theta = \pi/2$, which is exactly the original Dicke Hamiltonian $H_{\rm Dicke}$, \ Eq.~(\ref{Dicke_first}){}. This is illustrated in Fig. (\ref{EB04_Fig1.eps}), where the bifurcation at the critical point $\lambda_c$ of the Dicke model at $\theta = \pi/2$ separates the models $\theta <\pi/2$ and $\theta > \pi/2$, which have shifts corresponding to either the upper or the lower sign in $a \rightarrow a \pm \sqrt{\alpha}$ and $b \rightarrow b \mp \sqrt{\beta}$ for all coupling constants. This indicates that a phase transition in the spin-boson models, \ Eq.~(\ref{oham}, occurs only for `orthogonal' couplings $\theta = \pi/2$, which shows that the Dicke model is  unique within the whole class of Hamiltonians $H_{\theta}$. These findings are corroborated by a calculation of the excitation energy pairs $\varepsilon_{\pm}$ corresponding to the two collective modes of the diagonalized Hamiltonians for $j\to \infty$, cf. Fig. (\ref{EB04_Fig1.eps}) left, where critical behavior (the vanishing of $\varepsilon_-$) only occurs at $\theta=\pi/2$.  Furthermore, non-analyticities in the  ground-state energy, atomic inversion and  mean photon number of the ground-state at $\lambda=\lambda_c$, cf. Fig. (\ref{EB04_Fig5.eps}), is observed only at $\theta=\pi/2$ in agreement with these results.

\subsection{Quantum Phase Transitions and Entanglement}
Zero-temperature quantum phase transitions occur in models both from Quantum Optics and Condensed Matter Physics, and the relation between quantum entanglement and the singularities associated with the transition have been addressed in quite a large number of works recently. As the question of meaningful entanglement measures is non-trivial (in particular when it comes to, e.g., mixed states, infinite dimensional Hilbert spaces, or multi-partite systems), most of the research done so far deals with the entanglement entropy for bi-partite systems,  or the pairwise entanglement (concurrence) between two spin $1/2$s. One common feature of many of these works is the study of models that are exactly solvable in some limit, for example $XY$ models in one dimension and large-spin (boson) models, with some of the key topics being the role that entanglement in quantum phase transitions plays with respect to critical correlations \cite{Ostetal02,ON02}, in renormalization group theory \cite{Car04}, in conformal field theory \cite{VLRK03}, in finite-size scaling \cite{Ostetal02,RQJ04}, or in quantum chaos \cite{LEB04,HH04,FMT03,SS03,WGSH04}.

\subsubsection{Atom-Field Entanglement in the Dicke Model}
Lambert, Emary and Brandes \cite{LEB04} used the one-mode Dicke model $H_{\rm Dicke}$, \ Eq.~(\ref{Dicke_first}), for a study of quantum entanglement across a quantum phase transition, again combining analytical results for $j \to \infty$ with numerical diagonalizations at finite $j$. Defining $\hat{\rho}\equiv {\rm Tr_{spins}}|G\rangle \langle G| $ as the reduced density matrix of the field ($a^{\dagger}$) mode for the initially pure ground-state $|G\rangle$ of the total system, a measure for the entanglement between the atoms (i.e., the collection of $N$ identical two-level systems or spins, cf. \ Eq.~(\ref{Dicke_first}) ) and the field is given by the von Neumann entropy $S\equiv -{\rm Tr} \hat{\rho} \log_2 \hat{\rho}$. A peculiarity occurs in the superradiant phase with its two degenerate ground-states due to the broken parity symmetry $\hat{\Pi}$, leading to $S=S(\hat{\rho}_\pm)+1$ with $\hat{\rho}_\pm$ the reduced density matrix of either of the two macroscopically separated solutions. 

The calculation is most easily done in the $x$-$y$ representation, where in the normal phase the reduced density matrix is given by 
\begin{eqnarray}\label{rhoL}
  \rho_L(x,x')=c_L\int_{-\infty}^{\infty}dy f_L(y)\Psi^*({x,y})\Psi({x',y}),
\end{eqnarray}
where $c_L$ is a normalization constant and  $f_L(y)\equiv e^{-y^2/L^2}$ a cut-off function $f_L(y)\equiv e^{-y^2/L^2}$ introduced in order to discuss the effect of a partial trace over the atomic ($y$) modes. The density matrix \ Eq.~(\ref{rhoL}){} is identical to the density matrix of a single harmonic oscillator with frequency $\Omega_L$ in a canonical ensemble at an effective temperature $T\equiv1/\beta$ and can be obtained by simple Gaussian integration, yielding 
\begin{eqnarray}
\label{cosh}
  \cosh \beta \Omega_L = 1 + 2\frac{\varepsilon_-\varepsilon_+ + 4(\varepsilon_-\cos\gamma^2+\varepsilon_+
\sin\gamma^2)/L^2} {(\varepsilon_--\varepsilon_+)^2\cos\gamma ^2\sin\gamma^2},
\end{eqnarray}
where the angle $\gamma$ and the excitation energies were defined in \ Eq.~(\ref{Psi_normal}) and \ Eq.~(\ref{lcepspm}), respectively.

This leads to a simple expression for the  entropy (depending on the cut-off length parameter $L$),
\begin{eqnarray}\label{SLdef}
   S_L(\zeta) = \left[\zeta \coth \zeta - \ln (2\sinh \zeta)\right]/\ln 2,\quad \zeta\equiv \beta\Omega_L/2.
  \label{sT}
\end{eqnarray}
For $L=\infty$, the entropy undergoes a divergence at the critical point for the  approach to $\lambda_c$ from either side which is due to the vanishing of $\varepsilon_- \propto |\lambda-\lambda_c|^{2 \nu}$, \ Eq. (\ref{varepsilon_scaling}), cf. Fig. (\ref{LEB04.eps}). Together with  $S_\infty(\zeta)=[1-\ln (2\zeta)+\zeta^2/6]/\ln 2+O(\zeta^4)$, this yields the logarithmic divergence of  $S_{\infty}$,
\begin{eqnarray}
S_{\infty} \propto
- \nu \log_2 |\lambda-\lambda_c| = \log_2 \xi, \quad \nu = 1/4,
\end{eqnarray}
demonstrating that the entanglement between the atoms and field diverges with the same critical exponent as the characteristic length. For $\lambda\to\lambda_c$, the  parameter $\zeta=\hbar \Omega_{\infty}/k_B T$ of the fictitious thermal oscillator approaches zero, indicating that a {\em classical} limit is being approached, that can be interpreted either as the temperature $T$ going to infinity, or the frequency $\Omega_{\infty}$ approaching zero. An alternative is to keep  $\Omega$ fixed and compensate by introducing a squeezing parameter $\kappa$ that tends to $0$ at the critical point \cite{LEB04a}. In terms of the original parameters of the system, however, the dependence of the entropy is through the ratio of energies $\varepsilon_-/\varepsilon_+$ only. Although the entanglement is a genuine quantum property of the combined atom-field system, this highlights that in the limit of $N\to \infty$ atoms, the exact mapping of $H_{\rm Dicke}$ to two coupled oscillators is strongly connected to the corresponding (cusp) singularity and the vanishing of one of the eigenvalues of a quadratic form in the classical model.

As pointed out by Srednicky \cite{Sre93} in his discussion of entropy and area, the mapping onto a single harmonic oscillator density matrix is in general no longer possible for $N>2$ coupled oscillator modes, but the entropy of a sub-system of oscillators can still be expressed as a sum over single oscillator entropies $S_{\infty}$ , \ Eq.~(\ref{SLdef}){}, with the arguments $\zeta$ determined by eigenvalues of matrices. The Dicke model for $N\to \infty$ is in fact equivalent to a zero-dimensional field theory (there are only two degrees of freedom). However, the atomic ($y$) mode has an `internal structure' as it represents a collection of atoms (or pseudo spin $1/2$s ). Similar to Srednicky's tracing out of oscillator degrees of freedoms inside a finite volume, one can  ask what happens if the trace over the (atomic) $y$-coordinate is performed over only a finite region of size $L$ for the atomic wave function. Scaling of the entanglement {\em at the transition} $\lambda=\lambda_c$ is then calculated by keeping the parameter $L$ in \ Eq.~(\ref{SLdef}){} finite. At the transition $\varepsilon_-=0$, and the relevant dimensionless energy scale is now $\varepsilon_L\equiv 2/(L^2\varepsilon_+ c^2)$ such that the entanglement entropy diverges as 
\begin{eqnarray}\label{SLinf}
S_L\propto -(1/2)\log_2(2\varepsilon_L)= \log_2L,\quad L\to \infty.
\end{eqnarray}
The logarithmic divergence at the transition resembles the entropy of a sub-region of length $L$, $S_L\approx (c+\bar{c})/6\log_2 L + k$ in 1+1 conformal field theories with holomorphic and anti-holomorphic central charges $c$ and $\bar{c}$, as discussed by  Vidal, Latorre, Rico, and Kitaev \cite{VLRK03} in an analysis of entanglement in spin-$1/2$ chains.

\begin{figure}[t]
\centerline{
  \includegraphics[clip=false,width=0.5\textwidth]{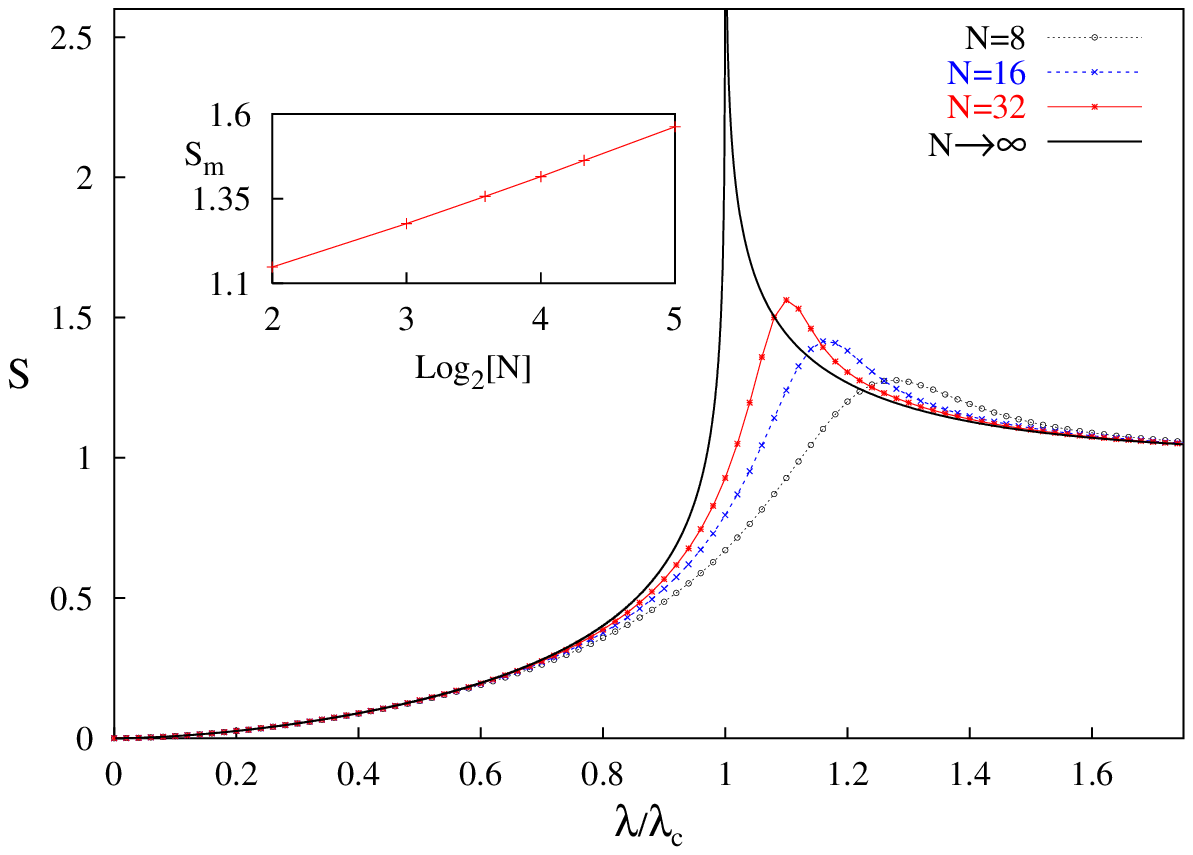}
  \includegraphics[clip=false,width=0.5\textwidth]{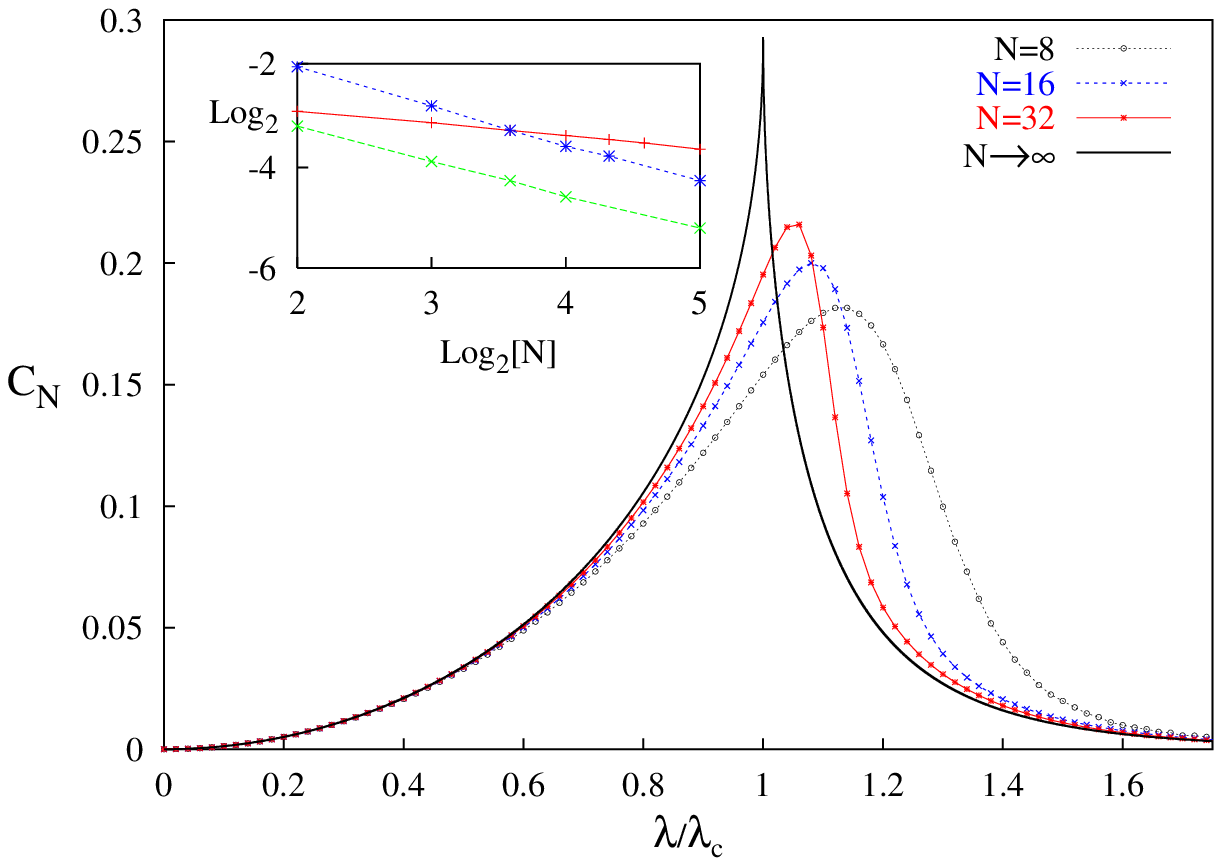} 
} \caption{\label{LEB04.eps}{\bf Left:} Entanglement of formation $S_{\infty}$
between atoms and field for both $N \rightarrow \infty$ and finite
$N$. Inset: Scaling of the value of the entanglement maximum as a
function of $\log_2N$. {\bf Right:} Scaled pairwise concurrence  $C_N=NC$
between two spins for both $N \rightarrow \infty$ and  finite $N$.
Inset:  Scaling of the value ($+$) and position ($\times$) of the
concurrence maximum, and the position of the entropy maximum ($*$)
as functions of $N$. The Hamiltonian is on scaled resonance
$\omega =\omega_0 = 1$. From \cite{LEB04}.}
\end{figure}

\subsubsection{Pairwise Entanglement and Squeezing in the Dicke Model}
As shown by Wooters \cite{Woo98}, the entanglement between any two spin-$1/2$s in a mixed state $\rho_{12}$ can be calculated from the  concurrence $C$ which in the Dicke model, \ Eq.~(\ref{Dicke_first}){} should be scaled by a factor $N$ in order to compensate for the $1/\sqrt{N}$ in the coupling energy,
\begin{eqnarray}\label{CN_def}
  C_N\equiv N C,\quad C\equiv{\rm max} \{0,\lambda_1-\lambda_2-\lambda_3-\lambda_4\}
\end{eqnarray}
where the $\lambda_i$ are the square roots of the eigenvalues (in descending order) of $\rho_{12}(\sigma_{1y}\otimes \sigma_{2y}) \rho_{12}^* (\sigma_{1y}\otimes \sigma_{2y})$.  
Wang and M{\o}lmer calculated the concurrence of pure Dicke states $|j=N/2,m\rangle$, \ Eq.~(\ref{eigenstates}){}, 
\begin{eqnarray}
  C=\frac{1}{2N(N-1)}\left\{N^2-4M^2 -\sqrt{(N^2-4M^2)[(N-2)^2-4M^2]}\right\},
\end{eqnarray}
using the $S_N$ permutation symmetry of the Dicke Hamiltonian $H_{\rm Dicke}$. Similarly, the mixed state $\rho_{12}$ as obtained from the ground state of $H_{\rm Dicke}$ has the form 
\begin{eqnarray}
\rho_{12}=
  \begin{pmatrix}
   \nu_+ & 0 & 0 & u \\
     0   & w & w & 0 \\
     0   & w & w & 0 \\
     u   & 0 & 0 & \nu_- 
\end{pmatrix},
\end{eqnarray}
where $\nu_\pm$, $u$, and $w$ can be expressed by the expectation values of the collective operators, $\langle J_z \rangle$, $\langle J_z^2 \rangle$, and $\langle J_+^2 \rangle$. For small coupling $\lambda$, perturbation theory yields an $N$-independent behavior of $C_N$,
\begin{eqnarray}
C_N(\lambda\to 0) \sim 2\alpha^2/(1+\alpha^2),\quad \alpha\equiv \lambda/(\omega+\omega_0).
\end{eqnarray}
At finite $N$, Wang and Sanders \cite{WS03} proved a quantitative relation between {\em spin squeezing} and pairwise entanglement valid for symmetric multi-qubit states for Hamiltonians with spin permutation symmetry. They considered a collective spin operator $S_\alpha\equiv \sum _{i=1}^N\sigma_{i\alpha}/2$, $\alpha=(x,y,z)$ and calculated the spin squeezing parameter $\xi$ that was first introduced by Kitawaga and Ueda \cite{KU93},
\begin{eqnarray}\label{KU_def}
  \xi^2 \equiv \frac{4}{N}(\Delta \vec{S}\vec{n} )^2 = 1-2(N-1)\left[|\langle \sigma_{i+} \otimes \sigma_{j+}\rangle| + \frac{|\langle \sigma_{iz} \otimes \sigma_{jz}\rangle|}{4}-\frac{1}{4}\right],
\end{eqnarray}
where the unit vector $\vec{n}$ is perpendicular to the mean spin $\langle \vec{S} \rangle$ and defines the direction of minimal variance $(\Delta S )^2$, and the spin correlation functions can be evaluated for any $i\ne j$. Messikh, Ficek, and Wahiddin \cite{MFW03} compared the Kitawaga-Ueda definition to another definition of spin squeezing by Wineland and co-workers \cite{WBIH94} for the two-atom Dicke model. They showed that the former is a better definition of entanglement. Wang and Sanders \cite{WS03a} discussed how both definitions coincide with bosonic squeezing for $N\to \infty$ and calculated the transfer of squeezing between the two modes in the RWA version of the Dicke Hamiltonian $H_{\rm Dicke}^{\rm RWA}$, \ Eq.~(\ref{Dicke_RWA}){}.

In the thermodynamic limit $N\to \infty$, the scaled concurrence in the Dicke model is expressed as
\begin{eqnarray}\label{Cs}
  C_{\infty} = (1+\mu)\left[\langle (d^{\dagger})^2 \rangle -
  \langle d^{\dagger}d\rangle \right]
+\frac{1}{2}(1-\mu),
\end{eqnarray}
where $\mu=1$ and $d^{\dagger}=b^{\dagger}$ in the normal phase ($\lambda<\lambda_c$),  whereas $\mu=(\lambda_c/\lambda)^2$ and $d^{\dagger}=b^{\dagger}+\sqrt{N(1-\mu)/2}$ in the superradiant phase ($\lambda>\lambda_c$). 
Recalling $b^{\dagger} = \sqrt{\omega_0/2}(y-ip_y/\omega_0)$, one can transform Eq.(\ref{Cs}) to establish a relation between the scaled concurrence and  the {\em momentum  squeezing} that occurs if the variance $(\Delta p_y) ^2 \equiv \langle p_y^2 \rangle - \langle p_y \rangle^2$ is less than $1/2$. Expressed in terms of $(\Delta p_y) ^2$, one obtains 
\begin{eqnarray}\label{C_infty}
C_{\infty} &=& (1+\mu)\left[\frac{1}{2} - (\Delta p_y)^2/\omega_0\right] + \frac{1}{2}(1-\mu),
\end{eqnarray}
where again, setting $\mu=(\lambda_c/\lambda)^2$ gives the superradiant phase equivalent.
The two quadrature variances $(\Delta x)^2$, $(\Delta p_x)^2$, $(\Delta y)^2$, and $(\Delta p_y)^2$ were calculated by Emary  and Brandes in \cite{EB03a}. For $\lambda\to \lambda_c$, the position variances $(\Delta x)^2$ and $(\Delta y)^2$ diverge whereas the momentum variances $(\Delta p_x)^2$, and $(\Delta p_y)^2$ show squeezing. 

Explicit analytical expressions for the concurrence $C_{\infty}$, \ Eq.~(\ref{C_infty}){}, were derived in \cite{LEB04} by using the mapping to the density matrix of a thermal oscillator as
\begin{eqnarray}
C_{\infty}&=& 1-({\mu \Omega}/{\omega_0})\coth (\beta \Omega/2),\quad   \cosh \beta \Omega = 1+2\varepsilon_-\varepsilon_+/D\nonumber\\
D&\equiv& [cs(\varepsilon_--\varepsilon_+)]^2,\quad 2\Omega/\sinh
\beta\Omega = D/(\varepsilon_-c^2+\varepsilon_+s^2).
\end{eqnarray}
Due to symmetry, these are the same parameters as for the reduced field ($x$) density matrix $\rho_{\infty}$ with $s=\sin \gamma$ and $c=\cos \gamma$ interchanged, and one obtains the simple result $ C_{\infty} = 1-\mu(\varepsilon_-s^2+\varepsilon_+c^2)/\omega_0$ which at resonance ($\omega=\omega_0$) reads
  \begin{eqnarray}\label{cusp1}
C_{\infty}^{x\le 1}&=&1-\frac{1}{2}\left[\sqrt{1+ x}+\sqrt{1- x}\right],\quad x\equiv \lambda/\lambda_c\\
C_{\infty}^{x\ge 1}&=&1-\frac{1}{\sqrt{2}x^2}\left[
\left(\sin^2 \gamma\right)\sqrt{1+x^4
- \sqrt{ \left(1-x^4\right)^2 + 4}}\right.\nonumber\\
&+&\left.\left(\cos^2 \gamma\right)\sqrt{1+x^4
+ \sqrt{ \left(1-x^4\right)^2 + 4}} \right],\quad 2\gamma=\arctan[2/(x^2-1)].
  \end{eqnarray}
The explicit expressions \ Eq.~(\ref{cusp1}){} reveal the  square-root non-analyticity of the scaled concurrence in the Dicke model near the critical point $\lambda_c$. Note that the concurrence assumes its finite maximum,
\begin{eqnarray}\label{ent_max_Dicke}
  C_{\infty}(\lambda_{c}) = 1-\frac{\sqrt{2}}{2}
\end{eqnarray}
{\em at}  the critical point $\lambda=\lambda_{c}$.

\subsubsection{Entanglement in Other Spin Models}
Osterloh, Amico, Falci, and Fazio \cite{Ostetal02}  presented a detailed {\em scaling analysis} of the concurrence $C(i)$ between two sites with distance $i$ in a spin-$1/2$ ferromagnetic chain in a  magnetic field (set to unity) with nearest-neighbor interactions ($XY$ models), 
\begin{eqnarray}
  H= -\lambda (1+\gamma) \sum_{i=1}^N \sigma^x_i\sigma^x_{i+1} - \lambda (1-\gamma) \sum_{i=1}^N \sigma^y_i\sigma^y_{i+1} -  \sum_{i=1}^N \sigma^z_i,
\end{eqnarray}
which for anisotropy parameters $0<\gamma\le 1$ belong to the Ising universality class with a quantum phase transition at $\lambda_c=1$ for $N\to \infty$. They obtained $C(i)$ for the case $\gamma=1$ using  correlation functions of the Ising model. For $N\to \infty$, the maximum of $C(1)$ does not coincide with the expected non-analyticity of $C(1)$ at $\lambda=\lambda_c$. The logarithmic divergence in the first derivative, $dC(1)/d\lambda=(8/3\pi^2) \ln |\lambda-\lambda_c| + {\rm const}$ for $N\to \infty$, can be related with its  precursors at $\lambda=\lambda_m$ for finite $N$ (with $\lambda_m-\lambda_c \propto N^{-1.86}$) by using a single-parameter scaling function $f(N^{1/\nu}(\lambda-\lambda_m))$ in order to analyze the data at different $N$. This analysis confirms the critical exponent $\nu=1$ known from the Ising model and demonstrates that scaling, as well as universality (by repeating the analysis for  $\gamma\ne 1$) works for the concurrence. Another interesting feature of this model is the fact that although the correlation length diverges at  the critical point, all concurrences $C(i)$ with $i>2$ vanish.

In an earlier calculation, Schneider and Milburn \cite{SM02} considered a driven, dissipative  large pseudo-spin model  described by the equation of motion for the atomic density operator $\rho(t)$,
\begin{eqnarray}
  \frac{\partial \rho}{\partial t}=-i\frac{\omega_0}{2}[J_++J_-,\rho]+\frac{\gamma_A}{2}\left(
2J_-\rho J_+ - J_+ J_- \rho - \rho J_+ J_- \right)
\end{eqnarray}
which is in an interaction picture within Markov and rotating wave approximation, where $\Omega$ is the  Rabi frequency and $\gamma_A$ the Einstein A coefficient (damping rate) of each atom. The model exhibits a (non-equilibrium) phase transition for $\Omega/j=\gamma_A$ and $\Omega,j\to \infty$. Schneider and Milburn calculated the unscaled two-atom concurrence for $j=N/2=1$ and found entanglement in the steady state, as well as entanglement maxima on the weak coupling side of the transition that moved closer to the critical point with increasing $j$. 

Vidal, Palacios and Mosseri \cite{VPM04} considered the Hamiltonian that was first introduced by Lipkin, Meshkov, and  Glick \cite{LMG65} in Nuclear Physics,
\begin{eqnarray}\label{Lipkin}
  H&\equiv& -\frac{\lambda}{N}\sum_{i<j}\left(\sigma_x^i\sigma_x^j + \gamma \sigma_y^i\sigma_y^j\right)
- \sum_i\sigma_z^i
= -\frac{2\lambda}{N}\left(J_x^2 + \gamma J_y^2\right) -2J_z + \frac{\lambda}{2}(1+\gamma),
\end{eqnarray}
which  displays a second-order, mean-field type quantum phase transition at $\lambda_c=1$ from a non-degenerate ground state to a doubly degenerate ground state for any anisotropy parameter $\gamma\ne 1$. They calculated the re-scaled concurrence $C_{N-1}$, \ Eq.~(\ref{CN_def}), for various $0\le \gamma \le 1$. $C_{N-1}$  develops a singularity at the critical point and for finite $N$ and $\gamma\ne 1$ scales like
\begin{eqnarray}\label{Lipkin_exp}
  1-C_{N-1}(\lambda_m)\sim N^{-0.33\pm 0.01},\quad \lambda_m-\lambda_c \sim N^{-0.66\pm 0.01},
\end{eqnarray}
where $\lambda_m$ is the value of $\lambda$ for which $C_{N-1}$ is maximum. As a further interesting feature, they found a vanishing of the concurrence for $\gamma\ne 0$ at a special value $\lambda_0(\gamma)$ that lead to a phase diagram in the $\lambda$-$\gamma$-plane separating ground states with $C_{N-1}\ne 0$ but zero spin squeezing ($\xi^2=1$, cf. \ Eq.~(\ref{KU_def}){}) for $\lambda>\lambda_0$ from ground states with spin squeezing, $\xi^2 = 1- C_{N-1}<1$ for  $\lambda<\lambda_0$. For $\gamma=0$, the ground state is always spin squeezed which is surprising since the $\gamma=0$ model belongs to the same universality class as the $\gamma\ne 0$ models.

Dusuel and Vidal \cite{DV04} used Wegner's continuous unitary transformation method in order to obtain finite-size scaling exponents, i.e., the $1/N$ corrections to the $N=\infty$ results for the Lipkin-Meshkov-Glick model. This allowed them to obtain analytical results for the exponents ($1/3$ and $2/3$ in \ Eq.~(\ref{Lipkin_exp}) ).
Furthermore, Latorre, Or\'{u}s, Rico, and Vidal \cite{LORV04} studied the entanglement entropy in this model and found a surprising similarity to the one-dimensional XY-model.

Reslen, Quiroga and Johnson \cite{RQJ04} used a second-order cumulant expansion  to derive an effective temperature-dependent Hamiltonian for the one-mode Dicke model $H_{\rm Dicke}$, \ Eq.~(\ref{Dicke_first}),
\begin{eqnarray}
  H_{\rm eff}(\beta) = \omega_0 J_z -\left[\frac{2\lambda}{\sqrt{N}}\right]^2
\left[1+\frac{2}{\beta(n(\beta)+1)}\right]J_x^2,\quad n(\beta) = (e^\beta -1)^{-1},
\end{eqnarray}
which they  used to calculate thermodynamic equilibrium expectation values at temperature $1/\beta$ for the {\em atomic} degrees of freedom.  For zero temperature, this corresponds to the an-isotropic Lipkin model, \ Eq.~(\ref{Lipkin}), with $\gamma=0$, $\omega_0=-2$, and $4\lambda^2\to 2\lambda$ (dropping the constant term). They found excellent agreement between the results ($N=15$) for the ground-state energy $E_G$ and the inversion $\langle J_z \rangle$ calculated with $H_{\rm Dicke}$ and with $H_{\rm eff}(\beta\to \infty)$. Furthermore, they analyzed the dependence on $N$ of the critical $\lambda$ and $C_N$, $\lambda_m-\lambda_c$ and $C_{\infty}(\lambda_c)-C_N(\lambda_m)$, cf. \ Eq.~(\ref{ent_max_Dicke}){}, and confirmed their respective scaling with the same exponents as for the Lipkin model, \ Eq.~(\ref{Lipkin_exp}), cf. however the discussion in  \cite{DV04}.


Levine and Muthukumar \cite{LM04} calculated the entanglement entropy for the spin-$1/2$ boson Hamiltonian, 
\begin{eqnarray}
  H = \Delta \sigma_x + \frac{\lambda}{\sqrt{2m\omega}}(a+a^{\dagger}) \sigma_z + \omega a^{\dagger}a,
\end{eqnarray}
which is canonically equivalent to the Rabi Hamiltonian, \ Eq.~(\ref{Dicke_Rabi}). They found a transition from zero to finite entropy at $\alpha\equiv \lambda^2/m\omega^2\Delta =1$ in the limit $\Delta/\omega\to \infty$. The corresponding bifurcation in the ground state was illustrated by Hines, Dawson, McKenzie, and Milburn \cite{HDMM04}, who analyzed a corresponding classical model (with a spinning top replacing the spin) and also confirmed the existence of the pitchfork bifurcation for the Dicke model, cf.  Fig. (\ref{EB04_Fig1.eps}). 

Finally, Verstraete, Martin-Delgado, and  Cirac \cite{VMC04} found the divergence of entanglement {\em without} a quantum phase transition in gapped quantum spin systems. They studied spin-1 Hamiltonians with Haldane gap such as the exactly solvable Affleck-Kennedy-Lieb-Tasaki model and calculated the so-called localizable entanglement. Remarkably, they  proved that the associated entanglement length $\xi_E$ can diverge, with the  correlation length $\xi_C$ remaining finite.


\section{Conclusion}
One of the motivations of this Review has been to establish connections between Quantum Optics and Condensed-Matter Physics, primarily in the area of electronic properties of mesoscopic systems. Activities at the interface between these fields, both theoretically and experimentally, have already started to grow rapidly, also driven by the search for elementary, scalable physical systems in which quantum mechanical operations (e.g., superposition and entanglement) can be controlled from the outside. As shown in the examples of (single or multiple) artificial two-level systems, the coupling to additional bosonic modes, to electronic reservoirs, or to dissipative environments very soon gives rise to an enormous complexity. 

Many theoretical problems still remain open, and many new problems will appear in the future. The description of the combined effects  of many-body interactions, non-equilibrium physics, and quantum coherence is a great challenge for a microscopic {\em transport theory}. For example, with regard to transport through single boson-mode models, a next step would be to fully understand the influence of non-linear oscillator-couplings and Kondo-type correlations on transport, frequency-dependent quantum noise and Full Counting Statistics. 

A further example is the understanding of {\em entanglement} in situations that go beyond the relatively simple models presented in the last section of this Review. Very little (if at all) is known about its role in many-body systems with quantum phase transitions such as, e.g., in disordered electronic systems. Another question is whether the relation between quantum chaos, entanglement, and the classical limit is in actual fact much deeper than it appears from the study of the single-mode Dicke or similar models.

\section*{Acknowledgments}
Many results presented in this work have been achieved in collaborations and discussions with colleagues from all over the world. In particular, I wish to acknowledge R. Aguado, A. Armour, M. Blencowe, R. H. Blick, M. B\"uttiker, Y.-N. Chen, S. Debald, A. Donarini, C. Emary, J. Inoue, A. Kawabata, J. Kotthaus, L. P. Kouwenhoven, B. Kramer, N. Lambert, J.-J. Liang, T. Novotn\'{y}, G. Platero, F. Renzoni, J. Robinson, B. Sanchez, A. Shimizu, J. Siewert,  W. van der  Wiel, T. Vorrath, and E. Weig. This research was supported by the Deutsche Forschungsgemeinschaft (DFG Br 1528/4-1), the United Kingdom EPSRC (GR/R44690/01), the EPSRC Quantum Circuits Network, the Nuffield Foundation, and the Wilhelm and Else Heraeus Foundation.

\appendix

\section{\bf Polaron-Transformed Master Equation}\label{appendix_pol}
This appendix provides details of the derivation of the `POL' Master equation \ Eq.~(\ref{eom3new})-
\ Eq.~(\ref{pdaggereqn}){}.
\subsection{Interaction picture}
The interaction picture for arbitrary operators $O$ and the $X$ operators, \ Eq.~(\ref{Xdef}){}, is defined by
\begin{eqnarray}\label{interact}
  \tilde{O}(t)\equiv e^{i\mathcal{H}_0t}\overline{O}e^{-i\mathcal{H}_0t},\quad
X_t\equiv e^{i\mathcal{H}_0t}{X}e^{-i\mathcal{H}_0t}.
\end{eqnarray}
In particular, one has $\tilde{n}_L(t)=n_L$, $\tilde{n}_R(t)=n_R$, and 
\begin{eqnarray}\label{operatorstime}
    \tilde{p}(t)&=&\hat{p}e^{i\varepsilon t}X_t,\quad
    \tilde{p^{\dagger}}(t)=\hat{p}^{\dagger}e^{-i\varepsilon t}X_t^{\dagger},\quad 
    \varepsilon\equiv \overline{\varepsilon_L}-\overline{\varepsilon_R}.
\end{eqnarray}
Furthermore, for the total density matrix $\chi(t)=e^{-i\mathcal{H}t}\chi_{t=0} e^{i\mathcal{H}t}$ one defines
\begin{equation}\label{intera1}
  \tilde{\chi}(t)\equiv e^{i\mathcal{H}_0t}\overline{\chi(t)}e^{-i\mathcal{H}_0t},\quad
  \overline{\chi(t)}\equiv e^{-i\overline{\mathcal{H}}t}\overline{\chi}_{t=0} e^{i\overline{\mathcal{H}}t}.
\end{equation}
The expectation value of any operator $O$ is given by
\begin{eqnarray}\label{expect}
  \langle O\rangle_t&\equiv &{\rm Tr} \left( \chi(t) O\right) 
={\rm Tr}  \left(\tilde{\chi}(t)\tilde{O}(t)\right),
\end{eqnarray}
for which equations of motions are derived from the 
equation of motion for $\tilde{\chi}$,
\begin{eqnarray}\label{writeother}
   \frac{d}{dt}\tilde{\chi}(t)&=& -i[\tilde{\mathcal{H}_T}(t)+\tilde{\mathcal{H}_V}(t),\tilde{\chi}(t)]
  = -i[\tilde{\mathcal{H}_T}(t),\tilde{\chi}(t)] - i[\tilde{\mathcal{H}_V}(t),
  \chi_0]\\
 &-&\int_0^tdt'
 [\tilde{\mathcal{H}_V}(t), 
 [\tilde{\mathcal{H}_T}(t')+\tilde{\mathcal{H}_V}(t'),  \tilde{\chi}(t')]].\nonumber
\end{eqnarray}
One defines the effective density operator of the system `dot+bosons',
\begin{equation}\label{rhotilde}
  \tilde{\rho}(t)\equiv{\rm Tr}_{\rm res}\tilde{\chi}(t)
\end{equation}
as the trace over the electron reservoirs (res) and assumes the second order Born approximation with respect to ${\mathcal{H}_V}$, 
\begin{equation}
\tilde{\chi}(t)\approx R_0 \otimes \tilde{\rho}(t),\quad t>0,
\end{equation}
where $R_0$ is the density matrix of the electron reservoirs. Then, terms linear in $\overline{\mathcal{H}}_V$ vanish and it remains
\begin{eqnarray}\label{master1a}
   \frac{d}{dt}\tilde{\rho}(t)&=& -i[\tilde{\mathcal{H}_T}(t),\tilde{\rho}(t)]
-{\rm Tr}_{\rm res}\int_0^tdt'
 [\tilde{\mathcal{H}_V}(t),[\tilde{\mathcal{H}_V}(t'),R_0 \otimes \tilde{\rho}(t')  ]].
\end{eqnarray}
Performing the commutators and using the free time evolution of the electron reservoir operators,
one finds
\begin{eqnarray}\label{master2}
\frac{d}{dt}\tilde{\rho}(t)&=& -i[\tilde{\mathcal{H}_T}(t),\tilde{\rho}(t)]\nonumber\\
&-&\sum_{k_i}\int_0^tdt'
g_{k_i}(t-t')\left\{\tilde{s}_L(t)\tilde{s}_L^{\dagger}(t')\tilde{\rho}(t')-
\tilde{s}_L(t')^{\dagger}\tilde{\rho}(t')\tilde{s}_L(t)\right\}\nonumber\\
&-&\sum_{k_i}\int_0^tdt'
\bar{g}_{k_i}(t'-t)\left\{\tilde{s}_L^{\dagger}(t)\tilde{s}_L(t')\tilde{\rho}(t')-
\tilde{s}_L(t')\tilde{\rho}(t')\tilde{s}_L^{\dagger}(t)\right\}\nonumber\\
&-&\sum_{k_i}\int_0^tdt'
g_{k_i}(t'-t)\left\{
\tilde{\rho}(t')\tilde{s}_L(t')\tilde{s}_L^{\dagger}(t)-
\tilde{s}_L^{\dagger}(t)\tilde{\rho}(t')\tilde{s}_L(t')\right\}\nonumber\\
&-&\sum_{k_i}\int_0^tdt'
\bar{g}_{k_i}(t-t')\left\{
\tilde{\rho}(t')\tilde{s}_L^{\dagger}(t')\tilde{s}_L(t)-
\tilde{s}_L(t)\tilde{\rho}(t')\tilde{s}_L^{\dagger}(t')\right\}\nonumber\\
g_{k_i}(\tau) &\equiv& |V_k^i|^2 f^i(\varepsilon_{k_i}) e^{i\varepsilon_{k_i}\tau},\quad
\bar{g}_{k_i}(\tau) \equiv |V_k^i|^2 [1-f^i(\varepsilon_{k_i}) ]e^{i\varepsilon_{k_i}\tau}, \quad i=L/R,
\end{eqnarray}
with the Fermi distributions $f^i(\varepsilon_{k_i})\equiv {\rm Tr}_{\rm res}(R_0c_{k_i}^{\dagger}c_{k_i}^{\phantom{\dagger}})$.
The sums over $k_i$ can be written as integrals, introducing the tunneling density of states
$\nu_i(\varepsilon)$ in lead $i$,
\begin{equation}
\sum_{k_i}|V_{k}^i|^2f^i(\varepsilon_{k_i})e^{i\varepsilon_{k_i}(t-t')}=
\int_{-\infty}^{\infty}d\varepsilon\nu_i(\varepsilon)f^i(\varepsilon)e^{i\varepsilon(t-t')},\quad
\nu_i(\varepsilon)\equiv \sum_{k_i}|V_{k}^i|^2\delta(\varepsilon-\varepsilon_{k_i}).
\end{equation}

\subsection{Markov Approximation}
In the infinite source-drain voltage limit $\mu_L\to \infty$ and $\mu_R\to -\infty$ introduced by 
Gurvitz and Prager \cite{GP96,Gur98}, and Stoof and Nazarov \cite{SN96},  the left Fermi function is one and the right Fermi function is zero. An additional simplification is obtained by assuming {\em constant tunneling densities of states}, $\nu_i(\varepsilon)=\nu_i=\Gamma_i/2\pi$, with constant tunnel rates
$\Gamma_i\equiv 2\pi\sum_{k_i}|V_k^i|^2\delta(\varepsilon-\varepsilon_{k_i})$,
$i=L/R$, cf. \ Eq.~(\ref{tunnel_rates_def}). 
This leads to Delta functions like
\begin{eqnarray}
\sum_{k_L}|V_k^L|^2 f^L(\varepsilon_{k_L})e^{i\varepsilon_{k_L}(t-t')}&=&
\Gamma_L\delta(t-t'),
\end{eqnarray}
and correspondingly for the other terms. In this {\em Markov limit}, 
the Master equation Eq.(\ref{master2}) becomes
\begin{eqnarray}\label{master3}
\tilde{\rho}(t)&=& \bar{\rho}_0-i\int_0^tdt'
[\tilde{H_T}(t),\tilde{\rho}(t)]\\
&-&\frac{\Gamma_L}{2}
\int_0^tdt'
\left\{\tilde{s}_L(t')\tilde{s}_L^{\dagger}(t')\tilde{\rho}(t')-
2\tilde{s}_L(t')^{\dagger}\tilde{\rho}(t')\tilde{s}_L(t')\right\}\nonumber\\
&-&\frac{\Gamma_L}{2}
\int_0^tdt'\left\{\tilde{\rho}(t')\tilde{s}_L(t')\tilde{s}_L^{\dagger}(t')\right\}
-\frac{\Gamma_R}{2}\int_0^tdt'\left\{
\tilde{s}_R^{\dagger}(t')\tilde{s}_R(t')\tilde{\rho}(t')\right\}\nonumber\\
&-&\frac{\Gamma_R}{2}
\int_0^tdt'\left\{-2 \tilde{s}_R(t')\tilde{\rho}(t')\tilde{s}_R^{\dagger}(t')
+\tilde{\rho}(t')\tilde{s}_R^{\dagger}(t')\tilde{s}_R(t')\right\}\nonumber,
\end{eqnarray}
where one integration from $0$ to $t$ was performed and $\int_0^t dt' \delta(t-t')f(t') =\frac{1}{2}f(t)$ was used.

\subsection{Equations of motions}
It is now convenient to derive the equations of motions for the expectation values of the dot
variables directly from the Master equation Eq.(\ref{master3}). One first calculates the commutators
\begin{eqnarray}\label{firstcomm}
[\tilde{n}_L(t),\tilde{\mathcal{H}}_T(t')]&=&-[\tilde{n}_R(t),\tilde{\mathcal{H}}_T(t')]=
T_c\left(\tilde{p}(t')-\tilde{p}^{\dagger}(t')\right)\nonumber\\
\left[\tilde{p}(t),\tilde{\mathcal{H}}_T(t')\right] &=& T_ce^{i\varepsilon(t-t')}
\left\{ \hat{n}_L X_t X_{t'}^{\dagger}-\hat{n}_RX_{t'}^{\dagger} X_t \right\}\nonumber\\
\left[\tilde{p}^{\dagger}(t),\tilde{\mathcal{H}}_T(t')\right]&=&T_ce^{-i\varepsilon(t-t')}
\left\{\hat{n}_RX_{t}^{\dagger}X_{t'}-\hat{n}_LX_{t'}X_t^{\dagger}\right\}
\end{eqnarray}
and uses the completeness relation
\begin{equation}
1=|0\rangle\langle 0|+\hat{n}_R+\hat{n}_L
\end{equation}
in the three-dimensional Hilbert space of the double dot to express
$\tilde{s}_L(t')\tilde{s}_L^{\dagger}(t')=|0\rangle\langle 0|=1-\hat{n}_R-\hat{n}_L.$
Multiplying Eq.(\ref{master3}) with $\hat{n}_L$, $\hat{n}_R$, $\hat{p}$, and $\hat{p}^{\dagger}$ and performing the trace with the three dot states, one obtains
\begin{eqnarray}\label{eom1}
\langle \hat{n}_L\rangle_t-\langle \hat{n}_L\rangle_0&=&-iT_c\int_0^tdt'\left\{
\langle \hat{p} \rangle_{t'}-\langle \hat{p}^{\dagger}\rangle_{t'} \right\} 
+\Gamma_L\int_0^tdt'(1-\langle \hat{n}_L\rangle_{t'} - \langle \hat{n}_R\rangle_{t'})\nonumber\\
\langle
\hat{n}_R\rangle_t-\langle \hat{n}_R\rangle_0&=&iT_c\int_0^tdt'\left\{ \langle
\hat{p}\rangle_{t'}-\langle \hat{p}^{\dagger}\rangle_{t'} \right\} 
-\Gamma_R\int_0^tdt'\langle \hat{n}_R\rangle_{t'} \nonumber\\ \langle
\hat{p}\rangle_t-\langle
\hat{p}\rangle_t^0&=&-\frac{\Gamma_R}{2}\int_0^tdt' e^{i\varepsilon(t-t')}  \langle
X_t^{\phantom{\dagger}}X_{t'}^{\dagger}\tilde{p}(t')\rangle_{t'}\\
&-&iT_c\int_0^tdt'e^{i\varepsilon(t-t')}
\left\{ \langle
\hat{n}_LX_t^{\phantom{\dagger}}X_{t'}^{\dagger}\rangle_{t'}- \langle
\hat{n}_RX_{t'}^{\dagger}X_t^{\phantom{\dagger}}\rangle_{t'}\right\}\nonumber\\
\langle \hat{p}^{\dagger}\rangle_t-\langle
\hat{p}^{\dagger}\rangle_t^0&=&-\frac{\Gamma_R}{2}\int_0^tdt' e^{-i\varepsilon(t-t')} \langle
\tilde{p}^{\dagger}(t')X_{t'}^{\phantom{\dagger}}X_{t}^{\dagger}\rangle_{t'}\nonumber\\
&+&iT_c\int_0^tdt'e^{-i\varepsilon(t-t')}
\left\{ \langle
\hat{n}_LX_{t'}^{\phantom{\dagger}}X_{t}^{\dagger}\rangle_{t'}- \langle
\hat{n}_RX_{t}^{\dagger}X_{t'}^{\phantom{\dagger}}\rangle_{t'}\right\}\nonumber.
\end{eqnarray}
Here, expectation values 
are defined as the trace over the dot {\em and} the boson system, and 
\begin{eqnarray}\label{pinitial}
\langle \hat{p}^{(\dagger)}\rangle_t^0\equiv {\rm Tr}\left( \bar{\rho}_0
(p e^{i\varepsilon t} X_t)^{(\dagger)}\right).
\end{eqnarray}
The time evolution of the expectation values $\langle \hat{p}^{(\dagger)}\rangle_t^0$ describes the decay of an initial polarization of the system and can be calculated exactly. This decay, however, plays no role for the stationary current, and one can safely assume zero inital expectation values  of $\hat{p}^{(\dagger)}$ whence 
$\langle \hat{p}^{(\dagger)}\rangle_t^0=0$ for all $t>0$. 

As can be recognized from Eq.(\ref{eom1}), the system of equations for the dot expectation values
is not closed since terms like $\langle \hat{n}_LX_t^{\phantom{\dagger}}X_{t'}^{\dagger}\rangle_{t'}$  contain products of dot and boson ($X$) operators. At this point, one invokes a physical argument to decouple the equations: if one is not interested in the backaction of the electron  onto the boson system, the latter can be assumed to be in thermal equilibrium all times, in particular when dealing with a continuum of infinitely many bosonic modes ${\bf Q}$. One {\em decouples} the reduced density matrix $\tilde{\rho}(t')$ according to
\begin{eqnarray}\label{dec}
\tilde{\rho}(t')\approx
\rho_B \otimes \tilde{\rho}_{\rm dot}(t'),\quad \tilde{\rho}_{\rm dot}(t') ={\rm Tr}_B \tilde{\rho}(t'),
\end{eqnarray}
cf. the discussion after \ Eq.~(\ref{decoupling_app}){}. This directly leads to  Eq.~(\ref{eom3new}).


\section{\bf{Calculation of the Boson Correlation Function $C_\varepsilon$}}\label{appendix_a}
Here, some explicit expressions for the Laplace transform of the  bosonic correlation function $C(t)$, \ Eq.~(\ref{Ct}),
\begin{eqnarray}\label{Czdefinition}
  \hat{C}(z) &\equiv& \int_{0}^{\infty}dt e^{-zt} C(t)\\
 C(t)&=& \exp\left\{\int_0^{\infty}d\omega
\frac{J(\omega)}{\omega^2} \left[ \left(1- \cos \omega t\right)
\coth \left(\frac{\beta \omega}{2}\right) + i \sin \omega t
\right]\right\}
\end{eqnarray}
are derived.

\subsection{Zero Temperature Ohmic Case}
For Ohmic dissipation with  $s=1$, \ Eq.~(\ref{Jomegageneric}), one has a boson spectral density $J(\omega)=2\alpha \omega \exp(-\omega{}/\omega{}_c)$.  At zero temperature ($1/\beta=T=0$), $C(t)=(1+i\omega_c t)^{-2\alpha}$, and one finds \cite{BAP04}
\begin{eqnarray}
  \hat{C}(z)&\equiv&\int_{0}^{\infty}dt e^{-zt}(1+i\omega_c t)^{-2\alpha}
=(i\omega_c)^{-2\alpha}z^{2\alpha-1}e^{-iz/\omega_c}\Gamma(1-2\alpha,-iz/\omega_c),
\end{eqnarray}
where Gradstein-Ryshik 3.382.4 was used and $\Gamma(\cdot,\cdot)$ denotes the incomplete Gamma function.
Measuring $\omega$ in units of the cut-off $\omega_c$ (setting $\omega_c=1$) simplifies the notation and one obtains
\begin{eqnarray}
  \hat{C}(-i\varepsilon)&=&-i\left(-\varepsilon\right)^{2\alpha-1}
e^{-\varepsilon}
\Gamma(1-2\alpha,-\varepsilon).
\end{eqnarray}
Note that $\varepsilon$ must have a small positive imaginary part  ${\mbox{Re}} z>0$ in the definition of
the Laplace transformation since the incomplete Gamma function $\Gamma(1-2\alpha,z)$ has a branch point at 
$z=0$. However, one can use the series expansion 
$ \Gamma(1-2\alpha,x)=\Gamma(1-2\alpha)-\sum_{n=0}^{\infty}{(-1)^nx^{1-2\alpha+n}}/[{n!(1-2\alpha+n)}]$ for $1-2\alpha\ne 0,-1,-2,...$ to obtain
\begin{eqnarray}
   \hat{C}(-i\varepsilon)&=&-i(-\varepsilon)^{2\alpha-1}e^{-\varepsilon}\Gamma(1-2\alpha)
+ie^{-\varepsilon}\sum_{n=0}^{\infty}\frac{\varepsilon^{n}}{n!(1-2\alpha+n)},\quad 2\alpha\ne 1,2,3,...
\nonumber
\end{eqnarray}
The second term is an analytic function  of $\varepsilon$. Now one writes
\begin{eqnarray}
  -i(-\varepsilon)^{2\alpha-1}&=&\left\{
  \begin{array}{cc}
-i|\varepsilon|^{2\alpha-1},&\varepsilon<0\\
\varepsilon^{2\alpha-1}e^{-\pi i(1/2+2\alpha-1)}      &\varepsilon>0.
  \end{array}\right.
  \end{eqnarray}
Recalling the reflection formula for the Gamma function,
$  \Gamma(1-z)=\frac{\pi}{\Gamma(z)\sin \pi z}$, this yields
\begin{eqnarray}
\hat{C}(-i\varepsilon)&=&
\frac{\pi}{\Gamma(2\alpha)}\varepsilon^{2\alpha-1}e^{-\varepsilon} + i\left[ \frac{\pi}{\Gamma(2\alpha)}\varepsilon^{2\alpha-1}e^{-\varepsilon} \cot 2\pi \alpha+
e^{-\varepsilon}\sum_{n=0}^{\infty}\frac{\varepsilon^{n}}{n!(1-2\alpha+n)}\right],\quad \varepsilon>0\nonumber\\
\hat{C}(-i\varepsilon)&=&
ie^{-\varepsilon}\left[-\frac{\pi}{\Gamma(2\alpha)\sin 2\pi\alpha}
|\varepsilon|^{2\alpha-1}+ \sum_{n=0}^{\infty}\frac{\varepsilon^{n}}{n!(1-2\alpha+n)}\right],\quad \varepsilon<0
\end{eqnarray}
From this, one reads off the real and the imaginary part of $\hat{C}(-i\varepsilon)$,
\begin{eqnarray}\label{PE}
  \mbox{Re} \hat{C}(-i\varepsilon)&\equiv& \pi P(\varepsilon)
=\frac{\pi}{\Gamma(2\alpha)}\varepsilon^{2\alpha-1}e^{-\varepsilon} \theta(\varepsilon)\\
   \mbox{Im} \hat{C}(-i\varepsilon)&\equiv&
e^{-\varepsilon}\left[\sum_{n=0}^{\infty}\frac{\varepsilon^{n}}{n!(1-2\alpha+n)}+
\frac{\pi|\varepsilon|^{2\alpha-1}}{\Gamma(2\alpha)\sin 2\pi\alpha}\cdot
\left\{
  \begin{array}{ll}
-1,&\varepsilon<0\\
\cos 2\pi \alpha ,&\varepsilon>0
  \end{array}\right\}
\right].
\end{eqnarray}

\subsection{`Structured Bath' with Oscillatory $J(\omega)$}
In the case of  more complicated spectral densities it is advantageous to split $J(\omega)$ into an Ohmic and a non-Ohmic part, e.g. for the the piezo-acoustic case, \ Eq.~(\ref{Jmicro}),
\begin{eqnarray}
  J(\omega) = J_{\rm ohm}(\omega) + J_{\rm osc}(\omega),\quad 
   J_{\rm ohm}(\omega)\equiv 2\alpha \omega e^{-\omega/\omega_c},
J_{\rm osc}(\omega) \equiv -2\alpha \omega_d \sin \left(\frac{\omega}{\omega_d}\right) e^{-\omega/\omega_c},
\end{eqnarray}
where $\omega_c$ is the high-frequency cut-off and $\omega_d=c/d$ the additional frequency scale of the bosonic bath, cf. \ Eq.~(\ref{Jmicro}). One writes 
\begin{eqnarray}
  C(t) &=& e^{-Q(t)},\quad Q(t) \equiv  Q_{\rm ohm}^{T=0}(t) +  Q_{\rm osc}^{T=0}(t) +
Q_{\rm ohm}^{T>0}(t) +  Q_{\rm osc}^{T>0}(t) \label{Csplitting}\\ 
Q_i^{T=0}(t) &\equiv& \int_{0}^{\infty}d\omega \frac{J_i(\omega)}{\omega^2}\left[1-\cos \omega t + i \sin \omega t \right]\\
Q_i^{T>0}(t) &\equiv& \int_{0}^{\infty}d\omega \frac{J_i(\omega)}{\omega^2}\left[\left( 1-\cos \omega t\right) \left(\coth (\beta\omega/2) -1 \right)\right],\quad i = {\rm ohm,osc},
\end{eqnarray}
thus separating the zero temperature contribution from the finite temperature contribution. 
\ Eq.~(\ref{Csplitting}) is convenient for a numerical evaluation of the Laplace transform, \ Eq.~(\ref{Czdefinition}), where one writes $z=-i\varepsilon +\delta$ and uses $e^{-z}= e^{-\delta}(\cos \varepsilon t + i \sin \varepsilon t )$ which is useful to take advantage of special routines for integrals over the semi-infinite, positive real axis with weight functions $\sin()$  and $\cos()$.

The zero temperature parts in \ Eq.~(\ref{Csplitting}) are
\begin{eqnarray}
Q_{\rm ohm}^{T=0}(t)&=& 2 \alpha \ln (1+i\omega_c t), \quad 
Q_{\rm osc}^{T=0}(t)= -2\alpha\frac{\omega_d}{\omega_c}\left[2 f\left(\omega_ct,\frac{\omega_c}{\omega_d}\right)+i
g\left(\omega_ct,\frac{\omega_c}{\omega_d}\right)\right]\\
f(x,y)&\equiv&\frac{1}{8}\Big\{y\ln\left[\frac{(1+(x+y)^2)(1+(x-y)^2)}{(1+y^2)^2}\right]
+x\ln\left[\frac{1+(x+y)^2}{1+(x-y)^2}\right]\nonumber\\
&+&2\arctan(x+y)+2\arctan(y-x) -4\arctan(y)\Big\}\nonumber\\
g(x,y)&\equiv&\frac{1}{2}\Big\{\frac{1}{2}\ln\left[\frac{1+(x+y)^2}{1+(x-y)^2}\right]
+(x+y)\arctan(x+y) -(y-x)\arctan(y-x) \Big\}.\nonumber 
\end{eqnarray}
Furthermore, the Ohmic finite temperature contribution is expressed in terms of Gamma functions of complex argument,
\begin{eqnarray}
  Q_{\rm ohm}^{T>0}(t) = -4\alpha \left\{ \ln
\left|\Gamma\left(1+\frac{1}{\beta\omega_c}+i\frac{t}{\beta}\right)\right| - 
\ln \left|\Gamma\left(1+\frac{1}{\beta\omega_c}\right)\right|\right\}.
\end{eqnarray}



\section{\bf{Memory Function Formalism for Quantum Wire in Magnetic Field}}\label{Appmemory}
The memory function formalism \cite{GW72} starts from the observation that it is more advantageous to perform an expansion of the inverse conductivity $\sigma ^{-1}$ rather than of $\sigma $ itself. The reason is that $\sigma \sim  \tau$, the transport time which (to lowest order) in turn is {\em inversely}  proportional to the square of the scattering potential matrix element. Therefore, one introduces a memory function which in the multichannel case becomes a matrix, 
\begin{equation}
\label{memoryfunction}
M(z)\equiv z\chi(z)[\chi^{0}-\chi(z)]^{-1}.
\end{equation}
Solving for the matrix 
\begin{equation}
\label{chi_matrix}
\chi(z)=[z+M(z)]^{-1}M\chi^{0}
\end{equation} 
and inserting into
Eq.~(\ref{sigmageneral}), with Eq.~(\ref{chi0}) one obtains
\begin{equation}
\label{sigmam}
\sigma (z)=ie^{2}\sum_{nm}\left([z+M]^{-1}\chi^{0}\right)_{nm}.
\end{equation}
Note that $M$ and $\chi^{0}$ are matrices so that in the multichannel case
a matrix inversion is required. 
The calculation is started by expanding Eq.~(\ref{chi_matrix}) in terms of the 
memory matrix, $z\chi=M\chi^{0}+...$.  
By calculating $M$ rather then $\chi$, a partial summation in the scattering potential (ladder diagrams) is already performed.
 
\subsection{Conductivity in a Multi-Channel System}
The equation of motion
\begin{equation}
z\langle\langle j_{n};j_{m}\rangle\rangle_{z} = L_s\langle [j_{n},j_{m}]\rangle
+\langle\langle A_{n};j_{m}\rangle\rangle_{z},\quad A_{n}\equiv[j_{n},H],
\end{equation}
together with $[j_{n},j_{m}]=0$ is used twice \cite{GW72} to obtain
an expression for $M$,
\begin{equation}
\label{mchi}
z(M\chi^{0})_{nm}= \phi_{nm}(z)-\phi_{nm}(0), \quad
\phi_{nm}(z) \equiv \langle\langle A_{n};A_{m}\rangle\rangle_{z}. 
\end{equation} 

The matrix $M(z)$ has a spectral representation and can be decomposed into 
real and imaginary part, $M(\omega +i0)=M'(\omega )+iM''(\omega )$ with real 
matrices $M'(\omega )=-M'(\omega )$ and $M''(\omega )=M''(-\omega )$.
For $\omega \to 0$, the real part $M'(0)=0$. Consequently, in the dc-limit
$z= \omega +i0\to 0+i0$,
\begin{equation}
\label{L}
M(z)\chi^{0}=\frac{\phi(z)-
\phi(0)}{z}
\to 
i{\rm 
Im}\,\left.\frac{\partial}{\partial\omega}\phi(\omega) \right|_{\omega 
=0}\equiv iL. 
\end{equation}
An expression for the ac conductivity can be obtained in the limit of frequencies $z$ so small that the dependence  of $M(z)$ on $z$ can be neglected. In the limit of $\hbar\omega\ll \varepsilon$, the energy dependence of the  scattering rates around the Fermi energy $\varepsilon_F$ is assumed to be negligible.  In terms of the $L$-matrix, $\sigma(z)$ can then be written as
\begin{equation}
  \label{eq:condl}
  \sigma(z)=ie^2\sum_{nm}\left(\chi^0[z\chi^0+iL]^{-1}\chi^0\right)_{nm}.
\end{equation}
The commutator $A_{n}$ is easily obtained as
\begin{equation}
\label{An}
A_{n}=\frac{1}{{L_{s}^2}}
\sum_{k,q,n'}\left[V_{nn'}(k,q)v_{nk}c^{+}_{nk}c_{n'k+q}-
V_{n'n}(k,q)v_{nk+q}c^{+}_{n'k}c_{nk+q}\right]. 
\end{equation}
Calculation of the matrix elements
$z(M(z)\chi^{0})_{nm}$, Eq.~(\ref{mchi}), requires the correlation function 
matrix elements which we denote by
\begin{equation}
<n,n';m,m'>\equiv\langle\langle 
c^{+}_{nk}c_{n'k+q};c^{+}_{mk'}c_{m'k'+q'}\rangle\rangle_{z}, 
\end{equation}
suppressing the indexes $k,k',q,q'$ which remain the same.
This leads to 
\begin{eqnarray}
\label{AA}
\phi_{nm}(z)&=&
\frac{1}{L_{s}^4}\sum_{n'm'kk'qq'}[
 V_{nn'}(k,q)V_{mm'}(k'q')v_{nk}v_{mk'}<n,n';m,m'>\nonumber\\
-&V_{nn'}(k,q)&V_{m'm}(k'q')v_{nk\phantom{+q}}v_{mk'+q'}<n,n';m',m>\nonumber\\
-&V_{n'n}(k,q)&V_{mm'}(k'q')v_{nk+q}v_{mk'\phantom{+q'}}<n',n;m,m'>\nonumber\\
+&V_{n'n}(k,q)&V_{m'm}(k'q')v_{nk+q}v_{mk'+q'}<n',n;m',m>].
\end{eqnarray}
The above equations constitute the 
general framework for the calculation of the conductivity in a multichannel 
system. To second order in the potential scattering, they are still 
completely general. For non-interacting electrons, one has
\begin{eqnarray}
\label{free}
<n,n';m,m'>&=&\delta _{q,-q'}\delta _{k',k+q}\delta _{nm'}\delta _{n'm}L_s
\varphi_{nm}(z),\quad 
\varphi_{nm}(z)\equiv
\frac{f(\varepsilon _{nk})-f(\varepsilon _{mk+q})}{z+\varepsilon 
_{nk}-\varepsilon _{mk+q}},
\end{eqnarray}
where we again suppressed the indexes $k$ and $k+q$. One obtains from  
Eq.~(\ref{free}) and Eq.~(\ref{AA})
\begin{eqnarray}
\label{AAfree}
\phi_{nm}(z)&=&
\frac{1}{L_{s}^3}\sum_{kq}|V_{nm}(q)|^{2}\left[v_{nk}v_{mk+q}\varphi_{nm}(z) 
+ v_{nk+q}v_{mk}\varphi_{mn}(z)\right]\\
&-&\delta _{nm}\frac{1}{L_{s}^3}\sum_{kqn'}
|V_{nn'}(q)|^{2}\left[v_{nk}v_{mk}\varphi_{nn'}(z) 
+ v_{nk+q}v_{mk+q}\varphi_{n'n}(z)\right]\nonumber.
\end{eqnarray}
In the limit of temperatures $k_{B}T,\hbar \omega\ll \varepsilon _{F}$, one has
\begin{eqnarray}
-{\rm Im}\,\varphi _{nm}(\omega )&=&\frac{\pi\omega 
}{v_{n}v_{m}}\left[\left.\left.\right.\right.\right.
\delta (k-k_{n})\left\{\delta (q+k_{n}-k_{m})+\delta (q+k_{n}+k_{m}) 
\right\}
\nonumber\\
&+&\delta (k+k_{n})\left.\left\{\delta (q-k_{n}-k_{m})+\delta (q-k_{n}+k_{m}) 
\right\}\right].
\end{eqnarray}
This leads to
\begin{eqnarray}
-{\rm Im}\,\phi_{nm}(\omega )
&=&
s\frac{4\pi\omega }{(2\pi)^{2}L_s}\left(
|V_{nm}(k_{n}-k_{m})|^{2}-|V_{nm}(k_{n}+k_{m})|^{2}\right)\\
&-&\delta _{nm}\sum_{n'}s
\frac{4\pi\omega}{(2\pi)^{2}L_s} \frac{v_{n}}{v_{n'}}\left(
|V_{nn'}(k_{n}-k_{n'})|^{2}+|V_{nn'}(k_{n}+k_{n'})|^{2}\right)\nonumber,
\end{eqnarray}
where $s=1$ or $s=2$ is the spin degeneracy.
Using Eq.~(\ref{chi0}) and Eq.~(\ref{L}), the matrix $L$ thus is
\begin{eqnarray}
\label{Lnmarray}
L_{nm}&=& 
\frac{s }{\pi L_s}\left(
|V_{nm}(k_{n}+k_{m})|^{2}-|V_{nm}(k_{n}-k_{m})|^{2}\right), \quad n\ne 
m\\
L_{nn}&=& \frac{s}{\pi L_s}\left[\sum_{n'\ne n}
\frac{v_{n}}{v_{n'}}
\left(|V_{nn'}(k_{n}-k_{n'})|^{2}+|V_{nn'}(k_{n}+k_{n'})|^{2}\right)
+2|V_{nn}(2k_{n})|^{2}\right]\nonumber.
\end{eqnarray}

\subsection{Potential Scattering Matrix Elements}\label{Appmatrix}
The momentum matrix element 
\begin{eqnarray}
  \label{eq:momentummatrix}
  \langle nk|e^{-i{\bf qx}}|n'k'\rangle=\delta_{k,k'+q_x}M^{q_x}_{nn'}(q_y)
\end{eqnarray}
reflects momentum conservation in $x$-direction. 
The matrix elements $M$ can be calculated exactly, their explicit expressions
for $n=0,1$ are
\begin{eqnarray}
  \label{eq:matrixexpl}
  |M^{q_x}_{00}(q_y)|^2&=&e^{-\frac{1}{2}\left(\xi^2+\eta^2\right)},\quad
  |M^{q_x}_{10}(q_y)|^2=e^{-\frac{1}{2}\left(\xi^2+\eta^2\right)}\frac{1}{2}
\left[\xi^2+\eta^2\right]\nonumber\\
  |M^{q_x}_{11}(q_y)|^2&=&e^{-\frac{1}{2}\left(\xi^2+\eta^2\right)}
\left[1-\frac{1}{2}(\xi^2+\eta^2)\right]^2,\quad
\xi = l_B\alpha q_x,\quad \eta=l_B q_y,
\end{eqnarray}
where we introduced the effective magnetic length $l_B$, the  cyclotron frequency $\omega_B$, and
the parameter $\alpha$ according to
\begin{equation}
  \label{eq:defimagnet}
   \alpha \equiv \frac{\omega_c}{\omega_B},\quad \omega_c \equiv\frac{eB}{m^*c},\quad
\omega_B\equiv\sqrt{\omega_0^2+\omega_c^2},\quad l_B\equiv\sqrt{\frac{\hbar}{m^*\omega_B}}.
\end{equation}
The matrix elements Eq. (\ref{eq:imaveragematrix}) can be evaluated explicitly
for Delta-scatterers with $u({\bf q})$ independent of ${\bf q}$. In this case,
\begin{equation}
  \label{eq:deltascatt}
  \left|u\left({\bf q}=(k-k',q_y)\right)\right|^2\equiv{V_0^2},
\end{equation}
The remaining sum 
$(1/L_s)\sum_{q_y}|M_{nn'}^{q_x}(q_y)|^2$
can be transformed into an integral and yields the result
\begin{eqnarray}
  \label{eq:Vfinalarray}
  \overline{|V_{00}(q)|^2}&=&\frac{n_iV_0^2L_s}{\sqrt{2\pi l_B^2}}e^{-\frac{1}{2}(l_B\alpha q)^2},\quad
  \overline{|V_{10}(q)|^2}=  \overline{|V_{00}(q)|^2}\frac{1}{2}\left[1+(l_B\alpha q)^2\right]\nonumber\\
  \overline{|V_{11}(q)|^2}&=&  \overline{|V_{00}(q)|^2}\left[\frac{3}{4}
-\frac{1}{2}(l_B\alpha q)^2 + \frac{1}{4}(l_B\alpha q)^4 \right]
\end{eqnarray}

\subsection{Explicit Expression for $\sigma(z)$}
The energy band-structure of a quantum wire with parabolic confinement
potential of strength $\hbar \omega_0$ in a perpendicular magnetic
field $B$ is
\begin{eqnarray}
  \label{eq:wireband}
  \varepsilon_{nk}=\left(n+\frac{1}{2}\right)\hbar\omega_B+\gamma_B\frac{\hbar^2}{2m^*}k^2,\quad
\gamma_B=\left(\frac{\omega_0}{\omega_B}\right)^2=\frac{1}{1+\left(\frac{\omega_c}{\omega_0}\right)^2},
\end{eqnarray}
i.e. a set of equidistant parabolas, labeled by the Landau band index $n$. 
Fixing the Fermi energy between the subbands $n=1$ and $n=2$,
i.e. $\varepsilon_F=2\hbar\omega_B$, the two subband Fermi wave vectors become
\begin{eqnarray}
  \label{eq:twofermiwave}
  k_0&=&\sqrt{\frac{2m^*}{\gamma_B\hbar^2}\left(\varepsilon_F-\frac{1}{2}
\hbar\omega_B\right)}=
\sqrt{\frac{3}{2}}\left(\frac{\omega_B}{\omega_0}\right)^{3/2}k_{F0}\nonumber\\
  k_1&=&\sqrt{\frac{2m^*}{\gamma_B\hbar^2}\left(\varepsilon_F-\frac{3}{2}\hbar\omega_B\right)}=
\sqrt{\frac{1}{2}}\left(\frac{\omega_B}{\omega_0}\right)^{3/2}k_{F0},\quad 
k_{F0}\equiv\sqrt{\frac{2m^*\omega_0}{\hbar}}.
\end{eqnarray}
Recognizing that $(l_B\alpha)^2\left(\frac{\omega_B}{\omega_0}\right)^{3}k_{F0}^2=2
\left(\frac{\omega_c}{\omega_0}\right)^{2}$,
the arguments $l_B\alpha q$ in the matrix elements become
\begin{eqnarray}
  \label{eq:sqrt3etc}
(l_B \alpha q)^2 &=& \left \{
  \begin{array}[l]{l}
\beta [\sqrt{3} + 1 ]^2 ,\quad q = k_0 + k_1\\
\beta [\sqrt{3} - 1 ]^2,\quad q = k_0 - k_1\\
\beta [2]^2,\quad q = 2k_0 \\
\beta [2\sqrt{3}]^2,\quad q = 2k_1\\
  \end{array}
\right.,\quad \beta \equiv \left(\frac{\omega_c}{\omega_0}\right)^2
\end{eqnarray}
for the four cases of intra-band backscattering $q=2k_0,2k_1$, 
inter-band backward ($q=k_0-k_1$) and inter-band forward scattering
($q=k_0+k_1$). 
The dependence on the magnetic field can be completely absorbed into the parameter 
$\beta$. We express the scattering matrix elements by the scattering rate $\tau^{-1}$ without
magnetic field,
\begin{eqnarray}
  \label{eq:taubzerodef}
  \tau^{-1}&\equiv&\frac{n_i^{2D}V_0^2}{\sqrt{2\pi l_0^2} v_{F_0}\hbar^2}
= \frac{n_iV_0^2m^*}{\sqrt{4\pi}\hbar^3}
\end{eqnarray}
with $v_{F_0}\equiv\hbar k_{F0}/{m^*}$ and $l_0 = \sqrt{\hbar/m^*\omega_0}$.
Then, one has $l_B = l_0(1+\beta)^{-1/4}$, and the conductivity can be written as
\begin{eqnarray}
  \label{eq:condlongeq}
  \sigma(z)&=&ie^2\frac{s}{\pi}v_{F0}\tau(1+\beta)^{-1/4} \nonumber\\
&\times&
\frac{z\tau
\left(\sqrt{\frac{3}{2}}+\sqrt{\frac{1}{2}} \right)
+i\left[ \sqrt{3}\tilde{L}_{11}+\frac{1}{\sqrt{3}} \tilde{L}_{00}-2\tilde{L}_{01}\right]}
{\left[z\tau + i \sqrt{\frac{2}{3}}\tilde{L}_{00}\right]
\left[z\tau + i \sqrt{\frac{1}{2}}\tilde{L}_{11}\right]
+\frac{2}{\sqrt{3}}\tilde{L}_{01}^2}\\
\tilde{L}_{00}&\equiv&\sqrt{1+\beta}\left\{\frac{\sqrt{3}}{2} \sum_{\sigma=\pm 1} 
\left\{\left( 1+\left[1+\sigma\sqrt{3}\right]^2\beta \right) e^{-\frac{1}{2}\beta
[1+\sigma\sqrt{3}]^2}\right\}+ 2e^{-6\beta} \right\}\nonumber\\
\tilde{L}_{11}&\equiv&\sqrt{1+\beta}\left\{\frac{1}{2\sqrt{3}} \sum_{\sigma=\pm 1} 
\left\{ \left( 1+\left[1+\sigma\sqrt{3}\right]^2\beta \right) e^{-\frac{1}{2}\beta
[1+\sigma\sqrt{3}]^2}\right\}\right\}\nonumber\\
&+&2\sqrt{1+\beta}\left(\frac{3}{4}-2\beta+4\beta^2\right)e^{-2\beta}\nonumber\\
\tilde{L}_{01}&\equiv&\sqrt{1+\beta}\frac{1}{2} \sum_{\sigma=\pm 1} \left\{
\sigma\left( 1+\left[1+\sigma\sqrt{3}\right]^2\beta \right) e^{-\frac{1}{2}\beta
[1+\sigma\sqrt{3}]^2}\right\},
\end{eqnarray}
where we used Eq.(\ref{eq:Liidef}),(\ref{eq:sqrt3etc}),
\begin{equation}
  \label{eq:tauuse}
  \frac{n_iV_0^2}{\sqrt{2\pi l_B^2}} =v_{F_0}\tau^{-1}\frac{l_0}{l_B}=
v_{F_0}\tau^{-1}(1+\beta)^{\frac{1}{4}},
\end{equation}
and dimensionless functions $\tilde{L}_{00}=\pi L_{00}/(sv_{F_0}\tau^{-1})$ etc.
The Fermi velocities $v_0$ and $v_1$ can be expressed by the 
Fermi velocity $v_{F_0}$ as
\begin{equation}
  \label{eq:vfermiexpress}
  v_0=v_{F_0}\sqrt{\frac{3}{2}}(1+\beta)^{-\frac{1}{4}},
\quad v_1=v_{F_0}\sqrt{\frac{1}{2}}(1+\beta)^{-\frac{1}{4}}.
\end{equation}


\end{document}